\documentclass[USletter,10pt]{article}
\usepackage[margin=1in,dvips]{geometry}
\usepackage{graphicx,cancel}
\usepackage{graphics}
\usepackage{color, soul}
\usepackage{amsmath, amsthm, slashed}
\usepackage{amssymb}
\usepackage{rotating}
\usepackage{mathrsfs}
\usepackage{epsfig}
\usepackage{verbatim}
\usepackage[affil-it]{authblk}
\usepackage{rotating}
\usepackage{graphicx}
\usepackage{accents}
\newtheorem{definition}{Definition}[section]
\newtheorem{theorem}{Theorem}[section]
 
\newtheorem*{conjecture*}{Conjecture}
\newtheorem{corollary}{Corollary}[section]
\newtheorem*{theorem*}{Theorem}
\newtheorem*{corollary*}{Corollary}
\newtheorem{proposition}{Proposition}[subsection]
\newtheorem{lemma}{Lemma}[subsection]
\newtheorem{remark}{Remark}[section]

\newtheorem{bigtheorem}{Theorem}
\newtheorem{introtheorem}{Theorem}
\newcommand{\Sf}{\accentset{\land}{\mathscr{S}}}
\newcommand{\Si}{\underaccent{\lor}{\mathscr{S}}}
\newcommand{\Gi}{\underaccent{\lor}{\mathscr{G}}}
\newcommand{\Gf}{\accentset{\land}{\mathscr{G}}}

\newcommand{\Kf}{\accentset{\land}{\mathscr{K}}}

\newcommand{\Olin}{\Omega^{-1}\accentset{\scalebox{.6}{\mbox{\tiny (1)}}}{\Omega}}
\newcommand{\Olino}{\accentset{\scalebox{.6}{\mbox{\tiny (1)}}}{\Omega}}
\newcommand{\glinh}{\accentset{\scalebox{.6}{\mbox{\tiny (1)}}}{\hat{\slashed{g}}}}
\newcommand{\glin}{\accentset{\scalebox{.6}{\mbox{\tiny (1)}}}{\slashed{g}}}
  \newcommand{\glinto}{\accentset{\scalebox{.6}{\mbox{\tiny (1)}}}{\sqrt{\slashed{g}}}}
\newcommand{\bmlin}{\accentset{\scalebox{.6}{\mbox{\tiny (1)}}}{b}}

\newcommand{\xblin}{\accentset{\scalebox{.6}{\mbox{\tiny (1)}}}{\underline{\hat{\chi}}}}
\newcommand{\xlin}{\accentset{\scalebox{.6}{\mbox{\tiny (1)}}}{{\hat{\chi}}}}
\newcommand{\xli}{\accentset{\scalebox{.6}{\mbox{\tiny (1)}}}{\chi}}
\newcommand{\eblin}{\accentset{\scalebox{.6}{\mbox{\tiny (1)}}}{\underline{\eta}}}
\newcommand{\elin}{\accentset{\scalebox{.6}{\mbox{\tiny (1)}}}{{\eta}}}
\newcommand{\otx}{\accentset{\scalebox{.6}{\mbox{\tiny (1)}}}{\left(\Omega tr \chi\right)}}
\newcommand{\otxb}{\accentset{\scalebox{.6}{\mbox{\tiny (1)}}}{\left(\Omega tr \underline{\chi}\right)}}
\newcommand{\olin}{\accentset{\scalebox{.6}{\mbox{\tiny (1)}}}{\omega}}
\newcommand{\olinb}{\accentset{\scalebox{.6}{\mbox{\tiny (1)}}}{\underline{\omega}}}

\newcommand{\ablin}{\accentset{\scalebox{.6}{\mbox{\tiny (1)}}}{\underline{\alpha}}}
\newcommand{\alin}{\accentset{\scalebox{.6}{\mbox{\tiny (1)}}}{{\alpha}}}
\newcommand{\pblin}{\accentset{\scalebox{.6}{\mbox{\tiny (1)}}}{\underline{\psi}}}
\newcommand{\plin}{\accentset{\scalebox{.6}{\mbox{\tiny (1)}}}{{\psi}}}
\newcommand{\Pblin}{\accentset{\scalebox{.6}{\mbox{\tiny (1)}}}{\underline{P}}}
\newcommand{\Plin}{\accentset{\scalebox{.6}{\mbox{\tiny (1)}}}{{P}}}
\newcommand{\bblin}{\accentset{\scalebox{.6}{\mbox{\tiny (1)}}}{\underline{\beta}}}
\newcommand{\blin}{\accentset{\scalebox{.6}{\mbox{\tiny (1)}}}{{\beta}}}
\newcommand{\rlin}{\accentset{\scalebox{.6}{\mbox{\tiny (1)}}}{\rho}}
\newcommand{\slin}{\accentset{\scalebox{.6}{\mbox{\tiny (1)}}}{{\sigma}}}
\newcommand{\Klin}{\accentset{\scalebox{.6}{\mbox{\tiny (1)}}}{K}}
\newcommand{\flinc}{\accentset{\scalebox{.6}{\mbox{\tiny (1)}}}{f}}

\newcommand{\Zlin}{\accentset{\scalebox{.6}{\mbox{\tiny (1)}}}{Z}}
\newcommand{\Ylin}{\accentset{\scalebox{.6}{\mbox{\tiny (1)}}}{Y}}
\newcommand{\Psilin}{\accentset{\scalebox{.6}{\mbox{\tiny (1)}}}{\Psi}}
\newcommand{\Psilinb}{\accentset{\scalebox{.6}{\mbox{\tiny (1)}}}{\underline{\Psi}}}

\DeclareMathAlphabet\mathbfcal{OMS}{cmsy}{b}{n}

\title{The linear stability of the Schwarzschild solution\\ to gravitational perturbations}

\author[1,2]{Mihalis Dafermos}
\author[3,1]{Gustav Holzegel}
\author[1]{Igor Rodnianski}

\affil[1]{\small Princeton University, Department of Mathematics, Fine~Hall,~Washington~Road,~Princeton,~NJ~08544,~United~States\vskip.2pc \ }

\affil[2]{\small University of Cambridge, Department of Pure Mathematics and Mathematical
Statistics, Wilberforce~Road,~Cambridge~CB3~0WA,~United~Kingdom\vskip.2pc \ }

\affil[3]{\small Imperial College London, Department of Mathematics, 
South~Kensington~Campus,~London~SW7~2AZ,~United~Kingdom\vskip.2pc \ }

\begin{document}

\maketitle

\begin{abstract}
We prove in this paper
the linear stability of the celebrated Schwarzschild family of  black holes
in general relativity: 
Solutions to the linearisation of 
the Einstein vacuum equations 
 (``linearised gravity'')
around a Schwarzschild metric arising
from regular initial data remain globally
bounded
on the black hole exterior and in fact decay to a linearised Kerr metric.
We express the equations in a suitable double null gauge.
To obtain decay, one must in fact add a residual pure gauge
solution which we prove to be itself
quantitatively controlled from
initial data. Our result \emph{a fortiori} 
includes decay statements for general
solutions of the 
Teukolsky equation (satisfied by  gauge-invariant 
null-decomposed curvature components).
These latter statements are in fact deduced  in the course of the proof 
by exploiting associated quantities
shown to satisfy
the Regge--Wheeler equation, for which appropriate decay  can be obtained
easily by adapting previous work on the linear scalar wave equation.
The bounds on the rate of decay to linearised Kerr are inverse 
polynomial, suggesting
that dispersion is sufficient to control the non-linearities of the Einstein equations
in a potential future
proof of nonlinear stability. This paper is self-contained and includes a
physical-space derivation
of the equations of linearised gravity around Schwarzschild
from the full non-linear Einstein vacuum equations
expressed in a double null gauge.
\end{abstract}

  \tableofcontents
  
\section{Introduction}
  \label{THEintroduction}
  The Schwarzschild family~\cite{schwarzschild1916}
  of spacetimes $(\mathcal{M},g_M)$,   expressed  in local coordinates as
  \begin{equation}
  \label{Scwintro}
 -(1-2M/r)dt^2 +(1-2M/r)^{-1}dr^2 +r^2(d\theta^2+\sin^2\theta d\phi^2),
  \end{equation}
  was discovered exactly 100 years before this writing and
   comprises 
  the most basic family of non-trivial solutions to the celebrated
 Einstein vacuum equations
 \begin{equation}
 \label{VACeq!}
 {\rm Ric}(g)=0
 \end{equation}
 of general relativity. Though originally geometrically
 obscured by the coordinate form $(\ref{Scwintro})$,
 the family has now long been 
 understood (see the textbook~\cite{Wald}) to yield, for  parameter values $M>0$, the  simplest
 examples of spacetimes containing a 
 so-called \emph{black hole}, with the coordinate range $r>2M$ corresponding
  to the  \emph{exterior}.

There is perhaps no question more fundamental to pose concerning  Schwarzschild than that
of the stability of its exterior:
\begin{quotation}
\noindent
{{\bf Fundamental question}:}~\emph{Is the Schwarzschild exterior metric $(\ref{Scwintro})$
stable as a solution to $(\ref{VACeq!})$?}
\end{quotation}
The whole tenability of the black hole notion rests on a positive answer to the above.
The question is further complicated by the fact that the Schwarzschild family sits
as a $1$-parameter subfamily of the more elaborate 2-parameter Kerr family
$(\mathcal{M},g_{M,a})$ discovered only much later~\cite{Kerr} in 1963.

One can distinguish between three formulations of the above fundamental question, each of 
increasing difficulty, beginning from the statements initially studied in the physics literature
and ending with the definitive formulation of the question as a problem of nonlinear stability
in the context of the Cauchy problem for $(\ref{VACeq!})$, in analogy with the
non-linear stability of Minkowski space, proven in the monumental work~\cite{ChristKlei} 
of Christodoulou and Klainerman.
\vskip1pc
1.~{\bf \emph{The formal mode analysis of the linearised
equations.}} The equations of gravitational perturbations  around Schwarzschild (i.e.~the
linearisation of $(\ref{VACeq!})$), ``linearised gravity'' for short, can be formally 
decomposed into modes by  associating $t$-derivatives with multiplication
by $i\omega$ and angular derivatives with multiplication by $\ell$.
 The formal 
study of fixed modes from the point of view of ``metric perturbations'' was initiated  in a seminal paper of Regge--Wheeler~\cite{Regge}. 
This study was completed by Vishveshwara~\cite{Vishveshwara} and Zerilli~\cite{Zerilli!}.
A gauge-invariant formulation of ``metric perturbations'' was then given by Moncrief~\cite{MoncriefSch}.
An alternative approach via     the  Newman-Penrose formalism~\cite{newmanpenrose} 
was conducted by Bardeen--Press~\cite{bardeen1973}.
This latter type of analysis was later extended to the Kerr family by 
Teukolsky~\cite{teukolsky1973}.
A highlight of this analysis was the discovery that various curvature components
in  a null frame satisfy a decoupled wave equation, the celebrated \emph{Teukolsky equation},
first discovered in the Schwarzschild case in~\cite{bardeen1973} 
and generalised  to the Kerr case in~\cite{teukolsky1973}.
The understanding of the problem in the early 1980's is summarised by the 
magisterial monograph of Chandrasekhar~\cite{Chandrasekhar}, who 
introduced (see also~\cite{Chandraschw}) 
an important transformation theory connecting solutions of  the two approaches.
A highly non-trivial result is the statement of \emph{mode stability} for the Teukolsky equation on
Kerr,
obtained in a seminal paper of Whiting~\cite{Whiting}.
\vskip1pc
2.~{\bf \emph{The  problem of linear stability of Schwarzschild.}} 
The true problem of linear stability concerns general solutions to the equations
of linearised gravity arising
from regular initial data, not
simply fixed modes.
 One can in fact distinguish between two linear stability statements:
\begin{enumerate}
\item[(2a)] the question of whether  all solutions to  
the linearised Einstein equations around Schwarzschild remain bounded for all time
by a suitable norm of their initial data
 and 
 \item[(2b)]
 the question of asymptotic linear stability--i.e.~whether all solutions
to the linearised equations asymptotically decay. In view of the existence of the Kerr family
and the gauge freedom of the equations,
the best result  would be that they decay to a linearised Kerr solution in some gauge.
\end{enumerate}
Note that the mode analysis corresponding to formulation
1.~described above yields necessary but not sufficient conditions for
either statements (2a) and (2b) of true
linear stability.\footnote{Thus, the mode analysis can be an effective
tool to show instability, but never, on its own, stability. For  instability results for
related problems proven via the existence of unstable modes, 
see~\cite{yakovinsta, Dold:2015cqa} and references therein.
See also discussion in~\cite{whitingsurvey}.} 
In the case of the linear \emph{scalar} wave equation
\[
\Box_g\varphi=0,
\] 
which can be thought of as a ``poor man's'' version of linearised gravity,
the analogue of (2a) for Schwarzschild was proven by Kay--Wald~\cite{KayWald},
and the analogue of (2a) and (2b) are shown now for the full subextremal
Kerr family in~\cite{withYakov},
following a host of recent
activity~\cite{DafRodKerr, Toha2, AndBlue, DafRodsmalla}.
See~\cite{Mihalisnotes, dafrodlargea} for a survey. 
See~\cite{blue2008decay, andersson2013uniform} 
for generalisations to the Maxwell equations and~\cite{Aretakis2}
for a discussion of the extremal case $|a|=M$.
Concerning the linearised Einstein
equations themselves,  work on the wave equation easily generalises to 
establish physical space decay on certain quantities,
for instance those gauge-invariant quantities satisfying the Regge--Wheeler equation on 
Schwarzschild~\cite{friedman2, blue2005wave, Donninger}. 
\emph{For the full system of linearised gravity however,
both problems (2a) and (2b) have remained open until today.}
We note explicitly that even the question of
uniform boundedness, let alone decay, for the gauge-invariant quantities satisfying the Teukolsky equation
on Schwarzschild
has remained open.

\vskip1pc

3.~{\bf \emph{The full nonlinear stability of Schwarzschild as a solution to the Cauchy
problem for the nonlinear Einstein vacuum equations $(\ref{VACeq!})$.}}
This is the definitive formulation of the fundamental question. See our previous~\cite{DHRscat} for 
a precise statement of the conjecture in the language of the Cauchy problem for $(\ref{VACeq!})$.
In analogy with 2.~above, one could distinguish between questions of (3a) orbital stability
and (3b) asymptotic stability. 
 Experience from 
non-linear problems, however, in particular the proof of the nonlinear stability
of Minkowski space~\cite{ChristKlei} referred to earlier (see also~\cite{Igor2, Lydia}), 
indicates that (3a) and (3b) are naturally
coupled.\footnote{This coupling arises from the super-criticality of the Einstein
vacuum equations $(\ref{VACeq!})$. Note that under spherical symmetry (where the vacuum equations must be replaced, however, by a suitable Einstein--matter system to restore
a dynamical degree of freedom) this super-criticality is broken in the presence of a black
hole. Orbital stability
can then be proven independently of asymptotic stability, cf.~\cite{Mihali1} 
with~\cite{DafRod}.
Similar symmetric reductions can be studied for the vacuum equations in higher
dimensions~\cite{formationofbh, HolBi9}.}
Since nonlinear stability is thus necessarily a question 
of asymptotic stability, the ``Schwarzschild'' problem is
 more correctly re-phrased as \emph{the nonlinear asymptotic
stability of the Kerr family in a neighbourhood of Schwarzschild}. 
For even if one restricts to small perturbations of Schwarzschild, it is expected that
generically, spacetime dynamically
asymptotes to a very slowly rotating Kerr solution with $a\ne 0$.
Since in the context of a non-linear stability proof,
one effectively must ``linearise'' around the solution one expects to approach,
this suggests that
 to resolve the full nonlinear formulation 3.~will require a solution of the linear formulation 2.~\emph{not just
for Schwarzschild, but for very slowly rotating
Kerr solutions with $|a|\ll M$}.
Though the non-linear stability problem is thus completely open,
in~\cite{DHRscat} we have proven, via a scattering theory construction,
the existence of a class of dynamical vacuum spacetimes,
without any symmetry assumptions, asymptotically
settling down to Kerr in accordance with the expectation of non-linear stability.
In view of the fast, exponential rate of approach which we impose in~\cite{DHRscat}, 
however, the class we construct is expected 
to be of infinite codimension in the space of all solutions.
\vskip1pc

The purpose of the present paper is to completely
resolve the \emph{linear} stability problem (i.e.~formulation 2.)~in the Schwarzschild case, 
in both its aspects (2a) and (2b).
A first version of our main result can be stated as follows:

\begin{theorem*}[Linear stability of Schwarzschild]
All solutions to the linearised vacuum Einstein equations (in double null gauge) around
Schwarzschild arising from regular asymptotically flat initial data 
\begin{itemize}
\item[(a)] {\bf\emph{remain uniformly bounded}} on the exterior and
\item[(b)] {\bf\emph{decay inverse polynomially}} 
(through a suitable foliation) to a standard
linearised Kerr solution
\end{itemize}
after adding a pure gauge
solution {which can itself be estimated by the size of the data}.
\end{theorem*}

See Theorems~\ref{boundednessintro} and~\ref{ILEDintro} of the detailed overview in 
Section~\ref{overviewsection}
as well as the more detailed later formulations in the bulk of the paper  referred to there.

A word about gauge is already in order. We will express the equations of linearised
gravity in a \emph{double null gauge}. This still allows, however, for a residual gauge freedom which
in linear theory  manifests itself in the existence of ``pure gauge solutions''  corresponding
to one parameter families of deformations of the ambient null foliation of Schwarzschild. 
To measure geometrically the initial data of a general solution of linearised gravity
so as to formulate the boundedness statement~(a), one ``normalises'' the solution on
initial data by adding an appropriate pure gauge solution which is computable explicitly
from the original solution's initial data.  Importantly, this gauge assures that the position
of the horizon
is  fixed and that the
 ``sphere
at infinity'' is round. We emphasise that at the level of natural
energy fluxes, we obtain a boundedness statement controlling the full normalised solution
without loss of derivatives.
An interesting aspect of our work is that to obtain the
decay statement~(b), we must add yet another pure gauge solution which effectively
re-normalises the gauge on the event horizon. It is fundamental that   
we can  quantitatively
control this new pure gauge solution
in terms of the geometry of initial data as expressed in the original normalisation, i.e.~the new
pure gauge solution, though not explicitly given from data, itself
satisfies a uniform boundedness statement.

In particular, our theorem 
is stronger than (and  thus includes \emph{a fortiori}) the statement that 
the gauge invariant linearised curvature components ($\alin$ and $\ablin$ in our notation)
satisfying the Teukolsky equation (discussed above in the context of formal mode analysis)
 themselves remain bounded and in fact
decay inverse
polynomially.  
We will in fact prove the boundedness and decay of these components as a preliminary
step to proving
the full theorem above.
More specifically, we will first prove estimates for certain \emph{higher order} 
gauge invariant quantities $\Plin$ 
and $\Pblin$ (at the level of $4$ derivatives of the linearised metric and $2$ derivatives
of $\alin$, $\ablin$), 
 which satisfy the \emph{Regge--Wheeler} equation. (This is the same equation
 which originally appeared in the ``metric perturbations''
approach discussed above in the context of formal mode analysis.)
 The significance of the Regge--Wheeler equation is that it can in fact be understood
using the methods developed for the scalar wave equation discussed above.
The quantities $\Plin$ and $\Pblin$ will serve as the key to unlocking the whole
system, leading first to control of $\alin$, $\ablin$, and then to the
entirety of the system.
 Let us state explicitly the following corollary of our proof of the main theorem:
\begin{corollary*}
All solutions of the Regge--Wheeler and Teukolsky 
equations on Schwarzschild arising from smooth
compactly supported data decay inverse-polynomially with respect to the time
function of a suitable foliation
of the exterior.
\end{corollary*}

See Theorems~\ref{gauge-invBintro} and~\ref{Teukintrotheorem} 
of Section~\ref{overviewsection}
and the more detailed later formulations  referred to there.

The  expression of $\Plin$ as a second order differential operator applied to  $\alin$ 
can be viewed as a physical-space version of the
fixed-frequency transformations of
Chandrasekhar~\cite{Chandrasekhar} referred to previously.
See also~\cite{sasaki1981regge}.
One of the main points of the present paper is that the physical space understanding
of the relation between $\alin$ and $\Plin$ 
clarifies the fact that the original $\alin$ can be recovered from $\Plin$ 
and \emph{initial data quantities} by
integrating \emph{transport equations} along light cones. 
Indeed, we succeed in estimating all quantities hierarchically,
gauge invariant and gauge dependent, from initial data, by appropriate estimates
of such equations, after control of $\Plin$ and $\Pblin$ via Regge--Wheeler.

We collect some additional references relevant for the problem.
Recent studies
of stability in the physics literature with an eye toward numerical implementation include~\cite{PhysRevD.64.024012, sarbach2001gauge, martelpoisson}.
For analysis of a model problem related to the axisymmetric reduction of the stability problem,
see~\cite{Ionescu:2014cta}.
We also note~\cite{finster2006decay, dotti2014nonmodal}.

Let us note finally that 
the inverse polynomial 
decay bounds shown in our theorem above
are in principle sufficiently strong so as to treat quadratic non-linearities of the type present
in~$(\ref{VACeq!})$ purely
by exploiting the dispersion embodied by our decay results.  This 
allows one already to try to address a restricted non-linear stability conjecture establishing
the existence of the full
 finite-codimension family of solutions which 
indeed asymptotically settle down to Schwarzschild. (See Section~\ref{restrictedconji} for a precise statement;
note that this conjecture would  include a fortiori various well-known symmetric reductions of the stability problem
which similarly impose a Schwarzschild end-state.)
If in future work the main theorem of the present paper can be extended to the Kerr case,
at least in the very slowly rotating regime $|a|\ll M$, then it indeed
opens the way for study
of formulation 3., and thus, for a definitive resolution of the stability question.

\paragraph{Acknowledgements.}
MD~acknowledges support through NSF grant DMS-1405291 and 
EPSRC grant \text{EP/K00865X/1}. GH~acknowledges support through an 
ERC Starting Grant. IR~acknowledges support through NSF grants 
DMS-1001500 and DMS-1065710.

\section{Overview}
\label{overviewsection}
We shall give in this section  a complete overview of our paper.

We begin in Section~\ref{replacementsec} with an introduction to the basic
properties of linearised gravity around Schwarzschild in a double
null gauge, corresponding to Sections~\ref{VEeqDNGsec}--\ref{newgaugesec} 
of
the body of the paper. We will then state rough versions of the main theorems
of the paper in Section~\ref{MTintro}, corresponding to the precise statements
of Section~\ref{Themainthesection} of the body. 
We make a brief aside (Section~\ref{CsweXINTRO}) to review the theory of the scalar
wave equation which is useful to have in mind before turning to the proofs.
We shall then return to outlining the present paper in
Section~\ref{THEOUTLINE}, where we shall review the proofs of the main theorems, following
closely Sections~\ref{sec:RW}--\ref{sec:gest1} of the body of the paper.
We finally give in Section~\ref{restrictedconji} a restricted version of the 
non-linear stability conjecture which in principle can be addressed using the results
of this paper.

We have included also a guide for reading
the paper (Section~\ref{odngos}) for the convenience of those who are only interested
in a subset of the results proven here.

\subsection{Linearised gravity around Schwarzschild in a double null gauge}
\label{replacementsec}
Our paper will employ a \emph{double null gauge} to express the equations of linearised gravity.
This will define an associated \emph{double null frame}. 
The setup is intimately related with the approach to this
problem via the Newman--Penrose formalism~\cite{newmanpenrose} 
studied in  the physics literature since~\cite{bardeen1973}. 
We note that double null gauges have figured prominently in the nonlinear analysis of
the Einstein vacuum equations~\cite{formationofbh, KlaiNic, LukRod, klailukrod} and
thus provide a promising setting for a future full non-linear analysis of the stability
problem. 
We describe here how the relevant
equations are obtained
in physical space, as well as their most basic properties, including their
initial value formulation and the issue of pure gauge solutions.

The subsections of this section follow closely 
Sections~\ref{VEeqDNGsec}--\ref{newgaugesec} of the body of the paper:
We shall review first in Section~\ref{evwmeva}
the form of the Einstein equations in double null gauge 
(cf.~Section~\ref{VEeqDNGsec}).
After reviewing the Schwarzschild manifold in Section~\ref{niceambience} (cf.~Section~\ref{prelimsection}), we shall derive
the equations of linearised gravity around Schwarzschild in Section~\ref{LGarSintro}
(cf.~Section~\ref{sec:lineq}). 
We shall then identify in Section~\ref{PGintro}    two special classes of solutions, pure gauge solutions
and a reference linearised Kerr family (cf.~Section~\ref{sec:specialsol}).
The presence of these special solutions motivate looking at a hierarchy of gauge
invariant quantities satisfying the Teukolsky and Regge--Wheeler equations; 
these will be introduced in Section~\ref{decoupledINTRO} (cf.~Section~\ref{sec:P}). 
We shall then discuss    
the characteristic initial value problem for linearised gravity in Section~\ref{introwellposedne}    (cf.~Section~\ref{sec:initialdata}), 
and the issue of gauge normalisation in   
Section~\ref{EDWgauges} (cf.~Section~\ref{newgaugesec}).

\subsubsection{The Einstein equations in a double null gauge}
\label{evwmeva}
We first review the general form of the Einstein vacuum equations
$(\ref{VACeq!})$ in a double null gauge, following Christodoulou~\cite{Christnotes,
formationofbh}.
This corresponds to  {\bf Section~\ref{VEeqDNGsec}}
of the body of the paper.

A double null gauge
is a coordinate system $\boldsymbol{u}$, $\boldsymbol{v}$, $\boldsymbol{\theta}^1$, 
$\boldsymbol{\theta}^2$ 
such that the metric takes the
form
\begin{equation}
\label{generalmetric}
\boldsymbol{g} = -4 {\boldsymbol{\Omega}}^2 \boldsymbol{d}\boldsymbol{u}\boldsymbol{d}\boldsymbol{v} + \boldsymbol{\slashed{g}}_{CD} 
\left(\boldsymbol{d}\boldsymbol{\theta}^C - \boldsymbol{b}^C \ \boldsymbol{d}\boldsymbol{v} \right) 
\left(\boldsymbol{d}\boldsymbol{\theta}^D - \boldsymbol{b}^D \ \boldsymbol{d}\boldsymbol{v} \right) \, .
\end{equation}
The hypersurfaces
of constant $\boldsymbol{u}$ and $\boldsymbol{v}$ are then manifestly null hypersurfaces. Moreover, the
coordinate vector field $\boldsymbol{\partial}_{\boldsymbol{u}}$ is in the direction of the null generator of
the constant-$\boldsymbol{v}$ hypersurfaces.


Associated to a double null gauge is a natural normalised null frame
\begin{equation}
\label{normnullintro}
\boldsymbol{e}_3 ={\boldsymbol{\Omega}}^{-1}\boldsymbol{\partial}_{\boldsymbol{u}}  , \qquad
\boldsymbol{e}_4={\boldsymbol{\Omega}}^{-1}
(\boldsymbol{\partial}_{\boldsymbol{v}} +{\boldsymbol{b}}^A{\boldsymbol{\partial}}_{\boldsymbol{\theta}^A})
\end{equation}
which, together with the choice of a local (not necessarily orthonormal)
frame $\{{\boldsymbol{e}}_1, {\boldsymbol{e}}_2\}$ with
$\boldsymbol{g}({\boldsymbol{e}}_4,{\boldsymbol{e}}_1)=0$, 
$\boldsymbol{g}({\boldsymbol{e}}_3, {\boldsymbol{e}}_2)=0$, 
 allows
one to decompose  components of the second fundamental form and curvature. 
In our notation this yields \emph{Ricci coefficients}
\begin{equation} \label{heretheRiccicoefs}
\begin{split}
\boldsymbol\chi_{AB} &= \boldsymbol{g} \left(\boldsymbol{\nabla}_A {\boldsymbol{e}}_4,{\boldsymbol{e}}_B\right) \textrm{ \ \ , \ \ \ \ } \underline{\boldsymbol\chi}_{AB} =  \boldsymbol{g} \left(\boldsymbol{\nabla}_A {\boldsymbol{e}}_3,{\boldsymbol{e}}_B\right) \, ,
\\
 \boldsymbol\eta_{A} &= -\frac{1}{2}  \boldsymbol{g} \left(\boldsymbol{\nabla}_{{\boldsymbol{e}}_3} {\boldsymbol{e}}_A,{\boldsymbol{e}}_4\right) \textrm{ \ \ , \ \ } \underline{\boldsymbol\eta}_{A} = -\frac{1}{2}  \boldsymbol{g} \left(\boldsymbol{\nabla}_{{\boldsymbol{e}}_4}{\boldsymbol{e}}_A,{\boldsymbol{e}}_3\right) \, ,
\\
\hat{\boldsymbol\omega} &= \frac{1}{2}  \boldsymbol{g} \left(\boldsymbol{\nabla}_{{\boldsymbol{e}}_4} {\boldsymbol{e}}_3,{\boldsymbol{e}}_4\right) \textrm{ \ \ \ \ , \ \ \ \ \ } \hat{\underline{\boldsymbol\omega}} = \frac{1}{2}  \boldsymbol{g} \left(\boldsymbol{\nabla}_{{\boldsymbol{e}}_3} {\boldsymbol{e}}_4,{\boldsymbol{e}}_3\right)  \,  ,
\end{split}
\end{equation}
%
%
as well as \emph{curvature components}
\begin{equation}  \label{curvCintro}
\begin{split}
\boldsymbol\alpha_{AB} &= \boldsymbol{R} \left( {\boldsymbol{e}}_A,  {\boldsymbol{e}}_4,  {\boldsymbol{e}}_B,  {\boldsymbol{e}}_4\right) \textrm{ \ \ \ ,  \ \  } \underline{\boldsymbol\alpha}_{AB} = \boldsymbol{R} \left( {\boldsymbol{e}}_A,  {\boldsymbol{e}}_3,  {\boldsymbol{e}}_B,  {\boldsymbol{e}}_3\right) \, , \\
\boldsymbol\beta_{A} &= \frac{1}{2} \boldsymbol{R} \left( {\boldsymbol{e}}_A,   {\boldsymbol{e}}_4,  {\boldsymbol{e}}_3,   {\boldsymbol{e}}_4\right) \textrm{ \ \ \ ,  \ \ \ } \underline{\boldsymbol\beta}_{A} = \boldsymbol{R} \left( {\boldsymbol{e}}_A,  {\boldsymbol{e}}_3,  {\boldsymbol{e}}_3,   {\boldsymbol{e}}_4\right) \, , \ \\
\boldsymbol\rho &= \frac{1}{4} \boldsymbol{R}\left(  {\boldsymbol{e}}_4,  {\boldsymbol{e}}_3, {\boldsymbol{e}}_4,  {\boldsymbol{e}}_3\right) \textrm{ \ \ \ , \ \ \ \ \ \  } \boldsymbol\sigma = \frac{1}{4} {}^\star \boldsymbol{R}\left(  {\boldsymbol{e}}_4,  {\boldsymbol{e}}_3, {\boldsymbol{e}}_4,  {\boldsymbol{e}}_3\right) \, .
\end{split}
\end{equation}

The content of the
 Einstein vacuum equations $(\ref{VACeq!})$ can  be expressed as a system of 
transport and elliptic equations for 
the metric $(\ref{generalmetric})$
and Ricci coefficients $(\ref{heretheRiccicoefs})$
coupled with Bianchi identities for the curvature $(\ref{curvCintro})$,
the latter capturing the essential hyperbolicity of the equations $(\ref{VACeq!})$.

Concerning the metric and Ricci coefficients, examples of transport equations are:
\begin{align}
\label{kiauto9elei}
\underline{\boldsymbol{D}}\boldsymbol{\slashed{g}} = 2 \boldsymbol{\Omega} \underline{\boldsymbol\chi} = 2 \boldsymbol{\Omega} \underline{\hat{\boldsymbol\chi}} + \boldsymbol{\Omega}\, \boldsymbol{tr} \hat{\boldsymbol\chi} \slashed{g} 
  \ \ \ ,  \ \ \ 
\boldsymbol{D} \boldsymbol{\slashed{g}} = 2 \boldsymbol{\Omega} \boldsymbol\chi = 2\boldsymbol{\Omega} \hat{\boldsymbol\chi} + \boldsymbol{\Omega} \boldsymbol{tr} \boldsymbol\chi \, \boldsymbol{\slashed{g}} \, ,
\end{align}
\begin{align}
\label{9eleivoumero}  
\boldsymbol{\slashed{\nabla}}_{\bf 3} \underline{\hat{\boldsymbol\chi}} + \boldsymbol{tr} \underline{\boldsymbol\chi} \ \hat{\underline{\boldsymbol\chi}} -\hat{\underline{\boldsymbol\omega}} \ \underline{\hat{\boldsymbol\chi}} =  -\underline{\boldsymbol\alpha} \ \ \ ,  \ \ \ 
\boldsymbol{\slashed{\nabla}}_{\bf 4}
 {\hat{\boldsymbol\chi}} + \boldsymbol{tr} {\boldsymbol\chi} \ \hat{\boldsymbol\chi} -\hat{{\boldsymbol\omega}} \ {\hat{\boldsymbol\chi}} =  {\boldsymbol\alpha} ,
\end{align}
\begin{align}
\label{9eleivoumero2}
\boldsymbol{\slashed{\nabla}}_{\bf 3} \left(\boldsymbol{tr} \underline{\boldsymbol\chi}\right) + \frac{1}{2}\left(\boldsymbol{tr} \underline{\boldsymbol\chi}\right)^2 - \underline{\hat{\boldsymbol\omega}} 
\boldsymbol{tr} \underline{{\boldsymbol\chi}} = - \left( \underline{\hat{\boldsymbol\chi}} , \underline{\hat{\boldsymbol\chi}}\right) \ \ , \ \ \boldsymbol{\slashed{\nabla}}_{\bf 4} \left(\boldsymbol{tr} \boldsymbol\chi \right) + \frac{1}{2}\left(\boldsymbol{tr} \boldsymbol\chi\right)^2 - \hat{\boldsymbol\omega} \boldsymbol{tr} {\boldsymbol\chi} = - \left( {\hat{\boldsymbol\chi}}  , {\hat{\boldsymbol\chi}}\right) , 
\end{align}
while an example of an elliptic equation is the Codazzi equation
\begin{equation}
\label{acodazz}
\boldsymbol{\slashed{div}} {\boldsymbol{\hat{\underline{\chi}}}} =  
  \frac{1}{2} \underline{\hat{\boldsymbol\chi}}^\sharp \cdot \left( \boldsymbol\eta - \underline{\boldsymbol\eta}\right) - \frac{1}{2} \boldsymbol{tr} \boldsymbol\chi\, {\boldsymbol\eta}  + \frac{1}{2\boldsymbol{\Omega}} \boldsymbol{\slashed{\nabla}} \left( \boldsymbol{\Omega} 
  \boldsymbol{tr} {\boldsymbol {\underline\chi}}\right) + \underline{\boldsymbol\beta} \, .
\end{equation}
Here $\hat{\boldsymbol\chi}$,
known
as the \emph{shear}, denotes the trace-free 
part of ${\boldsymbol\chi}$; the quantity $\boldsymbol{tr} {{\boldsymbol\chi}}$ is known
as the \emph{expansion}.
The operators $\boldsymbol{\underline{D}}$, $\boldsymbol{\slashed\nabla}_3$ (respectively, $\boldsymbol{D}$,
$\boldsymbol{\slashed\nabla}_4$) 
are geometric operators associated to $\boldsymbol{g}$ acting on
tensors differentiating
in the $\boldsymbol{e}_3$ (respectively, $\boldsymbol{e}_4$) directions, while $\boldsymbol{\slashed{div}}$
is a natural operator on the constant-$(\boldsymbol{u},\boldsymbol{v})$ spheres. 
Equations $(\ref{9eleivoumero2})$ are the celebrated \emph{Raychaudhuri equations}.
The full system of ``null structure'' equations satisfied by the metric and
Ricci coefficients is given in Section~\ref{sec:nse}.

Concerning the curvature components $(\ref{curvCintro})$, examples of Bianchi  identities are
\begin{align}
\label{examp1intro}
\boldsymbol{\slashed{\nabla}}_{\bf 3}  \boldsymbol\alpha + \frac{1}{2} 
\boldsymbol{tr} \underline{\boldsymbol\chi}\, \boldsymbol\alpha + 2 \underline{\hat{\boldsymbol\omega}} \boldsymbol\alpha &= -2 \boldsymbol{\slashed{\mathcal{D}}_2^\star }\boldsymbol\beta - 3 \hat{\boldsymbol\chi} \boldsymbol\rho- 3{}^\star \hat{\boldsymbol\chi} \boldsymbol\sigma  + \left(4\boldsymbol\eta + \boldsymbol\zeta\right) 
\boldsymbol{\hat{\otimes}} \boldsymbol\beta,  \\
\label{examp2intro}
\boldsymbol{\slashed{\nabla}}_{\bf 4} \boldsymbol\beta + 2 \boldsymbol{tr} \boldsymbol\chi \,\boldsymbol\beta - \hat{\boldsymbol\omega} \boldsymbol\beta &= \boldsymbol{\slashed{div}} \boldsymbol\alpha + \left(\underline{\boldsymbol\eta}^\sharp + 2 \boldsymbol\zeta^\sharp\right) \boldsymbol{\cdot} \boldsymbol\alpha ,
\end{align}
where the operator $\boldsymbol{\slashed{\mathcal{D}}_2^\star}$ is a suitable
adjoint of $\boldsymbol{\slashed{div}}$.
The full system is given in Section~\ref{bieq}.

\subsubsection{The ambient Schwarzschild metric}
\label{niceambience}
We note that 
any Lorentzian metric can \emph{locally} be put in the above form $(\ref{generalmetric})$.
The maximally extended 
Schwarzschild manifold and--less obviously--the globally hyperbolic region of
Kerr (see Pretorius and Israel~\cite{Pretorius})
can in fact both be \emph{globally} covered by such a coordinate system, where
$\theta^1,\theta^2$ are interpreted as coordinates on the sphere $\mathbb S^2$,
modulo the usual degeneration of spherical coordinates.
In the bulk of the paper, fixing an ambient Schwarzschild manifold and double null foliation 
will be the content of {\bf Section~\ref{prelimsection}}.

We summarise briefly here:
In the Schwarzschild case with parameter $M$, one easily derives
from the expression
$(\ref{Scwintro})$ a double null parametrisation of the exterior region $r>2M$.
Defining first
\begin{equation}
\label{r*defi}
r^*\doteq r+ 2M \log(r-2M) - 2M\log 2M,
\end{equation}
set
\begin{equation}
\label{udefiandvdefi}
u\doteq t-r^*, \qquad v\doteq t+r^*.
\end{equation}
Then $u$ and $v$ define null coordinates on the region $r>2M$, parametrising it 
as $(-\infty,\infty)\times (-\infty,\infty)\times \mathbb S^2$.
In the notation $(\ref{generalmetric})$, the Schwarzschild metric takes the form
\begin{equation}
\label{SchwarzschildHERE}
\Omega^2= 1- 2M/r ,\qquad \slashed{g} =  r^2\gamma , \qquad {b}^A=0
\end{equation}
where $\gamma$ denotes the  standard metric
of the unit sphere. Note that $r(u,v)$ is now defined implicitly by
$(\ref{r*defi})$ and $(\ref{udefiandvdefi})$.
In this context, the coordinates $u$ and $v$ are known as \emph{Eddington--Finkelstein
double null coordinates}.

Note that we distinguish Schwarzschild metric quantities in the above differentiable
structure by presenting them in regular type, 
 de-bolded compared with quantities associated to  
a general manifold and metric (\ref{generalmetric}). This notation is used in the remainder 
of the paper.

Upon rescaling the null coordinate $U=U(u)$ appropriately, one can extend the region on which the 
metric is defined to include the so-called event horizon $\mathcal{H}^+$, a null hypersurface
which then corresponds to $r=2M$. 
In the body of the paper, it is in fact this $U$-coordinate which we shall use to \emph{define} the
Schwarzschild manifold in Section~\ref{differentialstru}, with $u$ defined
by inverting the rescaling, and $t$ and $r^*$ by $(\ref{udefiandvdefi})$.
It is computationally useful however to work with the
irregular $u$ coordinate, which formally still parametrises $\mathcal{H}^+$ as $u=\infty$,
and to compensate for this by introducing renormalised quantities that are regular on the horizon.
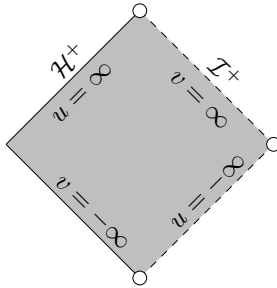
\[
\begin{picture}(0,0)%
\includegraphics{domaineverything.pstex}%
\end{picture}%
\setlength{\unitlength}{3158sp}%
\begingroup\makeatletter\ifx\SetFigFont\undefined%
\gdef\SetFigFont#1#2#3#4#5{%
  \reset@font\fontsize{#1}{#2pt}%
  \fontfamily{#3}\fontseries{#4}\fontshape{#5}%
  \selectfont}%
\fi\endgroup%
\begin{picture}(2174,2222)(3889,-5372)
\put(5476,-3661){\rotatebox{315.0}{\makebox(0,0)[lb]{\smash{{\SetFigFont{10}{12.0}{\rmdefault}{\mddefault}{\updefault}{\color[rgb]{0,0,0}$\mathcal{I}^+$}%
}}}}}
\put(4331,-3766){\rotatebox{45.0}{\makebox(0,0)[lb]{\smash{{\SetFigFont{10}{12.0}{\rmdefault}{\mddefault}{\updefault}{\color[rgb]{0,0,0}$\mathcal{H}^+$}%
}}}}}
\put(4321,-4101){\rotatebox{45.0}{\makebox(0,0)[lb]{\smash{{\SetFigFont{10}{12.0}{\rmdefault}{\mddefault}{\updefault}{\color[rgb]{0,0,0}$u=\infty$}%
}}}}}
\put(4276,-4561){\rotatebox{315.0}{\makebox(0,0)[lb]{\smash{{\SetFigFont{10}{12.0}{\rmdefault}{\mddefault}{\updefault}{\color[rgb]{0,0,0}$v=-\infty$}%
}}}}}
\put(5251,-4936){\rotatebox{45.0}{\makebox(0,0)[lb]{\smash{{\SetFigFont{10}{12.0}{\rmdefault}{\mddefault}{\updefault}{\color[rgb]{0,0,0}$u=-\infty$}%
}}}}}
\put(5176,-3736){\rotatebox{315.0}{\makebox(0,0)[lb]{\smash{{\SetFigFont{10}{12.0}{\rmdefault}{\mddefault}{\updefault}{\color[rgb]{0,0,0}$v=\infty$}%
}}}}}
\end{picture}%

\]
Thus, our basic double null foliation will be that defined by coordinates $u$ and $v$.

We note that the static Killing field $\partial_t$ whose existence is manifest from
$(\ref{Scwintro})$ is expressed in $(u, v)$ coordinates as 
\begin{equation}
\label{TDEFofintro}
T=\frac12(\partial_u+\partial_v).
\end{equation}
This vector field extends smoothly to a null vector on 
the horizon $\mathcal{H}^+$; in $(U,v)$ coordinates,
it takes the form $T=\frac12\partial_v$ on $\mathcal{H}^+$.

From $(\ref{SchwarzschildHERE})$, in 
the double null foliation defined by the $u$, $v$ coordinates, we compute
that,
with the notation of Section~\ref{evwmeva}, the non-vanishing 
Schwarzschild metric coefficients
are 
\begin{equation}
\label{Schvalintro1}
\Omega=\sqrt{1-2M/r},\qquad {\sqrt{\slashed{g}}}=
{r}^2 \sqrt{\gamma},
\end{equation}
and, with respect to the associated null frame:
\[
e_3={\Omega^{-1}}\partial_u,\qquad  e_4 = 
{\Omega^{-1}}\partial_v,
\]
the non-vanishing Ricci coefficients are
\begin{equation}
\label{Schvalintro2}
\begin{split}
{\chi}_{AB} =  r^{-1}\Omega {\slashed{g}}_{AB}  \  ,  \ {\underline{\chi}}_{AB} = - r^{-1}\Omega {\slashed{g}}_{AB} \ \ , \ \ 
{\hat{\omega}} = r^{-2} \Omega^{-1}M   \ , \ 
{\hat{\underline{\omega}}} =-r^{-2}\Omega^{-1}M \, ,
\end{split}
\end{equation}
while the only non-vanishing curvature component is 
\begin{align}
\label{Schvalintro3}
\rho = -\frac{2M}{r^3} \, .
\end{align}

\subsubsection{Linearised gravity around Schwarzschild}
\label{LGarSintro}

The equations of interest in this paper (the equations of gravitational perturbations around
Schwarzschild, or ``linearised gravity'' for short)
are those that arise from linearising
the system of Section~\ref{evwmeva}  around their Schwarzschild values $(\ref{Schvalintro1})$--$(\ref{Schvalintro3})$.
The equations are derived and presented in {\bf Section~\ref{sec:lineq}} of the body of
the paper. We give a brief 
outline here.

In deriving the equations of linearised gravity, we will fix the above background
differential structure $(\mathcal{M},g)$
with its Schwarzschild metric, and embed $(\ref{SchwarzschildHERE})$ 
in a $1$-parameter
family of metrics $\boldsymbol{g}$ of the form
$(\ref{generalmetric})$ satisfying the Einstein vacuum equations $(\ref{VACeq!})$.

The system of linearised gravity concerns linearised quantities  associated to the quantites
of Section~\ref{evwmeva}, namely \emph{linearised metric coefficients}
\begin{equation}
\label{metriccoefs}
\Olino\, , \, \glinto  \,  , \,
\glinh  \, ,  \, \bmlin \, ,
\end{equation}
\emph{linearised Ricci coefficients}
\begin{equation}
\label{Ricoedw}
\otx \, , \,
\otxb \, , \, \elin \, , \,
\eblin \, , \, \olin \, , \, \olinb
 \,  , \,
\xlin \, , \, \xblin \, ,
\end{equation}
as well as \emph{linearised curvature components}
\begin{equation}
\label{lincurvedw}
 \alin \, , \, \blin  \, , \, 
\rlin \, , \, \slin \, , \,
\bblin \, , \,  
\ablin \, .
\end{equation}

Our notation is motivated by the formal expansions
\begin{equation}
\label{formalexpy0}
\boldsymbol{\Omega}= \Omega+\epsilon \Olino  +O(\epsilon^2),
\end{equation}
\begin{equation}
\label{formalexpy}
\boldsymbol{\Omega}\boldsymbol{tr}{\boldsymbol\chi} = 
{\Omega tr\chi} +\epsilon 
\otx + O(\epsilon^2),
\end{equation}
\vspace{.005cm}
\begin{equation}
\label{formalexpy2}
\boldsymbol{\alpha}=0+\epsilon\alin +O(\epsilon^2),
\end{equation}
where we recall that the unbolded quantities without any 
superscripts denote the Schwarzschild values given
by $(\ref{Schvalintro1})$--$(\ref{Schvalintro3})$, and we have substituted that
${\alpha}=0$ in $(\ref{formalexpy2})$. See Section~\ref{hereformality} for details.

We note that because our frame $(e_3, e_4)$
is irregular\footnote{The 
linearisation process is frame covariant and thus coincides with the result of
linearising with respect to a regular frame. See Section~\ref{appendix:linco}.}
 (cf.~Section~\ref{niceambience}), some quantities require
${\Omega}$-weights so as to be regular on the horizon $\mathcal{H}^+$, 
e.g.~
\[
\Olin\, , \, \Omega^{-2} \otxb \, , \,
\Omega^2 \alin\, .
\]

The \emph{linearised Einstein equations}  thus
take the form of a linear system of equations
in the above quantities $(\ref{metriccoefs})$--$(\ref{lincurvedw})$.  
They can be derived from plugging in the expansions
of type $(\ref{formalexpy0})$--$(\ref{formalexpy2})$
into the equations of Section~\ref{evwmeva}
and collecting the linear terms in $\epsilon$ (see Section~\ref{sec:outlinelin}).
For instance, the linearised version of the first transport equation
$(\ref{kiauto9elei})$, decomposed into trace and trace-free part, is
\begin{equation}
\label{topros9esame}
\underline{D} \left(\frac{\glinto}{\sqrt{\slashed{g}}}\right)  = \otxb , \qquad 
\sqrt{\slashed{g}}\, \underline{D}\left( \frac{\glinh_{AB}}{\sqrt{\slashed{g}}} \right)   =2\Omega\, \xblin_{AB}.
\end{equation}
Here $\underline{D}$ is a projected Schwarzschild Lie derivative which on scalars
reduces simply to $\partial_u$. 
Concerning linearised Ricci coefficients,
the linearised versions of the  equations
$(\ref{9eleivoumero})$  are
\begin{equation}
\label{epitelousavafora}
\slashed{\nabla}_3  \left(\Omega^{-1}
\xblin \right)  + {\Omega^{-1}}
\left(tr \underline{\chi}\right) \xblin
= -{\Omega^{-1}} \ablin \, ,
\qquad 
\slashed{\nabla}_4  \left({\Omega^{-1}} \xlin \right) 
+  {\Omega^{-1}} 
{\left(tr \underline{\chi}\right)}\xlin= 
-{\Omega^{-1}}\alin \, ,
\end{equation}
of $(\ref{9eleivoumero2})$ is 
\[
{\Omega}\slashed{\nabla}_4\otxb =
{\Omega}^2 \left( 2 \slashed{div} \elin + 2 \rlin
 + 4{\rho}\, \Olin \right) - \frac{1}{2} {\left( \Omega tr \chi \right)} \left( \otxb - \otx \right)
\]
and of $(\ref{acodazz})$ is
\begin{equation}
\label{Lincodazz}
\slashed{div} \xblin = -\frac{1}{2} {\left(tr \underline{\chi}\right)} \elin 
+ \bblin + \frac{1}{2} {\Omega^{-1}}
 \slashed{\nabla}_A \otxb.  
\end{equation}
Here $\slashed{\nabla}_3$, $\slashed{\nabla}_4$ and $\slashed{div}$
are covariant differentiation operators defined with respect to the
Schwarzschild metric. The linearised version of $(\ref{examp1intro})$--$(\ref{examp2intro})$ is 
\begin{align}
\label{kiedw9adwsoume}
\slashed{\nabla}_3 \alin + \frac{1}{2} {(tr \underline{\chi})} \alin + 2 {\underline{\hat{\omega}}}\, \alin &= -2 \slashed{\mathcal{D}}_2^\star \blin - 3 
{\rho}\xlin \, ,  \\
\nonumber
\slashed{\nabla}_4 \blin + 2{(tr \chi)} \blin -{\hat{\omega}} \,\blin &= \slashed{div}\, \alin \, .\end{align}
Again, $\slashed{\mathcal{D}}_2^\star$ is an adjoint
of $\slashed{div}$  defined with respect to the Schwarzschild
background.
The complete system of linearised equations is given in Sections~\ref{=lmc}--\ref{=lcc}.

Let us make the following remark:
When linearising the  equations of Section~\ref{evwmeva} around Minkowski space,
the arising linearised Bianchi identities in fact \emph{decouple}, 
and this allows for them to be studied
independently of the full system of linearised gravity. 
Boundedness and decay results for this decoupled set of equations
(the so-called spin $2$ equations) were obtained in~\cite{asymppropCK} by Christodoulou and Klainerman
using robust vector field
methods, and this study played an important role as a preliminary
step for their later proof~\cite{ChristKlei} 
of the non-linear stability of Minkowski space.  In contrast, in our setting here,
 examining 
$(\ref{kiedw9adwsoume})$, we see immediately that
the linearised Bianchi identities  couple to the linearised null structure equations
through the appearance of the term  $-3\rho\xlin$.
As we shall see in Section~\ref{decoupledINTRO} below, however,
the quantity $\alin$  itself satisfies a second order decoupled wave equation.

We shall denote solutions of the above system by 
\begin{equation}
\label{denoteinintro}
\mathscr{S}=\left( \,\glinh, \, \glinto, \,
\Olino, \, \bmlin, \, \otx, \,
\otxb, \,  \xlin \, , \, \xblin \, , \, \elin \, , \, \eblin \, , \, \olin \, , \, \olinb \, , \,
\alin \, , \, \blin \,  , \, \rlin \, , \,  \slin \, , \, \bblin \,  , \, \ablin \, , \, \Klin\right).
\end{equation}
For convenience, in the above we have added an additional quantity, 
the linearised Gauss curvature $\Klin$, as an unknown.
We note that a solution $\mathscr{S}$ is completely determined by its linearised metric
coefficients $(\ref{metriccoefs})$, but nonetheless, we prefer to adjoin
all quantities as unknowns. 

Let us note that one can indeed quantitatively relate \emph{solutions} of this formal linearisation
to $1$-parameter families of \emph{solutions} of the actual vacuum 
Einstein equations $(\ref{VACeq!})$
as expressed by the system of Section~\ref{evwmeva}.\footnote{See~\cite{moncrieflinearizstab}
for subtleties that arise for this in the spatially compact case.}
In this paper, however, we will develop a self-contained theory of the
linearised system without
reference to an actual $1$-parameter family of solutions of the full nonlinear theory.

\subsubsection{Special solutions: pure gauge solutions and the linearised Kerr family}
\label{PGintro}
Let us discuss immediately two important
classes of special solutions of the above system; this corresponds to 
{\bf Section~\ref{sec:specialsol}} of the body of the paper.

For the first class, note that  the restriction  to coordinates of double null form $(\ref{generalmetric})$
is not sufficient to   uniquely determine them
on an abstract Lorentzian manifold.
There is residual gauge freedom:
 Change of coordinates that preserve
the form $(\ref{generalmetric})$, upon
linearisation, give rise to a special class of solutions
which we shall refer to as  
\emph{pure gauge solutions}.

An example of a pure gauge solution is one generated by a function $f(v,\theta^A)$:
\begin{align}
\label{EXpuregauge}
\mathscr{G}=\Bigg\{\, 
\Olino=\Omega^{-1}\partial_v \big(f {\Omega^2}\big) \ ,  \ {\bmlin}_A = -r^2 \slashed{\nabla}_A \big(\partial_v \left(r^{-1}{f}\right)\big)   \ , \ 
\glinto = \sqrt{\slashed{g}}\big(r^{-1}{\Omega^2} f +  r \slashed{\Delta}f\big) \  ,  \ \glinh 
= - 2r \slashed{\mathcal{D}}_2^\star \slashed{\nabla}_A f 
\Bigg\},
\end{align}
where we have shown explicitly only the metric perturbation from which all other geometric quantities can be determined. See Section~\ref{sec:ssgauge} where we shall classify all pure gauge solutions.
We will return to discuss pure gauge solutions in Section~\ref{EDWgauges}
when we discuss gauge-normalised solutions.

Another class of explicit solutions of    the linearised
system of Section~\ref{LGarSintro}   arises from linearising the Kerr family itself (in a convenient
coordinate representation) around
a given Schwarzschild solution.

Linearising  the Schwarzschild (i.e.~constant $a=0$)
sub-family with mass $\tilde{M}=M+\epsilon \mathfrak{m}$ (in a particular
double null coordinate
representation) gives  rise to 
a solution
\[
\mathscr{K}=\left\{
\Olino = - \frac12\Omega \mathfrak{m} \ \ \ , \ \ \ \glinto = -\sqrt{\slashed{g}}^{-1}\mathfrak{m} \ \ \ , \ \ \ \ \rlin
= -\frac{2M}{r^3} \cdot \mathfrak{m} \ \ \ , \ \ \ \Klin=\frac{\mathfrak{m}}{r^2}  \ \ \ , \ \ \ \textrm{rest}=0
\right\}
\]
while linearising the constant-$M$ mass Kerr subfamily
with rotation parameter
$\tilde{a}=0+\epsilon\mathfrak{a}$ gives rise to three linearly independent solutions, each of form:
\[
\mathscr{K}=\left\{ \Olino=0 \ \ \ , \ \ \  
\glinh = 0, \ \ \ , \ \ \ \glinto=0 
 \ \ \ , \ \ \   {\bmlin}_A= \frac{4M\mathfrak{a}}{r} \epsilon_{A}^{\phantom{B}C}\partial_C Y^1_m
\right\}
\]
with $Y^1_m$ ($m=-1,0,1$) being the three linearly independent $\ell=1$ spherical harmonics.
See Sections~\ref{sec:linss} and~\ref{sec:nontrivkerr}.
We will call the resulting $4$-dimensional subspace 
of solutions (parameterised by real
coefficients $\mathfrak{m}$, $s_{-1}$, $s_0$, $s_1$)
\emph{reference linearised Kerr solutions} and denote them
by $\mathscr{K}_{\mathfrak{m}, s_i}$.
We note that these solutions are supported entirely in angular modes $\ell=0$ and $\ell=1$.

In view of the existence of the solutions of this section, the best we can expect of
general solutions of our system of linearised gravity is that they decay
to a pure gauge solution $\mathscr{G}$ plus a reference linearised Kerr solution
$\mathscr{K}$.

\subsubsection{Hierarchy of gauge-invariant quantities}
\label{decoupledINTRO}
In view of the complication provided by the existence
of the solutions of Section~\ref{PGintro} 
above, it is useful to isolate quantities which vanish for all
  such solutions.
These are the so-called \emph{gauge-invariant quantities}.
In the body of the paper, these are discussed in {\bf Section~\ref{sec:P}}.

An example of such a quantity is the linearised curvature
component $\alin$ from $(\ref{lincurvedw})$:
As originally shown by Bardeen and Press~\cite{bardeen1973},
the component
$\alin$ in fact decouples from the full system and satisfies
the equation
\begin{align} \label{teukolskyintheintro}
\slashed{\nabla}_4 \slashed{\nabla}_3 \alin &+ \left(\frac{1}{2}{tr \underline{\chi}} 
+ 2 {\underline{\hat{\omega}}}
\right) \slashed{\nabla}_4 \alin +  \left(\frac{5}{2}{tr {\chi}} - {{\hat{\omega}}} \right) \slashed{\nabla}_3 \alin - \slashed{\nabla}^2 \alin \nonumber  \\
&+ \alin \left(5 {\underline{\hat{\omega}}} \, {tr \chi} - {\hat{\omega}}\, {tr \underline{\chi}}-{4\rho} + 2{K} + 
{tr \chi}\,{tr \underline{\chi}} -4
{\hat{\omega}}\,{\hat{\underline{\omega}}}\right) = 0 \, .
\end{align}
A similar equation (with the roles of the $3$ and $4$ directions and underlined and non-underlined
quantities reversed) is satisfied by $\ablin$.
These equations are known as the \emph{spin $\pm2$ 
Teukolsky equations}.\footnote{For the Schwarzschild case considered here,
these equations are also known as the \emph{Bardeen--Press equations}.
It was Teukolsky~\cite{teukolsky1973}
who showed 
that the structure allowing decoupled wave equations
for gauge invariant quantities survives when linearising the Einstein equations 
around Kerr. In that case, however,
the relevant gauge invariant quantities
are defined not with respect to the null
frame $(\ref{normnullintro})$, but with  respect to the so-called \emph{algebraically special}
frame. 
In Schwarzschild, the algebraically
special frame coincides with the  null frame $(\ref{normnullintro})$
associated to a double null foliation.}

We note already that the vanishing of \emph{both} 
$\alin$ and $\ablin$ identically
imply that a solution is a pure gauge solution plus a reference linearised Kerr,
provided that it be asymptotically flat (cf.~Section~\ref{introwellposedne}). 
We shall show this in Appendix~\ref{charofpuregauge}.

It turns out that $\alin$ and $\underline\alin$ are
best understood in the context of a \emph{hierarchy
of gauge-invariant quantities}.
We define
\begin{align}
\label{littlepsidef0}
\plin&\doteq-\frac{1}{2}r^{-1}{\Omega^{-2}}
\slashed{\nabla}_3\big(r{\Omega^2}\alin \, \big)&=
\slashed{\mathcal{D}}^\star_2\blin+\frac32{\rho}\,\xlin,
\end{align}
\begin{align}
\label{littlepsidef1}
\pblin&\doteq
\frac{1}{2}r^{-1}{\Omega^{-2}}
\slashed{\nabla}_4\big(r{\Omega^2}\underline\alin \, \big)
&=\slashed{\mathcal{D}}^\star_2\bblin-\frac32{\rho}\,\xblin
\end{align}
and 
\begin{align} \label{Pdefintro} 
\Plin &\doteq   r^{-3}{\Omega^{-1}}\slashed{\nabla}_3\big(\plin r^3\Omega\big) &=  
 \slashed{\mathcal{D}}_2^\star \slashed{\mathcal{D}}^\star_1 \left(-\rlin \, , \,  
 \slin\right) + \frac{3}{4} {\rho}\, {tr \chi}  \left(\xlin-\xblin\right),
 \\
\label{Pbardefintro}
\underline{\Plin} & \doteq  
-r^{-3}{\Omega^{-1}} \slashed{\nabla}_4\big(\underline\plin r^3 \Omega\big)
&=
 \slashed{\mathcal{D}}_2^\star \slashed{\mathcal{D}}^\star_1 \left(-\rlin \, , \,
  -\slin\right) + \frac{3}{4} {\rho}\,{tr \chi}  \left(\xlin-\xblin\right).
\end{align}
The second equalities in $(\ref{littlepsidef0})$--$(\ref{Pbardefintro})$ above are nontrivial 
and follow from the linearised Bianchi equations. It follows
from equation $(\ref{teukolskyintheintro})$ and the definitions 
$(\ref{littlepsidef0})$--$(\ref{Pbardefintro})$ alone that 
the quantities $\Plin$ (and similarly $\underline{\Plin}$)  satisfy the so-called
\emph{Regge-Wheeler equation}
\begin{align}   \label{im4INTRO}
\slashed{\nabla}_3 \slashed{\nabla}_4 \Plin &+ \slashed{\nabla}_4 \slashed{\nabla}_3 \Plin - 2 \slashed{\Delta} \Plin + \left(5 {tr \underline{\chi}} 
+ {\hat{\underline{\omega}}} \right) \cdot \slashed{\nabla}_4 \Plin + \left(5{tr {\chi}} + {\hat{\omega}}\right) \slashed{\nabla}_3 \Plin \nonumber \\ 
&+ \Plin \left(4{K} - \left(3 {tr \chi} + {\hat{\omega}}\right) 
2 {tr \chi} - 4 {\left(tr \chi\right)^2} + 2\slashed{\nabla}_3 {tr \chi}  -8{\hat{\omega}}\,{tr \chi} \right)
= 0 \, .
\end{align}
The relation defining quantity $\Plin$ from $\alin$ is a physical space interpretation of 
the fixed-frequency transformation theory of Chandresekhar~\cite{Chandrasekhar}.

In contrast to $(\ref{teukolskyintheintro})$, the equation $(\ref{im4INTRO})$ admits 
a positive
energy and can be understood    using the   methods developed for studying
the  scalar wave equation $\Box_g\varphi=0$ (see Section~\ref{CsweXINTRO} below).
This will allow us to view $\Plin$ and $\underline{\Plin}$ as the key to unlocking
the whole system (Section~\ref{tokleidi}).

It is remarkable that the  equation $(\ref{im4INTRO})$, 
which originally appeared as a quantity satisfied by
metric perturbations~\cite{Regge}, reappears in this context.

We note that there are solutions for which $\Plin$ and $\Pblin$ both vanish
identically and are \emph{not} pure gauge,
in particular the linearised Robinson--Trautman solutions discussed in 
Appendix~\ref{theRTrautsection}. Nonetheless,  as we shall see,
we can always estimate $\alin$ given control of $\Plin$
by integrating transport equations (see already Section~\ref{giqistoria} below), 
picking up also an initial data quantity
for $\alin$.

In the body of the paper, we 
will in fact give a self-contained theory of both the Regge--Wheeler 
equation~$(\ref{im4INTRO})$
and the Teukolsky equation $(\ref{teukolskyintheintro})$, defining these
in Section~\ref{def:teue}, proving a well-posedness theorem in Section~\ref{sec:ivprwteu},
and deriving $(\ref{im4INTRO})$ from $(\ref{teukolskyintheintro})$ in 
Section~\ref{sec:tratheo}. In these sections we naturally drop the superscript~$\accentset{(1)}{\ \ }$ from all quantities as we consider a general $P$ satisfying (\ref{im4INTRO}) and a general $\alpha$ satisfying (\ref{teukolskyintheintro}).
Only in
Section~\ref{sec:fullrel} do we derive that  
 $(\ref{teukolskyintheintro})$ 
is indeed satisfied by the component $\alin$
of a solution  to the full system of linearised gravity.   
The first two main theorems of this paper (Theorem~\ref{prop:summarypsi} 
and~\ref{theo:mtheogi} of Section~\ref{Themainthesection}) will prove
boundedness and decay for such 
general solutions of the Regge--Wheeler and
Teukolsky equations. For rough versions of the statements of these theorems, the reader can turn immediately to
Section~\ref{introstRWTK}.

\subsubsection{Characteristic initial data,  well-posedness and asymptotic flatness}
\label{introwellposedne}
The equations of linearised gravity described in Section~\ref{LGarSintro} 
above admit
 a \emph{well-posed initial value problem}.
 Though the structure which makes this possible can be viewed
 as inherited from the original non-linear system $(\ref{VACeq!})$, 
 the well-posedness allows us to develop
 a self-contained theory of solutions for the linear system \emph{without
 further reference to its origin}.
 In the body of the paper, this will be discussed in {\bf Section~\ref{sec:initialdata}}. We give 
 a brief treatment here.

 As is well known, initial data for the Einstein vacuum equations
 $(\ref{VACeq!})$ must satisfy
 \emph{constraints}. This feature is of course inherited by the linearisation.
As we are working in a double null gauge, it is more convenient
to discuss characteristic initial data. This has the advantage 
of reducing the constraints to ordinary differential equations, which can be solved by
integrating transport equations after prescribing--freely--suitable 
seed data.

We will introduce thus first  a notion of  a \emph{seed initial data set}
defined on two 
null cones $C_{u_0}$ and $C_{v_0}$ of the ambient Schwarzschild metric.
Refer to the diagram below:
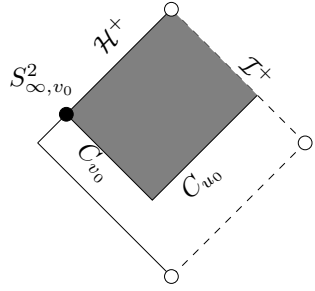
\[
\begin{picture}(0,0)%
\includegraphics{domainlinstab.pstex}%
\end{picture}%
\setlength{\unitlength}{3158sp}%
\begingroup\makeatletter\ifx\SetFigFont\undefined%
\gdef\SetFigFont#1#2#3#4#5{%
  \reset@font\fontsize{#1}{#2pt}%
  \fontfamily{#3}\fontseries{#4}\fontshape{#5}%
  \selectfont}%
\fi\endgroup%
\begin{picture}(2402,2222)(3661,-5372)
\put(5476,-3661){\rotatebox{315.0}{\makebox(0,0)[lb]{\smash{{\SetFigFont{10}{12.0}{\rmdefault}{\mddefault}{\updefault}{\color[rgb]{0,0,0}$\mathcal{I}^+$}%
}}}}}
\put(4423,-3575){\rotatebox{45.0}{\makebox(0,0)[lb]{\smash{{\SetFigFont{10}{12.0}{\rmdefault}{\mddefault}{\updefault}{\color[rgb]{0,0,0}$\mathcal{H}^+$}%
}}}}}
\put(4201,-4336){\rotatebox{315.0}{\makebox(0,0)[lb]{\smash{{\SetFigFont{10}{12.0}{\rmdefault}{\mddefault}{\updefault}{\color[rgb]{0,0,0}$C_{v_0}$}%
}}}}}
\put(5101,-4711){\rotatebox{45.0}{\makebox(0,0)[lb]{\smash{{\SetFigFont{10}{12.0}{\rmdefault}{\mddefault}{\updefault}{\color[rgb]{0,0,0}$C_{u_0}$}%
}}}}}
\put(3676,-3811){\makebox(0,0)[lb]{\smash{{\SetFigFont{10}{12.0}{\rmdefault}{\mddefault}{\updefault}{\color[rgb]{0,0,0}$S^2_{\infty, v_0}$}%
}}}}
\end{picture}%

\]
This data set will be described by symmetric traceless tensors
$\glinh_{\circ,out}$ and
$\glinh_{\circ,in}$, a one-form
$\bmlin_{\circ,out}$ 
and functions $\Olino_{\circ, out}$
and $\Olino_{\circ, in}$, each defined
on $C_{u_0}$, $C_{v_0}$ respectively, augmented
by certain additional geometric data
on  the event horizon   sphere $S^2_{\infty, v_0}$.
See Definition~\ref{def:seeddata}. 

We have the following foundational statement, which we summarise here as:
\begin{introtheorem}[Well-posedness of linearised gravity, rough formulation]
\label{LWPlgINTRO}
A smooth seed initial data set leads to a unique    
smooth solution $\mathscr{S}$
of the equations of linearised gravity
in the region $u_0\le u\le\infty$, $v_0\le v<\infty$.
\end{introtheorem}
The precise statement is given in the body of the paper as Theorem~\ref{theo:lwp}.

The boundedness and decay theorems of our paper will require that data,
in addition to smooth, be \emph{asymptotically flat}. (Note that in the full nonlinear
theory governed by $(\ref{VACeq!})$, this is necessary even for a local existence theorem with a $u$-time of existence  uniform in $v$, i.e.
up to future null infinity $\mathcal{I}^+$.) 
We will define asymptotic flatness in terms of seed data in Section~\ref{NEOKEF}, and
show  that it leads to a decay hierarchy for all quantities associated to a solution. 
These decay rates in fact propagate 
under evolution by Theorem~\ref{LWPlgINTRO}. See Theorem~\ref{prop:pwprop}.

\subsubsection{Gauge normalisation and final linearised Kerr}
\label{EDWgauges}
Before stating theorems which quantitatively estimate solutions, we must
confront the issue of gauge. In addition,
we can already identify the final linearised Kerr to which
our solution will eventually approach. In the body
of the paper, this is the content of {\bf Section~\ref{newgaugesec}}.

Two ``choices of gauge'', realised in our linear theory by the addition of two distinct pure gauge
solutions $\mathscr{G}$, will play an important role in this work.

1.~To formulate 
quantitative boundedness (cf.~(a) of the main theorem of Section~\ref{THEintroduction}), 
we first need a quantitative measure of the initial
data. For this, it is necessary to normalise the solution on the  initial data
hypersurface $C_{u_0}\cup C_{v_0}$  
by subtracting a pure gauge solution $\Gi$. 
Given a solution $\mathscr{S}$ asymptotically flat in
the sense of Section~\ref{NEOKEF}, then
 Theorem~\ref{prop:gaugeachieve} 
 (see Section~\ref{IDnGsec}) establishes that there indeed
exists a $\Gi$ normalising the solution on initial data, 
where
 addition of $\Gi$ ensures in particular that the ``location'' 
of the horizon is fixed and that the sphere at infinity is ``round''.
The resulting solution $\Si$ is known as  the 
 \emph{initial-data normalised solution}
\begin{equation}
\label{canform}
\Si \doteq \mathscr{S}+\Gi.
\end{equation}
Moreover, it is shown that
$\Gi$ is itself  asymptotically flat.

Eventually, our main \emph{boundedness} result
(Theorem~\ref{theo:mtheo} 
of Section~\ref{Themainthesection}) will uniformly bound  natural
energies (on both spheres and cones) for
$\Si$ 
from an
initial energy norm.
These integral bounds control all quantities restricted
to their $\ell\ge 2$ angular frequencies.
In~Theorem~\ref{etsilew} (see Section~\ref{vanishingsec}), 
we show that the projection of $\Si$ to its
$\ell=0$ and $\ell=1$ modes is precisely a linearised Kerr solution
$\mathscr{K}_{\mathfrak{m}, s_i}$. Thus, the energies of Theorem~\ref{theo:mtheo}
are coercive on $\Si^\prime=\Si-\mathscr{K}_{\mathfrak{m}, s_i}$.
As a simple corollary, we will obtain in particular pointwise uniform bounds on all quantities
associated to $\Si$
 (see Corollary~\ref{cor:pwe}) in terms of an initial energy and the Kerr parameters
 $\mathfrak{m}$, $s_i$ which can be read off explicitly
 (and thus in particular are bounded)
 from data.
Rough versions of the statement of the theorem and its corollary
are given in Section~\ref{BfsINTRO} below.

2.~To
prove decay (cf.~(b) of the main theorem of Section~\ref{THEintroduction}) 
of all quantities,
we need to ``choose a different
gauge'', re-normalised
at the event horizon $\mathcal{H}^+$. That is to say, we will add
yet another pure gauge  
solution $\accentset{\land}{\mathscr{G}}$
of the from $(\ref{EXpuregauge})$, where $f(v,\theta,\phi)$ 
is  determined from the metric component $\Olino$
of the initial data-normalised solution $\Si$ by solving the ODE:
\begin{equation}
\label{kaitaode}
\partial_vf +\frac{1}{2M}f = -\Olin
\end{equation}
 along $\mathcal{H}^+$, to obtain the \emph{horizon-renormalised solution} $\Sf$ defined by:
\begin{equation}
\label{definingthefuture}
\accentset{\land}{\mathscr{S}} \doteq \underaccent{\lor}{\mathscr{S}}  +\accentset{\land}{\mathscr{G}}.
\end{equation}
The equation $(\ref{kaitaode})$ ensures in particular that the (regular) linearised lapse
associated to $\Sf$ vanishes on the horizon
\begin{equation}
\label{normalisationstovori}
\Olin\left[\Sf\right]=0   \qquad \text{on the horizon}
\end{equation}
See Proposition~\ref{prop:hozrgauge} in Section~\ref{hngaugesec}. (For convenience, we in fact
alter the above definition
for the $\ell=0$ mode so that the linearised Kerr solutions $\mathscr{K}_{\mathfrak{m}, s_i}$ are
already correctly normalised.)

Eventually, our main \emph{decay} result
(Theorem~\ref{theo:mtheod} or Section~\ref{Themainthesection}) 
will give quantitative energy decay estimates for
$\accentset{\land}{\mathscr{S}}$ including the metric components $(\ref{metriccoefs})$ themselves. 
As with the boundedness theorem, these  estimates control the restriction of $\Sf$
to angular frequencies $\ell \ge2$, and thus, as above, they are
indeed coercive on $\Sf'=\Sf-\mathscr{K}_{\mathfrak{m}, s_i}$, i.e.~they
show decay of the solution to a reference linearised Kerr.
Moreover, we shall show (using Theorem~\ref{theo:mtheo}!)~that the pure gauge solution
$\accentset{\land}{\mathscr{G}}$ can itself be  uniformly
bounded (in  a weighted sense)
by the initial values of $\underaccent{\lor}{\mathscr{S}}$;
thus, the decay result is indeed
quantitative. We expect that this global quantitative 
control of the pure gauge $\Gf$ is potentially of fundamental
importance for 
 non-linear applications.
A rough version of the statement of Theorem~\ref{theo:mtheod} and its 
pointwise corollary is given in  Section~\ref{DFSintrsec} 
below.

We note that both 
the above normalised solutions $\Si$ and $\Sf$ enjoy various additional 
properties which will be useful later on. 
In particular, the so called horizon gauge conditions $(\ref{theypropag})$
hold globally on the horizon, not just the initial sphere $S_{\infty,v_0}^2$. 
The  roundness of the
sphere
at infinity and the 
good properties at the horizon  are captured by two quantities $\Ylin$ and $\Zlin$, respectively, the former
 defined by
the expression
\begin{equation}
\label{Ydefintro}
\Ylin :=  r \left(r^2 \slashed{\mathcal{D}}_2^\star  \slashed{div} \left({\Omega^{-1}}{r \, \xblin} \right)  - {\Omega^{-1}} {r^3 \underline{\plin}} \, \right)  \, .
\end{equation}
It is shown (see Section~\ref{sec:globcon})
that $\Ylin\left[\Si\right]$ is uniformly bounded along $C_{u_0}$.
In the course of the proof of Theorem~\ref{theo:mtheo}, we  shall  show that this uniform boundedness propagates. In Theorem~\ref{theo:mtheod}, 
we shall show that this boundedness holds also for $\Ylin\left[\Sf\right]$.
(This will imply, in particular, that the sphere at infinity remains round for the 
horizon-renormalised solution.)

\subsection{The main theorems}
\label{MTintro}
We give now our first rough
statements of the main theorems of this paper. These will correspond
to the more precise
statements given in the body of the paper in {\bf Section~\ref{Themainthesection}}.

\subsubsection{Boundedness and decay of gauge invariant quantities: the Regge--Wheeler and Teukolsky equations}\label{introstRWTK}
The first two main theorems
 correspond to boundedness and decay statements
for the gauge invariant quantities $\alpha$, $\psi$, and $P$ of Section~\ref{decoupledINTRO},
and their corresponding underlined quantities. We will state these as independent
statements
for general solutions of the Regge--Wheeler and Teukolsky equations.

The first statement concerns Regge--Wheeler. The precise statement is
Theorem~\ref{prop:summarypsi} of Section~\ref{sec:theorw}.
A rough formulation is as follows:
\begin{introtheorem}[rough version]
\label{gauge-invBintro}
Let $P$ be a solution of the Regge--Wheeler equation
 $(\ref{im4INTRO})$ 
 arising from regular data described in Section~\ref{decoupledINTRO},
 and define the rescaled quantity $\Psi=r^5P$. Then the following holds:

(a) The quantity $P$ remains uniformly bounded
with respect    to an $r$-weighted energy norm in terms of its initial flux
\begin{equation}
\label{avtekaitouto}
\mathbb{F}[\Psi] \lesssim \mathbb{F}_0 [\Psi].
\end{equation}

(b) The quantity $P$ decays to $0$ in the following quantitative senses:
An $r$-weighted integrated decay statement
\begin{equation}
\label{avtekaikeivo}
\mathbb{I}[\Psi]\lesssim \mathbb{F}_0[\Psi]
\end{equation}
holds, as well as polynomial decay of energy fluxes 
and pointwise polynomial decay.
\end{introtheorem}

The flux quantities $\mathbb F$ referred to in the above theorem are suprema over integrals
on constant $u$ and $v$ null cones ($C_u$ and $C_v$, respectively).
The integral $\mathbb I$  is a spacetime intergal over the shaded region
 of the diagram before Theorem~\ref{LWPlgINTRO}. 
 Both will be explained in 
Section~\ref{EGIQsec}.
 The  statements
 (a) and (b) above are  in fact just special cases of
 an $r^p$ hierarchy of flux bounds and integrated decay
 statements (cf.~Section~\ref{therphierforwave} below).
 Although the precise form of the statements
 obtained in Theorem~\ref{gauge-invBintro} is new, we again note previous decay-type results for
the Regge--Wheeler equation   in~\cite{friedman2, blue2005wave, Donninger}.

Using the above theorem, we can now obtain a result for
general solutions of the Teukolsky equation.
The precise statement is
Theorem~\ref{theo:mtheogi} of Section~\ref{GISsubsec}.
A rough formulation is as follows:
\begin{introtheorem}[rough version]
\label{Teukintrotheorem}
Let $\alpha$ be a solution of the spin $+2$ Teukolsky equation
 $(\ref{teukolskyintheintro})$ 
 arising from regular data described in Section~\ref{decoupledINTRO},
 and let $\psi$ and $\Psi=r^5P$ be the derived quantities 
 defined by $(\ref{littlepsidef0})$ and $(\ref{Pdefintro})$.
 Then the following holds:

(a) The triple $(\Psi, \psi, \alpha)$ remains uniformly bounded 
with respect    to an $r$-weighted energy norm in terms of its initial flux
\[
\mathbb{F}[\Psi, \psi, \alpha] \lesssim \mathbb{F}_0 [\Psi, \psi, \alpha].
\]

(b) The triple $(\Psi, \psi, \alpha)$ decays to $0$ in the following quantitative senses:
An $r$-weighted integrated decay statement
\[
\mathbb{I}[\Psi, \psi, \alpha]\lesssim \mathbb{F}_0[\Psi, \psi, \alpha]
\]
holds, as well as polynomial decay of suitable fluxes 
and pointwise polynomial decay.

A similar statement holds for solutions $\underline\alpha$ of
the spin $-2$ Teukolsky equation and its derived quantities $\underline\psi$
and $\underline{P}$.
\end{introtheorem}
Again, the precise versions of the
 quantities $\mathbb{F}$ and $\mathbb{I}$ referred to in the above theorem will be explained in 
Section~\ref{sec:teuenergy}.
As noted above, even a boundedness  statement  for the Teukolsky equation
on Schwarzschild was not previously known.

Applied to the full system of linearised gravity, the above yields the following statement
\begin{corollary*}
Let $\mathscr{S}$ be a solution of the full system of linearised gravity arising from regular, asymptotically flat initial data 
described in Section~\ref{introwellposedne}. 
Then Theorem \ref{theo:mtheogi} applies to yield 
boundedness and decay for the gauge invariant hierarchy
$\left(\Psilin,\plin, \alin\right)$ 
and $\left(\underline{\Psilin},\pblin, \ablin\right)$.
\end{corollary*}
See Corollary~\ref{cor:fully} for the precise statement.

We next  turn to the problem of
 boundedness for \underline{all} quantities $(\ref{denoteinintro})$ 
 associated to $\mathscr{S}$, not just the gauge-invariant ones.

\subsubsection{Boundedness of the full system of linearised gravity}\label{BfsINTRO}
The third main theorem is a (quantitative) boundedness statement
for the full system of linearised gravity, embodying (a)
of the main theorem of the introduction. 

The most
fundamental statement is again at the level of an energy flux, now
augmented by $L^2$ estimates on spheres,
 and must
be expressed with the help of the initial-data normalised solution $\Si$ discussed already in
Section~\ref{EDWgauges}.
The precise statement is formulated as Theorem~\ref{theo:mtheo} of 
Section~\ref{BIDGse}.
A  rough statement is as follows:
\begin{introtheorem}[rough version]
\label{boundednessintro}
Let $\mathscr{S}$ be a solution of the full system of linearised gravity arising from regular, asymptotically flat initial data 
described in Section~\ref{introwellposedne} and let 
$\Gi$ the pure gauge solution
such that
 \[
\Si= \mathscr{S} + \Gi
 \]
   defined
by $(\ref{canform})$ is normalised to initial data.
Then the solution  $\underaccent{\lor}{\mathscr{S}}$   remains uniformly
bounded  with respect to a weighted energy norm $\mathbb{F}$, augmented
by the supremum of a weighted $L^2$-norm on spheres (denoted $\mathbb D$), in terms of         
its initial norm:
\begin{equation}
\label{centrerdboundINTR}
\mathbb{D}\big[\underaccent{\lor}{\mathscr{S}}\big]+
\mathbb{F}\big[\underaccent{\lor}{\mathscr{S}}\big] \lesssim 
\mathbb{D}_0\big[\underaccent{\lor}{\mathscr{S}}\big]+
\mathbb{F}_0\big[\underaccent{\lor}{\mathscr{S}}\big].
\end{equation}
All
quantities $(\ref{denoteinintro})$ of $\Si$ are controlled  in $L^2$
on suitable null cones or spheres by the above norms up to their projections
to the $\ell=0$ and $\ell=1$ modes.

Moreover, there is a unique linearised Kerr solution $\mathscr{K}_{\mathfrak{m},s_i}$, 
computable explicitly from
initial data, such that $\Si^\prime=\Si-\mathscr{K}_{\mathfrak{m},s_i}$ 
has vanishing $\ell=0$ and $\ell=1$ modes and thus
the above $L^2$ control is coercive for {all} quantities  $(\ref{denoteinintro})$ associated
to $\Si^\prime$.
\end{introtheorem}

Again, the precise form of the flux $\mathbb{F}$ and the $L^\infty(L^2)$ norm
$\mathbb{D}$ is contained in  Section~\ref{remainingsec}. We emphasise
that the bound $(\ref{centrerdboundINTR})$ and resultant control of
 the restriction of the solution to angular frequencies $\ell \ge 2$
is obtained indepedently
of identifying the correct linearised Kerr solution $\mathscr{K}_{\mathfrak{m},s_i}$.

From the above flux bounds we immediately obtain by standard 
Sobolev inequalities pointwise bounds on all quantities $(\ref{denoteinintro})$ associated
to $\Si$
\begin{corollary*}
For sufficiently regular, asymptotically flat initial data for $\mathscr{S}$ as above, 
\underline{all} quantities $(\ref{denoteinintro})$
associated to $\Si$ are uniformly
bounded \underline{pointwise} in terms of an initial energy 
as on the right hand side of $(\ref{centrerdboundINTR})$, and the parameters
$\mathfrak{m}$, $s_i$ of $\mathscr{K}_{\mathfrak{m},s_i}$ (explicitly computable from--and thus also
bounded by--initial data).
\end{corollary*}
See Corollary~\ref{cor:pwe} for a precise statement.
Note that these pointwise bounds are again $r$-weighted bounds.

\subsubsection{Decay of the full system of linearised gravity in the future-normalised gauge}
\label{DFSintrsec}
The final part of our results is the statement of decay for the full system, 
embodying
(b) of the main theorem of the introduction. For this, we have already discussed
in Section~\ref{EDWgauges}
the necessity of adding a pure gauge solution $\Gf$
 normalised to the event horizon.

The precise decay theorem is formulated as
Theorem~\ref{theo:mtheod} of Section~\ref{DFNGseec}. 
A rough statement 
takes the following form:
\begin{introtheorem}[rough version]
\label{ILEDintro}
Let $\mathscr{S}$ 
be a solution of the full system of linearised gravity
arising from regular, asymptotically flat
initial data described in Section~\ref{introwellposedne}, let 
 $\Si$ be as in
 Theorem~\ref{boundednessintro}, and let 
 \[
\Sf=\Si  + \Gf
 \] 
defined by $(\ref{definingthefuture})$ be the solution normalised to the event horizon.
 Then the pure gauge solution $\Gf$, and thus,
 in view of $(\ref{centrerdboundINTR})$, also 
 $\accentset{\land}{\mathscr{S}}$,
 satisfy
 boundedness statements 
 \begin{equation}
 \label{Th4bondsta}
\mathbb{D}\big[\Gf\big]+ \mathbb{F}\big[\Gf \big] \lesssim\mathbb{D}_0\big[\Si\big]+ \mathbb{F}_0\big[\Si \big], \qquad 
\mathbb{D}\big[\Sf\big]+  \mathbb{F}\big[\Sf \big] \lesssim\mathbb{D}_0\big[\Si\big] +\mathbb{F}_0\big[\Si\big],
 \end{equation}
from which, as in Theorem~\ref{boundednessintro}, 
it follows that all quantities $(\ref{denoteinintro})$ of $\Gf$ (and thus $\Sf$) 
 are controlled in $L^2$ on suitable null cones or spheres,
 while 
  $\Sf $ moreover
 satisfies ``integrated local energy decay'',
schematically
\begin{equation}
\label{intendecFINALintro}
  \mathbb{I} \big[\Sf \big]
  	\lesssim \mathbb{D}_0\big[\Si\big] +\mathbb{F}_0 \big[\Si \big].
\end{equation}

From $(\ref{intendecFINALintro})$, a hierarchy of inverse-polynomial decay
estimates follows for \underline{all} quantities $(\ref{denoteinintro})$ associated to
$\Sf$. As in Theorem~\ref{boundednessintro}, these estimates control
the quantities $(\ref{denoteinintro})$ of $\Sf$ up to their projections to
the $\ell=0$ and $\ell=1$ modes. 

If $\mathscr{K}_{\mathfrak{m},s_i}$ is the linearised Kerr solution of Theorem~\ref{boundednessintro}, 
then $\Sf'=\Sf-\mathscr{K}_{\mathfrak{m},s_i}$
has vanishing $\ell=0$ and $\ell=1$ modes and the above control is indeed coercive
for all quantities associated to $\Sf'$. 
\end{introtheorem}

Again, for the precise form of the estimates, see the propositions referred to in the full
statement of the theorem
in Section~\ref{DFNGseec}.
In analogy with the Corollary of Theorem~\ref{boundednessintro}, 
from the above $L^2$ bounds we immediately
obtain pointwise decay estimates by standard Sobolev inequalities.

\begin{corollary*}
For sufficiently regular asymptotically flat data for $\mathscr{S}$ as above,
we have pointwise inverse polynomial decay of all quantities of $(\ref{denoteinintro})$ 
of $\Sf$ to those of $\mathscr{K}_{\mathfrak{m},s_i}$
in particular, quantitative inverse polynomial decay rates for
the linearised metric quantities $(\ref{metriccoefs})$
themselves.
\end{corollary*}
See Corollary~\ref{newcoroledw} for the precise statement.

\subsection{Aside: Review of the case of the scalar wave equation}
\label{CsweXINTRO}

The proofs of our main theorems build on recent advances in understanding the
much simpler problem of the linear scalar wave equation 
\begin{equation}
\label{LinScaEq}
\Box_g\varphi=0
\end{equation}
 on a fixed Schwarzschild background $(\mathcal{M},g)$,
 discussed already in the introduction.
 We interrupt the outline of the present paper to review the   
 definitive results for the scalar wave equation
$(\ref{LinScaEq})$ on Schwarzschild, following~\cite{DafRod2, Mihalisnotes, DafRodnew}. The reader
very familiar with this material can skip this section altogether.
 We will resume our
outline of the body of the paper in Section~\ref{THEOUTLINE} below.

Let us note at the outset that the results reviewed in the present section
can be viewed firstly as  precise scalar wave equation 
 prototypes 
for the statements in
Theorem~\ref{gauge-invBintro} 
of  Section~\ref{MTintro} concerning the Regge--Wheeler equation. 
In fact, 
we will see in Section~\ref{tokleidi} below
that the proof of Theorem~\ref{gauge-invBintro} indeed  follows closely from the results on the wave equation to be described below.   
Some of the phenomena, however, that enter will also explicitly
appear again in the proofs of  the remaining
Theorems~\ref{Teukintrotheorem}--\ref{ILEDintro}, in particular, the red-shift effect 
(see Section~\ref{BoundforWAVE} below), the notion of an integrated local energy estimate
(See Section~\ref{trappeddifficult} below)  and the $r^p$
hierarchy (see Section~\ref{therphierforwave} below).

\subsubsection{Boundedness: Conservation laws and the red-shift}
\label{BoundforWAVE}
The scalar wave equation prototype for statement $(\ref{avtekaitouto})$ of 
Theorem~\ref{gauge-invBintro} is again a  statement that the flux of
a non-degenerate, $r$-weighted energy associated to $\varphi$ is uniformly
bounded from initial energy. As we shall see, this 
statement naturally arises in stages.

We first consider unweighted, non-degenerate energy boundedness.
Explicitly, let  us define
\begin{align}
F[\varphi]&= \sup_{v} \int_{C_v} (  (\Omega^{-1}\partial_u\varphi)^2
+|\slashed\nabla\varphi|^2)\, r^2 \Omega^2\, du \, d\gamma   +\sup_u \int_{C_u}((\Omega^{-1}\partial_v\varphi)^2+|\Omega^{-1}\slashed\nabla\varphi|^2 )\,r^2\Omega^2 \, dv \, d\gamma\, ,
\label{EFluxforwave}
\end{align}
and $F_0$ to be the same quantity where $\sup_v\int_{C_v}$
is replaced by $\int_{C_{v_0}}$, and similarly for $u$. In regular coordinates, the integrands
above represent all tangential derivatives to the cones $C_v=\{u\ge u_0\}\times\{v\}\times S^2$ and $C_u=\{u\}\times\{v\ge v_0\}\times S^2$,
without degeneration, and correspond
to the energy flux with respect to the vector field $N$ to be defined below.
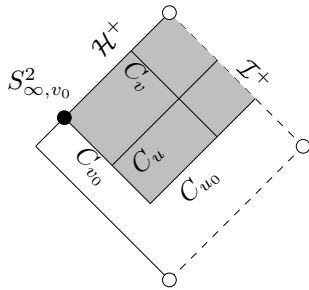
\[
\begin{picture}(0,0)%
\includegraphics{domainwfluxes.pstex}%
\end{picture}%
\setlength{\unitlength}{3158sp}%
\begingroup\makeatletter\ifx\SetFigFont\undefined%
\gdef\SetFigFont#1#2#3#4#5{%
  \reset@font\fontsize{#1}{#2pt}%
  \fontfamily{#3}\fontseries{#4}\fontshape{#5}%
  \selectfont}%
\fi\endgroup%
\begin{picture}(2402,2222)(3661,-5372)
\put(5476,-3661){\rotatebox{315.0}{\makebox(0,0)[lb]{\smash{{\SetFigFont{10}{12.0}{\rmdefault}{\mddefault}{\updefault}{\color[rgb]{0,0,0}$\mathcal{I}^+$}%
}}}}}
\put(4423,-3575){\rotatebox{45.0}{\makebox(0,0)[lb]{\smash{{\SetFigFont{10}{12.0}{\rmdefault}{\mddefault}{\updefault}{\color[rgb]{0,0,0}$\mathcal{H}^+$}%
}}}}}
\put(4201,-4336){\rotatebox{315.0}{\makebox(0,0)[lb]{\smash{{\SetFigFont{10}{12.0}{\rmdefault}{\mddefault}{\updefault}{\color[rgb]{0,0,0}$C_{v_0}$}%
}}}}}
\put(5101,-4711){\rotatebox{45.0}{\makebox(0,0)[lb]{\smash{{\SetFigFont{10}{12.0}{\rmdefault}{\mddefault}{\updefault}{\color[rgb]{0,0,0}$C_{u_0}$}%
}}}}}
\put(3676,-3811){\makebox(0,0)[lb]{\smash{{\SetFigFont{10}{12.0}{\rmdefault}{\mddefault}{\updefault}{\color[rgb]{0,0,0}$S^2_{\infty, v_0}$}%
}}}}
\put(4576,-3661){\rotatebox{315.0}{\makebox(0,0)[lb]{\smash{{\SetFigFont{10}{12.0}{\rmdefault}{\mddefault}{\updefault}{\color[rgb]{0,0,0}$C_v$}%
}}}}}
\put(4726,-4486){\rotatebox{45.0}{\makebox(0,0)[lb]{\smash{{\SetFigFont{10}{12.0}{\rmdefault}{\mddefault}{\updefault}{\color[rgb]{0,0,0}$C_u$}%
}}}}}
\end{picture}%

\]
The boundedness theorem for the unweighted energy $(\ref{EFluxforwave})$
for the scalar wave equation $(\ref{LinScaEq})$ then states 
\begin{theorem*}[\cite{DafRod2, Mihalisnotes}]
For solutions $\varphi$ of the wave equation $(\ref{LinScaEq})$ on Schwarzschild, 
we have
\begin{equation}
\label{1stanalogue}
F[\varphi]\lesssim F_0[\varphi].
\end{equation}
A similar non-degenerate higher order statement holds as well. (This implies
uniform pointwise estimates for $\psi$ and for all its derivatives up to any order.)
\end{theorem*}

The full statement of $(\ref{1stanalogue})$ can only be proven in conjunction
with the integrated decay to be shown in Section~\ref{trappeddifficult} 
which follows. We may already now prove,
however, a slightly weaker statement where we remove the $\Omega^{-1}$ weight
from the $\slashed\nabla\varphi$ term in the second integral of $(\ref{EFluxforwave})$. This estimate
is still non-degenerate on the cones $C_v$ and is thus sufficient to obtain pointwise estimates. 
We sketch the proof of this (slightly weaker) version of $(\ref{1stanalogue})$ 
below as it quite elementary and already illustrates two important features: 
\emph{conservation laws} and the \emph{red-shift}.

Recall the \emph{energy-momentum tensor} associated to $\varphi$ defined by
\begin{equation}
\label{QDefhere}
Q_{\mu\nu}[\varphi] = \partial_\mu\varphi\partial_\nu\varphi-\frac12 g_{\mu\nu} g^{\alpha\beta}\partial_\alpha
\varphi\partial_\beta\varphi,
\end{equation}
which for solutions of $(\ref{LinScaEq})$ satisfies
\begin{equation}
\label{DIVfree}
\nabla^\mu Q_{\mu\nu}[\varphi]=0.
\end{equation}
Contracting $(\ref{DIVfree})$ with the Schwarzschild Killing field $T$ defined
in $(\ref{TDEFofintro})$ one obtains
the conservation law 
\begin{equation}
\label{energyid}
\nabla^\mu J^T _\mu[\varphi] =0,
\end{equation}
where we use the notation $J^V_\mu[\varphi]=Q_{\mu\nu}[\varphi]V^\nu$ for an arbitrary vector field $V$. 
Integrating $(\ref{energyid})$ in a characteristic rectangle
bounded by the initial cones $C_{u_0}$, $C_{v_0}$ and two  later
cones $C_u$ and
$C_v$, one obtains a conservation law relating flux terms.
Using the fundamental positivity property of the tensor $(\ref{QDefhere})$:
\[
g(V,V)\le 0, \, \, g(V,W)\le0, \, \, g(W,W)\le 0 \implies
Q_{\mu\nu}[\varphi]V^\mu W^\nu\ge 0,
\]
it follows
that these  flux terms arising are nonnegative,
but degenerate at the horizon $\mathcal{H}^+$, where $T$ becomes null.
Thus, this conservation law yields
a version of the energy boundedness $(\ref{1stanalogue})$
but where $F$ and $F_0$
are replaced by fluxes $F^T$ and $F^T_0$
that degenerate at the event horizon $\mathcal{H}^+$, i.e.~\emph{a flux without
the $\Omega^{-1}$ factor on \underline{both} the $\partial_u$ and $\slashed\nabla$ terms 
in the definition $(\ref{EFluxforwave})$.}

This weaker, degenerate analogue of $(\ref{1stanalogue})$
can be thought of already as a statement of stability, but
it does \emph{not} allow one easily to infer uniform pointwise estimates up to
the horizon $\mathcal{H}^+$.\footnote{Here one should mention that the 
original Kay--Wald~\cite{KayWald} approach 
to boundedness on Schwarzschild obtained
pointwise estimates for $\varphi$ directly from this degenerate energy, applied to an auxilliary
solution $\tilde\varphi$ such that $\partial_t\tilde\varphi=\varphi$. 
The method of~\cite{KayWald} is fragile, however, and cannot, for instance, obtain boundedness for transversal derivatives. See the discussion in~\cite{Mihalisnotes}.}
It turns out, however, that 
given the above uniform degenerate energy
bound, one can than apply the so-called \emph{red-shift
energy identity},
 first introduced 
in~\cite{DafRod2},
 satisfied by a well-chosen  timelike vector field $N$, for
which the coercive property
\begin{equation}
\label{coercipr}
J^N_\mu[\varphi]N^\mu \lesssim \nabla^\mu J^N_\mu [\varphi]
\end{equation}
holds
near $\mathcal{H}^+$,
and upgrade the degenerate boundedness just obtained
to a version of $(\ref{1stanalogue})$, where 
the $\Omega^{-1}$ factor is now indeed obtained
in the $\partial_u$ term of the first integral in the definition~$(\ref{EFluxforwave})$
but must still be removed from the second
integral. (To obtain the $\Omega^{-1}$ factor
in the $\slashed\nabla$ term of the second integral over the $C_u$ cones, 
we must await for
Section~\ref{trappeddifficult}. We note that the flux terms of $(\ref{EFluxforwave})$ are precisely
the boundary terms that arise from integration of the divergence identity of $J^N_\mu$.)
The inequality $(\ref{coercipr})$ exploits the celebrated red-shift feature of the horizon (see
the discussion in~\cite{Mihalisnotes}).

To obtain a higher order analogue of $(\ref{1stanalogue})$, one can of course first
 commute the  wave equation $(\ref{LinScaEq})$
by the Killing vector field $T$ and obtain estimates for $T\varphi$. For a non-degenerate
statement, however,
we would like to obtain estimates for a \emph{strictly timelike} vector applied to $\varphi$.
For this, it turns out that one can additionally commute the wave
equation $(\ref{LinScaEq})$ by the ingoing null vector field $e_3$, 
and observe that the most dangerous new error
 term in the red-shift identity has a favourable sign (an enhanced red-shift),
 in fact for all $k\ge 0$, 
\begin{equation}
\label{STREd}
  J^N_\mu[e_3^k\varphi]N^\mu \lesssim (k+1) \, \nabla^\mu J^N_\mu [e_3^k\varphi] - \{\text{Controllable terms}\} \, .
\end{equation}
From $(\ref{STREd})$ and the fact that $T+e_3$ is timelike, 
a higher order analogue of $(\ref{1stanalogue})$ follows, nondegenerate on the $C_v$ 
cones. 
Pointwise bounds on $\varphi$ and all-order derivatives now 
follow from standard Sobolev-type inequalities.

We note already that inequality $(\ref{STREd})$ in fact ``strengthens'' the red-shift 
of $(\ref{coercipr})$
by an extra $k$ factor. This strengthening  will in fact play
an important role in the present paper.

\subsubsection{Trapped null geodesics and integrated local energy decay}
\label{trappeddifficult}
The scalar wave equation prototype for the  statement $(\ref{avtekaikeivo})$ of
Theorem~\ref{gauge-invBintro}
is again
a statement of weighted
 integrated local energy decay.
 We discuss in this section an \emph{unweighted} version of $(\ref{avtekaikeivo})$ for
 equation $(\ref{LinScaEq})$, deferring the question of
 proper $r$-weights to Section~\ref{therphierforwave} below.
 It is here already that we shall first encounter
 one of the fundamental aspects affecting quantitative
 control of decay: the existence
 of \emph{trapped null geodesics} associated to the photon sphere at $r=3M$. 
 
Let us define explicitly
\begin{equation}
\label{explitweig}
I [\varphi]=\int_{u_0}^\infty\int_{v_0}^\infty( r^{-3}(r-3M)( |T\varphi|^2+ |N\varphi|^2+ |\slashed\nabla \varphi|^2) +r^{-3}((\partial_u-\partial_v)\varphi)^2+   r^{-3}|\varphi|^2   \,) \, r^2 \Omega^2 du\, dv\, d\gamma.
\end{equation}
The degeneration of the first term in the integrand in $(\ref{explitweig})$ at $r=3M$ is related
precisely to trapping at the photon sphere.
We have 
\begin{theorem*}[\cite{DafRod2}]
For solutions $\varphi$ of the wave equation $(\ref{LinScaEq})$ on Schwarzschild, we have
the (unweighted, degenerate at $r=3M$) integrated local energy decay:
\begin{equation}
\label{iledintrowave}
I [\varphi]  \lesssim {F}_0 [\varphi].
\end{equation}
\end{theorem*}

To prove integrated local energy decay $(\ref{iledintrowave})$,  one
applies the energy identity arising from contracting $(\ref{DIVfree})$
with a well chosen vector field $X$ orthogonal to the constant-$r$ hypersurfaces:
The vector field $X$ is chosen
so that the bulk term of the identity
 $\nabla^\mu J^X_\mu$ has
coercivity properties
 allowing for the control of
the integrand of $I$, but with additional degeneration at the horizon.
The trapping constraint implies that to obtain nonnegativity of the bulk term, the 
vector field $X$ must necessarily vanish at $r=3M$, hence the degeneration
of $(\ref{explitweig})$ at $r=3M$. 
The boundary term fluxes of the energy identity, on the other hand,
are bounded by the fluxes of $J^T$.
Thus, in view of the trivial energy estimate for the fluxes of $J^T$, discussed already
in Section~\ref{BoundforWAVE}, 
one obtains a version of $(\ref{iledintrowave})$,
where one has additional degeneration of the integrand at the horizon and 
a correspondingly degenerate initial 
energy $F^T_0$ on the right hand side.

Let us note that 
a general result of Sbierski~\cite{sbierskigauss}
shows that the estimate $(\ref{iledintrowave})$ could not hold if some degeneration at $r=3M$
was not included in the definition $(\ref{explitweig})$.

In fact, the existence of the estimate $(\ref{iledintrowave})$ implicitly exploits the fact 
that trapping itself is ``unstable'' in the sense that geodesic flow is
hyperbolic.\footnote{If trapping is stable, then such an estimate cannot
hold and the rate of decay is only logarithmic. See~\cite{Keir:2014oka, Moschidis:2015wya,
gs:decay, newGustJac}.
It turns out that the good ``unstable'' Schwarzschild structure
of trapping is preserved in the Kerr case for the full sub-extremal range.
The analogous 
construction cannot, however, be done by a traditional vector field current; see~\cite{Mihalisnotes, AndBlue, Toha2} for the $|a|\ll M$ case and~\cite{withYakov} for the full $|a|<M$ case, where
in addition to the hyperbolicity of trapping, the fact that trapped frequencies are not superradiant also plays an important role.}
Let us note finally that the actual current applied is in fact more complicated than a pure
vector field current $J^X$, involving in fact $0$th order terms. See~\cite{DafRod2} for details. 
The analogous construction for $P$ in the proof of Theorem~\ref{gauge-invBintro} 
will in fact be simpler.

While the degeneration at $r=3M$ is necessary, the degeneration at the event horizon
is not.
One finally obtains the full $(\ref{iledintrowave})$ by adding the divergence
identity associated to the red-shift vector field $N$ of Section~\ref{BoundforWAVE},
in view again of $(\ref{coercipr})$.\footnote{Note however that for scattering results, it is useful
to have the original degenerate version of the estimate because it is time-reversible. 
See~\cite{Dafermos:2014jwa}.} It is at this point that we obtain also the
full $(\ref{1stanalogue})$, obtaining also the $\Omega^{-1}$ factor on the $\slashed\nabla$ term
of the second integral in the definition~$(\ref{EFluxforwave})$.

Let us note that by commuting the wave equation $(\ref{LinScaEq})$ with $T$, we may remove
the degeneration at $r=3M$ in the definition of $I$ in
$(\ref{iledintrowave})$ at the cost of requiring loss of differentiability in the
estimate, i.e.~replacing
$ F_0[\varphi]$ on the right hand side of $(\ref{iledintrowave})$
with $ F_0[\varphi]+ F_0[T\varphi]$.
In view also of $(\ref{STREd})$, we can further control an analogue of $I$ with all higher
derivatives   from a suitable higher order energy of initial data.

\subsubsection{The $r^p$ hierarchy, weighted estimates and inverse-polynomial decay bounds}
\label{therphierforwave}
To obtain the  scalar wave equation prototype for the full
\emph{weighted} boundedness $(\ref{avtekaitouto})$ 
and \emph{weighted} integrated local energy decay $(\ref{avtekaikeivo})$
of Theorem~\ref{gauge-invBintro},  
we must improve    the 
quantities $(\ref{EFluxforwave})$ and $(\ref{explitweig})$ with growing weights in $r$.
This is similar to--but more elaborate than--the improvement already
discussed which arises by adding the red-shift identity associated to 
the vector field $N$ and exploiting~$(\ref{coercipr})$.

The fundamental element is the following $r^p$ hierarchy of estimates:
\begin{theorem*}[\cite{DafRodnew}]
For $0\le p\le 2$, the 
following hierarchy of estimates holds for solutions of the wave equation
$(\ref{LinScaEq})$ on Schwarzschild
\begin{align}
\nonumber
\int_{\{r\ge R\} \cap \{u= u_2\}} r^p|\partial_v\varphi|^2+
\int_{\{ r\ge R\} \cap \{u_1\le u\le u_2\}}
r^{p-1}(p|\partial_v\varphi|^2+(2-p)|\slashed\nabla\varphi|^2)\\
\lesssim 
\int_{\{r\ge R\} \cap \{u= u_2\}} r^p|\partial_v\varphi|^2
+
\int_{\{ R-1\le r\le R\} \cap \{u_1\le u\le u_2 \}} |\partial_u\varphi|^2+|\partial_v\varphi|^2+|\slashed\nabla\varphi|^2.
\label{rpwenumber}
\end{align}
\end{theorem*}
We note that an analogue of the above identity holds for general asymptotically flat spacetimes.
See~\cite{Moschnewmeth}.

In particular, defining now 
\begin{equation}
\label{edwblackboard}
\mathbb F[\varphi]=F[\varphi]+\sup_u \int_{C_u\cap \{r\ge R\} } r^2|\partial_v\phi|^2,
\qquad
\mathbb I[\varphi]= I[\varphi] + \int_{r\ge R} r |\partial_v \varphi|^2,
\end{equation}
applying the above theorem in conjunction with the
theorems of Sections~\ref{BoundforWAVE} and~\ref{trappeddifficult} one
infers immediately  the following \emph{weighted} boundedness and integrated local
decay estimates
\begin{corollary*}
Let $\varphi$ be a solution of the scalar wave equation $(\ref{LinScaEq})$.
Then $\varphi$ satisfies the $r$-weighted uniform boundedness estimate:
\begin{equation}
\label{ari9mkiauto1}
\mathbb F[\varphi] \lesssim \mathbb F_0[\varphi]
\end{equation}
and the $r$-weighted integrated local energy estimate:
\begin{equation}
\label{ari9mkiauto2}
\mathbb I[\varphi] \lesssim \mathbb F_0[\varphi].
\end{equation}
\end{corollary*}

In $(\ref{ari9mkiauto1})$ and $(\ref{ari9mkiauto2})$, we have indeed now
obtained the scalar wave-equation prototypes of the estimates $(\ref{avtekaitouto})$ and
$(\ref{avtekaikeivo})$ of Theorem~\ref{gauge-invBintro}.

The above estimates  only represent the special $p=2$ case
of a hierarchy of estimates where the integrands
in the second terms of the right hand side of definitions $(\ref{edwblackboard})$ 
are replaced 
by the analogous integrands in $(\ref{rpwenumber})$ arising from a general $0\le p\le 2$.
It turns out that  from
repeated use of the above, exploiting the identity also for $p=0,1$ and the pigeonhole
principle,
one can obtain the following uniform $v$-decay for the flux of non-degenerate energy
\begin{equation}
\label{endecwaveintro}
\int_{C_u\cap\{v\ge \tilde{v}\}} (  (\Omega^{-1}\partial_v\varphi)^2+|\Omega^{-1}\slashed\nabla\varphi|^2)
\lesssim  \tilde{v}^{-2} (F_0[\varphi]+F_0[TT\varphi]),
\end{equation}
as well as similar bounds for fluxes on $C_v$ and for higher order energies.
Pointwise estimates 
such as 
\begin{equation}
\label{ptdecwaveintro}
|\varphi|\lesssim v^{-1}, \qquad  |r\varphi|\le u^{-1/2}, \qquad |\sqrt{r}\varphi|\lesssim u^{-1},
\end{equation}
the latter two for $r\ge r_0>2M$,
then follow from easy Hardy and Sobolev inequalities, where the implicit
constants depend on weighted $L^2$ norms of initial data.
The polynomial decay statements $(\ref{endecwaveintro})$--$(\ref{ptdecwaveintro})$
represent the precise scalar-wave equation
prototype for the polynomial decay of Theorem~\ref{gauge-invBintro}.

By commuting the wave equation by the vector field $e_4$ as well as the generators
of spherical symmetry, 
one can         apply an analogue of Theorem~\ref{gauge-invBintro} 
for higher values  $p\ge 2$. 
With this, one can further improve $(\ref{ptdecwaveintro})$
to decay rates of the form 
\begin{equation}
\label{survive}
|\varphi| \lesssim t^{-3/2},\qquad |\partial_t \varphi| \lesssim t^{-2}
\end{equation}
for fixed $r$, where the implicit constants
depend on even higher order weighted norms.
See~\cite{schlue2013decay} and the recent~\cite{Moschnewmeth}
for a definitive treatment. We will not obtain the analogue of such improvements here,
though the methods of~\cite{schlue2013decay, Moschnewmeth} easily generalise.

Recall that in Minkowski space, by the strong Huygens' principle,
solutions arising from
compactly supported initial data are in fact compactly
supported in $u$ and compactly supported in $t$ for fixed $r$. In the case
of Schwarzschild, on the other hand, even for
solutions arising from
compactly supported initial data, though the decay rate $(\ref{survive})$ for $\varphi$
can be indeed improved to  $t^{-3}$~\cite{DafRod, 
Donninger, tohaneanu}, it cannot, for 
generic initial data, be further improved~\cite{luk2015proof}. 
This is just one of a tower of $\ell$-dependent
linear obstructions which were first obtained heuristically by Price~\cite{Price:1971fb}, and
there is by now a large literature attached to them.
As opposed to  estimates of the form $(\ref{survive})$,
which indeed survive for  quasilinear problems without symmetry
(appearing in particular in
the proof of the stability of Minkowski space),
the relevance of sharper ``Price-law'' type decay estimates 
 for non-linear problems remains unclear.

\subsection{Outline of the proofs}
\label{THEOUTLINE}
Having recalled the theory of the scalar wave equation~$(\ref{LinScaEq})$
on Schwarzschild, 
we now return to the overview of our paper and 
give an outline of the proof of the main theorems of Section~\ref{MTintro}, following
Sections~\ref{sec:RW}--\ref{sec:gest1} of the body of the paper.

\subsubsection{Proof of Theorem~\ref{gauge-invBintro}: Boundedness and decay for Regge--Wheeler}
\label{tokleidi}
This will be the content of {\bf Section~\ref{sec:RW}} of the body of the paper.

The solution $P$ of the Regge--Wheeler equation can be estimated in direct analogy
with    our results for the wave equation~$(\ref{LinScaEq})$ 
described in Section~\ref{CsweXINTRO} above.
It is in fact natural to introduce from the beginning rescaled quantities
$\Psi= r^5P$.
The equation satisfied by $\Psi$ admits a conserved energy in analogy
with the $J^T$ energy discussed in Section~\ref{BoundforWAVE}. 
We immediately obtain in Section~\ref{sec:econs}
the global boundedness of a degenerate flux $F^T$ (Proposition~\ref{prop:consT}).

In Section~\ref{IDEforPhere} we obtain  our initial ``unweighted'' integrated
local energy decay, which moreover degenerates at the horizon.
This is in close analogy with the  estimate for the wave equation~$(\ref{LinScaEq})$
via the current associated to a vector field $X$ discussed in Section~\ref{trappeddifficult}. 
In fact, the construction
is easier than in the case of the wave equation, in particular since $P$, as
a symmetric traceless $S^2_{u,v}$ 2-tensor, is necessarily supported
only in angular frequencies $\ell\ge2$ (cf.~Section~\ref{sec:poinc}). 
The initial statement obtained
is the integrated local energy decay statement $(\ref{basmou})$, with
the degenerate fluxes on the right hand side.\footnote{In particular, it is here--and only here--that
the structure of trapping at the photon sphere appears directly  in this paper.} 
This is immediately improved in Section~\ref{sec:rshif} by     the analogue of  the 
red-shift estimate~$(\ref{STREd})$
to obtain non-degenerate boundedness (with an integrated decay which
does not degenerate on the horizon), and
in Section~\ref{sec:ini} by $(\ref{moraf})$, 
which is the analogue of the $r^p$-weighted estimate for the wave equation following from $(\ref{rpwenumber})$  with $p=2$.

Polynomial decay statements follow by adapting the arguments discussed in
Section~\ref{therphierforwave} for the wave equation~$(\ref{LinScaEq})$. This  is   
accomplished in Section~\ref{sec:decRW}. See Proposition~\ref{prop:decRW}.

\subsubsection{Proof of Theorem~\ref{Teukintrotheorem}: Boundedness and decay for Teukolsky}
\label{giqistoria}
This will be the content of {\bf Section~\ref{sec:hdgi}.}
We outline below.

Let $\alpha$ satisfy the spin $+2$ Teukolsky equation $(\ref{teukolskyintheintro})$,
and let $\psi$ and $P$ be defined by 
$(\ref{littlepsidef0})$ and $(\ref{Pdefintro})$, respectively. It follows that $P$ satisfies the Regge--Wheeler equation, and thus, Theorem~\ref{gauge-invBintro} applies to $P$.
The goal is to ascend the hierarchy,
 obtaining  estimates for $\psi$ and then $\alpha$ from estimates
for $P$ 
by integrating transport equations. (The results for a solution $\underline\alpha$
of the spin $-2$ Teukolsky equation are entirely analogous.)

Mutliplying $(\ref{Pdefintro})$ by $r^{2-\epsilon}\cdot \psi$
we obtain
\begin{equation}
\label{RHShereinthesketch}
\partial_u (|\psi|^2r^6 {\Omega} \cdot r^{2-\epsilon})
+ |\psi|^2r^6  {\Omega^4} r^{1-\epsilon} \lesssim 
r^{7+2-\epsilon}
|P|^2{\Omega^2}.
\end{equation}
We note that the right hand side of $(\ref{RHShereinthesketch})$ is indeed bounded
when integrated over spacetime by the weighted integrated decay estimate
for $P$ discussed above in Section~\ref{tokleidi}. Thus the first term on the left hand
side of $(\ref{RHShereinthesketch})$ gives an estimate for a flux on constant $u$ hypersurfaces,
while the second term gives a weighted integrated decay estimate. 
See $(\ref{heui})$ and $(\ref{kaivourgioovoma})$ of Proposition~\ref{prop:psie}.
Let us note that both the non-degeneracy at the horizon and the 
extra $r^p$ weights in our original estimate
for $P$ were fundamental for the numerology of $(\ref{RHShereinthesketch})$ to work out.
We thus indeed need the full strength of the weights in Theorem~\ref{gauge-invBintro}
to successfully estimate $\psi$.

Given now estimates for $\psi$, we similarly multiply $(\ref{littlepsidef0})$ 
by a weighted $r$-factor times $\alpha$ to obtain
\begin{equation}
\label{HERETO}
\partial_u( r^{4-\epsilon}\cdot r^2{\Omega^4}|\alpha|^2) 
+r^{3-\epsilon} r^2 {\Omega^6} |\alpha|^2 \lesssim r^{-1-\epsilon} 
{\Omega^4}|\psi|^2. 
\end{equation}
Again, the numerology is such that the right hand side can be estimated
by the integrated energy just controlled by Proposition~\ref{prop:psie}, again,
using in an essential way the non-degeneracy at the horizon and the $r^p$ weights.
Thus, upon integration over spacetime, $(\ref{HERETO})$ yields both an energy flux
and integrated energy decay statement for $\alpha$.
See Proposition~\ref{prop:alphae}.

Revisiting the equations, one can now  also 
estimate higher derivatives of $\psi$, $\underline\psi$, $\alpha$, $\underline\alpha$
from control of $P$ and $\underline P$ (Section~\ref{sec:higherder}). With this, one has all
the elements necessary to obtain polynomial decay estimates,
following the method of~\cite{DafRodnew} 
for the scalar wave equation discussed above 
in Section~\ref{therphierforwave}.  This is achieved in Section~\ref{sec:refinements}.

\subsubsection{Proof of Theorem~\ref{boundednessintro}: boundedness for linearised gravity}
\label{overviewbtp}
This will be the content of {\bf Section~\ref{sec:proofof3}} of the bulk of the paper. 
We outline the main points here.

\paragraph{Gauge invariant statements.}

Let $\Si$ be as in the statement of Theorem~\ref{boundednessintro}.
By the considerations
of Section~\ref{decoupledINTRO}, 
we have that $\alin$ and $\ablin$ satisfy the spin $\pm 2$
Teukolsky equations, and thus,
Theorem~\ref{Teukintrotheorem} applies, yielding boundedness
and decay estimates on
the hierarchy of gauge invariant quantities (see Section~\ref{collectthem}).
The goal is to promote this to  boundedness estimates on all quantities.

\paragraph{Fluxes on the horizon.}

The first task  (Section~\ref{sec:hdgk}) 
is to     estimate certain fluxes
on the horizon $\mathcal{H}^+$
which contain in their integrands non-gauge invariant quantities. 
For instance, from the boundedness of the $\plin$ flux on $\mathcal{H}^+$
and the second identity in~$(\ref{littlepsidef0})$,
we can immediately obtain a flux controlling  $\blin$ and $\xlin$,
in particular
\begin{equation}
\label{hatchifluxintro}
\int_{\mathcal{H}^+} |\xlin|^2.
\end{equation}
This is the content of Proposition~\ref{cor:chf}.
Using also the flux of $P$ we may obtain higher order fluxes associated
to the transversal derivative ${\Omega^{-1}}\slashed\nabla_3({\Omega} \xlin)$
\begin{equation}
\label{hatchifluxintro2}
\int_{\mathcal{H}^+} \left|{\Omega^{-1}}\slashed\nabla_3({\Omega} \xlin)\right|^2,
\end{equation}
see
Proposition~\ref{lem:gaugeta}, 
as well as the angular derivatives $\slashed{\mathcal{D}}^\star_2\slashed{\mathcal{D}}^\star_1(\, \rlin \, , \, \slin )$
and $\slashed{\mathcal{D}}^*_2\elin$. At this point, 
one can already obtain polynomial $v$-decay of various fluxes 
(Proposition~\ref{prop:decbh}).\footnote{It is worth noting here that although the integrands of the above fluxes $(\ref{hatchifluxintro})$
and $(\ref{hatchifluxintro2})$
are gauge-dependent,
the \emph{total fluxes} are gauge-invariant, given the horizon gauge conditions 
$(\ref{theypropag})$. 
The $\xlin$ flux $(\ref{hatchifluxintro})$
can in fact be directly related to an energy which arises from the Lagrangian
structure of the Einstein equations.  See Hollands and Wald~\cite{Waldstab}
for a general discussion of such fluxes and the upcoming~\cite{Holzegelfluxes} for
the relation to our setting.}

We remark that the above estimates  control only the projection of the above quantities
to angular frequencies $\ell\ge 2$. (Note that symmetric traceless $S^2_{u,v}$ 2-tensors
like $\xlin$ are necessarily supported only on $\ell\ge 2$; see Section~\ref{sec:poinc}.)
This will be a common feature of all estimates
obtained in the proof. 

\paragraph{Decay estimates for the outgoing shear $\hat\chi$.}
In analogy to how we estimated $\psi$ from $P$ and $\alpha$ from $\psi$ in Section~\ref{giqistoria}, one wishes to integrate the
transport equation $(\ref{epitelousavafora})$ for $\xlin$.
The problem is that the second term on the right hand side of $(\ref{epitelousavafora})$ has a bad
(i.e.~blue-shift)
sign at the horizon.
It turns out that, in analogy with the improved $k$ factor in $(\ref{STREd})$ for higher
transversal derivatives of the scalar wave equation $(\ref{LinScaEq})$ discussed
in Section~\ref{BoundforWAVE}, 
the bad term in $(\ref{epitelousavafora})$ can be killed by
applying suitable commutation. We see that the
quantity
\begin{equation}
\label{commutedpolu}
 \slashed{\nabla}_3 \left({\Omega^{-1}}
 \slashed{\nabla}_3\left(r^2 \xlin \, {\Omega} \, \right)\right)
\end{equation}
is in fact ``red-shifted''.  (This is apparent from equation $(\ref{dceq})$ of
Section~\ref{commutingitheresec}.)
We may now control  $\xlin$
in the darker shaded region $r\le r_1$ as follows:
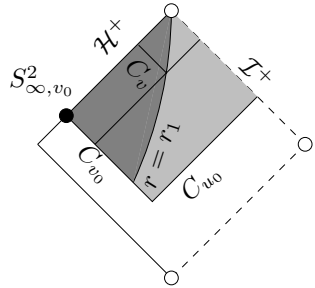
\[
\begin{picture}(0,0)%
\includegraphics{domainrsmall.pstex}%
\end{picture}%
\setlength{\unitlength}{3158sp}%
\begingroup\makeatletter\ifx\SetFigFont\undefined%
\gdef\SetFigFont#1#2#3#4#5{%
  \reset@font\fontsize{#1}{#2pt}%
  \fontfamily{#3}\fontseries{#4}\fontshape{#5}%
  \selectfont}%
\fi\endgroup%
\begin{picture}(2402,2222)(3661,-5372)
\put(5476,-3661){\rotatebox{315.0}{\makebox(0,0)[lb]{\smash{{\SetFigFont{10}{12.0}{\rmdefault}{\mddefault}{\updefault}{\color[rgb]{0,0,0}$\mathcal{I}^+$}%
}}}}}
\put(4423,-3575){\rotatebox{45.0}{\makebox(0,0)[lb]{\smash{{\SetFigFont{10}{12.0}{\rmdefault}{\mddefault}{\updefault}{\color[rgb]{0,0,0}$\mathcal{H}^+$}%
}}}}}
\put(4201,-4336){\rotatebox{315.0}{\makebox(0,0)[lb]{\smash{{\SetFigFont{10}{12.0}{\rmdefault}{\mddefault}{\updefault}{\color[rgb]{0,0,0}$C_{v_0}$}%
}}}}}
\put(5101,-4711){\rotatebox{45.0}{\makebox(0,0)[lb]{\smash{{\SetFigFont{10}{12.0}{\rmdefault}{\mddefault}{\updefault}{\color[rgb]{0,0,0}$C_{u_0}$}%
}}}}}
\put(4576,-3661){\rotatebox{315.0}{\makebox(0,0)[lb]{\smash{{\SetFigFont{10}{12.0}{\rmdefault}{\mddefault}{\updefault}{\color[rgb]{0,0,0}$C_v$}%
}}}}}
\put(3676,-3811){\makebox(0,0)[lb]{\smash{{\SetFigFont{10}{12.0}{\rmdefault}{\mddefault}{\updefault}{\color[rgb]{0,0,0}$S^2_{\infty, v_0}$}%
}}}}
\put(4825,-4622){\rotatebox{72.0}{\makebox(0,0)[lb]{\smash{{\SetFigFont{10}{12.0}{\rmdefault}{\mddefault}{\updefault}{\color[rgb]{0,0,0}$r=r_1$}%
}}}}}
\end{picture}%

\]
We couple red-shift estimates for 
$(\ref{commutedpolu})$ obtained by integrating along outgoing cones
$C_u$ in the manner of Section~\ref{giqistoria}
 to estimates for $\xlin$ and ${\Omega^{-1}}\slashed\nabla_3({\Omega} \xlin)$
 obtained by integrating (applying the fundamental theorem of calculus, twice)
 $(\ref{commutedpolu})$
along the cone
$C_v$
from the horizon, using our ``initial''
control of the horizon fluxes $(\ref{hatchifluxintro})$ and $(\ref{hatchifluxintro2})$.
This yields finally an integrated decay statement for
$\hat\chi$ restricted to the darker shaded region (Proposition~\ref{prop:nearhozchi}).
Finally, we may extend our estimates of $\xlin$ to the lighter shaded region above
by integrating directly $(\ref{epitelousavafora})$ along $C_u$ in the manner
of Section~\ref{giqistoria},
restricted to the lighter shaded
region, using our ``initial'' control along $r=r_1$ just obtained. 
This gives      a global integrated local energy decay statement for $\xlin$ (as
well as higher order derivatives). The estimate is stated as
Proposition~\ref{prop:chiall}.

\paragraph{Boundedness estimates on the ingoing shear $\hat{\underline\chi}$.}
\label{nondecayingsecIntro}
Having shown decay for the outgoing shear $\xlin$, we 
now turn to estimate the ingoing shear $\xblin$.
Here we will only be able to show boundedness, not decay.
First, we note that we are not able to estimate the evolution equation $(\ref{epitelousavafora})$
for $\xblin$  directly. Instead, 
we estimate $\xblin$ via the quantity $\Ylin$, defined by $(\ref{Ydefintro})$.

We have already noted that $\Ylin$ is  initially  bounded on the outgoing cone $C_{u_0}$,
capturing the roundness of the sphere at infinity which is ensured by
our gauge normalisation. It turns out that $\Ylin$ 
satisfies an evolution equation in the ingoing direction with 
gauge-invariant right hand side which is moreover integrable (see $(\ref{hopeid})$). 
This allows us to obtain the uniform boundedness of $\Ylin$ (Proposition~\ref{prop:Yest}).
From this one obtains the boundedness of angular derivatives of
$\xlin$ (Corollary~\ref{cor:crucialchibar})
as well as the boundedness of a flux of angular derivatives $\xblin$ on null cones
(Corollary~\ref{cor:xbar6}).

The $r$-weights above unfortunately do \underline{not}
allow us to
obtain integrated local energy decay for $\xblin$.
We can obtain, however, at this point such decay for the derivative
$\slashed\nabla_4({\Omega^{-1}}\xblin)$
(Proposition~\ref{prop:chibar4}) and upgrade it to a polynomial decay statement
on the event horizon (Proposition~\ref{prop:decdyad}).

\paragraph{Boundedness for the remaining quantities.}
Finally, we briefly discuss the remaining quantities (see Section~\ref{sec:bconclude}),
focussing on the order of the hierarchy and not the precise nature of the estimates.

We first note that the second identity of $(\ref{Pdefintro})$ relating $\Plin$
with the pair $(\rlin \, , \, \slin )$ and the shears
$\xlin$ and $\xblin$
allows us to obtain estimates for $\rlin$ and $\slin$.
Similarly, from  the second identities of $(\ref{littlepsidef1})$ and
$(\ref{littlepsidef0})$ we may now
obtain estimates for $\bblin$ and $\blin$ (see Section~\ref{titlosswstos}).
The quantities $\elin$ and $\eblin$ can also be controlled,
using however propagation equations which we have not introduced
in this overview.\footnote{Specifically,  equations $(\ref{tchi})$--$(\ref{chih3})$.
See Proposition~\ref{prop:etae}.}
We then estimate (see Section~\ref{edwmecodazzi})
 the quantity $\otxb$ via the Codazzi equation $(\ref{Lincodazz})$.
Finally, we may  
 estimate the metric components $(\ref{metriccoefs})$ themselves by integrating
their respective transport  equations. For instance, to estimate 
$\glinto$, we integrate $(\ref{topros9esame})$ using the bounds we have just obtained
on the right hand side (see Proposition~\ref{prop:meco1}).

At this point, we have estimated all quantities with the exception of $\otx$.
While this could easily be estimated via the analogue of $(\ref{Lincodazz})$ 
relating $\otx$ and $\xlin$, that method would not give us the quantitative boundedness  
as stated in  Theorem~\ref{boundednessintro},
as our previous estimates on $\xlin$ lose regularity. 
An alternative approach is thus given and it is here where the quantity
$\Zlin$ mentioned already in Section~\ref{EDWgauges} is in fact used.
We defer further discussion of this to Sections~\ref{sec:refha}--\ref{oxipolyrefined}.

By the above procedure, we 
obtain finally  $L^2$ estimates on all quantities $(\ref{denoteinintro})$
associated to $\Si$,
either on spheres or cones. As noted in the statement of
Theorem~\ref{boundednessintro}, these control only the restriction of all quantities
to angular frequencies $\ell\ge 2$.
We can now, however, apply Theorem~\ref{etsilew} to obtain that
the projection of $\Si$ to the $\ell=0$ and $\ell=1$ modes is given precisely
by a unique $\mathscr{K}_{\mathfrak{m}, s_i}$, and thus we have
truly coercive estimates on $\Si^\prime=\Si-\mathscr{K}_{\mathfrak{m}, s_i}$.
The pointwise estimates of the Corollary then follow by standard Sobolev inequalities  
applied to $\Si^\prime$
(see Section~\ref{sec:corpro}).

\subsubsection{Proof of Theorem~\ref{ILEDintro}: Decay for linearised gravity}
\label{DFNGintro}
This will be the content of {\bf Section~\ref{sec:gest1}}.
Again, we outline below:

\paragraph{Boundedness of the pure gauge solution.} 
Let $\Sf=\Si+\Gf$ be as in the statement of  Theorem~\ref{ILEDintro}.
The new normalisation $(\ref{normalisationstovori})$
allows us to derive the following evolution equation for the ingoing linearised shear on the horizon:
\[
\Omega \slashed{\nabla}_4 \left(\Omega^{-1} \xblin\left[\Sf\right]  \right) + \frac{1}{2M}   \Omega^{-1} 
\xblin\left[\Sf\right] = \frac{\Omega}{2M}  \xlin\left[\Sf\right]  - 2 
\slashed{\mathcal{D}}_2^\star \eblin\left[\Sf\right].
\]
See $(\ref{po})$. It follows that the regular quantity $\Omega^{-1} \xblin$ is ``red-shifted'',
in the sense that the sign of the factor multiplying $\xblin$ in the
second term on the the left hand side of the above identity is
positive, and moreover
the right hand side is suitably decaying.
This immediately allows for estimating  suitably normalised
$\xblin$
and $\otxb$
on the event horizon (see Proposition~\ref{lem:2dc} and its Corollaries). 
We can then translate this into bounds for derivatives
of the gauge function $f$ (see Proposition~\ref{cor:gfe2}), 
and thus, for $\Gf$ itself (see Proposition~\ref{prop:bndnew}).
This gives the boundedness statement for $\Gf$, and thus, in view of
Theorem~\ref{boundednessintro}, also for $\Sf$, i.e.~the statements
schematically represented as $(\ref{Th4bondsta})$.

We note that the reader may have expected that one must renormalise the gauge
also ``at null infinity'' in addition to the renormalisation carried out at the horizon.
It is indeed remarkable that  the horizon-renormalisation is on its own sufficient, at least
for the decay to be stated in the present paper.
This appears again to be fundamentally connected  to the red-shift at the horizon.

\paragraph{Integrated local energy decay.} 
Recall from our discussion in Section~\ref{overviewbtp} 
that it was precisely the quantity $\xblin$  which provided the first obstruction to showing
decay for $\Si$. Having obtained now decay for $\xblin\left[\Sf\right]$ 
on the horizon, we can
now upgrade this to a global integrated decay bound (Proposition~\ref{prop:chibar}), and from
this, repeating steps similar to those described in our discussion of
 the proof of Theorem~\ref{boundednessintro} in Section~\ref{overviewbtp}, 
we move up the hierarchy, to obtain now
integrated decay at each step for all other quantities (cf.~also the proof
of Theorem~\ref{Teukintrotheorem}). This corresponds to the statement
schematically represented as $(\ref{intendecFINALintro})$.
We defer further discussion to Section~\ref{kiedwILEDkiedw}.

\paragraph{Polynomial decay.}  
As in Theorems~\ref{gauge-invBintro} and~\ref{Teukintrotheorem},
 polynomial decay
can be obtained for all quantities by an adaptation of the method of~\cite{DafRodnew} described
in Section~\ref{therphierforwave} in the context of the scalar wave equation $(\ref{LinScaEq})$.
Refer to Section~\ref{sec:polyfinal}. (We only remark here that for
 reasons of length, since we have not derived integrated local energy decay
for the metric components themselves, we  obtain 
polynomial decay for these by directly 
integrating transport equations like $(\ref{topros9esame})$, using the decay already
obtained for the Ricci coefficients (see Section~\ref{sec:mecon}).)

As in the proof of Theorem~\ref{boundednessintro}, the above bounds in fact only
estimate the restriction of the solution to angular frequencies $\ell\ge 2$. In view of
Theorem~\ref{etsilew}, however, we again infer that we have indeed true decay
to $0$ for $\Sf^\prime=\Sf-\mathscr{K}_{\mathfrak{m},s_i}$ (without restriction
in angular frequency). In other words, all quantities $(\ref{denoteinintro})$ 
of $\mathscr{Sf}$ decay (in a suitable $L^2$ sense) inverse polynomially
to their $\mathscr{K}_{\mathfrak{m},s_i}$ values.
The pointwise polynomial decay estimates of the Corollary then follow
immediately by standard Sobolev inequalities.

\subsection{A restricted non-linear stability conjecture}\label{restrictedconji}

For background on the full non-linear problem, see the discussion in our
previous~\cite{DHRscat}, where we have in particular given a precise
formulation of the non-linear stability conjecture for Kerr. 
We end this overview section by noting that 
the linear stability around Schwarzschild proven in the present paper is in principle sufficient
to try to address the following restricted version of the full non-linear stability  conjecture:

\begin{conjecture*}[Full finite co-dimension non-linear stability of Schwarzschild]
Let $(\Sigma_M, \bar{g}_M, K_M)$ be the induced data 
on a spacelike asymptotically flat slice of the Schwarzschild solution of mass $M$ crossing the future horizon and bounded by a trapped surface. Then in the space of all nearby vacuum
data $(\Sigma, \bar{g}, K)$, in a suitable norm,
there exists a codimension-$3$ subfamily for which  the corresponding maximal vacuum
Cauchy development
$(\mathcal{M},g)$ contains a black-hole exterior region (characterized as the
past $J^-(\mathcal{I}^+)$ of a complete future null infinity $\mathcal{I}^+$), bounded by 
a non-empty future affine-complete event horizon $\mathcal{H}^+$,  
such that in $J^-(\mathcal{I}^+)$ 
(a) the metric remains
close to $g_M$  and moreover (b) asymptotically settles down to a nearby Schwarzschild metric 
$g_{\tilde{M}}$ at suitable inverse polynomial rates.
\end{conjecture*}

We emphasise that by dimensionality considerations, the above conjecture
would construct \underline{all}  spacetimes arising from data sufficiently near Schwarzschild
whose final state is again Schwarzschild.\footnote{The reader should compare this
with the result of~\cite{DHRscat} which gives a scattering construction of a family
of solutions asymptoting  to \emph{any} given particular subextremal Kerr $|a|<M$. The
solutions of~\cite{DHRscat}
 are constructed by prescribing exponentially decaying ``scattering'' data on the
 event horizon
and on null infinity and solving the vacuum Einstein equations $(\ref{VACeq!})$ 
\emph{backwards} in time. In view
of the strong, exponential rate of approach imposed, however,
this family is presumably very exceptional and in particular
of \emph{infinite} codimension in
the space of all solutions of the Einstein vacuum equations. Thus, when specialised
to $a=0$, the result of~\cite{DHRscat} is far from obtaining the above Conjecture.
See the comments in~\cite{DHRscat}
for futher discussion.}
In particular, we note 
that the above class of solutions would \emph{a fortiori} include the evolution of axisymmetric
initial data near Schwarzschild whose total angular momentum vanishes,
as these necessarily have a final Schwarzschild endstate (in view of the fact
that for the vacuum equations under axisymmetry, 
angular momentum does not radiate to null infinity).

\subsection{Guide to reading the paper}\label{odngos}
Though the paper has been written to be read linearly,
the reader interested only in certain results can skip various sections.
We give here a guide to various self-contained, more limited trajectories through the paper.

The reader only interested in results concerning the Regge--Wheeler and Teukolsky
equations
need only read Section~\ref{prelimsection} for notation and basic
differential operators defined on  Schwarzschild,  then
Section~\ref{sec:P}, up to Section~\ref{sec:tratheo},  for background on these equations, 
then
Sections~\ref{sec:theorw} and  \ref{GISsubsec} for the statement of   
Theorems~\ref{prop:summarypsi} and~\ref{theo:mtheogi}, and  finally
Sections~\ref{sec:RW} and \ref{sec:hdgi}
for their proofs.

The boundedness theorem (Theorem~\ref{theo:mtheo}) is independent of the decay theorem (Theorem~\ref{theo:mtheod})
and thus the reader interested only in the former can skip Section~\ref{hngaugesec}
concerning the horizon-normalised gauge, as well as Sections~\ref{DFNGseec} 
and~\ref{sec:gest1} giving
the statement and proof of Theorem~\ref{theo:mtheod}.

We note that the reader 
who does not want to concern themselves with some of the intricacies
of exploiting pure gauge solutions
can skip Section~\ref{IDnGsec},  and still understand the proof
of  Theorem~\ref{theo:mtheo}.

Finally, the reader interested in the full results but who is willing to take on faith
the system of linearised gravity can skip
Sections~\ref{VEeqDNGsec} and~\ref{hereformality}, 
though we note that the pure gauge solutions and    reference linearised Kerr solutions
of Section~\ref{sec:specialsol}
are more easily verified by applying the linearisation of Section~\ref{hereformality} than
by direct computation.

\section{The vacuum Einstein equations in a double null gauge}
\label{VEeqDNGsec}
In this section, we review the form of the vacuum Einstein equations $(\ref{VACeq!})$
written with respect to a natural null frame attached to a local double null
foliation of a Lorentzian manifold. See Christodoulou~\cite{formationofbh}.
It is these equations which we shall formally linearise in Section~\ref{hereformality} to obtain
the equations of linearised gravity.
The reader not interested in the derivation of the linearised system can skip immediately
to Section~\ref{prelimsection}.

An outline of the current section is as follows:
We begin in Section~\ref{sec:genmfld} with preliminaries, defining the notion of
double null gauge and associated notation.  Ricci coefficients and curvature
components are then defined in Section~\ref{sec:rccc}.  Finally, the vacuum Einstein equations
are presented in Section~\ref{nseq}.

\subsection{Preliminaries} \label{sec:genmfld}
Let $({\boldsymbol{\mathcal{M}},\boldsymbol{g}})$ be a $3+1$-dimensional 
Lorentzian manifold.

\subsubsection{Local double null gauge}
In a neighbourhood of any point ${\bf p}\in {\boldsymbol{\mathcal{M}}}$, we can introduce 
local coordinates $\boldsymbol{u},\boldsymbol{v}, \boldsymbol{\theta}^1,
\boldsymbol{\theta}^2$ such that the metric 
is
 expressed in ``canonical double-null form'':
\begin{equation} \label{metricdn}
\boldsymbol{g} = -4 {\boldsymbol\Omega}^2 \boldsymbol{d}\boldsymbol{u} 
\boldsymbol{d}\boldsymbol{v} + \boldsymbol{\slashed{g}}_{CD} \left(\boldsymbol{d}{\boldsymbol{\theta}}^C - 
{\boldsymbol{b}}^C \ \boldsymbol{d}{\boldsymbol{v}} \right) \left(\boldsymbol{d}{\boldsymbol{\theta}}^D - {\boldsymbol{b}}^D \ \boldsymbol{d}{\boldsymbol{v}} \right) \, 
\end{equation}
for a spacetime function $\boldsymbol{\Omega} : \boldsymbol{\mathcal{M}}\rightarrow \mathbb{R}$, 
an 
$\boldsymbol{S}_{\boldsymbol{u},\boldsymbol{v}}$-tangent vector ${\boldsymbol{b}}^A$ and a symmetric  $\boldsymbol{S}_{\boldsymbol{u},\boldsymbol{v}}$-tangent covariant symmetric $2$-tensor $\boldsymbol{\slashed{g}}_{AB}$. Here $\boldsymbol{S}_{\boldsymbol{u},\boldsymbol{v}}$ denotes the two-dimensional (Riemannian with metric $\boldsymbol{\slashed{g}}_{AB}$) manifold arising as the intersection of the hypersurfaces of constant $\boldsymbol{u}$ and $\boldsymbol{v}$.

\subsubsection{Normalised frames}
We can define a  normalised null frame associated to the above coordinates
as follows. We define
\begin{equation} \label{framedefl}
{\boldsymbol{e}}_3 = {\boldsymbol \Omega}^{-1} \boldsymbol{\partial}_{\boldsymbol u},
\qquad
{\boldsymbol{e}}_4 ={\boldsymbol \Omega}^{-1} \left( \boldsymbol{\partial}_{\boldsymbol v} + \boldsymbol{b}^A \boldsymbol{\partial}_{\boldsymbol\theta^A} \right), 
\qquad
{\boldsymbol{e}}_A = \boldsymbol{\partial}_{\boldsymbol\theta^A} \ \ \ \textrm{for $A=1,2$} \, ,
\end{equation}
for which we note the relations
\[
\boldsymbol{g}\left({\boldsymbol{e}}_3,{\boldsymbol{e}}_4\right) = -2 \ \ \ , \ \ \ \boldsymbol{g}\left({\boldsymbol{e}}_3,
{\boldsymbol{e}}_A \right) = 0 \ \ \ , \ \ \  \boldsymbol{g}\left({\boldsymbol{e}}_4, {\boldsymbol{e}}_A\right) = 0 \ \ \ , \ \ \ \boldsymbol{g} \left({\boldsymbol{e}}_A, {\boldsymbol{e}}_B \right) = \slashed{\boldsymbol{g}}_{AB} \, .
\] 
In particular, $\{\boldsymbol{e}_1,
\boldsymbol{e}_2\}$ 
constitutes a (local) coordinate frame field (not necessarily orthonormal) of the orthogonal complement of the span of ${\boldsymbol{e}}_3$, ${\boldsymbol{e}}_4$ (i.e.~in the tangent space of the submanifold $\boldsymbol{S}_{\boldsymbol{u},\boldsymbol{v}}$). In view of the above relations, we shall refer to the null frame 
$\boldsymbol{\mathcal{N}}=\left\{{\boldsymbol{e}}_3, {\boldsymbol{e}}_4, \boldsymbol{e}_1,\boldsymbol{e}_2\right\}$ as being \emph{normalised}.

\subsubsection{${\boldsymbol S}_{\boldsymbol{u},\boldsymbol{v}}$-tensor algebra}\label{tensalg}
In Section \ref{sec:rccc}, we will express the Ricci coefficients and curvature components of the metric (\ref{metricdn}) with respect to the null frame (\ref{framedefl}). These objects will then become ${\boldsymbol S}_{\boldsymbol{u},\boldsymbol{v}}$-tangent tensors, or ${\boldsymbol S}_{\boldsymbol{u},\boldsymbol{v}}$-tensors for short (see~\cite{formationofbh}).
Two types of such ${\boldsymbol S}_{\boldsymbol{u},\boldsymbol{v}}$-tensors will play a particularly important role: One-forms $\boldsymbol\xi$ and symmetric $2$-tensors $\boldsymbol\theta$, the latter being defined as satisfying
$\boldsymbol\theta_{AB} = \boldsymbol\theta_{BA}$ in any coordinate patch. A \emph{traceless} symmetric ${\boldsymbol S}_{\boldsymbol{u},\boldsymbol{v}}$
 $2$-tensor $\boldsymbol\theta$ satisfies in addition $\boldsymbol{\slashed{g}}^{AB}\boldsymbol\theta_{AB} = 0$.

Let $\boldsymbol\xi, \boldsymbol{\tilde \xi}$ be arbitrary ${\boldsymbol S}_{\boldsymbol{u},\boldsymbol{v}}$ one-forms and 
$\boldsymbol\theta, \boldsymbol{\tilde \theta}$ 
be arbitrary 
symmetric ${\boldsymbol S}_{\boldsymbol{u},\boldsymbol{v}}$ $2$-tensors. 

We denote by ${}^\star \boldsymbol\xi$ and ${}^\star \boldsymbol\theta$ the Hodge-dual (on $\left({\boldsymbol S}_{\boldsymbol{u},\boldsymbol{v}}, \boldsymbol{\slashed{g}}\right)$) of $\boldsymbol\xi$ and $\boldsymbol\theta$, respectively, and
denote by $\boldsymbol \theta^\sharp$ the tensor obtained from $\boldsymbol\theta$ by raising an index with $\slashed{\boldsymbol{g}}$. 

We define the contractions 
\[
\left(\boldsymbol\xi,\boldsymbol{\tilde \xi}\right):=\boldsymbol{\slashed{g}}^{AB}\boldsymbol\xi_A \boldsymbol{\tilde \xi}_B \textrm{ \ \ \  and \ \ \ } \left(\boldsymbol\theta,\boldsymbol{\tilde\theta}\right):=\boldsymbol{\slashed{g}}^{AB}\boldsymbol{\slashed{g}}^{CD}\boldsymbol\theta_{AC} \boldsymbol{\tilde\theta}_{BD},
\]
and
denote by $\boldsymbol\theta^\sharp \cdot \boldsymbol\xi$ the one-form $\boldsymbol\theta_A^{\phantom{A}B} \boldsymbol\xi_B$ arising from the contraction with $\slashed{\boldsymbol{g}}$. 

We finally define the $2$-tensors $\boldsymbol\theta \times \boldsymbol{\tilde\theta}$, 
$\boldsymbol\xi   \widehat{\otimes} \boldsymbol{\tilde \xi}$ 
and the scalar $\boldsymbol\theta  \wedge \boldsymbol{\tilde \theta}$  via 
\begin{align}
\left(\boldsymbol\theta \times \boldsymbol{\tilde\theta}\right)_{BC}&:=\slashed{\boldsymbol{g}}^{AD}\boldsymbol\theta_{AB}\boldsymbol{\tilde\theta}_{DC} \, ,\nonumber \\
\left(\boldsymbol\xi  \widehat{\otimes} \boldsymbol{\tilde \xi} \right)_{AB} 
&:= \boldsymbol\xi_A \boldsymbol{\tilde \xi}_B + \boldsymbol\xi_B \boldsymbol{\tilde \xi}_A - \slashed{\boldsymbol{g}}^{AB}\boldsymbol\xi_A \boldsymbol{\tilde \xi}_B  \, ,
\nonumber \\
\boldsymbol\theta  \wedge \boldsymbol{\tilde\theta} &:= \boldsymbol{\slashed{\epsilon}}^{AB} \slashed{\boldsymbol{g}}^{CD}\boldsymbol\theta_{AC} \boldsymbol{\tilde\theta}_{BD} \, , \nonumber
\end{align}
where $\boldsymbol{\slashed{\epsilon}}_{AB}$ denotes the components of the volume form associated with $\slashed{\boldsymbol{g}}$ on ${\boldsymbol S}_{\boldsymbol{u},\boldsymbol{v}}$. Note that $\boldsymbol\xi   \widehat{\otimes} \boldsymbol{\tilde \xi} $ is a symmetric traceless $2$-tensor.
\subsubsection{${\boldsymbol S}_{\boldsymbol{u},\boldsymbol{v}}$-projected Lie and
covariant derivates} \label{sec:pcd} 
We define the derivative operators $\underline{\boldsymbol D}$ and ${\boldsymbol D}$ to act on an ${\boldsymbol S}_{\boldsymbol{u},\boldsymbol{v}}$-tensor $\boldsymbol\phi$ as  
the projection onto ${\boldsymbol S}_{\boldsymbol{u},\boldsymbol{v}}$ of the Lie-derivative of $\boldsymbol\phi$ in the direction of $\boldsymbol \Omega{\boldsymbol{e}}_3$ and $\boldsymbol \Omega{\boldsymbol{e}}_4$ respectively. We hence have the following relations between the projected Lie-derivatives $\underline{\boldsymbol D}$ and ${\boldsymbol D}$ and the ${\boldsymbol S}_{\boldsymbol{u},\boldsymbol{v}}$-projected spacetime covariant derivatives $ \slashed{\nabla}_{\boldsymbol 3} = \slashed{\nabla}_{{\boldsymbol{e}}_3},\slashed{\nabla}_{\boldsymbol 4}= \slashed{\nabla}_{{\boldsymbol{e}}_4}$ in the direction ${\boldsymbol{e}}_3$ and ${\boldsymbol{e}}_4$ respectively: 
\begin{equation} \label{Dcovtransform}
\begin{split}
\boldsymbol {D} \boldsymbol f &= \boldsymbol{\Omega} {\slashed{\nabla}}_{\boldsymbol 4} \boldsymbol f \textrm{ \ \ \ on functions $\boldsymbol{f}$,} \\
\boldsymbol {D} \boldsymbol\xi &= \boldsymbol{\Omega} {\slashed{\nabla}}_{\boldsymbol 4}  \boldsymbol \xi + \boldsymbol{\Omega} \boldsymbol\chi^\sharp \cdot \boldsymbol\xi \textrm{ \ \ \ on one-forms $\boldsymbol \xi$,}
  \\
\boldsymbol {D} \boldsymbol\theta &=\boldsymbol{\Omega} {\slashed{\nabla}}_{\boldsymbol 4}  \boldsymbol\theta + \boldsymbol{\Omega} \boldsymbol\chi \times \boldsymbol\theta + \boldsymbol{\Omega} \boldsymbol\theta \times \boldsymbol\chi \textrm{ \ \ \ on symmetric $2$-tensors $\boldsymbol\theta$,} 
\end{split}
\end{equation}
and similarly for ${\slashed{\nabla}}_{\boldsymbol 3}$ replacing $\boldsymbol\chi$ by $\underline{\boldsymbol\chi}$ and ${\boldsymbol D}$ by $\underline{\boldsymbol D}$. 
See~\cite{formationofbh} for details.

\subsubsection{Angular operators on ${\boldsymbol S}_{\boldsymbol{u},\boldsymbol{v}}$} \label{sec:angop}
We employ the following notation (adapted from~\cite{ChristKlei}) for operators on the manifolds ${\boldsymbol S}_{\boldsymbol{u},\boldsymbol{v}}$.

Let $\boldsymbol\xi$ be an arbitrary one-form and $\boldsymbol\theta$ an arbitrary symmetric traceless $2$-tensor on ${\boldsymbol S}_{\boldsymbol{u},\boldsymbol{v}}$.
\begin{itemize}
\item $\boldsymbol{\slashed{\nabla}}$ denotes the covariant derivative associated with the metric $\slashed{\boldsymbol g}_{AB}$ on ${\boldsymbol S}_{\boldsymbol{u},\boldsymbol{v}}$. 
\item $\boldsymbol{\slashed{\mathcal{D}}}_1$ takes $\boldsymbol \xi$ into the pair of functions $\left(\slashed{\bf {div}} \boldsymbol \xi, \slashed{\bf {curl}} \boldsymbol \xi\right)$
where $\slashed{\bf {div}} \boldsymbol \xi = \slashed{\boldsymbol{g}}^{AB} \boldsymbol{\slashed{\nabla}}_A \boldsymbol\xi_B$ and $\slashed{\bf {curl}} \boldsymbol \xi = \boldsymbol{\slashed{\epsilon}}^{AB}  \boldsymbol{\slashed{\nabla}}_A \boldsymbol\xi_B$.
\item $\slashed{\mathbfcal{D}}_1^\star$, the formal\footnote{In our application,
the surfaces $\boldsymbol{S}_{\boldsymbol{u},\boldsymbol{v}}$ 
will be compact topological spheres and this will indeed define an adjoint on appropriate
spaces.} $L^2$-adjoint of $\boldsymbol{\slashed{\mathcal{D}}}_1$, takes any pair of scalars $\boldsymbol \rho, \boldsymbol \sigma$ into the ${\boldsymbol S}_{\boldsymbol{u},\boldsymbol{v}}$-one-form \\ $-\boldsymbol{\slashed{\nabla}}_A \boldsymbol \rho + \boldsymbol{\slashed{\epsilon}}_{AB} \boldsymbol{\slashed{\nabla}}^B \boldsymbol \sigma$.
\item $\slashed{\mathbfcal{D}}_2$ takes $\boldsymbol\theta$ into the ${\boldsymbol S}_{\boldsymbol{u},\boldsymbol{v}}$-one-form $\left(\slashed{\bf {div}} \boldsymbol\theta\right)_C=\slashed{\boldsymbol{g}}^{AB} \boldsymbol{\slashed{\nabla}}_A \boldsymbol\theta_{BC}$.
\item $\slashed{\mathbfcal{D}}_2^\star$, the formal $L^2$ adjoint of $\slashed{\mathbfcal{D}}_2$, takes $\boldsymbol\xi$ into \\
the symmetric traceless $2$-tensor $\left(\slashed{\mathbfcal{D}}^\star_2 \boldsymbol \xi\right)_{AB}=-\frac{1}{2} \left(\boldsymbol{\slashed{\nabla}}_B \boldsymbol\xi_A + \boldsymbol{\slashed{\nabla}}_A \boldsymbol\xi_B - \left(\slashed{\bf {div}} \boldsymbol\xi\right) \slashed{\boldsymbol{g}}_{AB}\right)$.
\end{itemize}

\subsection{Ricci coefficients and curvature components} \label{sec:rccc}
We now define the Ricci coefficients and curvature components associated to the metric (\ref{metricdn}) with respect to the normalised null frame $ \mathcal{N}=\left\{{\boldsymbol{e}}_3, {\boldsymbol{e}}_4, \boldsymbol{e}_1,\boldsymbol{e}_2\right\}$. 

For the Ricci coefficients, using the shorthand $\boldsymbol{\nabla}_A = \boldsymbol{\nabla}_{\boldsymbol{e}_A}$ we define
\begin{equation} \label{RicC}
\begin{split}
\boldsymbol\chi_{AB} &= \boldsymbol{g} \left(\boldsymbol{\nabla}_A {\boldsymbol{e}}_4,{\boldsymbol{e}}_B\right) \textrm{ \ \ , \ \ \ \ } \underline{\boldsymbol\chi}_{AB} =  \boldsymbol{g} \left(\boldsymbol{\nabla}_A {\boldsymbol{e}}_3,{\boldsymbol{e}}_B\right) \, ,
\\
 \boldsymbol\eta_{A} &= -\frac{1}{2}  \boldsymbol{g} \left(\boldsymbol{\nabla}_{{\boldsymbol{e}}_3} {\boldsymbol{e}}_A,{\boldsymbol{e}}_4\right) \textrm{ \ \ , \ \ } \underline{\boldsymbol\eta}_{A} = -\frac{1}{2}  \boldsymbol{g} \left(\boldsymbol{\nabla}_{{\boldsymbol{e}}_4}{\boldsymbol{e}}_A,{\boldsymbol{e}}_3\right) \, ,
\\
\hat{\boldsymbol\omega} &= \frac{1}{2}  \boldsymbol{g} \left(\boldsymbol{\nabla}_{{\boldsymbol{e}}_4} {\boldsymbol{e}}_3,{\boldsymbol{e}}_4\right) \textrm{ \ \ \ \ , \ \ \ \ \ } \hat{\underline{\boldsymbol\omega}} = \frac{1}{2}  \boldsymbol{g} \left(\boldsymbol{\nabla}_{{\boldsymbol{e}}_3} {\boldsymbol{e}}_4,{\boldsymbol{e}}_3\right)  \, , \\
\boldsymbol\zeta &= \frac{1}{2}  \boldsymbol{g} \left(\boldsymbol{\nabla}_A {\boldsymbol{e}}_4,{\boldsymbol{e}}_3\right) \, .
\end{split}
\end{equation}
%
%
Note that in view of $\boldsymbol \Omega^{-1} {\boldsymbol{e}}_3$  and $\boldsymbol \Omega^{-1} {\boldsymbol{e}}_4$ being geodesic vectorfields all other connection coefficients automatically vanish. It is natural to decompose ${\boldsymbol\chi}$ into its tracefree part
$\boldsymbol{\hat\chi}$, a symmetric traceless $\boldsymbol{S}_{\boldsymbol{u},
\boldsymbol{v}}$ 2-tensor and its trace $\boldsymbol{tr}\boldsymbol\chi$,
and similarly $\underline{\boldsymbol\chi}$.\footnote{This is of course unrelated to the $\hat{\ }$
in the case of the scalar quantity
$\boldsymbol{\hat\omega}$, which distinguishes it from other normalisations; 
we retain the 
notation $\boldsymbol{\hat\omega}$ to facilitate comparison with~\cite{formationofbh}.}

With $\boldsymbol{R}$ denoting the Riemann curvature tensor of (\ref{metricdn}), the null-decomposed curvature components  are defined as follows :
\begin{equation} \label{curvC}
\begin{split}
\boldsymbol\alpha_{AB} &= \boldsymbol{R} \left( {\boldsymbol{e}}_A,  {\boldsymbol{e}}_4,  {\boldsymbol{e}}_B,  {\boldsymbol{e}}_4\right) \textrm{ \ \ \ ,  \ \  } \underline{\boldsymbol\alpha}_{AB} = \boldsymbol{R} \left( {\boldsymbol{e}}_A,  {\boldsymbol{e}}_3,  {\boldsymbol{e}}_B,  {\boldsymbol{e}}_3\right) \, , \\
\boldsymbol\beta_{A} &= \frac{1}{2} \boldsymbol{R} \left( {\boldsymbol{e}}_A,   {\boldsymbol{e}}_4,  {\boldsymbol{e}}_3,   {\boldsymbol{e}}_4\right) \textrm{ \ \ \ ,  \ \ \ } \underline{\boldsymbol\beta}_{A} = \boldsymbol{R} \left( {\boldsymbol{e}}_A,  {\boldsymbol{e}}_3,  {\boldsymbol{e}}_3,   {\boldsymbol{e}}_4\right) \, , \ \\
\boldsymbol\rho &= \frac{1}{4} \boldsymbol{R}\left(  {\boldsymbol{e}}_4,  {\boldsymbol{e}}_3, {\boldsymbol{e}}_4,  {\boldsymbol{e}}_3\right) \textrm{ \ \ \ , \ \ \ \ \ \  } \boldsymbol\sigma = \frac{1}{4} {}^\star \boldsymbol{R}\left(  {\boldsymbol{e}}_4,  {\boldsymbol{e}}_3, {\boldsymbol{e}}_4,  {\boldsymbol{e}}_3\right) \, , 
\end{split}
\end{equation}
with ${}^\star \boldsymbol{R} $ denoting the Hodge dual on $({\boldsymbol{\mathcal{M}},\boldsymbol{g}})$ of $\boldsymbol{R}$. 
The above objects are $\boldsymbol{S}_{\boldsymbol{u},\boldsymbol{v}}$-tensors (functions, vectors, symmetric $2$-tensors) on $({\boldsymbol{\mathcal{M}},\boldsymbol{g}})$, cf.~\cite{ChristKlei}. Note also the relations
\begin{align}
\hat{\underline{\boldsymbol\omega}} = \frac{\boldsymbol{\partial}_u {\boldsymbol\Omega}}{\boldsymbol\Omega^2} \ \ , \ \ \hat{{\boldsymbol\omega}} = \frac{(\boldsymbol{\partial}_{\boldsymbol{v}} + \boldsymbol{b}^A \boldsymbol{\partial}_{\boldsymbol\theta^A}) {\boldsymbol\Omega}}{\boldsymbol\Omega^2} \ \ , \ \ 
\boldsymbol\eta_A = \boldsymbol\zeta_A + \boldsymbol{\slashed{\nabla}}_A \log \boldsymbol{\Omega}  \textrm{ \ \  ,  \ \ } \underline{\boldsymbol\eta}_A = -\boldsymbol\zeta_A + \boldsymbol{\slashed{\nabla}}_A \log \boldsymbol{\Omega}  .
\end{align}

\subsection{The Einstein equations} \label{nseq}
If $({\boldsymbol{\mathcal{M}},\boldsymbol{g}})$ satisfies the vacuum Einstein equations
\begin{equation}
\label{Evackiedw}
\boldsymbol{R}_{\mu \nu} \left[\boldsymbol{g}\right] = 0 \, ,
\end{equation}
the Ricci coefficients defined in (\ref{RicC}) and curvature components (\ref{curvC}) satisfy a system of equations, which is presented in this section.

\subsubsection{The null structure equations} \label{sec:nse} 
First, we have the important first variational formulae\footnote{Note that these formulae are equivalent to the statement that $\boldsymbol{\slashed{\nabla}}_3 \slashed{\boldsymbol{g}} = 0 = \boldsymbol{\slashed{\nabla}}_4\slashed{\boldsymbol{g}}$.}:
\begin{align} \label{firstvarf}
\boldsymbol{D} \slashed{\boldsymbol{g}} = 2 \boldsymbol{\Omega} \boldsymbol\chi = 2\boldsymbol\Omega \hat{\boldsymbol\chi} + \boldsymbol\Omega {\bf tr} \boldsymbol\chi \slashed{g} \textrm{ \ \ \ \ \ and \ \ \ \ \ } \underline{\boldsymbol{D}} \slashed{\boldsymbol{g}} = 2 \boldsymbol\Omega \underline{\boldsymbol\chi} = 2 \boldsymbol\Omega \underline{\hat{\boldsymbol\chi}} + \boldsymbol\Omega {\bf  tr} \hat{\boldsymbol\chi} \slashed{\boldsymbol{g}} \, .
\end{align}
Second,
\begin{align}  \label{chieq}
\boldsymbol{\slashed{\nabla}}_3 \underline{\hat{\boldsymbol\chi}} + {\bf tr} \underline{\boldsymbol\chi} \ \hat{\underline{\boldsymbol\chi}} -\hat{\underline{\boldsymbol\omega}} \ \underline{\hat{\boldsymbol\chi}} =  -\underline{\boldsymbol\alpha} \ \ \ ,  \ \ \ \boldsymbol{\slashed{\nabla}}_4 {\hat{\boldsymbol\chi}} + {\bf tr} {\boldsymbol\chi} \ \hat{\boldsymbol\chi} -\hat{{\boldsymbol\omega}} \ {\hat{\boldsymbol\chi}} =  {\boldsymbol\alpha} ,
\end{align}
\begin{align}
\boldsymbol{\slashed{\nabla}}_3 \left({\bf tr} \underline{\boldsymbol\chi}\right) + \frac{1}{2}\left({\bf tr} \underline{\boldsymbol\chi}\right)^2 - \underline{\hat{\boldsymbol\omega}} {\bf tr} \underline{{\boldsymbol\chi}} = - \left(\underline{\hat{\boldsymbol\chi}} , \underline{\hat{\boldsymbol\chi}}\right) \ \ , \ \ \boldsymbol{\slashed{\nabla}}_4 \left({\bf tr} \boldsymbol\chi \right) + \frac{1}{2}\left({\bf tr} \boldsymbol\chi\right)^2 - \hat{\boldsymbol\omega} {\bf tr} {\boldsymbol\chi} = - 
\left( {\hat{\boldsymbol\chi}}, {\hat{\boldsymbol\chi}}\right) . 
\end{align}
Note that the last two equations are the celebrated Raychaudhuri equations. We also have
\begin{align} \label{chieq2a}
\boldsymbol{\slashed{\nabla}}_3 {\hat{\boldsymbol\chi}} + \frac{1}{2} {\bf tr} \underline{\boldsymbol\chi} \ \hat{{\boldsymbol\chi}} +\hat{\underline{\boldsymbol\omega}} \ {\hat{\boldsymbol\chi}} = -2 \slashed{\mathbfcal{D}}_2^\star \boldsymbol{\eta} - \frac{1}{2} {\bf tr} \boldsymbol \chi \ \underline{\hat{\boldsymbol\chi}} + \left({\boldsymbol\eta} \widehat{\otimes} {\boldsymbol\eta}\right) \, ,
\end{align}
\begin{align}\label{chieq2b}
\boldsymbol{\slashed{\nabla}}_4 \underline{\hat{\boldsymbol\chi}} + \frac{1}{2} {\bf tr} {\boldsymbol\chi} \ \hat{\underline{\boldsymbol\chi}} +\hat{{\boldsymbol\omega}} \ \underline{\hat{\boldsymbol\chi}} = -2 \slashed{\mathbfcal{D}}_2^\star \underline{\boldsymbol\eta} - \frac{1}{2} {\bf tr} \underline{\boldsymbol \chi} \ \hat{\boldsymbol \chi} + \left(\underline{\boldsymbol\eta} \widehat{\otimes} \underline{\boldsymbol\eta}\right) \, ,
\end{align}

\begin{align}
\boldsymbol{\slashed{\nabla}}_3 \left({\bf tr} {\boldsymbol\chi}\right) + \frac{1}{2}\left({\bf tr} \underline{\boldsymbol\chi}\right) \left({\bf tr} {\boldsymbol\chi}\right)+ \underline{\hat{\boldsymbol\omega}} {\bf tr} {{\boldsymbol\chi}} = - \left(\underline{\hat{\boldsymbol\chi}}, {\hat{\boldsymbol\chi}}\right) + 2 \left( \boldsymbol\eta , \boldsymbol\eta\right) + 2\boldsymbol\rho +2\slashed{\bf{div}} \boldsymbol\eta \, ,
\end{align}
\begin{align} \label{endchi}
\boldsymbol{\slashed{\nabla}}_4 \left({\bf tr} \underline{\boldsymbol\chi}\right) + \frac{1}{2}\left({\bf tr} {\boldsymbol\chi}\right) \left({\bf tr} \underline{\boldsymbol\chi}\right)+ {\hat{\boldsymbol\omega}} {\bf tr} \underline{{\boldsymbol\chi}} = - \left( \underline{\hat{\boldsymbol\chi}}, {\hat{\boldsymbol\chi}} \right) + 2 \left( \underline{\boldsymbol\eta} ,\underline{\boldsymbol\eta}\right) + 2\boldsymbol\rho + 2\slashed{\bf{div}} \underline{\boldsymbol\eta} \, ,
\end{align}

\begin{align}
\boldsymbol{\slashed{\nabla}}_3 \underline{\boldsymbol\eta} = \underline{\boldsymbol\chi}^\sharp \cdot \left(\boldsymbol\eta - \underline{\boldsymbol\eta}\right) + \underline{\boldsymbol\beta} \ \ \ , \ \ \ 
\boldsymbol{\slashed{\nabla}}_4 {\boldsymbol\eta} = -{\boldsymbol\chi}^\sharp \cdot \left(\boldsymbol\eta - \underline{\boldsymbol\eta}\right) - {\boldsymbol\beta} \, ,
\end{align}
\begin{align}
\boldsymbol{D} \left(\boldsymbol\Omega \hat{\underline{\boldsymbol\omega}}\right) = \boldsymbol\Omega^2 \left[ 2 \left(\boldsymbol\eta, \underline{\boldsymbol\eta}\right) - |\boldsymbol\eta|^2 - \boldsymbol\rho \right] \textrm{ \ , \ } \underline{\boldsymbol{D}} \left(\boldsymbol\Omega \hat{\boldsymbol\omega} \right)= \boldsymbol\Omega^2 \left[ 2 \left(\boldsymbol\eta, \underline{\boldsymbol\eta}\right) - |\underline{\boldsymbol\eta}|^2 - \boldsymbol\rho \right] \, ,\nonumber
\end{align}
\begin{align} \label{bnle}
\boldsymbol{\partial}_{\boldsymbol{u}} 
\boldsymbol{b}^A = 2\boldsymbol{\Omega}^2 \left(\boldsymbol\eta^A-\underline{\boldsymbol\eta}^A\right) \, .
\end{align}
Finally, we have
\begin{align}
\slashed{\bf{curl}} \boldsymbol\eta = - \frac{1}{2} \boldsymbol\chi \wedge \underline{\boldsymbol\chi} + \boldsymbol\sigma \textrm{ \ \ \ \ \ and \ \ \ \ \ }\slashed{\bf {curl}} \underline{\boldsymbol\eta} = +\frac{1}{2} \boldsymbol\chi \wedge \underline{\boldsymbol\chi} - \boldsymbol\sigma \, ,
\end{align}
the Codazzi equations
\begin{align}
\slashed{\bf{ div}} \hat{\boldsymbol\chi} &=  -\frac{1}{2} \hat{\boldsymbol\chi}^\sharp \cdot \left( \boldsymbol\eta - \underline{\boldsymbol\eta}\right) + \frac{1}{4} {\bf tr} \boldsymbol\chi \left( \boldsymbol\eta - \underline{\boldsymbol\eta}\right)  + \frac{1}{2} \boldsymbol{\slashed{\nabla}} {\bf tr} {\boldsymbol \chi} - \boldsymbol\beta \nonumber \\
&=  -\frac{1}{2} \hat{\boldsymbol\chi}^\sharp \cdot \left( \boldsymbol\eta - \underline{\boldsymbol\eta}\right) - \frac{1}{2} {\bf tr} \boldsymbol\chi  \underline{\boldsymbol\eta}  + \frac{1}{2\boldsymbol\Omega} \boldsymbol{\slashed{\nabla}} \left( {\bf \Omega} {\bf tr} {\boldsymbol \chi}\right) - \boldsymbol\beta \, ,
\end{align}
\begin{align}
\slashed{\bf {div}} \underline{\hat{\boldsymbol\chi}} &=  \frac{1}{2} \underline{\hat{\boldsymbol\chi}}^\sharp \cdot \left( \boldsymbol\eta - \underline{\boldsymbol\eta}\right) - \frac{1}{4} {\bf tr} \underline{\boldsymbol\chi} \left( \boldsymbol\eta - \underline{\boldsymbol\eta}\right)  + \frac{1}{2} \boldsymbol{\slashed{\nabla}} {\bf tr} \underline{\boldsymbol \chi} + \underline{\boldsymbol\beta} \nonumber \\
&=  \frac{1}{2} \underline{\hat{\boldsymbol\chi}}^\sharp \cdot \left( \boldsymbol\eta - \underline{\boldsymbol\eta}\right) - \frac{1}{2} {\bf tr} \boldsymbol\chi {\boldsymbol\eta}  + \frac{1}{2\boldsymbol\Omega} \boldsymbol{\slashed{\nabla}} \left({\bf \Omega} {\bf tr} \underline{\boldsymbol \chi}\right) + \underline{\boldsymbol\beta} \, ,
\end{align}
and the Gauss equation
\begin{align}
\mathbf{K} = -\frac{1}{4} {\bf tr} \boldsymbol\chi {\bf tr} \underline{\boldsymbol\chi} + \frac{1}{2}\left( \hat{\boldsymbol\chi} , \hat{\underline{\boldsymbol\chi}} \right) - \boldsymbol\rho \, .
\end{align}

\subsubsection{The Bianchi equations} \label{bieq}
We finally turn to the equations satisfied by the curvature components of $({\boldsymbol{\mathcal{M}},\boldsymbol{g}})$, which are the well-known Bianchi equations:
\begin{align}
\boldsymbol{\slashed{\nabla}}_3  \boldsymbol\alpha + \frac{1}{2} {\bf  tr} \underline{\boldsymbol\chi} \boldsymbol\alpha + 2 \underline{\hat{\boldsymbol\omega}} \boldsymbol\alpha &= -2 {\bf \slashed{\mathbfcal{D}}_2^\star }\boldsymbol\beta - 3 \hat{\boldsymbol\chi} \boldsymbol\rho - 3{}^\star \hat{\boldsymbol\chi} \boldsymbol\sigma  + \left(4\boldsymbol\eta + \boldsymbol\zeta\right) \hat{\otimes} \boldsymbol\beta \, , \nonumber \\
\boldsymbol{\slashed{\nabla}}_4 \boldsymbol\beta + 2 {\bf  tr}\boldsymbol\chi \boldsymbol\beta - \hat{\boldsymbol\omega} \boldsymbol\beta &=  \slashed{\bf{div}}\boldsymbol\alpha + \left(\underline{\boldsymbol\eta}^\sharp + 2 \boldsymbol\zeta^\sharp\right) \cdot \boldsymbol\alpha 
 \, ,  \nonumber \\
\boldsymbol{\slashed{\nabla}}_3 \boldsymbol\beta + {\bf  tr} \underline{\boldsymbol\chi} + \underline{\hat{\boldsymbol\omega}} \boldsymbol\beta &= \slashed{\mathbfcal{D}}_1^\star \left(-\boldsymbol\rho, \boldsymbol\sigma\right) + 3 \boldsymbol\eta \boldsymbol\rho + 3{}^\star \boldsymbol\eta \boldsymbol\sigma + 2\hat{\boldsymbol\chi}^\sharp \cdot \underline{\boldsymbol\beta}
 \, ,  \nonumber \\
\boldsymbol{\slashed{\nabla}}_4 \boldsymbol\rho  + \frac{3}{2}{\bf  tr} \boldsymbol\chi\boldsymbol\rho &= \slashed{\bf{div}} \boldsymbol\beta + \left(2\underline{\boldsymbol\eta} + \boldsymbol\zeta, \boldsymbol\beta \right) - \frac{1}{2} \left(\underline{\hat{\boldsymbol\chi}}, \boldsymbol\alpha\right) 
 \, ,  \nonumber \\
\boldsymbol{\slashed{\nabla}}_4 \boldsymbol\sigma + \frac{3}{2} {\bf  tr} \boldsymbol\chi \boldsymbol\sigma &= -\slashed{\bf{curl}}\boldsymbol\beta - \left(2\underline{\boldsymbol\eta} + \boldsymbol\zeta \right) \wedge \boldsymbol\beta  + \frac{1}{2} \underline{\hat{\boldsymbol\chi}} \wedge \boldsymbol\alpha 
 \, ,  \nonumber
\end{align}
\begin{align} 
\boldsymbol{\slashed{\nabla}}_3 \boldsymbol\rho + \frac{3}{2} {\bf  tr} \underline{\boldsymbol\chi}\boldsymbol\rho &= -\slashed{\bf{div}}\underline{\boldsymbol\beta} - \left(2\boldsymbol\eta - \boldsymbol\zeta, \underline{\boldsymbol\beta}\right) - \frac{1}{2} \left(\hat{\boldsymbol\chi}, \underline{\boldsymbol\alpha}\right) 
 \, , \nonumber \\
\boldsymbol{\slashed{\nabla}}_3 \boldsymbol\sigma + \frac{3}{2} {\bf  tr} \underline{\boldsymbol\chi} \boldsymbol\sigma &= -\slashed{\bf{curl}} \underline{\boldsymbol\beta} 
- \left(2\boldsymbol\eta - \boldsymbol\zeta\right) \wedge \underline{\boldsymbol\beta} - \frac{1}{2} \hat{\boldsymbol\chi} \wedge \underline{\boldsymbol\alpha}
 \, ,  \nonumber \\
\boldsymbol{\slashed{\nabla}}_4 \underline{\boldsymbol\beta} + {\bf  tr} \boldsymbol\chi  \underline{\boldsymbol\beta} + \hat{\boldsymbol\omega} \underline{\boldsymbol\beta} &= \slashed{\mathbfcal{D}}_1^\star \left(\boldsymbol\rho, \boldsymbol\sigma\right) - 3 \underline{\boldsymbol\eta} \boldsymbol\rho + 3{}^\star \underline{\boldsymbol\eta} \boldsymbol\sigma  + 2 \underline{\hat{\boldsymbol\chi}}^\sharp \cdot \boldsymbol\beta 
 \, , \nonumber \\
\boldsymbol{\slashed{\nabla}}_3 \underline{\boldsymbol\beta} + 2 {\bf  tr} \underline{\boldsymbol\chi}  \underline{\boldsymbol\beta} - \hat{\underline{\boldsymbol\omega}} \underline{\boldsymbol\beta} &= - \slashed{\bf{div}} \underline{\boldsymbol\alpha} - \left(\boldsymbol\eta^\sharp - 2 \boldsymbol\zeta^\sharp\right) \cdot \underline{\boldsymbol\alpha}
 \, ,  \nonumber \\
\boldsymbol{\slashed{\nabla}}_4\underline{\boldsymbol\alpha} + \frac{1}{2}{\bf  tr} \boldsymbol\chi \underline{\boldsymbol\alpha} + 2 \hat{\boldsymbol\omega} \underline{\boldsymbol\alpha} &=  2 \slashed{\mathbfcal{D}}_2^\star \underline{\boldsymbol\beta} - 3 \underline{\hat{\boldsymbol\chi}} \boldsymbol\rho + 3{}^\star \underline{\hat{\boldsymbol\chi}} \boldsymbol\sigma - \left(4 \underline{\boldsymbol\eta} - \boldsymbol\zeta\right) \hat{\otimes} \underline{\boldsymbol\beta}
\, .
\nonumber
\end{align}
We note that the vacuum equations $(\ref{Evackiedw})$ further imply that the symmetric
tensors $\boldsymbol{\alpha}$
and $\underline{\boldsymbol{\alpha}}$ are in addition traceless.
The above equations encode the essential hyperbolicity of $(\ref{Evackiedw})$.
See~\cite{ChristKlei}.

\section{The Schwarzschild exterior background}
\label{prelimsection}

In this section, we shall introduce the Schwarzschild exterior metric
as well as  relevant background structure which will be useful in the paper.

We first fix in Section~\ref{differentialstru}
an ambient manifold-with-boundary
$\mathcal{M}$  on which we
define the Schwarzschild exterior metric $g$ with parameter $M$. 
We shall then pass to the computationally more
convenient Eddington--Finkelstein double null coordinates  $u$, $v$
and associated
null frames in Section~\ref{EFdnulldef}, computing the Ricci coefficients and
curvature components, commenting on issues associated
to  lack of regularity at the horizon. 
In Section~\ref{Sdiffopsandcon},
we shall introduce various natural differential operators associated to Schwarzschild,
specialising the definitions of Section~\ref{VEeqDNGsec}, 
and give some useful commutation formulas. Finally, in Section~\ref{elleivai01},
we recall some elementary properties of the classical $\ell=0, 1$
spherical harmonics and derive various important 
elliptic estimates on spheres.

\subsection{Differential structure and metric}
\label{differentialstru}
We define in this section the underlying differential structure and metric
in terms of a Kruskal coordinate system.

\subsubsection{An underlying Kruskal coordinate system}
\label{underunderlying}
Define the manifold with boundary
\begin{align} \label{SchwSchmfld}
\mathcal{M} := \mathcal{D} \times S^2 := \left(-\infty,0\right] \times \left(0,\infty\right) \times S^2
\end{align}
with coordinates $\left(U,V,\theta^1,\theta^2\right)$.
We will refer to these coordinates as \emph{Kruskal coordinates}.
 The boundary 
\[
\mathcal{H}^+:=\{0\}  \times \left(0,\infty\right) \times S^2
\]
will be referred to as the \emph{horizon}. 
We denote by $S^2_{U,V}$ the $2$-sphere $\left\{U,V\right\} \times S^2 \subset \mathcal{M}$ in $\mathcal{M}$.

\subsubsection{The Schwarzschild metric}
\label{officialdefsec}
We define the Schwarzschild metric on $\mathcal{M}$ as follows.

Fix a parameter $M>0$.
Let the function $r : \mathcal{M} \rightarrow \left[2M,\infty\right)$ be 
given implicitly as a function of the coordinates $U$ and $V$ by
\begin{align}
-UV = \frac{1-\frac{2M}{r}}{\frac{2M}{r}e^{-\frac{r}{2M}}} \, ,
\end{align}
and define also
\begin{align} \label{omkd}
\Omega_K^2 \left(U,V\right) = \frac{8M^3}{r\left(U,V\right)} e^{-\frac{r\left(U,V\right)}{2M}}  \ \ \ , \ \ \ \gamma_{AB} = \textrm{standard metric on $S^2$} \, .
\end{align}
Then the Schwarzschild metric $g$ with parameter $M$ is defined to be the metric:
\begin{align} \label{sskruskal}
g = -4 {\Omega}_K^2 \left(U,V\right) d{U} d{V} +   r^2 \left(U,V\right) \gamma_{AB} d{\theta}^A d{\theta}^B.
\end{align}
Note that the horizon $\mathcal{H}^+=\partial\mathcal{M}$ 
is a null hypersurface with respect to $g$. We will sometimes use the standard
spherical coordinates $(\theta^1,\theta^2)=(\theta, \phi)$, in which case
the metric $\gamma$ takes the explicit form
\begin{equation}
\label{gammaexplicit}
\gamma= d\theta^2+\sin^2\theta d\phi^2.
\end{equation}
 
The above metric  $(\ref{sskruskal})$
can in fact be extended to define the so-called maximally
extended Schwarzschild solution (see~\cite{synge, kruskal} and the textobook~\cite{Wald})
on the ambient manifold 
given by $(-\infty,\infty)\times (\infty,\infty)\times S^2$.
In this paper, however,
we  only need consider the manifold-with-boundary $\mathcal{M}$ as defined in $(\ref{SchwSchmfld})$.

\subsection{Eddington--Finkelstein double null coordinates $u$, $v$ and frames}
\label{EFdnulldef}
We have defined our manifold and metric as above so that its smoothness
is manifest. For computations,
however, it is much more convenient to rescale the null coordinates in such a way that quantities
are more symmetric.  The new coordinates are known as
\emph{Eddington--Finkelstein double null coordinates}.  Care must be taken
at the horizon $\mathcal{H}^+$, 
however, where these coordinates break down. We explain below.

\subsubsection{Eddington--Finkelstein double null coordinates}
\label{thedefofEF}
In this section we will define another double null coordinate system that covers 
$\mathcal{M}^o$, the interior of $\mathcal{M}$, modulo the degeneration of the 
angular coordinates. This coordinate system, 
$\left(u,v,\theta^1, \theta^2\right)$, will be referred to as \emph{Eddington--Finkelstein double null coordinates} and are defined via the relations
\begin{align} \label{UuVv}
U = -\exp\left(-\frac{u}{2M}\right) \ \ \ \textrm{and} \ \ \ \ V = \exp \left(\frac{v}{2M}\right) \, .
\end{align}
Using (\ref{UuVv}), we obtain the Schwarzschild metric on $\mathcal{M}^o$ in $\left(u,v,\theta^1, \theta^2\right)$-coordinates:
\begin{align} \label{ssef}
g =  - 4 {\Omega}^2 \left(u,v\right) \, d{u} \, d{v} +   r^2 \left(u,v\right) \gamma_{AB} d{\theta}^A d{\theta}^B  
\end{align}
with
\begin{align}
\label{officialOmegadef}
\Omega^2:= 1-\frac{2M}{r}
\end{align}
and the function $r: \left(-\infty,\infty\right) \times \left(-\infty,\infty\right) \rightarrow \left(2M,\infty\right)$ defined implicitly via the relation 
\begin{align}
\label{soimplicit}
e^{\frac{v-u}{2M}}=\left(\frac{r}{2M}-1\right)e^{\frac{r}{2M}}.
\end{align}

Note that setting $t=u+v$, we may rewrite the above metric in coordinates $(t, r, \theta, \phi)$
in the usual form $(\ref{Scwintro})$.

In $\left(u,v,\theta^1, \theta^2\right)$-coordinates, the horizon $\mathcal{H}^+$ can still be formally parametrised by $\left(\infty, v,\theta^2,\theta^2\right)$ with $v \in \mathbb{R}$, $\left(\theta^1,\theta^2\right) \in S^2$.
This will allow us to use these coordinates at $\mathcal{H}^+$, provided that we appropriately rescale all quantities
so as to be regular. We shall see this principle used already in the section below.

Let us also introduce the notation $S^2_{u,v}$ to denote the sphere $S^2_{U,V}$ 
where $U$ and $V$ are defined by $(\ref{UuVv})$. 
In the spirit of the remark of the previous paragraph, we shall write in addition
$S^2_{\infty, v}$ for the spheres of the horizon $\mathcal{H}^+$ (where we are to understand $U=0$).

Finally, we will often refer informally to the limit 
$v\to \infty$ as \emph{null infinity} $\mathcal{I}^+$,
which can  be parametrised as $\mathcal{I}^+=\{(u,\infty,\theta,\phi)\}$.

\subsubsection{Killing fields of the Schwarzschild metric} \label{sec:Killing}
It is natural at this point to already discuss the Killing fields associated to $g$.

We define the vectorfield $T$ to be the timelike Killing field $\partial_t$ of the $(r,t)$ 
coordinates (\ref{Scwintro}), which in Eddington--Finkelstein double null 
coordinates is given by 
\begin{equation}
\label{Tdefforreal}
T=\frac12(\partial_u + \partial_v).
\end{equation} 
The vector field extends to a smooth Killing field on the horizon $\mathcal{H}^+$,
which is  moreover
null and tangential to the null generator of $\mathcal{H}^+$. See 
Section~\ref{thenullframessection} below.

We can also define a basis of ``angular momentum operators'' $\mathnormal{\Omega}_i$, $i=1,2,3$,
for instance, fixing standard spherical coordinates $(\theta,\phi)$ on $S^2$ where
$\gamma$ takes the form $(\ref{sskruskal})$, by
\[
\mathnormal{\Omega}_1 = \partial_\phi , \qquad
\mathnormal{\Omega}_2 = -\sin \phi\partial_\theta -\cot\theta \cos\phi\partial_\phi ,\qquad
\mathnormal{\Omega}_3 =\cos\phi\partial_\theta-\cot\theta\sin\phi\partial_\phi.
\]

The Lie algebra of Killing fields of $g$ is then precisely that generated by
$T$ and $\mathnormal{\Omega}_i$, $i=1\ldots 3$.

\subsubsection{The null frames $\mathcal{N}_{EF}$, 
$\mathcal{N}_{EF^\star}$}
\label{thenullframessection}

 We define in this section two normalised null frames associated to Schwarzschild.

The most important one is the Eddington--Finkelstein frame $\mathcal{N}_{EF}$, which we define first. The vectorfields
\begin{align}
e_3 = \frac{1}{\Omega} \partial_u \ \ \ \ \ \ , \ \ \ \  \ \  e_4 = \frac{1}{\Omega} \partial_v 
\end{align}
defined with respect to $\left(u,v,\theta^1, \theta^2\right)$-Eddington--Finkelstein coordinates, together with a local frame field $(e_1, e_2)$ on $S^2_{u,v}$ provide a normalised frame on $\mathcal{M}^o$:
\[
\textrm{$\mathcal{N}_{EF}$ = normalised null frame $\left\{{e}_3, {e}_4, e_1, e_2\right\}$} \, .
\]

The above frame does not extend regularly to the horizon $\mathcal{H}^+$ (cf.~the comments
at the end of Section~\ref{thedefofEF}). However, it is easy to see that the rescaled null frame
\[
\textrm{$\mathcal{N}_{EF^\star}$ = normalised null frame $\left\{\Omega^{-1}{e}_3, \Omega {e}_4, e_1, e_2\right\}$} 
\]
does extend regularly to a non-vanishing null frame on
$\mathcal{H}^+$.  Though we shall always compute
with respect to $\mathcal{N}_{EF}$, passing to $\mathcal{N}_{EF^\star}$ will be useful
to understand which quantities are regular on the horizon.

Note finally that $2T=\Omega e_3+\Omega e_4 = \Omega^2 (\Omega^{-1}e_3) +\Omega e_4$,
from which it follows that on the horizon, $T$ corresponds up to a factor
with the null vector
of the $EF^*$ frame $T= \frac12\Omega e_4$.

\subsubsection{Connection coefficients and curvature components}
\label{hereforSchconcur}
We compute here the connection coefficients and curvature components with respect to
the two null frames of Section~\ref{thenullframessection}.

With respect to the Eddington--Finkelstein null frame $\mathcal{N}_{EF}=\left(\Omega^{-1} \partial_u, \Omega^{-1} \partial_v, e_1,e_2\right)$,  
we  compute the following non-vanishing Ricci coefficients (cf.~Section  \ref{sec:rccc}) on $\left(\mathcal{M}^o,g\right)$:
\begin{align} \label{ssvef} 
\chi_{AB} &:=g\left(-{e}_4 , \nabla_A e_B\right)  =  \frac{\Omega}{r}  r^2 \gamma_{AB}  , &\underline{\chi}_{AB} &:=g\left(-{e}_3 , \nabla_A e_B\right)  = -\frac{\Omega}{r}  r^2 \gamma_{AB}   \\
 \hat{\omega} &:=  \frac{1}{2} g\left(  \nabla_{{e}_4} {e}_3, {e}_4 \right) = \frac{M}{r^2 \Omega}  \textrm{ \ \ \ , \ \ \ }&  \hat{\underline{\omega}} &:= \frac{1}{2} g\left(  \nabla_{{e}_3} {e}_4, {e}_3 \right) =  -\frac{M}{r^2\Omega} \, .
\end{align}

\begin{remark}
Recall from the comments in Section~\ref{thenullframessection}
that the frame $\mathcal{N}_{EF}$ is not regular near the horizon $\mathcal{H}^+$ as is manifest from some of these quantities diverging at the horizon! Converting to the rescaled regular null frame $\mathcal{N}_{EF^\star}$ one easily finds $\chi=\frac{\Omega}{r} r^2 \gamma$, $\underline{\chi}=-r \gamma$, $\hat{\omega}=\frac{2M}{r^2}$ and $\underline{\hat{\omega}}=0$ which makes all components manifestly regular at $\mathcal{H}^+$.
\end{remark}

Turning to curvature, the only non-vanishing null curvature component on $\left(\mathcal{M},g\right)$ is 
\begin{align}
\rho := \frac{1}{4} Riem\left(  e_4,e_3,e_4,e_3\right) =  -\frac{2M}{r^3} \, ,
\end{align}
an identity
which holds with respect to both null frames $\mathcal{N}_{EF}$ and $\mathcal{N}_{EF^\star}$, as these only differ by a scaling of the two null-directions. We also introduce the notation
\begin{align} \label{GCdef}
K = \frac{1}{r^2}
\end{align}
for the Gauss curvature of the round $S^2_{u,v}$-spheres.

\subsection{Schwarzschild background  operators and commutation identities}
\label{Sdiffopsandcon}

In this section, we introduce a number of natural differential operators associated
to Schwarzschild. In Section~\ref{sec:ssdiffop}, we specialise the operators discussed
in Section~\ref{sec:genmfld} to Schwarzschild.
We then give in Section~\ref{sec:commutation}
some important commutation identities and
define the additional useful angular operators $\mathcal{A}^{[i]}$.

\subsubsection{The $S_{u,v}$-tensor analysis and natural  differential operators on Schwarzschild}
\label{sec:ssdiffop}
In this section we simply 
specialise some of the constructions of Section~\ref{sec:genmfld} to 
Schwarzschild. We explictly repeat
all definitions, however, so that this section can be read independently.

We recall the notion of an $S_{u,v}$ tensor from~\cite{formationofbh}. 
Specialised to Schwarzschild, these
are simply tensors which when expressed in the frame $\mathcal{N}_{EF}$ only have components
in the angular directions $e_1$ and $e_2$.

Particularly important for our purpose
will be one-forms $\xi$ and symmetric $2$-tensors $\theta$.
An $S_{u,v}$-$2$-tensor $\theta_{AB}$ is symmetric if  $\theta_{AB}=\theta_{BA}$ and traceless if
$\slashed{g}^{AB}\theta_{AB}=0$.

We quickly repeat the definitions from Section~\ref{tensalg}
for $\xi$, $\tilde\xi$ one forms and $\theta$, $\tilde\theta$ symmetric $2$-tensors.

We denote by ${}^\star \xi$ and ${}^\star \theta$ the Hodge duals (with respect to 
$\left({S}_{{u},{v}}, {\slashed{g}}\right)$) and
by $\theta^\sharp$ the tensor obtained from $\theta$ by raising an index with 
${\slashed{g}}$. 

We define the contractions 
\[
(\xi,\tilde\xi):={\slashed{g}}^{AB}\xi_A \tilde\xi_B , \qquad (\theta,\tilde\theta):={\slashed{g}}^{AB}
{\slashed{g}}^{CD}\theta_{AC} \tilde\theta_{BD}, \qquad \theta^\sharp \cdot \xi=\theta_A^{\phantom{A}B} \xi_B.
\]

We finally define the $2$-tensors $\theta \times \tilde\theta$, $\xi   \widehat{\otimes} \tilde\xi$ 
and the scalar $\theta \wedge \tilde\theta$  via 
\begin{align}
\left(\theta \times \tilde\theta \right)_{BC}&:=\slashed{{g}}^{AD}\theta_{AB}\tilde\theta_{DC} \, ,\nonumber \\
\left(\xi  \widehat{\otimes} \tilde \xi \right)_{AB} &:= \xi_A \tilde\xi_B + \xi_B \tilde\xi_A - \slashed{{g}}^{AB}\xi_A \tilde\xi_B  \, ,
\nonumber \\
\theta  \wedge \tilde\theta &:= {\slashed{\epsilon}}^{AB} \slashed{{g}}^{CD}\theta_{AC} \tilde\theta_{BD} \, , \nonumber
\end{align}
where ${\slashed{\epsilon}}_{AB}$ denotes the components of the volume form associated with $\slashed{{g}}$ on ${ S}_{{u},{v}}$, and where we note again 
that $\xi  \widehat{\otimes} \tilde\xi$ 
is a symmetric traceless $S^2_{u,v}$ $2$-tensor.

We now specialise the general definitions of 
the projected Lie and covariant differential operators in Section \ref{sec:pcd} to the case of the Schwarzschild manifold $\left(\mathcal{M},g\right)$, with its Eddington--Finkelstein 
double null coordinates $\left(u,v,\theta^1,\theta^2\right)$ and normalised null directions $e_3=\Omega^{-1}\partial_u$, $e_4 = \Omega^{-1}\partial_v$.

\begin{itemize}
\item 
The projection to the spheres $S^2_{u,v}$ of the Lie-derivative in the directions $\partial_u$ and $\partial_v$ is denoted by $\underline{D}$ and $D$ respectively. 
\end{itemize}
Hence if $\xi$ is an $S^2_{u,v}$ tensor of rank $n$ on $\left(\mathcal{M},g\right)$ we have in components\footnote{Observe that non-trivial terms would appear in the second formula if the background was not spherically symmetric.}
\begin{align} \label{lieinc}
\left(\underline{D} \xi\right)_{A_1...A_n} = \partial_u \left(\xi_{A_1...A_n}\right) \qquad \textrm{and}
\qquad \left(D \xi\right)_{A_1...A_n} = \partial_v \left(\xi_{A_1...A_n}\right) \, .
\end{align}
Similarly, 
\begin{itemize}
\item The projection to the spheres $S^2_{u,v}$ of the covariant derivative in the $e_3$-direction is denoted by $\slashed{\nabla}_3$ and that in the $e_4$-direction by $\slashed{\nabla}_4$. 
\end{itemize}
The relations (\ref{Dcovtransform}) now hold true ``un-bolding" all quantities. Since $\chi, \underline{\chi}$ only have a trace-component in $\left(\mathcal{M},g\right)$ one can deduce simplified formulae such as
\begin{align} \label{gotoco1}
\Omega \left( \slashed{\nabla}_3 \xi\right)_{A} = \partial_u \left( \xi_{A}\right) - \frac{1}{2} \Omega tr \underline{\chi} \xi_{A}  \ \ \ \textrm{for an $S^2_{u,v}$-$1$-form $\xi$}
\end{align}
and
\begin{align}\label{gotoco2}
\Omega \left( \slashed{\nabla}_4 \theta\right)_{AB} = \partial_v \left( \theta_{AB}\right) -  \Omega tr {\chi} \theta_{AB} \ \ \ \ \textrm{for a symmetric traceless $S^2_{u,v}$-2-tensor $\theta$}
\end{align}
which will be used below.

Finally, we specialise the definitons of  Section \ref{sec:angop}, i.e.~we introduce the 
notation:
\begin{itemize}
\item ${\slashed{\nabla}}$ denotes the covariant derivative associated with the metric $\slashed{g}$ on $S^2_{u,v}$ 
\item ${\slashed{\mathcal{D}}}_1$ takes any ${S}_{{u},{v}}$-tangent $1$-form $ \xi$ into the pair of functions $\left(\slashed{ {div}}  \xi, \slashed{ {curl}}  \xi\right)$
\item $\slashed{\mathcal{D}}_1^\star$, the $L^2$-adjoint (with respect to $\slashed{g}$) of $\slashed{\mathcal{D}}_1$, takes any pair of scalars $\rho, \sigma$ into the $S^2_{u,v}$-tangent $1$-form \\ $-{\slashed{\nabla}}_A  \rho + \slashed{\epsilon}_{AB}{\slashed{\nabla}}^B \sigma$ with $\slashed{\epsilon}_{AB}$ denoting the components of the volume form associated with $\slashed{g}$ on $S^2_{u,v}$.
\item $\slashed{\mathcal{D}}_2$ takes any $2$-covariant symmetric traceless
$S^2_{u,v}$-tensor $\xi$ into the $S^2_{u,v}$-tangent $1$-form $\slashed{div}\xi$.
\item $\slashed{\mathcal{D}}_2^\star$, the $L^2$ adjoint (with respect to $\slashed{g}$) of $\slashed{\mathcal{D}}_2$ takes any $S^2_{u,v}$-tangent $1$-form $\xi$ into \\
the $2$-form $-\frac{1}{2} \left({\slashed{\nabla}}_B \xi_A + {\slashed{\nabla}}_A\xi_B - \left(\slashed{ {div}} \xi\right) \slashed{{g}}_{AB}\right)$.
\end{itemize}
Recall that the spheres $S^2_{u,v}$ on the Schwarzschild manifold are equipped with the round metric $\slashed{g}_{AB} = r^2 \gamma_{AB}$, with $\gamma$ the metric on the unit sphere. We will use the notation $\epsilon_{AB}=r^{-2} \slashed{\epsilon}_{AB}$ for the components of the volume form on the unit sphere.

\subsubsection{Commutation formulae and the operators $\mathcal{A}^{[i]}$} \label{sec:commutation}
We have the following commutation formulae for projected covariant derivatives in Schwarzschild. If $\xi = \xi_{A_1...A_n}$ is an $n$-covariant $S^2_{u,v}$-tensor on the Schwarzschild manifold $\left(\mathcal{M},g\right)$, then
\begin{align}
\slashed{\nabla}_3 \slashed{\nabla}_B \xi_{A_1...A_n} - \slashed{\nabla}_B \slashed{\nabla}_3 \xi_{A_1...A_n} &= - \frac{1}{2} tr \underline{\chi}  \slashed{\nabla}_B \xi_{A_1...A_n} \, , \nonumber \\
\slashed{\nabla}_4 \slashed{\nabla}_B \xi_{A_1...A_n} - \slashed{\nabla}_B \slashed{\nabla}_4 \xi_{A_1...A_n} &= - \frac{1}{2} tr {\chi}  \slashed{\nabla}_B \xi_{A_1...A_n} \, , \\
\slashed{\nabla}_3 \slashed{\nabla}_4 \xi_{A_1...A_n} - \slashed{\nabla}_4 \slashed{\nabla}_3 \xi_{A_1...A_n} &= \hat{\omega} \slashed{\nabla}_3 \xi_{A_1...A_n} -\hat{\underline{\omega}} \slashed{\nabla}_4 \xi_{A_1...A_n} \, . \nonumber
\end{align}
In particular, we have
\begin{align}
\left[\slashed{\nabla}_4, r \slashed{\nabla}_A \right] \xi = 0 \ \ \ , \ \ \ \left[\slashed{\nabla}_3, r \slashed{\nabla}_A \right] \xi = 0 \ \ \ , \ \ \ \left[\Omega \slashed{\nabla}_3, \Omega \slashed{\nabla}_4 \right] \xi = 0 \, .
\end{align}

Finally, we introduce a shorthand notation for $i$ angular derivatives acting on a symmetric traceless $S^2_{u,v}$-tensor ($i\geq 1$). The crucial feature of these $\mathcal{A}^{[i]}$ is that they commute trivially with $\slashed{\nabla}_3$ and $\slashed{\nabla}_4$. We define
\begin{equation} \label{mathanot}
\mathcal{A}^{[2]} = r^2 \slashed{\mathcal{D}}_2^\star \slashed{\mathcal{D}}_2 \ \ \ \textrm{and inductively} \ \ \ \mathcal{A}^{[2i+1]} = r \slashed{\mathcal{D}}_2 \mathcal{A}^{[2i]}   \ \ \ , \ \ \ \mathcal{A}^{[2i+2]} = r^2 \slashed{\mathcal{D}}_2^\star \slashed{\mathcal{D}}_2 \mathcal{A}^{[2i]} \, .
\end{equation}

\subsection{The $\ell=0,1$ spherical harmonics and elliptic estimates on spheres}\label{elleivai01}
We collect in this final subsection some useful properties which require isolating
the $\ell=0,1$ angular frequencies of a tensor.
More specifically, 
after defining notation in Section~\ref{sec:notation},
we shall 
recall  in Section~\ref{herespherhar}
the classical $\ell=0,1$ spherical harmonics and
define what it means for  $S^2_{u,v}$ tensors of various types
to be supported on angular frequencies $\ell\ge2$.
This will then allow us to
infer in Section~\ref{sec:poinc} some
useful elliptic estimates  on spheres for such tensors.

\subsubsection{Norms on spheres}  \label{sec:notation}

Let $(\theta,\phi)$ denote standard spherical coordinates as in Section~\ref{officialdefsec}
where the spherical metric takes the form $(\ref{gammaexplicit})$.

We define the following pointwise norm for $S^2_{u,v}$-tensors of rank $n$, $\xi_{A_1...A_n}$:
\begin{align}
| \xi |^2 := \slashed{g}^{A_1 B_1} \ldots \slashed{g}^{A_n B_n} \xi_{A_1 \ldots A_n} \xi_{B_1 \ldots B_n} \, .
\end{align}
We also define the $L^2\left(S^2_{u,v}\right)$-norm
\begin{align} \label{normsp} 
\| \xi \|^2_{S^2_{u,v}} := \int_{S^2} r^2\left(u,v\right) \sin \theta d\theta d\phi |\xi|^2 \,
\end{align}
and note that\footnote{We will often write quantities in this form as it is easier to read off the decay.}
\begin{align}
\| r^{-1} \cdot \xi \|^2_{S^2_{u,v}} = \int_{S^2}  \sin \theta d\theta d\phi |\xi|^2 \, .
\end{align}

\subsubsection{The $\ell=0,1$ spherical harmonics and tensors supported on $\ell\ge 2$} \label{herespherhar}
Recall the well-known spherical harmonics $Y^{\ell}_m$ (where $\ell \in \mathbb{N}^0$ and $m \in \{-\ell, -\ell+1, ..., \ell-1,\ell \}$ admissible for fixed $\ell$) on the unit sphere.
The $\ell=0,1$ spherical harmonics are given explicitly by
\begin{equation}
\label{defY00}
Y^{\ell=0}_{m=0}=\frac{1}{\sqrt{4\pi}}
\end{equation}
and
\begin{equation}
\label{defYothers}
Y^{\ell=1}_{m=0} = \sqrt{\frac{3}{8\pi}} \cos \theta \ \ \ , \ \ \ Y^{\ell=1}_{m=-1} =\sqrt{\frac{3}{4\pi}} \sin \theta \cos \phi \ \ \ , \ \ \ Y^{\ell=1}_{m=1} = \sqrt{\frac{3}{4\pi}}\sin \theta \sin \phi \, .
\end{equation}
We have that the above family is orthogonal with respect to the standard
inner product on the sphere, and any arbitrary function 
$f\in L^2\left(S^2\right)$ can be expanded uniquely with respect to such a basis. 
\begin{definition} \label{def:supps}
We say a {\bf function} $f$ on $\mathcal{M}$ is {\bf supported on $\ell \ge 2$} if the projections
\[
\int \sin \theta d\theta d\phi \  f\cdot Y^{\ell}_m = 0
\]
vanish for $(\ref{defY00})$ and $(\ref{defYothers})$.
Any function $f$ can be uniquely decomposed orthogonally as
\[
f=  c(u,v) Y^{\ell=0}_{m=0} + \sum_{i=-1}^1 c_i(u,v) Y^{\ell=1}_{m=i} \left(\theta, \phi\right)   + f_{\ell \ge 2}
\]
where $f_{\ell \ge 2}$ is supported on $\ell \ge 2$. The functions $c(u,v)$ and $c_i(u,v)$
inherit regularity from $f$.
\end{definition}

Recall that an arbitrary one-form $\xi$ on $S^2$ has a unique representation $\xi = r\slashed{\mathcal{D}}_1^\star \left(f,g\right)$ in terms of two unique functions $f$ and $g$ on the unit sphere, both with vanishing mean. We can use this to define an analogous decomposition
for $S^2_{u,v}$ 1-forms on $\mathcal{M}$, we then have

\begin{definition} \label{def:supp1f}
We say a smooth $S^2_{u,v}$ {\bf one-form} $\xi$ on $\mathcal{M}$ 
is {\bf supported on $\ell\ge 2$} if the functions $f$ and $g$ in the unique representation
\[
\xi = r\slashed{\mathcal{D}}_1^\star \left(f,g\right)
\]
are supported on $\ell\ge 2$.
Any such one-form $\xi$ can be uniquely decomposed orthogonally
as 
\[
\xi = \xi_{\ell=1}+\xi_{\ell \ge 2}
\]
where the two scalar functions
\[
(r\slashed{div}\xi_{\ell=1}, r\slashed{curl}\xi_{\ell=1})
\]
are in the span of $(\ref{defYothers})$.
\end{definition}

For symmetric traceless $S^2_{u,v}$ 2-tensors, we have the following:
\begin{proposition}\label{syracelge2}
Let $\xi$ be a smooth symmetric traceless  $S^2_{u,v}$ 2-tensor. 
Then $\xi$ can be uniquely
represented as
\[
\xi=r^2\slashed{\mathcal{D}}^\star_2 \slashed{\mathcal{D}}_1^\star \left(f,g\right)
\]
where  $f$ and $g$ are supported on $\ell\ge 2$. In this sense,
{\bf any
symmetric traceless $2$-tensor on $S^2$ is supported on $\ell\geq 2$.}
\end{proposition}

Proposition~\ref{syracelge2} follows immediately by duality considerations
from the following lemma concerning 
the angular operator 
\begin{equation}
\label{definitionofopT}
\mathcal{T} = r^2 \slashed{\mathcal{D}}^\star_2 \slashed{\mathcal{D}}_1^\star \, ,
\end{equation}
which for fixed $u,v$ can be considered as an operator on the unit sphere\footnote{More precisely, it acts on the round spheres $S^2_{u,v}$ which have been rescaled (this is the reason for the factor $r^2$) to have unit radius.} which maps a pair of functions $\left(f_1(\theta,\phi),f_2(\theta,\phi)\right)$ to a symmetric traceless tensor on $S^2$.
Note that its adjoint, $r^2 \slashed{\mathcal{D}}_1 \slashed{\mathcal{D}}_2$, 
has trivial kernel in $L^2$.

For the computations in the following Lemma we regard $\mathcal{T}$ as an operator defined on pairs of smooth functions, which are dense in $L^2\left(S^2\right)$.

\begin{lemma} \label{lem:kernelexplore}
The kernel of $\mathcal{T}$ is finite dimensional. More precisely, if the pair of functions $\left(f_1, f_2\right)$ is in the kernel, then
\[
f_1 = c Y^{\ell=0}_{m=0} + \sum_{i=-1}^1 c_i Y^{\ell=1}_{m=i} \left(\theta, \phi\right) \ \ \ \textrm{and} \ \ \ f_2 = \tilde{c} Y^{\ell=0}_{m=0} + \sum_{i=-1}^1 \tilde{c}_i Y^{\ell=1}_{m=i} \left(\theta, \phi\right)
\]
for constants $c, c_i, \tilde{c}, \tilde{c}_i$ and $i=-1,0,1$ where $Y^\ell_m$ are the
spherical harmonics defined by $(\ref{defY00})$ and $(\ref{defYothers})$.
\end{lemma}

\begin{proof}
If $\left(f_1, f_2\right)$ is in the kernel then clearly
\[
\int_{S^2} \sin \theta d\theta d\phi \left[ \slashed{\mathcal{D}}^\star_2 \slashed{\mathcal{D}}_1^\star \left(f_1, f_2\right) \cdot \slashed{\mathcal{D}}^\star_2 \slashed{\mathcal{D}}_1^\star \left(f_1, f_2\right)\right] = 0 \, .
\]
Integrating by parts and using that $\slashed{\mathcal{D}}_2 \slashed{\mathcal{D}}_2^\star = -\frac{1}{2} \slashed{\Delta} - \frac{1}{2}K$ (with $K=\frac{1}{r^2}$ the Gauss curvature) as well as $\slashed{\mathcal{D}}^\star_1 \slashed{\mathcal{D}}_1 = -\slashed{\Delta} + K$ and $\slashed{\mathcal{D}}_1 \slashed{\mathcal{D}}_1^\star = -\slashed{\Delta}$, we find
\[
\int_{S^2} \sin \theta d\theta d\phi \left[ \frac{1}{2} f_1 \cdot \slashed{\Delta} \slashed{\Delta} f_1 + \frac{1}{r^2} f_1\slashed{\Delta}f_1 + \frac{1}{2} f_2 \cdot \slashed{\Delta} \slashed{\Delta}f_2 + \frac{1}{r^2} f_2 \slashed{\Delta}f_2\right] = 0
\]
and hence
\[
\int_{S^2} \sin \theta d\theta d\phi \left[ \frac{1}{2} | \slashed{\Delta} f_1 |^2 - \frac{1}{r^2} |\slashed{\nabla}f_1|^2 + \frac{1}{2} | \slashed{\Delta} f_2 |^2 - \frac{1}{r^2} |\slashed{\nabla}f_2|^2 \right] = 0 \, ,
\]
which can be written
\begin{align}
\int_{S^2} \sin \theta d\theta d\phi \Big[  \frac{1}{2} \Big| \slashed{\Delta} f_1 + \frac{2f_1}{r^2} \Big|^2 + \frac{1}{r^2} |\slashed{\nabla}f_1|^2 - \frac{2(f_1)^2}{r^4} 
 + \frac{1}{2} \Big| \slashed{\Delta} f_2 + \frac{2f_2}{r^2} \Big|^2 + \frac{1}{r^2}|\slashed{\nabla}f_2|^2 - 2\frac{(f_2)^2}{r^4} \Big] = 0 \, .
\end{align}
Clearly, the constant solutions $f_1=c$, $f_2=\tilde{c}$ satisfy this (and are obviously in the kernel). If we assume $f_1$ and $f_2$ to both have mean value zero, we see using the Poincar\'e 
inequality on the sphere that the only functions satisfying the above condition are the $\ell=1$ modes. Finally, one checks directly that the $\ell=1$ modes are indeed in the kernel: In components, the equation for $f_1$,
\[
\slashed{\nabla}_A \slashed{\nabla}_B Y^{\ell=1}_m + \slashed{\nabla}_B \slashed{\nabla}_A Y^{\ell=1}_m  = -2\slashed{g}_{AB} Y^{\ell=1}_m \, ,
\]
reads (using $\Gamma^{\theta}_{\phantom{\theta}\phi \phi} = -\sin \theta \cos \theta$ and $\Gamma^{\phi}_{\phantom{\phi}\theta \phi} = \frac{\cos \theta}{\sin \theta}$ in standard coordinates)
\begin{align}
\left(2\partial_\theta^2 + 2\right) Y^{\ell=1}_m &= 0  \\
\partial_\theta \partial_\phi Y^{\ell=1}_m - \frac{\cos \theta}{\sin \theta} \partial_\phi Y^{\ell=1}_m &= 0 \\
 \partial_\phi^2 Y^{\ell=1}_m+ \sin \theta \cos \theta \partial_\theta Y^{\ell=1}_m + \sin^2 \theta Y^{\ell=1}_m &= 0
 \end{align}
 and these identities are easily verified. The computation for $f_2$ is similar or can be inferred by duality.
\end{proof}

In particular, we have the following
\begin{corollary} \label{cor:nomode}
Let $\xi$ be a smooth
symmetric traceless $S^2_{u,v}$ 2-tensor on $\mathcal{M}$. Then
\[
\int_{S^2_{u,v}} \sin \theta d\theta d\phi \slashed{\mathcal{D}}_1 \slashed{\mathcal{D}}_2 \xi \cdot \left(c + c_i Y^{\ell=1}_{m=i}, \tilde{c}+\tilde{c}_i Y^{\ell=1}_{m=i}\right) = 0 \, .
\]
for any choice of constants $c, c_i, \tilde{c}, \tilde{c}_i$.
\end{corollary}

Note that this in particular means that if $\xi$ is a symmetric traceless
$S^2_{u,v}$ $2$-tensor, then the scalars
 $\slashed{div} \slashed{div} \xi$ and $\slashed{curl} \slashed{div} \xi$ are
 supported on $\ell \ge 2$.

\subsubsection{Elliptic estimates and positivity for angular operators on $S^2_{u,v}$-tensors}\label{sec:poinc}
We end with a discussion of
elliptic estimates giving positivity
for various angular operators acting on $S^2_{u,v}$
tensors supported on $\ell \ge2$.

We first give an estimate associated to the operator $\mathcal{T}$ from $(\ref{definitionofopT})$
acting on
pairs of scalar functions supported on $\ell \ge 2$.
\begin{proposition} \label{cor:elliptic12}
Let $\left(f_1, f_2\right)$ be a pair of functions on $S^2_{u,v}$
supported on $\ell \ge 2$. Then we have the elliptic estimate
\[
\sum_{i=0}^2 \int_{S^2_{u,v}} \sin \theta d\theta d\phi  \left[ |r^i \slashed{\nabla}^i f_1|^2 +  |r^i \slashed{\nabla}^i f_2|^2\right]  \lesssim \int_{S^2)_{u,v}} \sin \theta d\theta d\phi | r^2 \slashed{\mathcal{D}}^\star_2 \slashed{\mathcal{D}}_1^\star \left(f_1, f_2\right) |^2 \, .
\]
\end{proposition}
\begin{proof}
This follows immediately revisiting the computation of Lemma~\ref{lem:kernelexplore}.
\end{proof}

We next give identities (in the formulas below, recall that $K=\frac{1}{r^2}$) associated
to  operators acting on symmetric tracesless $2$-tensors and $1$-forms:
\begin{proposition} \label{lem:aist} 
Let $\xi$ be a smooth
symmetric traceless $S^2_{u,v}$ 2-tensor on $\mathcal{M}$. Then
\begin{align} \label{uid1}
 \int_{S^2_{u,v}} \sin \theta d\theta d\phi \left[ | \slashed{\nabla} \xi|^2 + 2K|\xi|^2 \right]= 2\int_{S^2_{u,v}}\sin \theta d\theta d\phi  | \slashed{div} \xi|^2 \, ,
\end{align}
\begin{align} \label{uid2}
\int_{S^2_{u,v}} \sin \theta d\theta d\phi  | \slashed{\mathcal{D}}_2^\star \slashed{div} \xi|^2 = \int_{S^2_{u,v}}\sin \theta d\theta d\phi  \left[ \frac{1}{4} | \slashed{\Delta} \xi |^2  +  K | \slashed{\nabla} \xi |^2 + K^2 |\xi|^2  \right]  \, .
\end{align}
Now let $\eta$ be a smooth $S^2_{u,v}$ one-form on $\mathcal{M}$. Then we have
\begin{align} \label{uid3}
\|\slashed{\mathcal{D}}^\star_2\slashed{\mathcal{D}}^\star_1\slashed{\mathcal{D}}_1\eta\|_{S^2_{u,v}}^2 = \|2 \slashed{\mathcal{D}}_2^\star \slashed{div} \slashed{\mathcal{D}}_2^\star \eta\|_{S^2_{u,v}}^2 + \|2K \slashed{\mathcal{D}}_2^\star \eta\|_{S^2_{u,v}}^2 + 8K\|\slashed{div} \slashed{\mathcal{D}}_2^\star \eta\|_{S^2_{u,v}}^2  .
\end{align}
\end{proposition}
\begin{proof}
See \cite{ChristKlei} for the first and note $\slashed{\mathcal{D}}_2^\star \slashed{div} \xi = \left( -\frac{1}{2} \slashed{\Delta}  + K\right) \xi$ for the second. 
For (\ref{uid3}) observe that
$
\slashed{\mathcal{D}}^\star_2\slashed{\mathcal{D}}^\star_1\slashed{\mathcal{D}}_1\eta =\slashed{\mathcal{D}}_2^\star \left(-\slashed{\Delta} + K \right) \eta=  2 \slashed{\mathcal{D}}_2^\star \slashed{div} \slashed{\mathcal{D}}_2^\star \eta +2K \slashed{\mathcal{D}}_2^\star \eta
$
and integrate the cross-term by parts.
\end{proof}

The identities (\ref{uid1}) and (\ref{uid2}) can be paraphrased as saying that the operator $\mathcal{A}^{[n]}$ defined in (\ref{mathanot}) acting on symmetric traceless $S^2_{u,v}$ 2-tensors is uniformly elliptic and positive definite. The identity (\ref{uid3}), on the other hand, when combined with Proposition~\ref{cor:elliptic12} and the identity (\ref{uid2}) leads to the following corollary, which can be 
thought of as an elliptic estimate
associated to the operator 
$\slashed{\mathcal{D}}^\star_2$ acting on $S^2_{u,v}$ $1$-forms $\eta$ supported on $\ell\ge 2$:

\begin{corollary} \label{cor:etavl1}
Let $\eta$ be a smooth $S^2_{u,v}$ one-form supported on $\ell\ge 2$. 
Then we have 
\[
\sum_{i=0}^3 \int_{S^2} \sin \theta d\theta d\phi   |r^i \slashed{\nabla}^i \eta|^2  \lesssim \int_{S^2} \sin \theta d\theta d\phi | \mathcal{A}^{[2]} \slashed{\mathcal{D}}^\star_2 \eta |^2 \, .
\]
The statement remains true replacing $3$ by $1$ in the sum on the left and removing $\mathcal{A}^{[2]}$ on the right hand side.
\end{corollary}

We also remark at this point already 
\begin{proposition} \label{lem:angularest}
Let $\xi$ be a smooth symmetric traceless $S^2_{u,v}$ $2$-tensor. Then we have the estimate
\begin{align}
- \int_{S^2} \sin \theta d\theta d\phi \left[\left( \slashed{\Delta}- \frac{4}{r^2}  \right) \xi \right]_{AB} \xi^{AB} \geq \frac{6}{r^2} \int_{S^2} \sin \theta d\theta d\phi |\xi|^2 \, .
\end{align}
\end{proposition}

\begin{proof}
We only outline the proof. The desired estimate follows from
\begin{equation}
- \int_{S^2} \sin \theta d\theta d\phi \slashed{\Delta} \xi \cdot \xi =  \int_{S^2} \sin \theta d\theta d\phi |\slashed{\nabla} \xi|^2 \geq \frac{2}{r^2}\int_{S^2} \sin \theta d\theta d\phi |\xi|^2 \, ,
\end{equation}
which holds for any symmetric traceless $S^2_{u,v}$ 2-tensor $\xi$. The latter can in turn be shown by representing the tensor $\xi$ as $\xi=\slashed{\mathcal{D}}_2^\star \slashed{\mathcal{D}}_1^\star \left(f,g\right)$ 
for unique functions $f$ and $g$ supported on $\ell\ge 2$
as in Proposition~\ref{syracelge2}, 
so in particular $-\int_{S^2} \sin \theta d\theta d\phi \slashed{\Delta}f \cdot f \geq 6r^{-2} \int_{S^2} \sin \theta d\theta d\phi |f|^2$ and the same estimate for $g$) and diligently integrating by parts using the properties of spherical harmonics (in particular, their orthogonality).
\end{proof}

\section{The equations of linearised  gravity around Schwarzschild} \label{sec:lineq}
In this section, we will present the equations of linearised gravity around Schwarzschild.
We begin in Section~\ref{hereformality} with a guide to the
formal derivation of this system from the equations
of Section~\ref{VEeqDNGsec}. The complete 
system of linearised gravity is then presented in
Section~\ref{sec:fulleq}.

\subsection{A guide to the formal derivation from the equations of Section~\ref{nseq}}
\label{hereformality}
We give in this section a formal derivation of the system from the equations of Section~\ref{VEeqDNGsec}.
The reader willing to take the system of linear gravity on faith can skip to Section~\ref{sec:fulleq}.

\subsubsection{Preliminaries} \label{sec:formprel}
We first identify the general manifold $\boldsymbol{\mathcal{M}}$ and its coordinates $\left(\boldsymbol{u}, \boldsymbol{v},\boldsymbol{\theta}^1, \boldsymbol{\theta}^2\right)$ of Section \ref{sec:genmfld}
with the interior of the Schwarzschild manifold $\mathcal{M}^\circ$ 
and its underlying Eddington--Finkelstein double null coordinates $(u,v,\theta^1,\theta^2)$.

On $\mathcal{M}^\circ$, we consider a one-parameter family of
Lorentzian metrics $\boldsymbol{g}\left(\epsilon\right)$
\begin{equation}
\label{familynewdef}
\boldsymbol{g}\left(\epsilon\right)
\doteq -4 {\boldsymbol\Omega}^2\left(\epsilon\right)  du dv + \slashed{\boldsymbol{g}}_{CD} \left(\epsilon\right)\left(d\theta^C -  {\boldsymbol{b}}^C \left(\epsilon\right) dv \right) \left(d\theta^D - {\boldsymbol{b}}^D\left(\epsilon\right) dv \right) \, ,
\end{equation}
such that $\boldsymbol{g}(0)=g_S$ expressed  in the Eddington--Finkelstein
double null form,
i.e.
\[
\boldsymbol{\Omega}^2(0)= \Omega^2 = (1-2M/r), \qquad \boldsymbol{b}(0)=0, \qquad
\boldsymbol{\slashed{g}}_{CD}= r^2\gamma_{CD}.
\]

We assume moreover that the family is \underline{smooth in the extended sense};
by this we mean it defines a smooth family of smooth metrics on the manifold $\mathcal{M}$
of Section~\ref{underunderlying}. 

This can be characterized explicitly in Eddington--Finkelstein double null coordinates
as follows: We require that  the function ${\boldsymbol\Omega}^2\left(\epsilon\right)$, the 
symmetric $S^2_{u,v}$ two-tensor $ \slashed{\boldsymbol{g}}_{CD} \left(\epsilon\right)$ and the $S^2_{u,v}$ one-form $\boldsymbol{b}^{C} \left(\epsilon\right)$ are smooth functions of 
the double null Eddington--Finkelstein coordinates on the interior $\mathcal{M}^o$ and that for any $n_1,n_2,n_3 \in \mathbb{N}_0$ and in any spherical coordinate chart the functions
\[
\left(e^\frac{u}{2M} \partial_u \right)^{n_1} \left(\partial_v\right)^{n_2} \left(\partial_{\theta_A}\right)^{n_3} \left( {\boldsymbol\Omega}^2\left(\epsilon\right)  e^{+\frac{u}{2M}}\right)
\]
\[
\left(e^\frac{u}{2M} \partial_u \right)^{n_1} \left(\partial_v\right)^{n_2} \left(\partial_{\theta_A}\right)^{n_3}  \slashed{\boldsymbol{g}}_{CD} \left(\epsilon\right)
\]
\[
\left(e^\frac{u}{2M} \partial_u \right)^{n_1} \left(\partial_v\right)^{n_2} \left(\partial_{\theta_A}\right)^{n_3}  {\boldsymbol{b}}^C\left(\epsilon\right)
\]
extend continuously to the boundary $\mathcal{H}^+$ (in particular, the limit $u \rightarrow \infty$ of the above quantities exists for any fixed $v,\theta_1,\theta_2$). 

We similarly define a function $\boldsymbol{f}$ of the Eddington--Finkelstein coordinates $\left({u}, {v},{\theta}^1, {\theta}^2\right)$ to be \ul{smooth in the extended sense} if $\boldsymbol{f}$
defines a smooth function on $\mathcal{M}$. Again, we can characterize this 
directly by requiring that $f$ 
 is a smooth function of its coordinates on $\mathcal{M}^o$ and moreover
\[
\left(e^\frac{u}{2M} \partial_u \right)^{n_1} \left(\partial_v\right)^{n_2} \left(\partial_{\theta_A}\right)^{n_3} \boldsymbol{f}
\]
extend continuously to the boundary $\mathcal{H}^+$. Symmetric tensorfields and one-forms which are smooth in the extended sense are defined completely analogously. We sometimes use the phrase that a function (or $S^2_{u,v}$ one-form or symmetric traceless $S^2_{u,v}$ 
tensor) \emph{extends regularly to $\mathcal{H}^+$} to mean that it is smooth in the extended sense.

In view of the general discussion in Section \ref{sec:genmfld} associated with (\ref{familynewdef}) is a family of null frames \begin{equation} \label{efframe}
\boldsymbol{\mathcal{N}}_{EF}=\left({\boldsymbol\Omega}^{-1}\left(\epsilon\right) \partial_u, {\boldsymbol\Omega}^{-1} \left(\epsilon\right)\left(\partial_v + \boldsymbol{b}^A\left(\epsilon\right) \partial_{\theta^A}\right), e_1,e_2\right).
\end{equation}
We can hence define the Ricci coefficients and curvature components for the family of metrics with respect to these frames as in Section \ref{sec:rccc} and formally expand them in powers of $\epsilon$.

Note that the frame (\ref{efframe}) does not itself
extend smoothly to the event horizon $\mathcal{H}^+$ in the sense that given a smooth (in the extended sense) function $\boldsymbol{f}$ of the Eddington--Finkelstein coordinates, the expression ${\boldsymbol\Omega}^{-1}\left(\epsilon\right) \partial_u \boldsymbol{f}$ does not extend continuously to $\mathcal{H}^+$. It is easily seen on the other hand that the rescaled frame 
\begin{equation} \label{efframe2}
\boldsymbol{\mathcal{N}}_{EF^\star}=\left({\boldsymbol\Omega}^{-2}\left(\epsilon\right) \partial_u,  \partial_v + \boldsymbol{b}^A\left(\epsilon\right) \partial_{\theta^A}, e_1,e_2\right)
\end{equation}
is smooth in the extended sense in that any element of the frame applied to a smooth (in the extended sense) function produces a smooth (in the extended sense) function.

\subsubsection{Outline of the linearisation procedure}  \label{sec:outlinelin}
We will now linearise the smooth one-parameter family of metrics (\ref{familynewdef}), i.e.~in particular we shall expand the equations of Sections \ref{sec:nse} and \ref{bieq} (with the Ricci and curvature components defined with respect to the family of frames (\ref{efframe})) to first order in $\epsilon$. 

We begin by recalling the derivative operators $\underline{\boldsymbol{D}}, \boldsymbol{D}$ associated with (\ref{familynewdef}) which (when acting on functions) read in coordinates:
\begin{align}
\underline{\boldsymbol{D}}  = \partial_{u} 
\ \ , \ \
{\boldsymbol{D}} = \partial_v + \boldsymbol{b}^A \left(\epsilon\right) \boldsymbol{e}_A
\ \ , \ \ 
\boldsymbol{e}_A = \frac{\partial}{\partial \theta^A} \, .
\end{align}

To formally linearise the full system of equations of Section \ref{nseq}, we invoke the following general notation: Geometric quantities defined with respect to the full metric (\ref{familynewdef}) are written in bold (e.g.~$\boldsymbol \chi$). Their Schwarzschild value (i.e.~the quantity defined with respect to the Schwarzschild metric) is written without any subscript and their linear perturbation with a superscript ``$(1)$". For instance, we write (recall $\Omega^2=1-\frac{2M}{r}$)
\begin{align} \label{expex}
\boldsymbol\Omega &\equiv \Omega + \epsilon \cdot \Olino \, , \nonumber \\
\slashed{\boldsymbol{g}}_{AB} &\equiv \left(\slashed{g}\right)_{AB} + \epsilon \cdot \glin_{AB} \, , \nonumber \\
\boldsymbol \Omega {\bf tr} \boldsymbol\chi &\equiv \left(\Omega tr \chi\right) + \epsilon \cdot \otx = +\frac{2}{r} \left(1-\frac{2M}{r}\right) +\epsilon \cdot \otx \, , \nonumber \\
\boldsymbol \Omega {\bf tr} \underline{\boldsymbol\chi} &\equiv \left(\Omega tr \underline{\chi}\right) + \epsilon \cdot \otxb = - \frac{2}{r} \left(1-\frac{2M}{r}\right) +\epsilon \cdot \otxb \, ,  \\
\boldsymbol\omega:=\boldsymbol\Omega \hat{\boldsymbol\omega} &\equiv \Omega \hat{\omega} + \epsilon \cdot \olin\, , \nonumber \\
\underline{\boldsymbol\omega}:=\boldsymbol\Omega \hat{\boldsymbol\omega} &\equiv \Omega \underline{\hat{\omega}} + \epsilon \cdot \olinb \, , \nonumber \\
\boldsymbol\rho &\equiv -\frac{2M}{r^3} + \epsilon \cdot \rlin \, , \nonumber
\end{align}
which covers all metric, Ricci and curvature coefficients which have non-trivial Schwarzschild values, cf.~(\ref{ssvef}), and
\begin{align}
\boldsymbol{b} \equiv 0 + \epsilon \cdot  \bmlin  \ \ \ , \ \ \ \hat{\boldsymbol\chi} \equiv 0 +\epsilon \cdot \xlin  \ \ \ , \ \ \ \underline{\hat{\boldsymbol\chi}} \equiv 0 +\epsilon \cdot \xblin
 \ \ \ , \ \ \ \boldsymbol\eta \equiv 0 + \epsilon \cdot \elin  \ \ \ , \ \ \ \underline{\boldsymbol\eta} \equiv 0 + \epsilon \cdot \eblin
\end{align}
as well as
\begin{align}
\boldsymbol\alpha \equiv 0 + \epsilon \cdot  \alin  \ \ \ , \ \ \ \underline{\boldsymbol\alpha} \equiv 0 + \epsilon \cdot \ablin \ \ \ , \ \ \ \boldsymbol\beta \equiv 0 + \epsilon \cdot  \blin   \ \ \ , \ \ \ \underline{\boldsymbol\beta} \equiv 0 + \epsilon \cdot \bblin \ \ \ , \ \ \ \boldsymbol\sigma \equiv 0 + \epsilon \cdot \slin 
\end{align}
for the coefficients which have vanishing Schwarzschild values. In the above $\equiv$ means
to first order in $\epsilon$.

The linearised equations are now obtained simply by expanding the equations of Section \ref{nseq} to order $\epsilon$ leading to the equations presented in Section \ref{sec:fulleq}.

To give a non-trivial example, we consider the second equation of (\ref{firstvarf}), which in view of formula for the projected Lie-derivative,
\begin{align}
\left( \boldsymbol{D} \slashed{\boldsymbol{g}}\right)_{AB} =  \partial_v \left( \slashed{\boldsymbol{g}}_{AB} \right) + \left(\boldsymbol{\slashed{\nabla}}_A \boldsymbol{b}_B + \boldsymbol{\slashed{\nabla}}_B \boldsymbol{b}_A \right) \, ,
\end{align} 
with $\partial_v$ acting on the components of $\slashed{\boldsymbol{g}}$ on the right, can be written as
\begin{align}
\partial_v \left( \slashed{\boldsymbol{g}}_{AB}\right) = 2 \boldsymbol\Omega \hat{\boldsymbol\chi}_{AB} + \slashed{\boldsymbol{g}}_{AB} \boldsymbol\Omega {\bf tr} \boldsymbol\chi + 2\left( \boldsymbol{\slashed{\mathcal{D}}}_2^\star \boldsymbol{b} \right)_{AB}- \slashed{\boldsymbol{g}}_{AB} \boldsymbol{\slashed{div}} \boldsymbol{b} \, .
\end{align}
Note that the operator $\partial_v$ on the left hand side coincides with the Schwarzschild differential 
operator $D$ introduced in Section~\ref{sec:ssdiffop}. 
Expanding in terms of powers of $\epsilon$ as above we find
\begin{align} \label{ixcn}
\partial_v \left( \glin_{AB} \right) = 2\Omega \xlin_{AB} + 2\left( \slashed{\mathcal{D}}_2^\star \bmlin\right)_{AB} + \slashed{g}_{AB} \left( \otx - \slashed{div} \bmlin\right) + \Omega tr \chi \cdot \glin_{AB} \, ,
\end{align}
where the unbolded operators are the Schwarzschild differential operators of
Section~\ref{sec:ssdiffop}.
We now decompose
\[
\glin_{AB}=  \glinh_{AB} + \frac{1}{2}\slashed{g}_{AB}  \cdot tr_{\slashed{g}} \glin \, ,
\]
where $ \glinh$ is traceless with respect to the round metric $\slashed{g}=r^2\gamma$. Note this implies to linear order
\begin{equation}
\boldsymbol\det \slashed{\boldsymbol{g}} \equiv  \det \slashed{g} \left(1+ \epsilon \cdot tr_{\slashed{g}} \glin \, \right) \textrm{ \ \ \ and we therefore define $\glinto:=\frac{1}{2} \sqrt{ \slashed{g}} \cdot tr_{\slashed{g}} \glin$,} 
\end{equation}
using the notation $\sqrt{\slashed{g}}:=\sqrt{\det \slashed{g}}$. Finally, upon contracting (\ref{ixcn}) with the inverse $\slashed{g}^{AB}$ we find
\begin{align}
\partial_v \left( tr_{\slashed{g}} \glin \, \right) = 2 \otx - 2\slashed{div} \bmlin \textrm{ \ \ and \ \ }
\sqrt{\slashed{g}} \partial_v \left( \, \glinh_{AB} \left( \sqrt{\slashed{g}}\right)^{-1} \right) &=2\Omega \xlin_{AB} + 2\left(\slashed{\mathcal{D}}_2^\star \bmlin \right)_{AB} \, , \nonumber
\end{align}
leading directly to equations (\ref{stos}) and (\ref{stos2}) of Section~\ref{=lmc} below. 
The linearisation of the first equation
 of (\ref{firstvarf}) is completely analogous.

The remaining equations for the metric components and Ricci coefficients are much simpler to linearise, as they are either scalar equations with spherically symmetric background values, or tensorial equations where the background quantity vanishes, in which case one can simply replace all operators and coefficients in the equation by their Schwarzschild ones (see Section \ref{Sdiffopsandcon}). 

We give two more examples: The non-linear Bianchi equation for $\boldsymbol\rho$ in the $4$-direction, which linearises via
\begin{align}
\partial_v \left(\rho + \epsilon \rlin \, \right) + \epsilon \bmlin^A \frac{\partial}{\partial \theta^A} \left(\rho + \epsilon \rlin \, \right) + \frac{3}{2} \left(\left(\Omega tr \chi\right) +\epsilon \otx \right) \left(\rho + \epsilon \rlin \, \right) =\epsilon \slashed{div} \blin+\mathcal{O}\left(\epsilon^2\right) \,  \nonumber
\end{align}
to produce the linearised equation (\ref{Bianchi4}) of
Section~\ref{=lcc} below and the Bianchi equation for $\beta$ in the $3$-direction, which using the linearisation formula
\[
 \boldsymbol{\slashed{\mathcal{D}}}_1^\star \left(-\boldsymbol \rho, \boldsymbol\sigma\right) \equiv -\frac{\partial}{\partial \theta^A} \left( \rho + \epsilon \cdot \rlin \, \right) +\boldsymbol{\slashed{\epsilon}}^{AB} \frac{\partial}{\partial \theta^B}\left(0+ \epsilon \cdot \slin \right) \equiv \epsilon \slashed{\mathcal{D}}_1^\star \left( -\rlin, \, \slin\right) \, ,
\]
leads to (\ref{Bianchi3}).

\begin{remark}
We emphasise that for the connection coefficients $\boldsymbol\chi$ and $\underline{\boldsymbol\chi}$ we linearise the equations for their tracefree and traceless parts $\hat{\boldsymbol{\chi}}, {\bf tr} \boldsymbol{\chi}$ and $\hat{\underline{\boldsymbol{\chi}}}, {\bf tr }\underline{\boldsymbol\chi}$ (taken with respect to the metric $\boldsymbol{g}$), cf.~(\ref{chieq})--(\ref{endchi}). This is different from linearising the connection coefficients $\underline{\boldsymbol{\chi}}$ and then splitting the resulting tensor into traceless and trace part with respect to the background spherical metric $\slashed{g}$ although the two are of course easily related. Consequently, when writing expressions like $\otx$, it is understood that this is a weighted linearised trace, not taking an actual trace of a tensor $\xli$.
\end{remark}

\subsubsection{Regular quantities at the horizon}
\label{sec:hozreg}

While it is indeed convenient
to perform the linearisation computation in Eddington--Finkelstein coordinates and with respect to the associated frame (\ref{efframe}) as indicated above, one should keep in mind that one will eventually need to consider \emph{regular} quantities. This will require rescaling appropriately some of the linearised quantities near the event horizon.

To understand the correct rescalings, 
as an example, 
consider the connection coefficient $\boldsymbol\chi$. 
Since the metric (\ref{familynewdef}) is smooth in the extended sense with respect to our Eddington--Finkelstein differential structure, it is $e^{\frac{-u}{4M}} \boldsymbol\chi$ which extends smoothly to the event horizon and hence it is the linearisation of this quantitity that one should consider. Equivalently, since $\boldsymbol{\Omega} \left(\epsilon\right)e^{\frac{u}{4M}}$ is a function which is smooth in the extended sense, we should 
consider the linearisation of  $\boldsymbol{\Omega}  \boldsymbol\chi$
near the horizon. 
Decomposing into 
the traceless part $\boldsymbol{\Omega}\boldsymbol{\hat\chi}$ and 
the trace ${\boldsymbol\Omega}{\bf tr}\boldsymbol{\chi}$,
we have
\[
\boldsymbol{\Omega} \hat{\boldsymbol{\chi}} \equiv 0 + \Omega \xlin 
\]
\[
\left(\boldsymbol{\Omega}{\bf tr} {\boldsymbol{\chi}}\right) \equiv \frac{2}{r}\left(1-\frac{2M}{r}\right) + \otx
\]
and hence the weighted linearised quantities $ \Omega \xlin $ as well as $\otx$ are the regular linearised quantities that we eventually have to estimate uniformly up to the horizon. Similarly for the metric component $\boldsymbol\Omega$, since it is $\boldsymbol\Omega e^{+\frac{u}{4M}}$ which extends smoothly to $\mathcal{H}^+$ we need to consider
the linearisation of
\[
\boldsymbol\Omega e^{+\frac{u}{4M}} \equiv \left(\Omega + \Olino\right) e^{+\frac{u}{4M}} = \left(1 + \Olin \right) \Omega e^{+\frac{u}{4M}} = \sqrt{\frac{2M}{r}} e^{\frac{r}{4M}} e^{-\frac{v}{4M}}\left(1+\Olin\right) \, .
\]
and hence the quantity $\Olin$. To give a final example, we have
\[
\frac{\boldsymbol\Omega {\bf tr} \underline{\boldsymbol\chi}}{\boldsymbol\Omega^2} \equiv \frac{\otxb}{\Omega^2} - 2\Olin \frac{\left(\Omega tr \underline{\chi}\right)}{\Omega^2} =   \frac{\otxb}{\Omega^2} +  \frac{4}{r}\Olin 
\]
which extends regularly to $\mathcal{H}^+$, so 
we need to consider 
 $\Omega^{-2} \otxb$ near the horizon. 
 
 The full list of rescaled quantities
is given by $(\ref{regq1})$ in Section~\ref{sec:unknowns} below.

\subsubsection{Aside: The relation between linearisation in the frames $\boldsymbol{\mathcal{N}}_{EF}$ and $\boldsymbol{\mathcal{N}}_{EF^\star}$} \label{appendix:linco}

The reader might wonder about the relation between the evolution equations of Section \ref{sec:fulleq} (which we recall arose from linearising metric,
Ricci and curvature components in the Eddington--Finkelstein frame 
$\boldsymbol{\mathcal{N}}_{EF}$~(\ref{efframe}) as indicated in the previous section) and the equations one would obtain if one defined these components 
using the regular frame $\boldsymbol{\mathcal{N}}_{EF^\star}$, cf.~(\ref{efframe2}), and then linearised the Einstein equations expressed with respect to that frame. 

We collect in this section the formulae allowing oneself to transform the equations from one to another leading to the notion of \emph{linearisation covariance}. To distinguish the components 
defined with respect to the two frames, we will use the subscripts ``$EF$" and ``$EF^\star$" below, which should not be confused with coordinate indices. 

For the curvature components we easily see
\begin{align}
\ablin_{EF^\star} = \Omega^{-2} \ablin_{EF} \ , \ \bblin_{EF^\star} = \Omega^{-1} \bblin_{EF} \ , \ \alin_{EF^\star} = \Omega^{2} \alin_{EF} \ , \ \blin_{EF^\star} = \Omega \blin_{EF}  \ , \  \rlin_{EF^\star} = \rlin_{EF} \ , \ \slin_{EF^\star} = \slin_{EF}.  \nonumber
\end{align}
For the Ricci coefficients we note
\begin{align}
\boldsymbol \chi_{EF^\star} = \boldsymbol\Omega \boldsymbol \chi_{EF} \ \ , \ \  \hat{\boldsymbol\chi}_{EF^\star} = \boldsymbol\Omega \hat{\boldsymbol \chi}_{EF} \ \ , \ \ tr \boldsymbol\chi_{EF^\star} = \boldsymbol\Omega tr \boldsymbol \chi_{EF} \, ,  \\
\underline{\boldsymbol \chi}_{EF^\star} = \boldsymbol\Omega^{-1} \underline{\boldsymbol\chi}_{EF} \ \ , \ \  \underline{\hat{\boldsymbol\chi}}_{EF^\star} = \boldsymbol\Omega^{-1} \underline{\hat{\boldsymbol \chi}}_{EF} \ \ , \ \ tr \underline{\boldsymbol\chi}_{EF^\star} = \boldsymbol\Omega tr \underline{\boldsymbol \chi}_{EF}  
\end{align}
and hence
\begin{align}
\xlin_{EF^\star} = \Omega \xlin_{EF} \ \ \ , \ \ \ \xblin_{EF^\star} = \Omega^{-1} \xblin_{EF} \, ,
\end{align}
\begin{align}
\accentset{(1)}{\left( tr \chi\right)}_{EF^\star} = \otx_{EF} \ \ \ , \ \ \ \accentset{(1)}{\left( tr \underline{\chi}\right)}_{EF^\star}= \Omega^{-2}\otxb_{EF} + \frac{4}{r} \Olin \, .
\end{align}
Also, since $\boldsymbol{\eta}_{EF^\star} = \boldsymbol{\eta}_{EF}$ and $\underline{\boldsymbol{\eta}}_{EF^\star} = \underline{\boldsymbol{\eta}}_{EF}$ we have
\begin{align}
\elin_{EF^\star} = \elin_{EF} \ \ \ , \ \ \ \eblin_{EF^\star} = \eblin_{EF} \ \ \ , \ \ \  \hat{\boldsymbol\omega}_{EF^\star} = 2 \boldsymbol\omega_{EF} = 2 \boldsymbol\Omega \hat{\boldsymbol\omega}_{EF} = \frac{2M}{r^2} + 2\olin_{EF} \ \ \ , \ \ \ \underline{\hat{\boldsymbol{\omega}}}_{EF^\star} = 0 \, ,
\end{align}
Note that the quantity $\olinb_{EF^\star}$ is hence automatically zero. 

Observe also that the linearised components in the frame $EF^\star$ are automatically regular at the horizon. This is of course consistent with our notion of regular quantities introduced in (\ref{regq1}), cf.~Section \ref{sec:hozreg}.
Using the formulae above one may easily reformulate the system of gravitational perturbations of Section \ref{sec:fulleq} as equations for linearised components in the frame $EF^\star$. This yields the same equations as linearising directly the full non-linear equations expressed in the frame $\boldsymbol{\mathcal{N}}_{EF^\star}$ (linearisation covariance).

\subsection{The full set of linearised equations} \label{sec:fulleq}
In the following subsections we present the equations arising from the  formal linearisation
(outlined in Section~\ref{hereformality}) 
of the equations of Section \ref{nseq}. 
These equations are physical space analogues of the 
equations appearing in Chandrasekhar's~\cite{Chandrasekhar}.
We stress that the system can be studied without reference to the full non-linear Einstein equations. In particular, the discussion below can be read independently of the formal derivation in 
Section~\ref{hereformality}.

\subsubsection{The complete list of unknowns} \label{sec:unknowns}

The equations will concern a set of quantities
\begin{align} \label{scollect}
\mathscr{S}=\left(\, \glinh \, , \, \glinto \, , \, \Olino \, , \,  \bmlin\, , \,  \otx \, , \,  \otxb\, , \,  \xlin\, , \, \xblin\, , \,  \eblin \, , \,  \elin \, , \, \olin \, , \,  \olinb \, , \,  \alin \, , \,  \blin \, , \,  \rlin \, , \,  \slin \, , \,  \bblin \, , \,  \ablin \, , \, \Klin \right)
\end{align}
of smooth (to be defined precisely below) functions, $S^2_{u,v}$-vectors and tensors defined
on domains of the Schwarzschild manifold $\left(\mathcal{M}, g\right)$.
Specifically, the quantities
\begin{itemize}
\item $\glinh\, , \,  \xlin\, , \,  \xblin\, , \,  \alin\, , \,  \ablin$ are symmetric trace free $S^2_{u,v}$ $2$-tensors
\item $\bmlin\, , \,  \elin\, , \,  \eblin\, , \,  \blin\, , \,  \bblin$ are $S^2_{u,v}$ $1$-forms 
\item $\Olino\, , \,  \glinto\, , \, \otx\, , \, \otxb\, , \, \olin \, , \, \olinb \, , \, \rlin \, , \, \slin \, , \, \Klin$  are scalar functions.\footnote{The quantity $\slin$ is actually a two-form on $S^2_{u,v}$ which we will identify with its scalar representative.} 
\end{itemize}
We will sometimes bundle some of these quantities and refer to   
\begin{itemize}
\item $\glinh\, , \,  \glinto\, , \,  \Olino\, , \,  \bmlin$ as the \emph{linearised metric components},
\item $\otx\, , \,  \otxb \, , \,  \xlin\, , \, \xblin\, , \,  \elin \, , \, \eblin \, , \,  \olin \, , \,  \olinb$ as the \emph{linearised Ricci coefficients} and 
\item $\alin\, , \,  \blin\, , \,  \rlin\, , \,  \slin\, , \,  \bblin\, , \,  \ablin$ as the \emph{linearised curvature components}.
\end{itemize}
We will also use the notation 
\[
\glin_{AB}\doteq\glinh_{AB}+\slashed g_{AB}(\sqrt{\slashed{g}})^{-1}\glinto.
\]

When we say that  $\mathscr{S}$ is smooth
on a set $\mathcal{D}\subset \mathcal{M}$, 
we mean--following the considerations of Section \ref{sec:hozreg} and our previous definition of smooth in the extended sense--that the above quantities $(\ref{scollect})$
are smooth functions of the Eddington--Finkelstein coordinates on $\mathcal{M}^o\cap\mathcal{D}$ and that the following weighted linearised quantities in $\mathscr{S}$ extend regularly to 
$\mathcal{H}^+\cap\mathcal{D}$:
\begin{gather}\label{regq1}
\glinh \, , \,   \glinto \, , \,  \bmlin \, , \,  \Olino \, , \, \otx \, , \,  \Omega^{-2} \otxb, \Omega \xlin \, , \,  \Omega^{-1} \xblin\, , \,  \elin\, , \,  \eblin\, , \,  \olin \, , \,  \Omega^{-2} \olinb \, , \,
\Omega^2 \alin\, , \,   \Omega \blin\, , \,  \rlin\, , \, \slin\, , \,  \Omega^{-1} \bblin\, , \,  \Omega^{-2} \ablin\, , \,  \Klin. 
\end{gather}
We recall that the latter means for any quantity $q$ from (\ref{regq1}) and any $n_1,n_2,n_3 \in \mathbb{N}_0$ we have that in any spherical coordinate patch
\[
\left(e^{\frac{u}{2M}} \partial_u\right)^{n_1} \left(\partial_v\right)^{n_2} \left(\partial_A\right)^{n_3} q_{\cdot \cdot}
\]
extends continuously to $\mathcal{H}^+$. Here $q_{\cdot \cdot}$ stands for the components of the quantity $q$, so it should be replaced by $q_{BC}$ for the symmetric traceless $S^2_{u,v}$-tensors in (\ref{regq1}), by $q_{B}$ for the $S^2_{u,v}$ one-forms and by $q$ for the scalars.

We say that a smooth $\mathscr{S}$ defined
on $\mathcal{D}\subset \mathcal{M}$
satisfies the equations of \emph{gravitational perturbations around Schwarzschild} (or \emph{linearised gravity})
if  (\ref{stos})--(\ref{Bianchi10}) to be given in the subsections below hold on $\mathcal{M}^o\cap
\mathcal{D}$.

 Note that given a solution of (\ref{stos})--(\ref{Bianchi10}), all quantities
 of $\mathscr{S}$ can in fact be reconstructed from knowing 
 just the ``metric perturbation" $\big(\Olino \, , \, \bmlin \, , \, \glin \, \big)$. 
Nonetheless, we shall view all quantities of $\mathcal{S}$ as unknowns.

\subsubsection{Equations for the linearised metric components}
\label{=lmc}
The equations for the metric components read 
\begin{align} \label{stos} 
\underline{D} \left(\frac{\glinto}{\sqrt{\slashed{g}}}\right)  = \otxb
\qquad , \qquad 
D \left(\frac{\glinto}{\sqrt{\slashed{g}}}\right) = \otx -  \slashed{div}\, \bmlin  \, ,
\end{align}
\begin{align} \label{stos2}
\sqrt{\slashed{g}}\, \underline{D}\left( \frac{\glinh_{AB}}{\sqrt{\slashed{g}}} \right)   =2\Omega\, \xblin_{AB}
\ \ \ ,  \ \ \
\sqrt{\slashed{g}}\, D\left( \frac{\glinh_{AB}}{\sqrt{\slashed{g}}} \right) &=2\Omega\, \xlin_{AB} + 2\left(\slashed{\mathcal{D}}_2^\star \bmlin \right)_{AB}  ,
\end{align}
\begin{align} \label{bequat} 
\partial_u \bmlin^A = 2 \Omega^2\left(\elin^A - \eblin^A\right) \, .
\end{align}
Note that the derivatives $D$ and $\underline{D}$ act on the components of $\glinh$ in (\ref{stos2}), cf.~formula (\ref{lieinc}).

\subsubsection{Equations for the linearised Ricci coefficients}
\label{=lRc}
We start with the equations for the weighted linearised traces of the second fundamental forms:

\begin{align} \label{dtcb}
D \otxb  = \Omega^2 \left( 2 \slashed{div}\, \eblin + 2\rlin + 4 \rho \, \Olin \right) - \frac{1}{2}  \Omega tr \chi \left( \otxb - \otx  \right) ,
\end{align}
\begin{align} \label{dbtc}
\underline{D} \otx  = \Omega^2 \left( 2 \slashed{div}\, {\elin} + 2 \rlin + 4\rho \, \Olin \right) - \frac{1}{2}  \Omega tr \chi \left( \otxb - \otx  \right) ,
\end{align}

\begin{align} \label{uray}
D \otx = - \left(\Omega tr \chi\right)\otx + 2 \omega \otx  + 2  \left(\Omega tr \chi \right) \olin ,
\end{align}
\begin{align} \label{vray}
\underline{D} \otxb = - \left(\Omega tr \underline{\chi}\right) \otxb  + 2 \underline{\omega} \otxb + 2  \left(\Omega tr \underline{\chi} \right) \olinb .
\end{align}
For the traceless parts we have
\begin{equation} \label{tchi} 
\begin{split}
\slashed{\nabla}_3  \left(\Omega^{-1} \xblin  \right)  +  \Omega^{-1} \left(tr \underline{\chi}\right) \xblin = -\Omega^{-1} \ablin \, , \\
\slashed{\nabla}_4  \left(\Omega^{-1} \xlin \right)  +  \Omega^{-1} \left(tr{\chi}\right) \xlin= -\Omega^{-1} \alin   \, ,
\end{split}
\end{equation}
\begin{align} \label{chih3}
\slashed{\nabla}_3  \left(\Omega \xlin \right)  + \frac{1}{2} \left(\Omega tr \underline{\chi}\right) \xlin + \frac{1}{2} \left( \Omega tr \chi\right) \xblin  &= -2 \Omega \slashed{\mathcal{D}}_2^\star \elin \, , \\
\slashed{\nabla}_4  \left(\Omega \xblin  \right) + \frac{1}{2} \left(\Omega tr \chi \right) \xblin  + \frac{1}{2} \left( \Omega tr \underline{\chi}\right) \xlin &= -2 \Omega \slashed{\mathcal{D}}_2^\star \eblin \, . \label{chih3b}
\end{align}
For $\elin, \eblin$ the equations read
\begin{align} \label{propeta}
\slashed{\nabla}_3 \eblin =  \frac{1}{2} \left(tr \underline{\chi}\right) \left( \elin - \eblin\right)  + \bblin
\textrm{ \ \ \ \ , \ \ \ \ }
\slashed{\nabla}_4 \elin =  -  \frac{1}{2} \left( tr {\chi}\right) \left( \elin - \eblin\right) - \blin .
\end{align}
The equations for the linearised lapse and its derivatives are given by 
\begin{align} \label{oml1}
D \olinb = -\Omega \left(\rlin + 2 \rho \Olin \right) \, ,
\end{align}
\begin{align} \label{oml2}
\underline{D}\olin = -\Omega \left(\rlin + 2 \rho \Olin \right) \, ,
\end{align}
\begin{align} \label{oml3}
\olin = {D} \left( \Olin \right) \textrm{ \ , \ }  \olinb = \underline{D} \left(\Olin\right) \textrm{ \  , \ }  \elin_A + \eblin_A = 2 \slashed{\nabla}_A \left(\Olin\right).
\end{align}
Finally we have the linearised Codazzi equations
\begin{equation}
\begin{split}
\slashed{div} \xblin = -\frac{1}{2} \left(tr \underline{\chi}\right)  \elin + \bblin + \frac{1}{2\Omega} \slashed{\nabla} \otxb , \\
\slashed{div} \xlin = -\frac{1}{2} \left( tr {\chi}\right) \eblin  -\blin + \frac{1}{2\Omega} \slashed{\nabla} \otx \label{ellipchi} \, ,
\end{split}
\end{equation}
and
\begin{align} \label{curleta}
\slashed{curl} \elin = \slin \ \ \ , \ \ \ \slashed{curl} \eblin = -\slin \, ,
\end{align}
as well as the linearised Gauss equation
\begin{equation} \label{lingauss}
\Klin = -\rlin - \frac{1}{4} \frac{tr {\chi}}{\Omega}\left( \otxb - \otx  \right) +\frac{1}{2}\Olin \left(tr \chi tr \underline{\chi} \right) \, .
\end{equation}

\subsubsection{Equations for linearised curvature components}
\label{=lcc}
We complete the system of linearised gravity with the linearised Bianchi equations:
\begin{align}
\slashed{\nabla}_3 \alin + \frac{1}{2} tr \underline{\chi}\alin + 2 \underline{\hat{\omega}} \alin &= -2 \slashed{\mathcal{D}}_2^\star \blin - 3 \rho\, \xlin \, ,  \label{Bianchi1} \\
\slashed{\nabla}_4 \blin + 2 (tr \chi) \blin - \hat{\omega} \blin &= \slashed{div}\, \alin \, , \label{Bianchi2} \\
\slashed{\nabla}_3 \blin + (tr \underline{\chi}) \blin + \underline{\hat{\omega}} \blin &= \slashed{\mathcal{D}}_1^\star \left(-\rlin \, , \, \slin \, \right) + 3\rho \, \elin \, ,   \label{Bianchi3}
\end{align}
\begin{align}
\slashed{\nabla}_4 \rlin + \frac{3}{2} (tr \chi) \rlin = \slashed{div}\, \blin - \frac{3}{2} \frac{\rho}{\Omega}  \otx \, , \label{Bianchi4}
\end{align}
\begin{align}
\slashed{\nabla}_3 \rlin + \frac{3}{2} (tr \underline{\chi}) \rlin = -\slashed{div}\, \bblin - \frac{3}{2} \frac{\rho}{\Omega} \otxb \, , \label{Bianchi5}
\end{align}
\begin{align}
\slashed{\nabla}_4 \slin + \frac{3}{2} (tr \chi) \slin&= -\slashed{curl}\, \blin \, , \label{Bianchi6} \\
\slashed{\nabla}_3 \slin + \frac{3}{2} (tr \underline{\chi}) \slin &= -\slashed{curl}\, \bblin \, , \label{Bianchi7} \\
\slashed{\nabla}_4 \bblin + (tr \chi)  \bblin + \hat{\omega} \bblin &= \slashed{\mathcal{D}}_1^\star \left(\rlin \, ,  \, \slin \, \right) - 3 \rho\, \eblin  \, ,  \label{Bianchi8} \\
\slashed{\nabla}_3 \bblin + 2 (tr \underline{\chi})  \bblin - \hat{\underline{\omega}} \bblin &= - \slashed{div}\, \ablin \, , \label{Bianchi9} \\
\slashed{\nabla}_4 \ablin + \frac{1}{2} (tr \chi) \ablin + 2 \hat{\omega} \ablin &=  2 \slashed{\mathcal{D}}_2^\star \bblin - 3 \rho\, \xblin \, .  \label{Bianchi10}
\end{align}
Note the coupling of the linearised Bianchi system
with the linearised Ricci coefficients
$\xlin$, $\elin$, $\eblin$ and $\xblin$, remarked already in
Section~\ref{LGarSintro}. (For comparison, recall that these terms do not arise in the Minkowski case
since the background curvature components vanish, in particular $\rho=0$.)

\section{Special solutions: pure gauge and linearised Kerr} \label{sec:specialsol}
In this section, we shall look at two
classes of special solutions of the system of linearised gravity given in
Section~\ref{sec:fulleq}. 
We begin in Section~\ref{sec:ssgauge} with a discussion of 
\emph{pure gauge solutions} followed by a presentation of a
$4$-dimensional family of  \emph{reference linearised Kerr solutions} in Section~\ref{4dimfam}.

\subsection{Pure gauge solutions $\mathscr{G}$}\label{sec:ssgauge}
As described already in Section~\ref{PGintro}, pure gauge solutions are those derived from
linearising the families of metrics that arise from applying to Schwarzschild 
smooth $1$-parameter families of  coordinate transformations which
preserve the double null form $(\ref{metricdn})$ of the metric.
We will classify such solutions here, deriving them from the formal linearisation
of Section~\ref{hereformality}
in order to best illustrate their geometric significance. (The reader can
alternatively simply directly verify that these 
solutions satisfy the system of linearised gravity of Section~\ref{sec:fulleq}.)

\subsubsection{General computations and formal developments} \label{sec:gencomp}

Let us fix $\Omega^2 \left(u,v\right) \cdot f_1 \left(u,\theta,\phi\right)$, $f_2 \left(v,\theta,\phi\right)$, $j_3\left(v,\theta,\phi\right)$, $j_4\left(v,\theta,\phi\right)$ functions on $\mathcal{M}^\circ$
extending smoothly to $\mathcal{M}$.

Consider a smooth one-parameter family of coordinates on $\mathcal{M}^\circ$ defined by
\begin{equation} \label{cotrafo}
\begin{split}
\tilde{\boldsymbol{u}} = \tilde{\boldsymbol{u}}_\epsilon  &:= u + \epsilon f_1 \left(u,\theta,\phi\right) \, ,  \\
\tilde{\boldsymbol{v}} = \tilde{\boldsymbol{v}}_\epsilon &:= v + \epsilon f_2 \left(v,\theta,\phi\right) \, , \\
\tilde{\boldsymbol\theta} = \tilde{\boldsymbol\theta}_\epsilon &:=\theta + \epsilon f_3\left(u,v, \theta,\phi\right) =  \theta +  \epsilon \frac{2}{r\left(u,v\right)} \left(f_2 \right)_\theta \left(v,\theta,\phi\right) + \epsilon j_3\left(v,\theta,\phi\right) \, , \\
\tilde{\boldsymbol\phi} = \tilde{\boldsymbol\phi}_\epsilon  &:= \phi +\epsilon f_4\left(u,v, \theta,\phi\right) = \phi +  \epsilon \frac{2}{r\left(u,v\right)} \frac{1}{\sin^2\theta} \left( f_2 \right)_\phi \left(v,\theta,\phi\right) +\epsilon j_4\left(v,\theta,\phi\right) \, .
\end{split}
\end{equation}

If we express the Schwarzschild metric in the form $(\ref{ssef})$:
\begin{equation} \label{schm} 
\boldsymbol{g}_S = -4\Omega^2  \left(\tilde{\boldsymbol{u}},\tilde{\boldsymbol{v}}\right) d\tilde{\boldsymbol{u}} d\tilde{\boldsymbol{v}}+ r^2 \left(\tilde{\boldsymbol{u}},\tilde{\boldsymbol{v}}\right) \left[ d\tilde{\boldsymbol\theta}^2 + \sin^2 \tilde{\boldsymbol\theta} d\tilde{\boldsymbol\phi}^2 \right] \, 
\end{equation}
where $\Omega^2$ and $r^2$ are $\gamma$ are defined by the expressions $(\ref{officialOmegadef})$,
$(\ref{soimplicit})$ and $(\ref{gammaexplicit})$, 
where however $u$, $v$, $\theta$, $\phi$ are replaced by  the new coordinates
$\tilde{\boldsymbol{u}}$, $\tilde{\boldsymbol{v}}$, $\tilde{\boldsymbol\theta}$,
$\tilde{\boldsymbol\phi}$,
then, in view of $(\ref{cotrafo})$, this defines with respect to the original 
fixed coordinates $u$, $v$, $\theta$, $\phi$ of Section~\ref{thedefofEF} a 
one-parameter family of metrics, whose first order $\epsilon$-dependence  can be expressed as 
\begin{align} \label{trafom} 
\boldsymbol{g}_S \left(\epsilon\right) \equiv & \ dudv \left[ -4 \Omega^2 + \epsilon \left( -\frac{8M}{r^2} \Omega^2 \left[f_2-f_1\right] - 4 \Omega^2 \left(f_2\right)_v -4 \Omega^2 \left(f_1\right)_u \right) \right] 
\nonumber \\ 
&+ dv d\theta \left(-4\Omega^2 \epsilon (f_1)_\theta + 2r^2 \epsilon (f_3)_v \right) 
+ dv d\phi \left(-4\Omega^2 \epsilon \left(f_1\right)_\phi + 2r^2 \sin^2\theta \epsilon (f_4)_v\right) 
\nonumber \\
&+ d\theta d\theta \left(r^2 + 2r\Omega^2 \epsilon \left[f_2-f_1\right] + 2r^2 \epsilon (f_3)_\theta\right) 
+ d\theta d\phi \left( 2\epsilon r^2 (f_3)_\phi + 2\epsilon r^2 \sin^2 \theta (f_4)_\theta  \right) 
\nonumber \\
&+ \sin^2 \theta d\phi d\phi \left[ r^2 + 2r\Omega^2 \epsilon \left[f_2-f_1\right] + 2r^2 \epsilon \left(f_4\right)_\phi + 2r^2 \frac{\cos \theta}{\sin \theta} \epsilon f_3 \right] .
\end{align}

Here, as before, ``$\equiv$" indicates that we are ignoring terms of order $\epsilon^2$ or higher and the subscripts ``$v$", ``$\theta$" etc.~indicating a partial derivative with respect to this variable. 
The smoothness assumptions on $f_1$, $f_2$, $j_3$ and $j_4$ ensure that
this defines in fact a smooth $1$-parameter family of metrics on $\mathcal{M}$, i.e.~including
the boundary $\mathcal{H}^+$.

Note that the right hand side of (\ref{trafom}) 
is of the double-null form (\ref{familynewdef}) with $\boldsymbol\Omega^2 \left(0\right)=\Omega^2$, $\boldsymbol{b}\left(0\right)=0$ and $\boldsymbol{\slashed{g}} \left(0\right)=\slashed{g}$. 
Since the right hand side of $(\ref{trafom})$ thus defines
a family diffeomorphic to Schwarzschild to first order in $\epsilon$,
it in particularly satisfies the Einstein equations
$(\ref{Evackiedw})$ to first order, and thus, still
gives rise to a solution of
linearised gravity, which can be read off as in Section~\ref{sec:outlinelin} 
from (\ref{trafom}) by collecting the terms at $\mathcal{O}\left(\epsilon\right)$.

Specifically, from (\ref{trafom}) we can read off
\begin{equation} \label{mpe1}
\bmlin^\theta =  \frac{1}{r^2} \left(2\Omega^2 (f_1)_\theta - r^2 (f_3)_v\right) \ \ \ \ , \ \ \ \ \bmlin^\phi = \frac{1}{r^2} \left(2\Omega^2 \frac{(f_1)_\phi}{\sin^2 \theta} - r^2 \left(f_4\right)_v\right) \, ,
\end{equation}
\begin{equation} \label{mpe2}
2\Olin =  \left(f_2\right)_v +  \frac{2M}{r^2} f_2 +  \left(f_1\right)_u -  \frac{2M}{r^2} f_1 \, ,
\end{equation}
\begin{equation} \label{mpe3}
\frac{\glinto}{\sqrt{\slashed{g}}} =  \frac{2 \Omega^2}{r} \left(f_2-f_1\right) +  \frac{1}{\sin \theta} \partial_\theta \left(\sin \theta f_3\right) +  (f_4)_\phi \, ,
\end{equation}
\begin{align} \label{mpe4}
\glinh_{\theta \theta} &= 2r^2 \left(f_3\right)_\theta - \frac{r^2}{\sin \theta} \partial_\theta \left(\sin \theta f_3\right) - r^2 \left(f_4\right)_\phi \, , \nonumber \\
\glinh_{\theta \phi} &= r^2 \left(f_3\right)_\phi + r^2 \sin^2 \theta \left(f_4\right)_\theta \, ,  \\
\sin^{-2} \theta \glinh_{\phi \phi} &= 2r^2 \left(f_4\right)_\phi + 2r^2 \frac{\cos \theta}{\sin \theta} f_3 - \frac{r^2}{\sin \theta} \partial_\theta \left(\sin \theta f_3\right) - r^2 \left(f_4\right)_\phi \nonumber
\, .
\end{align}

Note that since $\Omega^2 \left(u,v\right) \cdot f_1 \left(u,\theta,\phi\right)$, $f_2 \left(v,\theta,\phi\right)$, $j_3\left(v,\theta,\phi\right)$, $j_4\left(v,\theta,\phi\right)$ are functions smooth in the extended sense on $\mathcal{M}$, the perturbation $\left(\Olin \, , \, \bmlin \, , \, \glin \, \right)$ is smooth on $\mathcal{M}$.

All geometric quantities can now be computed from the above using the system of gravitational perturbations. For future reference we collect here the formulae for 
\begin{align} \label{otc}
\otx &= \partial_v \left(\frac{\glinto}{\sqrt{\slashed{g}}} \right) + \slashed{div} \bmlin = 
\partial_v \left(\frac{2\Omega^2}{r} \left(f_2-f_1\right) \right) + \frac{2\Omega^2}{r^2} \Delta_{S^2} f_1 \nonumber \\
&= \frac{2\Omega^2}{r} \partial_v f_2 + \frac{\Omega^2}{r^2} \left[ \left(2-4\Omega^2\right) \left(f_2-f_1\right) + 2 \Delta_{S^2} f_1\right] \, ,
\end{align}
and
\begin{align} \label{otc2}
\otxb &= \partial_u \left(\frac{\glinto}{\sqrt{\slashed{g}}}\right)  = 
\partial_u \left(\frac{2\Omega^2}{r} \left(f_2-f_1\right) \right) + \frac{2\Omega^2}{r^2} \Delta_{S^2} f_2 \nonumber \\
&= \frac{2\Omega^2}{r} \left(-\partial_u f_1\right) - \frac{\Omega^2}{r^2} \left[ \left(2-4\Omega^2\right) \left(f_2-f_1\right) - 2 \Delta_{S^2} f_2\right] \, ,
\end{align}
which are easily determined from (\ref{stos}). We conclude

\begin{proposition} \label{lem:pgs}
Let $\Omega^2 \cdot f_1 \left(u,\theta,\phi\right)$, $f_2 \left(v,\theta,\phi\right)$, $j_3\left(v,\theta,\phi\right)$, $j_4\left(v,\theta,\phi\right)$ be smooth functions on $\mathcal{M}$ and $f_3\left(u,v, \theta,\phi\right)$, $f_4\left(u,v, \theta,\phi\right)$ be defined through (\ref{cotrafo}). Then the metric perturbation with $\Olin$ defined as in (\ref{mpe2}), $\bmlin^A$ defined as in (\ref{mpe1}) and $\glin$ defined as in (\ref{mpe3}), (\ref{mpe4}) determines a smooth solution of the system of gravitational perturbations on $\mathcal{M}$. 

We call such a solution a {\bf pure gauge solution} and denote it by $\mathscr{G}$ or, to indicate how it is generated, by $\mathscr{G}\left(f_1,f_2,j_3,j_4\right)$.
\end{proposition}

Note that non-trivial $\left(f_1,f_2,j_3,j_4\right)$ can generate the trivial solution $\mathscr{G}=0$. For instance choosing $f_1=1$, $f_2=1$ and $j_3=j_4=0$ generate the zero solution. 

The validity of the above Proposition is clear by the above formal computations. One can also check explicitly that all equations of the system of gravitational perturbations (\ref{stos})--(\ref{Bianchi10}) are satisfied.

In the next three subsections we will look at the basic building blocks of pure gauge solutions arising from Proposition \ref{lem:pgs}. Specifically, we compute explicitly all Ricci coefficients and curvature quantities of three special gauge solutions produced by Proposition \ref{lem:pgs}:
\begin{itemize}
\item  pure gauge solutions arising from setting $\left(f_1=0,f_2=f\left(v,\theta,\phi\right),j_3=0,j_4=0\right)$:  Lemma \ref{lem:exactsol}
\item  pure gauge solutions arising from setting $\left(f_1=f\left(u,\theta,\phi\right),f_2=0,j_3=0,j_4=0\right)$: Lemma \ref{lem:exactsol2}
\item pure gauge solutions arising from setting $\left(f_1=0,f_2=0,j_3\left(v,\theta,\phi\right),j_4\left(v,\theta,\phi\right)\right)$: Lemma \ref{lem:exactsol3}
\end{itemize}
In view of linearity, the general pure gauge solution can be obtained from summing the three special ones.

We will also include the computation that all equations of (\ref{stos})--(\ref{Bianchi10}) are indeed satisfied for Lemma \ref{lem:exactsol} since we have omitted the lengthy but straightforward proof of Proposition \ref{lem:pgs}.

\subsubsection{Pure gauge solutions of the form $\left(f_1=0,f_2=f\left(v,\theta,\phi\right),j_3=0,j_4=0\right)$} \label{sec:famexplicit}

The following solution is the explicit form of the pure gauge solution $\left(f_1=0,f_2=f\left(v,\theta,\phi\right),j_3=0,j_4=0\right)$ arising from Proposition \ref{lem:pgs}:

\begin{lemma} \label{lem:exactsol} 
For any smooth function $f=f\left(v,\theta,\phi\right)$, the following is a pure gauge solution of the system of gravitational perturbations:
\begin{align}
2\Olin &= \frac{1}{\Omega^2} \partial_v \left(f \Omega^2\right)  , & \glinh&= - \frac{4}{r} r^2 \slashed{\mathcal{D}}_2^\star \slashed{\nabla}_A f  \ , &\frac{\glinto}{\sqrt{\slashed{g}}} &= \frac{2\Omega^2 f}{r} + \frac{2}{r} r^2 \slashed{\Delta}f  , \nonumber \\
\bmlin &= -2r^2 \slashed{\nabla}_A \left[ \partial_v \left(\frac{f}{r}\right)\right] , & \elin &= \frac{\Omega^2}{r^2} r \slashed{\nabla} f  , & \eblin &= \frac{1}{\Omega^2} r \slashed{\nabla} \left[\partial_v \left(\frac{\Omega^2}{r}f\right) \right]   \nonumber \, ,
\nonumber \\
\xblin &= -2\frac{\Omega}{r^2} r^2 \slashed{\mathcal{D}}_2^\star \slashed{\nabla} f  , & \otx &= 2 \partial_v \left(\frac{f \Omega^2}{r}\right)  , & \otxb &=  2\frac{\Omega^2}{r^2} \left[\Delta_{\mathbb{S}^2} f - f \left(1-2\Omega^2\right) \right]  , \nonumber \\
\rlin &= \frac{6M \Omega^2}{r^4} f  , & \bblin&= \frac{6M\Omega}{r^4}  r \slashed{\nabla} f , & \Klin &= -\frac{\Omega^2}{r^3}\left(\Delta_{\mathbb{S}^2} f + 2f\right) \nonumber
\end{align}
and
\[
\xlin = \alin = \ablin = 0 \ \ \ , \ \ \ \blin = 0 \ \ \ , \ \ \ \slin = 0 \nonumber \, .
\]
We will call $f$ a gauge function. 
\end{lemma}

\begin{remark}
Note that $\elin$, $\rlin$ and $\otx$ vanish on the horizon $\mathcal{H}^+$.
\end{remark}

\begin{remark} \label{rem:fv}
In Section~\ref{NEOKEF}, we will introduce the notion of asymptotically flat seed data (Definition \ref{def:afpeel}). In this language, note that the above pure gauge solution induces asymptotically flat seed data with weight $s$ to order $0$ on $C_{u_0} \cap C_{v_0}$, 
provided the gauge function $f$ satisfies  $|(r \slashed{\nabla})^n f | \lesssim v$ for $n\leq 2$, $|(r \slashed{\nabla})^n r^2 \partial_v \left(\frac{f}{r}\right)| \lesssim 1$ for $n\leq 1$ and $|r^{1+s} \partial_v \left( r^2 \partial_v \left(\frac{f}{r}\right)\right)\big| \lesssim 1$. In particular, $f$ does not have to be bounded. It is also easy to see what higher order assumptions on $f$ need to be imposed to guarantee that the seed data are asymptotically flat to higher order.
\end{remark}

\begin{proof}[Proof of Lemma \ref{lem:exactsol}] \label{sec:nolapem}
We verify some of the null stucture and all of the Bianchi equations below leaving the remaining equations to the reader. The left hand side of the (renormalised) (\ref{dtcb}) is
\begin{align}
D \left[r\otxb \right] = 2 \Delta_{\mathbb{S}^2} \left[ \partial_v \left( \frac{\Omega^2}{r}f\right)\right] - 2\partial_v \left(\frac{\Omega^2 \left(1-2\Omega^2\right)}{r}f \right) \, ,
\end{align}
while the right hand side is
\begin{align}
2 \Omega^2 r \left( \frac{1}{\Omega^2 r} \Delta_{\mathbb{S}^2} \left[  \partial_v \left( \frac{\Omega^2}{r}f\right) \right]  + \frac{6M}{r^4} \Omega^2 f -\frac{2M}{\Omega^2 r^3} \partial_ v \left(f \Omega^2\right) \right) + r \frac{\Omega^2}{r} 2 \partial_v \left(\frac{f \Omega^2}{r}\right) \nonumber \, .
\end{align}
The term involving the angular Laplacian cancels and since
\begin{align}
 - 2\partial_v \left(\frac{\Omega^2 \left(1-2\Omega^2\right)}{r}f \right) &= +2 \partial_v \left(\frac{\Omega^2}{r} f \right) - \partial_v \left( \frac{8M}{r^2} \Omega^2 f \right) \nonumber \\
 &= 2\Omega^2 \partial_v \left(\frac{\Omega^2}{r} f \right) + \frac{4M}{r} \partial_v \left(\frac{\Omega^2}{r} f \right)  - \partial_v \left( \frac{8M}{r^2} \Omega^2 f \right) \nonumber \\
& = 2\Omega^2 \partial_v \left(\frac{\Omega^2}{r} f \right) - \frac{4M}{r^2} \partial_v \left( f \Omega^2 \right) + \frac{12M}{r^3 } \Omega^4 f  \nonumber  \, ,
\end{align}
we have established that (\ref{dtcb}) holds. Let us check the two Codazzi equations. The one involving $\beta$ can be read off directly while the one involving $\underline{\beta}$ reads
\begin{align} \label{coval} 
+2 \Omega \slashed{\mathcal{D}}_2 \slashed{\mathcal{D}}_2^\star \slashed{\mathcal{D}}_1^\star \left( f , 0 \right) = 
+ \frac{\Omega}{r}  \frac{\Omega^2}{r^2} r \slashed{\nabla}_A f + \frac{6M\Omega}{r^4}  r \slashed{\nabla}_A f + \frac{1}{2\Omega} 2 \frac{\Omega^2}{r^2} \slashed{\nabla}_A \left[\Delta_{\mathbb{S}^2} f - f \left(1-2\Omega^2\right) \right] \, .
\end{align}
Now use the fact that
\begin{align}
\slashed{\mathcal{D}}_2 \slashed{\mathcal{D}}_2^\star \slashed{\mathcal{D}}_1^\star \left( f , 0 \right) = \left( -\frac{1}{2} \slashed{\Delta} - \frac{1}{2} K \right) \slashed{\mathcal{D}}_1^\star \left( f , 0 \right)
=\frac{1}{2} \slashed{\mathcal{D}}_1^\star \slashed{\mathcal{D}}_1 \slashed{\mathcal{D}}_1^\star \left( f , 0 \right) - K \slashed{\mathcal{D}}_1^\star \left( f , 0 \right) = \slashed{\nabla}_A \slashed{\Delta} f + \frac{1}{r^2} \nabla_A f
\end{align}
to validate (\ref{coval}). To validate (\ref{vray}), note that
\[
\partial_u \left[\otxb r^2 \Omega^{-2} \right] = -\frac{8M}{r^2} f \Omega^2 = -4 r \frac{1}{2} f \partial_u \left(\frac{2M}{r^2} \right) = -4r \olinb .
\]
The Bianchi equations (\ref{Bianchi1}) and (\ref{Bianchi2}) are trivially satisfied, (\ref{Bianchi3}) is easily checked by inspection. For (\ref{Bianchi4}) note that
\[
\partial_v \left(r^3 \rlin \, \right) = \partial_v \left(r^3 \frac{6M}{r^4} \Omega^2 f\right) = \frac{3M}{\Omega} 2 \partial_v \left(\frac{f\Omega^2}{r}\right) = \frac{3M}{\Omega r^3} r^3 \otx \, .
\]
For (\ref{Bianchi5}) note similarly
\[
\partial_u \left(r^3 \rlin \, \right) = f \partial_u \left(r^3 \frac{6M}{r^4} \Omega^2 \right) = -6M \Omega\left(\Delta_{\mathbb{S}^2} f \right) + \frac{3M}{\Omega} 2\frac{\Omega^2}{r^2} \left[\Delta_{\mathbb{S}^2} f - f \left(1-2\Omega^2\right) \right] \, ,
\]
which is readily verified through the identity $\partial_u \left(\frac{6M \Omega^2}{r}\right) = 6M \Omega^2 \left(\frac{\Omega^2}{r^2} - \frac{2M}{r^3}\right)$.
The Bianchi equations (\ref{Bianchi6}) and (\ref{Bianchi7}) hold trivially and so does (\ref{Bianchi10}). The equation (\ref{Bianchi9}) is also easily verified using that $f$ does not depend on $u$. It remains to verify (\ref{Bianchi8}). We renormalise it to
\[
\slashed{\nabla}_4 \left(r^2 \Omega \, \bblin \, \right) = -r^2 \Omega \slashed{\nabla}_A \rlin + \frac{6M}{r} \Omega \, \eblin \, .
\]
The left hand side is 
\[
\slashed{\nabla}_4 \left(r^2 \Omega \, \bblin \, \right) = \slashed{\nabla}_4 \left(\frac{6M \Omega}{r^2} r \slashed{\nabla}_A f\right) = \frac{6M r}{\Omega} \slashed{\nabla}_A \partial_v \left( \frac{f\Omega}{r^2}\right) \, ,
\]
while the right hand side is
\[
-r^2 \Omega \slashed{\nabla}_A \rlin + \frac{6M}{r} \Omega \, \eblin = -\frac{6M}{r^2}\Omega \left(1-\frac{2M}{r}\right) \slashed{\nabla}_A f + \frac{6M}{r} \frac{1}{\Omega^2} r \slashed{\nabla}_A \left[\partial_v \left(\frac{\Omega^2}{r}f\right) \right] \, .
\]
Equation (\ref{Bianchi8}) is now immediate.
\end{proof}

\subsubsection{Pure gauge solutions of the form $\left(f_1=f\left(u,\theta,\phi\right), f_2=0,j_3=0,j_4=0\right)$} \label{sec:famexplicit2}

Completely analogously one proves the following Lemma, which provides the explicit form of the pure gauge solution $\left(f_1=f\left(u,\theta,\phi\right),f_2=0,j_3=0,j_4=0\right)$ arising from Proposition \ref{lem:pgs}:

\begin{lemma} \label{lem:exactsol2} 
For any function $f=f\left(u,\theta,\phi\right)$, such that $\Omega^2 \cdot f\left(u,\theta,\phi\right)$ is smooth in the extended sense on $\mathcal{M}$ the following is a (pure gauge) solution of the system of gravitational perturbations 
\begin{align}
 2\Olin &= \frac{1}{\Omega^2} \partial_u \left(f \Omega^2\right), \, & \frac{\glinto}{\sqrt{\slashed{g}}} &= -\frac{2\Omega^2 f}{r} , & \bmlin &= 2\Omega^2 \slashed{\nabla} f, \nonumber \\
\xlin &= -2\frac{\Omega}{r^2} r^2 \slashed{\mathcal{D}}_2^\star \slashed{\nabla} f, & \elin &= \frac{1}{\Omega^2} r \slashed{\nabla} \left[\partial_u \left(\frac{\Omega^2}{r}f\right) \right] , & \eblin &= -\frac{\Omega^2}{r^2} r \slashed{\nabla} f , \nonumber \, \\
\otxb &= -2 \partial_u \left(\frac{f \Omega^2}{r}\right) , \ & \otx &= 2\frac{\Omega^2}{r^2} \left[\Delta_{\mathbb{S}^2} f - f \left(1-2\Omega^2\right) \right] , 
\nonumber  \\
\blin &= -\frac{6M\Omega}{r^4}  r \slashed{\nabla} f , &  \rlin &= -\frac{6M \Omega^2}{r^4} f  \, , & \Klin &= +\frac{\Omega^2}{r^3} \left(\Delta_{\mathbb{S}^2} f + 2 f\right) \nonumber
\end{align}
and
\[
0 = \glinh= \xblin = \alin = \ablin = 0 \ \ \ , \ \ \ \bblin = 0 \ \ \ , \ \ \ \slin = 0 \nonumber \, .
\]
We will call $f$ a gauge function.
\end{lemma}

\begin{remark} \label{rem:fu}
Recall that as long as $\Omega^2 \cdot f\left(u,\theta,\phi\right)$ is smooth in the extended sense and uniformly bounded the corresponding pure gauge solution is smooth in the extended sense all the way to the horizon. In addition, the linearised curvature components and Ricci coefficients of the above gauge solution satisfy the decay rates (\ref{decrete}) towards null infinity. Cf.~Remark \ref{rem:fv}.
\end{remark}

\subsubsection{Pure gauge solutions of the form $\left(f_1=0,f_2=0,j_3,j_4\right)$}

We finally consider pure gauge solutions of the form $\left(f_1=0,f_2=0,j_3,j_4\right)$. As we will see they will only generate non-trivial values for the metric components $\glinto$, $\glinh$ and $\bmlin$ while all other quantities of the solution vanish. To discuss these solutions it will be desirable to bring the formulae (\ref{mpe1})--(\ref{mpe4}) into a more geometric form. For this it is useful to think about the associated underlying coordinate transformation (\ref{cotrafo}) on the sphere as
\[
\tilde{\boldsymbol\theta}^A = \theta^A + \epsilon j^A \left(v,\theta,\phi\right) \ \ \ \textrm{with $\tilde{\boldsymbol\theta}^1=\tilde{\boldsymbol\theta}$, $\tilde{\boldsymbol\theta}^2=\tilde{\boldsymbol\phi}$ and ${\theta}^1=\theta$, ${\theta}^2={\phi}$}
\]
for a $v$-dependent vectorfield $j^A\left(v,\theta,\phi\right)$ on $S^2_{u,v}$ with components $j_3$ and $j_4$. We can then solve for each $u,v$ the equation
\[
j = r^2\slashed{\mathcal{D}}_1^\star \left(-q_1,-q_2\right) \ \ \textrm{or in components $j^A=\gamma^{AB} \partial_B q_1 - \epsilon^{AC} \partial_C q_2 $}
\]
for a pair of functions $\left(q_1 \left(v,\theta,\phi\right),q_2 \left(v,\theta,\phi\right)\right)$. Note that $q_1,q_2$ do not depend on $u$ as there is no $u$-dependence if the equation is written with indices upstairs. Note that $q_1, q_2$ are unique up to their spherical mean. The following lemma parametrises the pure gauge solutions of the title in terms of smooth functions $q_1 \left(v,\theta,\phi\right)$ and $q_2\left(v,\theta,\phi\right)$, which by the above considerations exploit the full freedom given by $j_3$ and $j_4$.

\begin{lemma} \label{lem:exactsol3}
For any smooth functions $q_1\left(v,\theta,\phi\right)$ and $q_2\left(v,\theta,\phi\right)$ the following is a pure gauge solution of the system of gravitational perturbations:
\begin{align}
\glinh  &=2r^2 \slashed{\mathcal{D}}_2^\star \slashed{\mathcal{D}}_1^\star \left(q_1,q_2\right) , &\frac{\glinto}{\sqrt{\slashed{g}}} &= r^2 \slashed{\Delta} q_1, & \bmlin &= r^2 \slashed{\mathcal{D}}_1^\star \left(\partial_v q_1, \partial_v q_2\right) \, , \nonumber 
\end{align}
while the linearised
metric coefficient $\Olino$ as well as all linearised connection coefficients and  curvature components vanish.
\end{lemma}
\begin{proof}
Note that since $\bmlin$ with indices upstairs does not depend on the variable $u$, (\ref{bequat}) indeed holds. The equations (\ref{stos}) and (\ref{stos2}) are also readily checked and the remaining equations hold trivially.
\end{proof}

\subsection{A $4$-dimensional reference linearised Kerr family $\mathscr{K}$}
\label{4dimfam}
The other class of interesting special solutions which we shall identify
corresponds to the $4$-dimensional family 
that arises  by linearising one-parameter representations of 
Kerr  (which of course 
solves the nonlinear equations $(\ref{Evackiedw})$) around Schwarzschild
in an appropriate coordinate system. We will present
such a family here, giving first in Section~\ref{sec:linss}  a $1$-dimensional linearised
Schwarzschild family and then in Section~\ref{sec:nontrivkerr}, a $3$-dimensional family
corresponding to Kerr with fixed mass $M$.

\subsubsection{Linearised Schwarzschild solutions} \label{sec:linss}
We begin by reminding the reader that in view of the pure gauge solutions identified in Section \ref{sec:ssgauge}, there is no unique way of identifying a $1$-parameter family of linearised Schwarzschild solutions. This \emph{uniqueness up to pure gauge solutions} is reflected in the choice of double null coordinates in which one linearises the one-parameter Schwarzschild family. A particularly simple such choice is given by writing the one-parameter Schwarzschild family in rescaled null coordinates
\begin{align} \label{ssrc}
g_{M} = 4M^2 \left(-4 \left(1-\frac{1}{x}\right)  d\hat{u} d\hat{v} + x^2 d\sigma_2\right) \textrm{ \ \ with $x$ defined via the relation $\left(x-1\right)e^x = e^{\hat{v}-\hat{u}}$}  \, . 
\end{align}
Note that setting $r=2Mx$ and $u=2M\hat{u}$ and $v=2M\hat{v}$ produces the metric in standard Eddington-- Finkelstein coordinates $u,v$. Since the $x$ in (\ref{ssrc}) does not depend on $M$ at all, the linearisation of (\ref{ssrc}) in the parameter $M$ is immediate.

One obtains thus  a proof of the following proposition (which can alternatively be proven  by directly verifying that the system of linearised gravity  is satisfied)
\begin{proposition} \label{lem:linss}
For every $\mathfrak{m} \in \mathbb{R}$, the following is a (spherically symmetric) solution of the system of gravitational perturbations (\ref{stos})--(\ref{Bianchi10}) in $\mathcal{M}$:
\begin{align}
 \glinh= \hat{\chi} = \xblin=\alin = \ablin = 0 \ \ \ , \ \  \   \bmlin = \elin = \eblin = \blin = \bblin =0 
\ \ \ , \ \ \ 
  \otx = \Omega^{-2} \otxb = 0   
\end{align}
and
\begin{align} 
2\Olin = - \mathfrak{m} \ \ \ , \ \ \ tr_{\slashed{g}} \glin= -2\mathfrak{m}  \ \ \ , \ \ \ \rlin= -\frac{2M}{r^3} \cdot \mathfrak{m}  \ \ \ , \ \ \ \Klin=\frac{\mathfrak{m} }{r^2}  \, .
\end{align}
We refer to the above one-parameter family as the {\bf reference linearised Schwarzschild solutions}.
\end{proposition}

As mentioned the above proposition exhibits the family of linearised Schwarzschild solutions in a particular gauge.
\begin{remark}
With respect to standard Eddington--Finkelstein coordinates $\left(u,v\right)$, the Schwarzschild family is given by
\begin{align} \label{ssrc2}
g_M = -4 \left(1-\frac{2M}{r_M}\right) du dv + (r_M)^2 d\sigma_2 \ \ \ \ \textrm{with $r_M$ defined via} \ \ \left(\frac{r_M}{2M}-1\right)e^{\frac{r_M}{2M}}=e^{\frac{v-u}{2M}} \, .
\end{align}
If one linearises (\ref{ssrc2}) with respect to the parameter $M$ fixing the $\left(u,v\right)$-differential structure, one obtains the sum of the family of Lemma \ref{lem:linss} and the pure-gauge transformation generated by $f_1=\frac{u}{2M}$ and $f_2=\frac{v}{2M}$ (and $f_3=f_4=j_3=j_4=0$) in (\ref{cotrafo}), the reason being that the coordinate transformation relating $\left(u,v\right)$ and $\left(\hat{u},\hat{v}\right)$ alluded to above depends on $M$ itself.
\end{remark}

\subsubsection{Linearised Kerr solutions leaving the mass unchanged} \label{sec:nontrivkerr}
Recall from the discussion in Section~\ref{niceambience}
that the Kerr family can globally be brought into the double null form (\ref{familynewdef}) in 
its exterior. This was achieved in \cite{Pretorius} (see also \cite{DHRscat}). One can linearise 
a one-parameter representation of the metric in this form with respect to the angular momentum parameter $a=\epsilon\mathfrak{a}$ to obtain what we shall call the (reference) linearised Kerr solution below. Alternatively one can take a shortcut and start from the Kerr metric expressed in standard Boyer--Lindquist coordinates ignoring all terms quadratic or higher in $a$:
\[
g_{Kerr} = - \left(1-\frac{2M}{r}\right) dt^2 + \frac{dr^2}{1-\frac{2M}{r}} + r^2 \left(d\theta^2 + \sin^2\theta d\phi^2\right) - \frac{4Ma}{r} \sin^2 \theta d\phi dt + \mathcal{O}\left(a^2\right) \, .
\]
One can now introduce the standard Eddington--Finkelstein coordinates $\left(u,v\right)$ for the Schwarzschild part and do a coordinate transformation $\phi \rightarrow \tilde{\phi} + f\left(v-u\right)$ for an appropriate function $f$ to bring the metric into the form (\ref{familynewdef}) to first order
in $\epsilon$
and still read off the metric perturbation. Either of these procedures leads to the ($m=0$ case of the) following 
proposition:

\begin{proposition} \label{lem:explicitkerr} 
Let $Y^{\ell=1}_m$ for $m=-1,0,1$ denote the spherical harmonics $(\ref{defYothers})$. 
For any $\mathfrak{a} \in \mathbb{R}$, the following is a smooth
solution of the system of gravitational perturbations (\ref{stos})--(\ref{Bianchi10}) on $\mathcal{M}$. The non-vanishing metric coefficients are
\begin{align}
\bmlin_A = \left(b^{Kerr}\right)^A=\frac{4M\mathfrak{a}}{r} \slashed{\epsilon}^{AB} \partial_B Y^{\ell=1}_m \, .
\end{align}
The non-vanishing Ricci coefficients are
\begin{align}
\elin^A = \left(\eta^{Kerr}\right)^A = \frac{3M\mathfrak{a}}{r^2}  \slashed{\epsilon}^{AB} \partial_B Y^{\ell=1}_m \ \ \ , \ \ \  \eblin= \underline{\eta}^{Kerr} = -\eta^{Kerr} \, .
\end{align}
The non-vanishing curvature components are
\begin{align}
\blin = \beta^{Kerr} = \frac{\Omega}{r} \eta^{Kerr} \ \ \ , \ \ \  \bmlin=\underline{\beta}^{Kerr} = -\beta^{Kerr} \ \ \ &,  \ \ \  \slin =\sigma^{Kerr} = \frac{6}{r^4} \mathfrak{a} M \cdot Y^{\ell=1}_m \, .
\end{align}
We will refer to this three parameter family spanned by
the above solutions with ($m=-1,0,1$) the  {\bf reference $\ell=1$ linearised Kerr solutions}.
\end{proposition}

Note that the above family may be parametrised by the $\ell=1$-modes of the curvature component $\slin$.\footnote{The scalar $\slin$ has no $\ell=0$ mode as it satisfies the equation 
$\slin = curl \elin$.}

\begin{proof}
We first note that $\slashed{div} b^{Kerr}=0$ and $\slashed{\mathcal{D}}_2^\star b^{Kerr}=0$ as well as
\[
\partial_u \left(b^{Kerr}\right)^A =  \partial_u \left(\frac{4Ma}{r^3} \gamma^{AB} \epsilon_A^{\phantom{A}B} \partial_B Y^{\ell=1}_m\right) = \frac{12Ma \Omega^2}{r^2} \slashed{\epsilon}^{AB} \partial_B Y^{\ell=1} = 2 \Omega^2 \left(\eta^{Kerr} - \underline{\eta}^{Kerr}\right)
\]
where we recall that $\slashed{g}=r^2 \gamma$ and that $\epsilon_A^{\phantom{A}B}=\slashed{\epsilon}_A^{\phantom{A}B}$ does not depend on $r$. Hence (\ref{stos})--(\ref{bequat}) all hold.

Since also $\slashed{div} \eta^{Kerr}=\slashed{div} \underline{\eta}^{Kerr}=0$ and $\slashed{\mathcal{D}}_2^\star \eta^{Kerr}=\slashed{\mathcal{D}}_2^\star \underline{\eta}^{Kerr}=0$ all null structure equations (\ref{dtcb})--(\ref{lingauss}) hold trivially except (\ref{propeta}) and (\ref{curleta}). Note that the Codazzi equations (\ref{ellipchi}) hold by definition of $\blin$ and $\bblin$ in terms of $\elin$ and $\eblin$. To verify (\ref{curleta}) we compute
\[
\slashed{curl} \eta^{Kerr} =\slashed{\epsilon}^{BA} \partial_B \eta^{Kerr}_A = \frac{3Ma}{r^2} \slashed{\epsilon}^{BA} \partial_B \slashed{\epsilon}_A^{\phantom{A}C}\partial_CY^{\ell=1}_m = -\frac{3Ma}{r^4}\Delta_{\mathbb{S}^2}Y^{\ell=1}_m =  \frac{6}{r^4} a M \cdot Y^{\ell=1}_m = \sigma^{Kerr}
\]
and similarly for $\underline{\eta}^{Kerr}$. To verify (\ref{propeta}) we compute
\begin{align}
(\Omega \slashed{\nabla}_4 r^2 \eta^{Kerr})_A &= \partial_v (r^2 \eta^{Kerr}_A) -\frac{\Omega^2}{r} (r^2\eta^{Kerr}_A) = -\Omega\beta^{Kerr}_A \, , \nonumber \\
(\Omega \slashed{\nabla}_3 r^2 \underline{\eta}^{Kerr})_A &= \partial_u (r^2 \underline{\eta}^{Kerr}_A) +\frac{\Omega^2}{r} (r^2\underline{\eta}^{Kerr}_A) = +\Omega\underline{\beta}^{Kerr}_A \, .
\end{align}
We finally turn to verifying the Bianchi equations (\ref{Bianchi1})--(\ref{Bianchi10}). We first note that the ones for $\alin$ and $\ablin$ as well as those for $\rlin$ are trivially satisfied. Also
\begin{align}
(\Omega \slashed{\nabla}_4( r^4 \Omega^{-1} \beta^{Kerr}))_A = (\Omega \slashed{\nabla}_4 (r^3\eta^{Kerr}))_A =0 \ \ \ ,  \ \ \ (\Omega \slashed{\nabla}_3 ( r^4 \Omega^{-1} \underline{\beta}^{Kerr}))_A = (\Omega \slashed{\nabla}_3 ( r^3\underline{\eta}^{Kerr}))_A =0
\end{align}
verifying (\ref{Bianchi2}) and (\ref{Bianchi9}). It remains to check that (\ref{Bianchi3}) and (\ref{Bianchi6})--(\ref{Bianchi8}) are satisfied. For the $\slin$ equations we note 
\[
\Omega\slashed{\nabla}_4 (r^3\sigma^{Kerr}) = -\frac{6}{r^2} aM \Omega^2 Y^{\ell=1}_m = 0 = -r^2 \sigma^{Kerr} \Omega^2 = -\Omega r^3 \slashed{curl} \beta^{Kerr}
\]
and similarly for equation (\ref{Bianchi7}). We finally verify (\ref{Bianchi3}) noting that (\ref{Bianchi8}) is verified analogously:
\begin{align}
\Omega\slashed{\nabla}_3 \left(r^2 \Omega \underline{\beta}^{Kerr} \right) &=\Omega \slashed{\nabla}_3 \left(r \Omega^2 \underline{\eta}^{Kerr} \right) =\partial_u \left(\frac{3Ma}{r}\Omega^2 \epsilon_{A}^{\phantom{A}B} \partial_B Y^{\ell=1}_m  \right)  + \frac{\Omega^2}{r}\left(\frac{3Ma}{r}\Omega^2 \epsilon_{A}^{\phantom{A}B} \partial_B Y^{\ell=1}_m  \right)  \nonumber \\ 
&= \frac{6Ma}{r^2} \Omega^2 \left(1-\frac{2M}{r}\right) \epsilon_{A}^{\phantom{A}B} \partial_B Y^{\ell=1}_m - \frac{3Ma}{r^3} 2M \Omega^2 \epsilon_{A}^{\phantom{A}B} \partial_B Y^{\ell=1}_m \nonumber \\
&= r^2 \Omega^2 \epsilon_{A}^{\phantom{A}B} \partial_B \sigma^{Kerr} - \frac{6M}{r} \Omega^2 \eta^{Kerr} \, . \nonumber
\end{align}
\end{proof}

The reader might wonder why the family in Proposition~\ref{lem:explicitkerr} is a $3$-parameter family of solutions, while the full Kerr metric with fixed mass is a $1$-parameter family. This can be explained as follows. When writing down the Kerr metric one fixes an axis of symmetry. Rotations of this axis in space correspond to the same Kerr metric expressed in different coordinates. At the linear level, if we linearised the metric at a \emph{non-trivial} ($a\neq0$) member of the Kerr family, this would manifest itself in the existence of non-trivial pure gauge solutions corresponding to a rotation of the axis. In contrast, here we are linearising with respect to the \emph{spherically symmetric} Schwarzschild metric. The associated pure gauge solutions of rotating the axis are then trivial in view of the isometry group of the round sphere. Hence, we must see three ``basis" Kerr metrics which cannot be connected by a pure gauge transformation. Note the aforementioned trivial pure gauge solution are seen as $\left(q_1=0, q_2=Y^1_m\right)$ generating the trivial solution in
Lemma~\ref{lem:exactsol3}.

Finally, let us  combine the 1-dimensional space of
reference linearised Schwarzschild  solutions and the 3-dimensional space
of reference $\ell=1$ linearised Kerrs in the following definition: 
\begin{definition} \label{def:kerrl}
Let $\mathfrak{m}$, $s_{-1}, s_0,s_1$ be four real parameters. We call the sum of the solution of Proposition~\ref{lem:linss} with parameter $\mathfrak{m}$ and the solution of 
Proposition~\ref{lem:explicitkerr} satisfying $\sigma^{Kerr} = \sum s_{m} Y^{\ell=1}_m$ the
{\bf reference linearised Kerr solution} with parameters $\left(\mathfrak{m},s_{-1},s_0,s_1\right)$ and denote it by $\mathscr{K}_{\mathfrak{m},s_i}$ or simply $\mathscr{K}$. 
\end{definition}

\section{The Teukolsky and Regge--Wheeler equations and the gauge invariant
hierarchy} \label{sec:P}

In this section, we 
will introduce the celebrated  spin $\pm 2$ Teukolsky equations and the Regge--Wheeler equation and explain the connection between the two
and their relation to
the full system of linearised gravity.

We begin in Section~\ref{sec:teurw} by
defining the above three equations,
considering them  as second order hyperbolic PDE's for independent
unknowns $\alpha$, $\underline\alpha$ and $P$,
independently that is  of the system of gravitational perturbations.
In Section~\ref{sec:ivprwteu}, we shall 
state 
a well-posedness
theorem for the characteristic initial value problem for these equations.
We then introduce in Section \ref{sec:tratheo}  a fundamental 
transformation mapping solutions
$\alpha$, $\underline\alpha$ of the spin $\pm 2$ Teukolsky equations to 
solutions $P$, $\underline{P}$ of the Regge-Wheeler equation. This
transformation will play a key role in understanding solutions to the Teukolsky equation itself. As was remarked in Section \ref{decoupledINTRO}, this transformation is a physical space version of transformations appearing in Chandrasekhar~\cite{Chandrasekhar} for fixed frequencies.
Note that Sections~\ref{sec:teurw}--\ref{sec:tratheo} are independent of 
Sections~\ref{sec:lineq} and~\ref{sec:specialsol}.

The relation of the above 
PDEs with the full system of linearised gravity 
is finally explained in Section \ref{sec:fullrel}, where we will see that given a smooth solution of the system of gravitational perturbations, the curvature components $\alin$ 
and $\ablin$ satisfy the Teukolsky equation.
By the above transformations, this gives rise to  
$\Plin$ and $\Pblin$ 
satisfying the Regge--Wheeler equation.
We shall see also (Proposition~\ref{prop:relfull}) 
how the latter can be re-expressed in various ways using the Bianchi identities.

\subsection{The spin $\pm2$ Teukolsky equations and the Regge--Wheeler equation} \label{sec:teurw}
The spin $\pm2$ Teukolsky equations
concern symmetric traceless $S_{u,v}$ 2-tensors which we shall denote by $\alpha$, $\underline\alpha$, 
in anticipation of Section~\ref{sec:fullrel}. 
For now let these be  defined on a subset $\mathcal{D}\subset \mathcal{M}$.
Note that with our normalisations, it
is natural to assume that the rescaled quantities
$\Omega^2\alpha$ and $\Omega^{-2}\underline\alpha$ are smooth on
$\mathcal{D}$ up to and including the horizon $\mathcal{D}\cap \mathcal{H}^+$.

\begin{definition} \label{def:teue}
Let ${\alpha}$ be a  symmetric traceless $S^2_{u,v}$ 2-tensor defined
on a subset $\mathcal{D}\subset \mathcal{M}$ such that $\Omega^2\alpha$ is smooth
on $\mathcal{D}$.
We say that $\alpha$ satisfies the {\bf Teukolsky equation of spin ${\bf +2}$} if $\alpha$ satisfies the following PDE:
\begin{align} \label{teukolsky}
\slashed{\nabla}_4 \slashed{\nabla}_3 \alpha &+ \left(\frac{1}{2} tr \underline{\chi} + 2 \underline{\hat{\omega}} \right) \slashed{\nabla}_4 \alpha +  \left(\frac{5}{2} tr {\chi} - {\hat{\omega}} \right) \slashed{\nabla}_3 \alpha - \slashed{\nabla}^2 \alpha \nonumber  \\
&+ \alpha \left[5 \underline{\hat{\omega}} tr \chi - \hat{\omega} tr \underline{\chi}-4\rho + 2K + tr \chi tr \underline{\chi} -4\hat{\omega} \hat{\underline{\omega}}\right] = 0 \, .
\end{align}
Let $\underline{\alpha}$ be a symmetric traceless $S^2_{u,v}$-tensor on $\mathcal{D}$ 
such that $\Omega^{-2}\underline{\alpha}$ 
is smooth on $\mathcal{D}$. 
We say that $\underline{\alpha}$ satisfies the {\bf Teukolsky equation of spin ${\bf -2}$} if $\underline{\alpha}$ satisfies the following PDE:
\begin{align} \label{teukolsky2}
\slashed{\nabla}_3 \slashed{\nabla}_4 \underline{\alpha} &+ \left(\frac{1}{2} tr {\chi} + 2 {\hat{\omega}} \right) \slashed{\nabla}_3 \underline{\alpha} +  \left(\frac{5}{2} tr \underline{\chi} - \underline{\hat{\omega}} \right) \slashed{\nabla}_4 \underline{\alpha} - \slashed{\nabla}^2 \underline{\alpha} \nonumber  \\
&+ \underline{\alpha} \left[5 {\hat{\omega}} tr \underline{\chi} - \underline{\hat{\omega}} tr {\chi}-4\rho + 2K + tr \chi tr \underline{\chi} -4\hat{\omega} \hat{\underline{\omega}}\right] = 0 \, .
\end{align}
\end{definition}

We note that the Teukolsky equation of spin $-2$ is obtained from that of spin $+2$ by interchanging $\slashed{\nabla}_3$ with $\slashed{\nabla}_4$ and underlined Schwarzschild quantities with non-underlined ones.

Let us repeat the explicit
characterization of  smoothness up to the horizon from  Section \ref{sec:lineq}
in terms of Eddington--Finkelstein double null coordinates $u$ and $v$:
An $S^2_{u,v}$-tensor $\Theta$ extends
smoothly to the horizon
if in the spherical coordinate chart the components $\Theta_{CD}$ are smooth functions of the double null Eddington--Finkelstein coordinates on the interior $\mathcal{M}^o$ and if 
for any $n_1,n_2,n_3 \in \mathbb{N}_0$ and $A\in \{1,2\}$ the functions
\[
\left(e^\frac{u}{2M} \partial_u \right)^{n_1} \left(\partial_v\right)^{n_2} \left(\partial_{\theta_A}\right)^{n_3} \Theta_{CD}
\]
extend continuously to the boundary $\mathcal{H}^+$.

According to Proposition~\ref{prop:relfull} of Section~\ref{sec:fullrel}, 
it will follow that given a solution $\mathscr{S}$ of linearised gravity,
then the quantities  $\alin$ and $\ablin$
satisfy the spin $\pm 2$ Teukolsky equations, respectively. For now, however,
we will study the Teukolsky equation in its own right, 
independently of the full system.

The other equation to be defined in this section is the Regge--Wheeler equation, to be
satisfied again by a symmetric traceless $S^2_{u,v}$-tensor $P$.

\begin{definition} \label{def:rwe}
Let $P$ be a  smooth, symmetric traceless $S^2_{u,v}$-tensor on $\mathcal{D}$.
We say that $P$ satisfies the {\bf Regge--Wheeler equation} if $P$ satisfies the following PDE:
\begin{align}   \label{im4}
\slashed{\nabla}_3 \slashed{\nabla}_4 P &+ \slashed{\nabla}_4 \slashed{\nabla}_3 P - 2 \slashed{\Delta} P + \left(5 tr \underline{\chi} + \hat{\underline{\omega}}\right) \cdot \slashed{\nabla}_4 P + \left(5tr {\chi} + \hat{\omega}\right) \slashed{\nabla}_3 P \nonumber \\ 
&+ P \left(4K - \left(3 tr \chi + \hat{\omega}\right) 2 tr \chi - 4 \left(tr \chi\right)^2 + 2\slashed{\nabla}_3 tr \chi  -8\hat{\omega} tr \chi \right)
= 0 \, .
\end{align}
\end{definition}
In Section~\ref{sec:tratheo}, 
we shall show that given solutions $\alpha$ and $\underline\alpha$ of
the spin $\pm2$ Teukolsky equations, respectively, we can derive
two solutions $P$ and $\underline{P}$, respectively, of the Regge--Wheeler equation.
In view of the above remarks, it follows that we can associate such solutions
to a solution $\mathscr{S}$ of the full system of linearised gravity. As with
$(\ref{teukolsky})$, $(\ref{teukolsky2})$,
however, for now we shall consider the equation $(\ref{im4})$ in its own
right. We again note 
that the Regge--Wheeler equation  was first derived in~\cite{Regge} in the context
of the theory of ``metric perturbations''. 

To bring (\ref{im4}) in a more familiar form we define the weighted symmetric traceless $S^2_{u,v}$-tensors
\begin{align} \label{PsiP}
\Psi = r^5 P  \ \ \ \ \ \textrm{and} \ \ \ \ \ \underline{\Psi} = r^5 \underline{P}
\end{align}
and conclude
\begin{corollary}
If $P$ satisfies the Regge--Wheeler equation, then the weighted symmetric traceless $S^2_{u,v}$-tensor $\Psi = r^5 P$ satisfies the equation
\begin{align} \label{rwro}
\Omega \slashed{\nabla}_3 \left(\Omega \slashed{\nabla}_4 \Psi \right) - \left(1-\frac{2M}{r}\right) \slashed{\Delta}\Psi + V \Psi = 0 \, \ \textrm{ \ with} \ \ V = \left[\frac{4}{r^2} - \frac{6M}{r^3}\right]\left(1-\frac{2M}{r}\right).
\end{align}
\end{corollary}
\begin{proof}
Direct computation using the Schwarzschild background values of Section \ref{hereforSchconcur}.
\end{proof}

In the context of the proof of Theorem \ref{prop:summarypsi} we will do estimates directly at the level of the tensorial equation (\ref{rwro}).
\begin{remark}
In the literature, the Regge-Wheeler equation is typically stated for a scalar function $f$ as
\[
\frac{1}{1-\frac{2M}{r}} \partial_u \partial_v f  - \slashed{\Delta}f  -\frac{6M}{r^3} f = 0 \, .
\]
It is not hard to see that if two smooth functions $f,g$ satisfy the scalar Regge-Wheeler equation, then the symmetric traceless tensor $\phi= r^2\slashed{\mathcal{D}}_2^\star \slashed{\mathcal{D}}_1^\star \left(f,g\right)$ satisfies (\ref{rwro}), the additional factor of $\frac{4}{r^2}$ appearing from the commutation of the angular operators with $\slashed{\Delta}$. Note also that one can reconstruct $f,g$ uniquely from $\phi$ up to the $\ell=0,1$ modes.
\end{remark}

\subsection{The characteristic initial value problem} \label{sec:ivprwteu}
For completeness, we state here a standard well-posedness theorem for both the Teukolsky and the Regge--Wheeler equation. In view of future applications we formulate it in the context of a characteristic problem: We fix a sphere $S^2_{u_0,v_0}$ in $\mathcal{M}$ and consider the outgoing Schwarzschild light cone $C_{u_0}=\{u_0\}\times\{v\ge v_0\}\times S^2$ and the ingoing light cone $C_{v_0}=\{u\ge u_0\}\times\{v_0\}\times S^2$ on which the data are being prescribed. 
In our convention, $C_{v_0}$ includes the horizon sphere $S^2_{\infty, v_0}$.

We begin with the spin $\pm2$ Teukolsky equations:
\begin{theorem}[Well-posedness for Teukolsky of spin $+2$]
 \label{theo:wpteu}
Given a sphere $S^2_{u_0,v_0}$ with corresponding null cones $C_{u_0}$ and $C_{v_0}$,
prescribe
\begin{itemize}
\item along $C_{v_0}$ a  symmetric traceless $S^2_{u,v}$-tensor $\alpha_{\circ, in}$,  such that $\Omega^2 \alpha_{\circ,in}$ is smooth
\item along $C_{u_0}$ a smooth symmetric traceless $S^2_{u,v}$-tensor $\alpha_{\circ, out}$ satisfying $\alpha_{\circ,out} = \alpha_{\circ,in}$ on $S^2_{u_0,v_0}$.
\end{itemize}
Then there exists a unique smooth symmetric traceless $S^2_{u,v}$
$2$-tensor $\Omega^2\alpha$
defined on $\mathcal{M} \cap \{u \geq u_0\} \cap \{v \geq v_0\}$ such that
\begin{itemize}
\item $\alpha$ satisfies the Teukolsky equation of spin $+2$ (\ref{teukolsky}) in $\mathcal{M} \cap \{u \geq u_0\} \cap \{v \geq v_0\}$,
\item $\Omega^2 \alpha \big|_{C_{u_0}}= \Omega^2\alpha_{\circ, in}$ and $\alpha \big|_{C_{v_0}}= \alpha_{\circ, out}$.
\end{itemize}
\end{theorem}
We emphasise that in our convention, 
the set $\mathcal{M} \cap \{u \geq u_0\} \cap \{v \geq v_0\}$ includes
$\mathcal{H}^+\cap\{v\geq v_0\}$.

\begin{theorem}[Well-posedness for Teukolsky of spin $-2$] 
\label{theo:wpteu2}
Theorem \ref{theo:wpteu} holds replacing all $\alpha$ by $\underline{\alpha}$, all $\Omega^2$ by $\Omega^{-2}$ and (\ref{teukolsky}) by (\ref{teukolsky2}) in the above statement.
\end{theorem}

The well-posedness statement for the Regge--Wheeler equation (\ref{im4}) is entirely analogous:

\begin{theorem}[Well-posedness for Regge--Wheeler] 
\label{theo:wprw}
Given a sphere $S^2_{u_0,v_0}$  with corresponding null cones $C_{u_0}$ and
$C_{v_0}$, prescribe
\begin{itemize}
\item along $C_{v_0}$ a smooth symmetric traceless $S^2_{u,v}$-tensor $P_{\circ, in}$,
\item along $C_{u_0}$ a smooth symmetric traceless $S^2_{u,v}$-tensor $P_{\circ, out}$ satisfying $P_{\circ,out} = P_{\circ,in}$ on $S^2_{u_0,v_0}$.
\end{itemize}
Then there exists a unique smooth symmetric traceless
$S_{u,v}$ 2-tensor $P$ defined on  $\mathcal{M} \cap \{u \geq u_0\} \cap \{v \geq v_0\}$
such that 
\begin{itemize}
\item $P$ satisfies the Regge--Wheeler equation
(\ref{im4}) in $\mathcal{M} \cap \{u \geq u_0\} \cap \{v \geq v_0\}$,
\item $P \big|_{C_{u_0}}= P_{\circ, in}$ and $P \big|_{C_{v_0}}= P_{\circ, out}$.
\end{itemize}
\end{theorem}

\subsection{The transformation theory: Definitions of $\psi$, \underline{$\psi$} \ and $P$, \underline{$P$}} \label{sec:tratheo}
We now describe a transformation theory relating solutions of the
Teukolsky equations to solutions of Regge--Wheeler.

Given a solution  $\alpha$ of the Teukolsky equation of spin $+2$, we can define the following \emph{derived} quantities:
\begin{align} 
\psi &:= -\frac{1}{2 r\Omega^2 } \slashed{\nabla}_3 \left(r\Omega^2 \alpha\right)\label{psidef} \, ,
 \\
P&:=\frac{1}{r^3 \Omega} \slashed{\nabla}_3 \left(\psi r^3 \Omega\right)  \label{evolpp} \, .
\end{align}
Similarly,  given 
a solution  $\underline{\alpha}$ of the Teukolsky equation of spin $-2$, we can define the following \emph{derived} quantities:
\begin{align} 
\underline{\psi} &:= \frac{1}{2 r \Omega^2} \slashed{\nabla}_4 \left(r\Omega^2 \underline{\alpha}\right) \, ,\label{psidef2} 
\\
\underline{P} &:= -\frac{1}{r^3 \Omega} \slashed{\nabla}_4 \left(\underline{\psi} r^3 \Omega\right)  \, .\label{evolpp2}
\end{align}
These quantities are again symmetric traceless $S^2_{u,v}$ $2$-tensors.

The following  proposition
can be proven by a straightforward computation:
\begin{proposition} \label{prop:rwt1}
Let $\alpha$ be a solution of the Teukolsky equation of spin $+2$ on $\mathcal{M} \cap \{u\geq u_0\} \cap \{v \geq v_0\}$ as arising from Theorem \ref{theo:wpteu}. Then the symmetric traceless tensor $P$ as defined through (\ref{psidef}) and (\ref{evolpp}) satisfies the Regge--Wheeler equation on $\mathcal{M} \cap \{u\geq u_0\} \cap \{v \geq v_0\}$. 

Now let $\underline{\alpha}$ be a solution of the Teukolsky equation of spin $-2$ on $\mathcal{M} \cap \{u\geq u_0\} \cap \{v \geq v_0\}$ as arising from Theorem \ref{theo:wpteu2}. Then the symmetric traceless tensor $\underline{P}$ as defined through (\ref{psidef2}) and (\ref{evolpp2}) satisfies the Regge--Wheeler equation on $\mathcal{M} \cap \{u\geq u_0\} \cap \{v \geq v_0\}$.
\end{proposition}

We note that Fourier space analogues of the above transformations were first
discovered by Chandrasekhar~\cite{Chandrasekhar}, who also discussed differential transformations
mapping solutions of Regge--Wheeler to solutions of Teukolsky.
In this paper, however, it is 
 the physical space structure of the above transformations which we shall exploit,
in particular, the fact that they can be interpreted as transport equations, which allow
$\alpha$ (respectively $\underline\alpha$) to be recovered from $P$ (respectively
$\underline P$) \emph{and initial data}.

\subsection{The connection with the system of gravitational perturbations} \label{sec:fullrel}
We will now finally relate the equations presented above to the full system of linearised
gravity.

Let  $\mathscr{S}$ be a smooth solution
of the system of gravitational perturbations and recall the quantities $\alin$
and $\ablin$ of Section~\ref{sec:unknowns}. Note that both these symmetric traceless 
$S^2_{u,v}$ 2-tensors are \emph{gauge invariant} 
in the sense that any of the pure gauge solutions discussed in Section \ref{sec:ssgauge} 
satisfies $\alin=\ablin=0$. The latter fact can 
be checked directly from (\ref{mpe1})--(\ref{mpe4}). 
We note also that $\alin$ and $\ablin$ vanish for the $4$-dimensional reference
linearised Kerr family  of Section~\ref{4dimfam}. We will in fact show in
Appendix~\ref{charofpuregauge} that, provided that $\mathscr{S}$ is asymptotically flat
(see Section~\ref{NEOKEF}), then the vanishing identically of
 $\alin$ and $\ablin$ implies that $\mathscr{S}$
is a sum of a pure gauge solution and a reference linearised Kerr.

Remarkably, as was first shown by Bardeen--Press~\cite{bardeen1973},
$\alin$ and $\ablin$ satisfy the Teukolsky equation of spin $+2$ and spin $-2$ respectively.
Combining this fact with results of the previous section yields:

\begin{proposition} \label{prop:relfull}
Let $\mathscr{S}$ be a smooth solution
of the system of gravitational perturbations on a domain $\mathcal{D}\subset \mathcal{M}$ 
and consider the curvature components $\alin$, $\ablin$ which are part of the solution $\mathscr{S}$. Then $\alin$ satisfies the Teukolsky equation of spin 2 and $\ablin$ satisfies the Teukolsky equation of spin $-2$. Moreover, the derived quantities (\ref{psidef})--(\ref{evolpp2}) that are defined for any solution of the Teukolsky equation, now denoted $\plin$, $\Plin$ and $\plin$, $\Plin$, can also be re-expressed in terms of derivatives of curvature components and Ricci coefficients using the Bianchi and null structure equations. We have
\begin{align} \label{psideffull}
\plin =\slashed{\mathcal{D}}^\star_2 \, \blin + \frac{3}{2} \rho \, \xlin  \ \ \ \ \ \textrm{and} \ \ \ \ \ \pblin = \slashed{\mathcal{D}}^\star_2 \, \bblin \, - \frac{3}{2} \rho\, \xblin \, ,
\end{align}
as well as
\begin{align} \label{Pdef} 
\Plin &=   \slashed{\mathcal{D}}_2^\star \slashed{\mathcal{D}}^\star_1 \left( -\rlin \, , \, \slin \, \right) + \frac{3}{4} \rho \, tr \chi  \left(\xlin - \xblin \, \right) \, ,  \\
\label{Pdefund}
\Pblin &=  \slashed{\mathcal{D}}_2^\star \slashed{\mathcal{D}}^\star_1 \left(  -\rlin \, , \,  -\slin \, \right) + \frac{3}{4} \rho \, tr \chi  \left(\xlin - \xblin \, \right) \, .
\end{align}
\end{proposition} 
\begin{proof}
The equation for $\alin$ is easily derived from taking a $4$-derivative of (\ref{Bianchi1}) and using (\ref{Bianchi2}). The equation for $\ablin$ follows from taking a $3$-derivative of (\ref{Bianchi10}) and using (\ref{Bianchi9}). The identities for $\plin$ and $\pblin$ are immediate from (\ref{Bianchi1}) and (\ref{Bianchi10}). To see the identity for $\Pblin$, one uses the Bianchi equation (\ref{Bianchi8}) and the null structure equation (\ref{chih3b}), as well as the formulae of Section \ref{sec:commutation} to obtain
\begin{align} \label{compbar}
\slashed{\nabla}_4 \pblin &= \slashed{\mathcal{D}}^\star_2 \left(  \slashed{\mathcal{D}}^\star_1 \left(\, \rlin \, , \, \slin \, \right) - 3 \, \eblin \, \rho - tr \chi \, \bblin - \hat{\omega} \, \bblin \right) - \frac{1}{2} tr \chi \slashed{\mathcal{D}}^\star_2\, \bblin \nonumber \\ 
&\qquad+ \frac{9}{4} \rho\,  tr \chi \, \xblin - \frac{3}{2} \rho \left( - \hat{\omega} \, \xblin - \frac{1}{2} tr \chi \, \xblin - \frac{1}{2} tr \underline{\chi}\, \xlin - 2\slashed{\mathcal{D}}^\star_2 \, \eblin \right) \nonumber \\
&=  \slashed{\mathcal{D}}_2^\star \slashed{\mathcal{D}}^\star_1 \left( \, \rlin \, , \, \slin \, \right) - \frac{3}{4} \rho tr \chi  \left(\, \xlin - \xblin \, \right) - \frac{3}{2} tr \chi \left( \slashed{\mathcal{D}}^\star_2 \, \bblin \, - \frac{3}{2} \rho\, \xblin \,\right) - \hat{\omega} \left( \slashed{\mathcal{D}}^\star_2 \, \bblin \, - \frac{3}{2} \rho\, \xblin \, \right) \ .
\end{align}
The computation for $\Plin$ is completely analogous.
\end{proof}

\begin{remark}
In view of the gauge invariance of $\alin$ and $\ablin$ in the sense above, 
it follows from the definitions (\ref{psidef})--(\ref{evolpp2}) that the quantities $\Plin$, $\Pblin$ and $\plin$, $\pblin$ are also manifestly gauge invariant.
We note however (see Appendix~\ref{theRTrautsection}) that there exist asymptotically flat 
solutions $\mathscr{S}$ which are \emph{not} pure gauge  
such that $\accentset{(1)}{P}$ and $\accentset{(1)}{\underline{P}}$ identically vanish.
 \end{remark}

The fact that $\accentset{(1)}{P}$ and $\accentset{(1)}{\underline{P}}$ above satisfy the Regge--Wheeler equation
$(\ref{im4})$, together
with the relations $(\ref{psidef})$--$(\ref{evolpp2})$ but also $(\ref{psideffull})$--$(\ref{Pdefund})$,
 will be the key to estimating the Teukolsky equations and
 unlocking the system of linearised gravity.

\section{Initial data and  well-posedness of linearised gravity} \label{sec:initialdata} 
We turn in this section to the well-posedness of the system of linearised gravity 
of Section~\ref{sec:fulleq}.

We first describe how to prescribe initial data in Section~\ref{seeddatasection} below.
Then we shall formulate the well-posedness theorem in Section~\ref{WPTsec}.
Finally, in Section~\ref{NEOKEF},
we define what it means for data to be asymptotically flat.

\subsection{Seed data on an initial double null cone}\label{seeddatasection}
In this section, we describe how to prescribe initial data for the system of gravitational perturbations derived in Section \ref{sec:lineq}.

The setting will be that of a characteristic initial value problem: We fix a sphere $S^2_{u_0,v_0}$ in $\mathcal{M}$ and consider the outgoing 
Schwarzschild
light cone $C_{u_0}=\{u_0\}\times\{v\ge v_0\}\times S_2$ 
and the ingoing light cone $C_{v_0}=\{u\ge u_0\}\times\{v_0\}\times S^2$ 
on which the data are being prescribed. 
Initial data are determined by so-called ``seed data'' that can be prescribed freely.
Recall that in our convention, $C_{v_0}$ includes the horizon sphere $S^2_{\infty, v_0}$.
The definition of seed data is given below. We remark that this is essentially
a linearised version of the prescription given in \cite{formationofbh}.

\begin{definition} \label{def:seeddata}
Given a sphere $S^2_{u_0,v_0}$ with corresponding null cones $C_{u_0}$, $C_{v_0}$, 
a {\bf smooth seed initial data set} consists of prescribing
\begin{itemize}
\item along $C_{v_0}$ a smooth symmetric traceless $S^2_{u,v}$ 2-tensor $\glinh_{\circ,in}\left(u,\theta,\phi\right)$ 
\item along $C_{u_0}$  a smooth symmetric traceless $S^2_{u,v}$ 2-tensor $\glinh_{\circ,out}\left(v,\theta,\phi\right)$ 
\item along $C_{v_0}$ a  
smooth function $\Omega^{-1}\Olino_{\circ,in}(u,\theta, \phi)$ \\
\item
along $C_{u_0}$ a smooth function $\Olino_{\circ,out}(v,\theta,\phi)$

\item along $C_{u_0}$ a smooth $S^2_{u,v}$-1-form $\bmlin_\circ \left(v,\theta,\phi\right)$

\item on the sphere $S^2_{\infty,v_0}$ a smooth function  $\Omega^{-2} \otxb_\circ(\theta,\phi)$

\item on the sphere $S^2_{\infty,v_0}$ a smooth function $tr_{\slashed{g}} \, \glin_\circ(\theta,\phi)$
\item on the sphere $S^2_{\infty,v_0}$ a smooth function  $ \otx_\circ(\theta,\phi)$ 
\item on the sphere $S^2_{\infty,v_0}$ a smooth 1-form $\elin_\circ(\theta,\phi)$.

\end{itemize}
\end{definition}
We emphasise that since
by our convention that $C_{v_0}$ includes $S^2_{\infty, v_0}$,
smoothness is to be understood above geometrically, up to and including
the horizon.

\subsection{The well-posedness theorem}\label{WPTsec}
We can now state the fundamental well-posedness theorem for
linearised gravity on Schwarzschild:
\begin{theorem}[Well-posedness] \label{theo:lwp} 
Fix a sphere $S^2_{u_0,v_0}$ and consider a smooth seed initial data set as in Definition \ref{def:seeddata}. Then there exists a unique smooth solution 
\[
\mathscr{S}=\left(\, \glinh \, , \, \glinto \, , \, \Olino \, , \,  \bmlin\, , \,  \otx \, , \,  \otxb\, , \,  \xlin\, , \, \xblin\, , \,  \eblin \, , \,  \elin \, , \, \olin \, , \,  \olinb \, , \,  \alin \, , \,  \blin \, , \,  \rlin \, , \,  \slin \, , \,  \bblin \, , \,  \ablin \, , \, \Klin \right)
\]
of linearised gravity defined in $\mathcal{M} \cap \{u \geq u_0\} \cap \{v\geq v_0\}$ which agrees with the seed data on $C_{u_0}$ and $C_{v_0}$.
\end{theorem}

\begin{remark}
Recall that from Section \ref{sec:unknowns} that our notion of smoothness includes the statement that the weighted quantities (\ref{regq1}) of $\mathscr{S}$ extend smoothly to 
$\mathcal{H}^+\cap
\{v\ge v_0\}$.
Note that our initial smoothness assumptions on the weighted quantities $\Olin_{\circ,in}$, 
etc., is indeed
consistent with this.
\end{remark}

\begin{proof}
We give a brief outline, exploiting Theorem~\ref{theo:wpteu}, leaving the details to the reader.

First, we show that the equations uniquely determine  from seed data all dynamical
quantities   on 
 $C_{u_0} \cup C_{v_0}$ such that all tangential equations are satisfied.
(This is in fact implicitly carried out in Appendix~\ref{appendix:datacon}.)
 
Now, since by Proposition~\ref{prop:relfull}, given
a solution $\mathscr{S}$, the quantities  $\alin$ and $\ablin$ satisfy the spin $\pm2$ Teukolsky equations, we can determine these globally from their initial values on $C_{u_0} \cup C_{v_0}$ by 
applying Theorem~\ref{theo:wpteu}. Once these are determined we may order a subset
of the remaining  equations hierarchically so all remaining quantities are 
determined by the previous by integrating transport equations or by taking derivatives. 
For instance,  given $\alin$ and $\ablin$, then
$(\ref{tchi})$ can be integrated as a linear o.d.e.~to determine $\xlin$ and $\xblin$ from
seed data. (Note that the projection to the $\ell=0,1$ modes behaves differently to the rest; c.f.~Section~\ref{vanishingsec}.) 
Finally, the fact that the tangential equations hold initially on  $C_{u_0} \cup C_{v_0}$  can be used to show that the complete
system of equations, i.e.~not just those used to construct the solution, is satisfied (``propagation of constraints'').
\end{proof}

\subsection{Pointwise asymptotic flatness}\label{NEOKEF}

As we shall see, our main boundedness theorem (Theorem~\ref{theo:mtheo}) 
is most
naturally formulated in terms of a solution $\Si$ arising from Theorem~\ref{theo:lwp}, 
where the seed data satisfy certain ``gauge normalisation conditions'' 
 (see Section~\ref{sec:terminology}) and such that 
 that certain weighted energy quantities 
are bounded (see the norms in Sections~\ref{EGIQsec} and~\ref{remainingsec}). 
A sufficient condition which ensures that given a general solution $\mathscr{S}$
arising from non-normalised seed data as in Theorem~\ref{theo:lwp}, 
a solution $\Si$ can be associated to $\mathscr{S}$ 
(by addition of a pure gauge solution)
satisfying the assumptions of Theorem~\ref{theo:mtheo} 
is to assume pointwise asymptotic flatness on the seed data.
(See the statement of Theorem~\ref{prop:gaugeachieve}.)
We give the relevant definition of this notion in this section.

To keep the notation concise we first define the following derived quantities along $C_{u_0}$ from a smooth seed initial data set as in  Definition \ref{def:seeddata}:
\begin{align}
\xlin_{\circ,out} &=  \slashed{\nabla}_4 \glinh_{\circ,out} \ \ \  \  \textrm{which is a symmetric traceless
$S^2_{u,v}$ 2-tensor, } \nonumber \\
\alin_{\circ, out} &=  \Omega \slashed{\nabla}_4 \left( r^2 \Omega^{-1} \slashed{\nabla}_4 \glinh_{\circ,out} -2 \Omega^{-2} r^2 \slashed{\mathcal{D}}_2^\star \, \bmlin_\circ \right)  \ \ \  \  \textrm{which is a symmetric traceless $S^2_{u,v}$ 2-tensor, } \,  \nonumber \\
\olin_\circ &= \partial_v \left(\Olin_{\circ,out} \right) \, \ \ \  \textrm{which is a scalar function.} \nonumber
\end{align}
Note these quantities are uniquely determined in terms of the seed data.

Let us also agree on a shorthand notation to handle higher derivatives. For $\xi$ an $S^2_{u,v}$-tensor of rank $n$ on $\mathcal{M}$, we define for any $n_1 \geq 0$, $n_2\geq 0$
\[
\mathfrak{D}_{n_1,n_2} \xi =  \left(r\slashed{\nabla}\right)^{n_1} \left(r \Omega \slashed{\nabla}_4 \right)^{n_2} \xi
\]
producing an $S^2_{u,v}$-tensor of rank $n+n_1$ on $\mathcal{M}$.

We may now state our pointwise notion  of asymptotic flatness:

\begin{definition} \label{def:afpeel}
We call a seed initial data set {\bf asymptotically flat} with weight $s$ to order $n$ provided the seed data satisfies the following estimates along $C_{u_0}$  for some  $0<s\leq 1$ and any $n_1 \geq 0$, $n_2\geq 0$ with $n_1+n_2\leq n$: 
\begin{align}\label{aoci1}
\big|\Olin_{\circ,out}\big| + \big| \mathfrak{D}_{n_1,n_2}  (r^{2+s} \olin_{\circ}) \big| \leq C_{\circ,n_1,n_2}
\end{align}
\begin{align}\label{aoci2}
 \big| \mathfrak{D}_{n_1,n_2}  (r \bmlin_{\circ}) \big| \leq C_{\circ,n_1,n_2}
\end{align}
\begin{align} \label{aoci}
\big|\glinh_{\circ,out}\big| +  \big| \mathfrak{D}_{n_1,n_2}  (r^2 \xlin_{\circ,out})\big| +  \big| \mathfrak{D}_{n_1,n_2}  (r^{3+s} \alin_{\circ,out}) \big| \leq C_{\circ,n_1,n_2}
\end{align}
for some constant $C_{\circ,n_1,n_2}$ depending on $n_1$ and $n_2$. We say that
the seed data are asymptotically  flat to all orders if the above bounds hold for any $n$.
\end{definition}

Observe that a trivial choice to construct asymptotically flat seed data is 
to choose $\glinh_{\circ,out}$ of compact support on $C_{u_0}$ and $\Olin_{\circ,out}\equiv 0$ and $\bmlin_{\circ}\equiv 0$ along $C_{u_0}$.

We will show in Theorem~\ref{theo:AFdata} of  Appendix~\ref{appendix:datacon} that asymptotically flat seed data
lead in particular to a hierarchy of decay for all quantities that moreover propagates under evolution
by Theorem~\ref{WPTsec}.

\section{Gauge-normalised solutions and identification of the Kerr parameters}
\label{newgaugesec}
In this section, we define the two gauge-normalisations which will play
a fundamental role in this paper and we  identify the correct linearised Kerr parameters of
a general asymptotically flat solution $\mathscr{S}$ to our system.

We first define in Section~\ref{sec:terminology} what it means for a solution~$\mathscr{S}$ 
to be initial-data normalised. (This condition can be read explicitly from the seed data.) 
Our main boundedness theorem (Theorem~\ref{theo:mtheo} of Section~\ref{BIDGse})
will then concern such normalised solutions.
 
We then show, in Section~\ref{IDnGsec}, that given a solution $\mathscr{S}$
arising from asymptotically flat
seed data in the sense of Definition~\ref{def:afpeel}  above, 
we can indeed associate to it
an initial data-normalised solution $\Si$, 
which is realised by adding to $\mathscr{S}$ a pure
gauge  solution $\Gi$. 
Importantly, $\Gi$ can be explicitly determined by
the  seed data of $\mathscr{S}$ and is itself asymptotically flat in the sense
of Definition~\ref{def:afpeel}.
This result is stated as Theorem~\ref{prop:gaugeachieve}.

We next define in Section~\ref{hngaugesec} a renormalised solution
$\Sf$
realised by the addition of an additional pure gauge solution $\Gf$ determined
by the behaviour of $\Si$ along the event horizon. 
We will call $\Sf$ the horizon-renormalised solution and it will  be the object
of our main decay theorem (Theorem~\ref{theo:mtheod} of Section~\ref{DFNGseec}).
As opposed to the  pure gauge solution   defining $\Si$,
the pure gauge solution  defining $\Sf$ is not explicitly computable
from the seed data of $\mathscr{S}$. Only in the final proof of the decay  theorem will we show that
the pure gauge solution defining $\Sf$
is itself 
bounded (with the appropriate weights near null infinity) by the seed data of $\Si$.
 
In Section \ref{sec:globcon}, we prove several global properties satisfied by the solutions $\Si$ and $\Sf$ that will be exploited later. 

Finally, in Section \ref{vanishingsec}, we 
show
that the projection of $\Si$  to the $\ell=0,1$ modes defines a unique linearised
Kerr solution $\mathscr{K}_{\mathfrak{m},s_i}$. This is the statement of Theorem~\ref{etsilew}.
In particular, the final Kerr parameters can indeed be read off from initial data.

\subsection{Initial-data normalised solutions $\protect\Si$} \label{sec:terminology}

In this section we define the notion of an initial-data normalised solution. 
As we will show in Theorem \ref{prop:gaugeachieve} below, given a seed initial data set and its
associated solution $\mathscr{S}$, we can find a pure gauge solution $\mathscr{G}$ such that the initial data for the sum $\mathscr{S}+\mathscr{G}$ satisfies all of these conditions.

\begin{definition} \label{def:gaugechoice}
Consider a seed data set as in Definition~\ref{def:seeddata} and let $\mathscr{S}$
be the resulting solution given by Theorem~\ref{theo:lwp}.
We say that the initial data satisfies  
\begin{itemize}
\item the {\bf lapse and shift condition} if 
\begin{align}
\begin{split}
\partial_u (\Olin )&= \olinb = 0 \ \ \  \ \ \textrm{along the null hypersurface $C_{v_0}$,}  \\
 \partial_v (\Olin ) &= \olin =  0 \ \ \  \ \ \textrm{along the null hypersurface $C_{u_0}$,}  \label{omchoice}  
  \end{split}
  \end{align}
\begin{align}
\bmlin^A &= \frac{2}{3} r^3 \slashed{\epsilon}^{AB} \partial_B \slin_{\ell=1} \ \ \  \ \ \textrm{along the null hypersurface $C_{u_0}$.} \label{shiftco} 
\end{align}

\item the {\bf round sphere condition at infinity} provided 
\begin{align}
\lim_{v \rightarrow \infty} r^2 \Klin_{\ell \geq 2} \left(u_0,v,\theta,\phi\right) = 0  \ \ \ \  \textrm{along the null hypersurface $C_{u_0}$,} \label{rsc}
\end{align}
\begin{align}
\lim_{v \rightarrow \infty} r^2 \slashed{\mathcal{D}}_2^\star \slashed{\mathcal{D}}_2 \glinh\left(u_0,v,\theta,\phi\right) = 0  \ \ \ \  \textrm{along the null hypersurface $C_{u_0}$.} \label{rscm}
\end{align}
\item the {\bf horizon gauge conditions} if the following hold on $S_\mathcal{H}:=S^2_{\infty,v_0}$:
\begin{align}
\otx&=0 \, ,  \label{mts}
\\
\rlin - \rlin_{\ell=0} + \slashed{div}\,  \elin &= 0 \, , \label{fchoi} \
\end{align}
\item the {\bf auxiliary gauge conditions} if the following holds on $S_\mathcal{H}:=S^2_{\infty,v_0}$:
\begin{align} 
2\Olin|_{\ell=0} &=4M^2 \rlin_{\ell=0} \, , \label{omchoicel1} \\
2\Olin|_{\ell=1} &=0  \label{omchoicel2} \, ,  \\
\frac{1}{\Omega^2}\otxb_{\ell=0,1}&=0 \, , \label{resi} \\
\glinto_{\ell=1}&=0   \label{hoc} \, . 
\end{align}
\end{itemize}
Finally, 
we call the solution $\mathscr{S}$ {\bf initial-data normalised} if it satisfies all gauge conditions (\ref{omchoice})--(\ref{hoc}) above.
We typically will denote such solutions by $\Si$.
\end{definition}

We note that the above conditions can all be written explicitly in terms of the seed
data.  The auxilliary gauge conditions are related to chosing  a centre of mass frame.

We also immediately note by straightforward computation
\begin{proposition} \label{prop:kerrnormalised}
The reference Kerr solutions $\mathscr{K}$ of Definition \ref{def:kerrl} are initial data normalised. 
\end{proposition}

\subsection{Achieving the initial-data normalisation for a general $\mathscr{S}$}\label{IDnGsec}
In this section, we prove the existence of 
a pure gauge solution $\Gi$ 
such that upon adding these to a given solution $\mathscr{S}$
arising from regular asymptotically flat seed data, the resulting
solution $\Si$ is generated by data satisfying all conditions
of Section~\ref{sec:terminology}.
This will define the  initial-data normalised solution.

\begin{theorem} \label{prop:gaugeachieve}
Consider a seed data set as in Definition~\ref{def:seeddata} and let ${\mathscr{S}}$
be the resulting solution given by Theorem~\ref{theo:lwp}. Assume the seed data are
asymptotically flat 
with weight $s$ to order $n \geq 10$ as in Definition \ref{def:afpeel}. 

Then there exists a pure gauge solution $\underaccent{\lor}{\mathscr{G}}$, explicitly computable from the seed
data of ${\mathscr{S}}$,
such that the sum
\[
\Si \doteq {\mathscr{S}}+\Gi
\]
is initial-data normalised, i.e.~all gauge conditions (\ref{omchoice})--(\ref{hoc}) of Definition \ref{def:gaugechoice} hold for
$\Si$. The pure gauge solution $\Gi$ is unique and arises itself from seed data which are
asymptotically flat to order $n-2$.
\end{theorem}

\subsubsection{Overview of the proof of Theorem \ref{prop:gaugeachieve}}
The proof of Theorem \ref{prop:gaugeachieve} requires a few preparatory propositions, collected in the four Sections \ref{sec:lapsegauge}--\ref{sec:resid}. The proof proper will then be carried out in Section \ref{sec:fipap}.
For the propositions, we make frequent use of Lemmas \ref{lem:exactsol}--\ref{lem:exactsol3}. Let us describe briefly what is achieved in each individual section:

\begin{enumerate}

\item In Section \ref{sec:lapsegauge} we prove that the families of Lemma \ref{lem:exactsol} and \ref{lem:exactsol2} can generate pure gauge solutions with arbitrary prescribed linearised lapse $\Olin$ along a double null-cone $C_{u_0} \cup C_{v_0}$ emanating from a fixed sphere $S^2_{u_0,v_0}$. Such a solution will clearly be useful to achieve (\ref{omchoice}), (\ref{omchoicel1}), (\ref{omchoicel2}).

\item The question of uniqueness of such pure gauge solutions is then addressed in Section \ref{sec:lapseconstraint}. In view of the linearity of the theory this is equivalent to understanding all gauge solutions generated by Lemma \ref{lem:exactsol} and \ref{lem:exactsol2} which do not change $\Olin$ on $C_{u_0} \cup C_{v_0}$. It turns out that uniqueness holds within the class of pure gauge solutions of Lemma \ref{lem:exactsol} and \ref{lem:exactsol2} \emph{up to specifying three free functions on a fixed sphere}. The reason for this freedom essentially arises from integration ``constants" when imposing the vanishing of (\ref{mpe2}) along $C_{u_0} \cup C_{v_0}$. These free functions can essentially (up to $\ell=0,1$ modes) be used to prescribe the horizon gauge conditions (\ref{mts}), (\ref{fchoi}) and the round sphere condition (\ref{rsc}). We remark that the $\ell=0$ modes require a special treatment as one needs to address the existence of the reference linearised Schwarzschild solutions, which requires special care in achieving (\ref{omchoicel1}) and (\ref{resi}).

\item Having fully exploited the special gauge solutions of Lemmas \ref{lem:exactsol} and \ref{lem:exactsol2} in the first two steps above we turn to Lemma \ref{lem:exactsol3}. Note that such pure gauge solutions only generate non-trivial values for $\glinh, \glinto$ and $\bmlin$, hence they do not interfere with the gauge conditions in the first two steps above. In Section \ref{sec:puremetricg} we construct a pure gauge solution which will allows us to achieve (\ref{shiftco}), the gauge solution being unique up to a pure gauge solution changing only $\glinh, \glinto$.

\item In Section \ref{sec:resid} we finally exploit the ``residual freedom" mentioned at the end of $3.$ to construct a pure gauge solutions allowing us to achieve (\ref{hoc}) and (\ref{rscm}). 
\end{enumerate}

\subsubsection{Pure gauge solutions with prescribed initial lapse} \label{sec:lapsegauge}

We show that we can use Lemma \ref{lem:exactsol} and Lemma \ref{lem:exactsol2} to obtain a pure gauge solution $\mathscr{G}$ with prescribed linearised lapse $\Olin$ on $C_{u_0} \cup C_{v_0}$:
\begin{proposition} \label{prop:lapsechoice}
Fix a sphere $S^2\left(u_0,v_0\right)$ in $\mathcal{M}$ with corresponding outgoing cone $C_{u_0}$ and ingoing cone $C_{v_0}$. Let $\Omega_{out}\left(v,\theta,\phi\right)$ be a bounded smooth function on $C_{u_0}$ and $\Omega_{in}\left(u,\theta,\phi\right)$ be a bounded function, smooth in the extended sense on $C_{v_0}$, such that $\Omega_{in} \left(u_0,\theta,\phi\right) = \Omega_{out}\left(v_0,\theta,\phi\right)$ holds on the sphere  $S^2\left(u_0,v_0\right)$. Then there exists a pure gauge solution $\mathscr{G}$ of the system of gravitational perturbations such that
\begin{align}
2\Olin \left(u_0,v,\theta,\phi\right) = \Omega_{out} \left(v,\theta,\phi\right) \ \ \ \textrm{and} \ \ \ \ 2\Olin \left(u,v_0,\theta,\phi\right) = \Omega_{in} \left(u,\theta,\phi\right) \, . \nonumber
\end{align}
\end{proposition}

\begin{proof}
Let $f_{out}$ be a function along $C_{u_0}$ determined as the solution to the ODE (for each $\theta, \phi$)
\begin{align} \label{ode1}
\partial_v f_{out} + \frac{2M}{r^2\left(u_0,v\right)} f_{out} = \Omega_{out} - \frac{1}{2} \Omega_{in} \left(u_0, \theta,\phi\right) \ \ \ \ , \ \  \ \ f_{out}\left(v_0,\theta,\phi\right) = 0 \, .
\end{align}
Let $f_{in}$ be a function along $C_{v_0}$ determined as the solution to the ODE
\[
\partial_u f_{in} - \frac{2M}{r^2\left(u,v_0\right)} f_{in} = \Omega_{in} - \frac{1}{2} \Omega_{out} \left(v_0,\theta,\phi\right) \ \ \ \ , \ \  \ \ f_{in}\left(u_0,\theta,\phi\right) = 0 \, .
\] 
We claim that the pure gauge solution generated by applying Lemma \ref{lem:exactsol} with $f_{out}$ added to the pure gauge solution generated by applying Lemma \ref{lem:exactsol2} with $f_{in}$, yields the desired solution. To see this, we compute
\begin{align}
2\Olin &= \partial_v f_{out} + \frac{2M}{r^2\left(u,v\right)} f_{out} +  \partial_u f_{in} - \frac{2M}{r^2\left(u,v\right)} f_{in}  \nonumber \\
&=  -\frac{2M}{r^2\left(u_0,v\right)} f_{out} + \Omega_{out} - \frac{1}{2}\Omega_{in}\left(u_0,\theta,\phi\right) + \frac{2M}{r^2\left(u,v\right)} f_{out} \nonumber \\
& \ \ \ +\frac{2M}{r^2\left(u,v_0\right)} f_{in} + \Omega_{in} - \frac{1}{2}\Omega_{out}\left(v_0,\theta,\phi\right) - \frac{2M}{r^2\left(u,v\right)} f_{in} \, , \nonumber
\end{align}
and hence
\[
2\Olin\left(u_0,v,\theta,\phi\right) = \Omega_{out}\left(v,\theta,\phi\right) 
\ \ \ , \ \ \ 
2\Olin\left(u,v_0,\theta,\phi\right) = \Omega_{in}\left(v_0,\theta,\phi\right)  \, .
\]
\end{proof}

Writing out the explicit solution of the ODE (\ref{ode1}) one also deduces from Lemma \ref{lem:exactsol} the following corollary (cf.~Remark \ref{rem:fv}):

\begin{corollary} \label{cor:np1}
If the function $\Omega_{out}$ on the outgoing cone $C_{u_0}$ arises as the quantity $\Olin_{\circ,out}$ of a seed data set which is asymptotically flat to order $n\geq 2$, i.e.~in particular (\ref{aoci1}) holds, then the pure gauge solution $\mathscr{G}$ constructed in Proposition \ref{prop:lapsechoice} arises itself from seed data which are asymptotically flat to order $n-2$.
\end{corollary}
\begin{proof}
For $f_{out}$ one easily checks the required estimates (\ref{aoci1})--(\ref{aoci}) on $C_{u_0}$ from Lemma \ref{lem:exactsol}, in particular that $|\frac{f_{out}}{r}|$ and $|r^2 \partial_v\frac{f_{out}}{r}|$ are uniformly bounded. For $f_{in}$ one uses Lemma \ref{lem:exactsol2} and the fact that $f_{in}$ is uniformly bounded with $f_{in}=0$ for $u=u_0$.
\end{proof}

For later purposes, we also note the following special case of Proposition \ref{prop:lapsechoice}:
\begin{corollary} \label{cor:lapsechoice}
Let $\mathfrak{m}\in \mathbb{R}$. The functions 
\[
f_{out} \left(v\right) = \frac{\mathfrak{m}}{2}  \Omega^{-2} \left(u_0,v\right) \left[ r\left(u_0,v\right) - r\left(u_0,v_0\right) \right]
\]
\[
f_{in} \left(u\right) = \frac{\mathfrak{m}}{2}  \Omega^{-2} \left(u,v_0\right) \left[ -r\left(u,v_0\right) + r\left(u_0,v_0\right) \right]
\]
generate a (spherically symmetric) pure gauge solution satisfying $2\Olin= \mathfrak{m}$ 
along both $C_{u_0}$ and $C_{v_0}$. It furthermore satisfies
 \begin{align} \label{polk}
 \rlin \left(\infty,v_0,\theta,\phi\right) &=-\frac{3}{(2M)^3} \frac{\mathfrak{m}}{2} \left(r\left(u_0,v_0\right) - 2M\right) \, ,
\nonumber \\
\otx \left(\infty,v_0,\theta,\phi\right) &= -\frac{1}{4M^2} \mathfrak{m} \left(r\left(u_0,v_0\right) - 2M\right) \, ,
 \nonumber \\
\Omega^{-2}\otxb \left(\infty,v_0,\theta,\phi\right) &= - \frac{1}{2M} \mathfrak{m} \cdot \frac{r\left(u_0,v_0\right)}{2M}\, .
 \end{align}
\end{corollary}
\begin{proof}
Direct computation using Lemma \ref{lem:exactsol} and Lemma \ref{lem:exactsol2}. For (\ref{polk}) recall that $f_{out}\left(v_0\right)=0$.
\end{proof}
\subsubsection{Pure gauge solutions with vanishing initial lapse} \label{sec:lapseconstraint}
In this section we explicitly parametrise the space of (the sum of) all special gauge transformations arising from Lemma \ref{lem:exactsol} and \ref{lem:exactsol2} which satisfy the condition
\begin{align} \label{gauck}
\textrm{ $\Olin=0$ along both $C_{u_0}$ and $C_{v_0}$.}
\end{align}

\begin{lemma} \label{lem:solgen} 
Let $h_1,h_2,h_3$ be smooth functions on the unit sphere and $R:=r\left(u_0,v_0\right)$. Then
\begin{align}
f_{1} \left(u,\theta,\phi\right) = \frac{1}{\Omega^2\left(u,v_0\right)} \left(h_2\left(\theta,\phi\right) r\left(u,v_0\right) + h_3\left(\theta,\phi\right)- \frac{2M}{r\left(u,v_0\right)}  h_1\left(\theta,\phi\right)\right)
\end{align}
\begin{align}
 f_{2} \left(v,\theta,\phi\right) 
= \frac{1}{\Omega^2\left(u_0,v\right)\Omega^2\left(u_0,v_0\right)} \Bigg(
&  \ \ \ h_2\left(\theta,\phi\right)  \left(r\left(u_0,v\right) \left[1- \frac{4M}{R}\right] - \frac{2M}{r\left(u_0,v\right)}R -R + 6M \right)  \nonumber \\
&+h_1\left(\theta,\phi\right) \left(\frac{2M}{R^2} r\left(u_0,v\right) + \frac{4M^2}{R r\left(u_0,v\right)}+1-\frac{6M}{R} \right) \nonumber \\
&+h_3\left(\theta,\phi\right) \left(-\frac{2M}{R^2}r\left(u_0,v\right) - \frac{2M}{r\left(u_0,v\right)}+\frac{4M}{R}\right) \Bigg) \nonumber
\end{align}
and $j_3\equiv j_4 \equiv 0$ generates a pure gauge solution satisfying (\ref{gauck}). Moreover, any pure gauge solution in Proposition \ref{lem:pgs} satisfying $j_3=j_4=0$ as well as  (\ref{gauck}) is of the form above.
\end{lemma}

Before we embark on the proof, note that the function $f_{1} \left(u,\theta,\phi\right) \cdot \Omega^2 \left(u,v\right)$ is indeed smooth in the extended sense on $\mathcal{M}$ since $\Omega^2\left(u,v\right) \cdot \Omega^{-2} \left(u,v_0\right) = \frac{r\left(u,v_0\right)}{r\left(u,v\right)} e^{(-r\left(u,v\right) + r\left(u,v_0\right))/2M} e^{(v-v_0)/2M}$ is smooth in the extended sense.

\begin{proof}
We note $f_{2}\left(v_0,\theta,\phi\right)= h_1\left(\theta,\phi\right) $ and $\partial_v(f_{2})\left(v_0,\theta,\phi\right) = h_2\left(\theta,\phi\right)$ \\ and $\frac{1}{\Omega^2\left(u,v\right)} \left( f_{1} \Omega^2\left(u,v\right)\right)_u \Big|_{v=v_0} = -\left(h_2\left(\theta,\phi\right) + h_1\left(\theta,\phi\right) \frac{2M}{r^2\left(u,v_0\right)}\right)$, which verifies $(\ref{mpe2})=0$ along $C_{v_0}$. Along $C_{u_0}$ we compute
\begin{align}
\frac{1}{\Omega^2\left(u,v\right)} \left( f_{2} \Omega^2\left(u,v\right)\right)_v \Big|_{u=u_0} = {\Omega^2\left(u_0,v_0\right)} \Bigg[ h_2\left(\theta,\phi\right) \left(1-\frac{4M}{R} + \frac{2M}{r^2\left(u_0,v\right)} R\right) \nonumber \\
+ h_1\left(\theta,\phi\right) \left(\frac{2M}{R^2} - \frac{4M^2}{Rr^2\left(u_0,v\right)} \right) + h_3\left(\theta,\phi\right) \left(-\frac{2M}{R^2} + \frac{2M}{r^2\left(u_0,v\right)}\right)\Bigg] \, , \nonumber
\end{align}
which verifies $(\ref{mpe2})=0$ along $C_{u_0}$ after observing that
\begin{align}
\left(f_{1}\right)_u \left(u_0,\theta,\phi\right) - \frac{2M}{r^2\left(u_0,v\right)} f_{1} \left(u_0,\theta,\phi\right) \nonumber \\
= \frac{1}{\Omega^2\left(u_0,v_0\right)} \Bigg\{ \left( -\frac{2M}{r^2\left(u_0,v\right)} + \frac{2M}{R^2}\right) \left[h_2\left(\theta,\phi\right)R +h_3\left(\theta,\phi\right) - \frac{2M}{R}h_1\left(\theta,\phi\right) \right] \nonumber \\
 - \left(h_2\left(\theta,\phi\right)+ \frac{2M}{R^2} h_1\left(\theta,\phi\right)\right)\left(1-\frac{2M}{R}\right) \Bigg\} \, . \nonumber
\end{align}
For the uniqueness assertion note that an arbitrary pure gauge solution $\left(f_1,f_2,j_3=0,j_4=0\right)$ satisfying (\ref{gauck}) is uniquely determined by specifying $f_{2}\left(v_0,\theta,\phi\right)$, $\partial_v f_{2}\left(v_0,\theta,\phi\right)$ and  $f_1\left(u_0,\theta,\phi\right)$ via elementary ODE theory applied to $(\ref{mpe2})=0$ along $C_{u_0}$ and $C_{v_0}$. On the other hand, the aforementioned values are seen to be in one-to-one correspondence with the functions $h_1,h_2,h_3$.
\end{proof}

\begin{corollary} \label{cor:np2}
The pure gauge solution of Lemma \ref{lem:solgen} induces seed initial data on $C_{u_0} \cup C_{v_0}$ which is asymptotically flat to any order.
\end{corollary}
\begin{proof}
This follows from carefully going through Lemma \ref{lem:exactsol} and \ref{lem:exactsol2}. The key is to note  that $f_1$ is uniformly bounded and that $f_2$ satisfies the estimates $|\frac{f_2}{r}|\lesssim 1$, $|r^2 \partial_v \left(\frac{f_2}{r}\right)| \lesssim 1$ and $|r^2 \partial_v \left( r^2 \partial_v \left(\frac{f_2}{r} \right)\right)| \lesssim 1$, the latter allowing one to obtain the estimates (\ref{aoci1}). The statements about higher derivatives in (\ref{aoci1})--(\ref{aoci}) are then straightforward.
\end{proof}

The gauge transformations of Lemma \ref{lem:solgen} can be used to prescribe additional geometric quantities on the horizon sphere $S^2_{\infty,v_0}$. We first state the three fundamental propositions before proving them.

\begin{proposition} \label{prop:hozgauge} 
Let $X^1,X^2$ be smooth functions on $S^2_{\infty,v_0}$ with vanishing projection to $\ell=0$ and $\ell=1$. There exists a pure gauge solution $\mathscr{G}\left(f_1,f_2,j_3=0,j_4=0\right)$ satisfying (\ref{gauck}), the round sphere condition (\ref{rsc}) and
\begin{align} \label{gsolp} 
\otx \left(\infty,v_0,\theta,\phi\right)  &= X^1 \, , \nonumber \\
\slashed{div}\, \elin+ \rlin \left(\infty,v_0,\theta,\phi\right) &= X^2 \, ,
\end{align}
Moreover, the functions $f_1$ and $f_2$ of the pure gauge solution $\left(f_1,f_2,j_3=0,j_4=0\right)$ are uniquely determined with their projection to $\ell=0$ and $\ell=1$ vanishing. Finally, the pure gauge solution induces asymptotically flat (to any order) seed data  on $C_{u_0} \cup C_{v_0}$.
\end{proposition}

For the $\ell=1$ modes we have an additional degree of freedom which stems from the fact the round sphere condition is always satisfied as the pure gauge solutions below cannot alter the linearised Gaussian curvature $\Klin$:
\begin{proposition} \label{prop:hozgauge2} 
Let $X^1,X^2,X^3$ be smooth functions on $S^2_{\infty,v_0}$ all supported on $\ell=1$ only. There exists a pure gauge solution $\mathscr{G}\left(f_1,f_2,j_3=0,j_4=0\right)$ satisfying (\ref{gauck}), which in addition satisfies
\begin{align} \label{gsolp2} 
\otx \left(\infty,v_0,\theta,\phi\right)  &= X^1 \, , \nonumber \\
\slashed{div} \, \elin + \rlin  \left(\infty,v_0,\theta,\phi\right) &= X^2 \, , \\
\Omega^{-2} \otxb  \left(\infty,v_0,\theta,\phi\right) &=X^3 \, . \nonumber
\end{align}
Moreover, the functions $f_1$ and $f_2$ of the pure gauge solution $\left(f_1,f_2,j_3=0,j_4=0\right)$ are uniquely determined and supported on $\ell=1$ only. Finally, the pure gauge solution induces asymptotically flat (to any order) seed data on $C_{u_0} \cup C_{v_0}$.
\end{proposition}

For the $\ell=0$ modes we cannot fix all three geometric quantities on the horizon independently because of a spherically symmetric degree of freedom that is \emph{not} pure gauge and corresponding to a linearised Schwarzschild solution,  discussed in Section \ref{sec:linss}. However, we can combine Lemma \ref{lem:solgen} with Corollary \ref{cor:lapsechoice} to prove:
\begin{proposition} \label{prop:hozgauge3} 
Let $\mathfrak{m}, X^1,X^3$ be constants on $S^2_{\infty,v_0}$. There exists a pure gauge solution $\mathscr{G}$ generated by $\left(f_1,f_2,j_3=0,j_4=0\right)$ with $f_1$ and $f_2$ being spherically symmetric which satisfies (\ref{omchoice}) and in addition 
\begin{align} \label{gsolp3} 
2 \Olin \left(\infty,v_0,\theta,\phi\right)  &= \mathfrak{m} \\ 
\otx \left(\infty,v_0,\theta,\phi\right)  &= X^1 \, ,\\
\Omega^{-2} \otxb  \left(\infty,v_0,\theta,\phi\right) &=X^3 \, .
\end{align}
Moreover, $f_1$ and $f_2$ are unique up to a constant $f_1=f_2=\lambda$ and hence the pure gauge solution is unique (as $\left(\lambda,\lambda,0,0\right)$ generates the zero solution). The solution also necessarily satisfies
\[
\rlin\left(\infty,v_0,\theta,\phi\right) = \frac{3}{4M}X^1 \, .
\]
Finally, the pure gauge solution induces asymptotically flat (to any order) seed data on $C_{u_0} \cup C_{v_0}$. 
\end{proposition}

\begin{proof}
We will prove Propositions \ref{prop:hozgauge}--\ref{prop:hozgauge3} all at the same time. We first compute from the general solution of Lemma \ref{lem:solgen} the following geometric quantities on the horizon:
\begin{align}
\otx \left(\infty,v_0,\theta,\phi\right) &= \frac{1}{2M^2} \left[ \left(\Delta_{\mathbb{S}^2} - 1\right) \left(2M \cdot h_2 \left(\theta,\phi\right)+h_3 \left(\theta,\phi\right)-h_1 \left(\theta,\phi\right)\right)\right] ,\nonumber \\
\left(\slashed{div} \,\elin + \rlin \right) \left(\infty,v_0,\theta,\phi\right) &=  \frac{1}{(2M)^3} \left( \Delta_{\mathbb{S}^2} \left(h_3\left(\theta,\phi\right) - 2h_1\left(\theta,\phi\right)\right) - 3 \left(2M \cdot h_2\left(\theta,\phi\right) + h_3 \left(\theta,\phi\right)- h_1\left(\theta,\phi\right)\right)\right),\nonumber \\ 
\Omega^{-2} \otxb \left(\infty,v_0,\theta,\phi\right) &= \frac{1}{2M^2} \left(\Delta_{\mathbb{S}^2}  h_1 \left(\theta,\phi\right) + h_1 \left(\theta,\phi\right) - h_3 \left(\theta,\phi\right)\right). \nonumber
\end{align}
To prove Proposition \ref{prop:hozgauge} we choose
\begin{align} \label{relrs}
h_3 - h_1 = h_2 \frac{R^2}{2M} \left(1-\frac{4M}{R}\right) \, .
\end{align}
One checks that with this choice both $f_1$ and $f_2$ (as well as angular derivatives thereof) are uniformly bounded, hence $r^3 \cdot \Klin$ is uniformly bounded and in particular the round sphere condition (\ref{rsc}) holds for any $h_1,h_2,h_3$ satisfying (\ref{relrs}). Plugging this relation into the first equation above and recalling $R>2M$ we see that the first equation uniquely determines $h_2$ to satisfy the condition in the proposition. Plugging (\ref{relrs}) into the second equation to isolate $h_1$, we see that we can uniquely solve for $h_1$ to determine the second condition in the proposition. Of course $h_3$ is determined by (\ref{relrs}).

To prove Proposition \ref{prop:hozgauge2} we can restrict to $\ell=1$. The three equations above then turn into a simple algebraic system with non-zero determinant which admits a unique solution for any $X_1,X_2,X_3$ prescribed.

To prove Proposition \ref{prop:hozgauge3} we project to $\ell=0$ to see the resulting algebraic system has one-dimensional kernel $h_2=0$, $h_1=h_3$. It is easy to see that such gauge solutions are trivial. We now add the pure gauge solution from Corollary \ref{cor:lapsechoice} to the aforementioned projection to obtain (setting $h_4=h_1-h_3$)
\begin{align}
\otx \left(\infty,v_0,\theta,\phi\right) &= -\frac{1}{2M^2} \left(2M \cdot h_2 - h_4\right) - \frac{1}{4M^2} \mathfrak{m} \left(R-2M\right) \, , \nonumber \\
\Omega^{-2} \otxb \left(\infty,v_0,\theta,\phi\right)  &= \frac{1}{2M^2} h_4 -\frac{1}{2M} \mathfrak{m} \frac{R}{2M} \, , \nonumber \\
 \rlin  \left(\infty,v_0,\theta,\phi\right) &= -\frac{3}{(2M)^3} \left(2M \cdot h_2 - h_4\right) - \frac{3}{(2M)^3} \frac{\mathfrak{m}}{2} \left(R-2M\right) \, .
\end{align}
Note that the right hand side of the third is a multiple of the first. It is immediate that we can solve the first two equations uniquely for $h_2, h_4$ given any left hand side and given any $\mathfrak{m}$. This provides the statement in the Proposition recalling that $h_2,h_4$ do not alter $\Olin$ on $C_{u_0} \cup C_{v_0}$. The uniqueness follows since (\ref{omchoice}) and (\ref{gsolp3}) at the horizon fix $\Olin$ on $C_{u_0} \cup C_{v_0}$, so the remaining gauge solution must be of the type of Lemma \ref{lem:solgen} projected to $\ell=0$. This solution is trivial if $\otx \left(\infty,v_0,\theta,\phi\right) =\Omega^{-2} \otxb \left(\infty,v_0,\theta,\phi\right)= 0$. Finally, the assertion about asymptotic flatness is a consequence of Corollary \ref{cor:np2}.
\end{proof}

\subsubsection{Pure gauge solutions with prescribed shift} \label{sec:puremetricg}

 In this section we exploit pure gauge solutions arising from 
 Lemma~\ref{lem:exactsol3}, which we recall only generate non-trivial metric components $\glinh$, $\glinto$ and $\bmlin$, 
 while all Ricci and curvature coefficients vanish. In particular, any such pure gauge solution will automatically satisfy all gauge conditions of Definition \ref{def:gaugechoice} except (\ref{shiftco}) and (\ref{hoc}).

\begin{proposition} \label{prop:zeroshift}
Let $\tilde{b}$ be a smooth $S^2_{u,v}$-valued 1-form 
prescribed along $C_{u_0}$. There exists a pure gauge solution of the type of Lemma \ref{lem:exactsol3}, which satisfies
\[
\bmlin = \tilde{b} \ \ \ \textrm{along $C_{u_0}$} 
\]
Moreover, except for $\glinh$ and $\frac{\glinto}{\sqrt{\slashed{g}}}$ which are potentially non-vanishing, all linearised Ricci and curvature components of this pure gauge solution vanish. The solution is unique up to a pure gauge solution generated by functions $q_1$ and $q_2$ depending only on the angular variables (Proposition \ref{prop:metricg}).

\end{proposition}

\begin{proof}
By Lemma \ref{lem:exactsol3} we need to determine $q_1$, $q_2$ solving
\begin{align} \label{detq}
\partial_v \left(\Delta_{S^2} q_1\right) = -\slashed{div}\, \tilde{b} \ \ \ , \ \ \ \partial_v \left(\Delta_{S^2} q_2\right) = -\slashed{curl}\, \tilde{b} \ \ \ \textrm{along $C_{u_0}$}
\end{align}
where $\Delta_{S^2}=r^2 \slashed{\Delta}$ is defined with respect to the round \emph{unit} sphere. We can solve these ODEs uniquely prescribing $\Delta_{S^2} q_1$ and $\Delta_{S^2} q_2$ freely (as functions with vanishing mean) initially at $v=v_0$ accounting for the non-uniqueness asserted in the proposition. The conclusions now follow from Lemma \ref{lem:exactsol3}. 
\end{proof}

\begin{corollary} \label{cor:np3}
If $\tilde{b}$ in Proposition \ref{prop:zeroshift} satisfies $|\left(r \slashed{\nabla}\right)^{n_1} \left(r \slashed{\nabla}_4\right)^{n_2} \tilde{b}|\lesssim v^{-1}$ along $C_{u_0}$ for $n_1+n_2 \leq n$ (as is the case when $\tilde{b}$ arises as the quantity $\bmlin_\circ$ of a seed data set which is asymptotically flat to order $n$, cf.~(\ref{aoci2})) then the pure gauge solution induces data on $C_{u_0} \cup C_{v_0}$ which are asymptotically flat to order $n$.
\end{corollary}

\subsubsection{Residual pure gauge solutions} \label{sec:resid}
We finally give an explicit parametrisation of the kernel in Proposition  \ref{prop:zeroshift} which we recall is generated by $q_1$ and $q_2$ being smooth function of the unit sphere the proof of which is immediate from Lemma \ref{lem:exactsol3}.

\begin{proposition} \label{prop:metricg}
Let $q_1$ and $q_2$ be smooth functions on the unit sphere. 
Then there exists a pure gauge transformation satisfying
\begin{align}
\glinh \left(u,v,\theta,\phi\right)= 2r^2 \slashed{\mathcal{D}}_2^\star \slashed{\mathcal{D}}_1^\star \left(q_1,q_2\right) \ \ \ , \ \ \ \frac{\glinto}{\sqrt{\slashed{g}}}=  r^2 \slashed{\Delta} q_1  \, ,
\end{align}
and with all other metric and Ricci coefficients and curvature components globally vanishing. In particular, the pure gauge solution is asymptotically flat to any order.
\end{proposition}
Note the above solutions in particular do not change the linearised
Gaussian curvature $\Klin$. 
We will use them below to bring the metric into ``standard form" on the sphere at infinity, i.e.~to achieve (\ref{rscm}) once we have (\ref{rsc}).

\subsubsection{Proof of Theorem \ref{prop:gaugeachieve}} \label{sec:fipap}
We can now prove Theorem \ref{prop:gaugeachieve}. Let $\mathscr{S}$ be as in the Proposition.

Applying Proposition \ref{prop:lapsechoice} with 
$\Omega_{out}= -\Olin_{\circ,out}$ and $\Omega_{in} = -\Olin_{\circ,in}$, we achieve that the sum of the original solution $\mathscr{S}$ and the pure gauge solution generated by Proposition \ref{prop:lapsechoice} satisfies $\Olin=0$ along $C_{u_0} \cup C_{v_0}$. We denote this solution by $\mathscr{S}_1$. 

The weighted geometric quantity $r^2 \Klin_{\ell \geq 2}$ of the solution $\mathscr{S}_1$ converges pointwise with at least $n-4$ angular derivatives $r\slashed{\nabla}$ to a smooth function $X_4\left(\theta,\phi\right)$ on the unit sphere along the cone $C_{u_0}$ as $v \rightarrow \infty$. This follows from (the first part of) Theorem \ref{prop:pwprop} in the appendix and the fact that $r^2\Klin$ has such a limit for the pure gauge solution applied in the previous step. Let $\bar{f}$ be the unique solution of the equation $\Delta_{\mathbb{S}^2} \bar{f} + 2\bar{f} = X_4$ on the unit sphere which has vanishing projection to $\ell=1$.

We now apply Proposition \ref{prop:lapsechoice} again, this time with $\Omega_{out}= \bar{f}$ and $\Omega_{in} = \bar{f}$ and add the resulting pure gauge solution to $\mathscr{S}_1$. The solution thus obtained will be denoted $\mathscr{S}_2$.
The solution $\mathscr{S}_2$ clearly satisfies (\ref{omchoice}) and also (\ref{rsc}), in fact it satisfies
\begin{align} \label{higherround}
\lim_{v\rightarrow \infty} \left( \left(r \slashed{\nabla}\right)^k r^2 \Klin_{\ell\geq 2} \left[\mathscr{S}_2\right] \left(u_0,v,\theta,\phi\right) \right)= 0
\end{align}
for $k\leq n-4$. To see the last claim, note that the $f_{out}$ associated with Proposition \ref{prop:lapsechoice} precisely cancels the weighted Gaussian curvature $r^2 \Klin$ at infinity of the solution $\mathscr{S}_1$ as can be seen directly from Lemma \ref{lem:exactsol}. On the other hand, the contribution from $f_{in}$ through Lemma \ref{lem:exactsol2} does not affect $r^2\Klin$ at infinity since $f_{in}$ is uniformly bounded.

Let now
\[
\otx|_{S^2_{\infty,v_0}}=X_1 \ ,  \  \left(\slashed{div} \,\elin +\rlin\right)\Big|_{S^2_{\infty,v_0}}|_{S^2_{\infty,v_0}}=X_2 \ , \  \Omega^{-2} \otxb |_{S^2_{\infty,v_0}}=X_3
\]
for $X_1,X_2,X_3$ smooth functions on the sphere $S^2_{\infty,v_0}$ denote the geometric quantities on the horizon for $\mathscr{S}_2$. We apply Proposition \ref{prop:hozgauge} to generate a pure gauge solution $\mathscr{G}_1$ satisfying (\ref{omchoice}) and (\ref{rsc}) and in addition $\otx|_{S^2_{\infty,v_0}}=X_1- \left(X_1\right)_{\ell=0,1}$ and $ \left(\slashed{div} \,\elin +\rlin \right)\big|_{S^2_{\infty,v_0}}=X_2-\left(X_2\right)_{\ell=0,1}$, where the notation indicates that the projection to $\ell=0$ and $\ell=1$ has been removed. We apply Proposition \ref{prop:hozgauge2} to generate a pure gauge solution $\mathscr{G}_2$ satisfying (\ref{omchoice}) and (\ref{rsc}) and in addition $\otx|_{S^2_{\infty,v_0}}= \left(X_1\right)_{\ell=1}$,  $\left(\slashed{div}\, \elin +\rlin\right)|_{S^2_{\infty,v_0}}=\left(X_2\right)_{\ell=1}$ and $\Omega^{-2}\otxb|_{S^2_{\infty,v_0}}=\left(X_3\right)_{\ell=1}$. We apply Proposition \ref{prop:hozgauge3} to generate a pure gauge solution $\mathscr{G}_3$ which satisfies $\otx|_{S^2_{\infty,v_0}}= \left(X_1\right)_{\ell=0}$ and $\Omega^{-2} \otxb|_{S^2_{\infty,v_0}}=\left(X_3\right)_{\ell=0}$ and $2\Olin|_{S^2_{\infty,v_0}}=4M^2 \left( -\left(X_2\right)_{\ell=0}+\frac{3}{4M} \left(X_1\right)_{\ell=0}\right)$, the last holding on all of $C_{u_0} \cup C_{v_0}$ by (\ref{omchoice}). By Proposition \ref{prop:hozgauge3} the solution necessarily satisfies $\rlin|_{S^2_{\infty,v_0}}=+\frac{3}{4M}\left(X_1\right)_{\ell=0}$.

If we now define the solution $\mathscr{S}_3:=\mathscr{S}_2 - \mathscr{G}_1 - \mathscr{G}_2 - \mathscr{G}_3$, then this solution satisfies (\ref{omchoice}), (\ref{rsc}) and also (\ref{omchoicel1}), (\ref{omchoicel2}) as well as the horizon gauge conditions and the auxiliary condition (\ref{resi}). One also checks directly that any pure gauge solution of the form $\left(f_1,f_2,j_3=0,j_4=0\right)$ with these properties is necessarily trivial.

The solution $\mathscr{S}_3$ satisfies $\bmlin=\tilde{b}$ along $C_{u_0}$ for some smooth $v$-valued $S^2_{u_0,v}$-one form $\tilde{b}$ along $C_{u_0}$. We apply Proposition \ref{prop:zeroshift} for $\tilde{b}^A-\frac{2}{3} r^3 \slashed{\epsilon}^{AB} \partial_B \slin_{\ell=1} \left[\mathscr{S}_3\right]$ and subtract (a representative of) the pure gauge solution generated by it from $\mathscr{S}_3$. We denote the resulting solution by  $\mathscr{S}_4$.

The solution $\mathscr{S}_4$ satisfies all of the desired gauge conditions except (\ref{hoc}) and (\ref{rscm}) and any such solution is determined up to a pure gauge solution of Proposition \ref{prop:metricg}. The geometric quantity $\glinh$ has a smooth limit along the cone $C_{u_0}$ and converges to a symmetric traceless $S^2_{u,v}$-tensor $\glinh_{\infty} \left(\theta,\phi\right)$ on the unit sphere. This follows from the assumptions in Proposition \ref{prop:pwprop} in conjunction with the constraint equations along $C_{u_0}$ and the fact that this is true for all pure gauge transformations applied so far. We solve the elliptic equation
\[
2r^2 \slashed{\mathcal{D}}_2^\star \slashed{\mathcal{D}}_1^\star \left(q_1,q_2\right)= \glinh_{\infty} \left(\theta,\phi\right)
\]
for $q_1, q_2$ on the unit sphere, which can be done uniquely up to $\ell=0$ and $\ell=1$ modes of $q_1$ and $q_2$. Noting that $\ell=0$-modes and the $\ell=1$ mode of $q_2$ generate trivial pure gauge solutions, we have determined $q_1,q_2$ up to trivial pure gauge solutions and the three $\ell=1$ modes for $q_1$. Let $X_5$ be the value of $\glinto_{\ell=1}$ on the horizon sphere $S^2_{\infty,v_0}$ of the solution $\mathscr{S}_4$ we determine $q_1$ uniquely by solving $\Delta_{S^2} q_1=X_4$. We now subtract the gauge solution generated by Proposition \ref{prop:metricg} for $q_1$ and $q_2$ as above from $\mathscr{S}_4$. We denote the resulting solution by $\mathscr{S}_5$. It is easy to see that all of the desired gauge conditions are now satisfied. The assertion about $\Gi$ (and hence $\Si$) being asymptotically flat follows from Corollaries \ref{cor:np1}, \ref{cor:np2} and \ref{cor:np3}.

\subsection{The horizon-renormalised solution $\protect\Sf$}
\label{hngaugesec}
As discussed already in Section~\ref{EDWgauges}, our main decay theorem, Theorem~\ref{theo:mtheod}, will require
passing to a new gauge normalised from the event horizon values of $\Si$. 
The following proposition defines and proves the existence of the horizon-renormalised 
solution $\Sf$:
\begin{proposition} \label{prop:hozrgauge}
Let $\Si$ be an initial-data normalised solution as in
Definition~\ref{def:seeddata}.
Then there exists a unique pure gauge solution $\Gf$ of the type of Lemma \ref{lem:exactsol},  computable from the
trace of $\Si$ on the event horizon $\mathcal{H}^+$,
such that the sum 
\begin{align}
\Sf \doteq \Si+\Gf
\end{align}
has the following properties:
\begin{enumerate}
\item The projection to $\ell\geq 2$ of the linearised lapse vanishes along the event horizon for $\Sf$, i.e.
\begin{align} \label{lapsevanish}
\Olin_{\ell\geq2} = 0 \ \ \ \textrm{holds along the event horizon $\mathcal{H}^+$}
\end{align}
\item The pure gauge solution $\Gf$ satisfies
\begin{align}
\Omega^{-2} \otxb \Big|_{S^2_{\infty,v_0}} = 0  \ \ \ \ \textrm{on the horizon sphere $S^2_{\infty,v_0}$. }
\end{align}
\item The function $f$ generating $\Gf$ has vanishing projection to $\ell=0,1$.
\end{enumerate}
We call $\Sf$ the  {\bf horizon-renormalised solution}. 
\end{proposition}

\begin{proof}
 Let $f$ be determined as the unique solution to the ODE
\begin{align} \label{odef}
\partial_v f + \frac{1}{2M} f = -2\Olin_{\ell \geq 2} \left[\Si\right] \left(\infty,v,\theta,\phi\right) \ \ \textrm{with} \ \ f\left(v_0, \theta, \phi\right)=0 \, 
\end{align}
where the right hand side of the ODE denotes the projection to $\ell \geq 2$ of the trace of the linearised lapse of $ \Si$ along the event horizon. Given $f$ as above, we define $\Gf$ to be the pure gauge solution associated by Lemma \ref{lem:exactsol} and $\Sf=\Si+\Gf$. It is easy to check that $\Sf$ satisfies (\ref{lapsevanish}). The uniqueness statement follows since satisfying the ODE is a necessary condition for (\ref{lapsevanish}) to hold and $f\left(v_0,\theta,\phi\right)=0$ is required by expression for $\Omega^{-2} \otxb \Big|_{S^2_{\infty,v_0}}$ in Lemma \ref{lem:exactsol}. 
\end{proof}

\begin{remark}
Since $f$ is supported for $\ell \geq 2$ only one easily sees from Lemma \ref{lem:exactsol} and Theorem \ref{etsilew} below that in fact one also has $\Olin_{\ell=1} = 0$ on $\mathcal{H}^+$ as this holds both for a reference linearised Kerr solution and for the solution $\Gf$.
\end{remark}

Note that $\Sf$ still satisfies the horizon gauge conditions (\ref{mts}), (\ref{fchoi}). Note also that if we apply the Proposition for a reference linearised Kerr solution, i.e.~with $\Si=\mathscr{K}$, then $\Sf=\Kf=\mathscr{K}$, so the reference Kerr is both in the initial data and in the horizon normalised gauge. Another way to say this is that the pure gauge solution $\Gf$ is not supported on $\ell=0,1$, the terminology being introduced below in Definition \ref{def:solsup} below.

In contrast to Theorem \ref{prop:gaugeachieve} concerning the initial-data normalised solution $\Si$ which
states asymptotic flatness for $\Gi$, at this point, we do not
know that $\Gf$ enjoys this property.
Thus, a priori the $\Sf$ defined by Proposition~\ref{prop:hozrgauge} 
may have  data which
are not asymptotically flat, even if the data corresponding to $\Si$ are
asymptotically flat.

While showing that $\Sf$ is asymptotically flat in this case
in the sense of Definition~\ref{def:afpeel} 
would
require an improvement of our polynomial decay bounds for gauge invariant quantities and decay estimates at all orders of derivatives, we will prove, in the context of the proof of Theorem~\ref{theo:mtheod}, \emph{weighted boundedness} estimates for $\Gf$. See Remarks \ref{rem:afrem} and \ref{rem:afrem2}.

\subsection{Global properties of the gauge-normalised solutions}
\label{sec:globcon}
In this section, we collect some global properties of  the system of linearised gravity that follow for the initial data-normalised and horizon-renormalised solutions
$\Si$ and $\Sf$.

\subsubsection{Propagation along the event horizon} \label{sec:hozprop}
We first deduce two conservation laws along the event horizon $\mathcal{H}^+$.
\begin{proposition} \label{prop:propagate} 
Consider a seed data set as in Definition~\ref{def:seeddata}, let $\mathscr{S}$
be the resulting solution given by Theorem~\ref{theo:lwp}. If $\mathscr{S}$ satisfies
the horizon gauge conditions (\ref{mts}) and (\ref{fchoi}) then we have
\begin{equation}
\label{theypropag}
\otx = 0 \textrm{ \ \ \ \ and \ \ \ \ }  \rlin\, - \rlin_{\ell =0} + \slashed{div}\, \elin = 0 \ \ \textrm{ \ \ pointwise along $\mathcal{H}^+$.}
\end{equation}
The assumption and hence the conclusion holds in particular for the solutions $\Si$ and $\Sf$ defined above. 
\end{proposition}

\begin{proof}
We write the linearised Raychaudhuri equation (\ref{vray}) as
\begin{align} \label{hoyu}
\partial_v \left(e^{-\frac{v}{2M}} r^2 \otx \right) = 0 \ \ \ \textrm{along} \ \mathcal{H}^+
\end{align}
and use the assumption that the quantity in brackets is zero on the initial sphere.
To show the second bound we note that using the first, we have the following propagation equation along $\mathcal{H}^+$:
\begin{align}
\slashed{\nabla}_4 \left( r^3 \rlin \, - r^3 \rlin_{\ell=0} + r^3 \slashed{div} \, \elin \right) = 0 \, 
\end{align}
and from (\ref{fchoi}) we conclude $r^3 \rlin \, - r^3 \rlin_{\ell=0} + r^3 \slashed{div} \, \elin =0$ pointwise on $\mathcal{H}^+$. The claim about $\Si$ is immediate as it satisfies (\ref{mts}) and (\ref{fchoi}) by definition. The claim about $\Sf$ follows since $\Gf$ satisfies $\elin=0$ and $\otx=\rlin=0$ on the horizon $\mathcal{H}^+$.
\end{proof}

\subsubsection{The geometric quantities $\protect\Ylin$ and $\protect\Zlin$}

In this section we will define two auxiliary quantities which will play a key role later in the analysis.  Specifically, we will later assume uniform boundedness of these quantities on the initial data and show that this is propagated in evolution. 

Besides the definition, we also prove two propositions, which show that the initial uniform boundedness of these quantities can in fact be \emph{deduced} for the solution $\Si$ arising from Theorem \ref{prop:gaugeachieve}.

 The quantities are defined as follows: 
 \begin{align} \label{Ydef}
\Ylin :=  r \left[r^2 \slashed{\mathcal{D}}_2^\star  \slashed{div} \left(\Omega^{-1} r\xblin \, \right)  - \Omega^{-1} r^3 \underline{\pblin}\right] =r^3 \slashed{\mathcal{D}}_2^\star \, \elin - 3Mr \xblin \Omega^{-1} + \frac{1}{2}\frac{r^4}{\Omega^2} \slashed{\mathcal{D}}_2^\star  \slashed{\nabla} \otxb \, ,
\end{align}
\begin{align} \label{Qquant}
\Zlin_A &:= \frac{r^3}{\Omega^2} \slashed{\nabla}_A \otx  - 2 r^2 \left(\elin_A + \eblin_A\right)=\frac{r^3}{\Omega^2} \slashed{\nabla}_A \left(\otx  - \frac{4}{r} \Omega^2 \Olin\right) \, ,
\end{align}
where the non-defining equalities hold for solutions for the system of gravitational perturbations.

We start with a proposition on $\Ylin$:
\begin{proposition} \label{prop:ybounded}
Consider a seed data set as in Definition~\ref{def:seeddata}, which is asymptotically flat to order $n\geq 12$. Let $\mathscr{S}$
be the resulting solution given by Theorem~\ref{theo:lwp} and let $\Si$
be as in Theorem \ref{prop:gaugeachieve}. Then the geometric quantity $\Ylin$ associated with $\Si$ is uniformly bounded along $C_{u_0}$. 
\end{proposition}
Before we prove the proposition let us remark that  we will eventually also prove that $\Ylin$ is bounded for $\Sf$ but this will require global boundedness estimates on the pure solution $\Gf$. See Theorem \ref{theo:mtheod}.

\begin{proof}
One first derives a propagation equation for the Gaussian curvature $\Klin$ along $C_{u_0}$ which follows by taking a $\partial_v$ derivative of (\ref{lingauss}). This reads schematically
\[
\partial_v \left( \left(r \slashed{\nabla}\right)^k r^2 \Klin \right) = \mathcal{Q}
\]
where $\mathcal{Q}$ satisfies $|\mathcal{Q}|\leq Cr^{-2}$ for $k\leq n-5$ from the seed data being asymptotically flat to order $n\geq 12$, cf.~Theorem \ref{theo:AFdata}. Using the round sphere condition at infinity, (\ref{rsc}), we obtain that $r^3 \Klin$ is uniformly bounded along $C_{u_0}$. Commuting with angular derivatives and using that the proof of Theorem \ref{prop:gaugeachieve} actually gave (\ref{higherround}) one obtains that $r^2 \slashed{\mathcal{D}}_2^\star \slashed{\nabla}_A r^3  \Klin$ is similarly uniformly bounded along $C_{u_0}$, cf.~(\ref{higherround}). One finally looks at the commuted linearised Gauss equation (\ref{lingauss}) which when combined with the decay rates  (\ref{decrete}) produces the boundedness of $\Ylin$.
\end{proof}
Remarkably, as we will see, the global uniform boundedness of $\Ylin$ is actually propagated by the equations, cf.~Proposition \ref{prop:Yest}. 

We now turn to the quantity $\Zlin$ above.
\begin{proposition} \label{prop:zbounded}
Consider a seed data set as in Definition~\ref{def:seeddata}, let $\mathscr{S}$
be the resulting solution given by Theorem~\ref{theo:lwp} and let $\Si$
be as in Theorem \ref{prop:gaugeachieve}. Then the geometric quantity $\Zlin\Omega^{-2}$ associated with $\Si$ is uniformly bounded along $C_{v_0}$. Moreover, since $\Zlin_A=-2r\Omega^2\slashed{\nabla}_A f$ near the horizon for pure gauge solutions of Lemma \ref{lem:exactsol} the boundedness statement holds equivalently for $\Zlin\Omega^{-2}$ associated with $\Sf$.
\end{proposition}

\begin{proof}
Note that along $C_{v_0}$ we have
\[
\partial_u \left(\otx - \frac{4}{r} \Omega^2 \Olin \right) =
 -\frac{1}{M^2} \Olin+ \frac{1}{M^2} \Olin + \mathcal{O}\left(\Omega^4\right) = \mathcal{O}\left(\Omega^4\right) \, ,
\]
where we have used the horizon gauge conditions and the lapse gauge condition. The same statement holds for arbitrary angular commutations. Since the quantity in brackets vanishes initially on the horizon $\mathcal{H}^+$ it actually vanishes  to order $\Omega^4$ by the estimate. The fact that $\Zlin_A=-2r\Omega^2\slashed{\nabla}_A f$ for pure gauge solution of Lemma \ref{lem:exactsol} is read off directly from this Lemma, so that the last statement follows from recalling that $\Gf=\Sf-\Si$ arises from Lemma \ref{lem:exactsol}.
\end{proof}

Remarkably, as we will see, the uniform boundedness of $\Zlin \Omega^{-2}$ near the 
horizon is again actually propagated by the equations, cf.~Proposition \ref{prop:zestimate}.

\subsection{The projection to the $\ell=0, 1$ modes and the Kerr parameters}
\label{vanishingsec}
The initial-data normalisation, as we have defined it, will allow to completely
understand the
 projection of solutions $\mathscr{S}$ to their $\ell=0, 1$ modes. 
The main result of
this section is the following
\begin{theorem}\label{etsilew}
Let $\mathscr{S}$  and $\Si$ be as in Theorem~\ref{prop:gaugeachieve}. 
Then  the projection of $\Si$ to its $\ell=0,1$ modes (See Definition~\ref{def:proj01}) is a reference
linearised Kerr solution $\mathscr{K}_{\mathfrak{m}, s_i}$ where the
parameters $\mathfrak{m}$, $s_i$ are given by:
\[
\mathfrak{m}=-4M^2 \rlin_{\ell=0}|_{S^2_{\infty,v_0}}      ,\qquad s_{-1}=\slin_{\ell=1,m=-1}|_{S^2_{\infty,v_0}}, \qquad s_0=\slin_{\ell=1,m=0}|_{S^2_{\infty,v_0}}   , \qquad s_1=\slin_{\ell=1,m=1}|_{S^2_{\infty,v_0}} \, .
\]
Here $\slin_{\ell=L,m=S}$ denotes the projection of $\slin$ to the spherical harmonic $Y^{L}_{S}$.
(Thus, in particular,
solutions $\mathscr{S}$ supported only on $\ell=0,1$  
are 
a linearised Kerr plus a pure gauge solution.)
\end{theorem}
 
\subsubsection{The projection to $\ell=0,1$}

We begin with the following definition
\begin{definition} \label{def:solsup}
We say that a solution $\mathscr{S}$ of the system of gravitational perturbations is {\bf supported only on $\ell=0,1$} if 
\begin{itemize}
\item all scalars $s$ in $\mathscr{S}$ are supported on $\ell=0,1$ only 
(cf.~Definition~\ref{def:supps}),
\item all one-forms $\xi$ in $\mathscr{S}$ are supported on $\ell=1$ only (cf.~Definition~\ref{def:supp1f}), 
\item all symmetric traceless tensors $\theta$ in $\mathscr{S}$ vanish (cf.~Proposition~\ref{syracelge2}).
\end{itemize}
Conversely, we define a solution $\mathscr{S}$ to have {\bf support outside $\ell=0,1$} if 
\begin{itemize}
\item all scalars $s$ in $\mathscr{S}$ are supported on $\ell\geq 2$ only
(cf.~Definition~\ref{def:supps}),
\item all one-forms $\xi$ in $\mathscr{S}$ are supported on $\ell\geq 2$ only
(cf.~Proposition~\ref{syracelge2}).
\end{itemize}
\end{definition}

Observe that the reference linearised Kerr solutions $\mathscr{K}$ 
are supported only on $\ell=0,1$. 

Note also that by Lemma \ref{lem:kernelexplore} a solution that is supported only on
$\ell=0,1$  satisfies that
\[
r^2 \slashed{\mathcal{D}}_2^\star \slashed{\nabla}_A s = 0 \textrm{ \  for all scalars $s$ in $\mathscr{S}$} \ \ \ \textrm{and} \ \ \  r \slashed{\mathcal{D}}_2^\star \xi = 0 \textrm{ \  for all one-forms $\xi$ in $\mathscr{S}$.}
\]
In general, it is easy to see that one has
\begin{lemma} \label{lem:udeco}
Let $\mathscr{S}$ be a smooth solution of the system of gravitational perturbations on $\mathcal{M} \cap \{u\geq u_0\} \cap \{v\geq v_0\}$. We have the unique decomposition
\[
\mathscr{S} = \mathscr{S}_{\ell=0,1} + \mathscr{S}_{\ell\ge 2}
\]
where $\mathscr{S}_{\ell=0,1}$ and $\mathscr{S}_{\ell\ge 2}$ are both solutions to the system of gravitational perturbations with $\mathscr{S}_{\ell=0,1}$ supported only on $\ell=0,1$
and $\mathscr{S}_{\ell \ge2}$ supported outside $\ell=0,1$. 
\end{lemma}

\begin{definition} \label{def:proj01}
We call the map $\mathscr{S} \rightarrow \mathscr{S}_{\ell=0,1}$ 
in Lemma \ref{lem:udeco} {\bf the projection of $\mathscr{S}$ to its $\ell=0,1$ modes}.
\end{definition}

\subsubsection{Proof of Theorem~\ref{etsilew}}
Let $\Si$ be as in Theorem \ref{prop:gaugeachieve}. The solution $\Si$ satisfies $\rlin_{\ell=0}\left[\Si\right]=-\frac{1}{4M^2} \mathfrak{m}$ and $\slin_{\ell=1}\left[\Si\right]=\sum_{i=-1}^1 s_i Y_i^{\ell=1}$ on the sphere $S^2_{\infty,v_0}$ for some real numbers $\mathfrak{m}$, $s_{-1}, s_0, s_1$. We can hence subtract from $\Si$ a reference linearised Kerr solution $\mathscr{K}_{\mathfrak{m},s_i}$ such that the projection to $\ell=0,1$ of the solution $\Si - \mathscr{K}_{R,S_i}$ satisfies in particular
\begin{itemize}
\item $\rlin_{\ell=0}=0$ and $\slin_{\ell=1}=0$ on  $S^2_{\infty,v_0}$,
\item $\Olin_{\ell=0,1}=0$ on $C_{u_0}$ and $C_{v_0}$, 
\item $\left( \slashed{div} \bmlin \, \right)_{\ell=1} =0$ and $\left( \slashed{curl} \bmlin \, \right)_{\ell=1}  =0$ along $C_{u_0}$, 
\item $\otx_{\ell=0,1} =0$ and $\left( \, \rlin + \slashed{div} \, \elin \, \right)_{\ell=1}=0$ on $S^2_{\infty,v_0}$,
\item $\Omega^{-2} \otxb_{\ell=0,1}=0$ and $\glinto_{\ell=1}=0$  on $S^2_{\infty,v_0}$, 
\end{itemize}
where we recall Proposition \ref{prop:kerrnormalised}. Note that the parameters $\mathfrak{m}$, $s_i$ are precisely the ones claimed in the theorem.
We will now show that this implies the following identities for $\Si - \mathscr{K}_{\mathfrak{m},s_i}$ globally on $\mathcal{M} \cap \{u \geq u_0\} \cap \{v \geq v_0\}$:
\begin{align}
\Olin \Big|_{\ell=0,1}= \glinto_{\ell=0,1} = \slin_{\ell=0,1} = \otx_{\ell=0,1}= \Omega^{-2} \otxb_{\ell=0,1} =  \rlin_{\ell=0,1}=0 \, ,
\nonumber \\
\left( \slashed{div} \, \bmlin \,  \right)_{\ell=0,1} = \Omega \left( \slashed{div} \, \elin \, \right)_{\ell=0,1} = \Omega^{-1}  \left( \slashed{div} \, \eblin \, \right)_{\ell=0,1} =  \Omega \left( \slashed{div} \ \blin \, \right)_{\ell=0,1} = \Omega^{-1}  \left( \slashed{div} \, \bblin \, \right)_{\ell=0,1} = 0 \, ,
 \nonumber \\
\left( \slashed{curl} \, \bmlin \,  \right)_{\ell=0,1}  = \Omega \left( \slashed{curl} \, \elin \right)_{\ell=0,1} = \Omega^{-1}  \left( \slashed{curl} \, \eblin \right)_{\ell=0,1} =  \Omega \left( \slashed{curl}\, \bblin \right)_{\ell=0,1} = \Omega^{-1}  \left( \slashed{curl} \, \blin \right)_{\ell=0,1} = 0 \nonumber \, .
\end{align}
which in turn implies that $\Si - \mathscr{K}_{\mathfrak{m},s_i}$ is supported outside $\ell=0,1$ providing the conclusion of the theorem. It is also easy to see that $\mathscr{K}_{\mathfrak{m},s_i}$ is unique as any other choice of parameters would make the solution non-trivial on the horizon sphere.

We first obtain the above identities on the initial null hypersurfaces and then 
show how the identities can be propagated globally.

\subsubsection*{The $\ell=0,1$ modes vanish on $C_{u_0}$ and $C_{v_0}$} \label{sec:boinh}

\begin{enumerate}
\item We first obtain additional identities on the sphere $S_{\infty,v_0}$ via elliptic equations. The linearised Gauss equation simplifies on the horizon $\mathcal{H}^+$ to
\begin{equation} \label{Gausshoz}
\Klin = -\rlin \, .
\end{equation}
Computing the linearised Gauss curvature in terms of $\glin$, we find
\begin{align} \label{Gaussinmet} 
2\Klin &= -\slashed{\Delta} \left( tr_{\slashed{g}}\glinh \, \right) + \slashed{div} \slashed{div} \, \glinh - tr_{\slashed{g}}  \, \glin  
=-\frac{1}{2} \slashed{\Delta} \left( tr_{\slashed{g}} \, \glin \right) + \slashed{div} \slashed{div} \, \glinh - \frac{1}{r^2} tr_{\slashed{g}} \glin \, .
\end{align}
Projecting on the $\ell=1$ modes we see that $\Klin_{\ell=1}=0$, cf.~Corollary \ref{cor:nomode}. By (\ref{Gausshoz}) also $\rlin_{\ell=1}=0$. From  $\left( \, \rlin + \slashed{div} \, \elin \, \right)_{\ell=1}=0$ and the fact that $\Olin_{\ell=1}=0$ implies $\slashed{div} \left( \elin+\eblin \, \right)_{\ell=1}=0$ we conclude $\left(\slashed{div} \eta \right)_{\ell=1}=\left(\slashed{div} \elin \right)_{\ell=1}=0$. Also $\slin_{\ell=1}=0$ implies $\left(\slashed{curl} \, \elin \, \right)_{\ell=1}=-\left(\slashed{curl} \, \eblin \, \right)_{\ell=1}=0$. Taking a divergence (and curl respectively) of the Codazzi equations (\ref{ellipchi}) we find $\left(\Omega \slashed{div} \, \blin \, \right)_{\ell=1}= \left( \Omega^{-1} \slashed{div} \, \bblin \, \right)_{\ell=1} = 0$ as well as $\left( \Omega \slashed{curl} \, \blin \,\right)_{\ell=1} = \left(\Omega^{-1} \slashed{curl} \, \bblin \, \right)_{\ell=1} = 0$ on $\mathcal{S}_{\mathcal{H}}$. Recall also that $\glinto_{\ell=0,1}=0$ on $S_{\infty, v_0}$ by (\ref{hoc}) and the fact that $\rlin_{\ell=0}$ implies $\glinto_{\ell=0}=0$ on $S_{\mathcal{H}}$ by combining (\ref{Gausshoz}) and (\ref{Gaussinmet}).

\item \label{item2} Commuting the Bianchi equation (\ref{Bianchi9}) with $\slashed{div}$ (and $\slashed{curl}$) we conclude $\left(\slashed{div} \, \bblin \, \right)_{\ell=1}=0$ and $\left( \slashed{curl} \, \bblin \, \right)_{\ell=1}=0$ hold along $v=v_0$. We then conclude $\left(\slashed{div} \, \eblin \, \right)_{\ell=1}=0$ and $\left( \slashed{curl} \, \eblin \, \right)_{\ell=1}=-\slin_{\ell=1}=0$ hold along $v=v_0$ from commuting (\ref{propeta}).

\item Propagating from the horizon outwards in the $3$-direction, we see  from (\ref{vray}) using $\Olin_{\ell=0,1}=0$ that 
\begin{align}
 \partial_u \left[ \otxb_{\ell=0,1} \Omega^{-2} r^2\right] = 0
\end{align}
and hence taking into account the projection of (\ref{mts}) 
\begin{equation}
\Omega^{-2}  \otxb_{\ell=0,1} = 0  \textrm{\ \  along $C_{v_0}$.}
\end{equation}
\item From (\ref{Bianchi5}) and (\ref{propeta}) we derive, using that $\slashed{div} \left(\, \elin + \eblin \, \right)_{\ell=1} = \slashed{curl} \left(\, \elin + \eblin \, \right)_{\ell=1}=0$ along the cone
\begin{align}
\frac{1}{\Omega} \slashed{\nabla}_3 \left(r^3 \rlin  + r \slashed{div} \left(r^2 \eblin \right) \right)|_{\ell=0,1} =0 \, ,
\end{align}
where the right hand side vanishes along $C_{v_0}$ by the previous step. Moreover, the quantity in brackets on the left is also zero initially on $S_{\mathcal{H}}$ when projected to $\ell=0$ and $\ell=1$. We conclude that 
\begin{align} \label{pkol} 
\rlin_{\ell=0} = 0 \ \ \ \textrm{and}\ \ \  \rlin_{\ell=1} + \left(\slashed{div} \, \eblin \, \right)_{\ell=1} = 0 \ \ \  \textrm{along $C_{v_0}$} \, .
\end{align}
By item \ref{item2} above this means that $\rlin_{\ell=1}= \left(\slashed{div} \, \eblin \, \right)_{\ell=1} =  \left(\slashed{div} \, \elin \, \right)_{\ell=1}=0$ individually on $C_{v_0}$.
We can also conclude $\glinto_{\ell=0}=0$ along $C_{v_0}$ from the projection to $\ell=0$ of (\ref{stos}) along $C_{v_0}$.
\item The previous step allows us to conclude that the condition $\otx_{\ell=0,1}=0$ is propagated along $v=v_0$. This follows by writing (\ref{dbtc}) as
\[
\frac{1}{\Omega^2} \partial_u \left(r \otx \right) = 2 r \slashed{div} \, \elin + 2 r \rlin - \frac{1}{2} \frac{r}{\Omega^2} \Omega tr \chi\otxb
\]
and noting that the right hand side vanishes on $C_{v_0}$ when projected on $\ell=0,1$.

\item The commuted (with $\slashed{div}$ and $\slashed{curl}$) Codazzi equation now gives $\left(\slashed{div} \, \blin\,\right)_{\ell=1}=0$ and $\left( \slashed{curl} \, \blin \, \right)_{\ell=1}=0$ along $v=v_0$.
With this, all required identities have been established on $C_{v_0}$.

\item We finally need to propagate the identities along the outgoing hypersurface $C_{u_0}$ 
from the sphere of intersection $S^2_{u_0,v_0}$ where all the desired identities have already been established. This follows analogously to what we have done before and will only be sketched. For $\otx_{\ell=0,1}$ this follows from the Raychaudhuri equation (\ref{uray}). For $\left(\slashed{div} \, \blin \, \right)_{\ell=1}$ and $\left(\slashed{curl} \, \blin \,  \right)_{\ell=1}$  this follows from commuting (with $\slashed{div}$ and $\slashed{curl}$) the Bianchi equation (\ref{Bianchi2}). Equation (\ref{propeta}) shows $\left(\slashed{div} \,\elin \, \right)_{\ell=1}=0$ and $\left(\slashed{curl} \, \elin \, \right)_{\ell=1}=0$ (and, by the lapse gauge condition, the $\eblin \, $-analogues). Finally, $\rlin_{\ell=1}=0$ and $\slin_{\ell=1}=0$ from their Bianchi equations in the $4$-direction and $\otxb_{\ell=0,1}=0$ from integrating the projection of (\ref{dtcb}).  Note that once we have that $\slin_{\ell=1}$ vanishes along $C_{u_0}$ we can conclude $\bmlin_{\ell=1}=0$.
\end{enumerate}

\subsubsection*{The $\ell=0,1$ modes vanish globally} \label{sec:boglo}
To obtain the identities of Theorem \ref{etsilew} globally we first observe that from the $\ell=1$-projection of the Bianchi equation (\ref{Bianchi2}) and (\ref{Bianchi9}) one concludes
\[
\Omega \left( \slashed{div} \blin \right)_{\ell=0,1} = \Omega^{-1} \left( \slashed{div} \bblin  \right)_{\ell=0,1} = 0  \ \ \ \textrm{and} \ \ \ \Omega  \left( \slashed{curl} \blin  \right)_{\ell=0,1} =\Omega^{-1} \left(  \slashed{curl} \bblin  \right)_{\ell=0,1}= 0 \, .
\]
Because these identities hold globally, the commuted Bianchi equations (\ref{Bianchi3}) and (\ref{Bianchi8}) yield the equations
\[
\ell \left(\ell+1\right) \rlin_{\ell=1} + 3\rho \left(\slashed{div} \, \elin \, \right)_{\ell=1} =0 \ \ \ \ \textrm{and} \ \ \  -\ell \left(\ell+1\right) \rlin_{\ell=1} - 3\rho \left(\slashed{div}\, \eblin \, \right)_{\ell=1} =0 \, ,
\]
from which we conclude $\left(\slashed{div} \, \elin \, \right)_{\ell=1} = \left(\slashed{div} \, \eblin \, \right)_{\ell=1}$ globally. Note also that the Bianchi equations for $\slin_{\ell=1}$ ensure that $\slin_{\ell=1}=0$ and hence $\left(\slashed{curl} \, \elin \,  \right)_{\ell=1} = \left(\slashed{curl} \, \eblin \, \right)_{\ell=1}$ globally. We obtain $\left(\slashed{div} \, \elin \,  \right)_{\ell=1} = 0 = \left(\slashed{div} \, \eblin \, \right)_{\ell=1}$ individually (and similarly for the $\slashed{curl}$) from the commuted (\ref{propeta}). We conclude from (\ref{oml3}) that $\Olin\Big|_{\ell=1}=0$. This allows to use Raychaudhuri (\ref{uray}), (\ref{vray}) to conclude $\otx_{\ell=1}=0= \Omega^{-2}\otxb_{\ell=1}$ globally. 
\vspace{.2cm}

It remains to show that the $\ell=0$ modes vanish globally. For this, note first that the $\ell=0$ projected linearised Raychaudhuri equations\footnote{We remark that the second holds without projection on $\ell=0$, while the first has a $\slashed{div} b$-term in general.} can be written
\begin{equation}
D \left( \frac{{\otx}_{\ell=0}}{\Omega^2} r - 4\Olin\Big|_{\ell=0} + \frac{\glinto_{\ell=0}}{\sqrt{\slashed{g}}} \right) = 0  \textrm{ \ \ and \ \ }
\underline{D} \left( \frac{\otxb_{\ell=0}}{\Omega^2} r + 4 \Olin\Big|_{\ell=0} - \frac{\glinto_{\ell=0}}{\sqrt{\slashed{g}}} \right) = 0 \, .  \nonumber
\end{equation}
Hence the quantities in brackets vanish identically and in particular, $\otx_{\ell=0} + \otxb_{\ell=0}=0$ globally. With this we can combine (\ref{dtcb}) and (\ref{oml1}) as well as (\ref{dbtc}) and (\ref{oml2}) as 
\[
D \left( r^2 \otxb_{\ell=0} - 2\olinb_{\ell=0}\right) = 0 \ \ \ , \ \ \ \underline{D} \left( r^2\otx_{\ell=0} - 2\olin_{\ell=0} \right) = 0 \, .
\]
The quantities in brackets vanish so in particular
\[
\left(\underline{D} + \underline{D} \right) \Olin \Big|_{\ell=0}  = 0
\]
and since $\Olin \Big|_{\ell=0}$ is zero on both $C_{u_0}$ and $C_{v_0}$ we can conclude global vanishing of $ \Olin \Big|_{\ell=0}$, hence of $
\olinb_{\ell=0}$ and $\olin_{\ell=0}$, hence of $\otx_{\ell=0}$ and $\otxb_{\ell=0}$ individually. Global vanishing of $ \glinto_{\ell=0}$ follows from (\ref{stos}).
\vspace{.2cm}

\section{Precise statements of the main theorems}
\label{Themainthesection}
In this section, we
present the precise statements
of the main theorems of this paper.
These will correspond to the rough statements already given in the overview
Section~\ref{MTintro}. 

Section~\ref{sec:theorw} will concern boundedness and decay statements for general solutions
$P$ of the Regge--Wheeler equation. The main result is Theorem \ref{prop:summarypsi}.

Section~\ref{GISsubsec} will concern boundedness and decay statements for general solutions $\alpha$, $\underline\alpha$ of the spin $\pm2$ Teukolsky equations. 
The main result is Theorem~\ref{theo:mtheogi}, while in Corollary~\ref{cor:fully},
we will apply this to linearised gravity and
infer boundedness and decay for the gauge invariant quantities 
$(\alin, \plin, \Psilin)$, $(\ablin, \pblin, \Psilinb)$
 characterizing
a solution $\mathscr{S}$
of the full system.

Section~\ref{BIDGse} will concern the boundedness of all quantities $(\ref{regq1})$
associated to an initial-data normalised solution $\Si$ of linearised gravity, not just the
gauge invariant quantities.
The main result is Theorem~\ref{theo:mtheo} and its pointwise Corollary~\ref{cor:pwe}.

Finally, Section~\ref{DFNGseec} will concern the decay theorem for the
future-renormalised solution $\Sf$. 
The main result is Theorem~\ref{theo:mtheod} and its pointwise Corollary~\ref{newcoroledw}.

The remainder of the paper will then concern the proofs of these theorems.   

\subsection{Theorem~\ref{prop:summarypsi}: Boundedness and decay for solutions to Regge--Wheeler} \label{sec:theorw}

Our first theorem (Theorem~\ref{prop:summarypsi})
is concerned purely with solutions of the Regge--Wheeler equation.
We will state the theorem in Section~\ref{diatupwsn9e1} below, 
after first defining in Section~\ref{EGIQsec}
the norms and energies  appearing in its formulation.

\subsubsection{Energies and norms} \label{EGIQsec}
 
We begin with the definition of various norms which will appear in Theorem~\ref{prop:summarypsi}.
Let $P$ below denote
a solution of the Regge--Wheeler equation as arising from Theorem \ref{theo:wprw}.

We define the following energies for the rescaled solution $\Psi=r^5 P$ defined in (\ref{PsiP}). 
The energy fluxes
\begin{align} \label{nuflu1}
F_u\left[\Psi\right] \left(v_1,v_2\right) = \int_{v_1}^{v_2} dv \sin \theta d\theta d\phi \left[ | \Omega \slashed{\nabla}_4 \Psi |^2  +   |\slashed{\nabla} \Psi|^2 + r^{-2} |\Psi|^2 \right] \, ,
\end{align}
\begin{align} \label{nuflu2}
F_v\left[\Psi\right] \left(u_1,u_2\right) = \int_{u_1}^{u_2} du \sin \theta d\theta d\phi \Omega^2 \left[ | \Omega^{-1} \slashed{\nabla}_3 \Psi |^2  + |\slashed{\nabla} \Psi|^2 + r^{-2} |\Psi|^2 \right]  \, ,
\end{align}
as well as the weighted (near infinity) fluxes
\begin{align} \label{fn1}
{F}^\mathcal{I}_u\left[\Psi\right] \left(v_1,v_2 \right) =
 \int_{v_1}^{v_2} d\bar{v} \left[ r^2 \| r^{-1} \Omega \slashed{\nabla}_4 \Psi\|_{S^2_{u,v}}^2 + \|r^{-1} \ \slashed{\nabla} \Psi \|_{S^2_{u,v}}^2 + r^{-2} \|r^{-1} \ \Psi\|_{S^2_{u,v}}^2 \right] \, ,
\end{align} 
\begin{align} \label{fn2}
{F}^\mathcal{I}_v \left[\Psi\right] \left(u_1,u_2\right) = \int_{u_1}^{u_2}  d\bar{u} \Omega^2 \left[ \| r^{-1} \ \Omega^{-1} \slashed{\nabla}_3 \Psi\|_{S^2_{u,v}}^2 + r^2 \| r^{-1} \ \slashed{\nabla} \Psi \|_{S^2_{u,v}}^2 + \| r^{-1} \ \Psi\|_{S^2_{u,v}}^2 \right] \, ,
\end{align}
where we recall the norms on the spheres $S^2_{u,v}$ defined in (\ref{normsp}). 
(The $\mathcal{I}$ superscript notation alludes to presence of non-trivial weights towards
future null infinity.) From these we define
\begin{align} \label{enreft}
\mathbb{F}\left[\Psi \right] =  \sup_u F^\mathcal{I}_{u} \left[\Psi \right] \left(v_0,\infty \right) + \sup_v F_{v}\left[\Psi\right] \left(u_0,\infty\right) 
\end{align}
with corresponding initial energies
\begin{align}  \label{enreft2}
\mathbb{F}_0\left[\Psi \right] =  F^\mathcal{I}_{u_0} \left[\Psi \right] \left(v_0,\infty \right) + F_{v_0}\left[\Psi\right] \left(u_0,\infty\right) 
\, . 
\end{align}

To estimate higher order energies we also introduce the following notation, taylored to the fact that the Regge--Wheeler equation (\ref{rwr}) commutes with $T$ and the angular momentum operators $\mathnormal{\Omega}_i$, cf.~Section \ref{sec:Killing}:
\begin{align} 
 \mathbb{F}^{n,T} \left[\Psi\right] := &\sum_{i=0}^n \sup_u F^\mathcal{I}_{u} \left[T^i \Psi\right] \left(v_0,\infty \right) + \sup_v \sum_{i=0}^n F^\mathcal{I}_{v}\left[T^i \Psi\right] \left(u_0,\infty\right)  \, , \label{do2Tg}
  \\
\mathbb{F}^{n,T,\slashed{\nabla}} \left[\Psi\right] :=& \sum_{i+j\leq n} \sup_u F^\mathcal{I}_{u} \left[T^i \left(r \slashed{\nabla}_A\right)^j \Psi\right] \left(v_0,\infty \right) + \sup_v  \sum_{i+j\leq n} F^\mathcal{I}_{v}\left[T^i \left(r \slashed{\nabla}_A\right)^j \Psi\right] \left(u_0,\infty\right) \label{do2TOg} \, ,
\end{align}
which initially become
\begin{align} 
 \mathbb{F}_0^{n,T} \left[\Psi\right] := &\sum_{i=0}^n F^\mathcal{I}_{u_0} \left[T^i \Psi\right] \left(v_0,\infty \right) + \sum_{i=0}^n F^\mathcal{I}_{v_0}\left[T^i \Psi\right] \left(u_0,\infty\right)  \, , \label{do2T}
  \\
\mathbb{F}_0^{n,T,\slashed{\nabla}} \left[\Psi\right] :=& \sum_{i+j\leq n} F^\mathcal{I}_{u_0} \left[T^i \left(r \slashed{\nabla}_A\right)^j \Psi\right] \left(v_0,\infty \right) + \sum_{i=0}^n F^\mathcal{I}_{v_0}\left[T^i \left(r \slashed{\nabla}_A\right)^j \Psi\right] \left(u_0,\infty\right) \label{do2TO} \, .
\end{align}

We also define spacetime energies, which will be used in the integrated local energy decay estimate. These will be denoted by the letter $\mathbb{I}$. We define
\begin{align}
\mathbb{I}_{deg} \left[\Psi\right] :=\int_{u_0}^\infty \int_{v_0}^\infty \int_{S^2_{\bar{u},\bar{v}}} d\bar{u} d\bar{v} \sin \theta d\theta d\phi  \Omega^2\Bigg[  \frac{1}{r^2}| \Omega \slashed{\nabla}_4 \Psi - \Omega \slashed{\nabla}_3 \Psi |^2   + \frac{1}{r^3} |\Psi|^2 \nonumber \\
+ \frac{(r-3M)^2}{r^3} \left( |\slashed{\nabla} \Psi|^2 + \frac{1}{r^2} | \Omega \slashed{\nabla}_4 \Psi  |^2 + \frac{\Omega^2}{r^2}| \Omega^{-1} \slashed{\nabla}_3 \Psi |^2 \right)  \Bigg] \, ,
\end{align}
which degenerates near the trapped set $r=3M$ and a weighted energy localised to $r\geq 4M$
\begin{align}
\mathbb{I}_{\mathcal{I},\epsilon} \left[\Psi\right] :=
\int_{u_0}^\infty \int_{v_0}^\infty \int_{S^2_{\bar{u},\bar{v}}} du dv \sin \theta d\theta d\phi \cdot \iota_{r\geq 4M} \Big[r | \Omega \slashed{\nabla}_4 \Psi|^2 + r^{-1-\epsilon} |\Omega \slashed{\nabla}_3 \Psi|^2  
+ r^{1-\epsilon} |\slashed{\nabla} \Psi|^2 + r^{-1-\epsilon} |\Psi|^2  \Big] ,\nonumber
\end{align}
for some $0<\epsilon<1/8$ now fixed once and for all and $\iota_{r\geq R}$ being the indicator function which equals $1$ for $r \geq R$ and is zero otherwise. The higher order analogues are defined in the obvious way:
\begin{align}
\mathbb{I}^{n,T,\slashed{\nabla}}_{\mathcal{I},\epsilon} \left[\Psi\right] := \sum_{i+j\leq n}^n \mathbb{I}_{\mathcal{I},\epsilon} \left[T^i \left(r \slashed{\nabla}_A\right)^j \Psi\right] 
\end{align}
and similarly for $\mathbb{I}^{n,T,\slashed{\nabla}}_{deg} \left[\Psi\right]$.

\subsubsection{Statement of the theorem}\label{diatupwsn9e1}
We are now ready to state the boundedness and decay theorem for solutions $P$
of the Regge--Wheeler equation.

\begin{bigtheorem} \label{prop:summarypsi}
Let $P$ be a solution of the Regge--Wheeler equation as arising from Theorem \ref{theo:wprw}. Then the weighted symmetric traceless $S^2_{u,v}$-tensor $\Psi = r^5 P$ satisfies equation (\ref{rwro}) and the following estimates hold provided the initial energies on the right hand sides are finite:
\begin{enumerate}
\item the basic boundedness and integrated decay estimates of Proposition \ref{prop:basicforrw} 
as well as the weighted boundedness estimate
\begin{align} \label{flux1}
\mathbb{F} \left[\Psi\right] \lesssim \mathbb{F}_0 \left[\Psi\right]  ,
\end{align}
\item the higher order estimates (for any integer $n\geq 0$)
\begin{align} \label{qxz}
\mathbb{F}^{n,T,\slashed{\nabla}} \left[\Psi\right] \lesssim \mathbb{F}^{n,T,\slashed{\nabla}} _0 \left[\Psi\right]  ,
\end{align}
\item the weighted integrated decay estimate (for any integer $n\geq 0$)
\begin{align} \label{bestintdec}
\mathbb{I}^{n,T,\slashed{\nabla}}_{\mathcal{I},\epsilon} \left[\Psi\right] + \mathbb{I}^{n,T,\slashed{\nabla}}_{deg} \left[\Psi\right] \lesssim \mathbb{F}_0^{n,T,\slashed{\nabla}}\left[\Psi\right]  \, . 
\end{align}

\item Finally, the polynomial decay estimates of Proposition \ref{prop:decRW} hold.
\end{enumerate}
\end{bigtheorem}

The proof of the above theorem will be the content of Section \ref{sec:RW}.

\subsection{Theorem~\ref{theo:mtheogi}: Boundedness and decay for solutions to  Teukolsky}
\label{GISsubsec}
Our second theorem is concerned purely with solutions to the spin $\pm2$ Teukolsky 
equations.  We define relevant energies and norms in Section~\ref{sec:teuenergy} below.
We state the theorem in Section~\ref{diatupwsn9e2}. We shall then infer an immediate application
of the result to the full system
of linearised gravity in Section~\ref{immedapp}.

\subsubsection{Energies and norms} \label{sec:teuenergy}
Let $\alpha$ be a solution to the Teukolsky equation of spin $+2$ as arising from Theorem \ref{theo:wpteu} and $\underline{\alpha}$ be a smooth solution to the Teukolsky equation of spin $-2$ as arising from Theorem \ref{theo:wpteu2}. 

Recall that associated with a solution to the Teukolsky equation of spin $+2$ are the derived quantities $\psi, P$ defined in (\ref{psidef}), (\ref{evolpp}) of Section \ref{sec:tratheo} and associated with a solution to the Teukolsky equation of spin $-2$ are the quantities $\underline{\psi}, \underline{P}$
defined in (\ref{psidef2}), (\ref{evolpp2}).

We define the following energies for the solution $\alpha$ and its derived quantities $\psi$, $P$ and the solution $\underline{\alpha}$ and its derived quantities $\underline{\psi}, \underline{P}$ 

\begin{align} 
\mathbb{F}\left[\Psi, \psi \right] =  \mathbb{F}\left[\Psi\right] + \sup_u \int_{v_0}^{\infty} dv \|r^{-1} \cdot \psi\|_{S^2_{u,v}}^2 r^{8-\epsilon} \Omega^2 
\ \ \ , \ \ \ 
\mathbb{F}\left[\underline{\Psi}, \underline{\psi} \right] =  \mathbb{F}\left[ \underline{\Psi} \right]  + \sup_v \int_{u_0}^{\infty} du \|r^{-1} \cdot \underline{\psi}\|_{S^2_{u,v}}^2 r^6 \, ,  \nonumber
\end{align}
with the obvious definition for $\mathbb{F}_0\left[\Psi, \psi \right]$, $\mathbb{F}_0\left[\underline{\Psi}, \underline{\psi} \right]$. Also 
\begin{align} 
\mathbb{F}\left[\Psi, \psi, \alpha \right] &= \mathbb{F}\left[\Psi, \psi \right]  + \sup_u \int_{v_0}^{\infty} dv \|r^{-1} \alpha\|_{S^2_{u,v}}^2 r^{6-\epsilon} \Omega^4  \, ,
\nonumber \\
\mathbb{F}\left[ \underline{\Psi}, \underline{\psi}, \underline{\alpha} \right] &= \mathbb{F}\left[ \underline{\Psi}, \underline{\psi} \right] + \sup_v \int_{u_0}^{\infty} du \|r^{-1} \underline{\alpha}\|_{S^2_{u,v}}^2 \Omega^{-2} \nonumber \, ,
\end{align}
again with the obvious definition for $\mathbb{F}_0\left[\Psi, \psi, \alpha \right]$, $\mathbb{F}_0\left[ \underline{\Psi}, \underline{\psi}, \underline{\alpha} \right]$. 
Finally, the higher order norms
\begin{align}
\mathbb{F}[\Psi,  \mathfrak{D}\psi, \alpha] &=  \mathbb{F}\left[\Psi, \psi, \alpha \right] + \sup_u \int_{v_0}^{\infty} dv \|r^{-1} \cdot \mathfrak{D} (\psi \Omega)\|_{S^2_{u,v}}^2 r^{8-\epsilon}   \, ,
\nonumber \\
\mathbb{F}[\underline{\Psi},  \mathfrak{D}\underline{\psi}, \underline{\alpha}] &=  \mathbb{F}\left[ \underline{\Psi}, \underline{\psi}, \underline{\alpha} \right] + \sup_v \int_{u_0}^{\infty} du \|r^{-1} \cdot \mathfrak{D} (\underline{\psi}\Omega^{-1})\|_{S^2_{u,v}}^2 \Omega^2 r^6 , \nonumber
\nonumber \\
\mathbb{F}[\Psi,  \mathfrak{D}\psi, \mathfrak{D}\alpha]  &= \mathbb{F}[\Psi,  \mathfrak{D}\psi, \alpha]  + \sup_u \int_{v_0}^{\infty} dv \|r^{-1} \cdot \mathfrak{D} (\alpha \Omega^2)\|_{S^2_{u,v}}^2 r^{6-\epsilon}  \, ,
\nonumber \\
\mathbb{F}[ \underline{\Psi}, \mathfrak{D}\underline{\psi}, \mathfrak{D}\underline{\alpha}] &= \mathbb{F}\left[ \underline{\Psi},\mathfrak{D} \underline{\psi}, \underline{\alpha} \right]  + \sup_v \int_{u_0}^{\infty} du \|r^{-1}\cdot\mathfrak{D} (\underline{\alpha} \Omega^{-2})\|_{S^2_{u,v}}^2 \Omega^{2} , \nonumber
\end{align}
where we have employed the short hand notation 
\[
\|r^{-1} \cdot \mathfrak{D} \xi\|_{S^2_{u,v}}^2:=\|r^{-1} \cdot r \slashed{\nabla}_A \xi\|_{S^2_{u,v}}^2+ \|r^{-1} \cdot \Omega^{-1} \slashed{\nabla}_3 \xi\|_{S^2_{u,v}}^2+ \|r^{-1} \cdot r \Omega \slashed{\nabla}_4 \xi\|_{S^2_{u,v}}^2 \, 
\]
for an $S^2_{u,v}$-tensor $\xi$. 

We also define a basic spacetime energy measuring some form of integrated decay:
\begin{align}
\mathbb{I}_{master} \left[\Psi, \mathfrak{D} \psi, \mathfrak{D}\alpha\right]=\ & \mathbb{I}_{deg} \left[\Psi\right] + \mathbb{I}^{\mathcal{I}}_{\epsilon} \left[\Psi\right]  \nonumber \\
  &+ \int_{u_0}^\infty  \int_{v_0}^\infty d\bar{u} d\bar{v} \Omega^2 \left[ r^{7-\epsilon} \| r^{-1} \cdot \mathfrak{D} \left(\Omega \psi\right)\|^2_{S^2_{\bar{u},\bar{v}}} + r^{5-\epsilon} \| r^{-1} \cdot \mathfrak{D} \left(\Omega \alpha\right)\|^2_{S^2_{\bar{u},\bar{v}}}\right] \, , \nonumber
\end{align}
\begin{align}
\mathbb{I}_{master} \left[\underline{\Psi}, \mathfrak{D} \underline{\psi}, \mathfrak{D}\underline{\alpha}\right]=\ & \mathbb{I}_{deg} \left[\underline{\Psi}\right] + \mathbb{I}^{\mathcal{I}}_{\epsilon} \left[\underline{\Psi}\right] \nonumber \\
& + \int_{u_0}^\infty  \int_{v_0}^\infty d\bar{u} d\bar{v} \Omega^2 \left[   r^{5-\epsilon} \| r^{-1} \cdot \mathfrak{D} \left(\Omega^{-1} \underline{\psi}\right)\|^2_{S^2_{\bar{u},\bar{v}}} +  r^{1-\epsilon} \| r^{-1} \cdot \mathfrak{D} \left(\Omega^{-1} \underline{\alpha}\right)\|^2_{S^2_{\bar{u},\bar{v}}}  \right] \, . \nonumber
\end{align}
The following higher order energies are then defined in the obvious way 
\begin{align}
\mathbb{F}^{n,T,\slashed{\nabla}} [ \underline{\Psi}, \mathfrak{D}\underline{\psi}, \mathfrak{D}\underline{\alpha}] \ , \ \mathbb{F}_0^{n,T,\slashed{\nabla}} [ \underline{\Psi}, \mathfrak{D}\underline{\psi}, \mathfrak{D}\underline{\alpha}] \ \ \ \textrm{and} \ \ \  \mathbb{I}^{n,T,\slashed{\nabla}}_{master} \left[ \underline{\Psi}, \mathfrak{D} \underline{\psi}, \mathfrak{D}\underline{\alpha}\right] \, ,
\end{align}
as are their non-underlined counterparts.

\subsubsection{Statement of the theorem}\label{diatupwsn9e2}
We are now ready to state the boundedness and decay theorem for solutions of the spin $\pm2$
Teukolsky equations.
\begin{bigtheorem} \label{theo:mtheogi} 
Let $\alpha$ be a solution of the spin $+2$ Teukolsky equation as arising from Theorem \ref{theo:wpteu}.  
Then the derived quantity $\Psi=r^5P$ with $P$ defined through (\ref{psidef}) and (\ref{evolpp}) satisfies the conclusions of Theorem \ref{prop:summarypsi}. Moreover, provided the initial energies on the right hand side of (\ref{teubo1})--(\ref{teubo3}) are finite, we have the following estimates:
\begin{enumerate}
\item the weighted boundedness estimate
\begin{align} \label{teubo1}
\mathbb{F}[\Psi, \psi, \alpha] \lesssim \mathbb{F}_0[\Psi, \psi, \alpha] \, ,
\end{align}
\item the higher order statements (for any integer $n\geq0$)
\begin{align} \label{teubo3}
 \mathbb{F}^{n,T, \slashed{\nabla}} [\Psi, \mathfrak{D}\psi, \mathfrak{D} \alpha] \lesssim  \mathbb{F}_0^{n,T, \slashed{\nabla}} [\Psi, \mathfrak{D}\psi, \mathfrak{D} \alpha] \, ,
\end{align}
\item the weighted integrated decay estimate (for any integer $n \geq 0$)
\begin{align} \label{masterid}
\mathbb{I}^{n,T,\slashed{\nabla}}_{master} \left[\Psi, \mathfrak{D} \psi, \mathfrak{D}\alpha\right] \lesssim \mathbb{F}_0^{n,T, \slashed{\nabla}} [{\Psi}, \mathfrak{D}{\psi}, \mathfrak{D} {\alpha}]  \, .
\end{align}
\item Finally, the polynomial decay estimates of Propositions~\ref{prop:refd}, \ref{prop:psibs2d2}, \ref{prop:psibs2d}, \ref{prop:as2d2} and the decay estimate of Proposition \ref{prop:l1est} hold.
\end{enumerate}
\noindent
Now let $\underline{\alpha}$ be a smooth solution of the spin $-2$ Teukolsky equation as arising from Theorem \ref{theo:wpteu2}. Then $\underline{\Psi}=r^5\underline{P}$, with $\underline{P}$ defined through (\ref{psidef}), (\ref{evolpp}), satisfies the conclusions of Theorem \ref{prop:summarypsi}. Moreover, the estimates $1.$-$4.$ above hold replacing the quantities $\alpha,\psi,\Psi$ by $\underline{\alpha}, \underline{\psi}, \underline{\Psi}$ provided the energies on the right hand side are finite.
\end{bigtheorem}

Note that the second sentence 
of Theorem \ref{theo:mtheogi}, that $\Psi$ satisfies the conclusions of Theorem \ref{prop:summarypsi}, is already immediate from Proposition \ref{prop:rwt1}. The same proposition
 immediately yields the analogous statement for $\underline{\Psi}$ claimed in the second part of the theorem.

The proof of Theorem \ref{theo:mtheogi} will be carried out in Section \ref{sec:hdgi}. Key to the proof is to exploit the transformation formulae of Section \ref{sec:tratheo}. 

\subsubsection{Application to the full system of linearised gravity: Boundedness and decay for the gauge invariant hierarchy}\label{immedapp}

In view of Proposition \ref{prop:relfull} we infer the following application to the full system
of linearised gravity:
\begin{corollary} \label{cor:fully}
Let $\mathscr{S}$ be a smooth solution of the system of gravitational perturbations arising from a smooth seed initial data set on $C_{u_0} \cup C_{v_0}$ through Theorem \ref{theo:lwp}. Then 
\begin{itemize}
\item the gauge invariant curvature component $\alin$ of the solution $\mathscr{S}$ satisfies the Teukolsky equation of spin $+2$ hence the first part of
Theorem \ref{theo:mtheogi} applies yielding boundedness and decay for $\left(\Psilin,\plin, \alin\right)$. 
\item the gauge invariant curvature component $\ablin$ of the solution $\mathscr{S}$ satisfies the 
Teukolsky equation of spin $-2$ and hence the second part of Theorem \ref{theo:mtheogi}  applies yielding boundedness and decay for $\left(\underline{\Psilin},\pblin, \ablin\right)$.
\end{itemize}
\end{corollary}
Let us note that we know more information about $\alin$ and $\ablin$ then
the statement that they satisfy the spin $\pm 2$ Teukolsky equations.
The solutions $\alin$ and $\ablin$ are in fact non-trivially related 
to each other through the system of linearised
Bianchi and null structure equations.
(For fixed frequency solutions, these relations are well known;
see~\cite{starobinskii1973amplification}.)
We stress that  the estimates inferred in Corollary~\ref{cor:fully} for $\mathscr{S}$
are derived without exploiting this relation.
The above corollary will be the starting point (see 
Section~\ref{collectthem}) for the proof of Theorem~\ref{theo:mtheo} to which we now turn.

\subsection{Theorem~\ref{theo:mtheo}:  Boundedness for solutions to the full system}
\label{BIDGse}
We now consider the full system of linearised gravity.
Our next theorem (Theorem~\ref{theo:mtheo}) asserts boundedness of  initial-data normalised
solutions $\Si$ as in Definition~\ref{def:gaugechoice}. In view
of Theorem~\ref{prop:gaugeachieve}, we will be able to apply
Theorem~\ref{theo:mtheo} to solutions $\mathscr{S}$ arising from general,
smooth asymptotically flat seed data. 
We first 
define some additional energies and norms in Section~\ref{remainingsec}
before stating the precise formulation of the theorem in Section~\ref{diatupwsn9e3}.

\subsubsection{Energies and norms}\label{remainingsec}
Let $\mathscr{S}$ be a solution of the system of gravitational perturbations as arising from Theorem \ref{theo:lwp}. 

Recall that 
by Proposition~\ref{prop:relfull},
the components $\alin$ and $\ablin$ of the solution satisfy the spin $\pm2$ Teukolsky equations
and thus, by Proposition~\ref{prop:rwt1},
the quantities $\Plin$ and $\Pblin$ derived from $\alin$ and $\ablin$ respectively  satisfy the Regge--Wheeler equation.
Thus, we may use the notations of Sections~\ref{EGIQsec}
and~\ref{sec:teuenergy} to denote energies associated
with these gauge-invariant quantities.
We will augment these with  the following combined notation
\[
\mathbb{F}[\Psilin, \Psilinb, \plin, \pblin,  \alin, \ablin]  :=\mathbb{F}[\Psilin,  \plin, \alin] 
+ \mathbb{F}[ \Psilinb, \pblin, \ablin ] \, .
\]

We proceed to define additional (gauge-dependent) energies.

We define the flux
\begin{align}
\Big\|(\slashed{\nabla}_3)^2 \, \xlin \, \Big\|^2_{L^\infty_v L^2(C_v)} = \sup_{v \geq v_0} \int_{u_0}^\infty d\bar{u} \int_{S^2\left(\bar{u},v\right)} \sin \theta d\theta d\phi \  \frac{\Omega^2}{r^{\epsilon}} \Bigg\{  \Big| \frac{1}{\Omega}\slashed{\nabla}_3 \left( \frac{1}{\Omega}\slashed{\nabla}_3\left(r^2 \, \xlin \Omega\right)\right) \Big|^2 \nonumber \\ 
+ \Big|\frac{1}{\Omega}\slashed{\nabla}_3\left(r^2 \,  \xlin \Omega\right) \Big|^2 +  \Big|r^2 \, \xlin \Omega\Big|^2 \Bigg\} \nonumber  \, .
\end{align}
Recall now the two auxiliary quantities $\Ylin$ and $\Zlin$ defined in (\ref{Ydef}) and (\ref{Qquant}). 
We define the following energy for the Ricci coefficients on spheres (the superscript $(5)$ stands for the fact that this energy is at the level of $5$ derivatives of the Ricci coefficients)
\begin{align} \label{D5def}
\mathbb{D}^{[5]} \big[\Ylin, \Zlin\big]  = \sup_{u,v} \|r^{-1} \cdot r^3  \slashed{\mathcal{D}}_2^\star  \slashed{div} \slashed{\mathcal{D}}_2^\star \Ylin \|^2_{S^2_{u,v}} +\sup_{u,v} \|r^{-1} \cdot r^2   \slashed{div} \slashed{\mathcal{D}}_2^\star r \Omega \slashed{\nabla}_4 \Ylin \|^2_{S^2_{u,v}} \nonumber \\
+\sup_{u,v} r^{2+\epsilon} \| r^{-1} \cdot \Omega^{-1} r^4 \slashed{div} \slashed{\mathcal{D}}_2^\star  \slashed{div}\slashed{\mathcal{D}}_2^\star \Zlin \|^2_{S^2_{u,v}} + \sup_{u,v} \Big\| r^{-1} \cdot r  \slashed{\nabla}_3 \left( r^4 \slashed{\mathcal{D}}_2^\star \slashed{div} \slashed{\mathcal{D}}_2^\star \slashed{\nabla} \big( r \Omega^{-2} \,  \otx \, \big)\right) \Big\|^2_{S^2_{u,v}} \, ,
\end{align}
which at the level of data is
\begin{align} \label{D5def0}
\mathbb{D}^{[5]}_0 \big[\Ylin, \Zlin\big]  = \sup_{v} \|r^{-1} \cdot r^3  \slashed{\mathcal{D}}_2^\star  \slashed{div} \slashed{\mathcal{D}}_2^\star \Ylin \|^2_{S^2_{u_0,v}} +\sup_{v} \|r^{-1} \cdot r^2  \slashed{div} \slashed{\mathcal{D}}_2^\star r \Omega \slashed{\nabla}_4 \Ylin \|^2_{S^2_{u_0,v}} \, \nonumber \\
+\sup_{u} r^{2+\epsilon} \| r^{-1} \cdot \Omega^{-1} r^4 \slashed{div} \slashed{\mathcal{D}}_2^\star  \slashed{div}\slashed{\mathcal{D}}_2^\star \Zlin \|^2_{S^2_{u,v_0}} + \sup_{u} \Big\|   \slashed{\nabla}_3\left( r^4 \slashed{\mathcal{D}}_2^\star \slashed{div} \slashed{\mathcal{D}}_2^\star \slashed{\nabla} \big( r \Omega^{-2} \,  \otx \, \big)\right) \Big\|^2_{S^2_{u,v_0}} \, .
\end{align}

Recall Propositions \ref{prop:zbounded} and \ref{prop:ybounded} which guarantee that the 
norm $\mathbb{D}^{[5]}_0 \big[\Ylin, \Zlin\big]$ is 
indeed finite for the initial data of the solution $\Si$ defined in Theorem \ref{prop:gaugeachieve}.

\begin{remark}
One should think of the last term in (\ref{D5def}) as the $\slashed{\nabla}_3$ derivative of $\Zlin$ but 
without the $\left(\elin+ \eblin\right)$-part. There is a small technical advantage in that the quantity in the energy satisfies a ``more decoupled" equation. Note also that for both $\Zlin$ 
and the last term we do not put the optimal weight near the horizon (which would allow another factor of $\Omega^{-1}$ in both terms of the second line, cf.~Proposition \ref{prop:zbounded}).
\end{remark}

Let us note finally that if $\mathscr{S}$ is supported on $\ell=0,1$ only, then
all the above energies manifestly vanish.
In particular, the above energies vanish for the reference linearised
Kerr solutions $\mathscr{K}_{\mathfrak{m},s_i}$.

\subsubsection{Statement of the theorem}\label{diatupwsn9e3}
We are now ready to state our boundedness theorem for the initial-data normalised solution
$\Si$ of the full system
of linearised gravity.

\begin{bigtheorem} \label{theo:mtheo} 
Let $\Si$ be a smooth solution of the system of gravitational perturbations arising from a smooth seed initial data set on $C_{u_0} \cup C_{v_0}$ through Theorem \ref{theo:lwp}, which is
moreover initial-data normalised according to Definition \ref{def:gaugechoice}.

(In particular, 
given a general smooth seed initial data set which is asymptotically
flat with weight $s$ to order $n \geq 10$ according to Definition~\ref{def:afpeel}, then defining
\[
\Si=\mathscr{S}+\underaccent{\lor}{\mathscr{G}}
\]
by applying Theorem \ref{prop:gaugeachieve},
it follows that $\Si$ satisfies the above assumption.)

Then the curvature quantities $\alpha$ and $\underline{\alpha}$ associated with $\underaccent{\lor}{\mathscr{S}}$ satisfy the conclusions of Theorem \ref{theo:mtheogi}.

We assume finiteness of the following initial energy, which is at the level of five derivatives of curvature and five derivatives of the Ricci coefficients
\begin{align}  \label{ughbo}
 \Big\|\slashed{\nabla}_3^2 \left(r^3 \slashed{div} \slashed{\mathcal{D}}_2^\star \slashed{div} \xlin \, \right)\Big\|^2_{ L^2 \left(C_{v_0}\right)} + \mathbb{D}^{[5]}_0 \big[\Ylin, \Zlin\big] 
+\mathbb{F}_0^{2,T,\slashed{\nabla}} \big[ \underline{\Psilin}, \mathfrak{D}\underline{\plin}, \mathfrak{D}\ablin \, \big]+\mathbb{F}_0^{2,T,\slashed{\nabla}} \big[ \Psilin ,  \mathfrak{D}\plin, \mathfrak{D}\alin \, \big] < \infty \, .
\end{align}
Then, we have the estimates
\begin{align} \label{ughbtriv}
\mathbb{F}_0^{2,T,\slashed{\nabla}} \big[ \underline{\Psilin}, \mathfrak{D}\underline{\plin}, \mathfrak{D}\ablin \, \big]+\mathbb{F}_0^{2,T,\slashed{\nabla}} \big[ \Psilin ,  \mathfrak{D}\plin, \mathfrak{D}\alin \, \big]   \lesssim \textrm{initial data energy (\ref{ughbo})}
\end{align}
and 
\begin{align}  \label{ughb}
\Big\|\slashed{\nabla}_3^2 \left(r^3 \slashed{div} \slashed{\mathcal{D}}_2^\star \slashed{div} \xlin \right)\Big\|_{L^\infty_v L^2 \left(C_v\right)} +    \mathbb{D}^{[5]} \big[\Ylin, \Zlin\big] \lesssim \textrm{initial data energy (\ref{ughbo})} \, ,
\end{align}
the first one being already immediate from Theorem \ref{theo:mtheogi}.

Moreover, the initial data energy (\ref{ughbo}) controls in addition:
\begin{enumerate}
\item Weighted $L^\infty_{u,v} L^2\left(S^2_{u,v}\right)$-norms for up to five angular derivatives of the metric coefficients \\ 
$\Big(\, \glinh \, , \,  \frac{\glinto}{\sqrt{\slashed{g}}} \, , \, \bmlin \, , \, \Olin \, \Big)$ as in Proposition \ref{prop:meco1}.

\item Weighted $L^\infty_{u,v}L^2\left(S^2_{u,v}\right)$-norms and weighted $L^2$-fluxes on null cones for 
\begin{itemize}
\item up to five angular derivatives of $\xblin$ as in Corollaries \ref{cor:crucialchibar} and \ref{cor:xbar6}
\item up to five angular derivatives of $\xlin$ as in Propositions \ref{prop:chiall}, \ref{prop:hioc}, \ref{prop:angc} and Proposition \ref{prop:5ca}

\item  up to five angular derivatives  of $\eblin$ as in Propositions \ref{prop:etae} and \ref{prop:etabe2}
\item up to five angular derivatives of $\elin$ as in Propositions \ref{prop:etae} and \ref{prop:etae2}
\item up to five angular derivatives for $\otx$ as in Corollary \ref{cor:crucialchibar}  and Corollary \ref{cor:im2}
\item up to five angular derivatives of  $\otxb$ as in Proposition \ref{prop:trxbflux} 
\item up to five angular derivatives of $\olin$ and $\olinb$ as in Proposition \ref{prop:omb}
\end{itemize}

\item Weighted $L^\infty_{u,v}L^2\left(S^2_{u,v}\right)$-norms for four angular derivatives and weighted flux estimates for five angular derivatives of the curvature components $\Big(\alin \, , \, \blin \, , \,\rlin  \, , \, \slin  \, , \, \bblin  \, , \,\ablin \, \Big)$ provided the non-degenerate  initial energies $\mathbb{F}^2_0\big[\Psilin\big]$ and $\mathbb{F}^2_0\big[\underline{\Psilin}\big]$ are finite and added (to (\ref{ughb})) on the right hand side.
See Propositions \ref{prop:fluv}, \ref{prop:fluu} and \ref{prop:5dera} for the flux estimates and Propositions \ref{prop:bc4} and \ref{prop:hioaa} for the $L^\infty_{u,v}L^2 \left(S^2_{u,v}\right)$ estimates.
\end{enumerate}

Finally, let $\mathscr{K}_{\mathfrak{m},s_i}$ be the initial data normalised Kerr solution
as in Theorem \ref{etsilew}
such that
$\Si'=\Si-\mathscr{K}_{\mathfrak{m},s_i}$ has support outside $\ell=0,1$.
Then the initial norm $(\ref{ughbo})$ coincides for $\Si^\prime$,
and the above statements of the theorem all hold 
applied to $\Si^\prime$ in place of $\Si$, 
where 
now the derived energy bounds are coercive
on all quantities $(\ref{regq1})$ of $\Si^\prime$.
\end{bigtheorem}

\subsubsection{Remarks and  uniform pointwise boundedness}
We give a number of remarks concerning the statement of Theorem~\ref{theo:mtheo}.

\begin{remark}
If $\Si$ in Theorem \ref{theo:mtheo} indeed arises through Theorem \ref{prop:gaugeachieve} from smooth asymptotically flat seed initial data of order $1/2\leq s\leq1$ and with $n \geq 15$, the finiteness of the initial energy (\ref{ughbo}) is seen to be a direct consequence of the estimates (\ref{decrete}) and Propositions \ref{prop:zbounded} and \ref{prop:ybounded}. 
\end{remark}

\begin{remark} \label{rem:indoflapse}
As we shall see, the boundedness and decay estimates proven in  Theorem \ref{theo:mtheo} can be proven also if the conditions (\ref{omchoice}), (\ref{shiftco}) did not hold for $\Si$. The only difference are additional boundary terms appearing on the right hand side in the estimates, cf.~Proposition \ref{prop:omb}. The horizon gauge conditions (\ref{hoc}), (\ref{resi}), the round sphere condition (\ref{rsc}) for $\Si$ and the finiteness of (\ref{ughbo}) are fundamental, however.
\end{remark}

\begin{remark} \label{rem:afrem}
The propagation of the weighted norm (\ref{ughbo}) in  (\ref{ughbtriv}) and (\ref{ughb}) and, intimately related with it, the propagation of the round sphere condition at infinity can be viewed as a version of propagation of asymptotic flatness for the solution $\Si$, which does not lose derivatives.
\end{remark}

\begin{remark}
In the course of the proof of Theorem~\ref{theo:mtheo}, we shall obtain several other estimates on various derivatives of the Ricci coefficients. We have not stated these estimates explicitly above but direct the reader to the body of Section~\ref{sec:proofof3}. Some of these estimates are needed to prove Corollary~\ref{cor:pwe} below. We also emphasise that the quantities $\blin$ and $\xlin$ can already be shown to decay to zero in time for $\Si^\prime$. 
This is not true for the other Ricci coefficients and curvature components (except, of course, for the gauge invariant quantities $\alin$ and $\ablin$ for which the conclusions of Theorem~\ref{theo:mtheogi} hold).
\end{remark}

\begin{remark}
The above statements focus on angular derivatives. A version of the above theorem can be obtained for \underline{all} derivatives of curvature and Ricci coefficients up to order five, provided appropriate quantities are assumed initially. As this is standard (but lengthy) we leave it to the reader.
\end{remark}

Simple Sobolev embedding on the spheres $S^2_{u,v}$ and using the fact that 
$\Si'$ of Theorem \ref{theo:mtheo} is supported outside of $\ell=0,1$, 
together with the boundedness of $\mathscr{K}_{\mathfrak{m},s_i}$
we obtain in particular (see Section \ref{sec:corpro})
\begin{corollary} \label{cor:pwe}
Let $\Si$, $\mathscr{K}_{\mathfrak{m},s_i}$ be as in the statement of Theorem~\ref{theo:mtheo}.
Then all quantities $(\ref{regq1})$ of  $\Si'=\Si-\mathscr{K}_{\mathfrak{m},s_i}$ 
are uniformly pointwise  bounded in that
\begin{align}
| r^{\frac{7}{2}-\epsilon}\Omega^2 \, \alin \, | + | \Omega r^{\frac{7}{2}-\epsilon} \, \blin \,| + | r^3 \rlin \,  | + |r^3 \slin \, | + | r^2\Omega^{-1} \, \bblin \,  | + | r \Omega^{-2}\, \ablin \, | &\lesssim \textrm{quantity (\ref{ughbo})}
\nonumber \\
|r^2 \Omega \, \xlin \,  | + | r \Omega^{-1} \, \xblin \,  | + | r \, \elin \, | + |r^2 \, \eblin \, | + |r^2 \Omega^{-2} \, \otx \, | + |r  \Omega^{-2} \otxb \, | + |r^{\frac{5-\epsilon}{2}} \olin| + |\Omega^{-2} \olinb| &\lesssim\textrm{quantity (\ref{ughbo})} \nonumber \\
 \Big|\glinh\Big| +   \Big| \frac{\glinto}{\sqrt{\slashed{g}}}\Big| + r^{\frac{1}{2}-\epsilon} | \, \bmlin \, | + \Big| \, \Olin  \, \Big| &\lesssim\textrm{quantity (\ref{ughbo})}  \, . \nonumber
\end{align}
The same bounds hold for $\Si$ in place of $\Si'$ if a constant depending on
$|\mathfrak{m}|+|s_{-1}|+|s_{0}|+|s_{1}|$
is added to the right hand side.
\end{corollary}

\begin{remark}
We indeed control $|r^2 \Omega^{-2} \otx |$ above because the regular quantity $ \otx$ vanishes linearly on the event horizon for the gauge $\Si$.
\end{remark}

\subsection{Theorem~\ref{theo:mtheod}: Decay for solutions to the full system in the future-normalised gauge}
\label{DFNGseec}
We may now state our final Theorem~\ref{theo:mtheod} giving quantitative decay, measured
in appropriate $L^2$ norms, for all quantities associated to the horizon-renormalised solution $\Sf$ defined in Proposition~\ref{prop:hozrgauge}. As a corollary we shall deduce 
in particular pointwise polynomial decay 
of the metric components of $\Sf$ to their linearised Kerr values given
by $\mathscr{K}_{\mathfrak{m}, s_i}$ 
of Theorem~\ref{etsilew}.
The norms appearing below have already been defined in 
Sections~\ref{EGIQsec},~\ref{sec:teuenergy} and~\ref{remainingsec}.

\begin{bigtheorem} \label{theo:mtheod}
Let $\Si$ be as in Theorem \ref{theo:mtheo}, in particular (\ref{ughbo}) holds initially. Let 
\[
\Sf = \Si +\Gf
\]
be the horizon-renormalised solution defined in Proposition \ref{prop:hozrgauge}. 
Then
\begin{enumerate}
\item The pure gauge solution $\Gf$ is uniformly bounded and controlled solely by the initial data energy (\ref{ughbo}) and the ingoing shear of the solution $\Si$ on the initial sphere $S^2_{\infty,v_0}$.  

In particular, the geometric quantities of $\Gf$ satisfy the weighted boundedness estimates of Proposition \ref{prop:bndnew}, which are
identical to the weighted boundedness estimates proven for $\Si$ up to four angular derivatives of all geometric quantities. 

In fact, except for a small loss\footnote{This loss can potentially be avoided with further work. See Remark \ref{rem:bndnew}.} of decay towards null infinity for the highest angular derivatives of some of the $\Gf$ quantities, any weighted quantity bounded in $\Si$ by Theorem \ref{theo:mtheo} is also bounded for $\Gf$ and hence, by linearity for $\Sf$.

\item The geometric quantities of $\Sf$ satisfy the integrated decay estimates of Propositions \ref{prop:chibar}, \ref{prop:chibar4}, \ref{prop:intdecay3rs}, \ref{prop:etaetab}, \ref{prop:trchib2an}, \ref{prop:beta5}. In particular, we obtain a degenerate (near $r=3M$) integrated decay estimate for five angular derivatives of the linearised curvature components $\left(\alin \, , \, \blin \, , \,\rlin  \, , \, \slin  \, , \, \bblin  \, , \,\ablin \, \right)$ and a non-degenerate estimate for four (or less) derivatives.

\item The geometric quantities of $\Sf$ satisfy the polynomial decay proven in Section \ref{sec:polyfinal}.
In particular, the metric coefficients satisfy the polynomial decay estimates
\begin{align} \label{mcd1}
\Big\| r^{-1} \cdot r^2  \slashed{\mathcal{D}}_2^\star \slashed{\nabla}_A \Olin \, \Big\|_{S^2_{u,v}} 
\lesssim \frac{1}{v} \left( \textrm{right hand side of (\ref{ughb})}\right) \, ,
\end{align}
\begin{align} \label{mcd4}
\Big\| r^{-1} \cdot  r \slashed{\mathcal{D}}_2^\star \bmlin \, \Big\|_{S^2_{u,v}} \lesssim \frac{1}{v^{1/2-\epsilon}} \left(\textrm{right hand side of (\ref{ughb})} \right)
\end{align} 
\begin{align} \label{mcd2}
\Big\| r^{-1} \cdot \mathcal{A}^{[2]} \glinh \, \Big\|_{S^2_{u,v}} 
\lesssim \frac{1}{v^{1/2-\epsilon/2}} \left( \textrm{right hand side of (\ref{ughb})}\right) \, ,
\end{align}
\begin{align} \label{mcd3}
\Big\|  r^{-1} \cdot r^2  \slashed{\mathcal{D}}_2^\star \slashed{\nabla}_A \frac{\glinto}{\sqrt{\slashed{g}}} \, \Big\|_{S^2_{u,v}} 
\lesssim \frac{1}{v^{1/2-\epsilon/2}} \left( \textrm{right hand side of (\ref{ughb})}\right)\, , 
\end{align}
to be proven in Section \ref{sec:mecon}.
\end{enumerate}

Finally, let $\mathscr{K}_{\mathfrak{m},s_i}$ be the reference linearised Kerr solution defined in Theorem \ref{theo:mtheo}. Then
$\Sf'=\Sf-\mathscr{K}_{\mathfrak{m},s_i}$ is supported away from $\ell=0,1$ and the above statements
of the theorem hold as stated for $\Sf'$. As in the final statement
of Theorem~\ref{theo:mtheo}, the energies are now coercive
on all quantities $(\ref{regq1})$ of $\Sf^\prime$.
\end{bigtheorem}

We append a remark  analogous to Remark~\ref{rem:afrem}  in Theorem~\ref{theo:mtheo}:

\begin{remark} \label{rem:afrem2}
The statement $1.$ in Theorem \ref{theo:mtheod} implies in particular the estimate (\ref{ughb}) for the geometric quantities of $\Sf$ on the left hand side. Therefore, analogous to Remark \ref{rem:afrem}, we can interpret the result as a propagation of asymptotic flatness for the solution $\Sf$.
\end{remark}

A simple application of the Sobolev embedding theorem on the round sphere to (\ref{mcd1})--(\ref{mcd3}) provides
\begin{corollary}\label{newcoroledw}
With $\Sf$ and $\mathscr{K}_{\mathfrak{m}, s_i}$ as in Theorem~\ref{theo:mtheod},
defining 
$\Sf^\prime\doteq\Sf-\mathscr{K}_{\mathfrak{m},s_i}$,
then the metric components $\Olino\left[\Sf^\prime\right]$,
$\bmlin\left[\Sf^\prime\right]$, $\glinto\left[\Sf^\prime\right]$, 
$\glinh\left[\Sf^\prime\right]$ of $\Sf^\prime$ 
satisfy 
the following uniform bounds  on $\mathcal{M} \cap \{u \geq u_0\} \cap \{v \geq v_0\}$:
\begin{align}
\label{pointwisedecayone}
| \Olin | \lesssim \frac{1}{v} \left( \textrm{right hand side of (\ref{ughb})}\right)
\end{align}
and
\begin{align}
\label{pointwisedecaytwo}
| \bmlin | + | \glinto| + |\glinh| \lesssim \frac{1}{v^{\frac{1}{2}-\epsilon}} \left( \textrm{right hand side of (\ref{ughb})}\right).
\end{align}
Thus, the metric components of $\Sf$ 
converge pointwise to the linearised
Kerr values of $\mathscr{K}_{\mathfrak{m}, s_i}$.
\end{corollary}
 Let us remark that one can also obtain pointwise bounds for all Ricci coefficients and curvature components from the bounds proven in this paper but we will note state these bounds explicitly here.

\section{Proof of Theorem~\ref{prop:summarypsi}} \label{sec:RW}
 
The present section contains the proof of Theorem \ref{prop:summarypsi}. 
As explained in Section~\ref{CsweXINTRO}, this proof follows 
closely previous work for the scalar wave equation~$(\ref{LinScaEq})$. 
The reader may wish to refer
to Section~\ref{CsweXINTRO} for comparison while reading the present section.

We begin in Section~\ref{sec:econs}   with the natural energy identity associated to the Regge--Wheeler equation.
We then show in Section~\ref{IDEforPhere} a version of integrated decay which
degenerates at $r=3M$, at the horizon $\mathcal{H}^+$ and at null infinity $\mathcal{I}^+$. 
The degeneration at $\mathcal{H}^+$
is completely removed in Section~\ref{sec:rshif} using the red-shift, whereas
the degeneration  at $\mathcal{I}^+$  is refined in Section~\ref{sec:ini} using an $r^p$ hierarchy.
Higher order estimates and polynomial decay estimates for the energy will be 
the content of Section \ref{sec:decRW}.

\subsection{Energy conservation for Regge--Wheeler} \label{sec:econs}
Let $\Psi$ be as in the statement of Theorem \ref{prop:summarypsi}.

From 
\begin{align} \label{rwr}
\Omega \slashed{\nabla}_3 \left(\Omega \slashed{\nabla}_4 \Psi \right) - \left(1-\frac{2M}{r}\right) \slashed{\Delta}\Psi + V \Psi = 0 \, \ \textrm{ \ with} \ \ V = \left(\frac{4}{r^2} - \frac{6M}{r^3}\right)\left(1-\frac{2M}{r}\right) \, ,
\end{align}
we easily derive the following identity:
\begin{align}
&\left[\Omega \slashed{\nabla}_3 + \Omega \slashed{\nabla}_4 \right] \int \sin \theta d\theta d\phi \Big\{ |\Omega \slashed{\nabla}_4 \Psi |^2 + |\Omega \slashed{\nabla}_3 \Psi |^2 +2 \frac{1-\frac{2M}{r}}{r^2} |r \slashed{\nabla} \Psi |^2 + 2V | \Psi|^2 \Big\} \nonumber \\
+ &\left[\Omega \slashed{\nabla}_3 - \Omega \slashed{\nabla}_4 \right]\int \sin \theta d\theta d\phi \Big\{ |\Omega \slashed{\nabla}_4 \Psi |^2 - |\Omega \slashed{\nabla}_3 \Psi |^2 \Big\} = 0 \label{eerwr} \, .
\end{align}
Using the notation $1-\mu=1-\frac{2M}{r}$, we define the null fluxes
\begin{align}
F^T_u\left[\Psi\right] \left(v_1,v_2\right) = \int_{v_1}^{v_2} dv \sin \theta d\theta d\phi \left[ | \Omega \slashed{\nabla}_4 \Psi |^2  +  \left(1-\mu\right) |\slashed{\nabla} \Psi|^2 + V |\Psi|^2 \right] \nonumber
\end{align}
\begin{align}
F^T_v\left[\Psi\right] \left(u_1,u_2\right) = \int_{u_1}^{u_2} du \sin \theta d\theta d\phi \left[ | \Omega \slashed{\nabla}_3 \Psi |^2  +  \left(1-\mu\right) |\slashed{\nabla} \Psi|^2 + V |\Psi|^2 \right] \nonumber
\end{align}
Note that $V= \left(1-\frac{2M}{r}\right) \left[ \frac{4}{r^2} - \frac{6M}{r^3} \right] \geq \frac{1}{r^2}\left(1-\frac{2M}{r}\right)$ and hence that these fluxes are manifestly coercive. 

Integrating (\ref{eerwr}) with respect to $dudv$ yields a conservation law:
\begin{proposition} \label{prop:consT}
For any $u\geq u_0$ and $v\geq v_0$ the $\Psi$ of Theorem \ref{prop:summarypsi} satisfies
\begin{align} \label{econs}
F^T_u\left[\Psi\right] \left(v_0,v\right) + F^T_v\left[\Psi\right] \left(u_0,u\right) = F^T_{v_0}\left[\Psi\right] \left(u_0,u\right) + F^T_{u_0}\left[\Psi\right] \left(v_0,v\right) \, .
\end{align}
\end{proposition}

The above is the precise analogue of the $T$-energy identity for solutions $\varphi$ 
of $(\ref{LinScaEq})$.\footnote{Let us note that one can indeed easily adapt the energy-momentum
tensor formalism of Section~\ref{BoundforWAVE} to 
Regge--Wheeler, but we here prefer to explicitly
integrate by parts.}

\subsection{Integrated decay estimate}
\label{IDEforPhere}
Let us define the operators
\[
T:= \frac{1}{2} \left[\Omega \slashed{\nabla}_3 + \Omega \slashed{\nabla}_4 \right] \ \ \ \  \textrm{and}  \ \ \ \ R^\star:= \frac{1}{2} \left[-\Omega \slashed{\nabla}_3 + \Omega \slashed{\nabla}_4 \right] \, .
\]
(We note that $T$ above coincides with Lie-differentiation $\mathcal{L}_T$ with respect to
the Killing field $T$ of Section~\ref{sec:Killing}, but the above form will be convenient
here.)

Let now $\mathfrak{f}$ be a function on $\mathcal{M}^\circ$
of $r^\star\doteq (v-u)$ only, i.e.~$T\left(\mathfrak{f}\right)=0$ and $f^\prime:=R^\star \left(\mathfrak{f}\right)$. We have the identity
\begin{align} \label{id1t}
&\left[\Omega \slashed{\nabla}_3 + \Omega \slashed{\nabla}_4 \right]  \left( \mathfrak{f} \Big\{  |\Omega \slashed{\nabla}_4 \Psi |^2 -  |\Omega \slashed{\nabla}_3 \Psi |^2 \Big\}  \right)  \\
& \left[\Omega \slashed{\nabla}_3 - \Omega \slashed{\nabla}_4 \right] \left( \mathfrak{f}\Big\{  |\Omega \slashed{\nabla}_4 \Psi |^2 +  |\Omega \slashed{\nabla}_3 \Psi |^2 -2\frac{1-\mu}{r^2} |r \slashed{\nabla} \Psi |^2 -2V|\Psi|^2 \Big\}  \right) \nonumber \\
&+2 \mathfrak{f}^\prime  \left(   |\Omega \slashed{\nabla}_4 \Psi |^2 +  |\Omega \slashed{\nabla}_3 \Psi |^2 \right) -4R^\star \left(\mathfrak{f} \frac{1-\mu}{r^2}\right) |r \slashed{\nabla} \Psi |^2 - 4R^\star \left(\mathfrak{f} V\right)|\Psi|^2 \equiv 0 \, , \nonumber
\end{align}
{\bf where $\equiv$ means that the above becomes an equality after integration over $\int \sin \theta d\theta d\phi$}, and the identity
\begin{align} \label{id2t}
&+\left[\Omega \slashed{\nabla}_3 + \Omega \slashed{\nabla}_4 \right]  \Big(\mathfrak{f}^\prime \Psi \cdot \left[\Omega \slashed{\nabla}_3 + \Omega \slashed{\nabla}_4 \right]  \Psi \Big) \nonumber \\
&- \left[\Omega \slashed{\nabla}_3 - \Omega \slashed{\nabla}_4 \right]  \Big(\mathfrak{f}^\prime \Psi \cdot \left[\Omega \slashed{\nabla}_3 - \Omega \slashed{\nabla}_4 \right]  \Psi +  \mathfrak{f}^{\prime \prime} |\Psi|^2 \Big) \nonumber \\
 & - 2\mathfrak{f}^{\prime \prime \prime} \|\Psi\|^2 - 4\mathfrak{f}^\prime \Omega\slashed{\nabla}_3 \Psi \cdot \Omega\slashed{\nabla}_4 \Psi + 4\mathfrak{f}^\prime \left(\frac{1-\mu}{r^2} |r \slashed{\nabla} \Psi |^2 +V|\Psi|^2  \right)\equiv 0 \, .
\end{align}
Adding (\ref{id1t}) and (\ref{id2t}) yields the identity
\begin{align} \label{mora}
0 \equiv \left[\Omega \slashed{\nabla}_3 + \Omega \slashed{\nabla}_4 \right]  \Big( \mathfrak{f} \Big\{  |\Omega \slashed{\nabla}_4 \Psi |^2 -  |\Omega \slashed{\nabla}_3 \Psi |^2 + \mathfrak{f}^\prime \Psi \cdot \left[\Omega \slashed{\nabla}_3 + \Omega \slashed{\nabla}_4 \right]  \Psi \Big\}\Big) \nonumber \\
+\left[\Omega \slashed{\nabla}_3 - \Omega \slashed{\nabla}_4 \right] \Bigg( \mathfrak{f} \Big\{  |\Omega \slashed{\nabla}_4 \Psi |^2 +  |\Omega \slashed{\nabla}_3 \Psi |^2 -2\frac{1-\mu}{r^2} |r \slashed{\nabla} \Psi |^2 -2V|\Psi|^2 \Big\} \nonumber \\
- \mathfrak{f}^\prime \Psi \cdot \left[\Omega \slashed{\nabla}_3 - \Omega \slashed{\nabla}_4 \right]  \Psi -  \mathfrak{f}^{\prime \prime} |\Psi|^2 \Bigg) \nonumber \\
+ 2\mathfrak{f}^\prime | \Omega \slashed{\nabla}_4 \Psi - \Omega \slashed{\nabla}_3 \Psi |^2 
+ |r \slashed{\nabla} \Psi |^2 \left[ -4 \mathfrak{f} \left(\frac{1-\mu}{r^2}\right)^\prime \right] + |\Psi|^2 \left(-4\mathfrak{f} V^\prime -2\mathfrak{f}^{\prime \prime \prime}\right) \, .
\end{align}
Note that after integration with respect to the measure $\int du dv \sin \theta d\theta d\phi$ the term in the last line is a spacetime term, while all others are boundary terms.

\subsubsection{The choice of $\mathfrak{f}$} \label{sec:fchoice}

The next Lemma shows that we can choose a function $\mathfrak{f}$ in the identity (\ref{mora}) such that the last line of the latter is a manifestly non-negative expression. The choice below has appeared before in \cite{Holzegelults}.

\begin{lemma} \label{lem:fchoice}
If we define
\begin{align} \label{fdef}
\mathfrak{f} = \left(1-\frac{3M}{r}\right) \left(1+\frac{M}{r}\right) \, ,
\end{align}
then there exists a constant $c$ such that the $\Psi$ in Theorem \ref{prop:summarypsi} satisfies
\begin{align} \label{bulkes}
\int \sin \theta d\theta d\phi \Bigg\{ &
   |r \slashed{\nabla} \Psi |^2 \left[ -\frac{1}{2} \frac{\mathfrak{f}}{1-\mu} \left(\frac{1-\mu}{r^2}\right)^\prime \right] 
    + |\Psi|^2 \left[-\frac{1}{2}\frac{V^\prime}{1-\mu} \mathfrak{f} - \frac{1}{4}\frac{\mathfrak{f}^{\prime \prime \prime}}{1-\mu}\right] \Bigg\} \geq \frac{c}{r^3}\int \sin \theta d\theta d\phi |\Psi|^2
\end{align}
for all $\left(u,v\right)$ with $r \in \left(2M,\infty\right)$.
\end{lemma}

\begin{remark}
With the above choice of $\mathfrak{f}$, the square bracket multiplying the term $|r \slashed{\nabla} \Psi|^2$ is non-negative.
Since (\ref{fdef}) implies $\frac{\mathfrak{f}^\prime}{1-\mu} \geq \frac{2M}{r^2}$, the last line of (\ref{mora}) is indeed non-negative.
\end{remark}

\begin{proof}
Since the term $|r \slashed{\nabla} \Psi|^2$ is non-negative, applying Lemma \ref{lem:angularest} shows that it suffices to establish
\begin{equation} \label{goali}
-\frac{1}{2}\frac{\left( V + \frac{2}{r^2}\left(1-\mu\right) \right)^\prime }{1-\mu} \mathfrak{f} - \frac{1}{4}\frac{\mathfrak{f}^{\prime \prime \prime}}{1-\mu} \geq \frac{c}{r^3} \, .
\end{equation}
We compute
\begin{align}
\mathfrak{f}_r &= \frac{3M}{r^2}\left(1+\frac{M}{r}\right) + \left(1-\frac{3M}{r}\right)\left(-\frac{M}{r^2}\right) = \frac{2M}{r^2} + \frac{6M^2}{r^3} \, , \nonumber \\
\mathfrak{f}_{rr} &= -\frac{4M}{r^3} - \frac{18M^2}{r^4} \, , \nonumber \\
\mathfrak{f}_{rrr} & = +\frac{12M}{r^4} + \frac{72M^2}{r^5} \, ,
\end{align}
hence 
\begin{align}
\mathfrak{f}^\prime &= \mathfrak{f}_r \left(1-\mu\right) = \left( \frac{2M}{r^2} + \frac{6M^2}{r^3} \right) \left(1-\mu\right) \, ,
\nonumber \\
\mathfrak{f}^{\prime \prime} &= \mathfrak{f}_{rr} \left(1-\mu\right)^2 + \mathfrak{f}_r \frac{2M}{r^2}\left(1-\mu\right) \, ,
\nonumber \\
\mathfrak{f}^{\prime \prime \prime} &= \mathfrak{f}_{rrr} \left(1-\mu\right)^3 + \mathfrak{f}_{rr} \frac{6M}{r^2}\left(1-\mu\right)^2 - 2M \mathfrak{f}_r \frac{2}{r^3}\left(1-\frac{3M}{r}\right) \left(1-\mu\right) \, ,
\end{align}
and therefore
\begin{align}
expr &= \frac{1}{2} \left(V+\frac{2}{r^2}\left(1-\mu\right)\right)^\prime \mathfrak{f} + \frac{1}{4}\mathfrak{f}^{\prime \prime \prime}  \nonumber \\
&= \left[ \frac{3M}{r^4} + \frac{18M^2}{r^5} \right] \left(1-\mu\right)^3 - \frac{3}{2}\frac{M}{r^2}\left(1-\mu\right)^2 \left(\frac{4M}{r^3} + \frac{18M^2}{r^4}\right) 
-\frac{M}{r^3} \left(\frac{2M}{r^2} + \frac{6M^2}{r^3} \right) \left(1-\mu\right) \left(1-\frac{3M}{r}\right) \nonumber \\
\nonumber
& \ \ \ + \frac{1}{2}\left(1-\frac{3M}{r}\right)\left(1+\frac{M}{r}\right)\left[ \frac{-2}{r^3}\left(1-\frac{3M}{r}\right) \left(1-\mu\right) \left(6- \frac{6M}{r}\right) + \frac{6M}{r^4} \left(1-\mu\right)^2\right] \, .
\end{align}
We claim that this expression is negative for $r \in  \left(2M,\infty\right)$. To see this, we write the expression as follows
\begin{align}
-\left(1-\mu\right)^{-1} expr  = -\left[ \frac{3M}{r^4} + \frac{18M^2}{r^5} \right] \left(1-\frac{4M}{r} + \frac{4M^2}{r^2}\right) \nonumber \\
+ \frac{3}{2}\frac{M}{r^2}\left(1-\frac{2M}{r}\right) \left(\frac{4M}{r^3} + \frac{18M^2}{r^4}\right) 
+\frac{M}{r^3} \left(\frac{2M}{r^2} + \frac{6M^2}{r^3} \right) \left(1-\frac{3M}{r}\right) \nonumber \\
+ \left(1-\frac{3M}{r}\right)\left(1+\frac{M}{r}\right)\left[ \frac{1}{r^3}\left(1-\frac{3M}{r}\right) \left(6 - \frac{6M}{r}\right) - \frac{3M}{r^4} \left(1-\frac{2M}{r}\right)\right] \nonumber \, ,
\end{align}
which is computed to be
\begin{align}
-\left(1-\mu\right)^{-1} expr  \nonumber \\ = -\left[ \frac{3M}{r^4}+ \frac{6M^2}{r^5} - \frac{60M^3}{r^6}+\frac{72M^4}{r^7} \right]
+ \frac{3}{2} \frac{M}{r^2}\left(\frac{4M}{r^3} +\frac{10M^2}{r^4} - \frac{36M^3}{r^5}\right) 
+\frac{M}{r^3} \left(\frac{2M}{r^2} - \frac{18M^3}{r^4} \right) \nonumber \\
+\frac{1}{r^3} \left(1-\frac{5M}{r} + \frac{3M^2}{r^2} + \frac{9M^3}{r^3}\right) \left(6 - \frac{6M}{r}\right)  -\frac{3M}{r^4} \left(1-\frac{4M}{r} + \frac{M^2}{r^2} + \frac{6M^3}{r^3} \right) \nonumber
\end{align}
and simplifies further to
\begin{align}
-\left(1-\mu\right)^{-1} expr \nonumber \\ \geq \frac{6}{r^3} + \frac{1}{r^4} \left(-3M -30M-6M-3M\right) 
+ \frac{1}{r^5} \left(-6M^2 + 6M^2 + 2M^2+ 18M^2 + 30M^2 + 12M^2 \right) \nonumber \\
+ \frac{1}{r^6} \left(60M^3 +15M^3+54M^3-18M^3-3M^3\right) 
+ \frac{1}{r^7} \left(-72M^4-54M^4-18M^4-54M^4-18M^4\right) \, . \nonumber
\end{align}
It thus suffices to establish positivity of the polynomial
\begin{align}
6 r^4 -42Mr^3+ 62M^2r^2+ 108M^3 r - 216
\end{align}
or, upon setting $r=2Mx$, positivity of
\begin{align}
p\left(x\right) = 12 x^4 -42 x^3 + 31 x^2 + 27x-27 \ \ \ for \ x \in \left(1,\infty\right)
\end{align}
which is easily established.
\end{proof}

\subsubsection{The basic estimate}
Upon integrating  (\ref{mora}) with respect to $dudv \sin \theta d\theta d\phi$ over any spacetime region $\left[u_0,u\right] \times \left[v_0,v\right] \times S^2_{\bar{u},\bar{v}}$ with $\mathfrak{f}$ as chosen in Lemma \ref{lem:fchoice}, we see ($\mathfrak{f}$ is uniformly bounded!) that we can estimate all boundary terms (null-fluxes) by the fluxes $F^T_{u} \left[\Psi\right] \left(v_0,v\right)$, $F^T_{v} \left[\Psi\right] \left(u_0,u\right)$, $F^T_{v_0} \left[\Psi\right] \left(u_0,u\right)$, $F^T_{u_0} \left[\Psi\right] \left(v_0,v\right)$, which by the conservation law (\ref{econs}) means all boundary terms are controlled by a constant times $F^T_{u_0} \left[\Psi\right] \left(v_0,v\right) + F^T_{u_0} \left[\Psi\right] \left(v_0,v\right)$ alone. Exploiting now the statement of Lemma \ref{lem:fchoice} for the term in the last line of (\ref{mora}) we obtain the basic Morawetz estimate
\begin{align} \label{mori}
\int_{u_0}^u \int_{v_0}^v \int_{S^2_{\bar{u},\bar{v}}} d\bar{u} d\bar{v} \sin \theta d\theta d\phi \left(1-\mu\right)  \left[ \frac{1}{r^2} | \Omega \slashed{\nabla}_4 \Psi - \Omega \slashed{\nabla}_3 \Psi |^2 + \frac{1}{r^3} |\Psi|^2 \right] \nonumber \\
\leq C \left[ F^T_{u_0} \left[\Psi\right] \left(v_0,v\right) + F^T_{v_0} \left[\Psi\right] \left(u_0,u\right) \right]   \, .
\end{align}

The above estimate (\ref{mori}) can be improved immediately. First of all, revisiting the proof of Lemma \ref{lem:fchoice} it is clear that the estimate (\ref{bulkes}) still holds if we borrow a little bit of the good angular term, so that the argument above actually establishes
\begin{align}
\int_{u_0}^u \int_{v_0}^v \int_{S^2_{\bar{u},\bar{v}}} d\bar{u} d\bar{v} \sin \theta d\theta d\phi \left(1-\mu\right) \Big[ \frac{1}{r^2} | \Omega \slashed{\nabla}_4 \Psi - \Omega \slashed{\nabla}_3 \Psi |^2 + \frac{(r-3M)^2}{r^3} |\slashed{\nabla} \Psi|^2 
+ \frac{1}{r^3} |\Psi|^2 \Big] \nonumber \\
\leq C \left[ F^T_{u_0} \left[\Psi\right] \left(v_0,v\right) + F^T_{v_0} \left[\Psi\right] \left(u_0,u\right) \right] . \nonumber
\end{align}
A standard argument allows to recover the missing derivative: For instance, integrating the identity (\ref{id1t}) with a bounded, monotonically increasing $\mathfrak{f}$ which vanishes to third order near $r=3M$, we obtain
\begin{align} \label{basmou}
\int_{u_0}^u \int_{v_0}^v \int_{S^2_{\bar{u},\bar{v}}} d\bar{u} d\bar{v} \sin \theta d\theta d\phi  \left(1-\mu\right) \Bigg[  \frac{1}{r^2}| \Omega \slashed{\nabla}_4 \Psi - \Omega \slashed{\nabla}_3 \Psi |^2   + \frac{1}{r^3} |\Psi|^2 \nonumber \\
 + \frac{(r-3M)^2}{r^3} \left( |\slashed{\nabla} \psi|^2 + \frac{1}{r^2} | \Omega \slashed{\nabla}_4 \Psi + \Omega \slashed{\nabla}_3 \Psi |^2 \right)  \Bigg] 
\leq C \left[ F^T_{u_0} \left[\Psi\right] \left(v_0,v\right) + F^T_{v_0} \left[\Psi\right] \left(u_0,u\right) \right]  \, . \end{align} 

The degeneration near $r=3M$ is the familiar trapping phenomenon and cannot be removed (although it can be improved to logarithmic loss, cf.~\cite{Toha1}). The degeneration at the horizon however can be removed by exploiting the redshift. The weights near infinity can also be improved.
We turn to these two refinements in Sections~\ref{sec:rshif} and~\ref{sec:ini} below.

\subsection{Improving the weights near the horizon $\mathcal{H}^+$: The redshift}
\label{sec:rshif}
Given Proposition~\ref{prop:consT} and estimate $(\ref{basmou})$, 
the argument exploiting the red-shift
identity as described in Sections~\ref{BoundforWAVE} and~\ref{trappeddifficult} 
for the scalar wave equation $(\ref{LinScaEq})$ (cf.~\cite{Mihalisnotes}), 
can be immediately
adapted to $\Psi$.

In particular,
one upgrades Proposition \ref{prop:consT} to the non-degenerate boundedness statement
\begin{align} \label{ncons}
F_u\left[\Psi\right] \left(v_0,v\right) + F_v\left[\Psi\right] \left(u_0,u\right) \lesssim F_{v_0}\left[\Psi\right] \left(u_0,u\right) + F_{u_0}\left[\Psi\right] \left(v_0,v\right) \, ,
\end{align}
where these are now \emph{non-degenerate} null-fluxes defined in (\ref{nuflu1}) and (\ref{nuflu2}),
and the estimate (\ref{basmou}) itself to the improved\footnote{in the sense that the regular transversal derivative $\frac{1}{\Omega}\slashed{\nabla}_3 \Psi$ is also 
controlled near the horizon--the degeneracy near $r=3M$ remains.} integrated decay estimate
\begin{align} \label{moraf} 
\int_{u_0}^u \int_{v_0}^v \int_{S^2_{\bar{u},\bar{v}}} d\bar{u} d\bar{v} \sin \theta d\theta d\phi  \Omega^2\Bigg[  \frac{1}{r^2}| \Omega \slashed{\nabla}_4 \Psi - \Omega \slashed{\nabla}_3 \Psi |^2   + \frac{1}{r^3} |\Psi|^2 \nonumber \\
 + \frac{(r-3M)^2}{r^3} \left( |\slashed{\nabla} \Psi|^2 + \frac{1}{r^2} | \Omega \slashed{\nabla}_4 \Psi  |^2 + \frac{\Omega^2}{r^2}| \Omega^{-1} \slashed{\nabla}_3 \Psi |^2 \right)  \Bigg] 
\lesssim F_{u_0} \left[\Psi\right] \left(v_0,v\right) + F_{v_0} \left[\Psi\right] \left(u_0,u\right)   \, . 
\end{align} 
Note that taking the limit $u,v \rightarrow \infty$ the left hand side is precisely $\mathbb{I}_{deg}\left[\Psi\right]$. Hence (\ref{ncons}) and (\ref{moraf}) prove 

\begin{proposition} \label{prop:basicforrw}
The $\Psi$ in Theorem \ref{prop:summarypsi} satisfies the boundedness estimate
\begin{align}
\sup_u F_u\left[\Psi\right] \left(v_0,\infty\right) + \sup_v F_v\left[\Psi\right] \left(u_0,\infty\right) \lesssim F_{v_0}\left[\Psi\right] \left(u_0,\infty\right) + F_{u_0}\left[\Psi\right] \left(v_0,\infty\right) \, ,
\end{align}
and the integrated decay estimate
\begin{align} \label{intdecweak}
\mathbb{I}_{deg} \left[\Psi\right]  \lesssim F_{v_0}\left[\Psi\right] \left(u_0,\infty\right) + F_{u_0}\left[\Psi\right] \left(v_0,\infty\right) \, ,
\end{align}
provided the initial energies on the right hand side are finite.
\end{proposition}
Note that higher order versions of the above Proposition are immediate from Lie differentation with the Killing fields of Section~\ref{sec:Killing}, i.e. $\mathcal{L}_T$ as well as $\mathcal{L}_{\mathnormal{\Omega}_i}$.

\subsection{Improving the weights near null infinity $\mathcal{I}^+$: The $r^p$ hierarchy} \label{sec:ini}
The $r^p$ hierarchy of~\cite{DafRodnew} recalled in Section~\ref{therphierforwave} in the context 
of the scalar wave equation~$(\ref{LinScaEq})$ 
can also now be adapted to $\Psi$.

From the Regge--Wheeler equation for $\Psi$ we derive the identity (for $1\leq p \leq 2$ and $k\geq 1$)
\begin{align} \label{nme} 
 \partial_u \left[ \frac{r^p}{\left(1-\mu\right)^k} |\Omega \slashed{\nabla}_4 \Psi|^2 \right] + \partial_v \left[ \frac{r^p}{\left(1-\mu\right)^{k-1}} |\slashed{\nabla} \Psi|^2\right] +  \partial_v \left[ r^p \frac{V}{\left(1-\mu\right)^k} |\Psi|^2 \right] \nonumber \\
 -\partial_u \left(\frac{r^p}{\left(1-\mu\right)^k}\right)   |\Omega \slashed{\nabla}_4 \Psi|^2    -\partial_v \left( \frac{Vr^p}{\left(1-\mu\right)^k} \right) |\Psi|^2  \nonumber \\
 + \left[ \left(2-p\right) r^{p-1} \left(1-\mu\right)^{1-k} + r^p\left(k-1\right) \left(1-\mu\right)^{-k}\frac{2M}{r^2}r_v \right] |\slashed{\nabla} \Psi|^2 \equiv 0  \, ,
\end{align}
where we have used 
\[
-2\slashed{\nabla}_4 \Psi \cdot \slashed{\Delta} \Psi \equiv \slashed{\nabla}_4 | \slashed{\nabla} \psi |^2 - 2\left[\slashed{\nabla}_4, \slashed{\nabla}\right] \Psi \cdot \slashed{\nabla}\Psi = \slashed{\nabla}_4 | \slashed{\nabla} \Psi |^2 + tr \chi |\slashed{\nabla} \Psi|^2 \, ,
\]
and $\equiv$ indicates that (\ref{nme}) becomes an identity after integration against $\sin \theta d\theta d\phi$.

For our current purposes it will be sufficient to integrate (\ref{nme}) for $1\leq  p \leq 2$ with respect to the measure $dudv\sin \theta d\theta d\phi$ in a region
\[
\mathcal{R} = \Big\{ \left(u,v\right) \in \mathcal{M} \  \Big|  \ r\left(u,v\right) \geq R \ \ \textrm{and} \ \  u_{final} \geq u\geq u_{data} \Big\}
\]
for sufficiently large $R$. Precisely, we choose $R$ sufficiently large such that
\[
 -\partial_u \left(\frac{r^p}{\left(1-\mu\right)^k}\right) \geq \frac{1}{2} r^{p-1}  \textrm{\ \ \  for all $1\leq p \leq 2$ and $k \leq 5$.}
\]
Note also that
\begin{align}
-\partial_v \left( \frac{Vr^p}{\left(1-\mu\right)^k} \right) = -\partial_v \left(\frac{4r^{p-2} - 6M r^{p-3}}{\left(1-\mu\right)^{k-1}} \right) \nonumber \\
= \frac{r_v}{\left(1-\mu\right)^k} \Big[ \left(4\left(2-p\right) r^{p-3}  - 6M\left(3-p\right) r^{p-4}\right)\left(1-\mu\right) 
+ \left(k-1\right) \frac{2M}{r^2} \left(4r^{p-2} - 6Mr^{p-3}\right)\Big] \nonumber \\
= \frac{r_v}{\left(1-\mu\right)^k}  \left[4\left(2-p\right) r^{p-3} + M r^{p-4} \left(8k-4 +3p-18\right) + M^2 r^{p-5} \left(...\right) \right]
 \end{align}
 holds, which means that given any $1\leq p \leq 2$ the choice $k=3$ ensures that also the estimate
 \[
 -\partial_v \left( \frac{Vr^p}{\left(1-\mu\right)^k} \right) \geq 2M r^{p-4}
 \]
holds in $\mathcal{R}$ for sufficiently large $R$. Therefore, integrating (\ref{nme}) for $p= 2$ with respect to the measure $dudv\sin \theta d\theta d\phi$ we obtain the estimate
\begin{align} \label{mainweighted}
\int_{\mathcal{D} \cap \{r\geq R\}}du dv \sin \theta d\theta d\phi \Big(r | \Omega \slashed{\nabla}_4 \Psi|^2 + r^{1-\epsilon} |\slashed{\nabla} \Psi|^2 + r^{-1-\epsilon} |\Psi|^2  \Big) \nonumber \\
\leq \int_{v_0}^{\infty}dv \int_{S^2} \sin \theta d\theta d\phi \left( r^2 | \Omega \slashed{\nabla}_4 \Psi|^2 \right) \left(u_0,v\right)+ C \Big( F_{u_0} \left[\Psi\right] \left(v_0,v\right) + F_{v_0} \left[\Psi\right] \left(u_0,u\right) \Big)  \, ,
\end{align}
where the last two terms on the right hand side account for the terms arising on the timelike hypersurface at $r=R$, which can be controlled by the Morawetz estimate (\ref{moraf}) after averaging in $R$. 

At the same time, we obtain also good boundary terms (fluxes) from the terms in the first line of (\ref{nme}). We have thus deduced both (\ref{flux1}) and the $n=0$ part of (\ref{bestintdec}) after recalling the shorthand notation (\ref{enreft}), (\ref{enreft2}) for the energies.

With these bounds established we deduce the following corollary, which is an immediate consequence of the fundamental theorem of calculus:
\begin{corollary} \label{cor:Pons2} 
Under the assumptions of Theorem \ref{prop:summarypsi} we also have  the estimate
\begin{align}
\sup_{u\geq u_0,v\geq v_0} \| \Psi\|^2_{S^2_{u,v}} \lesssim \sup_{u} \| \Psi\|^2_{S^2_{u,v_0}} + F^\mathcal{I}_{u_0} \left[\Psi\right] \left(v_0,\infty \right) + F^\mathcal{I}_{v_0}\left[\Psi\right] \left(u_0,\infty\right) \, .
\end{align}
\end{corollary}

We finally note that integrating (\ref{nme})  with $p=1$, $k=3$ (instead of $p=2$, $k=3$ as done to derive (\ref{mainweighted})) leads to additional estimates which together with the choice $p=0$, $k=0$ (for which the identity (\ref{nme}) also holds) constitute the Regge--Wheeler analogue of the $r^p$-hierarchy for the wave equation in \cite{DafRodnew}.

\subsection{Higher order estimates and polynomial decay}  \label{sec:decRW}
In this section we will extend the above weighted estimates to higher order and then infer
polynomial decay.

We note the trivial fact that the Regge--Wheeler equation~(\ref{rwr}) 
commutes with Lie differentation with the Killing fields of Section~\ref{sec:Killing}, i.e.
$\mathcal{L}_T$ as well as $\mathcal{L}_{\mathnormal{\Omega}_i}$.\footnote{As in \cite{DHRscat} we could alternatively commute ``tensorially" with $r \slashed{\nabla}_A$ and estimate the lower order terms inductively.}
Recalling the commuted energies (\ref{do2TOg}) we hence immediately conclude the following Corollary which provides the estimate (\ref{qxz}) and the $n>0$ part of the estimate  (\ref{bestintdec}) in Theorem \ref{prop:summarypsi}:
\begin{corollary} \label{cor:higherorderbasic}
If the $\Psi$ in Theorem \ref{prop:summarypsi} satisfies $\mathbb{F}^{n,T,\slashed{\nabla}}_0 \left[\Psi\right] < \infty$  for some integer $n\geq 0$, then we have the estimate
\begin{align}
\label{xwrisovomaoxipia}
\mathbb{I}^{n,T,\slashed{\nabla}}_{\mathcal{I},\epsilon} \left[\Psi\right] + \mathbb{I}^{n,T,\slashed{\nabla}}_{deg} \left[\Psi\right]  + \mathbb{F}^{n,T,\slashed{\nabla}} \left[\Psi\right] &\lesssim \mathbb{F}_0^{n,T,\slashed{\nabla}}\left[\Psi\right]  \, .
\end{align}
\end{corollary}

We can in fact show an analogue of the above for an $n$-th order
\emph{non-degenerate} energy, where 
higher derivatives have moreover additional weights in $v$.
This is a straightforward adaptation of the 
procedure       appearing in~\cite{Schlue, Moschnewmeth} to $\Psi$,
and follows
by commuting the equation with the redshift operator $\Omega^{-1} \slashed{\nabla}_3$ near the horizon and with the weighted operator $r \Omega \slashed{\nabla}_4$ near null infinity,
and observing that 
the terms not-controllable by $(\ref{xwrisovomaoxipia})$ occur with favourable signs.
We will simply state the estimate arising. 
Define the energy
\begin{align}
\mathbb{F}^n\left[\Psi\right] :=  &\sum_{i+j+k\leq n} \sup_u F^\mathcal{I}_{u} \left[\left(\Omega^{-1} \slashed{\nabla}_3\right)^i \left(r \Omega \slashed{\nabla}_4\right)^j \left(r \slashed{\nabla}_A\right)^k \Psi\right] \left(v_0,\infty \right)  \nonumber \\
+ &\sum_{i+j+k\leq n} \sup_v F^\mathcal{I}_{v} \left[\left(\Omega^{-1} \slashed{\nabla}_3\right)^i \left(r \Omega \slashed{\nabla}_4\right)^j \left(r \slashed{\nabla}_A\right)^k \Psi\right] \left(u_0,\infty\right) \label{do2}
\end{align}
with initial energy
\begin{align}
\mathbb{F}_0^n\left[\Psi\right] :=  &\sum_{i+j+k\leq n} F^\mathcal{I}_{u_0} \left[\left(\Omega^{-1} \slashed{\nabla}_3\right)^i \left(r \Omega \slashed{\nabla}_4\right)^j \left(r \slashed{\nabla}_A\right)^k \Psi\right] \left(v_0,\infty \right)  \nonumber \\
+ &\sum_{i+j+k\leq n} F^\mathcal{I}_{v_0} \left[\left(\Omega^{-1} \slashed{\nabla}_3\right)^i \left(r \Omega \slashed{\nabla}_4\right)^j \left(r \slashed{\nabla}_A\right)^k \Psi\right] \left(u_0,\infty\right). 
\label{do2i}
\end{align}
We have
\begin{corollary} \label{cor:higherorder}
If the $\Psi$ in Theorem \ref{prop:summarypsi} satisfies $\mathbb{F}^n_0\left[\Psi\right] < \infty$, then we have for any $n\geq 0$ and non-negative integers $i,j,k$ with $i+j+k\leq n$ the estimate
\begin{align}
\mathbb{F}^n\left[\Psi\right] \lesssim \mathbb{F}_0^n\left[\Psi\right] \, ,\nonumber \\
\mathbb{I}_{deg}\left[\left(\Omega^{-1} \slashed{\nabla}_3\right)^i \left(r \Omega \slashed{\nabla}_4\right)^j \left(r \slashed{\nabla}_A\right)^k \Psi\right] +\mathbb{I}_\epsilon^{\mathcal{I}} \left[\left(\Omega^{-1} \slashed{\nabla}_3\right)^i \left(r \Omega \slashed{\nabla}_4\right)^j \left(r \slashed{\nabla}_A\right)^k \Psi\right] \lesssim\mathbb{F}_0^n\left[\Psi\right]  \,  . \nonumber
\end{align}
\end{corollary}
We will in fact
only use Corollary \ref{cor:higherorder} later to optimise decay statements already obtained. \\

Exploiting the $r^p$-hierarchy for the Regge--Wheeler equation discussed in Section \ref{sec:ini}, 
polynomial decay estimates
can be obtained for $\Psi$ exactly as in~\cite{DafRodnew} for the case of the scalar wave
equation (cf.~the discussion in Section~\ref{therphierforwave}).
We only give the most elementary statement here. 

Let us fix $r_0 \geq 8M$ and denote by $u\left(v,r_0\right)$ the $u$-value corresponding to the sphere of intersection between the $r=r_0$ hypersurface and the constant $v$ hypersurface. Note $v \sim u\left(v,r_0\right)$ for large $v$. 
Applying step by step the method of~\cite{DafRodnew} 
 using Corollary~\ref{cor:higherorderbasic} in place of the analogous statement for the wave equation $(\ref{LinScaEq})$, 
we obtain the following result:
\begin{proposition} \label{prop:decRW}
Fix $r_0 \geq 8M$ and $v \geq v_0$ and suppose the $\Psi$ in Theorem \ref{prop:summarypsi} satisfies $\mathbb{F}_0^{2,T} \left[\Psi\right]<\infty$ initially.
Then for any $V \geq v$ and any $U \geq u\left(v,r_0\right)$ we have
\begin{align}
F_U\left[\Psi\right] \left(v,\infty\right) + F_V \left[\Psi\right] \left( u\left(v,r_0\right),\infty\right)  
  \lesssim \frac{1}{v^2} \cdot \mathbb{F}_0^{2,T}\left[\Psi\right] \, .
 \nonumber
\end{align}
Here the constant implicit in $\lesssim$ depends on the choice of $r_0$ and we recall the non-degenerate energy fluxes (\ref{nuflu1}) and (\ref{nuflu2}).
\end{proposition}

For more details      on the method of~\cite{DafRodnew}, 
see~\cite{Schlue, Moschnewmeth}.
\begin{corollary} \label{cor:dmdec}
Under the assumptions of the previous proposition we have the integrated decay estimate
\begin{align}
\int_{v}^\infty d\bar{v} \int_{u\left(v,r_0\right)}^\infty d\bar{u} \int_{S^2_{\bar{u},\bar{v}}} \sin \theta d\theta d\phi \frac{\Omega^2}{r^{3}} \left[ | \Psi |^2 + \left(1-\frac{3M}{r}\right)^2 \left(| r \slashed{\nabla} \Psi |^2 + |\Omega^{-1} \slashed{\nabla}_3 \Psi|^2 \right) \right] 
 \lesssim  \frac{1}{v^2} \cdot \mathbb{F}_0^{2,T}\left[\Psi\right]  \, . \nonumber
\end{align}
\end{corollary}
\begin{proof}
Apply Proposition \ref{prop:decRW} to (\ref{moraf}).
\end{proof}

Using a dyadic decomposition of the domain, we can also deduce the following spacetime $L^1$-estimate from Proposition \ref{prop:decRW} and Corollary \ref{cor:dmdec}, which will be useful later.
\begin{corollary} \label{cor:PL1}
Under the assumptions of the previous proposition we also have the $L^1$-estimates
\begin{align}
\int_{v_0}^\infty d\bar{v} \int_{u_0}^\infty d\bar{u} \int_{S^2_{\bar{u},\bar{v}}} \sin \theta d\theta d\phi \, \frac{\Omega^2}{r^2} \left[ | \Psi | + | r \slashed{\nabla} \Psi| + | \Omega^{-1} \slashed{\nabla}_3 \Psi |\right] \lesssim \sqrt{\mathbb{F}_0^{2,T}\left[\Psi\right]}  \nonumber \, .
\end{align}
\end{corollary}

\section{Proof of Theorem~\ref{theo:mtheogi}} \label{sec:hdgi} 

In this section we exploit the transformation formulae of Section \ref{sec:tratheo} together with
Theorem~\ref{prop:summarypsi}
 to prove Theorem \ref{theo:mtheogi}. The reader can refer to the overview of Section \ref{giqistoria}.

We begin in Section \ref{sec:9basic} with the most basic estimates for $\psi$, $\underline{\psi}$ and $\alpha$, $\underline{\alpha}$ that follow straight from the transport structure of equations (\ref{psidef}), (\ref{evolpp}) and  (\ref{psidef2}), (\ref{evolpp2}) in conjunction with Theorem \ref{prop:summarypsi}.

 In Section \ref{sec:higherder} we obtain higher derivative estimates for $\psi$, $\underline{\psi}$ and $\alpha$, $\underline{\alpha}$ that follow from commuting the aforementioned  transport equations. Sometimes pointwise algebraic identities (Section \ref{sec:usefulids}) can be used to avoid the loss of derivatives that is encountered in using the transport equations. Combining these results we finally complete the proof of the first three statements of Theorem \ref{theo:mtheogi} in Section \ref{sec:prot1}. The final subsection \ref{sec:refinements} presents some refinements of the previous results, including higher order estimates and polynomial decay statements for solutions of the Teukolsky equation, in particular the last statement in Theorem \ref{theo:mtheogi}.

\subsection{Ascending the hierarchy: basic transport estimates} \label{sec:9basic}
We will prove the statements of Theorem~\ref{theo:mtheogi}  regarding
spin $+2$ and $-2$ Teukolsky equations in parallel.
Let $\alpha$ and $\underline\alpha$ be as in the two parts of Theorem~\ref{theo:mtheogi}.

We first  define the derived quantities $\psi$ and $P$ from $\alpha$,
and $\underline{\psi}$ and $\underline{P}$ from $\underline\alpha$, from formulas
$(\ref{psidef})$--$(\ref{evolpp})$ and $(\ref{psidef2})$--$(\ref{evolpp2})$, respectively, of Section~\ref{sec:tratheo}.
By Proposition~\ref{prop:rwt1} it follows that both $P$ and $\underline{P}$
satisfy the Regge--Wheeler equation. We may thus apply Theorem~\ref{prop:summarypsi} to 
both $P$
and $\underline{P}$, noting that by the assumptions of Theorem~\ref{theo:mtheogi}, 
the  corresponding initial energies appearing in 
Theorem~\ref{prop:summarypsi} are finite.

In the subsections that follow, we will show how from control of $P$ and $\underline{P}$ we
can control $\psi$, $\underline\psi$, and then $\alpha$, $\underline\alpha$,
by estimating transport equations.

\subsubsection{Estimates for $\psi$ and \underline{$\psi$}} \label{sec:psi}
We begin by estimating $\psi$ and $\underline\psi$.
\begin{proposition} \label{prop:psie} 
The derived quantity $\psi$ associated with the solution $\alpha$ of Theorem \ref{theo:mtheogi} 
satisfies the following estimates. Along any null hypersurface of constant $u \geq u_0$, including the event horizon,
\begin{align} \label{heui} 
\int_{v_1}^{v_2} dv \|r^{-1} \psi\|_{S^2_{u,v}}^2 r^{8-\epsilon} \Omega^2 \left(u,v\right) \lesssim \mathbb{F}_0\left[\Psi,  \psi\right] \, .
\end{align}
In addition, we have for any $\epsilon>0$ the integrated decay estimate 
\begin{align}\label{kaivourgioovoma}
\int_{\mathcal{D}} du dv  \|r^{-1} \underline{\psi}\|_{S^2_{u,v}}^2 r^{5-\epsilon} \lesssim \mathbb{F}_0\left[\underline{\Psi}, \underline{\psi} \right] \, .
\end{align}
where the constant in $\lesssim$ depends on $\epsilon$.

The derived quantity $\underline{\psi}$ associated with the solution $\underline{\alpha}$ of Theorem \ref{theo:mtheogi} 
satisfies the following estimates. Along any null hypersurface of constant $v \geq v_0$, including in the limit on null infinity:
\begin{align} 
\int_{u_1}^{u_2} du \|r^{-1} \underline{\psi}\|_{S^2_{u,v}}^2 r^6 \left(u,v\right) \lesssim \mathbb{F}_0\left[ \underline{\Psi}, \underline{\psi} \right] \, .
\end{align}
In addition, we have for any $\epsilon>0$ the integrated decay estimate
\begin{align}\label{kaivourgioovoma2}
\int_{\mathcal{D}} du dv  \|r^{-1} \psi\|_{S^2_{u,v}}^2  r^{7-\epsilon} \Omega^4 \lesssim \mathbb{F}_0\left[\Psi, , \psi \right] \, .
\end{align}
where the constant in $\lesssim$ depends on $\epsilon$.
\end{proposition}

\begin{proof}
From (\ref{evolpp}) we derive
\begin{align} \label{bas1}
\partial_u \left[ |\psi|^2 r^6 \Omega^2 \right] = 2  r^6 \Omega^3 P \psi
\end{align}
or, multiplying by $r^n$ and using $r_u = - \Omega^2$,
\begin{align} 
\partial_u \left( |\psi|^2 r^6 \Omega^2 \cdot r^n \right) + |\psi|^2 r^6 \Omega^4 n r^{n-1}= 2  r^{6+n} \Omega^3 P \psi \leq \frac{1}{2}  |\psi|^2 r^6 \Omega^4 n r^{n-1} + \frac{2}{n} r^{7+n} |P|^2 \Omega^2 \nonumber
\end{align}
which simplifies to
\begin{align} \label{psi}
\partial_u \left( |\psi|^2 r^6 \Omega^2 \cdot r^n \right) + \frac{1}{2} |\psi|^2 r^6 \Omega^4 n r^{n-1}
\leq \frac{2}{n} r^{7+n} |P|^2 \Omega^2 \, .
\end{align}
The analogue of (\ref{bas1}) is
\begin{align} \label{bas2}
\partial_v \left[ |\underline{\psi}|^2 r^6 \Omega^2 \right] = -2  r^6 \Omega^3 \underline{P} \underline{\psi} \, .
\end{align}
We can multiply this by $\frac{1}{\Omega^2}$ which satisfies $\partial_v \Omega^{-2}=-\frac{1}{\Omega^2}\frac{2M}{r^2}$:
\begin{align} \label{psiba}
\partial_v \left[\frac{1}{\Omega^2} |\underline{\psi}|^2 r^6 \Omega^2 \right] + 2M r^4 |\underline{\psi}|^2 = -2  r^6 \Omega \underline{P} \underline{\psi}
\end{align}
and hence
\begin{align} \label{psibar}
\partial_v \left[\frac{1}{\Omega^2} |\underline{\psi}|^2 r^6 \Omega^2 \right] + M r^4 |\underline{\psi}|^2 \leq  \frac{1}{M} r^8 |\underline{P}|^2 \Omega^2 \, .
\end{align}
Multiplying instead by $\frac{1}{r^{\epsilon}}$ we find similarly
\begin{align} \label{psibar2}
\partial_v \left[\frac{1}{r^{\epsilon}} |\underline{\psi}|^2 r^6 \Omega^2 \right] +  \epsilon \cdot r^{-1-\epsilon} r^6 \Omega^4 |\underline{\psi}|^2   \leq C_\epsilon \cdot r^{7-\epsilon} |\underline{P}|^2 \Omega^2 \, .
\end{align}

For the right hand sides in (\ref{psi}) (applied with $n=2-\epsilon$  for $\epsilon>0$) as well as (\ref{psibar}) and (\ref{psibar2})  we already have an integrated decay estimate, i.e.~the right hand side remains controlled from initial data when integrated over any spacetime region with respect to the measure $du dv \sin \theta d\theta d\phi$, see (\ref{bestintdec}). Upon this integration, the left hand sides of (\ref{psi}), (\ref{psibar}) and (\ref{psibar2}) will provide estimates for fluxes of $\psi$ and $\underline{\psi}$, as well as integrated decay as stated in Proposition \ref{prop:psie}.
\end{proof}

\begin{corollary} \label{cor:psishb} 
In addition to the bounds of Proposition \ref{prop:psie} we have for fixed $u\geq u_0$ (including the horizon $u=\infty$) and any $v\geq v_0$
\begin{align}
\| r^{-1}\cdot \underline{\psi} \Omega^{-1} r^3 \|^2_{S^2_{u,v}} + \int_{v_0}^v d\bar{v} \frac{1}{r^2} \| r^{-1}\underline{\psi} \Omega^{-1} r^3 \|^2_{S^2_{u,\bar{v}}}  \lesssim \| r^{-1} \cdot \underline{\psi} \Omega^{-1} r^3\|^2_{S^2_{u,v_0}} + \mathbb{F}_0 \left[\underline{\Psi}\right]. \nonumber
\end{align}
\end{corollary}
\begin{proof}
Write (\ref{bas2}) as $
\partial_v \left[ |\underline{\psi}|^2 \Omega^{-2} r^6 \right] + \frac{4M}{r^2} r^6  |\underline{\psi}|^2 \Omega^{-2} = -2r^6 \underline{P} \Omega^{-1} \underline{\psi}$, integrate using Cauchy's inequality and control on the $\underline{P}$-flux from Theorem \ref{prop:summarypsi}. 
\end{proof}

\begin{corollary} \label{cor:psish} 
In addition to the bounds of Proposition \ref{prop:psie}  we have for fixed $v\geq v_0$  and any $u\geq u_0$ (including the horizon $u=\infty$)
\begin{align}
r^3 \| r^{-1}\cdot {\psi} \Omega r^3 \|^2_{S^2_{u,v}} + \int_{u_0}^u d\bar{u} \Omega^2 r^2 \| r^{-1}{\psi} \Omega r^3 \|^2_{S^2_{u,\bar{v}}}  \lesssim r^3 \| r^{-1} \cdot {\psi} \Omega r^3\|^2_{S^2_{u_0,v}} + \mathbb{F}_0 \left[\Psi\right]. \nonumber
\end{align}
\end{corollary}
\begin{proof}
Write (\ref{bas1}) as $
\partial_u \left[ r^3 |{\psi}|^2 \Omega^{2} r^6 \right] + 3\Omega^2 r^8  |{\psi}|^2 \Omega^{2} = 2r^9 {P} \Omega^{3} {\psi}$, integrate using Cauchy's inequality and control on the ${P}$-flux from Theorem \ref{prop:summarypsi}. 
\end{proof}

\subsubsection{Estimates for $\alpha$ and \underline{$\alpha$}} \label{sec:alpha}
Now let us obtain decay estimates for $\alpha$ and $\underline{\alpha}$. The following proposition (in conjunction with Proposition \ref{prop:psie}) proves the estimate (\ref{teubo1}) of Theorem \ref{theo:mtheogi}.

\begin{proposition} \label{prop:alphae} 
The solution $\alpha$ of Theorem \ref{theo:mtheogi} satisfies the following estimates: Along any null hypersurface of constant $u \geq u_0$, including the event horizon, we have
\begin{align}
\int_{v_1}^{v_2} dv \|r^{-1}\alpha\|_{S^2_{u,v}}^2 r^{6-\epsilon} \Omega^4 \left(u,v\right) \lesssim \mathbb{F}_0\left[\Psi,  \psi,  \alpha \right] \, .
\end{align}
The solution $\underline{\alpha}$ of Theorem \ref{theo:mtheogi} satisfies the following estimates: Along any null hypersurface of constant $v\geq v_0$, including in the limit on null infinity, we have
\begin{align}
\int_{u_1}^{u_2} du   \|r^{-1} \underline{\alpha}\|_{S^2_{u,v}}^2 r^2 \Omega^{-2} \left(u,v\right) \lesssim \mathbb{F}_0\left[\underline{\Psi}, \underline{\psi},  \underline{\alpha} \right] \, .
\end{align}
Finally, we have the integrated decay estimates
\begin{align}
\int_{\mathcal{D}} du dv  r^{1-\epsilon} \Omega^{-2} \|r^{-1} \underline{\alpha}\|_{S^2_{u,v}}^2 \lesssim \mathbb{F}_0\left[\underline{\Psi}, \underline{\psi}, \underline{\alpha} \right] \textrm{ \ \ \ and \ \ \ } \int_{\mathcal{D}} du dv  r^{5-\epsilon} \Omega^6 \|r^{-1}\alpha\|_{S^2_{u,v}}^2   \lesssim \mathbb{F}_0\left[\Psi,  \psi, \alpha \right]  . 
\end{align}
\end{proposition}

\begin{proof}
Observe that we can write
\begin{align}
\slashed{\nabla}_3 \left(r \Omega^2 \alpha \right) = -2 \psi \cdot r \Omega^2 \, .
\end{align}
It follows that
\begin{align}
\partial_u \left(r^2 \Omega^4 |\alpha|^2 \right) = -4 \psi \cdot r^2 \Omega^5 \alpha \, ,
\end{align}
which upon multiplication with $r^n$ becomes
\begin{align}
\partial_u \left(r^n \cdot r^2 \Omega^4 |\alpha|^2 \right) + n r^{n-1} r^2 \Omega^6 |\alpha|^2  = -4 \psi \cdot r^3 \Omega^5 \alpha r^{n-1}  \leq \frac{n}{2} r^{n-1} r^2 \Omega^6 |\alpha|^2 + \frac{4}{n} r^{n-1} r^4 \Omega^4 |\psi|^2 \, .\nonumber
\end{align}
From the resulting
\begin{align} \label{koi}
\partial_u \left(r^n \cdot r^2 \Omega^4 |\alpha|^2 \right) + \frac{n}{2} r^{n-1} r^2 \Omega^6 |\alpha|^2  \leq  \frac{4}{n} r^{n-1} r^4 \Omega^4 |\psi|^2
\end{align}
we see from (\ref{bestintdec}) that choosing $n=4-\epsilon$, after integration with respect to $du dv \sin \theta d\theta d\phi$, the right hand side is controlled by Proposition \ref{prop:psie}. Therefore we obtain the analogue of Proposition \ref{prop:psie} for $\alpha$.  The argument for $\underline{\alpha}$ is similar and we will only sketch it. From
\begin{align} \label{aevol} 
\slashed{\nabla}_4 \left(r \Omega^2 \underline{\alpha} \right) = 2 \underline{\psi} \cdot r \Omega^2
\end{align}
we derive ($\partial_v = \Omega \slashed{\nabla}_4$)
\begin{align} \label{auit} 
\partial_v \left(\Omega^{-6} | r\Omega^2 \underline{\alpha} |^2\right) + 3\frac{2M}{r^2} \Omega^{-6} | r\Omega^2 \underline{\alpha} |^2 = 4\underline{\psi} \cdot r \Omega^{-1} r \underline{\alpha} \, .
\end{align}
Multiplying by $\Omega^{-6}$, integrating and proceeding as above, in particular observing that
\[
\underline{\psi} r \frac{1}{\Omega} r\underline{\alpha} \leq \frac{r^2}{4M} | \underline{\psi} r|^2 + \frac{1}{\Omega^2} \frac{M}{r^2} |r \underline{\alpha} |^2 
\]
yields the result for $\underline{\alpha}$ with the weights claimed at the horizon but weaker ones at infinity. To obtain the optimal weights at infinity multiply (\ref{auit}) with $r^{-\epsilon}$ and apply Cauchy-Schwarz on the right.
\end{proof}
\subsection{Higher derivative estimates} \label{sec:higherder}
In the following subsections we estimate higher derivatives of the quantities $\psi$, \underline{$\psi$}, $\alpha$, \underline{$\alpha$} from our control on $P$ and $\underline{P}$.
\subsubsection{Some useful identities} \label{sec:usefulids}
We begin by computing some useful identities, which we shall write in regular form so that the behaviour at the horizon can be assessed directly.

\begin{lemma} \label{lem:compuseid}
Consider the solution $\alpha$ of Theorem \ref{theo:mtheogi} and the derived quantities $\psi, P$ defined via (\ref{psidef}), (\ref{evolpp}). The following identities hold true:
\begin{align} \label{poy2}
\slashed{\nabla}_4 \left(r\psi \Omega \right) +  \left( 2 tr {\chi} -2\hat{\omega}\right) \left(r \psi \Omega \right) =  r \Omega \slashed{\mathcal{D}}_2^\star \slashed{div} {\alpha}+ \frac{3M}{r^2} \Omega \alpha \, ,
\end{align}
\begin{align} \label{qoy1}
\Omega \slashed{\nabla}_4 \left(r^5 P \right) = - 2r^5 \slashed{\mathcal{D}}_2^\star \slashed{div} \left(\psi \Omega\right) + 6Mr^2 \psi \Omega - 2r^3 \psi \Omega + 3r M \Omega^2 \alpha \, ,
\end{align}
\begin{align} \label{roy2}
\Omega \slashed{\nabla}_4 \left(r^2 \Omega \slashed{\nabla}_4 \left(r^5 P \right) \right) &= r\Omega  \slashed{\nabla}_4 \left(r \Omega \slashed{\nabla}_4 \left(r^5 P \right) \right) + \Omega^2 \cdot r \Omega \slashed{\nabla}_4 \left(r^5 P \right) \nonumber \\
&= -2 r^4 \slashed{\mathcal{D}}_2^\star \slashed{div}\slashed{\mathcal{D}}_2^\star \slashed{div} \left(r^3 \alpha \Omega^2\right)-4M r^2 \slashed{\mathcal{D}}_2^\star \slashed{div} \left(r^3 \psi \Omega\right) -6M r^2 \slashed{\mathcal{D}}_2^\star \slashed{div} \left(r^2 \alpha \Omega^2\right) \nonumber \\
& \ \ + \left(-2+\frac{6M}{r}\right) \Omega \slashed{\nabla}_4 \left(r^5 \psi \Omega\right) - 18M \Omega^2 r \psi \Omega + 3M \Omega \slashed{\nabla}_4 \left(r^3 \Omega^2 \alpha\right).
\end{align}

Consider the solution $\underline{\alpha}$ of Theorem \ref{theo:mtheogi} and the derived quantities $\underline{\psi},\underline{P}$ defined via (\ref{psidef2}), (\ref{evolpp2}). Then the following identities hold true: 
\begin{align} \label{poy1} 
\slashed{\nabla}_3 \left(r \frac{\underline{\psi} }{\Omega} \right) + 2 tr \underline{\chi} \left(r \frac{\underline{\psi} }{\Omega} \right) = -r \Omega^{-1} \slashed{\mathcal{D}}_2^\star \slashed{div} \underline{\alpha} - \frac{3M}{r^2} \frac{\underline{\alpha}}{\Omega} \, ,
\end{align}
\begin{align} \label{qoy2}
\Omega^{-1} \slashed{\nabla}_3 \left(r^5 \underline{P} \right) = + 2r^5 \slashed{\mathcal{D}}_2^\star \slashed{div} \left(\underline{\psi} \Omega^{-1}\right) - 6Mr^2 \underline{\psi} \Omega^{-1} + 2r^3 \underline{\psi} \Omega^{-1} + 3r M \underline{\alpha} \, ,
\end{align}
\begin{align} \label{roy1}
\Omega^{-1} \slashed{\nabla}_3  \left( \Omega^{-1} \slashed{\nabla}_3 \left(r^5 \underline{P} \right)\right) = -2r^4 \slashed{\mathcal{D}}_2^\star \slashed{div}\slashed{\mathcal{D}}_2^\star \slashed{div}\left(\frac{r \underline{\alpha}}{\Omega^2}\right) + \frac{4}{r} r^2 \slashed{\mathcal{D}}_2^\star \slashed{div}\left(\frac{\underline{\psi}r^3}{\Omega}\right)- 6M r^2 \slashed{\mathcal{D}}_2^\star \slashed{div}\left(\frac{\underline{\alpha}r}{\Omega}\right)  \nonumber \\
 \left(-\frac{6M}{r} + 2\right)\Omega^{-1} \slashed{\nabla}_3 \left(\frac{\underline{\psi}r^3}{\Omega}\right) - \frac{6M}{r^2}\left(\frac{\underline{\psi}r^3}{\Omega}\right) + 3M  \Omega^{-1} \slashed{\nabla}_3 \left(r \underline{\alpha}\right) \, ,
\end{align}
where the last identity could be simplified further by reinserting (\ref{qoy2}).
\end{lemma}

\begin{proof}
Note that  (\ref{poy2}) and (\ref{poy1}) are just rewritings of the Teukolsky equation of spin $+2$ and spin $-2$ respectively. The identities (\ref{qoy1}) and (\ref{qoy2}) follow from inserting the definitions of $P$ and $\underline{P}$, commuting derivatives and inserting the relevant Teukolsky equation. The identities (\ref{roy2}) and (\ref{roy1}) follow by combining the estimates already obtained.
\end{proof}

\subsubsection{Angular derivatives of $\psi$ and \underline{$\psi$}}

From (\ref{qoy1}) and (\ref{qoy2}) we directly conclude using the bounds of Proposition \ref{prop:psie} and Theorem \ref{prop:summarypsi}
\begin{proposition} \label{prop:nif}
The quantity ${\psi}$ associated with the solution ${\alpha}$ of Theorem \ref{theo:mtheogi} through (\ref{psidef}) satisfies the estimate
\begin{align}
\sup_{u\geq u_0} \int_{v_0}^v d\bar{v} r^{8-\epsilon} \Big\|r^{-1}\cdot r^2   \slashed{\mathcal{D}}_2^\star \slashed{div} ({\psi} \Omega)  \Big\|_{S^2_{\bar{u},v}}^2  \lesssim \mathbb{F}_0\left[\Psi,{\psi},{\alpha}\right] \, ,
\end{align}
and the degenerate integrated decay estimate
\begin{align}
\int_{u_0}^\infty du \int_{v_0}^\infty dv \frac{\Omega^2}{r^3}\left(1-\frac{3M}{r}\right)^2 \Big\|r^{-1}\cdot r^2   \slashed{\mathcal{D}}_2^\star \slashed{div} (r^3 {\psi} \Omega)  \Big\|_{S^2_{u,v}}^2 \lesssim \mathbb{F}_0\left[{\Psi},{\psi},{\alpha}\right] \nonumber \, .
\end{align}

The quantity $\underline{\psi}$ associated with the solution $\underline{\alpha}$ of Theorem \ref{theo:mtheogi} through  (\ref{psidef2}) satisfies the estimate
\begin{align}
\sup_{v\geq v_0} \int_{u_0}^u d\bar{u}  \Big\|r^{-1}\cdot r^2   \slashed{\mathcal{D}}_2^\star \slashed{div} (r^3 \underline{\psi} \Omega^{-1})  \Big\|_{S^2_{\bar{u},v}}^2  \Omega^2 \left(\bar{u},v\right) \lesssim \mathbb{F}_0\left[\underline{\Psi},\underline{\psi},\underline{\alpha}\right] \, ,
\end{align}
and the degenerate integrated decay estimate
\begin{align}
\int_{u_0}^\infty du \int_{v_0}^\infty dv \frac{\Omega^2}{r^3}\left(1-\frac{3M}{r}\right)^2 \Big\|r^{-1}\cdot r^2   \slashed{\mathcal{D}}_2^\star \slashed{div} (r^3 \underline{\psi} \Omega^{-1})  \Big\|_{S^2_{u,v}}^2 \lesssim \mathbb{F}_0\left[\underline{\Psi},\underline{\psi},\underline{\alpha}\right] \nonumber \, .
\end{align}
\end{proposition}

\begin{remark}
Note that in view of (\ref{uid2}) we control in particular the flux of first angular derivatives of $\underline{\psi}$ on null infinity $\mathcal{I}^+$, a fact that will be exploited in the next proposition.
\end{remark}

We now look at the commuted equation
\begin{align}
\slashed{\nabla}_4 \left(r \slashed{div} \underline{\psi} r^3 \Omega\right) = - r^3 \Omega \cdot r \slashed{div} \underline{P} \, .
\end{align}
From this we derive
\begin{align}
& \Omega \slashed{\nabla}_4 \left( \left(\frac{1}{\Omega^2} - \frac{1}{\Omega^2}\Big|_{r=3M} \right)\| r \slashed{div}  \underline{\psi} r^3 \Omega \|^2 \right) +\frac{2M}{r^2}\| r \slashed{div}  \underline{\psi} r^3 \|^2 \nonumber \\
&= - 2r^3 \Omega^2 r \slashed{div}  \underline{P} \left(\frac{1}{\Omega^2} - \frac{1}{\Omega^2}\Big|_{r=3M} \right) r \slashed{div}  \underline{\psi} r^3 \Omega =  4r^3  \left(1-\frac{3M}{r}\right)r \slashed{div}  \underline{P} \cdot r \slashed{div}  \underline{\psi} r^3 \Omega  \nonumber \\
&\leq \frac{M}{r^2}\| r \slashed{div}  \underline{\psi} r^3 \|^2 + \frac{4}{M} \left(1-\frac{3M}{r}\right)^2 r^8 \| r \slashed{div}  \underline{P} \|^2 \Omega^2 \nonumber \, ,
\end{align}
which yields
\begin{align}
\Omega \slashed{\nabla}_4 \left( \left(\frac{1}{\Omega^2} - \frac{1}{\Omega^2}\Big|_{r=3M} \right)\| r \slashed{div}  \underline{\psi} r^3 \Omega \|^2 \right) +\frac{M}{r^2}\| r \slashed{div}  \underline{\psi} r^3 \|^2  
\leq \frac{4}{M} \left(1-\frac{3M}{r}\right)^2 r^8 \| r \slashed{div}  \underline{P} \|^2 \Omega^2\nonumber \, .
\end{align}
Integrating this over the spacetime region $\left[u_0, u\right] \times \left[v_0, v\right]$ for 
arbitrary $u \geq u_0$ and $v\geq v_0$, we observe that
\begin{itemize}
\item The right hand side is controlled by initial data from (\ref{moraf}) and (\ref{bestintdec}).
\item The future boundary term (flux) on the constant-$v$ hypersurface
is \emph{negative}. However, we already control it from initial data through Proposition \ref{prop:nif}.
\end{itemize}
We conclude
\begin{proposition} \label{prop:firstangularpsi}
The quantity $\psi$ associated with the solution $\alpha$ of Theorem \ref{theo:mtheogi} through (\ref{psidef}) satisfies the non-degenerate integrated decay estimate
\begin{align}
\int_{u_0}^\infty \int_{v_0}^\infty \int_{S^2_{\bar{u},\bar{v}}} d\bar{u} d\bar{v} \sin \theta d\theta d\phi \ r^{7-\epsilon} \Omega^4  \| r \slashed{div}  {\psi}  \|^2 \lesssim \mathbb{F}_0\left[\Psi, \mathfrak{D} \psi, \alpha \right] \, . \nonumber
\end{align}
The quantity $\underline{\psi}$ associated with the solution $\underline{\alpha}$ of Theorem \ref{theo:mtheogi} through (\ref{psidef2}) satisfies the non-degenerate integrated decay estimate
\begin{align}
\int_{u_0}^\infty \int_{v_0}^\infty \int_{S^2_{\bar{u},\bar{v}}} d\bar{u} d\bar{v} \sin \theta d\theta d\phi \ r^{5-\epsilon} \| r \slashed{div}  \underline{\psi} \|^2 \lesssim \mathbb{F}_0\left[\underline{\Psi}, \mathfrak{D} \underline{\psi}, \underline{\alpha} \right] \, .
\end{align}
\end{proposition}
\begin{proof}
The second bound follows directly from the computation above. (It follows with weight $4$ instead of $5-\epsilon$ but it is easy to see how to improve the weight a posteriori.) For the first, one repeats this argument for the commuted equation $\slashed{\nabla}_3 \left(r \slashed{div}  {\psi} r^3 \Omega\right) =  r^3 \Omega \cdot r \slashed{div}  {P}$, now multiplying with $\left[r^{2-\epsilon} -\left(3M\right)^{2-\epsilon}\right]\cdot r \slashed{div}  {\psi} r^3 \Omega$ and using that the flux arising on the horizon is a priori controlled by Proposition \ref{prop:nif}. 
\end{proof}

Similarly, we can directly integrate the equations
\[
\Omega \slashed{\nabla}_4 \Big| r \slashed{div}  \frac{\underline{\psi}r^3}{\Omega} \Big|^2 + \frac{2M}{r^2} \Big| r \slashed{div}  \frac{\underline{\psi}r^3}{\Omega} \Big|^2 = -2 r^3 r \slashed{div}  \underline{P} \cdot r \slashed{div}  \frac{\underline{\psi}r^3}{\Omega} 
\]
and (for any $1\geq \delta \geq 0$)
\[
\Omega \slashed{\nabla}_3 \left[ r^{3-\delta} | r\slashed{div}  (r^3 \Omega \psi) |^2 \right] + \frac{3-\delta}{r} \Omega^2  \left[ r^3 | r\slashed{div}  (r^3 \Omega \psi) |^2 \right] = r^{6-\delta} \Omega^2 r \slashed{div}  P \cdot r \slashed{div}  (\psi r^3 \Omega)
\]
from initial data (integrating also over the angular variables) to obtain, after applying 
Cauchy--Schwarz to the right hand side and using the $P$ and $\underline{P}$-fluxes

\begin{proposition} \label{prop:firstangularpsi2}
In addition to the estimates of Proposition \ref{prop:firstangularpsi} we have
\begin{align}
\sup_{u,v} \Big\|r^{-1} \cdot r \slashed{div}  \frac{\underline{\psi}r^3}{\Omega} \Big\|^2_{S^2_{u,v}} + \sup_u \int_{v_0}^\infty \frac{1}{r^2} \Big\|r^{-1} \cdot r \slashed{div}  \frac{\underline{\psi}r^3}{\Omega} \Big\|^2_{S^2_{u,v}} dv \lesssim \sup_{u} \Big\|r^{-1} \cdot r\slashed{div}  \frac{\underline{\psi}r^3}{\Omega} \Big\|^2_{S^2_{u,v_0}} + \mathbb{F}_0 \left[ \underline{\Psi}\right] \nonumber
\end{align}
and for any $1 \geq \delta \geq 0$
\begin{align}
 \sup_{u,v} \ r^{3-\delta} \Big\|r^{-1} \cdot r \slashed{div}  (\psi \Omega r^3) \Big\|^2_{S^2_{u,v}} + \sup_v \int_{u_0}^\infty \Omega^2 r^{2-\delta} \Big\|r^{-1} \cdot r\slashed{div}  (\psi \Omega r^3) \Big\|^2_{S^2_{u,v}}  du \nonumber \\
 \lesssim \sup_{v} \ r^{3-\delta} \Big\|r^{-1} \cdot r \slashed{div}  (\psi \Omega r^3) \Big\|^2_{S^2_{u_0,v}}+ \mathbb{F}_0 \left[\Psi\right] \nonumber \, .
\end{align}
\end{proposition}
\begin{corollary} \label{cor:1da}
In addition to the estimates of Proposition \ref{prop:firstangularpsi} we also have the estimates
\begin{align}
\sup_{u,v}  \| r^{-1}r\slashed{div} \underline{\alpha} \Omega^{-2} \|_{S^2_{u,v}}^2 \lesssim \sup_{u}  \| r^{-1}r\slashed{div} \underline{\alpha} \Omega^{-2} \|_{S^2_{u,v_0}}^2 +\sup_{u} \Big\|r^{-1} \cdot r\slashed{div}  \frac{\underline{\psi}r^3}{\Omega} \Big\|^2_{S^2_{u,v_0}} + \mathbb{F}_0 \left[ \underline{\Psi}\right] \nonumber \, ,
\end{align}
\begin{align}
\sup_{u,v} \ r^{7-\delta} \| r^{-1}r\slashed{div} \alpha \Omega^2 \|_{S^2_{u,v}}^2 \lesssim \sup_v \ r^{7-\delta} \| r^{-1}  r\slashed{div} \alpha \Omega^2 \|_{S^2_{u_0,v}}^2 + \sup_{v} \ r^{3-\delta} \Big\|r^{-1} r \slashed{div}  (\psi \Omega r^3) \Big\|^2_{S^2_{u_0,v}}+ \mathbb{F}_0 \left[\Psi\right] \nonumber \, .
\end{align}
\end{corollary}
\begin{proof}
For the second estimate, apply the once angular commuted (\ref{koi}) with $n=5-\delta$ and use the flux in the Proposition. For the first, use (\ref{auit}) (with an additional ($\Omega^{-2}$ weight)) and the flux of the Proposition.
\end{proof}

\subsubsection{Estimating all first derivatives of $\psi$, \underline{$\psi$}}
Note that we already control the $\slashed{\nabla}_3 \psi$ and $\slashed{\nabla}_4 \underline{\psi}$ derivatives directly from the transport equation they satisfy, (\ref{evolpp}). To estimate the remaining first derivative we commute these equations by $\slashed{\nabla}_{R^\star}:=\Omega \slashed{\nabla}_3 - \Omega \slashed{\nabla}_4$ and recall that $\slashed{\nabla}_{R^\star} \Psi$ satisfies a \emph{non-degenerate} (near $r=3M$) integrated decay estimate, cf.~(\ref{moraf}). We compute
\begin{align}
\Omega \slashed{\nabla}_4 \left(\slashed{\nabla}_{R^\star} \underline{\psi} r^3 \Omega\right) = -\slashed{\nabla}_{R^\star} \left(r^3 \Omega^2 \underline{P}\right)
\end{align}
since $\Omega \slashed{\nabla}_3$ and $\Omega \slashed{\nabla}_4$ commute. Because the right hand side again satisfies a \emph{non-degenerate} integrated decay estimate (\ref{moraf}) and (\ref{bestintdec}), we can actually repeat the estimate for the uncommuted equation (leading from (\ref{psi}) to (\ref{psibar})) to immediately obtain the analogue of Proposition \ref{prop:psie},\footnote{Note that this would not be possible if the right hand side satisfied an estimate which degenerated near $r=3M$. The reason is that such a right hand side would (just as for the angular derivatives in the previous section) force us to multiply with a weight that changes sign near $3M$ which gives the future boundary term the \emph{wrong} sign. In the angular case this flux of the wrong sign was controlled a priori. Here this flux is not available yet.} as the same considerations hold for the commuted $\slashed{\nabla}_3 \left(r^3 \Omega \psi\right)$-equation.

\begin{proposition} \label{prop:psie2} 
Consider the quantity $\psi$ associated with the solution $\alpha$ of Theorem \ref{theo:mtheogi} through (\ref{psidef}). Then along any null-hypersurface of constant $u$ in $\mathcal{D}$ (including the event horizon) we have
\begin{align} \label{heui2} 
\int_{v_1}^{v_2} dv \|r^{-1} \slashed{\nabla}_{R^\star} (\Omega \psi)\|_{S^2_{u,v}}^2 r^{8-\epsilon}  \left(u,v\right) \lesssim \mathbb{F}_0\left[\Psi,  \mathfrak{D}\psi \right]  \, .
\end{align}
In addition, we have the integrated decay estimate
\begin{align}
\int_{\mathcal{D}} du dv \|r^{-1} \slashed{\nabla}_{R^\star}(\Omega \psi)\|_{S^2_{u,v}}^2 \Omega^2  r^{7-\epsilon} 
\lesssim \mathbb{F}_0\left[\Psi, \mathfrak{D}\psi \right] \, .
\end{align}
Consider the quantity $\underline{\psi}$ associated with the solution $\underline{\alpha}$ of Theorem \ref{theo:mtheogi} through (\ref{psidef2}). Then along any null hypersurface of constant $v$ in $\mathcal{D}$ we have
\begin{align}  \label{heui2b}
\int_{u_1}^{u_2} du \|r^{-1} \slashed{\nabla}_{R^\star} (\Omega^{-1} \underline{\psi})\|_{S^2_{u,v}}^2 r^6 \Omega^2 \left(u,v\right) \lesssim \mathbb{F}_0\left[ \underline{\Psi}, \mathfrak{D}\underline{\psi} \right] \, .
\end{align}
In addition, we have the integrated decay estimate
\begin{align}
\int_{\mathcal{D}} du dv \|r^{-1} \slashed{\nabla}_{R^\star} (\Omega^{-1} \underline{\psi})\|_{S^2_{u,v}}^2 \Omega^2 r^{5-\epsilon} \lesssim \mathbb{F}_0\left[ \underline{\Psi}, \mathfrak{D}\underline{\psi} \right]  \, .
\end{align}
\end{proposition}
In view of 
\[
\Omega \slashed{\nabla}_4 (\Omega\psi) = \Omega \slashed{\nabla}_3 (\Omega \psi) - \slashed{\nabla}_{R^\star} (\Omega \psi) = \Omega^2 P + 3\frac{\Omega^2}{r} (\Omega \psi) - \slashed{\nabla}_{R^\star} (\Omega \psi) 
\]
(and similarly for $\underline{\psi}$) it is immediate that we can obtain a non-degenerate integrated decay estimate for 
all first derivatives of $\psi$ and $\underline{\psi}$. In particular, we can replace $\slashed{\nabla}_{R^\star}$ by both $\Omega \slashed{\nabla}_4$ or  $\Omega \slashed{\nabla}_3$ in the estimates of Proposition \ref{prop:psie2}. The estimate for $\Omega \slashed{\nabla}_4 (\Omega\psi)$ will then still be non-optimal in terms of $r$-weights at infinity and the estimate for $\Omega \slashed{\nabla}_3 \left(\Omega^{-1} \underline{\psi}\right) $ will not be optimal near the horizon. However, this is easily cured using $r$-weighted estimates and the redshift respectively. We indicate this for the $r$-weight before stating the final proposition. We have
\begin{align}
\Omega \slashed{\nabla}_3 \left[  \xi\left(r\right) r^{4-\epsilon} | \Omega\slashed{\nabla}_4 (r^3 \Omega \psi) |^2 \right] + \frac{4-\epsilon}{r} \Omega^2 \xi\left(r\right) \left[ r^{4-\epsilon} | \Omega\slashed{\nabla}_4 (r^3 \Omega \psi) |^2 \right] \nonumber \\
- \xi_r \Omega^2 \ r^{4-\epsilon} | \Omega\slashed{\nabla}_4 (r^3 \Omega \psi) |^2 = \xi r^{7-\epsilon} \Omega^2 \Omega\slashed{\nabla}_4 P \cdot  \Omega\slashed{\nabla}_4(\psi r^3 \Omega)
\end{align}
for a radial cut-off function $\xi$ which we choose to be $1$ for $r\geq 8M$ and $0$ for $r\leq 6M$. Using that one already has a non-optimal spacetime estimate for $r \leq 8M$, integrating the above over a spacetime region improves the weights.\footnote{Note again that this argument does not apply globally because $\slashed{\nabla}_4 P$ is only controlled in a degenerate norm near $r=3M$.}
We summarise

\begin{proposition} \label{prop:firstderpsi}
Consider the quantity $\psi$ associated with the solution $\alpha$ of Theorem \ref{theo:mtheogi} through (\ref{psidef}). We have 
the non-degenerate integrated decay estimate
\begin{align}
\int_{u_0}^\infty \int_{v_0}^\infty \int_{S^2_{\bar{u},\bar{v}}} d\bar{u} d\bar{v} \sin \theta d\theta d\phi \ r^{7-\epsilon}  \Big[ | \slashed{\nabla}_3 \left( \Omega \psi \right) |^2 + \Omega^2  |  r\cdot \Omega\slashed{\nabla}_4 \left(\Omega {\psi}\right) |^2\Big]  \lesssim \mathbb{F}_0\left[\Psi, \mathfrak{D} \psi\right] \, ,\nonumber
\end{align}
as well as the flux estimate
\begin{align}
\int_{v_0}^\infty dv \sin \theta d\theta d\phi  \ r^{4-\epsilon} | \Omega \slashed{\nabla}_4 \left(\Omega \psi r^3\right) |^2 
\lesssim  \mathbb{F}_0\left[\Psi, \mathfrak{D} \psi\right] \, .
\end{align}
Consider the quantity $\underline{\psi}$ associated with the solution $\underline{\alpha}$ of Theorem \ref{theo:mtheogi} through (\ref{psidef2}). We have the non-degenerate integrated decay estimate
\begin{align}
\int_{u_0}^\infty \int_{v_0}^\infty \int_{S^2_{\bar{u},\bar{v}}} d\bar{u} d\bar{v} \sin \theta d\theta d\phi \Big[ & r^4 \Omega^2 \left ( |  \Omega^{-1} \slashed{\nabla}_3 \left(\Omega^{-1} \underline{\psi}\right) |^2 +  |  r \Omega \slashed{\nabla}_4 \left(\underline{\psi}\Omega^{-1}\right) |^2\right)  \Big]
\lesssim \mathbb{F}_0\left[ \underline{\Psi}, \mathfrak{D} \underline{\psi} \right]\nonumber
\end{align}
as well as the flux estimate
\begin{align}
\int_{u_0}^\infty du \sin \theta d\theta d\phi \ \Omega^2  r^6 | \Omega^{-1} \slashed{\nabla}_3 (\Omega^{-1}\underline{\psi})|^2 
\lesssim  \mathbb{F}_0\left[\underline{\Psi}, \mathfrak{D} \underline{\psi}\right] .
\end{align}
\end{proposition}

\subsubsection{Estimating second angular derivatives of $\alpha$, \underline{$\alpha$}}
Proposition \ref{prop:firstderpsi} will directly imply control over two angular derivatives of $\alpha$ and $\underline{\alpha}$. This follows from the identities (\ref{poy1}) and (\ref{poy2}). 
We observe that by Propositions \ref{prop:firstderpsi} and \ref{prop:psie} we already control the flux (on constant $v$) of the left hand side of (\ref{poy1})  and the flux (on constant $u$) of the left hand side of (\ref{poy2}). Moreover, by the same Propositions we have integrated decay for the left hand sides of both (\ref{poy1}) and (\ref{poy2}). As $\alpha$ and $\underline{\alpha}$ itself are already controlled from Proposition \ref{prop:alphae} we immediately conclude:
\begin{proposition} \label{prop:twoangular}
Consider the solution $\alpha$ of Theorem \ref{theo:mtheogi}. We have the flux estimate
\begin{align} \label{fluxa} 
\sum_{i=0}^2 \int_{v_0}^\infty dv \sin \theta d\theta d\phi  \  r^{6+2i-\epsilon} |  \slashed{\nabla}^i (\alpha \Omega^2) |^2 
\lesssim \mathbb{F}_0\left[\Psi, \mathfrak{D} \psi, \alpha \right] \, 
\end{align}
for any hypersurface of constant $u$ and the integrated decay estimate
\begin{align}
\sum_{i=0}^2 \int_{v_0}^\infty \int_{u_0}^\infty du dv \int_{S^2_{u,v}} \sin \theta d\theta d\phi \Omega^2 \left[  \Omega^4 r^{5-\epsilon+2i} | \slashed{\nabla}^i \alpha |^2 \right]  
\lesssim \mathbb{F}_0\left[\Psi, \mathfrak{D} \psi, \alpha \right] \, . \nonumber
\end{align}
Consider the solution $\underline{\alpha}$ of Theorem \ref{theo:mtheogi}. We have the flux estimate
\begin{align}
\sum_{i=0}^2 \int_{u_0}^\infty du \sin \theta d\theta d\phi \ \Omega^2 r^{2+2i} |  \slashed{\nabla}^i (\underline{\alpha}\Omega^{-2}) |^2 
\lesssim  \mathbb{F}_0\left[\underline{\Psi}, \mathfrak{D} \underline{\psi}, \underline{\alpha} \right] \, 
\end{align}
on any constant $v$ hypersurface and the integrated decay estimate
\begin{align}
\sum_{i=0}^2 \int_{v_0}^\infty \int_{u_0}^\infty du dv \int_{S^2_{u,v}} \sin \theta d\theta d\phi \Omega^2 \left[ \Omega^{-4} r^{2i} | \slashed{\nabla}^i \underline{\alpha} |^2  \right]  
\lesssim \mathbb{F}_0\left[\underline{\Psi}, \mathfrak{D} \underline{\psi}, \underline{\alpha} \right] \, . \nonumber
\end{align}
\end{proposition}

\begin{remark} \label{rem:teuonly}
The estimates of Proposition \ref{prop:twoangular} still ``loses derivatives", in that the norm on the right hand side involves three derivatives of $\alpha$ (one derivative of $\Psi$ which itself is two derivatives of $\alpha$). An improved estimate which does not lose in this sense will be provided in Proposition \ref{prop:5dera} below.
\end{remark}

\subsubsection{Estimating all first derivatives of $\alpha$, \underline{$\alpha$}}
We now establish estimates for (all) \emph{first} derivatives of $\alpha$ and $\underline{\alpha}$. This is very straightforward as we can commute $\slashed{\nabla}_3 \left(r\Omega^2 \alpha\right) = -2\psi r \Omega^2$ and the corresponding $\slashed{\nabla}_4 \left(r\Omega^2 \underline{\alpha}\right) = 2\underline{\psi} r \Omega^2$ equation by any derivative from $r\slashed{\nabla}_4$, $\slashed{\nabla}_3$ and $r\slashed{\nabla}$ and observe that the right hand side satisfies a non-degenerate integrated decay estimate just as the uncommuted equation did (cf.~Propositions \ref{prop:firstangularpsi} and \ref{prop:firstderpsi}). Repeating the proof of Proposition \ref{prop:alphae} therefore immediately provides the analogue of that Proposition:

\begin{proposition} \label{prop:firstderalpha}
Consider the solution $\alpha$ of Theorem \ref{theo:mtheogi}. Along any null hypersurface of constant $u\ge u_0$, including the event horizon, we have for $v_2\ge v_1\ge v_0$
\begin{align}
\int_{v_1}^{v_2} dv \|r^{-1}\mathfrak{D} \alpha\|_{S^2_{u,v}}^2 r^{6-\epsilon} \Omega^4 \left(u,v\right) \lesssim \mathbb{F}_0\left[\Psi, \mathfrak{D}\psi,  \mathfrak{D}\alpha \right] \, .
\end{align}
In addition we have the integrated decay estimate
\begin{align}
 \int_{v_0}^\infty\int_{u_0}^\infty du dv r^{5-\epsilon} \Omega^6 \|r^{-1}\mathfrak{D} \alpha\|_{S^2_{u,v}}^2  
\lesssim \mathbb{F}_0 \left[\Psi, \mathfrak{D}\psi, \mathfrak{D} \alpha\right] . \nonumber
\end{align}
Consider the solution $\underline{\alpha}$ of Theorem \ref{theo:mtheogi}. Along any null hypersurface of constant $v\ge v_0$, including in the limit on null infinity, we have
\begin{align}
\int_{u_1}^{u_2} du   \|r^{-1} \mathfrak{D} \underline{\alpha}\|_{S^2_{u,v}}^2 r^2 \Omega^{-2} \left(u,v\right) \lesssim \mathbb{F}_0\left[\underline{\Psi}, \mathfrak{D}  \underline{\psi}, \mathfrak{D}\underline{\alpha} \right] \, .
\end{align}
In addition, we have the integrated decay estimate
\begin{align}
 \int_{v_0}^\infty\int_{u_0}^\infty du dv du dv  r^{1-\epsilon} \Omega^{-2} \|r^{-1} \mathfrak{D} \underline{\alpha}\|_{S^2_{u,v}}^2
\lesssim \mathbb{F}_0 \left[ \underline{\Psi}, \mathfrak{D} \underline{\psi}, \mathfrak{D} \underline{\alpha} \right]  . \nonumber
\end{align}
\end{proposition}
\begin{remark}
For $\alpha$ (and similarly for $\underline{\alpha}$) it should be possible to use the Teukolsky equation (\ref{teukolsky}) itself to control all second derivatives using the fact that two angular derivatives and all first derivatives are already under control by Propositions \ref{prop:firstderalpha} and \ref{prop:twoangular}. Such estimates can alternatively be derived directly from the Bianchi equations.
\end{remark}

\subsubsection{Completing the proof of  (\ref{teubo3}) and (\ref{masterid})} \label{sec:prot1}
We complete the proof of 
the statements 1.--3.~of
Theorem \ref{theo:mtheogi} by proving the estimates (\ref{teubo3}) and (\ref{masterid}),  recalling that (\ref{teubo1}) was proven already in Section \ref{sec:alpha}.

Combining the flux and integrated decay estimates of Proposition \ref{prop:firstderalpha}, Proposition \ref{prop:firstderpsi} and the estimates of Propositions \ref{prop:firstangularpsi2} and \ref{prop:nif}, we have established (\ref{teubo3}) and (\ref{masterid}) for $n=0$ (in fact also integrated decay and flux estimates on second angular derivatives of $\alpha, \underline{\alpha}$ in Proposition \ref{prop:twoangular} and flux estimates for \emph{second} angular derivatives of $\psi, \underline{\psi}$ in Proposition \ref{prop:nif}). Since the Regge--Wheeler equation commutes trivially with $T$ and angular momentum operators $\mathnormal{\Omega}_i$ (cf.~Section \ref{sec:Killing}), the higher order statements are immediate. 
\subsection{Refinements and polynomial decay estimates} \label{sec:refinements}
We now obtain the necessary refinements to obtain the final statement~4.~of
Theorem~\ref{theo:mtheogi}, namely Propositions~\ref{prop:refd}, \ref{prop:psibs2d2}, \ref{prop:psibs2d}, \ref{prop:as2d2} and the decay estimate of Proposition \ref{prop:l1est} below.

The statement of the first four Propositions is contained in 
Section~\ref{sec:decay} below. 
To prove them,
we shall first show certain higher order statements in Section~\ref{hosheredev3}. 
Finally,  in Section~\ref{auxdecestundpsi}, we shall prove Proposition \ref{prop:l1est}, which provides an $L^1$-estimate that we will be useful in the proof of Theorem \ref{theo:mtheo}.

\subsubsection{Top order statements}
\label{hosheredev3}
Using the previous bounds in conjunction with the identities (\ref{roy1}) and (\ref{roy2}) we can conclude a boundedness (of energy) statement for $\alpha$ and $\underline{\alpha}$ which does not lose derivatives in the sense that two derivatives of $\Psi = r^5 P$, $\underline{\Psi} = r^5 \underline{\Psi}$ control four derivatives of $\alpha, \underline{\alpha}$. This statement does however require the higher order estimates of Corollary \ref{cor:higherorder}, which we recall include the commuted redshift near the horizon and weighted commutation near null infinity. 
Recall the notation (\ref{mathanot}) and the ellipticity of these operators, cf.~(\ref{uid2}).

\begin{proposition} \label{prop:5dera}
Consider the solution $\alpha$ of Theorem \ref{theo:mtheogi}. Then we have the flux bound
\begin{align}
\sup_{v\geq v_0} \int_{u_0}^\infty du \sin \theta \theta d\phi \ \Omega^2 \Big| \mathcal{A}^{[4]} \left(\frac{r \underline{\alpha}}{\Omega^2}\right) \Big|^2 \lesssim \mathbb{F}_0^{1} \left[\underline{\Psi}\right] + \mathbb{F}_0 \left[\underline{\Psi}, \mathfrak{D}\underline{\psi}, \mathfrak{D} \underline{\alpha}\right] \, .
\end{align}
Consider the solution $\underline{\alpha}$ of Theorem \ref{theo:mtheogi}. Then we have the flux bound
\begin{align}
\sup_{u \geq u_0} \int_{v_0}^\infty dv \sin \theta \theta d\phi \ r^{6-\epsilon} \Big| \mathcal{A}^{[4]} \left(\alpha \Omega^2\right) \Big|^2 \lesssim \mathbb{F}_0^{1} \left[{\Psi}\right] + \mathbb{F}_0 \left[{\Psi}, \mathfrak{D}{\psi}, \mathfrak{D} {\alpha}\right] \, .
\end{align}
\end{proposition}
\begin{proof}
Apply the flux bounds of Propositions \ref{prop:nif}, \ref{prop:firstderpsi}, \ref{prop:twoangular} and Corollary \ref{cor:higherorder} to the identities (\ref{roy1}) and (\ref{roy2}).
\end{proof}

\begin{proposition} \label{prop:a5id}
Consider the solution $\alpha$ of Theorem \ref{theo:mtheogi}. We have for $i\geq 4$ the integrated decay estimate
\begin{align}
\int_{u_0}^\infty du \int_{v_0}^\infty dv \frac{\Omega^2}{r^{1+\epsilon}} \left(1-\frac{3M}{r}\right)^{2} r^{-1} \Big\|r^{-1}\cdot  \mathcal{A}^{[i]} (r^3 \alpha \Omega^2)  \Big\|_{S^2_{u,v}}^2  \lesssim 
  \mathbb{F}_0^{i-3} \left[{\Psi}\right] + \mathbb{F}^{i-4, T, \slashed{\nabla}}_0 \left[{\Psi}, \mathfrak{D}{\psi}, \mathfrak{D} {\alpha}\right]
\end{align}
Consider the solution $\underline{\alpha}$ of Theorem \ref{theo:mtheogi}. 
\begin{align}
\int_{u_0}^\infty du \int_{v_0}^\infty dv \frac{\Omega^2}{r^{1+\epsilon}} \left(1-\frac{3M}{r}\right)^{2}  \Big\|r^{-1}\cdot  \mathcal{A}^{[i]} (r \underline{\alpha} \Omega^{-2}) \Big\|_{S^2_{u,v}}^2  
\lesssim \mathbb{F}_0^{i-3} \left[\underline{\Psi}\right]+  \mathbb{F}^{i-4,T,\slashed{\nabla}}_0 \left[\underline{\Psi}, \mathfrak{D}\underline{\psi}, \mathfrak{D} \underline{\alpha}\right] \, .
\end{align}
Moreover, for both estimates, if for any $i \geq 5$ we replace $\mathcal{A}^{[i]}$ by $\mathcal{A}^{[i-1]}$ on the left hand side (while keeping the right hand side fixed), the degeneration factor $(1-\frac{3M}{r})^2$ can be dropped.
\end{proposition}
\begin{proof}
Apply the integrated decay estimates of Propositions \ref{prop:nif}, \ref{prop:firstderpsi}, \ref{prop:twoangular}, \ref{prop:firstangularpsi} and Corollary \ref{cor:higherorder} to the identities (\ref{roy1}) and (\ref{roy2}).
\end{proof}

We also derive some boundedness estimates on the spheres $S^2_{u,v}$:
\begin{proposition} \label{prop:hioaa}
The solution $\alpha$ of Theorem \ref{theo:mtheogi} satisfies for any $u\geq u_0$ and $v\geq v_0$ the estimates
\begin{align}
\| r^{-1} \cdot \mathcal{A}^{[2]} \Psi \|^2_{S^2_{u,v}} \lesssim \mathbb{F}_0^{2} \left[{\Psi}\right]   \, ,
\end{align}
\begin{align}
r^3 \| r^{-1} \cdot \mathcal{A}^{[3]} r^3 \psi \Omega \|^2_{S^2_{u,v}}  +  r\| r^{-1} \cdot \mathcal{A}^{[4]} r^3 {\alpha} \Omega^{2} \|^2_{S^2_{u,v}} &\lesssim \mathbb{F}_0^{2} \left[{\Psi}\right] +  \mathbb{F}^{2,T,\slashed{\nabla}}_0 \left[{\Psi}, \mathfrak{D}{\psi}, \mathfrak{D} {\alpha}\right] \, .
\end{align}
The solution $\underline{\alpha}$ of Theorem \ref{theo:mtheogi} satisfies for any $u\geq u_0$ and $v\geq v_0$ the estimates
\begin{align}
 \| r^{-1} \cdot \mathcal{A}^{[2]} \underline{\Psi} \|^2_{S^2_{u,v}} \lesssim \mathbb{F}_0^{2} \left[\underline{\Psi}\right] \, ,
\end{align}
\begin{align}
  \| r^{-1} \cdot \mathcal{A}^{[3]} r^3 \underline{\psi} \Omega^{-1} \|^2_{S^2_{u,v}} +  \| r^{-1} \cdot \mathcal{A}^{[4]} r \underline{\alpha} \Omega^{-2} \|^2_{S^2_{u,v}}&\lesssim \mathbb{F}_0^{2} \left[\underline{\Psi}\right] +  \mathbb{F}^{2,T,\slashed{\nabla}}_0 \left[\underline{\Psi}, \mathfrak{D}\underline{\psi}, \mathfrak{D} \underline{\alpha}\right] \, .
\end{align}
\end{proposition}
\begin{proof}
The first estimate follows from one-dimensional Sobolev embedding and the boundedness statement of Corollary \ref{cor:higherorder}. The statements for $\psi$ and $\underline{\psi}$ follow from the once-angular commuted (\ref{qoy1}) and (\ref{qoy2}) respectively, after using again one-dimensional Sobolev embedding and the boundedness statement of Corollary  \ref{cor:higherorder}. Similarly, the bounds on $\alpha$ and $\underline{\alpha}$ follow from the identities (\ref{roy1}) and (\ref{roy2}) and the previous estimates (observe that the last term in (\ref{roy1}) only requires a bound on $S^2_{u,v}$ for $T\frac{r \underline{\alpha}}{\Omega^2}$ and $\frac{r \underline{\alpha}}{\Omega^2}$, which is again easily derived in terms
of the right hand side from the aforementioned transport equations.
\end{proof}

\subsubsection{Polynomial decay for $\psi$, $\alpha$ and \underline{$\psi$}, \underline{$\alpha$}} \label{sec:decay}

We finally record how the refined decay estimates for $\Psi$ of Section \ref{sec:decRW} are inherited by the derived quantities $\psi, \underline{\psi}, \alpha, \underline{\alpha}$. We remark that we are not aiming for the optimal or exhaustive statement here in terms of rates or regularity.

\begin{proposition} \label{prop:refd}
Fix $r_0$ as in Proposition \ref{prop:decRW}, let $v \geq v_0$ and recall the notation $u\left(v,r_0\right)$. Consider the solution $\alpha$ of Theorem \ref{theo:mtheogi} and the derived quantity $\psi$ defined via (\ref{psidef}). Then $\psi$ and $\alpha$ satisfy the following integrated decay estimates
\begin{align}
\int_v^\infty d\bar{v} \int^\infty_{u\left(v,r_0\right)} d\bar{u} \left( \|r^{-1} \cdot \psi \Omega\|_{S^2_{\bar{u},\bar{v}}}^2 \Omega^2 r^{5-\epsilon}  + \|r^{-1} \cdot \alpha \Omega^2\|_{S^2_{\bar{u},\bar{v}}}^2 \Omega^2 r^{3-\epsilon}  \right) 
 \lesssim \frac{1}{v^2} \left(\mathbb{F}_0^{2,T}\left[\Psi\right] + \mathbb{F}_0 \left[\Psi, \psi, \alpha\right] \right) \, . \nonumber
\end{align}
Consider the solution $\underline{\alpha}$ of Theorem \ref{theo:mtheogi} and the derived quantity $\underline{\psi}$ defined via (\ref{psidef2}). Then $\underline{\psi}$ and $\underline{\alpha}$ satisfy the following integrated decay estimates
\begin{align}
 \int_v^\infty d\bar{v} \int^\infty_{u\left(v,r_0\right)} d\bar{u} \left( \|r^{-1} \cdot \underline{\psi} \Omega^{-1}\|_{S^2_{\bar{u},\bar{v}}}^2 \Omega^2 r^{5-\epsilon} + \|r^{-1} \cdot \underline{\alpha} \Omega^{-2}\|_{S^2_{\bar{u},\bar{v}}}^2 \Omega^2 r^{-1-\epsilon} \right)   \lesssim \frac{1}{v^2} \left(\mathbb{F}_0^{2,T}\left[\underline{\Psi}\right] + \mathbb{F}_0 \left[\underline{\Psi},  \underline{\psi}, \underline{\alpha}\right] \right) . \nonumber
\end{align}
\end{proposition}

\begin{proof}
The proof is fairly simple but outlined in the following two subsections. 
\end{proof}

\subsubsection*{Polynomial decay for $\psi$ and $\alpha$}
We pick a dyadic $v$-sequence, $v_{i+1}=2v_i$, with associated $u$-sequence, $u_{i} = u\left(v_i,r_0\right)$. The spacetime integral of Proposition \ref{prop:psie} allows us to find in each dyadic interval $\left[u_i,u_{i+1}\right]$ a slice $\tilde{u}_i$ with
\[
\int_{v_0}^\infty d\bar{v} r^{7-\epsilon} \Omega^2  \|r^{-1} \cdot \psi \Omega \|_{S^2_{\tilde{u}_i,\bar{v}}}^2 \lesssim \frac{1}{\tilde{u}_i}\cdot \mathbb{F}_0 \left[\Psi,  \psi, \alpha\right]
\]
hence in particular
\[
\int_{\tilde{v}_i}^\infty d\bar{v} r^{7-\epsilon}  \|r^{-1} \cdot \psi \Omega \|_{S^2_{\tilde{u}_i,\bar{v}}}^2 \lesssim \frac{1}{\tilde{v}_i}\cdot \mathbb{F}_0 \left[\Psi,  \psi, \alpha\right] \, ,
\]
where
$\tilde{v}_i$ defined implicitly by $\tilde{u}_i=u(\tilde{v}_i,r_0)$.
Fix now an arbitrary $u \geq u_0$. Pick the $\tilde{u}_i$ such that $\tilde{u}_i < u \leq \tilde{u}_{i+1}$. Applying the inequality (\ref{psi}) with $n=1-\epsilon$ and integrated over the spacetime region $\left[\tilde{u}_i, u\right] \times \left[\tilde{v}_i,\infty\right] \times S^2$,
we export the decay to the slice $C_u$ and finally obtain
\[
\int_{v}^\infty d\bar{v} r^{7-\epsilon}  \|r^{-1} \cdot \psi \Omega \|_{S^2_{u,v}}^2 +  \int_v^\infty d\bar{v} \int^\infty_{u\left(v,r_0\right)} d\bar{u} \|r^{-1} \cdot \psi \Omega\|_{S^2_{\bar{u},\bar{v}}}^2 \Omega^2 r^{6-\epsilon} 
 \lesssim \frac{1}{v} \left(\mathbb{F}_0^{2,T}\left[\Psi\right] + \mathbb{F}_0 \left[\Psi,  \psi, \alpha\right] \right) \, 
\]
for any $v\geq v_0$ and $u \geq u\left(v,r_0\right)$. 
The method to establish decay for $\alpha$ given the decay for $\psi$ is entirely analogous to the one seen above establishing decay for $\psi$ from $P$. Hence
\begin{align} 
\int_{v}^\infty d\bar{v} r^{5-\epsilon}  \|r^{-1} \cdot \alpha \Omega^2 \|_{S^2_{u,v}}^2
+  \int_v^\infty d\bar{v} \int^\infty_{u\left(v,r_0\right)} d\bar{u} \|r^{-1} \cdot \alpha \Omega^2\|_{S^2_{\bar{u},\bar{v}}}^2 \Omega^2 r^{4-\epsilon} 
 \lesssim \frac{1}{v} \left(\mathbb{F}_0^{2,T}\left[\Psi\right] + \mathbb{F}_0 \left[\Psi,  \psi, \alpha\right] \right) \, \nonumber
\end{align}
for any $ v\geq v_0$ and $u \geq u\left(v,r_0\right)$. It is clear that repeating the iteration once more one obtains the first estimate claimed in the proposition. Note that the above proof also generates control of various fluxes, in particular
\begin{corollary} \label{cor:hozflu}
Consider the solution $\alpha$ of Theorem \ref{theo:mtheogi} and the derived quantity $\psi$ defined via (\ref{psidef}). Then on the event horizon $\mathcal{H}^+$ we have the flux bound
\begin{align}
\int_{v}^\infty d\bar{v} \left[ \| \alpha \Omega^2\|^2_{S^2_{\infty,\bar{v}}} + \|\psi \Omega \|^2_{S^2_{\infty,\bar{v}}} \right] \lesssim \frac{1}{v^2} \left(\mathbb{F}_0^{2,T}\left[\Psi\right] +\mathbb{F}_0 \left[\Psi,  \psi, \alpha\right]\right) \, .
\end{align}
\end{corollary}

\subsubsection*{Polynomial decay for \underline{$\psi$} and \underline{$\alpha$}}
For the underlined quantities one proceeds analogously, except that in this direction we do not have to lose powers of $r$. We again only carry out the argument for $\underline{\psi}$ and simply state the estimate for $\underline{\alpha}$ afterwards.

From the spacetime estimate of Proposition \ref{prop:psie} we find in each dyadic interval $\left[v_i,v_{i+1}\right]$ a $\tilde{v}_i$ such that
\[
\int_{u_0}^\infty d\bar{u} r^{5-\epsilon} \Omega^2  \|r^{-1} \cdot \underline{\psi} \Omega^{-1} \|_{S^2_{\bar{u},\tilde{v}_i}}^2 \lesssim \frac{1}{\tilde{v}_i}\cdot \mathbb{F}_0 \left[ \underline{\Psi}, \underline{\psi}, \underline{\alpha}\right]  \, .
\]
Fix now an arbitrary $v \geq v_0$. Pick the $\tilde{v}_i$ such that $\tilde{v}_i < v \leq \tilde{v}_{i+1}$. Integrating  (\ref{psibar}) over the region $\left[\tilde{u}_i, \infty\right] \times \left[v_i,v\right] \times S^2 \cap \{ r\leq 8M\}$ (where weights in $r$ are irrelevant) and using Corollary \ref{cor:dmdec} eventually yields
\begin{align}
&\int_{u\left(v,r_0\right)}^\infty d\bar{u}  \Omega^{2} \|r^{-1} \cdot \underline{\psi} \Omega^{-1} \|_{S^2_{u,V}}^2 \  \iota_{r\leq 8M} \nonumber \\
+  &\int_v^\infty d\bar{v} \int^\infty_{u\left(v,r_0\right)} d\bar{u} \ \Omega^2 \|r^{-1} \cdot \underline{\psi} \Omega^{-1}\|_{S^2_{\bar{u},\bar{v}}}^2 \iota_{r\leq 8M} 
 \lesssim \frac{1}{v}\left(\mathbb{F}_0^{2,T}\left[\underline{\Psi}\right] + \mathbb{F}_0 \left[ \underline{\Psi}, \underline{\psi}, \underline{\alpha}\right]\right) \,  \nonumber
\end{align}
for any $V \geq v\geq v_0$ and $u \geq u\left(v,r_0\right)$ with $\iota_{r\leq 8M}$ denoting the indicator function. Repeating the dyadic argument finally improves the power on the right hand side from $\frac{1}{v}$ to $\frac{1}{v^2}$. To extend the estimate to the region $r \geq 8M$ we use 
the estimate (\ref{psibar2}) in connection with the previous bound and again Corollary \ref{cor:dmdec}. This proves the estimate for $\underline{\psi}$ claimed in the Proposition

For $\underline{\alpha}$ one repeats the above argument now (instead of (\ref{psibar})) applied to the equation
\[
\Omega \slashed{\nabla}_4 \left(r \underline{\alpha} \Omega^{-2}\right) + \frac{4M}{r^2}  \left(r \underline{\alpha} \Omega^{-2}\right) = 2 \frac{r \underline{\psi}}{\Omega}
\]
and using that polynomial decay has already been established for the right hand side in the previous step. Note that the above proof also produces polynomial decay of various fluxes. In particular:
\begin{corollary} \label{cor:fluxbpsib}
Consider the solution $\underline{\alpha}$ of Theorem \ref{theo:mtheogi} and the derived quantity $\underline{\psi}$ defined via (\ref{psidef2}). Then for any $V \geq v \geq v_0$ we have the flux bound
\begin{align}
\int_{u\left(v,r_0\right)}^\infty d\bar{u}  \frac{\Omega^{2}}{r^\epsilon} \left( \|r^{-1} \cdot r^3 \underline{\psi} \Omega^{-1} \|_{S^2_{u,V}}^2 + \|r^{-1} \cdot \underline{\alpha} r \Omega^{-2} \|_{S^2_{u,V}}^2 \right) \  \lesssim \frac{1}{v^2} \left(\mathbb{F}_0^{2,T}\left[\Psi\right] + \mathbb{F}_0 \left[ \underline{\Psi}, \underline{\psi}, \underline{\alpha}\right] \right) \, .\nonumber
\end{align}
\end{corollary}
\begin{remark}
The $\epsilon$ could be removed with further work but we will not pursue this as the decay rate is also non-optimal.
\end{remark}

\subsubsection*{Polynomial decay for $\psi$ and \underline{$\psi$} on spheres $S^2_{u,v}$}
For later purposes we also note estimates for $\psi$ and $\underline{\psi}$ on spheres $S^2_{u,v}$.
\begin{proposition} \label{prop:psibs2d2}
Consider the solution $\alpha$ of Theorem \ref{theo:mtheogi} and the derived quantity $\psi$ defined via (\ref{psidef}). The following estimate holds for any $u\geq u_0$, $v\geq v_0$ including the sphere $S^2_{\infty,v_0}$:
\begin{align}
\|r^{-1} \cdot \Omega \psi r^3 \|^2_{S^2_{u,v}}  \lesssim \frac{1}{v^2} \left(\mathbb{F}_0^{2,T}\left[\Psi\right] + \sup_v \|r^{-1} \cdot \Omega \psi r^4 \|^2_{S^2_{u_0,v}}  \right) \, .
\end{align}
\end{proposition}

\begin{proof}
Fix $v \geq v_0\geq 2u_0$. From the defining equation (\ref{evolpp}) we derive the estimate
\[
\| r^{-1} \cdot \Omega r^3 \psi \|_{S^2_{u,v}} \lesssim \| r^{-1} \cdot \Omega r^3 \psi \|_{S^2_{u_0,v}} + \int_{u_0}^{u=\frac{1}{2}v} du \Omega^2 \| r^{-1} \cdot P r^3 \|_{S^2_{u,v}} +  \int_{u=\frac{1}{2}v}^{\max\left(v,\frac{1}{2}v\right)} du \Omega^2 \| r^{-1} \cdot P r^3 \|_{S^2_{u,v}}.
\]
Applying Cauchy--Schwarz on the integrals we obtain
\[
\| r^{-1} \cdot \Omega r^3 \psi \|_{S^2_{u,v}} \lesssim \frac{1}{v} \| r^{-1} \cdot \Omega r^4 \psi \|_{S^2_{u_0,v}} +  \sqrt{\int_{u_0}^{u=\frac{1}{2}v} du \Omega^2 \| r^{-1} \cdot \Psi \|^2_{S^2_{u,v}}} \frac{1}{v^\frac{3}{2}} + \sqrt{\int_{u=\frac{1}{2}v}^{\max\left(v,\frac{1}{2}v\right)} du \frac{\Omega^2}{r^2} \| r^{-1} \cdot \Psi \|^2_{S^2_{u,v}}}
\]
which yields the result after observing that the first integral is bounded and the second controlled by Proposition \ref{prop:decRW}.
\end{proof}

\begin{proposition} \label{prop:psibs2d}
Consider the solution $\underline{\alpha}$ of Theorem \ref{theo:mtheogi} and the derived quantity $\underline{\psi}$ defined via (\ref{psidef2}). Then, for $\left(u,v\right) \times S^2_{u,v}$ in $\mathcal{M} \cap \{u\geq u_0\} \cap \{v \geq v_0\} \cap \{r\leq r_0\}$ including the event horizon we have 
\begin{align}
 \|r^{-1} \cdot \Omega^{-1} r^3 \underline{\psi} \|^2_{S^2_{u,v}} \lesssim \frac{1}{v^2} \left( \mathbb{F}_0^{2,T}\left[\underline{\Psi}\right]  + \sup_{u} \|r^{-1} \cdot \Omega^{-1} r^3 \underline{\psi} \|^2_{S^2_{u,v_0}} \right) \, ,
\end{align}
while for $\left(u,v\right) \times S^2_{u,v}$ in $\mathcal{M} \cap \{u\geq u_0\} \cap \{v \geq v_0\} \cap \{r\geq r_0\}$ the above estimate holds replacing $\frac{1}{v^2}$ by $\frac{1}{u^2}$ on the right hand side.
\end{proposition}
\begin{proof}
For the region $r\geq r_0$ one integrates the estimate (\ref{psibar}) and uses a dyadic argument together with the fact that the flux arising on the right hand side of  (\ref{psibar}) satisfies a polynomial decay estimate from Proposition \ref{prop:decRW}. For the region $r \geq r_0$ one notes that weights of $\Omega$ can be ignored and that $u\sim v$ on $r=r_0$. One then 
integrates as 
above using again the polynomial decay estimate from Proposition \ref{prop:decRW}.
\end{proof}

\subsubsection*{Polynomial decay for \underline{$\alpha$} and $\alpha$ on spheres $S^2_{u,v}$}
Completely analogously to Propositions~\ref{prop:psibs2d2} and~\ref{prop:psibs2d}  one proves the following proposition, now starting from the defining equations for $\alpha$ and $\underline{\alpha}$, equations (\ref{psidef}) and (\ref{psidef2}):

\begin{proposition} \label{prop:as2d2}
Consider the solution $\alpha$ of Theorem \ref{theo:mtheogi} and the derived quantity $\psi$ defined via (\ref{psidef}). The following estimate holds for any $u\geq u_0$, $v\geq v_0$ including the sphere $S^2_{\infty,v_0}$:
\begin{align}
\|r^{-1} \cdot \Omega^2 \alpha r^2 \|^2_{S^2_{u,v}}  \lesssim \frac{1}{v^2} \left(\mathbb{F}_0^{2,T}\left[\Psi\right] + \sup_v \|r^{-1} \cdot \Omega^2 \alpha r^3 \|^2_{S^2_{u_0,v}}  \right) \, .
\end{align}
Consider now the solution $\underline{\alpha}$ of Theorem \ref{theo:mtheogi}. For $\left(u,v\right) \times S^2_{u,v}$ in $\mathcal{M} \cap \{u\geq u_0\} \cap \{v \geq v_0\} \cap \{r\leq r_0\}$ including the event horizon we have 
\begin{align} \label{ifa}
 \|r^{-1} \cdot \Omega^{-2} r \underline{\alpha} \|^2_{S^2_{u,v}} \lesssim \frac{1}{v^2} \left( \mathbb{F}_0^{2,T}\left[\underline{\Psi}\right]  + \sup_{u} \|r^{-1} \cdot \Omega^{-2} r \underline{\alpha} \|^2_{S^2_{u,v_0}} \right) \, ,
\end{align}
while for $\left(u,v\right) \times S^2_{u,v}$ in $\mathcal{M} \cap \{u\geq u_0\} \cap \{v \geq v_0\} \cap \{r\geq r_0\}$ (\ref{ifa}) holds replacing $\frac{1}{v^2}$ by $\frac{1}{u^2}$ on the right hand side.
\end{proposition}

Let us note that from the above estimates and additional commutations by 
$\mathnormal{\Omega}_i$, one obtains trivially also \emph{pointwise} 
decay estimates on $\alpha$,
with the corresponding commuted initial data norm on the right hand side of the estimate.

\subsubsection{Some auxiliary decay estimates}\label{auxdecestundpsi}
We collect here an auxiliary decay estimate for the derived quantity $\underline{\psi}$, which will be useful later when we consider the full system of gravitational perturbations.
\begin{proposition} \label{prop:l1est}
Consider the solution $\underline{\alpha}$ of Theorem \ref{theo:mtheogi} and the derived quantity $\underline{\psi}$ defined via (\ref{psidef2}). We define the $L^1$-fluxes
\[
A_1 \left(v\right) = \int_{u_0}^\infty du \Omega^2 \Big\|r^{-1} \cdot r \slashed{div} \frac{r^3\underline{\psi}}{\Omega}\Big\|_{S^2_{u,v}} \ \ \ \ \ \textrm{and} \ \ \ \ A_2 \left(v\right) =\int_{u_0}^\infty du \Omega^2 \Big\|r^{-1} \cdot \Omega^{-1} \slashed{\nabla}_3 \frac{r^3\underline{\psi}}{\Omega}\Big\|_{S^2_{u,v}} \, .
\]
We then have the estimates
\[
A_i\left(v\right) \lesssim A_i \left(v_0\right) + \sqrt{\mathbb{F}_0^{2,T}\left[\underline{\Psi}\right]} \lesssim \sqrt{\mathbb{F}_0\left[\underline{\Psi},\mathfrak{D}\underline{\psi}, \underline{\alpha}\right]} + \sqrt{\mathbb{F}_0^{2,T}\left[\underline{\Psi}\right]} \, ,
\]
with the second inequality being an immediate application of Cauchy--Schwarz.
\end{proposition}
\begin{proof}
From $\Omega \slashed{\nabla}_4 \left(\underline{\psi} r^3 \Omega\right) = -\underline{P}r^3 \Omega^2$ we derive
 \begin{align}
 \Omega \slashed{\nabla}_4 \left(r \slashed{div} \underline{\psi} r^3 \Omega\right) = -r \slashed{div} \underline{P}r^3 \Omega^2 \ \ \ \ \textrm{and} \ \ \ \
  \Omega \slashed{\nabla}_4 \left(\Omega\slashed{\nabla}_3 (\underline{\psi} r^3 \Omega)\right) = -\Omega \slashed{\nabla}_3 (\underline{P}r^3 \Omega^2)
 \end{align}
 From this it is manifest that the estimate for $A_1$ holds if we can uniformly bound the spacetime integral
\begin{align} \label{l1need}
\int_{v_0}^v d\bar{v} \int_{u_0}^\infty du \|r^{-1}\cdot r \slashed{div} r^5 \underline{P}\|_{S^2_{\bar{u},v}}  \frac{\Omega^2}{r^2}  ,
\end{align}
which we do by Corollary \ref{cor:PL1}. Applying the same argument to the other equation yields a bound for
the quantity $\int_{u_0}^\infty du \Omega^2 \|r^{-1} \cdot \Omega^{+1} \slashed{\nabla}_3 \frac{r^3\underline{\psi}}{\Omega}\|_{S^2_{u,v}}$ instead of $A_2\left(v\right)$, i.e.~the bound degenerates near the horizon. However, for $r\leq 8M$ we have the simple estimate
\begin{align}
\int_{u_0}^\infty du\, \iota_{r\geq 8M} \Omega^2 \Big\|r^{-1} \cdot \Omega^{-1} \slashed{\nabla}_3 \frac{r^3\underline{\psi}}{\Omega}\Big\|_{S^2_{u,v}} \lesssim \sqrt{\mathbb{F}_0\left[\underline{\Psi},\mathfrak{D}\underline{\psi}, \underline{\alpha}\right]} 
\end{align}
from the Cauchy--Schwarz inequality.
\end{proof}
\begin{corollary} \label{cor:l1est}
Under the assumptions of the previous proposition we also have the estimate
\begin{align}
\int_{u_0}^\infty du \Omega^2 \Big( \Big\|r^{-1} \cdot \frac{r^3\underline{\psi}}{\Omega}\Big\|_{S^2_{u,v}}
 + \Big\|r^{-1} \cdot r^2 \slashed{\mathcal{D}}_2^\star \slashed{div} \frac{r\underline{\alpha}}{\Omega^2}\Big\|_{S^2_{u,v}} \Big) \lesssim \sqrt{\mathbb{F}_0\left[\underline{\Psi},\mathfrak{D}\underline{\psi}, \underline{\alpha}\right]} + \sqrt{\mathbb{F}_0^{2,T}\left[\underline{\Psi}\right]} \, .
\end{align}
\end{corollary}
\begin{proof}
This follows from (\ref{uid1}) and (\ref{poy1}). Note that by (\ref{uid2}) this controls the $L^1$-flux of all angular derivatives of $\underline{\alpha}$ up to order two.
\end{proof}
\section{Proof of Theorem~\ref{theo:mtheo}} \label{sec:proofof3}
In this section we shall prove Theorem~\ref{theo:mtheo}. The reader can refer to the overview of Section \ref{overviewbtp}.

The first ingredient in the proof will be  the bounds on
gauge invariant quantities which follow from applying Theorem~\ref{theo:mtheogi}  
to 
the curvature components $\alin$ and $\ablin$, respectively, of our solution
to the system of linearised gravity.
We shall  collect these gauge invariant estimates in Section~\ref{collectthem}. 
We shall then obtain estimates for certain
fluxes on the horizon in Section~\ref{sec:hdgk}.
These, together with a red-shift commutation argument, 
will be used in Section~\ref{sec:chihatest} to obtain
decay estimates for the outgoing shear $\xlin$. We then obtain boundedness
estimates
for the ingoing shear $\xblin$ in Section~\ref{sec:bndchi}, 
and finally boundedness
for all remaining quantities in Section~\ref{sec:bconclude}.

\subsection{Gauge-invariant estimates from Theorem~\ref{theo:mtheogi}}
\label{collectthem}
Let 
\[
\Si=
\left(\, \glinh \, , \, \glinto \, , \, \Olino \, , \,  \bmlin\, , \,  \otx \, , \,  \otxb\, , \,  \xlin\, , \, \xblin\, , \,  \eblin \, , \,  \elin \, , \, \olin \, , \,  \olinb \, , \,  \alin \, , \,  \blin \, , \,  \rlin \, , \,  \slin \, , \,  \bblin \, , \,  \ablin \, , \, \Klin \right)
\]
be as in the statement of Theorem~\ref{theo:mtheo}.

The linearised curvature components $\alin$  and $\underline\alin$ 
satisfy the spin $\pm2$ Teukolsky equations, respectively,
by Proposition~\ref{prop:relfull}  of Section~\ref{sec:fullrel}. 
Thus, Theorem~\ref{theo:mtheogi} applies to yield Corollary \ref{cor:fully}. Note also that the assumption (\ref{ughbo}) of Theorem~\ref{theo:mtheo} implies that the initial norms of Theorem~\ref{theo:mtheogi},
applied to $\alin$  and $\underline\alin$, 
respectively, are indeed finite for $n=2$.
 
From Proposition~\ref{prop:relfull}, 
 we see also that the gauge-invariant estimates on $\plin$, $\pblin$,
 $\Plin$, and $\Pblin$ arising from Theorem~\ref{theo:mtheogi} 
 can immediately
 be re-interpreted as estimates for   the right hand sides of 
$(\ref{psideffull})$--$(\ref{Pdefund})$. These estimates will also be useful in what follows.

\subsection{Fluxes on the horizon $\mathcal{H}^+$} \label{sec:hdgk}
In this section we exploit the horizon gauge conditions (\ref{mts}) and (\ref{fchoi}) to control additional (not necessarily gauge invariant) fluxes \emph{in terms of the gauge invariant quantities $\Psilin$, $\plin$ and $\alin$}. 

In Section \ref{sec:hoz1} we obtain bounds for the linearised outgoing shear $\xlin$ 
itself, derived from the fact that $\plin$ is controlled on the horizon. In Section \ref{sec:gebo} we obtain similar bounds for the transversal derivative $\Omega^{-1} \slashed{\nabla}_3(\Omega \xlin)$, derived from the fact that $\Psilin$ is controlled on the horizon. Higher derivative fluxes are obtained in Section \ref{sec:hozf} from the fact that the energy of $\Psilin$ (cf.~Theorem \ref{prop:summarypsi}) controls also $\slashed{\nabla} \Psilin$ and $\slashed{\nabla}_4 \Psilin$ on the horizon. Finally, polynomial decay bounds for $\xlin$ and its transversal derivatives (which are inherited from polynomial decay bounds of $\plin$ and $\Psilin$) are stated in Section \ref{sec:hozpd}.

\subsubsection{Obtaining the $\protect\xlin$-flux on $\mathcal{H}^+$} \label{sec:hoz1}
\begin{proposition} \label{cor:chf} 
The geometric quantities associated with $\Si$ satisfy on any sphere on the horizon
\begin{align}
\int_{S_{\infty,v}^2} \sin \theta d\theta d\phi \left[ |\slashed{\mathcal{D}}^\star_2 \blin|^2 + \frac{9}{4} \rho^2 |\xlin |^2 +\frac{6M}{r^3}| \blin |^2 + |\slashed{\mathcal{D}}^\star_2 \slashed{div} \xlin |^2 \right] 
 \lesssim \sup_v   \| r^{-1/2} \cdot {\plin} \Omega r^3\|^2_{S^2_{u_0,v}}  + \mathbb{F}_0\left[\Psilin \right] \, .\nonumber
\end{align}
We also control the horizon flux
\begin{align}
\int_{\mathcal{H}\left(v_0,\infty\right)} \left[ |\slashed{\mathcal{D}}^\star_2 \blin |^2 + \frac{9}{4} \rho^2 |\xlin |^2 +\frac{6M}{r^3}|\blin |^2 + |\slashed{\mathcal{D}}^\star_2 \slashed{div} \xlin |^2 \right] \Omega^2 dv \sin \theta d\theta d\phi  \lesssim \mathbb{F}_0\left[\Psilin,  \plin \right] \nonumber .
\end{align}
\end{proposition}

\begin{proof}
The bounds follow from Proposition \ref{prop:psie} and Corollary \ref{cor:psish}. We compute
\[
|\plin|^2 = |\slashed{\mathcal{D}}_2^\star \blin |^2 + 3\rho \xlin \cdot \slashed{\mathcal{D}}_2^\star \blin + \frac{9}{4} \rho^2 | \xlin |^2 \, ,
\]
which since the adjoint of $\slashed{\mathcal{D}}_2^\star$ is $\slashed{div}$ yields
\begin{align} \label{psiidh}
\int \sin \theta d\theta d\phi | \plin |^2 =\int \sin \theta d\theta d\phi \Big( |\slashed{\mathcal{D}}_2^\star \blin  |^2 -\frac{6M}{r^3} \slashed{div} \xlin \cdot \blin + \frac{9}{4} \rho^2 |\xlin |^2 \Big) \, .
\end{align}
Inserting the Codazzi equation (\ref{ellipchi}) restricted to the horizon ($\slashed{div} \xlin  = -\blin$; here we use that $\otx=0$ on $\mathcal{H}^+$ for $\Si$, cf.~Proposition \ref{prop:propagate}!) we obtain the first bound from Corollary \ref{cor:psish} and the second from (\ref{heui}). 
\end{proof}

\begin{remark}
Note that the above provides a bound on $\blin$ itself. The underlying reason is that by 
Theorem~\ref{etsilew}, $\blin_{\ell=1}$ actually vanishes on the horizon since it vanishes 
there for any reference Kerr solution $\mathscr{K}$.
\end{remark}

\begin{remark}
By an elliptic estimate (cf.~(\ref{uid2})) one obtains Proposition \ref{cor:chf} for all angular derivatives up to order $2$ of $\xlin$.\end{remark}

\subsubsection{Obtaining the $\frac{1}{\Omega} \slashed{\nabla}_3 \left(\Omega \protect\xlin\right)$-flux on $\mathcal{H}^+$} \label{sec:gebo}

\begin{proposition} \label{lem:gaugeta}
The geometric quantities associated with $\Si$ satisfy on any sphere on the horizon
\begin{align}
\int_{S^2_{\infty,v}} \sin \theta d\theta d\phi \left[   |\slashed{\mathcal{D}}^\star_2 \elin|^2 
+  |\slashed{div} \slashed{\mathcal{D}}^\star_2 \elin |^2 + |\slashed{\mathcal{D}}^\star_2 \slashed{div} \slashed{\mathcal{D}}^\star_2 \elin |^2 + \Big| \frac{1}{\Omega} \slashed{\nabla}_3 \left(\Omega \xlin\right) \Big|^2+ \Big| \slashed{\mathcal{D}}^\star_2 \slashed{div}\frac{1}{\Omega} \slashed{\nabla}_3 \left(\Omega \xlin \right) \Big|^2 \right] 
\nonumber \\ 
\lesssim \sup_v   \| r^{-1/2} \cdot {\plin} \Omega r^3\|^2_{S^2_{u_0,v}}  + \mathbb{F}_0\left[\Psilin \right]  . \nonumber
\end{align}
We also control the horizon flux
\begin{align}
\int_{\mathcal{H}\left(v_0,\infty \right)} dv \sin \theta d\theta d\phi \left[   |\slashed{\mathcal{D}}^\star_2 \elin |^2 +  |\slashed{div} \slashed{\mathcal{D}}^\star_2 \elin |^2 + |\slashed{\mathcal{D}}^\star_2 \slashed{div} \slashed{\mathcal{D}}^\star_2 \elin |^2 + \Big| \frac{1}{\Omega} \slashed{\nabla}_3 \left(\Omega \xlin \right) \Big|^2 +  \Big| \slashed{\mathcal{D}}^\star_2 \slashed{div}\frac{1}{\Omega} \slashed{\nabla}_3 \left(\Omega \xlin \right) \Big|^2 \right] \nonumber \\
\lesssim  \mathbb{F}_0\left[\Psilin,  \plin \right] \, . \nonumber
\end{align}
\end{proposition}

\begin{remark}
Note that the above implies that we control $\elin$ up to its $\ell=1$ modes, cf.~Corollary \ref{cor:etavl1}.
\end{remark}

\begin{proof}
By our gauge conditions on $\Si$ we have that $div \elin = -\rlin+\rlin_{\ell=0}$ 
holds on the horizon (cf.~Proposition \ref{prop:propagate}). This implies that
\begin{align} \label{pjk}
\|\slashed{\mathcal{D}}_2^\star \slashed{\mathcal{D}}_1^\star \left( \slashed{div} \, \elin ,
\slashed{curl} \, \elin \right)\|_{S^2_{\infty,v}}^2 = \| \slashed{\mathcal{D}}_2^\star \slashed{\mathcal{D}}_1^\star \left(-\rlin,\slin \right)\|_{S^2_{\infty,v}}^2 \lesssim \| \Plin \|_{S^2_{\infty,v}}^2 + \|\Omega \xlin\|_{S^2_{\infty,v}}^2
\end{align}
by the definition of $P$.
Using Proposition \ref{cor:chf} as well as Theorem \ref{prop:summarypsi} on the right hand side and Lemma \ref{lem:aist} on the left, we derive the desired estimates except for the $\slashed{\nabla}_3 \left(\Omega \hat{\chi}\right)$-terms. To obtain the latter, note that equation (\ref{chih3}) when restricted to the horizon, yields
\begin{align} \label{pjk2}
\frac{1}{\Omega} \slashed{\nabla}_3 \left(\Omega \xlin \right) = -2 \slashed{\mathcal{D}}^\star_2 \elin  + \frac{1}{2M} \Omega \xlin \, ,
\end{align}
so that control on $\xlin$ (cf.~again Proposition \ref{cor:chf}) and the $\slashed{\mathcal{D}}^\star_2 \elin$-fluxes implies control on the normal derivative.
\end{proof}

\subsubsection{Obtaining higher order fluxes on $\mathcal{H}^+$} \label{sec:hozf} 
We collect some additional flux estimates on the horizon that follow directly from the horizon fluxes on $\Plin$ and $\Pblin$. 

\begin{proposition} \label{prop:angularbetahoz}
The geometric quantities associated with $\Si$ satisfy the following estimates on the horizon for any $v\geq v_0$:
\begin{equation}
\sum_{i=0}^3 \int_{v_0}^v d\bar{v}  \| \slashed{\nabla}^i (\Omega \blin ) \|_{S^2\left(\infty,\bar{v}\right)}^2   \lesssim \mathbb{F}_0\left[\Psilin, \plin, \alin \right]  \, .
\end{equation}
The same estimate holds with $\xlin$ replacing $\blin$ and $3$ replacing $4$ in the sum. Moreover,
\begin{equation} \label{lyu1} 
 \int_{v_0}^v d\bar{v}   \| \slashed{div} \slashed{\mathcal{D}}_2^\star \slashed{\mathcal{D}}_1^\star \left(\rlin,\slin\right) \|_{S^2\left(\infty,\bar{v}\right)}^2  \lesssim  \mathbb{F}_0\left[\Psilin \right] \, ,
\end{equation}
as well as
\begin{equation} \label{lyu2}
 \int_{v_0}^v d\bar{v}   \|  \mathcal{A}^{[3]}\slashed{\mathcal{D}}_2^\star \elin \|_{S^2\left(\infty,\bar{v}\right)}^2  \lesssim  \mathbb{F}_0\left[\Psilin \right]\, .
\end{equation}
\end{proposition}

\begin{proof}
Note that we control the horizon flux
\begin{align} \label{fut} 
 \int_{v_0}^v d\bar{v} \left( \| \Omega \slashed{\nabla}_4 \Plin \|_{S^2\left(\infty,\bar{v}\right)}^2 +  \| \slashed{\nabla}_A \Plin \|_{S^2\left(\infty,\bar{v}\right)}^2 +  \|  \Plin \|_{S^2\left(\infty,\bar{v}\right)}^2\right)   \lesssim \mathbb{F}_0\left[\Psilin \right]
\end{align}
from Theorem \ref{prop:summarypsi}. Now \emph{restricted to the horizon} we have
\[
\Omega \slashed{\nabla}_4 \Plin = \slashed{\mathcal{D}}_2^\star \slashed{\mathcal{D}}_1^\star \left(-\slashed{div} \Omega \blin, -\Omega \slashed{curl} \blin\right) + \frac{3M}{(2M)^4} \Omega^2 \alin - \frac{6M}{(2M)^6} \Omega \xlin
\]
which in view of Proposition \ref{cor:chf} and Proposition \ref{prop:alphae} yields the first estimate after using elliptic estimates on $S^2$ and the fact that the zeroth order term of $\blin$ is controlled from Proposition \ref{cor:chf}. The claim about $\xlin$ is then an immediate consequence of the $\slashed{div} \xlin = -\blin$ holding on $\mathcal{H}^+$. For (\ref{lyu1}), note that restricted to the horizon
\[
\slashed{div} \Plin = \slashed{div}\slashed{\mathcal{D}}_2^\star \slashed{\mathcal{D}}_1^\star \left(-\rlin, \slin\right)  -\frac{3M}{(2M)^4} \slashed{div} \Omega \xlin \, ,
\]
and use (\ref{fut}) and Proposition \ref{cor:chf}. Finally, for (\ref{lyu2}) recall Proposition \ref{prop:propagate} which implies
\[
\slashed{\mathcal{D}}_2^\star \slashed{\mathcal{D}}_1^\star \left(\rlin + \slashed{div}\elin, 0\right)  =0 \ \ \ \textrm{on $\mathcal{H}^+$.}
\]
Using the identity (\ref{uid3}) and the previous bound (\ref{lyu1}) and Proposition \ref{lem:gaugeta}, the desired estimate follows.
\end{proof}

\subsubsection{Obtaining polynomial decay bounds on $\mathcal{H}^+$} \label{sec:hozpd}
Using the decay statements for the horizon fluxes in Proposition \ref{prop:decRW} and Corollary \ref{cor:hozflu} we can obtain
\begin{proposition} \label{prop:decbh}
We have the following flux bounds along the event horizon $\mathcal{H}^+$:
\begin{align}
\int_{v}^\infty d\bar{v} \Big[ \| \mathcal{A}^{[4]} \xlin \Omega \|^2_{S^2_{\infty,\bar{v}}} + \| \mathcal{A}^{[3]}\slashed{\mathcal{D}}_2^\star\elin \|^2_{S^2_{\infty,\bar{v}}} &+ \|\mathcal{A}^{[3]} \Omega^{-1} \slashed{\nabla}_3 \left(\Omega \xlin \right) \|_{S^2_{\infty,\bar{v}}} \Big] \nonumber \\
& \lesssim \frac{1}{v^2} \left(\mathbb{F}_0^{2,T}\left[\Psilin\right] + \mathbb{F}_0 \left[\Psilin, \Psilinb, \plin,\pblin, \alin, \ablin \right] \right) \, .
\end{align}
\end{proposition}
\begin{proof}
From Corollary \ref{cor:hozflu}, we already have the flux bound for $\plin$ and $\alin$. 
The identity (\ref{psiidh}) restricted to the horizon $\mathcal{H}^+$ already produces the desired statement for $\mathcal{A}^{[2]}\xlin\Omega$ instead of $\mathcal{A}^{[4]}\xlin\Omega$. Repeating the argument of Proposition \ref{lem:gaugeta} yields the flux for three derivatives of $\elin$. Repeating the proof of Proposition \ref{prop:angularbetahoz} finally provides the statement for $\mathcal{A}^{[4]}\xlin$ and four derivatives of $\elin$. The remaining estimate now follows from (\ref{pjk2}).
\end{proof}

\begin{proposition} \label{prop:decbh2}
We have the following bounds along the event horizon $\mathcal{H}^+$:
\begin{align}
 \| \mathcal{A}^{[2]} \xlin \Omega \|^2_{S^2_{\infty,v}} + \| \mathcal{A}^{[2]}\slashed{\mathcal{D}}_2^\star\elin  \|^2_{S^2_{\infty,v}} &+ \|\mathcal{A}^{[2]} \Omega^{-1} \slashed{\nabla}_3 \left(\Omega \xlin \right) \|_{S^2_{\infty,v}} \nonumber \\
& \lesssim \frac{1}{v^2} \left(\mathbb{F}_0^{2,T}\left[ \Psilin \right] + \mathbb{F}_0 \left[\Psilin, \Psilinb, \plin, \pblin, \alin, \ablin \right]+ \sup_v \|\Omega \plin \|^2_{S^2_{u_0,v}} \right) \, .
\end{align}
\end{proposition}
\begin{proof}
We use Proposition \ref{prop:psibs2d2} and the identity (\ref{psiidh}) restricted to the horizon to obtain the estimate on $\xlin$. For the estimate on $\elin$ 
we use the identities (\ref{pjk}) and (\ref{eac}) in conjunction with the estimate just obtained and the estimate of Proposition \ref{prop:decRW}. The remaining estimate now follows directly from (\ref{pjk2}).
\end{proof}
\subsection{Decay estimates for the outgoing shear $\protect\xlin$} \label{sec:chihatest} 
The main result of this section is
\begin{proposition} \label{prop:chiall} 
Consider the solution $\Si$ of Theorem \ref{theo:mtheo}. The following estimate holds for any $\epsilon>0$, and any $u\ge u_0$, $v\ge v_0$:
\begin{align}
\int_{v_0}^v d\bar{v} \int_{u_0}^u d\bar{u} \int_{S^2\left(\bar{u},\bar{v}\right)} \sin \theta d\theta d\phi \  \frac{\Omega^2}{r^{1+\epsilon}} \Bigg\{  \Big| \frac{1}{\Omega}\slashed{\nabla}_3 \left( \frac{1}{\Omega}\slashed{\nabla}_3\left(r^2 \xlin \Omega\right)\right) \Big|^2
+ \Big|\frac{1}{\Omega}\slashed{\nabla}_3\left(r^2 \xlin \Omega\right) \Big|^2 +  \Big|r^2 \xlin
\Omega\Big|^2 \Bigg\}   \nonumber \\
\lesssim  \Big\|(\slashed{\nabla}_3)^2 \xlin \Big\|^2_{L^2(C_{v_0})} +  \mathbb{F}_0\left[\Psilin, \plin, \alin \right] . \nonumber
\end{align}
We also control the flux
\begin{align} \label{mtheogam1}
 \int_{u_0}^\infty d\bar{u} \int_{S^2\left(\bar{u},v\right)} \sin \theta d\theta d\phi \  \frac{\Omega^2}{r^{0+\epsilon}} \Bigg\{  \Big| \frac{1}{\Omega}\slashed{\nabla}_3 \left( \frac{1}{\Omega}\slashed{\nabla}_3\left(r^2 \xlin \Omega\right)\right) \Big|^2 
+ \Big|\frac{1}{\Omega}\slashed{\nabla}_3\left(r^2 \xlin \Omega\right) \Big|^2 +  \Big|r^2 \xlin \Omega\Big|^2 \Bigg\} \nonumber \\  \lesssim  \Big\|(\slashed{\nabla}_3)^2\xlin\Big\|^2_{L^2(C_{v_0})} +  \mathbb{F}_0\left[\Psilin, \plin, \alin \right] 
\end{align}
for any $v\geq v_0$.
\end{proposition}

\begin{remark} \label{rem:improwd}
We will see from the proof that the $\epsilon$ can actually be removed in the estimate (\ref{mtheogam1}) if the term $ \big|r^2 \xlin \Omega\big|^2$ is removed from the curly bracket. 
\end{remark}

We have already discussed in Section~\ref{overviewbtp} the main difficulty of estimating
$\xlin$ directly:
In the equation $(\ref{tchi})$, there is a  blue-shift factor. As explained 
in Section~\ref{overviewbtp},
this difficulty can be corrected by commuting twice  by the  ``red-shift'' operator
$\frac{1}{\Omega} \slashed{\nabla}_3$ (exploiting the improvement discussed
in Section~\ref{BoundforWAVE} in the context of the scalar wave equation), coupled with our a priori bound on the 
flux of the transversal derivative, $\frac{1}{\Omega} \slashed{\nabla}_3 \left(\Omega \xlin \right)$, on the horizon--now established in Proposition \ref{lem:gaugeta}.

We give a brief outline of this section.
In Section~\ref{commutingitheresec}, we will derive the commutation formulas with the
operator $\frac{1}{\Omega} \slashed{\nabla}_3$.
The proof of Proposition \ref{prop:chiall} will then be carried out in 
Sections~\ref{sec:hose} (which will obtain bounds near the horizon) 
and~\ref{sec:hose2} (which will extend the bounds globally).
Some higher order estimates which follow from commuting and repeating the proof of Proposition \ref{prop:chiall} will be stated in Section \ref{sec:hios}.

\subsubsection{Commuting the $\slashed{\nabla}_4 \protect\xlin$-equation}
\label{commutingitheresec}
We  write the transport equation for $\xlin$, 
equation (\ref{tchi}), as
\begin{align} \label{basty}
\Omega \slashed{\nabla}_4 \left(r^2 \xlin \Omega\right) -2\Omega\hat{\omega} \left(r^2 \xlin \Omega\right) = - \alin \Omega^2 r^2 \, .
\end{align}
Note that $r^2 \xlin \Omega$ is regular both at the horizon and null 
infinity and that the second term in the above is a blue--shift term, i.e.~its sign is negative. We easily deduce from Section \ref{sec:commutation} the commutation formulae
\begin{align}
\left[ \slashed{\nabla}_3, \Omega \slashed{\nabla}_4\right] = +\hat{\omega} \Omega \slashed{\nabla}_3 
\textrm{ \ \ \ as well as  \ \ }
\left[\frac{1}{\Omega} \slashed{\nabla}_3, \Omega \slashed{\nabla}_4\right] = +2\hat{\omega} \Omega \cdot  \frac{1}{\Omega} \slashed{\nabla}_3 \, .
\end{align}
Hence commuting (\ref{basty}) with $\frac{1}{\Omega}\slashed{\nabla}_3$ removes the blue-shift term in (\ref{basty}) and we obtain
\begin{align}
\Omega \slashed{\nabla}_4 \left( \frac{1}{\Omega}\slashed{\nabla}_3\left(r^2 \xlin \Omega\right)\right) -\frac{1}{\Omega}\slashed{\nabla}_3 \left( 2\Omega\hat{\omega} \right) \left(r^2 \xlin \Omega\right) = - \frac{1}{\Omega}\slashed{\nabla}_3 \left( \alin \Omega^2 r^2 \right) \, ,
\end{align}
which simplifies to
\begin{align} \label{commu1} 
\Omega \slashed{\nabla}_4 \left( \frac{1}{\Omega}\slashed{\nabla}_3\left(r^2 \xlin \Omega\right)\right) -\frac{4M}{r^3} \left(r^2 \xlin \Omega\right) = 2\frac{1}{r} \plin r^3 \Omega + \alin r \Omega^2  \, .
\end{align}
Let us commute again with $\slashed{\nabla}_3$ to obtain
\begin{align} \label{dceq}
\Omega \slashed{\nabla}_4 \left( \slashed{\nabla}_3 \left( \frac{1}{\Omega}\slashed{\nabla}_3\left(r^2 \xlin \Omega\right)\right)\right) + \hat{\omega}\Omega \left( \slashed{\nabla}_3 \left( \frac{1}{\Omega}\slashed{\nabla}_3\left(r^2 \xlin \Omega\right)\right)\right) \nonumber \\
-\frac{4M}{r^3} \slashed{\nabla}_3 \left(r^2 \xlin \Omega\right) - \frac{12M}{r^4} \Omega \left(r^2 \xlin \Omega\right)  = - \slashed{\nabla}_3 \left( \frac{1}{\Omega}\slashed{\nabla}_3 \left( \alin \Omega^2 r^2 \right) \right) =2r^2 \Omega \Plin \, .
\end{align}
To derive the above we have used the identities
\begin{align} \label{nebrec}
- \frac{1}{\Omega}\slashed{\nabla}_3 \left( \alin \Omega^2 r \cdot r \right) = -\frac{r}{\Omega} \slashed{\nabla}_3 \left(\alin r \Omega^2 \right) + \alin r \Omega^2 = 2\frac{1}{r} \plin r^3 \Omega + \alin r \Omega^2 \, ,
\end{align}
and 
\[
 - \slashed{\nabla}_3 \left( \frac{1}{\Omega}\slashed{\nabla}_3 \left( \alin \Omega^2 r^2 \right) \right) = 2\plin \Omega^2 r +2r^2 \Omega \Plin -2\plin r \Omega^2 =2r^2 \Omega \Plin \, .
\]

\subsubsection{The main estimate near the horizon} \label{sec:hose} 
We shall first prove an unconditional estimate for $\xlin$ 
in a region $r \leq r_1$ for some $r_1>2M$ close to the horizon, which one may think of as Proposition \ref{prop:chiall} restricted to a region near the horizon. Refer to the diagram
of Section~\ref{overviewbtp}.
\begin{proposition} \label{prop:nearhozchi} 
Consider the solution $\Si$ of Theorem \ref{theo:mtheo}. There exists an 
$r_1>2M$ such that the following estimate holds for any $v\ge v_0$.
\begin{align} \label{ao}
\int_{v_0}^v d\bar{v} \int_{u\left(r_1,\bar{v}\right)}^{\infty} d\bar{u} \int_{S^2\left(\bar{u},\bar{v}\right)} \sin \theta d\theta d\phi \  \Omega^2 \Bigg\{  \Big| \frac{1}{\Omega}\slashed{\nabla}_3 \left( \frac{1}{\Omega}\slashed{\nabla}_3\left(r^2 \xlin \Omega\right)\right) \Big|^2 
+ \Big|\frac{1}{\Omega}\slashed{\nabla}_3\left(r^2 \xlin \Omega\right) \Big|^2 +  \Big|r^2 \xlin\Omega\Big|^2 \Bigg\}  \nonumber \\ \lesssim\mathbb{F}_0\left[\Psilin, \plin \right] 
+  \frac{1}{2} \int_{u\left(r_1,v_0\right)}^{\infty} d\bar{u}\int_{S^2\left(\bar{u},v_0\right)} \sin \theta d\theta d\phi \  \Omega^2 \Big| \frac{1}{\Omega}\slashed{\nabla}_3 \left( \frac{1}{\Omega}\slashed{\nabla}_3\left(r^2 \xlin \Omega\right)\right) \Big|^2 \, .
\end{align}
Here $u\left(r,v\right)$ denotes the $u$-value of the intersection of the hypersurfaces of constant $v$ and those of constant $r$.
Moreover,
the same estimate holds replacing $\int_{v_0}^v d\bar{v}$ by $\sup_{v \in \left(v_0,\infty\right)}$.
\end{proposition}

\begin{proof}[Proof of Proposition~\ref{prop:nearhozchi}]
We consider the region $r \geq r_1\geq 2M$, $v \geq v_0$ for some $r_1$ close to $2M$ chosen below. We let $u\left(r_1,v\right)$ denote the $u$-value where the hypersurface of constant $v$ intersects $r=r_1$ and similarly for $v\left(u,r_1\right)$.

The following Lemma expresses the fact that in the region $r\leq r_1$, we control the spacetime integral of $| \Omega r^2 \xlin |^2$ in a neighborhood of the horizon (``a small region in physical space") by an $\epsilon$ times the horizon flux and the spacetime integral
of the transversal $\Omega^{-1}\slashed\nabla_3$-derivative.
\begin{lemma} \label{auxlem}
Let $\xlin$ be a symmetric traceless $S^2_{u,v}$-tensor. The following estimate holds in $\mathcal{M} \cap \{r\leq r_1\}$:
\begin{align} \label{finiu}
\int_{v_0}^v d\bar{v} \int_{u\left(r_1,\bar{v}\right)}^{\infty} d\bar{u} \int_{S^2\left(u,\bar{v}\right)} \sin \theta d\theta d\phi \  \Omega^2  | \xlin \Omega r^2|^2 \left(\bar{u},\bar{v},\theta,\phi\right) \nonumber \\
\leq 2 |r_1-2M| \int_{v_0}^v d\bar{v}  \int_{S^2\left(\infty,\bar{v}\right)} \sin \theta d\theta d\phi  | \xlin \Omega r^2|^2 \left(\infty,\bar{v},\theta,\phi\right) \nonumber \\
+4 | r_1-2M|^2\int_{v_0}^v d\bar{v} \int_{u\left(r_1,\bar{v}\right)}^{\infty} d\bar{u} \int_{S^2\left(u,\bar{v}\right)} \sin \theta d\theta d\phi \  \Omega^2   \Big|\frac{1}{\Omega} \slashed{\nabla}_3 \left( \xlin\Omega r^2\right)\Big|^2 \left(\bar{u},\bar{v},\theta,\phi\right) \, .
\end{align}
\end{lemma}
\begin{proof}
By the fundamental theorem of calculus
\begin{align}
\int_{S^2\left(u,v\right)} | \xlin \Omega r^2|^2 \left({u},v,\theta,\phi\right) \sin \theta d\theta d\phi = \int_{S^2\left(\infty,v\right)} | \xlin \Omega r^2|^2 \left(\infty,v,\theta,\phi\right) \sin \theta d\theta d\phi \nonumber \\
 - \int_u^{\infty} d\bar{u}\int_{S^2\left(\bar{u},v\right)} \Omega\slashed{\nabla}_3 \left( |\xlin
 \Omega r^2|^2\right) \left(\bar{u},v,\theta,\phi\right) \sin \theta d\theta d\phi \nonumber \, .
\end{align}
We can estimate the last term for any $\lambda>0$ by
\begin{align}
|LT| &\leq  \frac{1}{\lambda} \int_u^{\infty} d\bar{u}\int_{S^2\left(\bar{u},v\right)} \Omega^2   |\xlin \Omega r^2|^2 \left(\bar{u},v,\theta,\phi\right) \sin \theta d\theta d\phi \nonumber \\
& \ \ \ \ + {\lambda} \int_u^{\infty} d\bar{u}\int_{S^2\left(\bar{u},v\right)} \Omega^2   \Big|\frac{1}{\Omega} \slashed{\nabla}_3 \left( \xlin \Omega r^2\right)\Big|^2 \left(\bar{u},v,\theta,\phi\right) \sin \theta d\theta d\phi \nonumber \\
& \leq \frac{|r_1-2M|}{\lambda} \sup_{\bar{u} \in (u(r_1,v),\infty)} \int_{S^2\left(\bar{u},v\right)}  | \xlin \Omega r^2|^2 \left(\bar{u},v,\theta,\phi\right)  \nonumber \\
& \ \ \ \ +{\lambda} \int_u^{\infty} d\bar{u}\int_{S^2\left(\bar{u},v\right)} \Omega^2   \Big|\frac{1}{\Omega} \slashed{\nabla}_3 \left( \xlin \Omega r^2\right)\Big|^2 \left(\bar{u},v,\theta,\phi\right) \sin \theta d\theta d\phi \, .
\end{align}
Choosing $\lambda =2 |r_1-2M|$ yields for any $u\left(r_1,v\right) \leq u \leq \infty$
\begin{align}
\int_{S^2\left(u,v\right)} | \xlin \Omega r^2|^2 \left({u},v,\theta,\phi\right) \sin \theta d\theta d\phi  
\leq 2 \int_{S^2\left(\infty,v\right)} | \xlin \Omega r^2|^2 \left(\infty,v,\theta,\phi\right) \sin \theta d\theta d\phi \nonumber \\
+4 | r_1-2M| \int_u^{\infty} d\bar{u}\int_{S^2\left(\bar{u},v\right)} \Omega^2   \Big|\frac{1}{\Omega} \slashed{\nabla}_3 \left( \xlin \Omega r^2\right)\Big|^2 \left(\bar{u},v,\theta,\phi\right) \sin \theta d\theta d\phi .  \nonumber
\end{align}
Multiplying this with $\Omega^2=-r_u$ and integrating in $u$ from the horizon, we deduce also
\begin{align}
\int_{u}^{\infty} d\bar{u} \ \Omega^2 & \int_{S^2\left(u,v\right)} | \xlin \Omega r^2|^2 \left(\bar{u},v,\theta,\phi\right) \sin \theta d\theta d\phi  \nonumber \\
\leq & \ \ \  2 |r_1-2M| \int_{S^2\left(\infty,v\right)} | \xlin \Omega r^2|^2 \left(\infty,v,\theta,\phi\right) \sin \theta d\theta d\phi \label{huyt} \\
&+4 | r_1-2M|^2 \int_u^{\infty} d\bar{u}\int_{S^2\left(\bar{u},v\right)} \Omega^2   \Big|\frac{1}{\Omega} \slashed{\nabla}_3 \left( \xlin \Omega r^2\right)\Big|^2 \left(\bar{u},v,\theta,\phi\right) \sin \theta d\theta d\phi \nonumber
\end{align}
and after integration in $v$ we obtain (\ref{finiu}).
\end{proof}

We now obtain the estimate (\ref{ao}) from the doubly commuted equation (\ref{dceq}).
Upon multiplication of (\ref{dceq}) with $\left[ \slashed{\nabla}_3 \left( \frac{1}{\Omega}\slashed{\nabla}_3\left(r^2 \xlin \Omega\right)\right)\right] $ and integration over the region $r\leq r_1$, the terms in the first line yield:
\begin{align} \label{poli}
FL = \frac{1}{2} \int_{u\left(v,r_1\right)}^{\infty} d\bar{u}\int_{S^2\left(\bar{u},v\right)} \sin \theta d\theta d\phi \  \Omega^2 \Big| \frac{1}{\Omega}\slashed{\nabla}_3 \left( \frac{1}{\Omega}\slashed{\nabla}_3\left(r^2 \xlin \Omega\right)\right) \Big|^2 \nonumber \\
+ \textrm{pos.~term on $r=r_1$} \nonumber \\
-  \frac{1}{2} \int_{u\left(v_0,r_1\right)}^{\infty} d\bar{u}\int_{S^2\left(\bar{u},v_0\right)} \sin \theta d\theta d\phi \  \Omega^2 \Big| \frac{1}{\Omega}\slashed{\nabla}_3 \left( \frac{1}{\Omega}\slashed{\nabla}_3\left(r^2 \xlin \Omega\right)\right) \Big|^2 \nonumber \\
+  \int_{v_0}^v d\bar{v} \int_{u\left(r_1,\bar{v}\right)}^{\infty} d\bar{u} \int_{S^2\left(u,\bar{v}\right)} \sin \theta d\theta d\phi \  \Omega^2 \hat{\omega}\Omega   \Big| \frac{1}{\Omega}\slashed{\nabla}_3 \left( \frac{1}{\Omega}\slashed{\nabla}_3\left(r^2 \xlin \Omega\right)\right) \Big|^2 \left(\bar{u},\bar{v},\theta,\phi\right) \, .
\end{align}
Recall $\hat{\omega} \Omega = + \frac{M}{r^2}$. The right hand side of (\ref{dceq}) after multiplication by $\slashed{\nabla}_3 \left( \frac{1}{\Omega}\slashed{\nabla}_3\left(r^2 \xlin \Omega\right)\right)$ can be estimated
\[
2r^2 \Omega \Plin \cdot \left( \slashed{\nabla}_3 \left( \frac{1}{\Omega}\slashed{\nabla}_3\left(r^2 \xlin \Omega\right)\right)\right) \leq \frac{1}{16} \frac{M}{r^2} \Bigg\|\slashed{\nabla}_3 \left( \frac{1}{\Omega}\slashed{\nabla}_3\left(r^2 \xlin \Omega\right) \right)\Bigg\|^2 +
 16\Omega^2  \frac{r^6}{M}|\Plin|^2 \, ,
\]
so that after integration over the region $r\leq r_1$ the first term can be absorbed by the good term in (\ref{poli}), while the last term is controlled by $\mathbb{F}\left[\Psilin \right]$ from the integrated decay estimate for $\Plin=r^{-5}\Psilin$, Theorem \ref{prop:summarypsi}.

The two remaining terms arising from multiplication of (\ref{dceq}) can be
controlled as follows. For the second term in the second line of (\ref{dceq}) we simply note
\begin{align} \label{poik}
- \frac{12M}{r^4} \Omega \left(r^2 \xlin \Omega\right)\cdot \left( \slashed{\nabla}_3 \left( \frac{1}{\Omega}\slashed{\nabla}_3\left(r^2 \xlin \Omega\right)\right)\right)
& \leq \frac{\hat{\omega} \Omega}{2} \Big| \slashed{\nabla}_3 \left( \frac{1}{\Omega}\slashed{\nabla}_3\left(r^2 \xlin \Omega\right)\right)\Big|^2 + \frac{1}{2} \frac{144}{\hat{\omega}\Omega} \frac{M^2}{r^8} \cdot \Omega^2 |r^2 \xlin\Omega|^2 \nonumber \\
&\leq \frac{\hat{\omega}\Omega}{2} \Big| \slashed{\nabla}_3 \left( \frac{1}{\Omega}\slashed{\nabla}_3\left(r^2 \xlin \Omega\right)\right)\Big|^2 +  \frac{72M^2}{r^6} \cdot \Omega^2 |r^2 \xlin \Omega|^2 \, .
\end{align}
The first term on the right hand side of (\ref{poik}) will again be absorbed by the good terms in (\ref{poli}) while for the second we will eventually apply Lemma \ref{auxlem}.
For the first term in the second line of (\ref{dceq}) we note that
\begin{align} \label{tyu}
-\frac{4M}{r^3}\Omega \frac{1}{\Omega} \slashed{\nabla}_3 \left(r^2 \xlin \Omega\right)\cdot\left( \slashed{\nabla}_3 \left( \frac{1}{\Omega}\slashed{\nabla}_3\left(r^2 \xlin \Omega\right)\right)\right) \\
= - \frac{2M}{r^3} \partial_u \left|  \frac{1}{\Omega}\slashed{\nabla}_3\left(r^2 \xlin \Omega\right)\right|^2 
= -  \partial_u \left( \frac{2M}{r^3} \Big| \frac{1}{\Omega}\slashed{\nabla}_3\left(r^2 \xlin  \Omega\right)\Big|^2 \right) + \frac{6M}{r^4} \Omega^2  \Big| \frac{1}{\Omega}\slashed{\nabla}_3\left(r^2 \xlin\Omega\right)\Big|^2 \, . \nonumber
\end{align}
Upon integration over the spacetime region the second term has a good sign while the first has a bad sign on the horizon and a good sign on the timelike boundary $r=r_1$.

We summarise the resulting estimate as
\begin{align} \label{poli2}
& \ \ \ \ \frac{1}{2} \int_{u\left(v,r_1\right)}^{\infty} d\bar{u}\int_{S^2\left(\bar{u},v\right)} \sin \theta d\theta d\phi \  \Omega^2 \Big| \frac{1}{\Omega}\slashed{\nabla}_3 \left( \frac{1}{\Omega}\slashed{\nabla}_3\left(r^2 \xlin \Omega\right)\right) \Big|^2 \nonumber \\
&\ \ +\int_{v_0}^v d\bar{v} \int_{u\left(r_1,\bar{v}\right)}^{\infty} d\bar{u} \int_{S^2\left(u,\bar{v}\right)} \sin \theta d\theta d\phi \  \Omega^2\frac{1}{4}\frac{M}{r^2}   \Big| \frac{1}{\Omega}\slashed{\nabla}_3 \left( \frac{1}{\Omega}\slashed{\nabla}_3\left(r^2 \xlin \Omega\right)\right) \Big|^2  \nonumber \\
&\ \ + \int_{v_0}^v d\bar{v} \int_{u\left(r_1,\bar{v}\right)}^{\infty} d\bar{u} \int_{S^2\left(u,\bar{v}\right)} \sin \theta d\theta d\phi \  \Omega^2 \frac{6M}{r^4}   \Big|\frac{1}{\Omega}\slashed{\nabla}_3\left(r^2 \xlin \Omega\right) \Big|^2
 \nonumber \\
&\leq  \frac{1}{2} \int_{u\left(v_0,r_1\right)}^{\infty} d\bar{u}\int_{S^2\left(\bar{u},v_0\right)} \sin \theta d\theta d\phi \  \Omega^2 \Big| \frac{1}{\Omega}\slashed{\nabla}_3 \left( \frac{1}{\Omega}\slashed{\nabla}_3\left(r^2 \xlin \Omega\right)\right) \Big|^2 \nonumber \\
&\ \ +\int_{v_0}^{v} d\bar{v}\int_{S^2\left(\infty,v\right)} \sin \theta d\theta d\phi \  \frac{2M}{r^3} \Big|\left( \frac{1}{\Omega}\slashed{\nabla}_3\left(r^2 \xlin \Omega\right)\right) \Big|^2 \nonumber \\
&\ \ +  \int_{v_0}^v d\bar{v} \int_{u\left(r_1,\bar{v}\right)}^{\infty} d\bar{u} \int_{S^2\left(u,\bar{v}\right)} \sin \theta d\theta d\phi \  \frac{72M^2}{r^6} \cdot \Omega^2 |r^2 \xlin \Omega|^2 + C\cdot \mathbb{F}_0\left[\Psilin \right] \, .
\end{align}
Finally, applying Lemma \ref{auxlem} will allow us to absorb the last term of (\ref{poli2}) by the term in the third line (for $r_1-2M$ sufficiently small; we now \emph{fix} $r_1$ (depending only on $M$) such that this is possible) at the cost of another flux-term on the horizon.
\begin{align} \label{poli3}
 \frac{1}{2} \int_{u\left(v,r_1\right)}^{\infty} d\bar{u}\int_{S^2\left(\bar{u},v\right)} \sin \theta d\theta d\phi \  \Omega^2 \Big| \frac{1}{\Omega}\slashed{\nabla}_3 \left( \frac{1}{\Omega}\slashed{\nabla}_3\left(r^2 \xlin \Omega\right)\right) \Big|^2 \nonumber \\
+  \int_{v_0}^v d\bar{v} \int_{u\left(r_1,\bar{v}\right)}^{\infty} d\bar{u} \int_{S^2\left(u,\bar{v}\right)} \sin \theta d\theta d\phi \  \Omega^2\frac{1}{4}\frac{M}{r^2}   \Big| \frac{1}{\Omega}\slashed{\nabla}_3 \left( \frac{1}{\Omega}\slashed{\nabla}_3\left(r^2 \xlin \Omega\right)\right) \Big|^2  \nonumber \\
+  \int_{v_0}^v d\bar{v} \int_{u\left(r_1,\bar{v}\right)}^{\infty} d\bar{u} \int_{S^2\left(u,\bar{v}\right)} \sin \theta d\theta d\phi \  \Omega^2 \frac{3M}{r^4}   \Big|\frac{1}{\Omega}\slashed{\nabla}_3\left(r^2 \xlin \Omega\right) \Big|^2
 \nonumber \\
\leq  \frac{1}{2} \int_{u\left(v_0,r_1\right)}^{\infty} d\bar{u}\int_{S^2\left(\bar{u},v_0\right)} \sin \theta d\theta d\phi \  \Omega^2 \Big| \frac{1}{\Omega}\slashed{\nabla}_3 \left( \frac{1}{\Omega}\slashed{\nabla}_3\left(r^2 \xlin \Omega\right)\right) \Big|^2 + C \mathbb{F}_0\left[\Psilin\right] \nonumber \\
+\int_{v_0}^{v} d\bar{v}\int_{S^2\left(\infty,v\right)} \sin \theta d\theta d\phi \left[ \frac{1}{4M^2} \Big|\left( \frac{1}{\Omega}\slashed{\nabla}_3\left(r^2 \xlin \Omega\right)\right) \Big|^2  + |r^2 \Omega \xlin|^2 \right] \, .
\end{align}
To obtain the estimate of Proposition \ref{prop:nearhozchi}, note that the flux term on the horizon in the last line is controlled from Lemma \ref{lem:gaugeta}.

To obtain the last statement of  Proposition \ref{prop:nearhozchi} concerning
the $\sup$, we use the positive first term in (\ref{poli3}) for the highest derivative flux. For the lower order terms we use this flux together with the estimate (\ref{huyt}), where in the latter we replace $\xlin \Omega r^2$ by $\Omega^{-1} \slashed{\nabla}_3 \left( \xlin \Omega r^2\right)$ such that the horizon term in (\ref{huyt}) is controlled by Lemma \ref{lem:gaugeta} while the flux term is controlled by the first term in (\ref{poli3}). This gives the $\Omega^{-1} \slashed{\nabla}_3 \left( \xlin \Omega r^2\right)$ flux on any $v \geq v_0$. To obtain the $\xlin \Omega r^2$-flux on any $v \geq v_0$ one again uses (\ref{huyt}) now in its original form and Lemma \ref{lem:gaugeta}.
\end{proof}
\subsubsection{Completing the proof of Proposition \ref{prop:chiall}} \label{sec:hose2}
Proposition \ref{prop:nearhozchi} provides integrated decay (and fluxes) in $r\leq r_1$ from quantities purely at the level of initial data. One can now generalise this to integrated decay (and fluxes) globally. From $\slashed{\nabla}_4 \left(\frac{ \xlin }{\Omega} r^2\right) = \frac{\alin}{\Omega}r^2$ we obtain
\begin{equation} \label{dirb} 
\partial_v \left(f \Big| \frac{ \xlin }{\Omega} r^2 \Big|^2 \right) - \partial_v f \Big| \frac{ \xlin }{\Omega} r^2 \Big|^2= 2 \frac{ \alin }{\Omega}r^2 \cdot \frac{ \xlin }{\Omega} r^2 \Omega f \, .
\end{equation}
Choosing $f$ equal to $\frac{1}{r^\epsilon}$ in $\left[r_1,\infty\right)$ and equal to zero in $\left[2M, 2M + \frac{r_1-2M}{2}<r_1\right)$  and integrating over a spacetime region $\left[u_0,u\right] \times \left[v_0,v\right] \times S^2$ we obtain, noting that the spacetime error in $r\leq r_1$ can be absorbed by Proposition \ref{prop:nearhozchi}, that the boundary term on constant $v$ is positive\footnote{This term vanishes in the limit on null infinity. 
To control the $\xlin$-flux on null infinity one should choose $f\sim 1$ near infinity. However, in this case one needs to know decay in $u$ of $\alpha$ in order to control the right hand side. One can obtain $L^\infty_{u,v} L^2 \left(S^2\right)$ bounds with $f=1$, however. See Corollary \ref{cor:l2sb} below.} and that near infinity ($\Omega^2 \approx 1$)
\[
 2 \frac{ \alin }{\Omega}r^2 \cdot \frac{ \xlin }{\Omega} r^2 r^{-\epsilon} \leq \frac{\epsilon}{2} \cdot r^{-1-\epsilon}\Omega^2 \Big| \frac{ \xlin }{\Omega} r^2 \Big|^2 + C_\epsilon \cdot r^{1-\epsilon} r^4 |\alin |^2
\]
the estimate
\begin{align}
\sup_{\bar{v} \in \left(v_0,v\right)} \int_{u_0}^{u} d\bar{u} \int_{S^2\left(\bar{u},\bar{v}\right)} \sin \theta d\theta d\phi \ \frac{\Omega^2}{r^{\epsilon}}  \Big|r^2  \xlin  \Omega\Big|^2
+ \int_{v_0}^v d\bar{v} \int_{u_0}^{u} d\bar{u} \int_{S^2\left(\bar{u},\bar{v}\right)} \sin \theta d\theta d\plin \ \frac{\Omega^2}{r^{1+\epsilon}}  \Big|r^2 \xlin \Omega\Big|^2  \nonumber \\
\lesssim \mathbb{F}_0\left[\Psilin, \plin, \alin \right]  + \Big\|(\slashed{\nabla}_3)^2 \xlin \Big\|^2_{L^2(C_{v_0})} \, . \nonumber 
\end{align}
Now that we have a global integrated decay (and flux) estimate on $ \xlin $ 
we can revisit (\ref{commu1}) and repeat the above argument (either with exactly the same $f$, which leads to the estimate of the Proposition or with $f=1+\frac{1}{r^\epsilon}$, which would give the estimate alluded to in Remark \ref{rem:improwd}; the improvement exploits the fact that the right hand side of (\ref{commu1}) decays better in $r$ than that of (\ref{basty})) using that zeroth order terms in the equation are already controlled and that the right hand side of (\ref{commu1}) already obeys an integrated decay estimate. Finally, we apply the argument to the twice commuted equation (\ref{dceq}) using again the control on the lower order terms from the previous steps and the integrated decay estimate for $P$ on the right hand side. This completes the proof of Proposition \ref{prop:chiall}.

\begin{corollary} \label{cor:l2sb}
Consider the solution $\Si$ of Theorem \ref{theo:mtheo}. We also have $L^\infty_{u,v} L^2 \left(S^2\right)$ estimates. In particular, for any $u\geq u_0$, $v\geq v_0$
\begin{align}
\int_{S^2_{u,v}} \sin \theta d\theta d\phi \left[ | \Omega \xlin r^2 |^2 + \Big| \frac{1}{\Omega} \slashed{\nabla}_3 (\Omega \xlin  r^2) \Big|^2 \right]  \leq \mathbb{F}_0\left[\Psilin, \plin, \alin \right] +\sup_v \|r^{-1} \plin \Omega r^3\|^2_{S^2_{u_0,v}} +  \Big\|(\slashed{\nabla}_3)^2\xlin \Big\|^2_{L^2(C_{v_0})} . \nonumber
\end{align}
\end{corollary}

\begin{proof}
(Sketch) It is straightforward to obtain these bounds in a region $r\leq 4M$ (or globally with weaker $r$-weights) using the $L^2_{\infty,v} \left(S^2\right)$ bounds on these quantities on the horizon (cf.~Proposition \ref{cor:chf} and Proposition \ref{lem:gaugeta}, acounting for the second term on the right) and the fluxes Proposition \ref{prop:chiall} together with the fundamental theorem of calculus. For $r\geq 4M$ one integrates the transport equations $\slashed{\nabla}_4 \left(\frac{ \xlin }{\Omega} r^2\right) = \frac{\alin}{\Omega}r^2$ and (\ref{commu1}) respectively towards infinity using the fluxes on $\alin$ and $\plin$.
\end{proof}

Using the horizon flux of $\xlin$ 
(Proposition \ref{cor:chf}) in conjunction with the integrated decay estimate on $\Big| \frac{1}{\Omega} \slashed{\nabla}_3 (\Omega \xlin r^2) \Big|^2$ of Proposition \ref{prop:chiall} one also proves
\begin{corollary} \label{cor:l2sc}
Consider the solution $\Si$ of Theorem \ref{theo:mtheo}. For any $u \geq u_0$
\begin{align}
\int_{v_0}^\infty dv \int_{S^2_{u,v}} \sin \theta d\theta d\phi r^{-1-\epsilon} | \Omega \xlin r^2 |^2  
 \leq \mathbb{F}_0\left[\Psilin, \plin, \alin \right]  +  \Big\|(\slashed{\nabla}_3)^2 \xlin \Big\|^2_{L^2(C_{v_0})} \, .
\end{align}
\end{corollary}

\subsubsection{Higher order estimates and summary} \label{sec:hios}
The arguments of the previous sections can be repeated for angular commuted equations. We have in particular

\begin{proposition} \label{prop:hioc}
Proposition \ref{prop:chiall} holds replacing $\xlin$ on the left by $\mathcal{A}^{[3]}\xlin$ 
and replacing the right hand side by  $\Big\|\slashed{\nabla}_3^2 \left(r^3 \slashed{div} \slashed{\mathcal{D}}_2^\star \slashed{div}\, \xlin \right)\Big\|_{L^2 \left(C_{v_0}\right)} + \mathbb{F}_0^{2,T,\slashed{\nabla}} [{\Psilin}, \mathfrak{D}{\plin}, \mathfrak{D}{\alin}]$.
\end{proposition}

\begin{proof}
For \emph{two} angular commutations, i.e.~replacing $\mathcal{A}^{[3]}$ by $\mathcal{A}^{[2]}$ on the left, the statement follows immediately by the fact that the angular derivatives commute trivially.  We sketch the argument (by revisiting the proof of Proposition \ref{prop:chiall}) that we can actually estimate three angular derivatives. Observe that commuting (\ref{dceq}) three times yields
\begin{align} \label{dceqc}
\Omega \slashed{\nabla}_4 \left[ \slashed{\nabla}_3 \left( \frac{1}{\Omega}\slashed{\nabla}_3 \left(r^3 \slashed{div} \slashed{\mathcal{D}}_2^\star \slashed{div}\left(r^2 \xlin \Omega\right)\right)\right)\right] + \hat{\omega}\Omega  \slashed{\nabla}_3 \left( \frac{1}{\Omega}\slashed{\nabla}_3\left(r^3 \slashed{div} \slashed{\mathcal{D}}_2^\star \slashed{div}\left(r^2 \xlin \Omega\right)
\right)\right) \nonumber \\
-\frac{4M}{r^3} \slashed{\nabla}_3\left(r^3 \slashed{div} \slashed{\mathcal{D}}_2^\star \slashed{div} \left(r^2\xlin \Omega\right)\right) - \frac{12M}{r^4} \Omega \left(r^3 \slashed{div} \slashed{\mathcal{D}}_2^\star \slashed{div} \left(r^2\xlin \Omega\right)\right)  =2r^2 \Omega r^3 \slashed{div} \slashed{\mathcal{D}}_2^\star \slashed{div} \Plin \, .
\end{align}
The key ingredients to derive Proposition \ref{prop:chiall} from this identity were 1) a non-degenerate (near $3M$) integrated decay estimate of the right hand side and 2) control of the fluxes of $\mathcal{A}^{[3]}(\xlin \Omega)$ and $\mathcal{A}^{[3]}\Omega^{-1} \slashed{\nabla}_3(\xlin\Omega)$ on the event horizon. The second ingredient is present by the angular commuted version of Propositions \ref{cor:chf} and \ref{lem:gaugeta}. The first ingredient is \emph{not} present a priori because three angular derivatives of $\Plin$ cannot be estimated non-degeneratedly by $\mathbb{F}_0^{2,T,\slashed{\nabla}} [{\Psilin}, \mathfrak{D}{\plin}, \mathfrak{D}{\alin}]$. However, we note that the right hand side of (\ref{dceqc}) can be written,  using the Regge--Wheeler equation for 
$\Plin$, as
\begin{align}
2r^2 \Omega r^3 \slashed{div} \slashed{\mathcal{D}}_2^\star \slashed{div} \Plin &= 2r^3 \Omega \slashed{div} \left[ 2 \slashed{\nabla}_4 \left(\Omega \slashed{\nabla}_3 \Plin \right) + a_1 \Omega \slashed{\nabla}_3 \Plin + a_2 \Omega \slashed{\nabla}_4 \Plin + a_3 \Plin \right] \nonumber \\
&= 2\Omega \slashed{\nabla}_4 \left(r^3 \slashed{div} \Omega \slashed{\nabla}_3 \Plin \right) + \textrm{l.o.t.}
\end{align}
The first term can be ``renormalised" into the derivative term on the left hand side of (\ref{dceqc}), after which at most second derivatives of $P$, one of them always being angular, appear on the right hand side. For such terms a non-degenerate integrated decay estimate is available in terms of $\mathbb{F}^{2,T,\slashed{\nabla}}\left[\Psilin\right]$ (see Corollary \ref{cor:higherorderbasic}; note in particular that the $\slashed{div} R^\star \Psilin$-derivative is controlled non-degenerately already in $\mathbb{F}^{1,T,\slashed{\nabla}}\left[\Psilin\right]$ while the $\slashed{div} T \plin$-derivative is non-degenerately controlled in $\mathbb{F}^{2,T,\slashed{\nabla}}\left[\Psilin\right]$).
\end{proof}

The analogues of Corollary \ref{cor:l2sb} and \ref{cor:l2sc} are now deduced just as before. Note that from the commuted version of Proposition \ref{cor:chf} and \ref{lem:gaugeta}  controlling $\mathcal{A}^{[3]} \xlin \Omega$ and $\mathcal{A}^{[3]}\Omega^{-1} \slashed{\nabla}_3 \left(\xlin \Omega\right)$ in $L^\infty$ on the horizon (or the fluxes on the horizon respectively) requires only one angular derivative of $\psi$ and one derivative of $P$ which is clearly controlled by $\mathbb{F}_0^{2,T,\slashed{\nabla}} [{\Psilin}, \mathfrak{D}{\plin}, \mathfrak{D}{\alin}]$. This proves

\begin{proposition} \label{prop:angc}
We have for any $u\geq u_0$, $v\geq v_0$
\begin{align}
\Big\| r^{-1} \cdot \mathcal{A}^{[3]}\Omega \xlin r^2 \Big\|_{S^2_{u,v}}^2 + \Big\|r^{-1} \cdot \mathcal{A}^{[3]}\frac{1}{\Omega} \slashed{\nabla}_3 (\Omega \xlin r^2) \Big\|_{S^2_{u,v}}^2   \lesssim \Big\|\slashed{\nabla}_3^2 \left(\mathcal{A}^{[3]} \xlin \right)\Big\|_{L^2 \left(C_{v_0}\right)} + \mathbb{F}_0^{2,T,\slashed{\nabla}} [{\Psilin}, \mathfrak{D}{\plin}, \mathfrak{D}{\alin}] \, , \nonumber
\end{align}
and for any $u \geq u_0$
\begin{align}
\int_{v_0}^\infty dv \int_{S^2_{u,v}} \sin \theta d\theta d\phi r^{-1-\epsilon} |\mathcal{A}^{[3]} \Omega \xlin r^2 |^2    \lesssim \Big\|\slashed{\nabla}_3^2 \left(\mathcal{A}^{[3]} \xlin \right)\Big\|_{L^2 \left(C_{v_0}\right)} + \mathbb{F}_0^{2,T,\slashed{\nabla}} [{\Psilin}, \mathfrak{D}{\plin}, \mathfrak{D}{\alin}] \, .
\end{align}
\end{proposition}

\subsection{Boundedness estimates for the ingoing shear $\protect\xblin$} \label{sec:bndchi}

In this section we will prove \emph{boundedness} of the quantity $\xblin$ and derivatives thereof. Key to the boundedness proof is the auxiliary geometric quantity $\Ylin$ 
introduced in (\ref{Ydef}), which satisfies a propagation equation with integrable gauge invariant right hand side.

\begin{remark}
With the redshift estimates on $\xlin$ 
of Section \ref{sec:chihatest} in mind, one might hope that a similar argument near 
null infinity will produce \emph{decay} estimates for the geometric quantity $\xblin$ 
of $\Si$. However, this is not the case and we shall only be able to prove boundedness. It is only for the horizon-renormalised  $\Sf$, where the geometric quantity $\xblin$ decays. See 
Theorem~\ref{theo:mtheod}.
\end{remark}

\subsubsection{Control on angular derivatives of $\protect\xblin$}

We recall the quantity $\Ylin$, a symmetric traceless $S^2_{u,v}$-tensor, from (\ref{Ydef}).

\begin{lemma}
We have the propagation equations
\begin{align} \label{hopeid} 
\slashed{\nabla}_3 \Ylin  = \frac{1}{2} r tr \underline{\chi} \frac{r^3 \pblin}{\Omega} + \frac{3M \ablin{r}}{\Omega} \, ,
\end{align}
and 
\begin{align} \label{hopeid1b}
\slashed{\nabla}_3 \left( \mathcal{A}^{[2]} \Ylin \right) = \slashed{\nabla}_3 \left(\frac{1}{2}  {\Psilinb} - M r^3 {\pblin} \Omega^{-1} \right) -r^3 {\pblin}  + \frac{11M}{r} {\blin} r^3 - \frac{3M}{2} \Omega {\alin} r - 9M^2\frac{{\alin}}{\Omega} \, ,
\end{align}
\begin{align} \label{hopeid1c}
\slashed{\nabla}_3 \left( \mathcal{A}^{[4]} \Ylin \right) = \slashed{\nabla}_3 \left(\frac{1}{2} r^2 \slashed{\mathcal{D}}_2^\star \slashed{div} {\Psilinb} - M r^5 \slashed{\mathcal{D}}_2^\star \slashed{div} {\pblin} \Omega^{-1} +\frac{1}{2}  {\Psilinb} - M r^3 {\pblin} \Omega^{-1}\right) + J_2 \, ,
\end{align}
where $J_2$ denotes a linear combination of terms $r^3 \pblin $ and $\frac{r \ablin}{\Omega}$ with uniformly bounded coefficients depending only on $M$.
\end{lemma}

\begin{remark}
Recall that the operators $\mathcal{A}^{[2]} = r^2 \slashed{\mathcal{D}}_2^\star \slashed{div}$ and $\mathcal{A}^{[4]} = r^4 \slashed{\mathcal{D}}_2^\star \slashed{div}\slashed{\mathcal{D}}_2^\star \slashed{div}$ acting on symmetric traceless tensors are uniformly elliptic, cf.~(\ref{uid2}).
\end{remark}

\begin{proof}
To derive the above, note that by (\ref{poy1})
\begin{align} \label{poy3} 
\slashed{\nabla}_3 \left(r^4 \frac{\pblin }{\Omega} \right) + \frac{1}{2} tr \underline{\chi} \left(r^4 \frac{\pblin }{\Omega} \right) = -r^4 \Omega^{-1} \slashed{\mathcal{D}}_2^\star \slashed{div} \ablin - 3Mr \frac{\ablin}{\Omega} \, ,
\end{align}
while
\begin{align}
\slashed{\nabla}_3  \left( \frac{r^2 \xblin}{\Omega}\right) = - \frac{r^2}{\Omega} \ablin \ \ \ \textrm{hence} \ \ \ \slashed{\nabla}_3  \left( \frac{r^4 \slashed{\mathcal{D}}_2^\star  \slashed{div} \xblin}{\Omega}\right) = - \frac{r^4}{\Omega}\slashed{\mathcal{D}}_2^\star  \slashed{div} \ablin  \, .
\end{align}
Subtracting (\ref{poy3}) from the last identity yields (\ref{hopeid}). For the second identity the computation is lengthy but straightforward. The identity (\ref{poy1}) is key as well as the fact that
\begin{align}
r^5 \slashed{\mathcal{D}}_2^\star \slashed{div}\pblin = \frac{r^2}{2}\slashed{\nabla}_4 \left(r^3 \slashed{\mathcal{D}}_2^\star \slashed{div} \ablin\right) + \frac{M}{r} \Omega^{-1} r^3\slashed{\mathcal{D}}_2^\star \slashed{div} \ablin \, ,
\end{align}
which upon inserting the identity (\ref{poy1}) allows to write the top order term as a $\slashed{\nabla}_3$-derivative of $\Psilinb$.
\end{proof}
The point of the identity (\ref{hopeid1b}) is that the highest derivatives on the right hand side appear as a boundary term while the remaining terms are essentially as in (\ref{hopeid}), i.e.~loosely speaking commuting with two angular derivatives doesn't ``lose" regularity.

\begin{proposition} \label{prop:Yest}
Consider the solution $\Si$ of Theorem \ref{theo:mtheo}. We have for any $u \geq u_0$ and $v \geq v_0$ the estimate
\begin{align}
\|r^{-1} \cdot \Ylin\|_{S^2_{u,v}} \lesssim \|r^{-1} \cdot \Ylin\|_{S^2_{u_0,v}} + \sqrt{\mathbb{F}_0^{2,T}\left[\Psilinb\right]} + \sqrt{\mathbb{F}_0 \left[\Psilinb, \pblin, \ablin\right]}
\end{align}
and
\begin{align} \label{plo1}
\|r^{-1} \cdot \mathcal{A}^{[3]} \, \Ylin\|_{S^2_{u,v}} \lesssim & \ \|r^{-1} \cdot  \mathcal{A}^{[3]}\,  \Ylin\|_{S^2_{u_0,v}} + \sqrt{ \mathbb{F}_0^{2,T, \slashed{\nabla}}\left[\Psilinb\right]} + \sqrt{\mathbb{F}_0 \left[\Psilinb, \pblin, \ablin\right]} \nonumber \\
& +  \| r^{-1} \cdot r\slashed{div} \pblin \Omega^{-1} r^3\|_{S^2_{u,v_0}}  +  \| r^{-1} \cdot r\slashed{div} \Psilinb \Omega^{-1}\|_{S^2_{u,v_0}} 
\end{align}
as well as 
\begin{align} \label{plo2}
\|r^{-1} \cdot  \mathcal{A}^{[4]}\,  \Ylin \|_{S^2_{u,v}} \lesssim & \ \|r^{-1} \cdot  \mathcal{A}^{[4]}  \Ylin \|_{S^2_{u_0,v}} + \sqrt{ \mathbb{F}_0^{2,T,\slashed{\nabla}}\left[\Psilinb\right]} + \sqrt{\mathbb{F}_0 \left[\Psilinb, \pblin, \ablin\right]} \nonumber \\
& +  \| r^{-1} \cdot \mathcal{A}^{[2]} \pblin \Omega^{-1} r^3\|_{S^2_{u,v_0}}  +  \| r^{-1} \cdot  \mathcal{A}^{[2]} \Psilinb \Omega^{-1}\|_{S^2_{u,v_0}} 
\end{align}
where we recall (\ref{do2T}) for the definition of $\mathbb{F}_0^{2,T}\left[\Psilinb\right]$. Note that by assumption (\ref{ughbo}) all right hand sides are finite.
\end{proposition}
\begin{remark}
We can actually drop the last term on the right hand side of (\ref{plo1}) and (\ref{plo2}) as it is controlled by $\mathbb{F}_0^{2,T, \slashed{\nabla}}\left[\Psilinb\right]$. We can also drop the penultimate term in (\ref{plo1}) and (\ref{plo2}) provided we replace $\mathbb{F}_0 \left[\Psilinb, \pblin, \ablin\right]$ by $\mathbb{F}^{2,T,\slashed{\nabla}}_0 \left[\Psilinb, \mathfrak{D} \pblin, \ablin\right]$. This is also a direct consequence of one-dimensional Sobolev embedding and the definition of the norms.
\end{remark}

\begin{proof}
It is clear from (\ref{hopeid}) that the first estimate would follow from boundedness for the $L^1$-fluxes
\[
 \int_{u_0}^\infty du \Omega^2 \Big\|r^{-1} \cdot \frac{r^3\pblin}{\Omega}\Big\|_{S^2_{u,v}} \ \ \ \ \ \textrm{and} \ \ \ \  \int_{u_0}^\infty du \Omega^2 \Big\|r^{-1} \cdot \frac{r\ablin}{\Omega^2}\Big\|_{S^2_{u,v}} \, .
\]
But this is immediate by Propositions \ref{prop:l1est} and Corollary \ref{cor:l1est}. For the second estimate we commute (\ref{hopeid1b}) with $r \slashed{div}$ and estimate the quantity $\|r^3 \slashed{div} \slashed{\mathcal{D}}_2^\star \slashed{div}\,\Ylin - \frac{1}{2} r^5 r\slashed{div} \Pblin + M r\slashed{div} r^3 \pblin \Omega^{-1} \|_{S^2_{u,v}}$ by integrating the transport equation using again Propositions \ref{prop:l1est} and Corollary \ref{cor:l1est}. Afterwards we use Corollary \ref{cor:Pons2} and Proposition \ref{prop:firstangularpsi2} to deduce the desired estimate. The third estimate is similar.
\end{proof}

Combining Proposition \ref{prop:Yest} with the definition of $\Ylin$ and Corollary \ref{cor:psishb} we deduce
\begin{corollary} \label{cor:crucialchibar}
Consider the solution $\Si$ of Theorem \ref{theo:mtheo}. We have for any $u\geq u_0$ and $v\geq v_0$ and $i=0,2,3$
\begin{align}
\Big\|r^{-1} \cdot\mathcal{A}^{[i+2]} \left(\frac{r\xblin}{\Omega}\right) \Big\|_{S^2_{u,v}} \lesssim & \ \|r^{-1} \cdot \mathcal{A}^{[i]}  \Ylin\|_{S^2_{u_0,v}} + \sqrt{ \mathbb{F}_0^{2,T, \slashed{\nabla}}\left[\Psilinb\right]} + \sqrt{\mathbb{F}_0 \left[\Psilinb, \pblin, \ablin\right]}  +  \| r^{-1} \cdot \mathcal{A}^{[i]} \pblin \Omega^{-1} r^3\|_{S^2_{u,v_0}}  \nonumber \, .
\end{align}
\end{corollary}

Using that we can multiply both the square of (\ref{plo1}) and (\ref{plo2}) by $\frac{\Omega^2}{r^2}$ and integrate in $u$ we find using the (twice angular commuted) fluxes of Proposition \ref{prop:nif}
\begin{corollary} \label{cor:xbar6}
Consider the solution $\Si$ of Theorem \ref{theo:mtheo}. We have for any $v\geq v_0$ the flux estimates
\begin{align}
\int_{u_0}^\infty du \Omega^2 \Big\|r^{-1} \cdot\mathcal{A}^{[5]} \left(\frac{r\xblin}{\Omega}\right) \Big\|^2_{S^2_{u,v}} \lesssim & \ \|r^{-1} \cdot \mathcal{A}^{[3]} \Ylin\|_{S^2_{u_0,v}} +  \mathbb{F}_0^{2,T, \slashed{\nabla}}\left[\Psilinb\right] +\mathbb{F}^{2,T,\slashed{\nabla}}_0 \left[\Psilinb, \pblin, \ablin\right] \, ,  \nonumber
\end{align}
\begin{align}
\int_{u_0}^\infty du \Omega^2 \Big\|r^{-1} \cdot\mathcal{A}^{[6]} \left(\frac{r\xblin}{\Omega}\right) \Big\|^2_{S^2_{u,v}} \lesssim & \ \|r^{-1} \cdot \mathcal{A}^{[4]} \Ylin\|_{S^2_{u_0,v}} +  \mathbb{F}_0^{2,T, \slashed{\nabla}}\left[\Psilinb\right] + \mathbb{F}^{2,T,\slashed{\nabla}}_0 \left[\Psilinb, \pblin, \ablin\right] \,.   \nonumber
\end{align}

\end{corollary}

\subsubsection{Control on angular derivatives of $\slashed{\nabla}_4\protect\xblin$}
Commuting (\ref{hopeid}) with $\Omega \slashed{\nabla}_4$ we derive
\begin{align} \label{hopeid2} 
\slashed{\nabla}_3 \left(\Omega \slashed{\nabla}_4 \Ylin \right) = r^3 \Omega \Pblin + 3M  \left(2r\pblin - 2 \hat{\omega} \ablin r \right) \, 
\end{align}
using (\ref{evolpp}) and (\ref{aevol}). 

\begin{proposition} \label{prop:chibar4} 
Consider the solution $\Si$ of Theorem \ref{theo:mtheo}. We have for any $u\geq u_0$, $v\geq v_0$ the estimate
\begin{align}  \label{4Ya}
 \|r^{-1} \cdot r \Omega \slashed{\nabla}_4 \Ylin  \|^2_{S^2_{u,v}}  \lesssim \sup_v \|r^{-1} \cdot r \Omega \slashed{\nabla}_4 \Ylin \|^2_{S^2_{u_0,v}}   +  \mathbb{F}_0\left[ \Psilinb, \pblin,\ablin\right] 
\end{align}
as well as for any $v\geq v_0$ the flux estimate
\begin{align} \label{4Yb}
\int_{u_0}^\infty d\bar{u} \Omega^2 r^{-1} \|r^{-1} \cdot r \Omega \slashed{\nabla}_4 \Ylin \|^2_{S^2_{\bar{u},v}}   \lesssim  \sup_v \|r^{-1} \cdot r \Omega \slashed{\nabla}_4 \Ylin \|^2_{S^2_{u_0,v}}   +  \mathbb{F}_0\left[ \Psilinb, \pblin,\ablin\right]  \, .
\end{align}
We also have the following flux and integrated decay estimate for any $v\geq v_0$, $u\geq u_0$
\begin{align} \label{4Yc}
 \int_{v_0}^v d\bar{v} r^{-1-\epsilon} \|r^{-1} \cdot r \Omega \slashed{\nabla}_4 \Ylin \|^2_{S^2_{u,\bar{v}}} + \int_{v_0}^v d\bar{v} \int_{u_0}^u d\bar{u} r^{-2-\epsilon} \Omega^2 \|r^{-1} \cdot r \Omega \slashed{\nabla}_4 \Ylin \|^2_{S^2_{\bar{u},\bar{v}}}  
\nonumber \\ 
\lesssim  \frac{1}{\epsilon} \sup_v \|r^{-1} \cdot r \Omega \slashed{\nabla}_4 \Ylin \|^2_{S^2_{u_0,v}}   +  \mathbb{F}_0\left[ \Psilinb, \pblin,\ablin\right]  \, .
\end{align}
\end{proposition}
\begin{proof}
From (\ref{hopeid2}), we deduce for $\gamma>0$ the estimate
\begin{align} \label{cvb}
\frac{1}{2} \Omega \slashed{\nabla}_3 \left[ \left( \Omega \slashed{\nabla}_4 \Ylin \right)^2 r^\gamma\right] + \frac{1}{4} \gamma r^{\gamma-1} \Omega^2 \left( \Omega \slashed{\nabla}_4 \Ylin \right)^2 \leq \Omega^2 r^{\gamma-3} \left(|\Psilinb|^2 + | r^3 \pblin \Omega^{-1}|^2 + |\ablin r \Omega^{-2}|^2 \right) 
\end{align}
The estimates of the proposition now follow from direct integration over the angular variables and either the $u$ direction (for the first two estimates, $\gamma=2$) or both $u$ and $v$ (for the last estimate, $\gamma=1-\epsilon$).
\end{proof}
\begin{remark} \label{rem:improveswY4}
Stronger $r$-weighted norms are propagated (in particular, we could apply (\ref{cvb}) with $\gamma=3$ and $\gamma=2-\epsilon$ respectively) but we will not make use of this here.
\end{remark}
A simple commutation yields (using that $r \slashed{\nabla}_A$ commutes trivially)
\begin{corollary} \label{cor:comcorj}
Consider the solution $\Si$ of Theorem \ref{theo:mtheo}. Let $i \in \{1,2,3\}$. The three estimates (\ref{4Ya})--(\ref{4Yc}) hold replacing 
\begin{align}
\slashed{\nabla}_4 \Ylin \textrm{ \  by  \ } \mathcal{A}^{[i]} \slashed{\nabla}_4 \Ylin \textrm{\ everywhere} \ \ \ \ \textrm{and} \ \ \ \ \mathbb{F}_0\left[ \Psilinb, \pblin,\ablin\right] \ \textrm{by} \  
\mathbb{F}_0^{2,T, \slashed{\nabla}}\left[   \Psilinb, \mathfrak{D} \pblin,\mathfrak{D} \ablin\right] \textrm{\ on the right.} \nonumber
\end{align}
\end{corollary}
Note that it is indeed $\mathbb{F}_0^{2,T, \slashed{\nabla}}\left[   \Psilinb, \mathfrak{D} \pblin,\mathfrak{D} \ablin\right]$ appearing on the new right hand side which stems from the fact that the estimate invoked in the proof only requires the fluxes of $\Psilinb$, 
and $\mathbb{F}_0^{2,T, \slashed{\nabla}}\left[   \Psilinb, \mathfrak{D} \pblin,\mathfrak{D} \ablin\right]$ by definition already contains three angular derivatives of $\Psilinb$.

Given the estimates for $\slashed{\nabla}_4 \Ylin$ we can obtain estimates for $\slashed{\nabla}_4 \xblin$ from the easily verified identity 
\begin{align} \label{tryu} 
\Omega \slashed{\nabla}_4 \Ylin = \frac{\Omega^2}{r} \Ylin + r \cdot r^2 \slashed{\mathcal{D}}_2^\star  \slashed{div} \Omega \slashed{\nabla}_4 \left(\frac{r \xblin}{\Omega} \right)-r^4 \Pblin - 2M r^2 \frac{\pblin}{\Omega} \, .
\end{align}
In particular, multiplying the above by $r$, squaring and integrating over the angular variable we deduce
\begin{corollary}
Consider the solution $\Si$ of Theorem \ref{theo:mtheo}. We have for any $u\geq u_0$, $v\geq v_0$ the estimate
\begin{align}
\Big\| r^{-1} \cdot r^2 \cdot r^2 \slashed{\mathcal{D}}_2^\star  \slashed{div} \Omega \slashed{\nabla}_4 \left(\frac{r \xblin}{\Omega} \right) \Big\|_{S^2_{u,v}}^2 \lesssim  & \sup_v \|r^{-1} \cdot r \Omega \slashed{\nabla}_4 \Ylin \|^2_{S^2_{u_0,v}}+ \sup_v \|r^{-1} \cdot  \Ylin \|^2_{S^2_{u_0,v}} +\mathbb{F}_0\left[ \Psilinb, \pblin,\ablin\right] \nonumber \\
&+ \sup_u  \| r^{-1} \cdot \pblin \Omega^{-1} r^3\|^2_{S^2_{u,v_0}}  + \sup_u \| r^{-1} \cdot \Psilinb \Omega^{-1}\|^2_{S^2_{u,v_0}}  \, .
\end{align}
\end{corollary}
Note that the last three terms on the right hand side could be replaced by $\mathbb{F}_0\left[ \Psilinb, \mathfrak{D}\pblin,\ablin\right]$ using Sobolev embedding in dimension 1. Using Corollary \ref{cor:comcorj} we also have the commuted version:
\begin{corollary} \label{cor:crucialchibar4}
Consider the solution $\Si$ of Theorem \ref{theo:mtheo}. We have for any $u\geq u_0$, $v\geq v_0$ the estimate
\begin{align}
\Big\| r^{-1} \cdot r^2 \cdot \mathcal{A}^{[4]} \Omega \slashed{\nabla}_4 \left(\frac{r \xblin}{\Omega} \right) \Big\|_{S^2_{u,v}}^2 \lesssim  & \sup_v \|r^{-1} \cdot r^3\slashed{\mathcal{D}}_2^\star  \slashed{div} \Omega \slashed{\nabla}_4 \Ylin \|^2_{S^2_{u_0,v}}+ \sup_v \|r^{-1} \cdot r^2 \slashed{\mathcal{D}}_2^\star  \slashed{div}\, \Ylin \|^2_{S^2_{u_0,v}} \nonumber \\
&+\mathbb{F}_0^{2,T, \slashed{\nabla}}\left[   \Psilinb, \mathfrak{D} \pblin,\ablin\right]  \, .
\end{align}
\end{corollary}
The horizon flux estimate of the next corollary follows directly from the (twice angular commuted) flux of (\ref{4Yc}) (applied with $u \rightarrow \infty$) and suggests that while $\xblin\Omega$ itself does not decay, applying a $T$-derivative gives rise to a decaying quantity:
\begin{corollary} \label{cor:dog}
Consider the solution $\Si$ of Theorem \ref{theo:mtheo}. We control the horizon flux
\begin{align}
\int_{v_0}^\infty d\bar{v} \| \mathcal{A}^{[4]} \Omega\slashed{\nabla}_4 (r \xblin \Omega^{-1})\|^2_{S^2_{\infty,\bar{v}}}\|^2 \lesssim  \sup_v \|r^{-1} \cdot \mathcal{A}^{[2]} r \Omega \slashed{\nabla}_4 \Ylin \|^2_{S^2_{u_0,v}}   +  \mathbb{F}^{2,T,\slashed{\nabla}}_0\left[ \Psilinb, \pblin,\ablin\right]\, .
\end{align}
\end{corollary}

\subsubsection{A polynomial decay estimate for $\slashed{\nabla}_4 (\Omega^{-1} \protect\xblin)$ on the horizon}
For later purposes we derive here a simple polynomial decay estimate for the quantity $\slashed{\nabla}_4 (\Omega^{-1}\xblin)$. 
\begin{proposition} \label{prop:decdyad}
Consider the solution $\Si$ of Theorem \ref{theo:mtheo}. We have along the event horizon the estimate
\[
\| \mathcal{A}^{[2]} \Omega\slashed{\nabla}_4 (r \xblin \Omega^{-1})\|^2_{S^2_{\infty,\bar{v}}} \lesssim \frac{1}{v^{2}} \left( \textrm{right hand side of (\ref{ughb})}\right)
\]
\end{proposition}

\begin{proof}
Starting from the defining equation (\ref{hopeid2}), one repeats the argument (splitting integrals when integrating the transport equations) of the proof of Proposition \ref{prop:psibs2d2} using also the flux bound of Corollary \ref{cor:fluxbpsib}. This yields in particular
\begin{align}
\|  \Omega\slashed{\nabla}_4\Ylin \|^2_{S^2_{\infty,\bar{v}}} \lesssim \frac{1}{v^{2}} \left( \textrm{right hand side of (\ref{ughb})}\right) \,.
\end{align} 
One now revisits the identity (\ref{tryu}) evaluated on the horizon and uses the fact that $v^{-2}$-decay for $\pblin$ is implied by Proposition \ref{prop:psibs2d}, while $v^{-2}$-decay for $\Psilinb$ (on spheres on the horizon) is an immediate consequence of Proposition \ref{prop:decRW}.
\end{proof}

\subsection{Boundedness for all remaining quantities} \label{sec:bconclude}

In this section we conclude the proof of Theorem \ref{theo:mtheo} by exploiting the estimates derived on the outgoing linearised shear $\hat{\chi}$ and the ingoing 
linearised shear $\xblin$ in the previous two subsections to bound all remaining linearised
Ricci and curvature components of the solution $\Si$.

\subsubsection{$L^\infty_{u,v}$-estimates on $S^2_{u,v}$ and fluxes on constant $v$-hypersurfaces}
\label{titlosswstos}
The estimates obtained thus far are sufficient to obtain flux bounds for five angular derivatives of the curvature components $\bblin, \left(\rlin,\slin\right)$, Note that we already control the flux of five angular derivatives of $\ablin$ by Proposition \ref{prop:5dera} as well as six angular derivatives of $\xblin$ from Corollary \ref{cor:xbar6}.
\begin{proposition} \label{prop:fluv}
Consider the  solution $\Si$ of Theorem \ref{theo:mtheo}.  We have the following flux estimates
\begin{align}
\sup_v \int_{u_0}^\infty du \Omega ^2 \| r^{-1} \cdot \mathcal{A}^{[3]}r^2  \slashed{\mathcal{D}}_2^\star \slashed{\mathcal{D}}_1^\star\left(r^3 \rlin,r^3 \slin\right) \|_{S^2_{u,v}}^2  \lesssim \textrm{right hand side of (\ref{ughb})} \, , 
\end{align}
\begin{align}
\sup_v \int_{u_0}^\infty du \Omega ^2 \| r^{-1} \cdot \mathcal{A}^{[4]}r  \slashed{\mathcal{D}}_2^\star \left(r^2 \bblin\Omega^{-1}\right) \|_{S^2_{u,v}}^2\lesssim \textrm{right hand side of (\ref{ughb})} \, .
\end{align}
\end{proposition}
\begin{proof}
The estimates follow from the identities 
\begin{align} \label{rsid2}
\mathcal{A}^{[i]} r^2  \slashed{\mathcal{D}}_2^\star \slashed{\mathcal{D}}_1^\star\left(r^3 \rlin,r^3 \slin\right) &= \mathcal{A}^{[i]} (r^5 \Plin) +3M\Omega r \left(\mathcal{A}^{[i]}\xlin - \mathcal{A}^{[i]}\xblin\right) \, ,
\end{align}
\begin{align} \label{betuid}
\mathcal{A}^{[i]} \slashed{\mathcal{D}}_2^\star (\bblin \Omega^{-1})= \mathcal{A}^{[i]} (\Omega^{-1}\pblin) + \frac{3}{2} \rho\mathcal{A}^{[i]} (\Omega^{-1}\xblin) \, ,
\end{align}
the flux-estimates on $P$ obtained in Theorem \ref{prop:summarypsi} and the estimates on $\xlin$ and $\xblin$ obtained in Corollary \ref{cor:xbar6} and Proposition \ref{prop:angc}. 
\end{proof}

The next proposition concerns $L^2$-estimates on spheres. Note that below for $\blin$ and $\bblin$ we need the redshift (and $r\slashed{\nabla}_4$-) commuted energy $\mathbb{F}_0^2\left[\Psilin\right]$ and $\mathbb{F}_0^2\left[\Psilinb\right]$ on the right hand side. The same is true for the $L^\infty$ estimates for $\alin$ and $\ablin$ obtained previously  in Proposition \ref{prop:hioaa}.
\begin{proposition} \label{prop:bc4}
Consider the  solution $\Si$ of Theorem \ref{theo:mtheo}. We have for any $u\geq u_0$ and any $v\geq v_0$ the estimates
\begin{align}
\| r^{-1} \cdot \mathcal{A}^{[2]}r^2  \slashed{\mathcal{D}}_2^\star \slashed{\mathcal{D}}_1^\star\left(r^3 \rlin,r^3 \slin\right) \|_{S^2_{u,v}}^2 \lesssim \textrm{right hand side of (\ref{ughb})}
\end{align}
and
\begin{align}
\| r^{-1} \cdot \mathcal{A}^{[3]}r  \slashed{\mathcal{D}}_2^\star \left(r^2 \bblin\Omega^{-1}\right) \|_{S^2_{u,v}}^2 +
\| r^{-1} \cdot \mathcal{A}^{[3]}r  \slashed{\mathcal{D}}_2^\star \left(r^{7/2} {\blin}\Omega\right) \|_{S^2_{u,v}}^2 
\lesssim \textrm{right hand side of (\ref{ughb})} + \mathbb{F}_0^2\left[\Psilin, \Psilinb\right] \, , \nonumber
\end{align}
provided $ \mathbb{F}_0^2\left[\Psilin, \Psilinb\right]<\infty$.
Moreover, if on the left hand side of the second estimate $\mathcal{A}^{[3]}$ is replaced by $\mathcal{A}^{[2]}$ and the exponent $7/2$ by $7/2-\epsilon$, then the last term on the right hand side can be dropped.
\end{proposition}
\begin{proof}
Use the identities (\ref{rsid2}), (\ref{betuid}), (\ref{betid}) now with Corollary~\ref{cor:crucialchibar} applied with $i=0$ and $i=2$, Proposition~\ref{prop:angc} and Proposition~\ref{prop:hioaa}. For the final remark use the once angular commuted Proposition~\ref{prop:firstangularpsi2} with $\delta=\epsilon$ and observe that the right hand side of the latter is controlled by the $\textrm{right hand side of (\ref{ughb})}$ alone via one-dimensional Sobolev embedding.
\end{proof}

\subsubsection{Estimates for four angular derivatives of $\protect\elin$ and $\protect\eblin$}
With the estimates at our disposal we can already prove
\begin{proposition} \label{prop:etae}
Consider the  solution $\Si$ of Theorem \ref{theo:mtheo}. For any $u\geq u_0$ and $v\geq v_0$ we have for $i=1,2,3$ the estimate
\begin{align}
\int_{S^2_{u,v}} \sin \theta d\theta d\phi \cdot  \left[ r^6 |\mathcal{A}^{[i]} \slashed{\mathcal{D}}_2^\star \eblin |^2 + r^4 |\mathcal{A}^{[i]} \slashed{\mathcal{D}}_2^\star {\elin} |^2 \right] \nonumber \\ 
\lesssim \Big\|\slashed{\nabla}_3^2 \left(r^3 \slashed{div} \slashed{\mathcal{D}}_2^\star \slashed{div} \, \xlin \right)\Big\|_{ L^2 \left(C_{v_0}\right)} + \mathbb{D}^{[5]}_0 \left[\Ylin, \Zlin\right] +\mathbb{F}_0^{2,T,\slashed{\nabla}} [ \Psilinb, \mathfrak{D}\pblin, \mathfrak{D}\ablin]+\mathbb{F}_0^{2,T,\slashed{\nabla}} [ \Psilin, \mathfrak{D}{\plin}, \mathfrak{D}{\alin}] \nonumber \\
\lesssim \textrm{right hand side of (\ref{ughb})}.
\end{align}
For $\elin$ we have in addition the flux estimate\footnote{Note this estimate loses $r^\epsilon$ compared with the previous but gains an integration in $u$.}
\begin{align} \label{fluh}
\int_{u_0}^\infty du \int_{S^2_{u,v}} \sin \theta d\theta d\phi  \frac{\Omega^2}{r^\epsilon} r^4 |\mathcal{A}^{[3]} \slashed{\mathcal{D}}_2^\star {\elin} |^2 \lesssim \textrm{right hand side of (\ref{ughb})}. \, 
\end{align}
\end{proposition}
\begin{proof}
Consider the estimate
\begin{align} \label{fipo}
r^2 |\mathcal{A}^{[i]} r \slashed{\mathcal{D}}_2^\star \elin|^2 \lesssim  \Big| \mathcal{A}^{[i]} \frac{1}{\Omega}\slashed{\nabla}_3\left(r^2 \xlin \Omega\right) \Big|^2 + \frac{1}{r^2} | \mathcal{A}^{[i]} r^2 \xlin \Omega|^2 + |\mathcal{A}^{[i]}r \xblin \Omega|^2 \, ,
\end{align}
which is an easy consequence of (\ref{chih3}) and trivial angular commutation with the $\mathcal{A}^{[i]}$ defined in (\ref{mathanot}), and
\begin{align} \label{fipo2}
r^4 |\mathcal{A}^{[i]} r \slashed{\mathcal{D}}_2^\star \eblin|^2 \lesssim \Big| \mathcal{A}^{[i]} r^2 \Omega \slashed{\nabla}_4\left(\frac{r\xblin}{\Omega} \right) \Big|^2 + | \mathcal{A}^{[i]} r^2 \xlin \Omega|^2 + |\mathcal{A}^{[i]} r\xblin \Omega^{-1}|^2 \, ,
\end{align}
which is easily derived from (\ref{chih3b}). For $i=1,2,3$ the terms on the right hand sides are controlled by Proposition \ref{prop:angc} and Corollary \ref{cor:crucialchibar}  and Corollary \ref{cor:crucialchibar4}. For the flux estimate we need to use in addition Proposition \ref{prop:hioc}. 
\end{proof}

\begin{remark} \label{rem:improvefluh}
The $\epsilon$ in (\ref{fluh}) can be removed if Remark \ref{rem:improwd} and Proposition \ref{prop:hioc} are taken into account.
\end{remark}

\subsubsection{Control on five angular derivatives of $\protect\olin$ and $\protect\olinb$} 
We turn to estimates for $\olin$ and $\olinb$:
\begin{proposition}\label{prop:omb}
Consider the  solution $\Si$ of Theorem \ref{theo:mtheo}. For any $u\geq u_0$ and $v\geq v_0$ we have the estimate
\begin{align}
\| r^{-1} \cdot \mathcal{A}^{[3]} r^2 \slashed{\mathcal{D}}_2^\star \slashed{\nabla}_A \olin \|_{S^2_{u,v}} \lesssim 
r^{-\frac{5-\epsilon}{2}} \left(u,v\right)\cdot \left(\textrm{right hand side of (\ref{ughb})}\right) \, . 
\end{align}
We also have the estimate
\begin{align}
\| r^{-1} \cdot \mathcal{A}^{[2]} r^2 \slashed{\mathcal{D}}_2^\star \slashed{\nabla}_A \olinb \Omega^{-2} \|_{S^2_{u,v}} \lesssim 
\textrm{right hand side of (\ref{ughb})} \, .
\end{align}
\end{proposition}
\begin{proof}
Commuting equation (\ref{oml1}) we can write for $i=0,1,2,3$
\[
\Omega \slashed{\nabla}_3 \left(\mathcal{A}^{[i]} r^2 \slashed{\mathcal{D}}_2^\star \slashed{\nabla}_A \olin \right) = \Omega^2 \left(\mathcal{A}^{[i]} r^2 \Plin - \mathcal{A}^{[i]} r^2 \Pblin + \frac{6M}{r^2} \mathcal{A}^{[i]} \left(\Omega \xlin - \Omega \xblin\right) \right) + \Omega^2 \frac{2M}{r} \mathcal{A}^{[i]}  \slashed{\mathcal{D}}_2^\star \left(\elin + \eblin\right) \,.
\]
Integrating this transport equation using the Cauchy--Schwarz inequality in conjunction with Theorem \ref{prop:summarypsi} and Corollary \ref{cor:higherorderbasic}, Proposition  \ref{prop:angc} and Corollary \ref{cor:xbar6} as well as Proposition \ref{prop:etae} yields the result. Note that the initial term on $u=u_0$ vanishes in view of $\Si$ satisfying the gauge condition (\ref{omchoice}). It is easy to see how to incorporate a non-vanishing boundary term if (\ref{omchoice}) did not hold on the data, cf.~Remark \ref{rem:indoflapse}.

For $\olinb$ the argument is similar, now integrating in the $4$-direction and starting from (\ref{oml2}) written in the red-shifted form
\[
\Omega \slashed{\nabla}_4 \left(\olinb \Omega^{-2}\right) + \frac{2M}{r^2} \olinb \Omega^{-2} = -\rlin + \frac{4M}{r^3} \Olin \, 
\]
from which after angular commutation as above, one proves the boundedness statement.
\end{proof}

\subsubsection{Control on five angular derivatives of $\protect\otxb$}
\label{edwmecodazzi}
The control obtained on four angular derivatives of $\elin$ is sufficient to estimate all angular derivatives of $\otxb$. For this we write the Codazzi equation (\ref{ellipchi}) as
\begin{align} \label{eqhc2} 
\mathcal{A}^{[i]}\slashed{\mathcal{D}}_2^\star \slashed{div} \Omega^{-1} \xblin = +\frac{1}{r} \mathcal{A}^{[i]}\slashed{\mathcal{D}}_2^\star {\elin} + \mathcal{A}^{[i]} \Omega^{-1} \pblin + \frac{3}{2}\rho \mathcal{A}^{[i]} \xblin\Omega^{-1} + \frac{1}{2\Omega^2} \mathcal{A}^{[i]} \slashed{\mathcal{D}}_2^\star \slashed{\nabla}_A \otxb \, .
\end{align}

\begin{proposition} \label{prop:trxbflux}
Consider the  solution $\Si$ of Theorem \ref{theo:mtheo}. For any $u\geq u_0$ and $v \geq v_0$ we have the estimate
\begin{align}
\Big\| r^{-1} \cdot r^3 \mathcal{A}^{[3]} \slashed{\mathcal{D}}_2^\star \slashed{\nabla}_A \frac{\otxb}{\Omega^2} \Big\|^2_{S^2_{u,v}} \lesssim \textrm{right hand side of (\ref{ughb})} \, .
\end{align}
Moreover, we have the flux estimate
\begin{align}
\sup_v \int_{u_0}^\infty du \Omega^2 \Big\| r^{-1} \cdot r^3 \mathcal{A}^{[3]} \slashed{\mathcal{D}}_2^\star \slashed{\nabla}_A \frac{\otxb}{\Omega^2} \Big\|^2_{S^2_{u,v}} \lesssim \textrm{right hand side of (\ref{ughb})} \, .
\end{align}
\end{proposition}
\begin{proof}
This is a consequence of the identity (\ref{eqhc2}) and Corollaries \ref{cor:crucialchibar}, \ref{cor:xbar6}, Proposition \ref{prop:etae} and the estimates on $\pblin$ obtained in Propositions \ref{prop:nif} and \ref{prop:hioaa} (the latter only for the first bound).
\end{proof}

\subsubsection{Boundedness of the metric components}
\begin{proposition} \label{prop:meco1}
Consider the  solution $\Si$ of Theorem \ref{theo:mtheo}. We have for any $v\geq v_0$ and $u\geq u_0$ and $i=0,1,2,3$
\begin{align} \label{nirs}
\Big\| r^{-1} \cdot \sqrt{r} \mathcal{A}^{[i]} r^2 \slashed{\mathcal{D}}_2^\star \slashed{\nabla} \frac{\glinto}{\sqrt{\slashed{g}}} \Big\|_{S^2_{u,v}}^2 \lesssim  \Big\| r^{-1} \cdot \sqrt{r} \mathcal{A}^{[i]} r^2 \slashed{\mathcal{D}}_2^\star \slashed{\nabla}  \frac{\glinto}{\sqrt{\slashed{g}}} \Big\|_{S^2_{u_0,v}}^2 + \textrm{right hand side of (\ref{ughb})}
\end{align}
and also
\begin{align} \label{nirs2}
 \| r^{-1} \cdot \sqrt{r} \mathcal{A}^{[i]} \glinh \, \Big\|_{S^2_{u,v}}^2 \lesssim  \| r^{-1} \cdot \sqrt{r} \mathcal{A}^{[i]} \glinh \, \Big\|_{S^2_{u_0,v}}^2 + \textrm{right hand side of (\ref{ughb})} \, .
 \end{align}

For the metric component $\bmlin$ we have
\begin{align} \label{beq1}
r^{1-\epsilon} \Big\| r^{-1} \cdot \mathcal{A}^{[i]} r \slashed{\mathcal{D}}_2^\star \bmlin \, \Big\|_{S^2_{u,v}}^2 \lesssim 
 \textrm{right hand side of (\ref{ughb})} \, .
\end{align}
For $\Olin$ we have
\begin{align} \label{omeq1}
 \Big\| r^{-1} \cdot \mathcal{A}^{[i]} r^2 \slashed{\mathcal{D}}_2^\star \slashed{\nabla}\, \Olin\, \Big\|_{S^2_{u,v}}^2 \lesssim  \textrm{right hand side of (\ref{ughb})}.
\end{align}
The last two estimates can be improved to $i=4$ with further work once higher angular derivatives of $\elin, \eblin$ are estimated. See Section \ref{sec:refha} and Corollary \ref{cor:hot}.  
Note also that the initial term vanishes for $(\ref{beq1})$.
\end{proposition}

We remark that the first term on right hand side of (\ref{nirs}) and (\ref{nirs2}) is in fact also controlled by the right hand side of (\ref{ughb}) and could hence be dropped. This follows easily from the round sphere conditions (\ref{rsc}) and (\ref{rscm}) and the boundedness of the initial energy $(\ref{ughb})$: Integrate (\ref{stos}) and (\ref{stos2}) from infinity.

The estimates above hence show in particular that the round sphere conditions (\ref{rsc}) and (\ref{rscm}) are preserved in evolution.

\begin{proof}
From (\ref{stos}) we derive for any $n \in \mathbb{R}$
\[
\frac{1}{2} \partial_u \left[r^n \Big| \mathcal{A}^{[i]} r^2 \slashed{\mathcal{D}}_2^\star \slashed{\nabla}  \frac{\glinto}{\sqrt{\slashed{g}}}\Big|^2  \right] + \frac{\Omega^2 n}{2r} \left[r^n \Big| \mathcal{A}^{[i]} r^2 \slashed{\mathcal{D}}_2^\star \slashed{\nabla}  \frac{\glinto}{\sqrt{\slashed{g}}}\Big|^2  \right] = r^n  \mathcal{A}^{[i]} r^2 \slashed{\mathcal{D}}_2^\star \slashed{\nabla} \otxb \cdot \mathcal{A}^{[i]} r^2 \slashed{\mathcal{D}}_2^\star \slashed{\nabla}  \frac{\glinto}{\sqrt{\slashed{g}}}\, ,
\]
which we apply with $n=1$ from initial data, use Cauchy--Schwarz on the right hand side and the flux of Proposition \ref{prop:trxbflux}. Similarly from the equation below (\ref{stos}) we have
\[
\frac{1}{2} \partial_u \left(r^n | \mathcal{A}^{[i]}\glinh |^2 \right) + \frac{1}{2} \frac{n \Omega^2}{r} \left(r^n | \mathcal{A}^{[i]}\glinh |^2 \right) = 2r^{n-1} \Omega^2 \mathcal{A}^{[i]}\left(r \xblin \Omega^{-1}\right)\cdot \mathcal{A}^{[i]}\glinh \, ,
\]
which we apply with $n=1$ from initial data, use Cauchy-Schwarz on the right hand side and the flux of Corollary \ref{cor:xbar6}.
From the transport equation (\ref{bequat}) we derive
\begin{align} \label{bidm}
\frac{1}{2} \partial_u \left(r^n \Big| \frac{\mathcal{A}^{[i]} r \slashed{\mathcal{D}}_2^\star \bmlin}{r}\Big|^2 \right) + \frac{n \Omega^2}{2r} \left(r^n \Big| \frac{\mathcal{A}^{[i]}r \slashed{\mathcal{D}}_2^\star \bmlin}{r}\Big|^2\right) = 2\Omega^2\frac{1}{r} \mathcal{A}^{[i]} r \slashed{\mathcal{D}}_2^\star \left(\elin - \eblin\right) \cdot \frac{\mathcal{A}^{[i]} r \slashed{\mathcal{D}}_2^\star \bmlin}{r} \cdot r^n \, ,
\end{align}
which we apply with $n=3-\epsilon$, use Cauchy--Schwarz on the right hand side and Proposition \ref{prop:etae} the flux estimate (\ref{fluh}). The last estimate of the Proposition follows directly from (\ref{oml3}).
\end{proof}

\begin{remark}
The proof shows that the $\epsilon$ in (\ref{beq1}) can be removed if Remark \ref{rem:improvefluh} is taken into account.
\end{remark}

\subsubsection{Top order estimates for angular derivatives of $\protect\xlin$ 
and $\protect\otx$: The role of $\protect\Zlin$} \label{sec:refha}
At this point we have estimated all metric and Ricci coefficients except $\otx$. 
Estimates for the latter could be obtained directly from the estimates on $\xlin$ 
of Section \ref{sec:chihatest} (cf.~Propositions \ref{prop:hioc} and \ref{prop:angc}) and the Codazzi equation (\ref{eqhc}). However, the estimates of Section \ref{sec:chihatest} ``lose" $\slashed{\nabla}_3$ derivatives, in the sense that we could only estimate $\xlin$ 
after commutation (twice!)~with the redshift vectorfield. In this section we show how to avoid this loss of derivatives and how to estimate five angular derivatives of $\xlin$. 

Key to the argument is the auxiliary quantity $\Zlin$ 
defined in (\ref{Qquant}) which essentially allows to prove that, given estimates on 
$\elin$ and $\eblin$, the quantity $\otx$ 
can be estimated without a loss near the horizon. The Codazzi equation (\ref{eqhc}) can then be used to estimate angular derivatives of $\xlin$. The weights near infinity obtained in the process are not optimal. However, once estimates for angular derivatives of $\xlin$ are available near the horizon, one can use the angular commuted transport equation for $\xlin$, (\ref{tchi}), to optimise the weights for (angular derivatives of) $\xlin$ and then use once again  (\ref{eqhc}) to optimise them for (angular derivatives of)  $\otx$.

\paragraph{Angular derivatives of  $\protect\otx$.}
Observe that we can write
\begin{align} \label{rewrite} 
D \left(\otx \frac{r^2}{\Omega^2} - 4 r \Olin\right) = - 4\Omega^2\Olin \, .
\end{align}
Recall the  quantity $\Zlin=\frac{r^3}{\Omega^2} \slashed{\nabla}_A \otx  - 2 r^2 \left(\elin + \eblin\right)$ defined in (\ref{Qquant}). Commuting,  we can write
\begin{align}
\Omega \slashed{\nabla}_4 \left(\mathcal{A}^{[i]}r \slashed{\mathcal{D}}_2^\star \Zlin\right) = - 2\Omega^2 \mathcal{A}^{[i]} r^2 \slashed{\mathcal{D}}_2^\star \left(\elin + \eblin\right) \, ,  \label{rewrite3} 
\end{align}
where we recall the definition of $\mathcal{A}^{[i]}$ in (\ref{mathanot}).
Contracting (\ref{rewrite3}) with $\frac{1}{\Omega^2} \mathcal{A}^{[i]} r \slashed{\mathcal{D}}_2^\star \Zlin$ (with the appropriate $i$) we find\footnote{In principle we could contract with $\frac{1}{\Omega^4}\mathcal{A}^{[i]} r \slashed{\mathcal{D}}_2^\star \Zlin$ to estimate a stronger norm but since in our gauge $\elin$ and $\eblin$ do not decay along the event horizon we need an additional $\Omega^2$ do perform the $v$ integration in the estimate below.}
\[
\partial_v \left[\frac{|\mathcal{A}^{[i]} r \slashed{\mathcal{D}}_2^\star \Zlin|^2}{r^n \Omega^2}\right] +\left(\frac{n\Omega^2 }{r^{1}} + \frac{2M}{r^{2}} \right) \frac{|\mathcal{A}^{[i]} r \slashed{\mathcal{D}}_2^\star 
\Zlin|^2}{r^n \Omega^2} = - 2 r^{2-n} \left( \mathcal{A}^{[i]} \slashed{\mathcal{D}}_2^\star  \left(\elin + \eblin\right), \mathcal{A}^{[i]} r \slashed{\mathcal{D}}_2^\star \Zlin\right) \, .
\]
Integrating and using Cauchy--Schwarz (note that $r \slashed{\mathcal{D}}_2^\star \Zlin$ vanishes on the sphere $S^2_{\infty,v_0}$ in our gauge) one finds

\begin{proposition} \label{prop:zestimate}
Consider the  solution $\Si$ of Theorem \ref{theo:mtheo}. We have for $n\geq 0$, any $i \in \mathbb{N}$ and any $u\geq u_0$ the estimate
\begin{align}
\int_{S^2_{u,v}} \sin \theta d\theta d\phi \frac{|\mathcal{A}^{[i]} r \slashed{\mathcal{D}}_2^\star \Zlin|^2}{r^n \Omega^2} + \int_{v_0}^v d\bar{v} \int_{S^2_{u,\bar{v}}} \sin \theta d\theta d\phi \frac{1}{r} \cdot \frac{|\mathcal{A}^{[i]} r \slashed{\mathcal{D}}_2^\star \Zlin|^2}{r^n \Omega^2} \nonumber \\
\lesssim  \int_{S^2_{u,v_0}} \sin \theta d\theta d\phi \frac{|\mathcal{A}^{[i]} r \slashed{\mathcal{D}}_2^\star \Zlin|^2}{r^n\Omega^2} 
+ \int_{v_0}^v d\bar{v}  \int_{S^2_{u,\bar{v}}} \sin \theta d\theta d\phi r^{5-n} \Omega^2  \Big|\mathcal{A}^{[i]} \slashed{\mathcal{D}}_2^\star  \left(\elin + \eblin\right) \Big|^2\nonumber .
\end{align}
\end{proposition}

Applying the Proposition we can reinsert the definition of $\Zlin$ and obtain an estimate for $i$ angular derivatives of $\otx$ in terms of $i-1$ angular derivatives of $\elin +\eblin$:

\begin{corollary}  \label{cor:5for4}
Consider the  solution $\Si$ of Theorem \ref{theo:mtheo}. We have for $n\geq 0$, any $i \in \mathbb{N}$ and any $\left(u\geq u_0,v\geq v_0\right)$ the estimate
\begin{align}
\int_{S^2_{u,v}} \sin \theta d\theta d\phi r^{8-n}  \Big|\mathcal{A}^{[i]}\slashed{\mathcal{D}}_2^\star \slashed{\nabla}_A \frac{\otx}{\Omega^2}\Big|^2 \nonumber \\ \lesssim \int_{S^2_{u,v_0}} \sin \theta d\theta d\phi \frac{|\mathcal{A}^{[i]} r \slashed{\mathcal{D}}_2^\star \Zlin|^2}{r^n\Omega^2} +\sup_v \int_{S^2_{u,v}} \sin \theta d\theta d\phi r^{6-n}  \Big|   \mathcal{A}^{[i]}\slashed{\mathcal{D}}_2^\star  \left(\elin + \eblin\right) \Big|^2  \nonumber \\
+ \int_{v_0}^v d\bar{v} \int_{S^2_{u,\bar{v}}} \sin \theta d\theta d\phi r^{5-n} \Omega^2  \Big|   \mathcal{A}^{[i]} \slashed{\mathcal{D}}_2^\star  \left(\elin + \eblin\right) \Big|^2 \, . \nonumber
\end{align}
\end{corollary}

This estimate is crucial as it allows us to estimate $i+1$ angular derivatives of  $\otx$ in terms of $i$ angular derivatives of $\elin$ and $\eblin$. We now apply Corollary \ref{cor:5for4} with $i=3$ and $n=2+\epsilon$ (to make the right hand side integrable) using the bounds of Proposition \ref{prop:etae}
to conclude
\begin{align} \label{dush}
\sup_{u,v} r^{2-\epsilon} \Big\| r^{-1} \cdot  \frac{1}{\Omega^2} \cdot \mathcal{A}^{[3]} r^2 \slashed{\mathcal{D}}_2^\star \slashed{\nabla}_A\otx\Big\|_{S^2_{u,v}}^2 \lesssim \textrm{right hand side of (\ref{ughb})} \, .
\end{align}
Note the factor of $\Omega^{-2}$. The estimate is clearly not optimal near infinity, as $2-\epsilon$ should be replaced by $4$. This will be achieved below.
\paragraph{Higher angular derivatives of  $\protect\xlin$}
We now write the Codazzi equation (\ref{ellipchi}) as
\begin{align} \label{eqhc} 
\mathcal{A}^{[i]}\slashed{\mathcal{D}}_2^\star \slashed{div} \Omega \xlin = -\frac{\Omega^2}{r} \mathcal{A}^{[i]}\slashed{\mathcal{D}}_2^\star \eblin - \mathcal{A}^{[i]} \Omega \plin - \frac{3}{2}\rho \mathcal{A}^{[i]} \xlin\Omega + \frac{1}{2} \mathcal{A}^{[i]} \slashed{\mathcal{D}}_2^\star \slashed{\nabla}_A \otx
\end{align}
and use the estimates available for all terms on the right hand side we will be able to conclude
\begin{proposition} \label{prop:5ca}
Consider the  solution $\Si$ of Theorem \ref{theo:mtheo}. We have for any $u\geq u_0$, $v \geq v_0$ the estimate
\begin{align} \label{qaz}
 \| r^{-1} \cdot \mathcal{A}^{[2]} r^2 \slashed{\mathcal{D}}_2^\star \slashed{div} \left(\Omega \xlin r^2\right)\|^2_{S^2_{u,v}}  \lesssim \textrm{right hand side of (\ref{ughb})}  \, ,
\end{align}
and
\begin{align} \label{qaz2}
\sup_{u\geq u_0} \int_{v_0}^\infty dv r^{-1-\epsilon} \| r^{-1}\cdot \mathcal{A}^{[2]} r^2 \slashed{\mathcal{D}}_2^\star \slashed{div} \left(\Omega \xlin r^2\right)\|^2_{S^2_{u,v}}  \lesssim \textrm{right hand side of (\ref{ughb})} \, .
\end{align}
Suppose now also that the initial norm $ \mathbb{F}_0^2\left[\Psilin\right]<\infty$. Then we can replace $\mathcal{A}^{[2]}$ by $\mathcal{A}^{[3]}$ in (\ref{qaz2}) provided we add the expression $\mathbb{F}_0^2\left[\Psilin\right]$ on the right. Assuming moreover that $\sup_v r^3 \| r^{-1} \mathcal{A}^{[3]} \Omega \plin r^3 \|^2_{S^2_{u_0,v}} + \mathbb{F}_0^2\left[\Psilin\right]<\infty$ we can replace $\mathcal{A}^{[2]}$ by $\mathcal{A}^{[3]}$ on the left in (\ref{qaz}), provided we add the latter expression on the right.
\end{proposition}
\begin{proof}
A weaker version of (\ref{qaz}), namely with weight $r^{-2-\epsilon}$ on the left hand side, follows directly from the identity (\ref{eqhc}) after applying the estimates (\ref{dush}), Proposition \ref{prop:angc}, Propositions \ref{prop:firstangularpsi2} and \ref{prop:nif} and \ref{prop:etae}. To optimise the weight near infinity one integrates the angular commuted transport equation $\Omega \slashed{\nabla}_4 \left(\xlin \Omega^{-1} r^2\right) = \alin r^2$ starting \emph{away} from the horizon, where the estimate has already been proven and uses the (twice angular commuted) flux bounds on $\alin$ in (\ref{fluxa}). To obtain the statement with $\mathcal{A}^{[3]}$ one follows the same argument, except that now one requires the flux of (the once angular commuted) Proposition \ref{prop:5dera} in the last step, which accounts for the additional term. The proof of (\ref{qaz2}) is similar.
\end{proof}
Revisiting again (\ref{eqhc}) immediately yields
\begin{corollary} \label{cor:im2}
Consider the  solution $\Si$ of Theorem \ref{theo:mtheo}. We have
\begin{align} \label{dushh}
\sup_{u,v} \Big\| r^{-1} \cdot \Omega^{-2} \cdot \mathcal{A}^{[2]} r^4 \slashed{\mathcal{D}}_2^\star \slashed{\nabla}_A \otx\Big\|_{S^2_{u,v}}^2 \lesssim \textrm{right hand side of (\ref{ughb})} \, .
\end{align}
Moreover, provided the initial norm $\sup_v r^3 \| r^{-1} \mathcal{A}^{[3]} \Omega \plin r^3 \|^2_{S^2_{u_0,v}} + \mathbb{F}_0^2\left[\Psilin\right]$ is finite, the estimate remains true if we replace $\mathcal{A}^{[2]}$ by $\mathcal{A}^{[3]}$ on the left and add the aforementioned initial norm on the right.
\end{corollary}
Proposition \ref{prop:5ca} without the improvement mentioned in the second part of its statement is already sufficient to prove the analogue of Proposition \ref{prop:fluv}, i.e.~flux estimates on constant $u$-hypersurfaces for the quantities $\rlin$ and $\blin$:
\begin{proposition} \label{prop:fluu}
Consider the  solution $\Si$ of Theorem \ref{theo:mtheo}. We have the following flux estimates
\begin{align}
\sup_u \int_{v_0}^\infty dv \frac{1}{r^2} \| r^{-1} \cdot \mathcal{A}^{[3]}r^2  \slashed{\mathcal{D}}_2^\star \slashed{\mathcal{D}}_1^\star\left(r^3 \rlin,r^3 \slin\right) \|_{S^2_{u,v}}^2  \lesssim \textrm{right hand side of (\ref{ughb})} \, ,
\end{align}
\begin{align}
\sup_u \int_{v_0}^\infty dv\frac{1}{r^2} \| r^{-1} \cdot \mathcal{A}^{[4]}r  \slashed{\mathcal{D}}_2^\star \left(r^3 {\blin}\Omega\right) \|_{S^2_{u,v}}^2  \lesssim  \textrm{right hand side of (\ref{ughb})} \, .
\end{align}
\end{proposition}
\begin{proof}
The estimates follow from the identities 
\begin{align}
\mathcal{A}^{[i]} r^2  \slashed{\mathcal{D}}_2^\star \slashed{\mathcal{D}}_1^\star\left(r^3 \rlin,r^3 \slin\right) &= \mathcal{A}^{[i]} (r^5 \Plin) +3M\Omega r \left(\mathcal{A}^{[i]}\xlin - \mathcal{A}^{[i]}\xblin\right). \nonumber
\end{align}
\begin{align} \label{betid}
\mathcal{A}^{[i]} \slashed{\mathcal{D}}_2^\star ({\blin} \Omega)= \mathcal{A}^{[i]} (\Omega{\plin}) - \frac{3}{2} \rho\mathcal{A}^{[i]} (\Omega{\xlin})
\end{align}
and the estimate on $\xlin$ obtained in (\ref{qaz2}) as well as Corollary \ref{cor:crucialchibar}.
\end{proof}

\subsubsection{Refined estimates for higher angular derivatives}
\label{oxipolyrefined}

We finally prove some refined estimates for higher angular derivatives which will eventually allows us to prove the estimate (\ref{ughbo}), i.e.~to show the propagation of the $\mathbb{D}^{[5]}$-norm in Section \ref{sec:corpro} below.

\paragraph{Higher angular derivatives of $\slashed{\nabla}_3 \protect\otx$ and $\slashed{\nabla}_3 (\protect\xlin \Omega)$}
Now that we have estimated five angular derivatives of $\xlin$ we can apply the identity (\ref{fipo2}) with $i=4$ to estimate five angular derivatives of $\eblin$. This is Proposition \ref{prop:etae2} below. To estimate five derivatives of $\elin$ we first estimate $ \mathcal{A}^{[4]}\Omega^{-1} \slashed{\nabla}_3 \otx$ and then revisit (\ref{eqhc}) with $i=2$ and one $\Omega \slashed{\nabla}_3^{-1}$ derivative applied to it to obtain a bound on $ \mathcal{A}^{[4]}\Omega^{-1} \slashed{\nabla}_3 \xlin$. Finally, revisiting (\ref{fipo}) now with $i=4$ will then control $5$ derivatives of $\elin$.

\begin{proposition} \label{prop:otxe}
Consider the  solution $\Si$ of Theorem \ref{theo:mtheo}. We have for $i=2$ and $i=3$ the estimate
\begin{align}
\sup_{u,v} \Big\| r^{-1} \cdot r  \slashed{\nabla}_3 \left(\mathcal{A}^{[i]} r^2 \slashed{\mathcal{D}}_2^\star \slashed{\nabla}_A \frac{r\otx }{\Omega^2} \right) \Big\|^2_{S^2_{u,v}} \lesssim  \textrm{right hand side of (\ref{ughb})}  .
\end{align}
Moreover,
\begin{align} \label{qaz2b}
\sup_{u,v} r^{-\epsilon} \| r^{-1}\cdot  \Omega^{-1} \slashed{\nabla}_3 \left(\mathcal{A}^{[2]}  r^2 \slashed{\mathcal{D}}_2^\star \slashed{div} \left(\Omega \xlin r^2\right)\right)\|^2_{S^2_{u,v}}  \lesssim \textrm{right hand side of (\ref{ughb})} \, .
 \end{align}
\end{proposition}

\begin{proof}
Starting from (\ref{dbtc})  and using (\ref{uray}), (\ref{dtcb}), (\ref{propeta}) and (\ref{Bianchi4}) we derive
\begin{align}
\Omega \slashed{\nabla}_4 \left(r \Omega^{-1} \slashed{\nabla}_3 \left[\mathcal{A}^{[i]} r^2 \slashed{\mathcal{D}}_2^\star \slashed{\nabla}_A \frac{r\otx}{\Omega^2} \right] \right) + \frac{2M}{r^2} \left(r \Omega^{-1} \slashed{\nabla}_3 \left[\mathcal{A}^{[i]} r^2 \slashed{\mathcal{D}}_2^\star \slashed{\nabla}_A \frac{r\otx }{\Omega^2} \right] \right) \nonumber \\
= \frac{8M}{r^2} \mathcal{A}^{[i]} r^2 \slashed{\mathcal{D}}_2^\star \left(\elin +\eblin\right) - 4 \mathcal{A}^{[i]} r^3  \slashed{\mathcal{D}}_2^\star \slashed{\nabla}_A \rlin + \left(\frac{2M}{r} - \Omega^2\right)\mathcal{A}^{[i]} r^2 \slashed{\mathcal{D}}_2^\star \slashed{\nabla}_A \frac{\otx}{\Omega^2} \, 
\end{align}
from which the redshift is manifest. Since for $\Si$ the quantities $\elin$ and $\eblin$ are not expected to decay, we multiply the above by $\Omega^2 \cdot  \left(r \Omega^{-1} \slashed{\nabla}_3 \left[\mathcal{A}^{[i]} r^2 \slashed{\mathcal{D}}_2^\star \slashed{\nabla}_A \frac{r\otx}{\Omega^2} \right] \right)$. Using the flux of Proposition \ref{prop:fluu} and $L^\infty_{u,v}L^2 \left(S^2_{u,v}\right)$ bounds on $\elin$ and $\eblin$ of Proposition \ref{prop:etae} as well as the bound (\ref{dush}) we conclude the first estimate. The second now follows from (\ref{eqhc}) with $i=2$ and $\frac{1}{\Omega}\slashed{\nabla}_3$ applied to this identity.
\end{proof}
The above bounds are still not optimal in terms of the $r$-weights. However, away from the horizon, we can simply integrate the commuted transport equation $\Omega \slashed{\nabla}_4 \left(\xlin \Omega^{-1} r^2\right) = \alin r^2$ (both in $u$ and $v$ or just in $v$) to optimize the weights in $r$. For later purposes we only state the flux bound which can be derived along those lines.

\begin{proposition}
Consider the  solution $\Si$ of Theorem \ref{theo:mtheo}. We have the flux bound
\begin{align}
\sup_{v\geq v_0} \int_{u_0}^\infty du  \| r^{-1}\cdot  \Omega^{-1} \slashed{\nabla}_3 \left(\mathcal{A}^{[2]}  r^2 \slashed{\mathcal{D}}_2^\star \slashed{div} \left(\Omega \xlin r^2\right)\right)\|^2_{S^2_{u,v}}  \lesssim \textrm{right hand side of (\ref{ughb})} + \mathbb{F}_0^2\left[\Psilin\right] \, ,
\end{align}
provided $\mathbb{F}_0^2\left[\Psilin\right] <\infty$.
\end{proposition}

\paragraph{Top order angular derivatives of $\protect\elin$ and $\protect\eblin$} 
\label{had-sec}
$\phantom{X}$\\
With the results above we immediately obtain estimates for the highest derivatives of $\elin$ and $\eblin$:

\begin{proposition} \label{prop:etabe2}
Consider the  solution $\Si$ of Theorem \ref{theo:mtheo}. For any $u\geq u_0$ and $v\geq v_0$ we have 
\begin{align}
\int_{S^2_{u,v}} \sin \theta d\theta d\phi \cdot  r^{6} |\mathcal{A}^{[4]} \slashed{\mathcal{D}}_2^\star \eblin |^2  \lesssim \textrm{right hand side of (\ref{ughb})}.
\end{align}
\end{proposition}

\begin{proof}
Apply (\ref{fipo2}) with $i = 4$  and use previous bounds.
\end{proof}

\begin{proposition} \label{prop:etae2}
Consider the  solution $\Si$ of Theorem \ref{theo:mtheo}. For any $u\geq u_0$ and $v\geq v_0$ we have the estimate
\begin{align}
\int_{S^2_{u,v}} \sin \theta d\theta d\phi \cdot  \left[ r^{4} |\mathcal{A}^{[4]} \slashed{\mathcal{D}}_2^\star {\elin} |^2 \right]  
\lesssim \textrm{right hand side of (\ref{ughb})} .
\end{align}
For $\elin$ we have in addition the flux estimate
\begin{align} \label{fluh2}
\int_{u_0}^\infty du \int_{S^2_{u,v}} \sin \theta d\theta d\phi  \Omega^2 r^4 |\mathcal{A}^{[4]} \slashed{\mathcal{D}}_2^\star {\elin} |^2 \lesssim \textrm{right hand side of (\ref{ughb})} . \, 
\end{align}
\end{proposition}
\begin{proof}
Apply (\ref{fipo}) with $i=4$ and use previous bounds.
\end{proof}

\begin{corollary} \label{cor:hot}
The estimates (\ref{beq1}) and (\ref{omeq1}) hold also for $i=4$.
\end{corollary}
\begin{proof}
Repeat the proof of the last two estimates of Proposition \ref{prop:meco1} now using the higher order estimates on $\elin$ and $\eblin$ above. 
\end{proof}

\subsubsection{Proof of (\ref{ughb}) and Corollary \ref{cor:pwe}} \label{sec:corpro}
The statement (\ref{ughb}) in the boundedness theorem now follows from (\ref{plo1}) and Corollary \ref{cor:comcorj} with $i=2$ for the $\Ylin$-part in the $\mathbb{D}\left[\Ylin,\Zlin\right]$-norm, from Proposition \ref{prop:zestimate} applied with $n=2+\epsilon$ for the $\Zlin$-part in the $\mathbb{D}\left[\Ylin,\Zlin\right]$-norm and finally Proposition \ref{prop:otxe} applied with $i=2$ for the remaining part in (\ref{D5def}).

For Corollary \ref{cor:pwe} one uses the classical Sobolev embedding on $S^2_{u,v}$ in conjunction with the estimates of Corollary \ref{cor:crucialchibar} for $\hat{\underline{\chi}}$, Proposition \ref{prop:angc} for $\xlin$, Proposition \ref{prop:etae} for $\elin, \eblin$ (recall that the $\ell=0, 1$ modes vanish for $\Si^\prime$), Proposition \ref{prop:meco1} for all metric quantities, Proposition \ref{prop:trxbflux} for $\otxb$, Corollary \ref{cor:im2} for $\otx$, Proposition \ref{prop:bc4} for $\rlin,\slin, \blin, \bblin$ and Corollary \ref{cor:1da} for $\alin, \ablin$. The pointwise
bounds 
for $\Si$ itself follow from the identity $\Si=\Si^\prime+\mathscr{K}_{\mathfrak{m},s_i}$,
together with the fact that, as is checked by direct computation,
 reference Kerr solutions $\mathscr{K}_{\mathfrak{m},s_i}$  indeed
satisfy the boundedness property of
the Corollary with right hand side controlled by a constant depending only
on the parameters $\mathfrak{m}, s_i$.

\section{Proof of Theorem~\ref{theo:mtheod}} \label{sec:gest1} 
In this final section of the paper, we turn to the proof of
Theorem \ref{theo:mtheod}. The reader can again
refer to the overview in Section \ref{DFNGintro}.

In Section~\ref{sec:controlgauge}, we shall show
that the pure gauge
solution $\Gf$, and thus also $\Sf$, satisfies a uniform boundedness statement and
an asymptotic flatness statement.
This gives statement 1.~of the Theorem~\ref{theo:mtheod}.
In Section~\ref{kiedwILEDkiedw}, we obtain statement 2.~of the theorem  concerning
integrated local energy decay.
Finally, we prove the final statement~3.~of the theorem concerning polynomial decay
in Section~\ref{sec:polyfinal}.

\subsection{Boundedness of the pure gauge solution $\protect\Gf$} \label{sec:controlgauge}
Let $\Gf$ denote the pure gauge solution in the statement
of Theorem \ref{theo:mtheod}.  The goal of this section is to prove the boundedness
of $\Gf$, from which a similar statement will follow for $\Sf=\Si+\Gf$,
in view of Theorem~\ref{theo:mtheo} applied to $\Si$.

We will begin in Section~\ref{horizfluxesinfour} below with certain preliminary
estimates for $\Sf$ on the horizon. We shall then use these in Section~\ref{elegxwelegxw} 
to infer estimates for the function $f$ defining $\Gf$.  
The precise boundedness statements that follow will be given in Section~\ref{kaibkaiaf}.

To distinguish between the quantities $(\ref{scollect})$ 
associated to $\Si$ or $\Sf$ we agree on the 
following {\bf convention}: The geometric quantities of the solution $\Si$ will from now on be denoted with an additional $\left[{\Si}\right]$ next to them while those of $\Sf$ will appear without any additional notation, unless there is potential confusion, in which case we add $\left[\Sf\right]$. The general rationale is to always write an estimate for a quantity of $\Sf$ on the left in terms of initial quantities of $\Si$ on the right.

For the geometric quantities associated to the pure gauge solution
$\Gf$, we shall always add $\left[\Gf\right]$.

\subsubsection{Decay bounds on the ingoing shear $\protect\xblin$ at the horizon}\label{horizfluxesinfour}

Recall from Section~\ref{sec:bndchi} 
of the proof of Theorem~\ref{theo:mtheo} that the ``first'' obstruction
to proving decay for $\Si$ arose from the quantitity $\xblin[\Si]$.  We will show in
this section  that our choice of $\Gf$ ensures that $\xblin=\xblin[\Sf]$ does indeed
decay along the event horizon $\mathcal{H}^+$.
The estimates obtained will then allow us in the next section to infer bounds for the
gauge function $f$ defining $\Gf$.

First some preliminary remarks:
We already note that the pure gauge solution $\Gf$ has vanishing linearised shear 
$\xlin[\Gf]=0$. Therefore, in addition to the estimates on the gauge invariant quantities, also the estimates on $\xlin$ proven in Section \ref{sec:chihatest} remain valid as stated for $\xlin=\xlin\left[\Sf\right]$. We also observe that $\Omega^{-2}\otxb [\Gf]=0$, 
$\Omega^{-1}\xblin[\Gf]=0$ on $S^2_{\infty,v_0}$
and thus
\begin{equation}
\label{kaloilogariasmoi}
\Omega^{-2}\otxb [\Sf]=\Omega^{-2}\otxb [\Si], \qquad 
\Omega^{-1}\xblin[\Sf]=\Omega^{-1}\xblin[\Si] \qquad {\rm on}\qquad S^2_{\infty,v_0}.
\end{equation}

We have the following  flux bounds on the horizon:
\begin{proposition} \label{lem:2dc} 
On the horizon $\mathcal{H}^+$, the geometric quantities of $\Sf$ in Theorem \ref{theo:mtheod} satisfy for $i\geq 3$
\begin{align}
 \Big\| \mathcal{A}^{[i]} \Omega^{-1} \xblin \Big\|^2_{S^2_{\infty,V}}
+ \int_{v_0}^V d\bar{v}  \Big\| \mathcal{A}^{[i]} \Omega^{-1} \xblin\Big\|_{S^2_{\infty,\bar{v}}}^2  
\lesssim   \Big\| \mathcal{A}^{[i]}  \Omega^{-1}\xblin\left[{\Si}\right] \Big\|_{S^2_{\infty,v_0}}^2 
 + \mathbb{F}^{i-3, T, \slashed{\nabla}}_0 \left[\Psilin, \plin, \alin \right] \, , \nonumber
\end{align}
\begin{align}
 \Big\|  \mathcal{A}^{[i-2]} \slashed{\mathcal{D}}_2^\star \slashed{\nabla}_A \frac{\otxb}{\Omega^2} \Big\|^2_{S^2_{\infty,V}}
+ \int_{v_0}^V d\bar{v}   \Big\| \mathcal{A}^{[i-2]} \slashed{\mathcal{D}}_2^\star \slashed{\nabla}_A \frac{\otxb}{\Omega^2} \Big\|_{S^2_{\infty,\bar{v}}}^2 
\nonumber \\ \lesssim   \Big\|  \mathcal{A}^{[i-2]} \slashed{\mathcal{D}}_2^\star \slashed{\nabla}_A \frac{\otxb \left[{\Si}\right] }{\Omega^2} \Big\|^2_{S^2_{\infty,v_0}}+  \mathbb{F}^{i-3,T, \slashed{\nabla}}_0 \left[\Psilin \right] \, .
\end{align}
\end{proposition}

\begin{proof}
Restricting the angular commuted (\ref{chih3b}) to the horizon we have on $\mathcal{H}^+$ 
\begin{equation} \label{po}
\Omega \slashed{\nabla}_4 \left(\mathcal{A}^{[i]} \frac{\xblin}{\Omega}\right) + \frac{1}{2M}  \mathcal{A}^{[i]}  \frac{\xblin}{\Omega} = \frac{\Omega}{2M} \mathcal{A}^{[i]}  \xlin - 2 \mathcal{A}^{[i]}   \slashed{\mathcal{D}}_2^\star \eblin \, .
\end{equation}
Hence in particular
\begin{align} \label{orig}
\frac{1}{2} \partial_v \Big| \mathcal{A}^{[i]}  \frac{\xblin}{\Omega}\Big|^2 + \frac{1}{4M} \Big| \mathcal{A}^{[i]}  \frac{\xblin}{\Omega}\Big|^2 \lesssim  | \mathcal{A}^{[i]}  ( \Omega\xlin )|^2 + | \mathcal{A}^{[i]}  \slashed{\mathcal{D}}_2^\star \eblin |^2 \, .
\end{align}
Taking into account   $(\ref{kaloilogariasmoi})$
on the sphere $S^2_{\infty,v_0}$ and that the quantities therefore agree for $\Si$ and $\Sf$ on that sphere, integration yields
\begin{align}
 \Big\| \mathcal{A}^{[i]}  \frac{ \xblin }{\Omega} \Big\|^2_{S^2_{\infty,V}}
+ \int_{v_0}^V d\bar{v}  \Big\|\mathcal{A}^{[i]} \frac{ \xblin }{\Omega} \Big\|_{S^2_{\infty,\bar{v}}}^2  \nonumber \\
\lesssim  \Big\|\mathcal{A}^{[i]}   \frac{ \xblin \left[\Si\right]}{\Omega} \Big\|_{S^2_{\infty,v_0}}^2
+ \int_{v_0}^V d\bar{v}  \left[  \|\mathcal{A}^{[i]}  (\Omega \xlin)\|_{S^2_{\infty,\bar{v}}}^2 + \| \mathcal{A}^{[i]} \slashed{\mathcal{D}}_2^\star \eblin \|_{S^2_{\infty,\bar{v}}}^2\right]  .\nonumber
\end{align}
Observing that on the horizon 
$\eblin\left[\Sf\right]=-\elin\left[\Sf\right]=-\eblin\left[\Sf\right]$ and $\xlin\left[\Sf\right]=\xlin\left[\Si\right]$ and using Proposition~\ref{prop:angularbetahoz} (recall $\slashed{div} \,\xlin=-\blin$ holds on $\mathcal{H}^+$ from (\ref{ellipchi})) we obtain the first estimate.

For the second, we proceed similarly. We write (\ref{dbtc}) on the horizon $\mathcal{H}^+$ as 
\begin{align}
\label{asbaloumekiedw}
\partial_v \left[\mathcal{A}^{[i-3]}\slashed{\mathcal{D}}_2^\star \slashed{\nabla}_A 
\frac{\otxb}{\Omega^2} \right] + \frac{1}{2M}\mathcal{A}^{[i-3]} \slashed{\mathcal{D}}_2^\star \slashed{\nabla}_A \frac{ \otxb }{\Omega^2} = 4\mathcal{A}^{[i-3]}\slashed{\mathcal{D}}_2^\star \slashed{\nabla}_A \rlin \, ,
\end{align}
where we have used 
that $\Olin \left[\Sf\right] \left(\infty,v,\theta,\phi\right)=0$ and 
$\slashed{div} \elin + \rlin= 0$ on $\mathcal{H}^+$ (for both $\Si$ and $\Sf$). 
Integrating the identity
$(\ref{asbaloumekiedw})$ as in the previous case yields the second estimate after applying 
Proposition~\ref{prop:angularbetahoz} to control the flux on the right hand side.
\end{proof}

\begin{corollary} \label{cor:cb4a}
We also have for $i\geq 3$
\begin{align}
 \int_{v_0}^\infty d\bar{v}  \Big\|\Omega \slashed{\nabla}_4 \mathcal{A}^{[i]}  \frac{\xblin}{\Omega}\Big| \Big\|_{S^2_{\infty,\bar{v}}}^2  
\lesssim  \Big\|\mathcal{A}^{[i]} \frac{\xblin\left[\Si\right]}{\Omega} \Big\|_{S^2_{\infty,v_0}}^2  +\mathbb{F}^{i-3,T,\slashed{\nabla}}_0 \left[\Psilin, \plin, \alin \right]  \, .
\end{align} \nonumber
\end{corollary}
\begin{proof}
Follows directly from (\ref{po}) recalling that  $\eblin\left[\Sf\right]=-\elin \left[\Sf\right]$ along $\mathcal{H}^+$ and using the flux bounds of Proposition~\ref{lem:2dc} and 
Proposition~\ref{prop:angularbetahoz}.
\end{proof}

\begin{corollary} \label{cor:cb4}
We have in addition the $L^\infty_v L^2 \left(S^2_{\infty,v}\right)$-bound for $i\geq 2$
\begin{align}
 \sup_v \Big\|\Omega \slashed{\nabla}_4 \mathcal{A}^{[i]}  \frac{\xblin}{\Omega} \Big\|_{S^2_{\infty,v}}^2 \lesssim \Big\|\mathcal{A}^{[i]}  \frac{\xblin\left[\Si\right]}{\Omega}  \Big\|_{S^2_{\infty,v_0}}^2 + \sup_v   \| r^{-1/2} \cdot \mathcal{A}^{[i-2]}{\plin} \Omega r^3\|^2_{S^2_{u_0,v}}
 + \mathbb{F}^{i-2,T,\slashed{\nabla}}_0 \left[\Psilin, \plin, \alin \right]  . \nonumber
\end{align}
\end{corollary}

\begin{proof}
Revisit (\ref{po}) and use the $L^\infty_{u,v}$-bound of Proposition~\ref{lem:2dc}, the 
$L^\infty_{u,v}$-bound on $ \mathcal{A}^{[i]} \xlin$ of Proposition \ref{cor:chf} and the $L^\infty_{u,v}$ bound on $\mathcal{A}^{[i]}\slashed{\mathcal{D}}_2^\star \eblin$ arising from the fact that on the event horizon we have from (\ref{ellipchi})
\begin{align}
\Omega \mathcal{A}^{[2]}\xlin = \Omega\slashed{\mathcal{D}}_2^\star \slashed{div}\xlin = -\Omega\slashed{\mathcal{D}}_2^\star  \blin = -\Omega \plin + \frac{3}{2} \rho \Omega \xlin
\end{align}
and
\begin{align} \label{eac}
\| \mathcal{A}^{[2]} \slashed{\mathcal{D}}_2^\star { \elin }\|^2 \leq \|\slashed{\mathcal{D}}_2^\star \slashed{\mathcal{D}}_1^\star \slashed{\mathcal{D}}_1 { \elin }\|^2 \lesssim \|\slashed{\mathcal{D}}_2^\star \slashed{\mathcal{D}}_1^\star \left(-\rlin,\slin \right)\| \leq \|\Psilin\|^2 + \|\Omega \xlin \|^2
\end{align}
using (\ref{uid3}) and the gauge condition (\ref{fchoi}). Finally, use $\| \Psilin \|_{S^2_{\infty,v_0}} \lesssim \mathbb{F}_0 \left[\Psilin\right]$.
\end{proof}

If we use the polynomial decay estimates of Proposition \ref{prop:decRW} and Proposition \ref{prop:refd} we also have using Proposition~\ref{lem:2dc} in conjunction with a pigeonhole principle 
\begin{corollary}\label{cor:fnewg}
We have along the event horizon $\mathcal{H}^+$ the decay estimate
\begin{align} 
 \Big\| \mathcal{A}^{[3]} \frac{\xblin}{\Omega} \Big\|^2_{S^2_{\infty,v}} + 
  \Big\| \mathcal{A}^{[2]} \Omega \slashed{\nabla}_4 \frac{\xblin}{\Omega} \Big\|^2_{S^2_{\infty,v}} \lesssim \frac{1}{v^2} \left ( \Big\| \mathcal{A}^{[3]} \frac{\xblin\left[\Si\right]}{\Omega} \Big\|^2_{S^2_{u_0,v}} + \mathbb{F}_0^{2,T}\left[\Psilin\right] + \mathbb{F}_0 \left[\Psilin, \Psilinb, \plin, \pblin, \alin, \ablin \right]    \right). \nonumber
 \end{align}
\end{corollary}
\begin{proof}
For the bound on $\xblin$ we combine Proposition~\ref{lem:2dc} with a simple dyadic argument. In particular, we use the fact that the fluxes appearing on the horizon on the right hand side of (\ref{orig}) (after integration) satisfy the polynomial decay estimates of Proposition \ref{prop:decbh}. To derive the bound for $\slashed{\nabla}_4 \xblin$ we revisit the identity (\ref{po}) with $i=2$ and show that all other terms have the desired decay. The bound for two angular derivatives of $\xblin$ has just been obtained. From Proposition \ref{prop:psibs2d2} and the identity (\ref{psiidh}) restricted to the horizon we see we have the decay bound for two angular derivatives of $\xlin$ on $S^2_{\infty,v}$. Finally, to estimate the term involving three derivatives of $\eblin = -{\elin} = -\eblin\left[\Si\right]$ in (\ref{po}) we use the identity (\ref{eac}) and the fact that an estimate for $\Psi$ on the horizon follows from Proposition \ref{prop:decRW}.
\end{proof}

\subsubsection{Controlling the gauge function}\label{elegxwelegxw}
With Proposition~\ref{lem:2dc} 
controlling $\xblin$ on the horizon (from data in $\Si$) and Corollary \ref{cor:crucialchibar} controlling  $\xblin\left[\Si\right]$ on the horizon (also from data in $\Si$) we can infer boundedness of the gauge function:
\begin{proposition} \label{cor:gfe2} 
The gauge function $f=f\left(v,\theta,\phi\right)$ associated with the pure gauge solution $\Gf$ in Theorem \ref{theo:mtheod} via Proposition \ref{prop:hozrgauge} satisfies for $i=5$
\begin{align}  \label{piuy}
\int_{S^2} \sin \theta d\theta d\phi  |\mathcal{A}^{[i]} r^2 \slashed{\mathcal{D}}_2^\star  \slashed{\nabla}_A f |^2 
\lesssim  \Big\|\mathcal{A}^{[i]}  \Omega^{-1} \xblin \left[\Si\right] \Big\|_{S^2_{\infty,v_0}}^2 +  \textrm{right hand side of (\ref{ughb})} 
\end{align}
\begin{align}  \label{piuy2}
\int_{S^2} \sin \theta d\theta d\phi  |\mathcal{A}^{[i-1]} r^2 \slashed{\mathcal{D}}_2^\star  \slashed{\nabla}_A \partial_v f |^2 
\lesssim  \Big\|\mathcal{A}^{[i-1]}  \Omega^{-1} \xblin \left[\Si\right] \Big\|_{S^2_{\infty,v_0}}^2 +  \textrm{right hand side of (\ref{ughb})} .
\end{align}
We also have the flux bound
\begin{align} \label{piuy3}
\int_{v_0}^\infty d\bar{v} \int_{S^2} \sin \theta d\theta d\phi  |\mathcal{A}^{[i-1]} r^2 \slashed{\mathcal{D}}_2^\star  \slashed{\nabla}_A \partial_v f |^2 
\lesssim  \Big\|\mathcal{A}^{[i-1]} \Omega^{-1} \xblin \left[\Si\right]  \Big\|_{S^2_{\infty,v_0}}^2 +  \textrm{right hand side of (\ref{ughb})} 
\end{align}
and the decay bound
\begin{align} \label{piuy4}
v^2 \int_{S^2} \sin \theta d\theta d\phi \ |\mathcal{A}^{[2]} r^2 \slashed{\mathcal{D}}_2^\star  \slashed{\nabla}_A \partial_v f |^2 \lesssim \Big\|\mathcal{A}^{[3]} \Omega^{-1} \xblin \left[\Si\right] \Big\|_{S^2_{\infty,v_0}}^2 +  \textrm{right hand side of (\ref{ughb})} .
\end{align}
\end{proposition}

\begin{proof}
We have from Lemma \ref{lem:exactsol}
\begin{equation} \label{lem32d}
r \cdot \Omega^{-1} r \, \xblin \left[\Sf\right]   - r \cdot \Omega^{-1} r \, \xblin \left[\Si\right] = r \cdot \Omega^{-1} r \, \xblin \left[\Gf\right] = -2 r^2  \slashed{\mathcal{D}}_2^\star  \slashed{\nabla}_A f \, ,
\end{equation}
which when restricted to the horizon $u=\infty$ (where $r=2M$) leads to (\ref{piuy}) after using Proposition~\ref{lem:2dc} and Corollary \ref{cor:crucialchibar}. For the second estimate, we commute the defining equation (\ref{odef}) with $\mathcal{A}^{[i-1]} r^2 \slashed{\mathcal{D}}_2^\star  \slashed{\nabla}_A$, and estimate $f$ by (\ref{piuy}) and the quantity $\mathcal{A}^{i-1} r \slashed{\mathcal{D}}_2^\star \left(\elin + \eblin\right)$ from Proposition \ref{prop:etae2} and (the twice angular commuted $S^2_{\infty,v}$-estimate of) Proposition \ref{lem:gaugeta}. For the third estimate we use again Lemma \ref{lem:exactsol} to conclude that on the horizon $\mathcal{H}^+$
\[
\Omega \slashed{\nabla}_4 \left(r\xblin \left[\Sf\right] \Omega^{-1}\right) - \Omega \slashed{\nabla}_4 \left(r \xblin \left[\Si\right] \Omega^{-1}\right) = -\frac{2}{2M}r^2  \slashed{\mathcal{D}}_2^\star  \slashed{\nabla}_A \partial_v f \, .
\]
The flux estimates of Corollary \ref{cor:cb4a} (with $i=4$) and Corollary \ref{cor:dog} produce (\ref{piuy3}). Combining Proposition \ref{prop:decdyad} with Corollary \ref{cor:fnewg} yields the bound (\ref{piuy4}).
\end{proof}

\subsubsection{Boundedness of the pure gauge and horizon-renormalised solution}\label{kaibkaiaf}

Combining the estimates of Proposition \ref{cor:gfe2} with Lemma \ref{lem:exactsol} we can easily deduce the uniform boundedness of $\Gf$ in Theorem \ref{theo:mtheod} as well as deduce a uniform boundedness statement for  $\Sf$ from the estimate on $\Si$ and $\Gf$:
\begin{proposition} \label{prop:bndnew}
The curvature components $\rlin \, , \, \slin\, , \,  \blin\, , \,  \bblin$ of the pure gauge solution $\Gf$  in Theorem \ref{theo:mtheod} satisfy the same boundedness estimates as these quantities for $\Si$ in Proposition \ref{prop:bc4}, provided the term 
\[ 
\big\|\mathcal{A}^{[5]}\Omega^{-1} \xblin  \left[\Si\right] \big\|_{S^2_{\infty,v_0}}^2
\]
is added on all right hand sides of that Proposition. Furthermore, the Ricci and metric
coefficients of the solution $\Gf$ satisfy for all $u$ and $v$:
\begin{align}
 r\| r^{-1} \cdot \mathcal{A}^{[5]} r \slashed{\mathcal{D}}_2^\star \eblin \, \|_{S^2_{u,v}}+ r^2\| r^{-1} \cdot \mathcal{A}^{[2]} r \slashed{\mathcal{D}}_2^\star \eblin \, \|_{S^2_{u,v}}  
+  r \| r^{-1} \cdot \mathcal{A}^{[4]} r \slashed{\mathcal{D}}_2^\star \elin \, \|_{S^2_{u,v}} \nonumber \\
+ \| r^{-1} \cdot \mathcal{A}^{[5]} r \Omega^{-1} \xblin \|_{S^2_{u,v}} +\Big\| r^{-1} \cdot r^3 \mathcal{A}^{[3]} \slashed{\mathcal{D}}_2^\star \slashed{\nabla}_A \Omega^{-2} \otxb \Big\|^2_{S^2_{u,v}}  
+\Big\| r^{-1} \cdot \Omega^{-2} \mathcal{A}^{[2]} r^4 \slashed{\mathcal{D}}_2^\star \slashed{\nabla}_A\otx \Big\|_{S^2_{u,v}}^2 \nonumber \\
\Big\| r^{-1} \cdot \sqrt{r} \mathcal{A}^{[3]} r^2 \slashed{\mathcal{D}}_2^\star \slashed{\nabla} \frac{\glinto}{\sqrt{\slashed{g}}} \Big\|_{S^2_{u,v}}^2 +  \| r^{-1} \cdot \sqrt{r} \mathcal{A}^{[5]} \glinh \, \Big\|_{S^2_{u,v}}^2+r^{1-\epsilon} \Big\| r^{-1} \cdot \mathcal{A}^{[4]} r \slashed{\mathcal{D}}_2^\star \bmlin \, \Big\|_{S^2_{u,v}}^2 \nonumber \\
  \lesssim  \big\|\mathcal{A}^{[5]}\Omega^{-1} \xblin  \left[\Si\right] \big\|_{S^2_{\infty,v_0}}^2 +  \textrm{right hand side of (\ref{ughb})} \, .
 \end{align}
\end{proposition}

\begin{proof}
Use Lemma \ref{lem:exactsol} in conjunction with Corollary \ref{cor:gfe2} and
\begin{itemize}
\item Propositions \ref{prop:etae} and \ref{prop:etabe2} for $\eblin$,
\item Propositions \ref{prop:etae} and \ref{prop:etae2} for $\elin$,
\item Corollary \ref{cor:crucialchibar} for $\otx$,
\item Proposition \ref{prop:trxbflux} for $\otxb$,
\item Corollary \ref{cor:im2} for $\otx$,
\item Proposition \ref{prop:meco1} for the metric coefficients. Note that the initial terms appearing in (\ref{nirs}) and (\ref{nirs2}) can be estimated by  the right hand side of (\ref{ughb}) using the round sphere condition (\ref{rsc}) and (\ref{rscm}).
\end{itemize}
\end{proof}
We leave stating the estimate for five angular derivatives of $\otx$ arising from Corollary \ref{cor:im2} and more refined estimates for the metric coefficients to the reader. Note that the estimate for $\xlin$ is unchanged as $\Gf$ has $\xlin=0$.

In view of $\Sf = \Si + \Gf$ and the way the above bounds were derived we immediately conclude:
\begin{corollary} \label{cor:bndnew}
The estimates of Proposition \ref{prop:bndnew} hold also for the solution $\Sf$.
\end{corollary}

\begin{remark} \label{rem:bndnew}
Proposition \ref{prop:bndnew} and Corollary \ref{cor:bndnew} can be paraphrased by saying that the solution in the future gauge $\Sf=\Si + \Gf$ satisfies the same boundedness estimates as the 
solution  $\Si$ in the initial data gauge. In particular, there is no loss of derivatives at the level of
flux bounds. It is important to note, 
however, that the estimates we obtain for five angular derivatives of $\eblin$, $\bmlin$ and $\Olin$) are slightly weaker in terms of their $r$-decay towards null infinity than those for $\elin\left[\Si\right]$, $\bmlin \left[\Si\right]$ and $\Olin \left[\Si\right]$. This is because the estimate establishing decay of $f_v$ (\ref{piuy3}) is not optimal in terms of regularity. Compare the estimates (\ref{piuy2}) and (\ref{piuy4}).
\end{remark}

\subsection{Integrated local energy decay}\label{kiedwILEDkiedw}
We now turn to show 
integrated decay statements for the quantities
associated to $\Sf$.

Recall from our comments above that in addition to the estimates for the gauge-invariant quantities,
the results of Section~\ref{sec:chihatest}, in particular the integrated decay statement
Proposition~\ref{prop:chiall}, remains valid for~$\xlin\left[\Sf\right]$.

The first quantity for which one could not obtain integrated decay in the initial data normalised
gauge was the quantity~$\xblin\left[\Si\right]$. In Section~\ref{epitelousdecay} 
below, we will  succeed to obtain an integrated local energy
decay statement for~$\xblin\left[\Sf\right]$, obtaining also a similar estimate for 
 $\slashed\nabla_4\xblin$ in Section~\ref{kaistotes}. 
 We will then finally unravel the decay
 hierarchy, proving successively integrated local energy decay for 
  $\rlin$, $\slin$ and $\bblin$ in Section~\ref{tamesea}, 
  $\elin$ and $\eblin$ in Section~\ref{taetaria}   and
  $\otxb$ in Section~\ref{kaigiaautoedw}.  
  Finally, we will obtain various refined statements for higher angular
  derivatives in Section~\ref{polyrefined} which allow to infer integrated
  local energy decay for $\blin$.

\subsubsection{Integrated decay for angular derivatives of $\protect\xblin$}\label{epitelousdecay}
Using the new horizon fluxes obtained in the previous section, we will now obtain global control on angular derivatives of $\xblin$. 

\begin{proposition} \label{prop:chibar} 
We have the following integrated decay estimate for the ingoing shear $\hat{\underline{\chi}}$ of the solution $\Sf$ in Theorem \ref{theo:mtheod}. For any $v\geq v_0$
\begin{align}
\int_{v_0}^v d\bar{v} \int_{u}^\infty d\bar{u} \int_{S^2\left(u,\bar{v}\right)} \sin \theta d\theta d\phi \frac{\Omega^2}{r^{1+\epsilon}} \Big|\mathcal{A}^{[5]} \left(\frac{r\xblin}{\Omega}\right)\Big|^2
\lesssim \Big\|\mathcal{A}^{[3]}\frac{\pblin}{\Omega}\Big\|_{S^2_{\infty,v_0}}^2 + \textrm{right hand side of (\ref{ughb})} . \nonumber
\end{align}
\end{proposition}

\begin{proof}
We multiply the once more angular commuted (\ref{hopeid1b}) by $r^{-2-\epsilon} \mathcal{A}^{[3]}\Ylin$, to obtain
\begin{align}
\frac{1}{2} \partial_u \left[ -r^{-2-\epsilon} |\mathcal{A}^{[3]} \Ylin|^2 \right]  + \frac{2+\epsilon}{2r^{3+\epsilon}} \Omega^2 |\mathcal{A}^{[3]} \Ylin|^2
\leq \frac{1}{2} \frac{1}{r^{3+\epsilon}} \Omega^2 |\mathcal{A}^{[3]}\Ylin|^2 
 \nonumber \\
 + C  \cdot \frac{\Omega^2}{r^{1+\epsilon}} \Big( |\Omega^{-1}\slashed{\nabla}_3 (r \slashed{div} \Psilinb) |^2 +  |\Omega^{-1}\slashed{\nabla}_3 (r\slashed{div} \Omega^{-1} \pblin r^3) |^2 + |r\slashed{div} \Omega^{-1} \pblin r^3|^2 + |r\slashed{div} \Omega^{-2} \ablin r|^2\Big) \, .
\end{align}
Absorbing the first term on the right by the left hand side and using the previous bounds on the term in the second line (the integrated decay estimate on $\pblin$ of Proposition \ref{prop:psie} and that on $\ablin$ of Proposition \ref{prop:alphae}), we obtain an integrated decay estimate for $\mathcal{A}^{[3]}\Ylin$ after integrating the estimate over spacetime, observing that the flux term on the horizon has the wrong sign but is controlled from Proposition \ref{prop:firstangularpsi2} and Proposition~\ref{lem:2dc}. We then use the definition of $\Ylin$ (\ref{Ydef}) and control on $\mathcal{A}^{[3]}\pblin$ (from the twice angular commuted Proposition \ref{prop:firstangularpsi}) to descend to the desired estimate for angular derivatives of $\xblin$.
\end{proof}

\subsubsection{Integrated decay for angular derivatives of $\slashed{\nabla}_4\protect\xblin$}\label{kaistotes}

\begin{proposition} \label{prop:chibar4id} 
We have the following integrated decay estimate for the quantity $\slashed{\nabla}_4\hat{\underline{\chi}}$ of the solution $\Sf$ in Theorem \ref{theo:mtheod}. For any $v\geq v_0$:
\begin{align}
 \int_{v_0}^v d\bar{v} \int_{u}^\infty d\bar{u} \int_{S^2\left(u,\bar{v}\right)} \sin \theta d\theta d\phi  \Omega^2 r^{-1-\epsilon} \Big|\mathcal{A}^{[4]} r \Omega \slashed{\nabla}_4 \left(\frac{r\xblin}{\Omega}\right)\Big|^2 
& \lesssim  \Big\|\mathcal{A}^{[4]} \frac{\xblin\left[\Si\right] }{\Omega} \Big\|_{S^2_{\infty,v_0}}^2 +  \textrm{right hand side of (\ref{ughb})}  \, . \nonumber
\end{align}
\end{proposition}

\begin{proof}
We recall (\ref{hopeid2}),
\begin{align} 
\slashed{\nabla}_3 \left(\mathcal{A}^{[2]}\Omega \slashed{\nabla}_4 \Ylin\right) = r^3 \Omega \Pblin + 3M  \left(2r\pblin - 2 \hat{\omega} \ablin r \right) \, 
\end{align}
which we multiply with $r^{-\epsilon}\mathcal{A}^{[2]}\Omega \slashed{\nabla}_4 \Ylin$, to obtain
\begin{align}
\frac{1}{2} \partial_u \left[ -r^{-\epsilon} |\mathcal{A}^{[2]} \Omega \slashed{\nabla}_4 \Ylin|^2 \right]  + \frac{\epsilon}{2r^{1+\epsilon}} \Omega^2 |\mathcal{A}^{[2]} \Omega \slashed{\nabla}_4 \Ylin|^2 \leq \frac{\epsilon}{4} \frac{1}{r^{1+\epsilon}} \Omega^2 |\mathcal{A}^{[2]}\Omega \slashed{\nabla}_4 \Ylin|^2 \nonumber \\
+ C_\epsilon \Omega^2 r^{1-\epsilon} r^{-4} \left( |\mathcal{A}^{[2]} \Psilinb|^2 + |\mathcal{A}^{[2]}\pblin r^3 \Omega^{-1}|^2 + |\mathcal{A}^{[2]} \ablin r \Omega^{-2}|^2\right) \, . \nonumber
\end{align}
We absorb the first term on the right by the left hand side. Upon integration over a spacetime region, for the last term on the right  we use the integrated decay estimate on $\Psilinb$ of Theorem \ref{prop:summarypsi}, that on $\pblin$ of Proposition \ref{prop:psie} and that on $\ablin$ of Proposition \ref{prop:alphae}. The flux term on the horizon (which has a bad sign) arising from the first term on the left is controlled using the commuted (\ref{tryu}), which reads
\[
\mathcal{A}^{[2]} \Omega \slashed{\nabla}_4 \Ylin = \frac{\Omega^2}{r} \mathcal{A}^{[2]}  \Ylin + r \cdot \mathcal{A}^{[4]}  \Omega \slashed{\nabla}_4 \left(\frac{r \xblin}{\Omega} \right)-r^4 \mathcal{A}^{[2]}  \Pblin - 2M r^2 \mathcal{A}^{[2]} \frac{\pblin}{\Omega} \, ,
\]
to which Corollary \ref{cor:cb4} can be applied. This gives integrated decay for $\slashed{\nabla}_4 \Ylin$. To descend to $\xblin$ we use again the above identity in conjunction with the integrated decay estimate of Proposition \ref{prop:chibar}.
\end{proof}

\subsubsection{Integrated decay for $\protect\rlin$, $\protect\slin$ and $\protect\bblin$}
\label{tamesea}

With the integrated decay estimates on $\xblin$ we can easily show decay of all curvature components. Recall that for $\alin$ and $\ablin$ we already have these statements from Proposition \ref{prop:a5id}. The estimate for $\blin$ will be proven in Proposition \ref{prop:beta5} after we have estimated four angular derivatives of $\xlin$. (The $i=2$ non-degenerate version for $\blin$ can be proven at this point already, cf.~the proof of Proposition \ref{prop:beta5}.)

\begin{proposition} \label{prop:intdecay3rs}
For $i=3$ the following holds for the geometric quantities of $\Sf$ in Theorem \ref{theo:mtheod}:
\begin{align} 
\int_{u_0}^\infty du \int_{v_0}^\infty dv \frac{\Omega^2}{r^{1+\epsilon}} \left(1-\frac{3M}{r}\right)^{2} \Bigg(& \Big\|r^{-1}\cdot  \mathcal{A}^{[i]} \slashed{\mathcal{D}}_2^\star \slashed{\mathcal{D}}_1^\star\left(r^3 \rlin,r^3 \slin\right) \Big\|_{S^2_{u,v}}^2 \nonumber \\
 &+\Big\|r^{-1}\cdot  \mathcal{A}^{[i+1]}\slashed{\mathcal{D}}_2^\star \left(\bblin r^2 \Omega^{-1}\right) \Big\|_{S^2_{u,v}}^2 \Bigg)   
\lesssim   \textrm{right hand side of (\ref{ughb})}   . \nonumber
\end{align}
For $i=2$ this estimate holds without the degenerating factor of $(1-\frac{3M}{r})^2$.
\end{proposition}

\begin{proof}
This is a direct consequence of the identities (\ref{rsid2}) and (\ref{betuid}), the identities  (\ref{qoy1}) and (\ref{qoy2}) and the estimates already established.
\end{proof}

\subsubsection{Integrated decay for $\slashed{\mathcal{D}}_2^\star \protect\elin$ and $\slashed{\mathcal{D}}_2^\star \protect\eblin$}\label{taetaria}

\begin{proposition} \label{prop:etaetab}
For any $v\geq v_0$ we have the integrated decay estimate on the solution $\Sf$ in Theorem \ref{theo:mtheod}:
\begin{align}
 \int_{v_0}^v d\bar{v} \int_{u_0}^\infty d\bar{u} \Omega^2 r^{1-\epsilon} \Big\{ \Big\|r^{-1}  \cdot r \slashed{\mathcal{D}}_2^\star \eblin \Big\|_{S^2_{\bar{u},\bar{v}}}^2 + \Big\|r^{-1}  \cdot r \slashed{\mathcal{D}}_2^\star {\elin} \Big\|_{S^2_{\bar{u},\bar{v}}}^2 \Big\} 
\lesssim \textrm{right hand side of (\ref{ughb})}  \, . \nonumber
\end{align}

\end{proposition}

\begin{proof}
This follows directly from (\ref{fipo}) and (\ref{fipo2}) using the integrated decay estimates of Proposition \ref{prop:chibar}, \ref{prop:chibar4} and \ref{prop:hioc}.
\end{proof}

\subsubsection{Integrated decay for angular derivatives of $\protect\otxb$}
\label{kaigiaautoedw}

\begin{proposition} \label{prop:trchib2an} 
We have the following integrated decay estimate for the solution $\Sf$ in Theorem \ref{theo:mtheod}:
\begin{align}
 \int_{v_0}^\infty dv \int_{u_0}^\infty du \frac{\Omega^2}{r^{1+\epsilon}} \Big\| r^{-1} \frac{\mathcal{A}^{[3]}r^3\slashed{\mathcal{D}}_2^\star \slashed{\nabla}_A \otxb}{\Omega^2} \Big\|^2_{S^2_{u,v}} \nonumber 
 \lesssim  \textrm{right hand side of (\ref{ughb})}  \, .
\end{align}
\end{proposition}

\begin{proof}
Follows directly from (\ref{eqhc2}) using the integrated decay estimates of Propositions \ref{prop:chibar}, \ref{prop:etaetab} and (the commuted version of) \ref{prop:firstangularpsi}.
\end{proof}

\subsubsection{Refined estimates for higher angular derivatives and integrated decay for $\protect\blin$}
\label{polyrefined}

We easily derive the following analogue of Proposition \ref{prop:zestimate} by integrating both in $u$ and $v$ (and not in $v$ only as in the derivation of Proposition  \ref{prop:zestimate}):
\begin{proposition} \label{prop:zestimate2}
We have for $n\geq 0$, any $i \in \mathbb{N}$ and any $u\geq u_0$ the estimate ($d\omega=\sin \theta d\theta d\phi$)
\begin{align}
\int_{u_0}^\infty d\bar{u} \int_{S^2_{\bar{u},v}} d\omega \frac{|\mathcal{A}^{[i]} r \slashed{\mathcal{D}}_2^\star \Zlin|^2}{r^n \Omega^2} + \int_{u_0}^\infty d\bar{u} \int_{v_0}^v d\bar{v} \int_{S^2_{\bar{u},\bar{v}}} d\omega \frac{1}{r} \cdot \frac{|\mathcal{A}^{[i]} r \slashed{\mathcal{D}}_2^\star \Zlin|^2}{r^n \Omega^2} \nonumber \\
 \lesssim
  \int_{u_0}^\infty d\bar{u} \int_{S^2_{\bar{u},v_0}}d\omega \frac{|\mathcal{A}^{[i]} r \slashed{\mathcal{D}}_2^\star \Zlin \left[\Si\right]|^2}{r^n\Omega^2} 
+\int_{u_0}^\infty d\bar{u} \int_{v_0}^v d\bar{v}  \int_{S^2_{\bar{u},\bar{v}}}d\omega r^{5-n} \Omega^2  \Big|\mathcal{A}^{[i]} \slashed{\mathcal{D}}_2^\star  \left(\elin+ \eblin\right) \Big|^2\nonumber 
\end{align}
for the solution $\Sf$ in Theorem \ref{theo:mtheod}. 
\end{proposition}
We remind the reader that following our notation, $\elin+ \eblin$ on the right denote the geometric quantities of $\Sf$.
In the above Proposition we have used that $\Zlin\left[\Sf\right]=\Zlin\left[\Si\right]$ holds on $v=v_0$ (recall $\Zlin\left[\Gf\right]=0$ on $v=v_0$ by (\ref{odef})). Recall also that $\Zlin \sim \Omega^2$ holds on $v=v_0$ near the horizon by Proposition \ref{prop:zbounded} so the first term in the second line is indeed finite for smooth data. We can reinsert the definition of $\Zlin$ for the second term on the left to obtain and an estimate for $i$ angular derivatives of $\otx$ in terms of $i-1$ angular derivatives of $\elin +\eblin$:

\begin{corollary}  \label{cor:5for42}
We have for $n\geq 0$, any $i \in \mathbb{N}$ and any $\left(u\geq u_0,v\geq v_0\right)$ the estimate
\begin{align}
 \int_{u_0}^\infty d\bar{u} \int_{v_0}^v d\bar{v} \int_{S^2_{\bar{u},\bar{v}}} d\omega r^{7-n} \Omega^2 \Big|\mathcal{A}^{[i]}\slashed{\mathcal{D}}_2^\star \slashed{\nabla}_A \frac{\otx}{\Omega^2}\Big|^2  \nonumber \\
 \lesssim
  \int_{u_0}^\infty d\bar{u} \int_{S^2_{\bar{u},v_0}} d\omega \frac{|\mathcal{A}^{[i]} r \slashed{\mathcal{D}}_2^\star \Zlin\left[\Si\right]|^2}{r^n\Omega^2} 
+\int_{u_0}^\infty d\bar{u} \int_{v_0}^v d\bar{v}  \int_{S^2_{\bar{u},\bar{v}}} d\omega r^{5-n} \Omega^2  \Big|\mathcal{A}^{[i]} \slashed{\mathcal{D}}_2^\star  \left(\elin + \eblin\right) \Big|^2\nonumber 
\end{align}
for the solution $\Sf$ in Theorem \ref{theo:mtheod}. 
\end{corollary}
Again, consistent with our notation, $\elin+ \eblin$ above denotes the geometric quantities of $\Sf$.
We conclude
\begin{proposition} \label{prop:xtrxf}
We have the integrated decay estimate
\begin{align} 
 \int_{u_0}^\infty d\bar{u} \int_{v_0}^\infty d\bar{v} \ \frac{\Omega^2}{r^{3+\epsilon}} \Bigg( \Big\| r^{-1} \cdot \mathcal{A}^{[i]}r^2\slashed{\mathcal{D}}_2^\star \slashed{\nabla}_A \frac{r^2\otx}{\Omega^2}\Big\|_{S^2_{\bar{u},\bar{v}}}^2 +   \| r^{-1} \cdot \mathcal{A}^{[3]} r^2 \slashed{\mathcal{D}}_2^\star \slashed{div} \left(\Omega \xlin r^2\right)\|^2_{S^2_{\bar{u},\bar{v}}}  \Bigg) \nonumber \\
  \lesssim \textrm{right hand side of (\ref{ughb})} + \int_{u_0}^\infty d\bar{u} \int_{S^2_{\bar{u},v_0}} d\omega \frac{|\mathcal{A}^{[i]} r \slashed{\mathcal{D}}_2^\star \Zlin\left[\Si\right]|^2}{r^n\Omega^2} \nonumber
\end{align}
for the geometric quantities of $\Sf$ in Theorem \ref{theo:mtheod}.
\end{proposition}

\begin{proof}
Apply Corollary \ref{cor:5for42} with $n=2+\epsilon$ and use  Proposition \ref{prop:etaetab} to obtain the first part of the estimate. Then use identity (\ref{eqhc}) to obtain the second part.
\end{proof}

\begin{remark}
The weights near infinity are far from optimal. The weight near infinity for $\mathcal{A}^{[5]}\xlin$ can be improved a posteriori from the transport equation (\ref{tchi}) and the (degenerate near $3M$) integrated decay estimate for  $\mathcal{A}^{[5]}\alin$. See also Corollary \ref{cor:im2}.
\end{remark}

The weights are sufficient to prove the integrated decay estimate of Proposition \ref{prop:intdecay3rs} for the missing curvature component $\blin$:

\begin{proposition} \label{prop:beta5}
For the geometric quantities of $\Sf$ in Theorem \ref{theo:mtheod} we have the following integrated decay estimate for $i=3$ 
\begin{align} 
\int_{u_0}^\infty du \int_{v_0}^\infty dv \frac{\Omega^2}{r^{1+\epsilon}} \left(1-\frac{3M}{r}\right)^{2} \Bigg( \Big\|r^{-1}\cdot  \mathcal{A}^{[i+1]}r \slashed{\mathcal{D}}_2^\star \left({\blin} r^3 \Omega^{-1}\right) \Big\|_{S^2_{u,v}}^2 \Bigg)   
\nonumber \\
  \lesssim \textrm{right hand side of (\ref{ughb})} + \int_{u_0}^\infty d\bar{u} \int_{S^2_{\bar{u},v_0}} \sin \theta d\theta d\phi \frac{|\mathcal{A}^{[i]} r \slashed{\mathcal{D}}_2^\star \Zlin\left[\Si\right]|^2}{r^n\Omega^2} \, . \nonumber \nonumber
\end{align}
For $i=2$ this estimate holds without the degenerating factor of $(1-\frac{3M}{r})^2$ on the left and without the last term on the right.
\end{proposition}

\begin{proof}
For $i=3$ this follows from the identity (\ref{betid}) and application of Proposition \ref{prop:xtrxf}. For $i=2$ this also follows from (\ref{betid}) but observing that now Proposition \ref{prop:hioc} already estimates three angular derivatives of $\xlin$, while (the twice angular commuted) Proposition \ref{prop:firstangularpsi} estimates (non-degenerately) three angular derivatives of $\plin$.
\end{proof}

Finally, repeating the arguments\footnote{All that is required is to insert an additional $u$-integration everywhere.} leading to Proposition \ref{prop:otxe} and Propositions \ref{prop:etabe2} and \ref{prop:etae2} we can prove integrated decay for five derivatives of $\elin$ and $\eblin$ of the solution $\Sf$, i.e.~the estimate
\begin{align} \label{fluh2o}
\int_{v_0}^\infty d\bar{v} \int_{u_0}^\infty d\bar{u} \int_{S^2_{\bar{u},\bar{v}}} \sin \theta d\theta d\phi  \Omega^2 r^{3-\epsilon} \left( |\mathcal{A}^{[4]} \slashed{\mathcal{D}}_2^\star {\elin} |^2 + |\mathcal{A}^{[4]} \slashed{\mathcal{D}}_2^\star \eblin|^2 \right) \nonumber \\
 \lesssim \textrm{right hand side of (\ref{ughb})} + \int_{u_0}^\infty d\bar{u} \int_{S^2_{\bar{u},v_0}} \sin \theta d\theta d\phi \frac{|\mathcal{A}^{[i]} r \slashed{\mathcal{D}}_2^\star \Zlin \left[\Si\right]|^2}{r^n\Omega^2}. \, 
\end{align}

\subsection{Polynomial decay estimates and conclusions} \label{sec:polyfinal}
Finally, 
in this section we prove appropriate 
$L^2$ polynomial decay of all quantites associated to $\Sf$.
This corresponds to statement~3.~of Theorem~\ref{theo:mtheod}. 
We shall then infer Corollary~\ref{newcoroledw} giving pointwise estimates.

We will consider first Ricci coefficients in Section~\ref{sec:somepd}, and
then the metric components themselves in Section~\ref{sec:mecon}.
We shall treat Corollary~\ref{newcoroledw} in Section~\ref{proofofcorsecedw}.

\subsubsection{Polynomial decay for $\protect\xlin$, $\protect\xblin$, $\protect\elin$ and $\protect\eblin$} \label{sec:somepd}
\begin{proposition} \label{prop:pold}
Fix $r_0$ as in Proposition \ref{prop:decRW} and let $v \geq v_0$ and recall the notation $u\left(v,r_0\right)$. We have the following decay estimates for the geometric quantities of $\Sf$ (and equivalently $\Si$) in Theorem \ref{theo:mtheod}:
\begin{align}
 \int^\infty_{u\left(v,r_0\right)} d\bar{u} \Bigg( \frac{\Omega^2}{r^{2+\epsilon}} \|r^{-1} \cdot r^2 \xlin \Omega\|_{S^2_{\bar{u},V}}^2   + \frac{\Omega^2}{r^{\epsilon}} \|r^{-1} \cdot \Omega^{-1} \slashed{\nabla}_3 \left( r^2 \xlin\Omega\right) \|_{S^2_{\bar{u},V}}^2    \nonumber \\
+ \frac{\Omega^2}{r^{\epsilon}} \|r^{-1} \cdot \Omega^{-1} \slashed{\nabla}_3 \left(\Omega^{-1} \slashed{\nabla}_3\left( r^2 \xlin \Omega\right)\right) \|_{S^2_{\bar{u},V}}^2\Bigg) \lesssim \frac{1}{v^2} \left(\mathbb{F}_0^{2,T}\left[\Psilin\right] + \mathbb{F}_0 \left[\Psilin, \Psilinb, \plin, \pblin, \alin, \ablin\right] \right)
\end{align}
and
\begin{align}
\int_v^\infty d\bar{v} \int^\infty_{u\left(v,r_0\right)} d\bar{u} \Bigg( \frac{\Omega^2}{r^{3+\epsilon}} \|r^{-1} \cdot r^2 \xlin\Omega\|_{S^2_{\bar{u},\bar{v}}}^2   + \frac{\Omega^2}{r^{1+\epsilon}} \|r^{-1} \cdot \Omega \slashed{\nabla}_3 \left( r^2 \xlin\Omega\right) \|_{S^2_{\bar{u},\bar{v}}}^2  \nonumber \\ + \frac{\Omega^2}{r^{1+\epsilon}} \|r^{-1} \cdot \Omega^{-1} \slashed{\nabla}_3 \left(\Omega^{-1} \slashed{\nabla}_3\left( r^2 \xlin \Omega\right)\right) \|_{S^2_{\bar{u},V}}^2\Bigg) 
 \lesssim \frac{1}{v^2} \left(\textrm{right hand side of (\ref{ughb})} \right) \, .
\end{align}
\end{proposition}

\begin{proof}
Recall that $\xlin \left[\Sf\right]=\xlin\left[\Si\right]$. By Proposition \ref{prop:decbh} we have on the horizon
\[
\int_{v}^\infty d\bar{v} \left( \| \Omega \xlin\|_{S^2_{\infty,\bar{v}}}^2 + \| \Omega^{-1} \slashed{\nabla}_3 \left( \Omega \xlin\right) \|_{S^2_{\infty,\bar{v}}}^2 \right) \lesssim \frac{1}{v^2} \left(\mathbb{F}_0^{2,T}\left[\Psilin\right] + \mathbb{F}_0 \left[\Psilin, \Psilinb, \plin, \pblin, \alin, \ablin\right] \right) \, .
\]
The boundedness of these fluxes was crucial in the proof of Proposition \ref{prop:chiall}. Now that we have decay for them we can repeat the proof (localised to dyadic regions and starting from ``good" slices generated from a pigeonhole argument) and
use the above flux and the integrated decay estimates on $\Psi$, $\plin$ and $\alin$ of Proposition \ref{prop:refd} and Corollary \ref{cor:dmdec} to obtain the statements of the proposition.
The only surprising fact is that the estimates on $\Omega^{-1} \slashed{\nabla}_3 \left( \Omega \xlin\right)$ and $\Omega^{-1} \slashed{\nabla}_3 \Omega^{-1} \slashed{\nabla}_3 \left( \Omega \xlin\right)$ do not lose $r$-weights. This is because the right hand side of (\ref{commu1}) and (\ref{dceq}) has improved decay in $r$ allowing one to avoid a loss of $r$-weights in the pigeonhole argument. Finally, recall that $\Gf$ has $\xlin=0$ globally, so indeed the statement holds equivalently for $\Sf$ and $\Si$.
\end{proof}
Repeating the proof of Corollary \ref{cor:l2sb} using now the decay estimates of Proposition \ref{prop:decbh2} we also have
\begin{corollary} \label{cor:pold}
Fix $v\geq v_0$. Then for all $u \geq u\left(v,r_0\right)$ and $V \geq v$ we have
\begin{align}
 \frac{1}{r^{1+\epsilon}} \| \Omega r^2\xlin\|_{S^2_{u,V}}^2 + \| \Omega^{-1} \slashed{\nabla}_3 \left( r^2 \Omega \xlin\right) \|_{S^2_{u,V}}^2 \lesssim \frac{1}{v^2} \left(\textrm{right hand side of (\ref{ughb})} \right) \, 
 \end{align}
for the geometric quantities of $\Sf$ (and equivalently $\Si$) in Theorem \ref{theo:mtheod}.
\end{corollary}

For the shear in the ingoing direction, $\xblin$, we obtain

\begin{proposition} \label{prop:pold2}
Fix $r_0$ as in Proposition \ref{prop:decRW} and let $v \geq v_0$ and recall the notation $u\left(v,r_0\right)$. We have the following decay estimate for the geometric quantities of $\Sf$ in Theorem \ref{theo:mtheod}:
\begin{align}
 \int_v^\infty d\bar{v} \int^\infty_{u\left(v,r_0\right)} d\bar{u} \frac{\Omega^2}{r^{1+\epsilon}} \left( \|r^{-1} \cdot r \xblin \Omega^{-1}\|_{S^2_{\bar{u},\bar{v}}}^2  + \|r^{-1} \cdot  r \Omega \slashed{\nabla}_4 \left( r \xblin\Omega^{-1}\right)\|_{S^2_{\bar{u},\bar{v}}}^2  \right)  \nonumber \\
\lesssim \frac{1}{v^2} \left(\textrm{right hand side of (\ref{ughb})} \right) \, .
\end{align}
\end{proposition}
\begin{proof}
We first note that Proposition~\ref{lem:2dc}, Corollary \ref{cor:cb4a} and Proposition \ref{prop:firstangularpsi2} can be combined with the decay estimate of Proposition \ref{prop:refd} to produce the following estimate for the horizon flux:
\begin{align} \label{hogu}
\int_{v}^\infty d\bar{v} \left( \| \mathcal{A}^{[3]} \Omega^{-1} \xblin\|_{S^2_{\infty,\bar{v}}}^2 + \|  \mathcal{A}^{[3]} \Omega \slashed{\nabla}_4 \left( \Omega^{-1} \xblin\right) \|_{S^2_{\infty,\bar{v}}}^2 + \| \Omega^{-1} \pblin\|_{S^2_{\infty,\bar{v}}}^2 \right) \nonumber \\
 \lesssim \frac{1}{v^2} \left(\mathbb{F}_0^{2,T}\left[\Psilin\right] + \mathbb{F}_0^{2,T}\left[\Psilinb\right] + \mathbb{F}_0 \left[\Psilin, \Psilinb, \plin, \pblin, \alin, \ablin\right] \right) \, .
\end{align}
This in turn allows to repeat the proof of Proposition \ref{prop:chibar} and Proposition \ref{prop:chibar4} to deduce the bound claimed in the Proposition.
\end{proof}

We also directly prove
\begin{proposition} \label{prop:pold3}
Fix $r_0$ as in Proposition \ref{prop:decRW} and fix $v \geq v_0$. We have for all $V \geq v$ the estimate
\begin{align} \label{metb1}
 \int^\infty_{u\left(v,r_0\right)} d\bar{u} \frac{\Omega^2}{r^\epsilon} \|r^{-1} \cdot r^2 \slashed{\mathcal{D}}_2^\star \slashed{div} \left( r \xblin \Omega^{-1}\right)\|_{S^2_{\bar{u},V}}^2   \lesssim \frac{1}{v^2} \left(\mathbb{F}_0^{2,T}\left[\Psilin\right] + \mathbb{F}_0^{2,T}\left[\Psilinb\right] + \mathbb{F}_0 \left[\Psilin, \Psilinb, \plin, \pblin, \alin, \ablin\right] \right)
\end{align}
for the geometric quantities of $\Sf$ in Theorem \ref{theo:mtheod}. In addition, for all $V \geq v$ and any $u \geq u \left(v,r_0\right)$
\[
\|r^{-1} \cdot r^2 \slashed{\mathcal{D}}_2^\star \slashed{div} \left( r \xblin\Omega^{-1}\right)\|_{S^2_{u,V}}^2 \lesssim  \frac{1}{v^2} \left(\mathbb{F}_0^{2,T}\left[\Psilin\right] + \mathbb{F}_0^{2,T}\left[\Psilinb\right] + \mathbb{F}_0 \left[\Psilin, \Psilinb, \plin, \pblin, \alin, \ablin\right] \right) \, .
\]
\end{proposition}

\begin{remark}
The $\epsilon$ can be removed with further work.
\end{remark}

\begin{proof}
We first have from (\ref{hopeid})
\begin{align} \label{ftp}
-\frac{1}{2} \partial_u \frac{|\Ylin|^2}{r^{1+\epsilon}} + \frac{1}{4} \Omega^2 \frac{|\Ylin|^2}{r^{2+\epsilon}} \lesssim \frac{\Omega^2}{r^\epsilon}\left( |\ablin r \Omega^{-2}|^2 + | \pblin r^3 \Omega^{-1}|^2\right) \, ,
\end{align}
which we would like to integrate from the horizon. An easy pigeonhole argument using (\ref{hogu}) and the propagation equations for $\pblin$ and $\xblin$ along the horizon yields that on the horizon 
\[
\int_{S^2_{\infty,v}} |\Ylin|^2 \sin \theta d\theta d\phi \lesssim \frac{1}{v^2} \left(\mathbb{F}_0^{2,T}\left[\Psilin\right] + \mathbb{F}_0^{2,T}\left[\Psilinb\right] + \mathbb{F}_0 \left[\Psilin, \Psilinb, \plin, \pblin, \alin, \ablin\right] \right) \, ,
\]
so that integration of (\ref{ftp}) in $u$ from the horizon yields a decay estimate for $Y$ after using the decay of the fluxes 
on the right hand side from Corollary \ref{cor:fluxbpsib}. The global estimate for $\pblin$
of Proposition \ref{prop:psibs2d} now allows to go from $\Ylin$ to $r^2 \slashed{\mathcal{D}}_2^\star \slashed{div} ( r \xblin \Omega^{-1})$.
\end{proof}

\begin{corollary} \label{cor:doil}
Fix $r_0$ as in Proposition \ref{prop:decRW} and let $v \geq v_0$ and recall the notation $u\left(v,r_0\right)$. We have the following integrated decay estimate for the geometric quantities of $\Sf$ in Theorem \ref{theo:mtheod}:
\begin{align}
\int_v^\infty d\bar{v} \int^\infty_{u\left(v,r_0\right)} d\bar{u} \frac{\Omega^2}{r^{1+\epsilon}} \left(  \|r^{-1} \cdot r^2  \slashed{\mathcal{D}}_2^\star \elin \|_{S^2_{\bar{u},\bar{v}}}^2   +  \|r^{-1} \cdot r^2  \slashed{\mathcal{D}}_2^\star \eblin \|_{S^2_{\bar{u},\bar{v}}}^2   \right) 
 \lesssim \frac{1}{v^2} \left(\textrm{right hand side of (\ref{ughb})} \right) \, . \nonumber
\end{align}
We also have the flux estimates for any $V \geq v$
\begin{align}
 \int^\infty_{u\left(v,r_0\right)} d\bar{u} \frac{\Omega^2}{r^\epsilon} \|r^{-1} \cdot r^2 \slashed{\mathcal{D}}_2^\star \elin\|_{S^2_{\bar{u},V}}^2 \lesssim  \frac{1}{v^2} \left(\textrm{right hand side of (\ref{ughb})} \right) \, ,
 \end{align}
\begin{align} \label{metb2}
 \int^\infty_{u\left(v,r_0\right)} d\bar{u} \frac{\Omega^2}{r^\epsilon}  \Big\| r^{-1} \cdot r^3  \slashed{\mathcal{D}}_2^\star \slashed{\nabla}_A \frac{\otxb}{\Omega^2} \Big\|^2_{S^2_{u,V}}\lesssim \frac{1}{v^2} \left(\textrm{right hand side of (\ref{ughb})} \right) \, .
\end{align}
Finally, for any $V \geq v$ and $u\geq u\left(v,r_0\right)$ we have
\begin{align}
\|r^{-1} \cdot r^2 \slashed{\mathcal{D}}_2^\star \elin\|_{S^2_{\bar{u},V}}^2 \lesssim  \frac{1}{v^2} \left(\textrm{right hand side of (\ref{ughb})} \right) \, .
\end{align}
\end{corollary}
\begin{proof}
The first two and the last estimate follow directly from Propositions \ref{prop:pold}, Propositions \ref{prop:pold2} and \ref{prop:pold3} and the identities (\ref{fipo}) and (\ref{fipo2}).  The remaining estimate follows similarly from (\ref{eqhc2}) and those Propositions.
\end{proof}

We finally derive an estimate for $\eblin$ on the spheres:
\begin{proposition} \label{prop:doil2}
Fix $v\geq v_0$. For any $V \geq v$ and $u\geq u\left(v,r_0\right)$ we have
\begin{align}
\|r^{-1} \cdot r^2 \slashed{\mathcal{D}}_2^\star \eblin\|_{S^2_{\bar{u},V}}^2 \lesssim  \frac{1}{v^2} \left(\textrm{right hand side of (\ref{ughb})} \right) \, 
\end{align}
for the geometric quantity $\eblin$ of $\Sf$ in Theorem \ref{theo:mtheod}.
\end{proposition}
\begin{proof}
Recalling that $\elin+\eblin=0$ on the horizon $\mathcal{H}^+$ for $\Sf$ we obtain
\[
\| \slashed{\mathcal{D}}_2^\star \eblin\|^2_{\infty,v}  \lesssim \frac{1}{v^2} \left(\mathbb{F}_0^{2,T}\left[\Psilin\right] + \mathbb{F}_0^{2,T}\left[\Psilinb\right] + \mathbb{F}_0 \left[\Psilin, \Psilinb, \plin, \pblin, \alin, \ablin\right] \right) \, .
\]
Using this decay estimate we can integrate
\begin{align}
\Omega \slashed{\nabla}_3 \left(r^2 \slashed{\mathcal{D}}_2^\star \eblin\right) = -\Omega^2 r \slashed{\mathcal{D}}_2^\star {\elin} + r^2 \Omega^2 \left(\pblin \Omega^{-1} + \frac{3}{2} \rho \xblin\Omega^{-1}\right) 
\end{align}
from the horizon and use the previous bounds.
\end{proof}

\subsubsection{Polynomial decay of the metric coefficients: Proof of (\ref{mcd1})--(\ref{mcd3})} \label{sec:mecon}
In this final subsection we prove 
the estimates  (\ref{mcd1})--(\ref{mcd3}) on the metric quantities of $\Sf$.

The estimate (\ref{mcd1}) is a direct consequence of Corollary \ref{cor:doil}, Proposition \ref{prop:doil2} and the definition (\ref{oml3}). 

For (\ref{mcd2}) we present the proof without the (trivial) commutation with $\mathcal{A}^{[2]}$ which can be inserted into all formulae below. We write (\ref{stos2}) in the form
\begin{align}
\Omega \slashed{\nabla}_3 \hat{\slashed{g}}^{(1)}= 2 \Omega \xblin
\end{align}
and derive for any fixed $u\geq u_0$, $v \geq v_0$
\begin{align}
\big\| r^{-1} \cdot \hat{\slashed{g}}^{(1)}\big\|_{S^2_{u,v}} \lesssim
\big\| r^{-1} \cdot \glinh \big\|_{S^2_{u_0,v}} + \int_{u_0}^u d\bar{u} \| r^{-1} \cdot \xblin \Omega \|_{S^2_{\bar{u},v}} \, .
\end{align}
Consider first the first term on the right hand side. Recall that the solution $\Si$ had $\bmlin=0$ on $C_{u_0}$. By Lemma \ref{lem:exactsol} and the estimates on the gauge function in Corollary \ref{cor:gfe2} the pure gauge solution $\Gf$ generates a $b$ satisfying $|\slashed{\mathcal{D}}_2^\star \bmlin |\lesssim r^{-2}$ along $C_{u_0}$. Moreover, again by the boundedness estimates on $f$ in Corollary \ref{cor:gfe2}, the pure gauge solution $\Gf$ also satisfies the round sphere condition (\ref{rscm}) at infinity. Therefore, integrating equation (\ref{stos2}) from infinity along $C_{u_0}$ using Corollary \ref{cor:l2sc} and the aforementioned bound on $\bmlin$ yields:
\[
\big\| r^{-1} \cdot \glinh\big\|_{S^2_{u_0,v}} \lesssim \frac{1}{r\left(u_0,v\right)} \cdot \sqrt{\textrm{right hand side of (\ref{ughb})}} \, .
\]
For the second term we define $u_\star:=\min \left(u,\frac{3}{4}v\right)$ and split the integral as
\[
\int_{u_0}^u d\bar{u} \| r^{-1} \cdot \xblin \Omega \|_{S^2_{\bar{u},v}} = \int_{u_0}^{u_\star} d\bar{u} \| r^{-1} \cdot \xblin \Omega \|_{S^2_{\bar{u},v}} + \int_{u_\star}^u d\bar{u} \| r^{-1} \cdot \xblin \Omega \|_{S^2_{\bar{u},v}} \, .
\]
Now for the first integral we have
\begin{align}
\int_{u_0}^{u_\star} d\bar{u} \| r^{-1} \cdot \xblin \Omega \|_{S^2_{\bar{u},v}} &\lesssim
\sqrt{\int_{u_0}^{u_\star} d\bar{u} \frac{\Omega^2}{r^\epsilon} \| r^{-1} \cdot \left( r \xblin \Omega^{-1}\right) \|^2_{S^2_{\bar{u},v}}} \sqrt{\int_{u_0}^{u_\star} d\bar{u} \frac{\Omega^2}{r^{2-\epsilon}}} \nonumber \\
& \lesssim 
 \sqrt{\textrm{right hand side of (\ref{ughb})}} \left[r\left(\frac{3}{4}v,v\right)\right]^{-1/2+\epsilon/2}.
\end{align}
For the second integral (which vanishes if $u\leq \frac{3}{4}v$) we have
\begin{align}
\int_{u_\star}^{u} d\bar{u} \| r^{-1} \cdot \xblin\Omega \|_{S^2_{\bar{u},v}} &\lesssim
\sqrt{\int_{3/4v}^{\infty} d\bar{u} \frac{\Omega^2}{r^\epsilon} \| r^{-1} \cdot \left( r \xblin \Omega^{-1}\right) \|^2_{S^2_{\bar{u},v}}} \sqrt{\int_{u_0}^{\infty} d\bar{u} \frac{\Omega^2}{r^{2-\epsilon}}} \nonumber \\
& \lesssim 
 \sqrt{\textrm{right hand side of (\ref{ughb})}} \cdot v^{-1} \, ,
\end{align}
where we have used the decay estimate of Proposition \ref{prop:pold3} and boudnedness of the integral in the second square root. Combining the estimates yields (\ref{mcd2}).

The argument to prove (\ref{mcd3}) is similar now starting from (\ref{stos}), which reads
\begin{align}
\Omega \slashed{\nabla}_3\frac{\glinto}{\sqrt{\slashed{g}}}= \otxb \, .
\end{align}
Using the bound (\ref{metb2}), the estimate on $\bmlin$ mentioned above and the fact that the round sphere condition (\ref{rsc}) in conjunction with (\ref{rscm}) implies that $\frac{\glinto}{\sqrt{\slashed{g}}}$ vanishes for $\ell\geq 2$ at infinity (cf.~\ref{Gaussinmet}) we derive (\ref{mcd3}) following the argument above.

To derive (\ref{mcd4}) we first note that (\ref{beq1}) yields
\begin{align}
 \big\| r^{-1} \cdot  r \slashed{\mathcal{D}}_2^\star \bmlin\big\|_{S^2_{u,v}}^2 \lesssim \frac{1}{v^{1-\epsilon}} \left(\textrm{right hand side of (\ref{ughb})} \right) \ \ \ \ \ \textrm{for $r\left(u,v\right) \gtrsim \frac{1}{4} v$} \, .
\end{align}
Using the estimates on the gauge function of Corollary \ref{cor:gfe2} and Lemma \ref{lem:exactsol} we conclude
\begin{align} \label{zd}
 \big\| r^{-1} \cdot  r \slashed{\mathcal{D}}_2^\star \bmlin \big\|_{S^2_{u,v}}^2 \lesssim \frac{1}{v^{1-\epsilon}} \left(\textrm{right hand side of (\ref{ughb})} \right) \ \ \ \ \ \textrm{for $r\left(u,v\right) \gtrsim \frac{1}{4} v$} \, .
\end{align}
Fix now $v \geq v_0$. Then on the hypersurface $\bar{u}=u\left(r_0,\frac{3}{4}v\right)$ we have for all $V\geq v$ that $r\left(\bar{u},V\right) \gtrsim \frac{1}{4}v$. Applying (\ref{bidm}), this time for the geometric quantities of $\Sf$ with $n=2-\epsilon$ and integrating from $u=\bar{u}$ towards the horizon for any $V\geq v$ we obtain upon inserting the bound (\ref{zd}) and the bounds on $\elin$ and $\eblin$ (Corollary \ref{cor:doil} and Proposition~\ref{prop:doil2}) that
\begin{align} \label{zd2}
r^{-\epsilon} \big\| r^{-1} \cdot r \slashed{\mathcal{D}}_2^\star \bmlin \big\|_{S^2_{u,V}}^2 \lesssim \frac{1}{v^{1-\epsilon}} \left(\textrm{right hand side of (\ref{ughb})} \right) \ \ \ \ \ \textrm{for $V\geq v$ and $u\geq u \left(r_0,v\right)$} \, .
\end{align}
This implies that in $r\geq r_0$ we have 
\begin{align} 
r^{-\epsilon} \big\| r^{-1} \cdot  r \slashed{\mathcal{D}}_2^\star \bmlin \big\|_{S^2_{u,v}}^2 \lesssim \frac{1}{u^{1-\epsilon}} \left(\textrm{right hand side of (\ref{ughb})} \right) \ \ \ \ \ \textrm{for $r\geq r_0$ }
\end{align}
Interpolation with (\ref{zd}) (note $u \sim v$ and $r \lesssim u$ in the region $r_0 \leq r \lesssim \frac{1}{4}v$) yields the desired bound.

\subsubsection{Proof of Corollary \ref{newcoroledw}}\label{proofofcorsecedw}
The proof of Corollary \ref{newcoroledw} is now immediate: 
We consider $\Sf^\prime=\Sf-\mathscr{K}_{\mathfrak{m},s_i}$.
Using the bounds (\ref{mcd1})--(\ref{mcd3}), we
apply the classical Sobolev embedding on the spheres $S^2_{u,v}$ to the left hand side. Note that the quantities $\glinto$, $\glinh$, $\bmlin$ and $\Olin$ associated
to $\Sf^\prime$ are supported on $\ell \geq 2$ and that Proposition \ref{cor:elliptic12} and Corollary \ref{cor:etavl1} guarantee that all second order derivatives are indeed controlled in $L^2\left(S^2_{u,v}\right)$. The estimates $(\ref{pointwisedecayone})$--$(\ref{pointwisedecaytwo})$ follow immediately.

\appendix

\section{Construction of data and propagation of asymptotic flatness} \label{appendix:datacon} 
In this appendix we construct and estimate from a smooth seed initial data set (Definition \ref{def:seeddata}) which is asymptotically flat with weight $s$ to order $n$ (Definition \ref{def:afpeel}) all quantities of the solution $\mathscr{S}$ associated with the data set through Theorem \ref{theo:lwp}, first on the initial cones $C_{u_0} \cup C_{v_0}$ and then globally in the spacetime.  The main result is the following

\begin{theorem} \label{theo:AFdata} \label{prop:pwprop}
Consider a smooth seed initial data set which is asymptotically flat with weight $s$ to order $n \geq 11$ and the corresponding smooth solution $\mathscr{S}$ arising from Theorem \ref{theo:lwp}. For $\xi$ an element of $\mathscr{S}$ we denote
\[
|\mathfrak{D}^k \xi |= \sum_{0\leq j_1+j_2+j_3 \leq k} | \left(\Omega^{-1} \slashed{\nabla}_3 \right)^{j_1} \left(r \slashed{\nabla}\right)^{j_2} \left(r \Omega \slashed{\nabla}_4 \right)^{j_3} \xi| \, .
\]
The solution $\mathscr{S}$ has the following property: On the initial cones 
$C_{u_0} \cup C_{v_0}$, the  estimates 
\begin{equation}\label{decrete} 
\begin{split}
 |\mathfrak{D}^k (r^{3+s} \alin \Omega^2 \, )| + |\mathfrak{D}^k( r^{3+s} \blin \Omega )| + |\mathfrak{D}^k (r^3 \rlin )\, | +  |\mathfrak{D}^k (r^3 \slin )| +  |\mathfrak{D}^k (r^2\bblin \Omega ^{-1} )|  \leq C_k  \, ,   \\
 |\mathfrak{D}^k (r \ablin \Omega^{-2} )| +  |\mathfrak{D}^k (r^2\Klin ) | +  |\mathfrak{D}^k(r^2 \xlin \Omega )| +  |\mathfrak{D}^k (r \xblin \Omega^{-1} )| +  | \mathfrak{D}^k (r\elin \, )| + |\mathfrak{D}^k (r^2\eblin \, )| \leq C_k \, , \\
  |\mathfrak{D}^k (r^{2+s} \olin) | + |\mathfrak{D}^k (\Omega^{-2} \olinb ) |  +  | \mathfrak{D}^k (r^2 \otx) | + | \mathfrak{D}^k (r \Omega^{-2} \otxb) |   \leq C_k  \, ,  \\ 
 \Big| \mathfrak{D}^k (\Olin \,) \Big| + |\mathfrak{D}^k \glinh| + |\mathfrak{D}^k ( tr_{\slashed{g}} \glin \, )|  + |\mathfrak{D}^k (r \bmlin \,)| \leq C_k  \, 
\end{split}
\end{equation}
hold for any $k \leq n-3$ and a constant $C_k$ which can be computed explicitly and depends only on finitely many constants $C_{\circ, n_1,n_2}$ appearing in Definition \ref{def:afpeel}. In particular, the constants $C_{\circ,n_1,n_2}$ with $n_1+n_2 \leq k+3$ are sufficient. For any $k\leq n-4$ the quantities on the left hand side with of (\ref{decrete}) with $s=0$ have well-defined limits on null infinity.

Moreover, given any $u_0 < U<\infty$, the estimates (\ref{decrete}) actually hold 
hold for any $k \leq n-10$ in any spacetime region $\mathcal{M} \cap \{u_0 \leq u\leq U\}$ where $C_k$ now also depends on the choice of $U$ and the constants $C_{\circ,n_1,n_2}$ with $n_1+n_2 \leq k+10$ appearing in Definition \ref{def:afpeel}. For any $k\leq n-11$ the quantities on the left hand side of (\ref{decrete}) with $s=0$ have well-defined limits on null infinity.
\end{theorem}

We remark that the conditions $n_1+n_2 \leq k+3$ in the first part and $n_1+n_2 \leq k+10$ in the second part of the theorem account for the loss of derivatives in the characteristic initial value problem and losses from applying Sobolev embedding. It is of course not optimal. We also remark that we have stated (\ref{decrete}) for quantities \emph{regular} at the horizon (cf.~(\ref{regq1})). Since $C_k$ is allowed to depend on $U$ in the second part of the theorem this is not essential.

The proof of Theorem \ref{prop:pwprop} will proceed in two steps. In Section \ref{sec:holdondata} we prove that the estimates (\ref{decrete}) hold on the initial cone $C_{u_0} \cup C_{v_0}$ by constructing from an asymptotically flat seed initial data set of Definition \ref{def:afpeel} all quantities of the solution $\mathscr{S}$ on $C_{u_0} \cup C_{v_0}$. In the second step, we show that these bounds are in fact propagated by the evolution. The statement about the limit at null 
infinity will follow from the fact that the $\Omega\slashed{\nabla}_4$-derivative of any of the quantities in the round brackets of (\ref{decrete}) is always integrable in $v$. Note that $\alin$ and $\olin$ are part of the seed data and we gain integrability from taking the limit of $r^3\alin$ and $r^2\olin$.

\begin{remark} \label{ftnt:cv}
Observe that all quantities in $\mathscr{S}$ except $\olinb$ decay at least as fast towards null infinity as their background Schwarzschild value. For $\olinb$ we can only propagate boundedness, while $\underline{\omega}$ decays like $r^{-2}$ towards null infinity. This exceptional behaviour is rooted in our choice of null frame for the linearisation. If we compare the components linearised 
with respect to the frame $\boldsymbol{\mathcal{N}}_{EF^\star}$, where $\underline{\boldsymbol\omega}=0$ and hence $\olinb=0$ identically (cf.~Section \ref{appendix:linco}) all linearised quantities decay as fast as their Schwarzschild value.
\end{remark}

\subsection{Proof of the first part: Constructing the data} \label{sec:holdondata}
In this section we shall prove the first part of Theorem \ref{theo:AFdata}, i.e.~the statement
that the solution is determined from seed data 
with  the bounds (\ref{decrete}) holding  on on $C_{u_0}\cup C_{v_0}$. 
We will focus on establishing the latter bounds for $k=0$. The statement for arbitrary angular commutations is then easily obtained from the fact that $r\slashed{\nabla}$ commutes with $\Omega\slashed{\nabla}_3$ and $\Omega\slashed{\nabla}_4$ and has good commutation properties with angular derivatives in the sense that
\[
| \left[r\slashed{\nabla}_A, r\slashed{\nabla}_B\right] \xi | \leq C |\xi| \, .
\]
To obtain the remaining tangential derivatives and the transversal derivatives we will use the null structure and Bianchi equations directly in conjunction with an inductive procedure which is outlined below.

For the remainder of the proof, we will allow ourselves to drop the $\circ$ subscript  from all quantities as well as the ``in" and "out" from $\glin$ as it will be clear from the context which cone we are on.

\subsubsection*{Elliptic equations on the  horizon sphere $S^2_{\infty,v_0}$}
We first note that the seed data determines on the horizon sphere:
\begin{itemize}
\item $\slin$ uniquely from (\ref{curleta})
\item $\Klin$ uniquely from the fact that $\glinh$ and $\frac{\glinto}{\sqrt{\slashed{g}}}$ are part of the seed data and (\ref{Gaussinmet})
\item $\rlin$ uniquely from (\ref{lingauss})
\item $\bblin$ uniquely from (\ref{ellipchi})
\item $\elin$ uniquely from $\eblin=-\elin+ 2 \slashed{\nabla}_A \Olin$, equation (\ref{oml3}).
\end{itemize}

\subsubsection*{Transport equations along $C_{v_0}$: Part I}

We now integrate our seed data from $S^2_{\infty,v_0}$ along the cone $C_{v_0}$. 

Along $C_{v_0}$ the tensor $\glinh$ is part of the seed data. Note that this determines uniquely along $C_{v_0}$ the quantities $\Omega^{-1}\xblin$  via (\ref{stos}) and $\Omega^{-2} \ablin$ via (\ref{tchi}).

Next, from (\ref{uray}) we have
\[
\partial_u \left[ \frac{r^2}{\Omega^2} \otxb\right] =  -4r \olinb \, .
\]
Since $\olinb$ is prescribed along $C_{v_0}$ as part of the seed data as is the value of $\Omega^{-2} r^2  \otxb$ on $S^2_{\infty,v_0}$, the above ODE determines $\otxb$ uniquely along $C_{v_0}$. Note that now $\frac{\glinto}{\sqrt{\slashed{g}}}$ is now uniquely determined along $C_{v_0}$ by (\ref{stos}).

Since $\bblin$ was already uniquely determined on $S^2_{\infty,v_0}$ above, we can integrate the Bianchi equation (\ref{Bianchi9}) written as
\[
\Omega \slashed{\nabla}_3 \left[ r^4 \Omega^{-1} \bblin \right] = - r^4 \slashed{div} \ablin  \, ,
\]
along $C_{v_0}$ which, since the right hand side is uniquely determined from seed data,
determines $\bblin$ uniquely on $C_{v_0}$. 
The Bianchi equations (\ref{Bianchi4}) and (\ref{Bianchi7}) read as ODEs along $C_{v_0}$ now uniquely determine $\rlin$ and $\slin$ since the initial value of these quantities on $S^2_{\infty,v_0}$ were already determined by seed data above. Similarly $\eblin$ is determined uniquely from (\ref{propeta}). Using (\ref{oml3}) we conclude that $\elin$ is also uniquely determined along $C_{v_0}$. We finally use (\ref{dbtc}) written as
\[
\partial_u \left(r \otx \right) = \mathcal{Q}
\]
with $\mathcal{Q}$ uniquely determined on $C_{v_0}$ to determine uniquely $r \otx$ along $C_{v_0}$. 
Noting that $\Klin$ is uniquely determined by (\ref{lingauss}) along $C_{v_0}$ we conclude that we have determined all geometric quantities of a solution $\mathcal{S}$ uniquely along $C_{v_0}$ except $\xlin$, $\blin$, $\alin$ and $\bmlin$, $\olin$ along $C_{v_0}$. 

\subsubsection*{Estimates on the sphere $S^2_{u_0,v_0}$}
We note that since $\glinh$, $\Olin$ and $\bmlin$ are part of the seed data on $C_{u_0}$:
\begin{itemize}
\item the quantity $\xlin$ is determined uniquely on $S^2_{u_0,v_0}$ by (\ref{stos2}), 
\item the quantity $\alin$ is determined uniquely on $S^2_{u_0,v_0}$ by (\ref{tchi}),
\item the quantity $\olin$ is determined uniquely on $S^2_{u_0,v_0}$ by (\ref{oml3}).
\end{itemize}
Hence by (\ref{ellipchi}) and the fact that we already determined $\eblin$ and $\otx$ uniquely on $S^2_{u_0,v_0}$ above, the quantity $\blin$ is also uniquely determined on $S^2_{u_0,v_0}$. 

\subsubsection*{Transport equations along $C_{v_0}$: Part II}
We can now determine the missing quantities $\xlin$, $\blin$, $\alin$ and $\bmlin$, $\olin$ along $C_{v_0}$ recalling that they have been determined uniquely on $S^2_{u_0,v_0}$: Use (\ref{bequat}) to determine $\bmlin$, (\ref{chih3}) to determine $\xlin$, (\ref{Bianchi3}) to determine $\blin$, (\ref{Bianchi1}) to determine $\alin$ and finally $\olin$ from (\ref{oml2}) all uniquely along $C_{v_0}$.

We have determined all geometric quantities in terms of seed data and uniform bounds for all quantities which extend smoothly to the horizon $\mathcal{H}^+$ along $C_{v_0}$.

\subsubsection*{Transport equations along $C_{u_0}$}
We finally turn to the conjugate cone $C_{u_0}$. Recall that all geometric quantities have been determined uniquely on the sphere $S^2_{u_0,v_0}$ and that moreover the seed data $\glinh$ along $C_{u_0}$ determines uniquely 
$\xlin$ by (\ref{stos2}) and $\alin$ by (\ref{tchi}) along $C_{u_0}$.

We now determine all quantities along $C_{u_0}$. Starting with (\ref{uray}) we have
\[
\partial_v \left( \frac{r^2}{\Omega^2} \otx \right) = 4r \olin
\]
We see that the right hand side is integrable by the asymptotic flatness condition therefore producing the bound
\[
|r^2\otx| < C \, ,
\]
where $C$ can be computed explicitly from the seed data (and depends on $0<s\leq 1$). From (\ref{stos}) and the asymptotic flatness condition on $\bmlin$ we immediately conclude
\[
\Big|\frac{\glinto}{\sqrt{\slashed{g}}}\Big| < C \, .
\]
Of course, the same bounds hold for arbitrary many angular derivatives $r\slashed{\nabla}$ of these quantities.
\begin{remark}
One also sees that the quantities $\glinh$ and $\frac{\glinto}{\sqrt{\slashed{g}}}$ (as well as angular derivatives $r\slashed{\nabla}$ of these quantities) have smooth limits at null infinity.
By this we mean that there exists a symmetric $2$-tensor  $\glinh_{\infty}$ and a scalar $\frac{\glinto_{\infty}}{\sqrt{\slashed{g}}}$ on $\mathcal{M}$ satisfying in any spherical coordinate patch
\[
\partial_u \frac{\Big(\glinh_{\infty}\Big)_{AB}}{\sqrt{\slashed{g}}}=\partial_v \frac{\Big(\glinh_{\infty}\Big)_{AB}}{\sqrt{\slashed{g}}}= 0 \ \ \ , \ \ \ \partial_u \frac{\glinto_{\infty}}{\sqrt{\slashed{g}}}=\partial_v\frac{\glinto_{\infty}}{\sqrt{\slashed{g}}}=0
\]
and
\[
\lim_{v\rightarrow \infty} \frac{\glinh_{AB}}{\sqrt{\slashed{g}}} = \frac{\Big(\glinh_{\infty}\Big)_{AB}}{\sqrt{\slashed{g}}} \ \ \ , \ \ \ \lim_{v\rightarrow \infty}  \frac{\glinto}{\sqrt{\slashed{g}}}=\frac{\glinto_{\infty}}{\sqrt{\slashed{g}}}
\]
in the limit along the cone $C_{u_0}$.
\end{remark}

We now note that $\blin$ is uniquely determined from (\ref{Bianchi2}) producing the uniform bound
\[
|r^{3+s} \blin | < C \, .
\]
Similarly we can determine $\elin$ and $\eblin$. For this we note the equations (following from (\ref{propeta}) and (\ref{oml3})):
\begin{align} \label{erty}
\slashed{\nabla}_4 (r^2 \eblin) &= \slashed{\nabla}_4 \left(r^2 \elin\right) - \slashed{\nabla}_4 \left( r^2 \eblin + r^2 \elin\right) = 2\frac{\Omega}{r} \left(\elin +\eblin \right) -r^2\blin - 2\slashed{\nabla}_4 \slashed{\nabla}_A  r^2\left(\Olin \right)=-r^2\blin -2r^2 \slashed{\nabla} \olin \, . 
\end{align}
Note that the right hand side is uniquely determined along $C_{u_0}$ and integrable leading to $\eblin$ being uniquely determined along $C_{u_0}$ with the uniform bound
\[
|r^2 \eblin | < C \, .
\]
We also have, by the relation (\ref{oml3}) that $\elin$ is uniquely determined along $C_{u_0}$ with the uniform bound
\[
|r \elin| < C \, .
\]
We turn to (\ref{Bianchi4}) and (\ref{Bianchi6}), which clearly determine $\rlin$ and $\slin$ uniquely along $C_{u_0}$ with the uniform bounds
\[
|r^3 \rlin | + |r^3 \slin | < C \, .
\]
In fact, since we can repeat the above procedure commuting with angular derivatives, we also have in particular
\[
| r \slashed{\nabla} (r^3 \rlin )| + |r \slashed{\nabla}  (r^3 \slin)| < C \, .
\] 
It is now easy to see that (\ref{Bianchi8}) determines $\bblin$ uniquely with the uniform bound
\[
|r^2 \bblin | < C \, ,
\]
and that (\ref{chih3b}) determined $\xblin$ uniquely with the uniform bound
\[
|r \xblin | < C \, .
\]
Similarly, (\ref{Bianchi10}) determines $\ablin$ uniquely with the uniform bound
\[
|r \ablin | < C \, .
\]
For the remaining components note that (\ref{oml2}) determines $\olinb$ uniquely with the uniform bound
\[
| \olinb | < C \, .
\]
The quantity $\otx$ is determined uniquely from (\ref{dbtc}) with the bound
\[
|r \otx| < C \, ,
\]
where we have used that we can write (\ref{dbtc}) as
\[
\partial_v \left(r \otx\right) = RHS
\]
with the right hand side uniquely determined and integrable along $C_{u_0}$. Finally, equation (\ref{lingauss}) determines $\Klin$ uniquely along $C_{u_0}$ with the uniform bound
\[
|r^2 \Klin | < C \, .
\]
In fact by the remark above and the fact that the right hand side in $\slashed{\nabla}_4 (r^2 \Klin)=RHS$ is integrable, the weighted linearised Gaussian curvature $r^2 \Klin$ extends 
smoothly to null infinity.

We note once more that the same bounds hold for arbitrarily many angular derivatives $r\slashed{\nabla}$ of the quantities estimated either by trivial commutation with 
angular momentum operators $\mathnormal{\Omega}_i$ or,
if the reader prefers, tensorial commutation
with $r\slashed{\nabla}$ and inductively estimating lower order terms.

We conclude the proof by estimating the remaining weighted tangential derivatives $r\slashed{\nabla}_4$ and the transversal derivative $\slashed{\nabla}_3$ on the cone $C_{u_0}$. The procedure on $C_{v_0}$ is analogous (but easier since there are no weights at null infinity) and is hence omitted. 

For the remaining weighted tangential derivative $r \slashed{\nabla}_4$ we use 
\begin{itemize}
\item equation (\ref{Bianchi2}) pointwise for $r \slashed{\nabla}_4 \blin$
\item equation (\ref{Bianchi4}) pointwise for $r \slashed{\nabla}_4 \rlin$
\item equation (\ref{Bianchi4}) pointwise for $r \slashed{\nabla}_4 \slin$
\item equation (\ref{Bianchi8}) pointwise for $r \slashed{\nabla}_4 \bblin$
\item equation (\ref{Bianchi10}) pointwise for $r \slashed{\nabla}_4 \ablin$
\end{itemize}
which estimates the first derivative of all linearised curvature components. Similarly one can use the null structure equations to exchange a $4$-derivative by an angular derivative to estimate all linearised Ricci coefficients and the linearised metric components. This estimates all (first) derivatives tangential to the cone and it is easy to see how to continue inductively to estimate tangential derivatives of arbitrary high order. 

To estimate the transversal derivatives on $C_{u_0}$ one follows a similar procedure, now using the null structure and Bianchi equations in the $\slashed{\nabla}_3$-direction: More specifically, one
\begin{itemize}
\item uses the null structure equations, which express the transversal derivatives of all Ricci coefficients in terms of angular derivatives or curvature components that have already been obtained.
\item uses the Bianchi equations, which express the transversal derivatives of all curvature components in term of angular derivatives of curvature and Ricci coeffcients that have already been obtained. For instance, (\ref{Bianchi1}) for $\alin$, (\ref{Bianchi3}) for $\blin$, (\ref{Bianchi5}) for $\rlin$, (\ref{Bianchi7}) for $\slin$ and (\ref{Bianchi9}) for the transversal derivatives of $\bblin$ along $C_{u_0}$. Finally for $\slashed{\nabla}_3 \alin$ one needs to commute (\ref{Bianchi10}) and use the fact that the transversal derivative of $\underline{\beta}$ has just been obtained.
\end{itemize}
A simple induction allows to estimate all derivatives of all quantities on $C_{u_0} \cup C_{v_0}$. Finally, counting derivatives in the above procedure one observes that the bounds claimed in Theorem \ref{theo:AFdata} hold on the hypersurface $C_{v_0} \cup C_{u_0}$ for a constant $C_k$ which only depends on the constants $C_{\circ,n_1,n_2}$ with $n_1+n_2 \leq k +3$ in the definition of an asymptotically flat seed initial data set and the size of the data on the compact hypersurface 
${C_{v_0}}$. One also sees that applying a $\Omega \slashed{\nabla}_4$-derivative to any quantity in the round brackets of (\ref{decrete}) the right hand side is integrable, which ensures the existence of the limit at null infinity. This generally loses a derivative, e.g.~(\ref{erty}), explaining the $k\leq n-4$.

\subsection{Proof of the second part: Propagation of decay} \label{sec:propagateib} 
Knowing that the desired bounds hold on $C_{v_0} \cup C_{u_0}$ we continue with the proof of Theorem \ref{theo:AFdata}.

As noted in Sections \ref{sec:tratheo} and \ref{sec:fullrel} the derived quantities $\Plin$ and $\Pblin$ of the solution $\mathscr{S}$, which can be expressed through (\ref{Pdef}) and (\ref{Pdefund}), satisfy the Regge--Wheeler equation. It is easy to see that for asymptotically flat seed initial data with weight $s$ of order $n\geq 10$, the initial energies of Corollary \ref{cor:higherorder}, $\mathbb{F}^k_0\left[\Psilin=r^5 \Plin\right]$, $\mathbb{F}^k_0\left[\underline{\Psilin}=r^5 \Pblin\right]$ are indeed finite for every $k\leq n-6$ with the bound depending only constants $C_{\circ,n_1,n_2}$ in the definition of asymptotically flat seed data with $n_1+n_2 \leq k + 6$. Corollary \ref{cor:Pons2} and standard Sobolev embedding hence yield the bound
\[
| \mathfrak{D}^k (\Plin r^5) | + | \mathfrak{D}^k (\Pblin r^5) | < C_k \left(U\right)
\]
in any spacetime region $\mathcal{M} \cap \{u_0 \leq u\leq U\}$ with $C_k\left(U\right)$ as claimed. We now use the definition of $\Plin$, $\Pblin$ via the transport equations (\ref{psidef})--(\ref{evolpp2}) to obtain the bounds
\[
| \mathfrak{D}^k (\plin r^{4+s})) | + | \mathfrak{D}^k (\pblin r^3) | + | \mathfrak{D}^k (\alin r^{3+s}) | + | \mathfrak{D}^k (\ablin r) |  < C_k \left(U\right) 
\]
in any spacetime region $\mathcal{M} \cap \{u_0 \leq u\leq U\}$ with $C_k\left(U\right)$ as claimed. 
Indeed, these bounds hold initially on $C_{u_0} \cup C_{v_0}$ and are propagated by the transport equations. Note that weights near the horizon are irrelevant as $U<\infty$ and the constants are allowed to depend on $U$.
We can now use the equations (\ref{tchi}) to obtain
\[
| \mathfrak{D}^k (\xlin r^{2})) | + | \mathfrak{D}^k (\xblin r) |  < C_k \left(U\right) 
\]
and the expressions for $\plin$, $\pblin$ in (\ref{psideffull}) to deduce
\[
| \mathfrak{D}^k (\slashed{\mathcal{D}}_2^\star \blin r^{4+s})) | + | \mathfrak{D}^k (\slashed{\mathcal{D}}_2^\star \bblin r^3) |  < C_k \left(U\right) \, ,
\]
both in any spacetime region $\mathcal{M} \cap \{u_0 \leq u\leq U\}$ with $C_k\left(U\right)$ as claimed. 
To derive the above note that $r\left(u_0,v\right) \leq r\left(U,v\right) + C \left(U\right)$ for $U<\infty$. Using (\ref{chih3}) pointwise and then (\ref{propeta}) in the $3$-direction, as well as (\ref{Bianchi7}) in the $3$-direction and the definition of $\Plin$ we obtain the estimates
\[
| \mathfrak{D}^k (\slashed{\mathcal{D}}_2^\star \elin \, r^{2})) | + | \mathfrak{D}^k (\slashed{\mathcal{D}}_2^\star \eblin \, r^{3})) |  + | \mathfrak{D}^k (r^5 (\slashed{\mathcal{D}}_2^\star \slashed{\nabla}_A \left(\rlin, \slin\right) \, ) | < C_k \left(U\right) \, .
\]
The Codazzi equations (\ref{ellipchi}) and equation (\ref{oml3}) now provide the bounds
\[
| \mathfrak{D}^k (r^3 \slashed{\mathcal{D}}_2^\star \slashed{\nabla}_A \otxb ) | + | \mathfrak{D}^k (r^4 \slashed{\mathcal{D}}_2^\star \slashed{\nabla}_A \otx) | + | \mathfrak{D}^k (r^2 \slashed{\mathcal{D}}_2^\star \slashed{\nabla}_A \Olin \, ) |  < C_k \left(U\right) \, 
\]
in any spacetime region $\mathcal{M} \cap \{u_0 \leq u\leq U\}$ with $C_k\left(U\right)$ as claimed. Note this implies already the desired bound for $\olinb= \partial_u (\Olin)$. For $\olin$ we use (\ref{oml2}) to obtain the improved bound
\[
 | \mathfrak{D}^k (r^{4+s} \slashed{\mathcal{D}}_2^\star \slashed{\nabla}_A \olin \, ) |  < C_k \left(U\right) \,.
\]
Finally, we use (\ref{stos}), (\ref{stos2}) in the $3$-direction and (\ref{bequat}) to deduce also
\[
 |\mathfrak{D}^k \glinh| + |\mathfrak{D}^k \left( r^2 \slashed{\mathcal{D}}_2^\star \slashed{\nabla}_A ( tr_{\slashed{g}} \glin \, ) \right) |  + |\mathfrak{D}^k (r^2 \slashed{\mathcal{D}}_2^\star  \bmlin \,)| \leq C_k \left(U\right)
 \]
in any spacetime region $\mathcal{M} \cap \{u_0 \leq u\leq U\}$ with $C_k\left(U\right)$ as claimed.

In view of Corollaries \ref{cor:elliptic12} and \ref{cor:etavl1}, we have proven Theorem \ref{theo:AFdata} except for the $\ell=0$ and $\ell=1$ modes of the solution $\mathscr{S}$. The latter have been understood in detail in Section \ref{newgaugesec}, Theorem \ref{etsilew}. Specifically, we can now add to $\mathscr{S}$ a pure gauge solution $\mathscr{G}$ (generated by a gauge function supported on $\ell=0,1$ only) and a member of the Kerr family $\mathscr{K}$ with the following properties
\begin{itemize}
\item both $\mathscr{G}$ and $\mathscr{K}$ satisfy the desired bounds of Theorem \ref{theo:AFdata}
\item both $\mathscr{G}$ and $\mathscr{K}$ do not alter any of the bounds proven in this section
\item $\mathscr{S} + \mathscr{G} + \mathscr{K}$ is supported on $\ell \geq 2$ only
\end{itemize}
This finishes the proof up to the claim concerning the limits on null infinity. 
These follow by the argument given in the first part of the proof, 
which can be repeated on any cone $C_{u}$ with $u \leq U$.

\subsection{Propagation of roundness at infinity}
In this section we state a corollary to Theorem~\ref{theo:AFdata}, which can be understood as the propagation of the round sphere condition (\ref{rsc}) at null infinity. It states that if the linearised Gaussian curvature $\Klin$ behaves like $r^{-3}$ on the outgoing cone $C_{u_0}$, then $\Klin r^{3}$ remains bounded on any cone which is a finite $u$ distance away. 

Note that Theorem \ref{prop:gaugeachieve} stated in particular that, given any solution $\mathscr{S}$ as in Theorem \ref{theo:AFdata}, we can construct a pure gauge solution $\mathscr{G}$ such that the sum $\mathscr{S}+\mathscr{G}$ satisfies on $C_{u_0}$ the stronger bounds in the Corollary below.\footnote{At the non-linear level this can be interpreted as refoliating the cone such that the sphere at infinity becomes round.} The corollary then shows that these stronger bounds are propagated. Of course this is directly related to the propagation of uniform boundedness for the quantity $\Ylin$ in our Theorems \ref{theo:mtheo} and \ref{theo:mtheod}.

\begin{corollary}
With the assumptions of Theorem \ref{theo:AFdata}, assume in addition that
\begin{align} \label{rcend}
| \left( r \slashed{\nabla}\right)^m \left( r^3 \Klin \right) |\leq C_m
\end{align}
holds for all $m \leq n-9$ on $C_{u_0}$. Then, given any $u_0 < U<\infty$, the estimate (\ref{rcend}) actually 
holds for any $m \leq n-10$ in any spacetime region $\mathcal{M} \cap \{u_0 \leq u\leq U\}$ where $C_m$ now also depends on the choice of $U$ and the constants $C_{m}$ appearing in Theorem \ref{theo:AFdata}.
\end{corollary}

\begin{proof}
Compute from (\ref{lingauss})
\[
\slashed{\nabla}_3 \left( \left(r \slashed{\nabla}\right)^m \Klin r^3\right) = \mathcal{Q}_m
\]
and deduce that $\mathcal{Q}_m$ is pointwise uniformly bounded using the bounds (\ref{decrete}) of Theorem \ref{theo:AFdata}. 
\end{proof}

We finally remark that a similar corollary is easily deduced for the quantities $\left( r \slashed{\nabla}\right)^m \frac{\glinto}{\sqrt{g}}$ and $\left( r \slashed{\nabla}\right)^m \glinh$.
\section{Characterizing the vanishing of gauge invariant quantities} \label{sec:robtraut} 

In this appendix, we discuss solutions characterized by the vanishing of gauge
invariant quantities. 

We first show in Appendix~\ref{charofpuregauge} that
the vanishing of $\alin$ and $\ablin$ identically implies that a solution $\mathscr{S}$ is
the sum $\mathscr{S}=\mathscr{G}+\mathscr{K}$ of a pure gauge solution and
a linearised Kerr solution, provided that that $\mathscr{S}$ is asymptotically flat.
We then consider in Appendix~\ref{theRTrautsection} 
the larger class of solutions such that $\Plin$ and $\Pblin$ vanish
identically. We shall show that this class corresponds precisely
to the linearised Robinson--Trautman
solutions, again up to the addition of a pure gauge solution.

\subsection{$\protect\alin=\protect\ablin=0\implies \mathscr{S}=\mathscr{G}+\mathscr{K}$}
\label{charofpuregauge}
In this section we prove that any solution $\mathscr{S}$
which is asymptotically flat and satisfies $\alin=\ablin\equiv 0$ globally 
is necessarily
a pure gauge solution $\mathscr{G}$ plus a
reference linearised Kerr solution $\mathscr{K}$. 

\begin{theorem} \label{theo:aabgauge}
Let $\mathscr{S}$ be a smooth solution of the full system of linearised gravity 
arising from a smooth seed initial data set on $C_{u_0} \cup C_{v_0}$ through Theorem \ref{theo:lwp}. Assume that the data are asymptotically flat with weight $s$ as in Definition \ref{def:afpeel}. Assume further that
\begin{align} \label{globvanishaa}
\alin = \ablin = 0
\end{align}
holds globally  on $\mathcal{M} \cap \{u\geq 0\} \cap \{v\geq 0\}$. Then $\mathscr{S}$ is 
the sum of a pure gauge solution $\mathscr{G}$ 
and a reference linearised Kerr solution $\mathscr{K}$.
\end{theorem}

We remark that the assumption (\ref{aoci}) in Definition \ref{def:afpeel} can actually be deduced from (\ref{globvanishaa}), so the assumptions (\ref{aoci1}) and (\ref{aoci2}) on the data suffice in conjunction with (\ref{globvanishaa}).

\begin{proof}
We let $\Si$ be as in Theorem~\ref{prop:gaugeachieve}, i.e.~we put the solution $\mathscr{S}$ in the initial data gauge. Moreover,
let us subtract $\mathscr{K}_{\mathfrak{m},s_i}$ of Theorem~\ref{etsilew}
so that $\Si^\prime\doteq \Si-\mathscr{K}_{\mathfrak{m},s_i}$
is supported outside of $\ell=0,1$. Below, we shall consider quantities
associated to $\Si^\prime$.

 From the fact that $\Ylin$ is bounded for asymptotically flat initial data and $\pblin=0$ globally we deduce, noting that $\ablin=0$ implies $\pblin=0$, the bound
\[
|r^2 \xblin| \leq C  \ \ \ \textrm{along $C_{u_0}$}
\]
for a constant determined purely by the seed initial data. Let $f_{out}\left(v,\theta,\phi\right)$ be a solution of the elliptic equation along $C_{u_0}$:
\[
 r^2 \slashed{\mathcal{D}}_2^\star \slashed{\nabla}_A f_{out}\left(v,\theta,\phi\right)  = \frac{r^2}{2\Omega^2} \left(u_0,v\right) \cdot \xblin \left(u_0,v,\theta,\phi\right) \, .
\]
The solution is uniquely determined once we insist that $f_{out}$ has vanishing $\ell=0$ and $\ell=1$ modes. It is also clear that $f_{out}$ is uniformly bounded. 

We add the pure gauge solution generated by Lemma \ref{lem:exactsol} through $f_{out}$ to 
$\Si^\prime$ and call the resulting solution $\mathscr{S}_1$. The solution $\mathscr{S}_1$ satisfies $\xblin=0$ along $C_{u_0}$ and hence globally on $\mathcal{M} \cap \{u\geq 0\} \cap \{v\geq 0\}$ by the transport equation (\ref{tchi}). Importantly, since $f_{out}$ is uniformly bounded, the solution $\mathscr{S}_1$ also still satisfies the round sphere condition (\ref{rsc}) and (\ref{rscm}).

Let now $\Omega^2 f_{in}\left(u,\theta,\phi\right)$ be determined by the following elliptic equation along $C_{v_0}$
\[
 r^2 \slashed{\mathcal{D}}_2^\star \slashed{\nabla}_A f_{in}\left(u,\theta,\phi\right) \Omega^2 \left(u,v_0\right) = \frac{\Omega r^2}{2} \left(u,v_0\right) \xlin \left[\mathscr{S}_1\right] \left(u,v_0,\theta,\phi\right).
\]

We add the pure gauge solution generated by Lemma \ref{lem:exactsol3} through $f_{in}$ to $\mathscr{S}_1$  and call the resulting solution $\mathscr{S}_2$. The solution $\mathscr{S}_2$ satisfies  $\xlin=0$ on $C_{v_0}$ which implies $\xlin=0$ globally through the transport equation (\ref{tchi}). Importantly, since $\Omega^2 f_{in}$ is uniformly bounded, the solution $\mathscr{S}_2$ also still satisfies the round sphere condition (\ref{rsc}) and (\ref{rscm}). Note also that the pure gauge solution added through  Lemma \ref{lem:exactsol3} has $\xblin=0$ so $\mathscr{S}_2$ satisfies 
\begin{align} \label{almosthere}
\alin = \ablin = \xlin = \xblin = 0 \ \ \ \textrm{on $\mathcal{M} \cap \{u\geq 0\} \cap \{v\geq 0\}$.}
\end{align}
Note that both pure gauge transformation added to $\Si$ do not change the conclusions of Theorem \ref{etsilew} as $f_{out}$ and $f_{in}$ both have vanishing projection to $\ell=0,1$. From (\ref{chih3}) and (\ref{chih3b}) one now concludes that $\slashed{\mathcal{D}}_2^\star \elin=\slashed{\mathcal{D}}_2^\star \eblin=0$ and hence since $\mathscr{S}_2$ satisfies $\left( \slashed{div} \elin \, \right)_{\ell=1}= \left( \slashed{curl} \elin \, \right)_{\ell=1}=0$ (and similarly for $\eblin$) that in fact $\elin=\eblin=0$ on $\mathcal{M} \cap \{u\geq 0\} \cap \{v\geq 0\}$. Since $\alin=\ablin=0$ implies $\plin=\pblin=0$, we have from the formulae (\ref{psideffull}) that $\blin=\bblin=0$ after using the fact that $\left( \slashed{div} \blin \, \right)_{\ell=1}= \left( \slashed{curl} \blin \, \right)_{\ell=1}=0$ (and similarly for $\bblin$). The Codazzi equations (\ref{ellipchi}) and the fact that the $\ell=0,1$ modes of $\otx$ and $\otxb$ vanish then yield $\otx=\otxb=0$ identically. Equation (\ref{oml3}) and the vanishing of the $\ell=0,1$ modes 
allows the conclusion for $\Olin$ and hence $\olin$ and $\olinb$.
Finally, global vanishing of $\rho$ and $\sigma$ is now a consequence of (\ref{dtcb}) and (\ref{curleta}).

So far, we have shown that all linearised curvature and all linearised Ricci coefficients of the solution vanish for $\mathscr{S}_2$. To conclude also the vanishing of the linearised metric components we need to add another pure gauge solution. 

The geometric quantity $\bmlin$ of the solution $\mathscr{S}_2$, while having globally vanishing projection to $\ell=1$,\footnote{Recall that by this we mean that $\left( \slashed{div} \bmlin \, \right)_{\ell=1}= \left( \slashed{curl} \bmlin \, \right)_{\ell=1}=0$ hold.} has a potentially non-vanishing trace on $C_{u_0}$, which we denote by $\tilde{b}$. It satisfies the bound $|\slashed{\nabla} \tilde{b}| \lesssim v^{-2}$ along $C_{u_0}$ following from the fact that $\mathscr{S}$ satisfies it and that all pure gauge solutions added so far do. 

Propositions \ref{prop:zeroshift} and \ref{prop:metricg} construct a pure gauge solution $\mathscr{G}$, generated by bounded $q_1\left(v,\theta,\phi\right)$ and $q_2\left(v,\theta, \phi\right)$ having vanishing projection to $\ell=0,1$ with the property that $\mathscr{G}$ satisfies the round sphere condition (\ref{rscm}) and moreover $\bmlin \left[\mathscr{G}\right]=-\tilde{b}$ along $C_{u_0}$. Finally, all linearised Ricci coefficients and curvature components vanish for $\mathscr{G}$. 

It is easy to see that the solution $\mathscr{S}_3=\mathscr{S}_2 + \mathscr{G}$ is the trivial solution: One uses the propagation equation (\ref{stos2}) from infinity to first conclude that $\glinh=0$ identically along $C_{u_0}$ and then that $\glinh=0$ identically on $\mathcal{M} \cap \{u\geq 0\} \cap \{v\geq 0\}$ by using the propagation in the $3$-direction of (\ref{stos2}).

We have shown thus that $\Si^\prime$ is the sum of  pure
gauge solutions  and thus the original $\mathscr{S}=\Si^\prime -\Gi +\mathscr{K}$ 
is the sum of a pure gauge solution 
and a reference linearised Kerr.
\end{proof}

\subsection{$\protect\Plin=\protect\Pblin=0\implies$
linearised Robinson--Trautman}
\label{theRTrautsection}

While global vanishing of $\alin$ and $\ablin$ together with asymptotic flatness implies that the solution is the sum of
a pure gauge solution and
 linearised Kerr solution by Theorem \ref{theo:aabgauge}, one may ask whether vanishing of the derived quantities $P$ and $\Pblin$  is sufficient for this conclusion. 
As we shall see in this section, this is \emph{not} the case. The  non-trivial solutions arising
can however be completely described: They are given by the linearisation of a family of algebraically special solutions to the Einstein vacuum equations, the celebrated \emph{Robinson--Trautman metrics}~\cite{RobTraut}. 
These vacuum metrics can be characterized geometrically by the fact that they 
admit a shear-free congruence of null-geodesics which is also hypersurface orthogonal.
See \cite{RobTraut, singleton} and also Section 10 of \cite{DHRscat} for an introduction to this family.

We shall only sketch here the relevant computations.

\subsubsection{The Calabi equation}
Suppose first $\Plin=0$ \emph{or} $\Pblin=0$. Then we have one of
\[
r^2 \slashed{\mathcal{D}}_2^\star \slashed {\mathcal{D}}_1^\star \left( r^3 \rlin,\mp r^3 \slin \, \right)-3M r \xblin + \mathcal{W} =0 \, ,
\]
with the upper sign in case $\Plin=0$ and the lower sign if $\Pblin=0$, where $\mathcal{W}$ indicates a term that vanishes in the limit on null infinity by asymptotic flatness. 
Taking a $3$-derivative yields
\[
r^2\slashed{\mathcal{D}}_2^\star \slashed {\mathcal{D}}_1^\star \left(-r^3 \slashed{div} \bblin, \pm r^3 \slashed{curl} \bblin \right)+3M r \ablin+ \mathcal{W}  = 0 \, .
\]
Application of another $3$-derivative yields the equation
\[
r^4 \slashed{\mathcal{D}}_2^\star \slashed {\mathcal{D}}_1^\star  \left(\slashed{div} \slashed{div} \left( r \ablin \right),\mp \slashed{curl} \left( \slashed{div} r \ablin \right) \right)+3M \slashed{\nabla}_3 \left( r\ablin \right) + \mathcal{W}  = 0 \, .
\]
Therefore, if $\Pblin=0$ then the fourth order parabolic equation (cf.~the Calabi equation in \cite{RobTraut})
\begin{align} \label{calabie} 
 \slashed{\nabla}_3 \left( r \ablin \right)^{\mathcal{I}} = -\frac{1}{3M} r^4  \slashed{\mathcal{D}}_2^\star \slashed {\mathcal{D}}_1^\star  \slashed{\mathcal{D}}_1   \slashed{\mathcal{D}}_2  \left( r\ablin\right)^{\mathcal{I}}  \, 
\end{align}
has to hold along null infinity. Here $ \left( r \ablin \right)^{\mathcal{I}}$ is the symmetric traceless tensor obtained as the limit at null infinity $v\rightarrow \infty$ of the quantity $r \ablin \left(u,v,\theta\right)$ measured in an orthonormal frame on the spheres $S^2_{u,v}$. We can interpret $\left( r \ablin \right)^{\mathcal{I}}$ either as a symmetric traceless spacetime $S^2_{u,v}$-tensor whose components in an orthonormal frame do not depend on $v$ or as a symmetric traceless $S^2_u$-tensor defined on the cylinder $\left[u_0,\infty\right) \times S^2$ equipped with the metric $-du^2+d\theta^2 + \sin^2\theta d\phi^2$. Taking the latter point of view and considering  $r^4 \slashed{\mathcal{D}}_2^\star \slashed {\mathcal{D}}_1^\star  \slashed{\mathcal{D}}_1   \slashed{\mathcal{D}}_2$ in (\ref{calabie}) as an operator on the unit sphere, equation (\ref{calabie}) becomes a parabolic equation on the cylinder $\left[u_0,\infty\right) \times S^2$ whose solution is uniquely determined if data are prescribed on the ``initial" sphere $S^2_{u_0}$. 

 A priori it seems we have the full freedom of specifying a symmetric traceless tensor. However, if in addition $P=0$, then $ \slashed{\mathcal{D}}_2^\star \slashed {\mathcal{D}}_1^\star\left(0, \slin\right)=\frac{1}{2} \left(\Plin - \Pblin\right) = 0$ everywhere and hence $\slin\equiv0$ globally provided the $\ell=0,1$ modes of $\slin$ also vanish. This implies that $ r^4\slashed{\mathcal{D}}_2^\star \slashed{\mathcal{D}}_1^\star  \slashed{curl} \slashed{div} \left( r \ablin \right)^{\mathcal{I}}=0$ on null infinity. It follows that we can prescribe $ \left( r \ablin \right)^{\mathcal{I}}$ initially on one sphere subject to the condition $ r^4\slashed{\mathcal{D}}_2^\star \slashed{\mathcal{D}}_1^\star  \slashed{curl} \slashed{div} \left( r \ablin \right)^{\mathcal{I}}=0$ (which then propagates), which reduces the number of degrees of freedom to one function on the initial sphere, just as it is the case for the Robinson-Trautman class \cite{RobTraut}. 
 
 It is useful to scalarise equation (\ref{calabie}) by setting $ \left( r \ablin \right)^{\mathcal{I}} = r^2 \slashed{\mathcal{D}}_2^\star \slashed {\mathcal{D}}_1^\star \left(f, 0\right)$ determining uniquely  (up to projection to $\ell=0,1$) a function $f$ on $\left[u_0,\infty\right) \times S^2$. If  $\left( r \ablin \right)^{\mathcal{I}}$ satisfies (\ref{calabie}) then $f$ satisfies
\begin{align} \label{fecalabi}
\partial_u f = -\frac{1}{6M} \left(\Delta_{\mathbb{S}^2} \Delta_{\mathbb{S}^2} + 2 \Delta_{\mathbb{S}^2} f\right) \, .
\end{align}
This should be compared with equation (2.4), (2.5) in \cite{singleton} which upon linearisation $f=1+\epsilon \flinc + \mathcal{O}\left(\epsilon^2\right)$ and setting $M=2m$ yields (\ref{fecalabi}).

We can solve (\ref{fecalabi}) mode by mode obtaining the solution
\[
f_{\ell,m} \left(u\right) = \exp\left(-\frac{u}{2M} \left(\ell-1\right) \frac{\ell \left(\ell+1\right) \left(\ell+2\right)}{3} \right) Y^\ell_m \, .
\]
Note that for fixed $v$ this behaves like an integer power of $\Omega^2 \left(u,v\right)\sim \exp\left(-u/2M\right)$ near the event horizon.

\subsubsection{Constructing the full solution in the horizon-normalised gauge}
We now outline the argument that assuming $\Plin=\Pblin=0$ and specifying such an $ \left( r \ablin \right)^{\mathcal{I}}$ initially on the sphere $S^2_{u_0,\infty}$ determines a solution of the system of gravitational perturbations  which is unique up to pure gauge solutions. It is important to note that (\ref{calabie}) will produce solutions $ \left( r \ablin \right)^{\mathcal{I}}$ which decay exponentially in $u$.

Suppose we have an asymptotically flat seed initial data set for which $\Plin$ and $\Pblin$ are zero on $C_{u_0} \cap C_{v_0}$.  Since $\Plin$ and $\Pblin$ satisfy the Regge-Wheeler equation, $\Plin \equiv 0$ and $\Pblin \equiv 0$ in $\mathcal{M} \cap \{ u \geq u_0 \} \cap \{ v \geq v_0 \}$. We construct the full solution $\Sf'=\Sf-\mathscr{K}_{\mathfrak{m},s_i}$ 
directly in the horizon-normalised 
gauge of Theorem \ref{theo:mtheod}.
\begin{enumerate}
\item From the discussion above we have
\[
\slin = 0 \, \ \ \ \textrm{in $\mathcal{M} \cap \{ u \geq u_0 \} \cap \{ v \geq v_0 \}$} .
\]
 \item On the horizon $\mathcal{H}^+$ we have 
\begin{align}
0=\int_{v_0}^\infty dv \int_{S^2} \sin \theta d\theta d\phi |\Plin|^2 = \int_{v_0}^v dv \int_{S^2} \sin \theta d\theta d\phi |\slashed{\mathcal{D}}_2^\star \slashed{\mathcal{D}}_1^\star \left(\rlin ,0 \, \right)+\frac{3}{8M^3} \Omega\xlin|^2 \nonumber \\
= \int_{v_0}^\infty dv \int_{S^2} \sin \theta d\theta d\phi \left[  |\slashed{\mathcal{D}}_2^\star \slashed{\mathcal{D}}_1^\star \left(\rlin ,0\right)|^2 +\frac{9|\Omega\xlin |^2}{64M^6}  + \frac{3}{4M^3} \mathcal{D}_1^\star \left(\rlin ,0\right) \left(-\Omega \blin \right)\right] \nonumber \\
= \int_{v_0}^\infty dv \int_{S^2}\sin \theta d\theta d\phi \left[  |\slashed{\mathcal{D}}_2^\star \slashed{\mathcal{D}}_1^\star \left(\rlin ,0\right)|^2 +\frac{9}{64M^6} |\Omega \xlin |^2 + \frac{3}{4M^3} \left( -\frac{1}{2} \right)\partial_v | \rlin |^2 \right] 
\nonumber
\end{align}
and since $\rlin \rightarrow 0$ as $v \rightarrow \infty$ by Theorem \ref{theo:mtheod} (recall Corollary \ref{cor:pold}, the restriction of $\Plin$ on the horizon and the fact that $\rlin_{\ell=0,1}=0$) we conclude $\Omega \xlin = 0$, $\Omega \blin = 0$ (Codazzi) and $\rlin=0$ along $\mathcal{H}^+$. Therefore also $\Omega^2 \alin=0$, $\Omega \plin = 0$ on $\mathcal{H}^+$. We also have $\elin=0$ on $\mathcal{H}^+$ since $\slashed{curl} \elin=0$ holds from $\slin=0$ and $\slashed{div} \elin=0$ by the horizon gauge condition (\ref{fchoi}). Note that these bounds hold both for the solution $\Sf$ of Theorem \ref{theo:mtheod} and the solution $\Si$ of Theorem \ref{theo:mtheo}.
\item The equations $\Omega^{-1}\slashed{\nabla}_3 \left(\plin r^2 \Omega\right)=0$ and $\Omega^{-1}\slashed{\nabla}_3 \left(\alin r \Omega^2\right)=0$ following from (\ref{psidef}), (\ref{evolpp}) now imply 
\begin{align} \label{trgh}
\plin \Omega = \alin \Omega^2 \equiv 0 \, \ \ \ \textrm{in $\mathcal{M} \cap \{ u \geq u_0 \} \cap \{ v \geq v_0 \}$} .
\end{align}
\item On null infinity we know that $r \ablin= \left(r \ablin \right)^\mathcal{I} \left(u,\theta\right)$ is entirely determined by the parabolic equation and exponentially decaying. The solution decays at least like $e^{-\frac{4}{M}u}$ (as follows from the fact that $\alpha$ has at least $\ell \geq 2$, see also \cite{Chruscielexistence} and compare with $\Omega^2 \sim e^{-\frac{u}{2M}}$). Therefore (recalling $\slin=0$)
\begin{align}
\lim_{v\rightarrow \infty} \left(r^2 \bblin\right) =\left(r^2 \bblin\right)^\mathcal{I}  \left(u,\theta\right) &= \int_{u}^\infty r \slashed{div}   \left( r \ablin\right)^\mathcal{I}  \left(\bar{u},\theta\right) d\bar{u} , \nonumber \\
\lim_{v\rightarrow \infty} \left( r^3\rlin \, \right) = \left( r^3\rlin \, \right)^\mathcal{I}  \left(u,\theta\right) &= \int_{u}^\infty r \slashed{div} \left(r^2 \bblin\right)^\mathcal{I} \left(\bar{u},\theta\right) d\bar{u} ,
 \nonumber \\
\lim_{v\rightarrow \infty} \left( r^4 \blin \right) = \left( r^4 \blin \right)^\mathcal{I}  \left(u,\theta\right) &= -\int_{u}^\infty r \ \slashed{\mathcal{D}}_1^\star \left(- \left( r^3\rlin \, \right)^\mathcal{I} , 0 \right) \left(\bar{u},\theta\right) d\bar{u} 
\end{align}
are all determined on null infinity. The existence of these limits and their vanishing as $u \rightarrow \infty$ is a consequence of Theorem \ref{theo:mtheod} (Propositions \ref{prop:psibs2d} and \ref{prop:pold3}) for $\left(r^2 \bblin\right)^\mathcal{I}$ and $\left(r^3\rlin \, \right)^\mathcal{I}$. For $ \left( r^4 \blin \right)^\mathcal{I}$ it follows from $\alin$ globally vanishing, (\ref{Bianchi2}) and the fact that $\blin\Omega$ vanishes on the horizon. We also see from (\ref{tchi})
\[
\lim_{v\rightarrow \infty} \left( r \xblin \right)  = \left(r \xblin\right)^\mathcal{I} =  \int_{u}^\infty  \left( r \ablin\right)^\mathcal{I}   \left(\bar{u},\theta\right) d\bar{u} \, .
\]
\item Integrating backwards from null infinity (\ref{Bianchi4}) now yields from (\ref{trgh})
\begin{align}
\blin \left(v, u,\theta\right) &= \frac{\Omega}{r^4} \left[ \left(r^4 \blin\right)^\mathcal{I}  \left(u,\theta\right) \right] \, ,\nonumber \\
\xlin \left(v, u,\theta\right) &= \frac{4M}{3}\Omega r^3 \slashed{\mathcal{D}}_2^\star \blin =  \frac{4M}{3}r^{-2} \Omega \cdot r \slashed{\mathcal{D}}_2^\star \left( \left(r^4 \blin\right)^\mathcal{I}   \left(u,\theta\right)\right) \, , \nonumber
\end{align}
valid in an orthonormal frame on the spheres $S^2_{u,v}$.\footnote{Recall that the statement that the spacetime tensor $\left(r^4 \blin\right)^\mathcal{I}$ does not depend on $v$ is true only in an orthonormal frame. Otherwise factors of $r$ appear from raising and lowering indices.}

\item Recall now the equations
\[
\slashed{\nabla}_4 \left(\pblin r^3 \Omega\right) = -r^3 \Omega \underline{\Plin} \ \ \ \ \ \textrm{and} \ \ \ \ \ \slashed{\nabla}_4 \left(r \Omega^2 \ablin \right) = 2  r \Omega^2 \pblin \, , 
\]
from which we conclude
\[
r^3 \Omega \pblin \left(v,u,\theta\right) = \int_{u}^\infty r^2 \slashed{\mathcal{D}}_2^\star \slashed{div} \left( r \ablin\right)^\mathcal{I}  \left(\bar{u},\theta\right) d\bar{u}
\]
and
\begin{align} \label{arep}
r\Omega^2 \ablin\left(v,u,\theta\right) = \left( r \ablin \right)^\mathcal{I}  \left({u},\theta\right) - \frac{2}{r\left(u,v\right)}  \int_{u}^\infty r^2 \slashed{\mathcal{D}}_2^\star \slashed{div}  \left( r \ablin \right)^\mathcal{I}  \left(\bar{u},\theta\right) d\bar{u} \, ,
\end{align}
again valid in an orthonormal frame. Note that $ \left( r \ablin \right)^\mathcal{I}  \left({u},\theta\right)$ needs to decay at least as fast as $\Omega^{4}$ towards the event horizon for $\ablin$ to extend regularly to $\mathcal{H}^+$.

\item Since in the horizon-normalised gauge we also have $\eblin=0$ on $\mathcal{H}^+$ we conclude using (\ref{chih3b}) and the fact that $\Omega^{-1} \xblin \rightarrow 0$ along 
$\mathcal{H}^+$ that $\Omega^{-1}\xblin=0$ on $\mathcal{H}^+$. We now determine $\xblin$ globally from (\ref{tchi}) using (\ref{arep}) as
\[
\xblin r^2 \Omega^{-1} \left(v,u, \theta\right) = \int_u^\infty  r^2 \ablin \left(v,\bar{u},\theta\right) d\bar{u} \, .
\]
With $\xlin$ and $\xblin$ determined globally,  (\ref{chih3}) and (\ref{chih3b}) allow us to obtain $\elin$ and $\eblin$ since the $\ell=0,1$ modes of all quantities vanish for $\Sf$ (cf.~Corollary \ref{cor:etavl1}). Codazzi (\ref{ellipchi}) implies expressions for $\otx$ and $\Omega^{-2} \otxb$. Finally (\ref{oml3}) implies an expression for $\Olin$. 

\item To determine the remaining metric quantities $\left(\bmlin, \glinto, \glinh \right)$ it is most convenient to add to $\Sf^\prime$ 
another pure gauge solution which achieves $\bmlin=0$ on $C_{u_0}$ while preserving the condition (\ref{rscm}) and not changing any of the quantities discussed in $1.-7.$ above. The existence of such a solution follows from Proposition \ref{prop:zeroshift}. One can then use (\ref{stos}), (\ref{stos2}) and (\ref{bequat}) to determine explicit formulae for  $\left(\bmlin, \glinto, \glinh \right)$.
\end{enumerate}

\subsubsection{Regularity}
To determine the smoothness properties of the solution constructed it suffices to check its regularity at the level of the seed initial data, i.e.~whether the following quantities extend to the horizon on $C_{v_0}$:
\begin{align}
\left( e^{-\frac{u}{2M}} \partial_u\right)^n \left(\xblin r^2 \Omega^{-1}\right) \ \ \ , \ \ \ \left( e^{-\frac{u}{2M}} \partial_u\right)^n \left(\otxb \Omega^{-2}\right) \ \ \ , \ \ \ \left( e^{-\frac{u}{2M}} \partial_u\right)^n \left(\Olin\right) \, .
\end{align}
Since all quantities of the solution determined above behave like $\left[\Omega^{2}\left(u,v_0\right)\right]^k$ for some integer $k$, the solution is smooth in the extended sense.

Interestingly, the non-smoothness observed in~\cite{singleton} 
for the class of Robinson--Trautman metric seems to be a feature of the non-linear terms in the parabolic equation and is not seen at the linearised level.

\bibliographystyle{hsiam}
\bibliography{schwarrefs}

\end{document}